\documentclass[aps,prd,onecolumn,groupedaddress,showpacs,nofootinbib,amssymb]{revtex4}
\usepackage[dvips]{graphicx}
\usepackage{amssymb}
\usepackage{amsmath}
\usepackage{graphicx,,color}
\usepackage{amsfonts}
\usepackage{bm}
\usepackage{cancel}
\usepackage{comment}
\usepackage{floatflt}
\usepackage{slashed}
\usepackage{appendix}
\usepackage{array}
\usepackage{tabularx}
\usepackage[a4paper,margin=2.5cm]{geometry}
\usepackage{enumitem}
\usepackage{xcolor}
\newcommand\be{\begin{equation}}
\newcommand\ee{\end{equation}}

\allowdisplaybreaks[4]

\begin{document}

\title{Thermodynamic Origin of the Tully-Fisher Relation in Dark Matter Dominated Galaxies: A Theoretical-Empirical Derivation}
\author{V.K. Oikonomou$^{1,2,3}$}\email{voikonomou@gapps.auth.gr,v.k.oikonomou1979@gmail.com}
\affiliation{$^{1)}$Department of Physics, Aristotle University of
Thessaloniki, Thessaloniki 54124, Greece} \affiliation{$^{2)}$L.N.
Gumilyov Eurasian National University - Astana, 010008,
Kazakhstan} \affiliation{$^{3)}$Laboratory for Theoretical
Cosmology International Center of Gravity and Cosmos\\ Tomsk State
University of Control Systems and Radioelectronics
(TUSUR) 634050\\
 Tomsk, Russian Federation}

\tolerance=5000

\begin{abstract}
In this work we introduce the concept of self-interacting dark
matter with scale-dependent equation of state, in the context of
which dark matter is collisional and its equation of state is
radius-dependent and has the form
$P(r)=K(r)\left(\frac{\rho(r)}{\rho_{\star}}\right)^{\gamma(r)}$.
We confronted the effectively 2-parameter model with 174 galaxies
from the SPARC data, and we found that the rotation curves of 100
galaxies can be perfectly fitted by the model. These galaxies are
dark matter dominated, mostly dwarfs, low-luminosity and
low-surface-brightness spiral galaxies. We demonstrate that
scale-dependent self-interacting dark matter solves the cusp-core
issue for dark matter dominated galaxies. More importantly, the
structure of the scale-dependent SIDM model produces in a
semi-theoretically and semi-empirically way the canonical
Tully-Fisher and the baryonic Tully-Fisher relations when these
100 viable dwarfs, low-surface-brightness and low-luminosity
galaxies are taken into account. The behavior of the entropy
function $K(r)$ is assumed to be
$K(r)=K_0\times\left(1+\frac{r}{r_c}\right)^{-p}$. The perfect
fits of the rotation curves come for a nearly isothermal and
virialized dark matter halo, which naturally predicts the
correlation $K_0\sim V_{\mathrm{max}}^2$. This correlation holds
true empirically as confirmed by the data and we also find
empirically $L\sim K_0^2$ from the data, thus the canonical
Tully-Fisher relation is reproduced semi-theoretically and
semi-empirically. We perform the same task and we find
theoretically, for dark matter dominated galaxies, that $K_0\sim
V_{\mathrm{flat}}^2$ which is also confirmed empirically from the
data, along with the correlation $K_0\sim M_b^{0.5}$, hence the
baryonic Tully-Fisher law naturally emerges in a semi-theoretical
and semi-empirical manner.
\end{abstract}


\maketitle

\section{Introduction}

Galaxies, galactic and extra-galactic physics are still mysterious
to scientists and many phenomena and even the very own
classification of galaxies are still vague questions with no
decisive answer. Questions why the Universe is filled with
galaxies instead of a uniform sea of stars, why are most stars
assembled in galaxies with luminosity $L\sim 3\times
10^{10}L_{\odot}$, what is the reason for the existence of a
fundamental plane in elliptic galaxies and the Tully-Fisher
relation in spiral galaxies, remain essentially unanswered. The
main purpose of structure formation theory is to explain why stars
are assembled to galaxies, galaxies to clusters and
super-clusters, and these originate by the inhomogeneities created
by the primordial perturbations in the Universe. Although we have
a clear-cut perspective of how the large scale structures emerged
in the Universe, no direct proof of these phenomena can be given.
Dark matter (DM) plays a fundamental role in large scale
structure. Primordial perturbations are neither spherically
symmetric nor isolated in the Universe, and dark matter is
believed to collapse to these primordial inhomogeneities,
undergoing a violent relaxation to a virialized structure known as
a halo. It is exactly on these halo potentials that baryons
undergo a similar relaxation process and protogalaxies and
galaxies are formed. There exist constraints on the DM halo shape,
structure, composition, mass, local density and size, but no
direct evidence exists to date to validate the above features.
Spiral galaxies show a characteristic pattern similar to all the
spirals, the rotation curves are either flat or rising at large
radii. The fact that different Hubble types of galaxies show a
similar pattern in their rotation curves imply that the rotation
velocities at the outer skirts of the galaxies are determined by a
DM halo. DM was firstly proposed by Zwicky in 1933 who observed
the mass-to-light ratio in the Coma cluster and compared it with
the mass-to-light ratio of the luminous part of spiral galaxies as
measured by the rotation velocities of their outer skirts. He
concluded that there was at least 400 times more dark mass
compared to the luminous mass in the Coma cluster. Galaxy
formation is believed to be a hierarchical process and DM plays a
crucial role to galaxy formation.

Although cold DM (CDM) in the context of the
$\Lambda$-Cold-Dark-Matter ($\Lambda$CDM) model is challenged,
there is no alternative theory that can explain the Cosmic
Microwave Background (CMB), the spin problem of galaxies, baryon
acoustic oscillations and pre-recombination baryon perturbations
and numerous other observational features of galaxies, such as
gravitational lensing and so on. Modified Newtonian Dynamics
(MOND), that is considered by a minority of scientists to be an
alternative to particle DM, lacks of a concrete explanation of the
CMB, the spin problem of galaxies, baryon acoustic oscillations
and pre-recombination baryon perturbations and gravitational
lensing and numerous other DM successes. MOND also lacks of a
solid and consistent relativistic formulation compatible with the
GW170817 event. The most important challenges for CDM in the
context of the $\Lambda$CDM model are the cusp-core problem, the
too-big-to-fail problem and the diversity problem
\cite{Tulin:2017ara}. The cusp-core problem refers to dwarf
galaxies and low-surface-brightness spirals which show flat or
cored central density profiles. The too-big-to-fail problem is
sourced to the fact that real satellites of the Milky Way have
shallower gravitational potentials than expected. The diversity
problem is sourced to the fact that galaxies with similar halo
masses show observationally different rotation curve shapes,
especially in their inner region, and it is difficult to reproduce
them with existing CDM models. To all the above models
self-interacting DM (SIDM) offer a consistent explanation. Dwarfs
irregulars and low-surface-brightness and low luminosity spirals
are expected to be DM dominated and these are perfect laboratories
to study DM models since baryons do not affect their dynamics
significantly.

In this work we shall assume that DM is self interacting and has a
scale-dependent equation of state (EoS) of the form
$P(r)=K(r)\left(\frac{\rho}{\rho_{\star}}\right)^{\gamma(r)}$,
with
\begin{align}\label{iniDMpres}
    \gamma(r) &= \gamma_0 - \delta_\gamma\,\tanh\!\left(\frac{r - r_\gamma}{2.0\mathrm{Kpc}}\right), \\
    K(r) &= K_0\,\left(1 + \frac{r}{r_c}\right)^{-p}\, .
\end{align}
Thus DM has a radius dependent polytropic EoS. In the literature
there exist various frameworks that used a polytropic EoS for DM,
but in our article we shall assume that the DM fluid has a radius
dependent polytropic EoS, a generalization of the polytropic EoS.
For a mainstream of articles and reviews on SIDM and DM with
polytropic EoS see
\cite{Slepian:2011ev,Saxton:2014swa,Alonso-Alvarez:2024gdz,Kaplinghat:2015aga,Ahn:2004xt,Saxton:2012ja,Saxton:2010jk,Saxton:2016ozz,Arbey:2003sj,
Heikinheimo:2015kra,Wandelt:2000ad,Spergel:1999mh,Loeb:2010gj,Ackerman:2008kmp,Goodman:2000tg,Arbey:2006it}.
In this article we shall investigate the phenomenological
implications of such a scale-dependent polytropic EoS, and as we
shall see we shall be able to mimic the rotation curves of dwarfs,
irregular galaxies, low-surface-brightness spirals and
low-luminosity spirals, with a two parameter SIDM model given in
Eq. (\ref{iniDMpres}). Also we shall show that our DM model of
scale-dependent polytropic EoS can produce the Tully-Fisher law
for spirals theoretically and empirically from the results and
also the model yields cored DM profiles. We examine 174 galaxies
from the SPARC data \cite{Lelli:2016zqa} and we find that our
scale-dependent SIDM model can fit perfectly 100 galaxies, 27
galaxies with marginal fit while it cannot fit 47 in total
galaxies. The 100 viable galaxies are mostly dwarfs, irregular
galaxies, and low-surface-brightness or low-luminosity spiral
galaxies. For these galaxies we are able to provide
semi-theoretical and semi-empirical proofs for the canonical
Tully-Fisher relation and the baryonic Tully-Fisher relations.

We shall make no assumption for the SIDM, and we shall not connect
it to some model of SIDM. The only assumptions made about the SIDM
is that it may be composed by atoms, ions, radiation and stable
particles. However the results of the galactic dynamics and the
related physical outcomes are not related to some specific model
of SIDM. These are universal results and model-independent and are
derived solely by the scale-dependent EoS of Eq.
(\ref{iniDMpres}).

However, we find quite fascinating the mirror DM perspective
firstly developed in \cite{Kobzarev:1966qya} and later studied in
Refs. \cite{Hodges:1993yb,Foot:2004pa,Berezhiani:2003wj} and also
in Refs.
\cite{Silagadze:2008fa,Foot:2000tp,Chacko:2005pe,Berezhiani:2000gw,Blinnikov:2009nn,Tulin:2017ara,Mohapatra:2001sx,
Blinnikov:1982eh,Blinnikov:1983gh,Foot:2016wvj,Foot:2014osa,Foot:2014uba,
Foot:2004pq,Foot:2001ft,Foot:2004dh,Foot:1999hm,Foot:2001pv,Foot:2001ne,Foot:2000iu,Pavsic:1974rq,Foot:1993yp,Ignatiev:2000yy,Ignatiev:2003js,
Ciarcelluti:2004ik,Ciarcelluti:2004ip,Ciarcelluti:2010zz,Dvali:2009fw,Foot:2013msa,Foot:2013vna,Cui:2011wk,Foot:2015mqa,Foot:2014mia,Cline:2013zca,Ibe:2019ena,
Foot:2018qpw,Howe:2021neq,Cyr-Racine:2021oal,Armstrong:2023cis,Ritter:2024sqv,Mohapatra:1996yy,Mohapatra:2000qx,Goldman:2013qla,Berezhiani:1995am,Oikonomou:2024geq}.
The alignment between the scale-dependent EoS SIDM and mirror DM,
comes due to the fact that for a combination of SIDM particles and
atoms, nature does not need a unique EoS to describe the
particles. The processes involved in a galaxy determine the actual
EoS, but we need a description to model what we see.
Scale-dependent EoS does exactly this work. It might be possible
that a galaxy is exactly like a neutron star, meaning that a
piecewise polytropic EoS is needed to accurately describe it. A
piecewise polytropic EoS might be needed, with the difference
being that a galaxy is by far more complex compared to a neutron
star and SIDM might be quite distinct than nuclear matter. Many
energetic astrophysical processes at the center of the galaxy
might drastically affect the DM EoS. Also if SIDM is
multi-component, composed by atoms, molecules, particles and
radiation, its EoS will change as a function of the radius,
starting from the center of the galaxy. Galaxies form in parts,
not simultaneously, first the bulge, next the outer layers, so the
actual DM EoS is a quantity that forms dynamically and changes as
time passes, while the galaxy is formed. Our scale-dependent EoS
approach applies at well formed galaxies. We need to note though
that the galactic scale radius-dependent EoS for DM approach might
hold at galactic and at cluster and super cluster scales, not at
cosmological scales of the order 100Mpc and larger. Thus this
scale-dependent approach for DM is assumed to hold at galactic and
cluster scales.

For the presentation of galactic rotation curves, we chose a small
number of several viable and non-viable galaxies from the total of
174 galaxies of the SPARC data. The complete behavior of the rest
of the galaxies is  presented in the extensive Appendix indented
only for the arXiv version of the article, not the journal version
of it.

\section{Motivation for a Scale-dependent Polytropic SIDM: Atomic DM vs non-atomic DM at Galactic Scales and their Implications for Galaxies}

As we mentioned in the introduction, in this work we shall assume
that DM is self interacting and that it has a scale-dependent
(radius dependent) equation of state (EoS) of the form
$P(r)=K(r)\left(\frac{\rho}{\rho_{\star}}\right)^{\gamma(r)}$,
with
\begin{align}\label{ScaledependentEoSDM}
    \gamma(r) &= \gamma_0 - \delta_\gamma\,\tanh\!\left(\frac{r - r_\gamma}{2.0\mathrm{Kpc}}\right), \\
    K(r) &= K_0\,\left(1 + \frac{r}{r_c}\right)^{-p}\, .
\end{align}
Now we shall discuss the motivation for having a scale-dependent
EoS for DM. In thermodynamics, the polytropic coefficient \(K(r)\)
in the EoS
$P(r)=K(r)\left(\frac{\rho}{\rho_{\star}}\right)^{\gamma(r)}$ can
be interpreted to be related to the specific entropy of the fluid,
especially in the adiabatic case. So a varying $K(r)$ spatially is
equivalent to having non-isentropic evolution, which basically
corresponds to entropy gradients of the system. Now regarding the
question whether this variation of $K(r)$ is consistent with DM
physics, from a physics standpoint it is, if the microphysics of
DM allows for non-adiabatic processes like dissipation and energy
transport, if local self-interactions exist that may lead to local
thermal equilibrium without global constant entropy. Thus the
varying $K(r)$ is consistent with SIDM halos.

Regarding a varying polytropic index $\gamma(r)$, let us discuss
the motivation for using this. The polytropic index in a fluid
measures the way that pressure responds to changes in the density
of the fluid. It basically encodes the degrees of freedom of the
fluid. The same applies for the SIDM which recall that it may be
comprised by atoms, ions, radiation and even elementary dark
particles, like copies of ordinary electrons in the context of
mirror DM. Thus in general, the thermodynamic state of this dark
plasma and gas and fluid of SIDM may vary with the environment.
The microphysical mechanisms that may induce a radius varying
polytropic index might be higher density and temperatures at
galactic centers, with fewer degrees of freedom available at the
centers. At larger galactic radii, lower densities are found, and
thus cooling by thermal Bremsstrahlung among DM particles, or
ion-mirror electron recombination is applicable. Hence, at larger
radii, DM atoms may be formed, and more degrees of freedom due to
atoms are excited, like vibrational, rotational and so on. Hence,
at different radii, distinct degrees of freedom occur, thus at a
galactic scale, the variation of the polytropic index is in
principle allowed, if SIDM is comprised by atoms, ions, dark
radiation and dark elementary particles. Just to mention, mirror
DM captures all these features in a very well motivated particle
physics framework. Having a radius dependent polytropic index may
induce cored inner profiles for the density instead of cusps, and
thus no modification of gravity is needed to comply with the
rotation curves, since the radius-dependent EoS replaces the
modified gravity-MOND-like effects. In simple words, MOND
phenomenology emerges directly from DM sector microphysics, not
from modified dynamics. In a section much later, we shall further
elaborate on the link between SIDM with scale-dependent EoS and
MOND effects. Of course, if dark stars can be formed, the
functions $K(r)$ and $\rho(r)$ might be affected, through outflows
and radiation heat, which can directly affect for example $K(r)$
but we will not discuss these issues here, see for example
\cite{Silagadze:2008fa,Foot:2000tp,Chacko:2005pe,Berezhiani:2000gw,Blinnikov:2009nn,Tulin:2017ara,Mohapatra:2001sx,
Blinnikov:1982eh,Blinnikov:1983gh,Foot:2016wvj,Foot:2014osa,Foot:2014uba,
Foot:2004pq,Foot:2001ft,Foot:2004dh,Foot:1999hm,Foot:2001pv,Foot:2001ne,Foot:2000iu,Pavsic:1974rq,Foot:1993yp,Ignatiev:2000yy,Ignatiev:2003js,
Ciarcelluti:2004ik,Ciarcelluti:2004ip,Ciarcelluti:2010zz,Dvali:2009fw,Foot:2013msa,Foot:2013vna,Cui:2011wk,Foot:2015mqa,Foot:2014mia,Cline:2013zca,Ibe:2019ena,
Foot:2018qpw,Howe:2021neq,Cyr-Racine:2021oal,Armstrong:2023cis,Ritter:2024sqv,Mohapatra:1996yy,Mohapatra:2000qx,Goldman:2013qla,Berezhiani:1995am,Oikonomou:2024geq}.
Thus a radius-dependent polytropic index $\gamma(r)$ encodes the
nature of DM particles and their degrees of freedom, while a
radius-dependent $K(r)$, encodes the thermodynamic state of the DM
fluid, that is, its entropy and its thermodynamic history.

\section{Analysis of Simple Rotation Curve Simulation with SIDM: Strategy for Solving the Hydrodynamic Equations and Equilibrium}

\subsection{Rigorous mathematical, physical and numerical analysis
of the implemented model}

Now at this point we shall analyze in detail our numerical
approach to solve the hydrodynamic equilibrium equations for
scale-dependent EoS DM. We fix the dimensions and also we discuss
in detail the numerical details of our study. We will also
introduce the temperature parameter
$T(r)=K(r)\,(\rho(r)/\rho_{\star})^{\gamma(r)-1}$ and discuss its
physical significance for the scale-dependent DM.

We model a spherically symmetric DM halo with scale-dependent EoS,
in hydrostatic equilibrium under its own gravity. The
scale-dependent EoS for SIDM  is a radius-dependent, normalized
polytropic law as follows,
\begin{equation}\label{mainequationofstate}
P(r) \;=\;
K(r)\,\left(\frac{\rho(r)}{\rho_\star}\right)^{\gamma(r)} ,
\end{equation}
where $\rho(r)$ is the mass density with units of $M_\odot\,{\rm
Kpc}^{-3}$, also $K(r)$ is a pressure-scale coefficient with fixed
physical units which we define shortly, $\gamma(r)$ is a smoothly
varying polytropic index, and $\rho_\star$ is a fixed reference
density introduced so that the exponent acts on a dimensionless
ratio. We will take $\rho_\star=1$$M_\odot\,{\rm Kpc}^{-3}$ for
simplicity, as it is an auxiliary parameter, not a parameter with
significant physical importance. The hydrostatic equilibrium in
spherical symmetry reads,
\begin{equation}
\frac{dP}{dr} \;=\; -\,\frac{G\,M(r)\,\rho(r)}{r^{2}},
\end{equation}
where the enclosed mass is
\begin{equation}
M(r)\;=\;\int_0^r 4\pi r'^{2}\,\rho(r')\,dr'.
\end{equation}
We differentiate $P(r)$ with respect to $r$ using the product and
chain rules,
\begin{align}
 &\frac{dP}{dr}
=
\frac{d}{dr}\!\left[K(r)\left(\frac{\rho}{\rho_\star}\right)^{\gamma(r)}\right]
\\ \notag &
 = \left(\frac{\rho}{\rho_\star}\right)^{\gamma}\frac{dK}{dr}
   +
   K\,\frac{d}{dr}\!\left[\left(\frac{\rho}{\rho_\star}\right)^{\gamma(r)}\right]\,
   ,
\end{align}
and the second term expands as,
\begin{equation}
\frac{d}{dr}\!\left[\left(\frac{\rho}{\rho_\star}\right)^{\gamma(r)}\right]
 = \left(\frac{\rho}{\rho_\star}\right)^{\gamma}
   \left(\frac{d\gamma}{dr}\ln\!\frac{\rho}{\rho_\star}
         + \frac{\gamma}{\rho}\frac{d\rho}{dr}\right).
\end{equation}
Hence,
\begin{equation}
\frac{dP}{dr} =
\left(\frac{\rho}{\rho_\star}\right)^{\gamma}\frac{dK}{dr}
 + K\left(\frac{\rho}{\rho_\star}\right)^{\gamma}\!\left(\frac{d\gamma}{dr}\ln\!\frac{\rho}{\rho_\star}\right)
 + K\,\gamma\left(\frac{\rho}{\rho_\star}\right)^{\gamma-1}\!\frac{d\rho}{dr}\frac{1}{\rho_\star}.
\end{equation}
The hydrostatic equilibrium requires
\begin{equation}
\frac{dP}{dr} \;=\; -\,\frac{G\,M(r)\,\rho}{r^{2}}\, ,
\end{equation}
so upon substituting the derivative of $P$ and solving
algebraically for $d\rho/dr$ gives,
\begin{align}
K\,\gamma\!\left(\frac{\rho}{\rho_\star}\right)^{\gamma-1}\frac{d\rho}{dr}\frac{1}{\rho_\star}
&=\; -\,\frac{G\,M(r)\,\rho}{r^{2}}
 - \left(\frac{\rho}{\rho_\star}\right)^{\gamma}\frac{dK}{dr}
 - K\left(\frac{\rho}{\rho_\star}\right)^{\gamma}\ln\!\frac{\rho}{\rho_\star}\frac{d\gamma}{dr},
\\ \notag &
\frac{d\rho}{dr}=\frac{-\,\dfrac{G\,M(r)\,\rho}{r^{2}}
      - \left(\dfrac{\rho}{\rho_\star}\right)^{\gamma}\dfrac{dK}{dr}
      - K\left(\dfrac{\rho}{\rho_\star}\right)^{\gamma}\ln\!\dfrac{\rho}{\rho_\star}\,\dfrac{d\gamma}{dr}}
     {K\,\gamma\left(\dfrac{\rho}{\rho_\star}\right)^{\gamma-1}\frac{1}{\rho_\star}}.
\end{align}
The corresponding mass differential equation is the standard
relation,
\begin{equation}
\frac{dM}{dr} = 4\pi r^{2}\,\rho(r).
\end{equation}
Hence the complete set of differential equations that we will
solve numerically is the following:
\begin{align}\label{differentialequationsmainset}
\frac{dP}{dr} &= -\rho(r)\frac{G M(r)}{r^2}, \\
P(r) &= K(r) \left(\frac{\rho(r)}{\rho_{\star}}\right)^{\gamma(r)}, \\
\frac{dM}{dr} &= 4\pi r^2 \rho(r).
\end{align}
We aim to integrate these equations outward from the galactic
center, modelling a SIDM halo from the core to the outskirts of
the galaxy. Now regarding the initial conditions, we will start
the integration not from a distance $r_0=0$ but from $r_0 \sim
10^{-4}~\mathrm{Kpc}$ with an initial density $\rho_0$ in the
range $\rho_0 \sim 10^6 - 10^9~M_\odot/\mathrm{Kpc}^3$  well
motivated by the physics of galaxies and for numerical reasons.
For $\rho_0$ the choice is in the range $\rho_0 \sim 10^6 -
10^9~M_\odot/\mathrm{Kpc}^3$ because in low-surface-brightness and
dwarfs we have densities in the core of the order
$0.01$-$0.1\,M_\odot/\mathrm{pc}^3$ $\rightarrow$
$10^6$-$10^7~M_\odot/\mathrm{Kpc}^3$. Thus a small but finite
$r_0$ and physical $\rho_0$ avoids the singularities in the
pressure gradient and the gravitational term at $r = 0$. Also
$\rho_0$ sets the normalization for the whole halo mass and size
in polytropic models. Now regarding the mass $M_0$ at the radius
$r_0 \sim 10^{-4}~\mathrm{Kpc}$ we will take $M(r_0) = 0$, which
ensures consistency with the mass continuity equation
\[
\frac{dM}{dr} = 4\pi r^2 \rho(r).
\]
Note that if we used $r = 0$ instead of $r_0$ as the starting
point of the integration, we would have $\frac{GM}{r^2}$ and
$\frac{dP}{dr}$ singular at $r=0$. Now we will solve the set of
the differential equations (\ref{differentialequationsmainset}) to
find numerically the density profile $\rho(r)$ and the enclosed DM
galactic mass $M(r)$. From these, we will find the rotation curve
$v_c(r)=\sqrt{\frac{GM(r)}{r}}$. As we stated earlier, we shall
choose the following model for scale-dependent SIDM,
\begin{align}\label{tanhmodel}
    \gamma(r) &= \gamma_0 - \delta_\gamma\,\tanh\!\left(\frac{r - r_\gamma}{2.0\,\mathrm{Kpc}}\right),
    \\\notag &
    K(r)= K_0\,\left(1 + \frac{r}{r_c}\right)^{-p}\, ,
\end{align}
The model is essentially a two parameter model since only $K_0$
and $\rho_0$ will be allowed to vary, while the rest of the
parameters are fixed to have the following parameters:
\begin{align}\label{modelparametervalues}
  & \gamma_0=1.0001\\ \notag &
  \delta_\gamma=1.2\times 10^{-9}\\ \notag &
  r_c=0.5\,\mathrm{Kpc}\\ \notag &
  p=0.01\\ \notag &
  r_\gamma=1.5 \,\mathrm{Kpc}
\end{align}
As we will show, the above two-parameter model (virtually one
parameter model since $\rho_0$ is an initial condition not a model
parameter value) successfully reproduces the observed transition
from cored to cusp density profiles and the approximately flat
rotation curves of dwarfs, low-luminosity and
low-surface-brightness galaxies. Now we shall confront the model
(\ref{tanhmodel}) against the galactic data contained in the SPARC
database \cite{Lelli:2016zqa}, using the simple model approach in
which the galactic rotation speed is affected only by DM, without
taking into account the disk, gas and if applicable, bulge
velocity. Our results indicate that low-luminosity,
low-surface-brightness and dwarf galaxies are perfectly described
by the scale-dependent EoS DM profile, but larger spirals require
the inclusion of gas, bulge and disk components. In this case, we
will dub the study as extended model, and the rotation curve
velocity will be,
\[
V_{\text{total}}^2(r) = V_{\text{disk}}^2(r) +
V_{\text{bulge}}^2(r) + V_{\text{gas}}^2(r) + V_{\text{DM}}^2(r)\,
.
\]
Each term represents the squared contribution to the circular
velocity from a distinct mass distribution. For the extended
model, when used, usually in large spirals, we shall fix $K_0$,
$\rho_0$ and vary only $\gamma_0$ and $\delta_\gamma$, thus the
model is again a two parameter model. Regarding the units, the
pressure has units,
\begin{equation}
[P] = [\rho][v^2] = M_\odot\,{\rm Kpc}^{-3} \, ({\rm km/s})^2.
\end{equation}
In the normalized EoS, we have,
\[
P(r) = K(r)\left(\frac{\rho(r)}{\rho_\star}\right)^{\gamma(r)},
\]
and the ratio $\rho/\rho_\star$ is dimensionless. Thus,
\begin{equation}
[K(r)] = [P] = M_\odot\,{\rm Kpc}^{-3} \, ({\rm km/s})^2\, .
\end{equation}
For the numerical integration we used a stiff differential
equation solver, the Radau. For each galaxy studied, apart from
$\rho(r)$ we will also present the behavior of the temperature
parameter defined as,
\begin{equation}
T(r) = K(r)\left(\frac{\rho(r)}{\rho_\star}\right)^{\gamma(r)-1}\,
.
\end{equation}
Differentiating we get,
\begin{equation}
\frac{d\ln T}{dr} = \frac{d\ln K}{dr} +
(\gamma-1)\frac{d\ln\rho}{dr} +
\ln\!\frac{\rho}{\rho_\star}\frac{d\gamma}{dr}\, ,
\end{equation}
thus typically we expect that all contributions make $T(r)$ to
decrease with radius. Thus the integration must lead to a
declining profile in the temperature parameter. For the simple
model analysis of SIDM, we shall also include well known DM
profiles, such as the Navarro-Frenk-White (NFW) profile
\cite{Navarro:1996gj},
\[
\rho(r) = \frac{\rho_s}{\left(\frac{r}{r_s}\right)\left(1 +
\frac{r}{r_s}\right)^2}\, ,
\]
the Burkert profile \cite{Burkert:1995yz}
\[
\rho(r) = \frac{\rho_0^B}{\left(1 + \frac{r}{r_0}\right) \left(1 +
\left(\frac{r}{r_0}\right)^2 \right)},
\]
with \(\rho_0^B\) being the central density for this profile and
\(r_0\) being the core radius. The Burkert DM profile is a cored
density profile often used to fit dwarf and low-surface-brightness
galaxy halos. Finally, we will include the Einasto DM profile
\cite{Baes:2022pbc},
\[
\rho(r) = \rho_e \, \exp\!\left[-\frac{2}{\alpha} \left(
\left(\frac{r}{r_e}\right)^{\alpha} - 1 \right)\right]\, .
\]

\section{Simulations with Real Spiral Galaxies Data from SPARC for Collisional Dark Matter}

In this section we shall present some characteristic galaxies from
the SPARC data \cite{Lelli:2016zqa} with which the SIDM model with
scale-dependent EoS (\ref{mainequationofstate}), with $K(r)$ and
$\rho(r)$ chosen as in  Eq. (\ref{tanhmodel}). The model
(\ref{tanhmodel}) is essentially a one parameter model since only
$K_0$ is allowed to vary along with the initial condition for the
central density $\rho_0$, and the rest parameters are given in Eq.
(\ref{modelparametervalues}). We examined in total 174 galaxies,
and the model produces perfectly compatible results for 100 of
them, marginally viable results for 27 and non-viable results for
47. For the non-viable and the marginally viable galaxies we also
included the extended rotation velocity, including the disk, gas
and if applicable bulge velocity. We found that some of these
extended versions are viable. For the extended galactic rotation
curves of SIDM we assumed that only $\gamma_0$ and $\delta_\gamma$
vary, so in this case too the model is a two parameter model.

The galaxies which are viable and compatible with the SIDM model
are: \textbf{CamB, D512-2, D564-8,D631-7, D631-7,DDO064, DDO154,
DDO170, ESO444-G084, F561-1, F563-1, F563-V1, F563-V2,F565-V2,
F567-2, F568-1, F568-3, F568-V1, F571-V1, F574-1, F574-2, F579-V1,
F583-1, F583-4, IC2574, KK98-251, NGC0024, NGC0055, NGC0100,
NGC1705, NGC2366, NGC2683, NGC2915, NGC2976, NGC3109, NGC3769,
NGC3877, NGC3893, NGC3917, NGC3949, NGC3953, NGC3972, NGC4068,
NGC4085, NGC4088, NGC4157, NGC4183, NGC4217, NGC4389, NGC6789,
PGC51017, UGC00634, UGC00891, UGC01230, UGC01281, UGC02023,
UGC04325, UGC04483, UGC04499, UGC05005, UGC05414, UGC05750,
UGC05829, UGC05918, UGC05986, UGC05999, UGC06399, UGC06628,
UGC06667, UGC06917, UGC06923, UGC06930, UGC06983, UGC07089,
UGC07151, UGC07232 , UGC07261 , UGC07559, UGC07577, UGC07603,
UGC07608, UGC07690, UGC07866, UGC08837, UGC09037, UGC09992,
UGC10310, UGC11557, UGC11914, UGC12632, UGCA281, UGCA444}.

The galaxies that are marginally compatible with the SIDM model
are: \textbf{DDO168, DDO161, IC4202, NGC0300, NGC1090,NGC2903,
NGC3198, NGC3521, NGC3726, NGC3992, NGC4051, NGC4138, NGC5985,
NGC6674, NGC7814, UGC00191, UGC03205, UGC03546, UGC04278,
UGC04305, UGC05721, UGC05764, UGC06446, UGC07125, UGC08286,
UGC08490, UGC08550, UGC11455, UGCA442,ESO563-G021,NGC0801,
NGC0891}.

The galaxies that are incompatible in general with the SIDM model
are:  \textbf{ESO079-G014, ESO116-G012, F571-8, NGC0247, NGC0289,
NGC1003,NGC2403, NGC2841, NGC2955, NGC2998, NGC3741, NGC4010,
NGC4013, NGC4100, NGC4214, NGC4559, NGC5005, NGC5033, NGC5055,
NGC5371, NGC5585, NGC5907, NGC6015, NGC6195 , NGC6503, NGC6946,
NGC7331, NGC7793, UGC00128, UGC00731, UGC02259, UGC02487,
UGC02885, UGC02916, UGC02953, UGC03580, UGC05253, UGC05716,
UGC06614, UGC06786 , UGC06787, UGC06818, UGC06973, UGC07323,
UGC07399, UGC07524, UGC08699, UGC09133, UGC11820, UGC12506,
UGC12732}.

In the following subsections we shall present some characteristic
viable, marginally viable and non-viable models. The rest of the
results are contained in the Appendix of the arXiv version of this
article. Most of the models use $K_0$ and $\rho_0$ as free
parameters, but for some galaxies we also considered fixing $K_0$
and we varied $\gamma_0$ and $\delta_{\gamma}$ just to show the
flexibility of the model in modelling galactic rotation curves.
The data are available upon request.

\subsection{Analysis and Simulation of SPARC Galaxy Data and Fitting with two-parameter Simple SIDM Model: A Sample of Viable Galaxies}

\subsubsection{The Galaxy NGC2915}

\begin{figure}[h!]
\centering
\includegraphics[width=20pc]{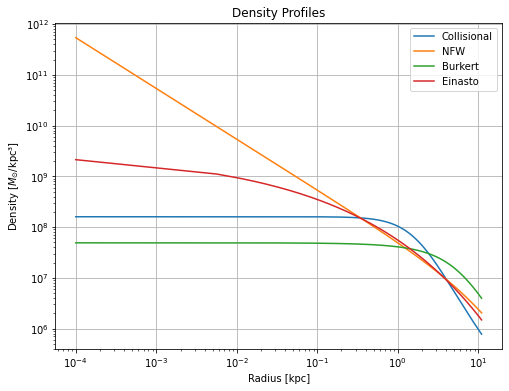}
\caption{The density of the collisional DM model (\ref{tanhmodel})
for the galaxy NGC2915, as a function of the radius.}
\label{NGC2915dens}
\end{figure}
\begin{figure}[h!]
\centering
\includegraphics[width=20pc]{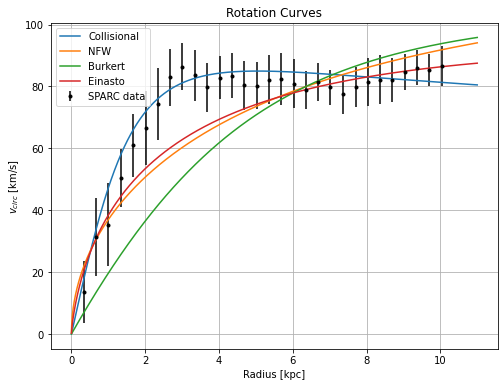}
\caption{The predicted rotation curves after using an optimization
for the collisional DM model (\ref{tanhmodel}), versus the SPARC
observational data for the galaxy NGC2915. We also plotted the
optimized curves for the NFW model, the Burkert model and the
Einasto model.} \label{NGC2915}
\end{figure}
\begin{figure}[h!]
\centering
\includegraphics[width=20pc]{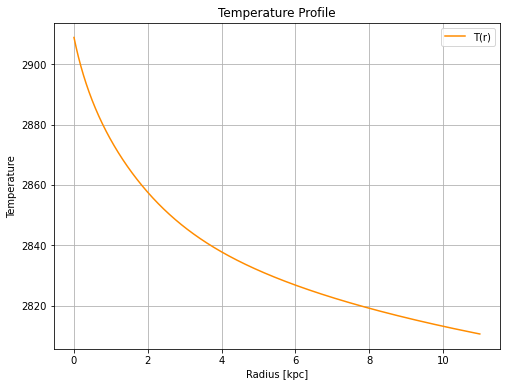}
\caption{The temperature as a function of the radius for the
collisional DM model (\ref{tanhmodel}) for the galaxy NGC2915.}
\label{NGC2915temp}
\end{figure}

For this galaxy, we shall choose $\rho_0=1.6\times
10^8$$M_{\odot}/\mathrm{Kpc}^{3}$. Galaxy NGC\,2915 is a nearby
blue compact dwarf (optically compact, star-forming core) embedded
in an extremely extended, low surface-brightness HI disk with
well-defined spiral structure. Dynamically it behaves like a
late-type/spiral system and is dark-matter dominated at nearly all
radii. Its distance is $D \sim 3.78\ \mathrm{Mpc}$. In Figs.
\ref{NGC2915dens}, \ref{NGC2915} and \ref{NGC2915temp} we present
the density of the collisional DM model, the predicted rotation
curves after using an optimization for the collisional DM model
(\ref{tanhmodel}), versus the SPARC observational data and the
temperature parameter as a function of the radius respectively. As
it can be seen, the SIDM model produces viable rotation curves
compatible with the SPARC data. Also in Tables \ref{collNGC2915},
\ref{NavaroNGC2915}, \ref{BuckertNGC2915} and \ref{EinastoNGC2915}
we present the optimization values for the SIDM model, and the
other DM profiles. Also in Table \ref{EVALUATIONNGC2915} we
present the overall evaluation of the SIDM model for the galaxy at
hand. The resulting phenomenology is viable.
\begin{table}[h!]
  \begin{center}
    \caption{Collisional Dark Matter Optimization Values}
    \label{collNGC2915}
     \begin{tabular}{|r|r|}
     \hline
      \textbf{Parameter}   & \textbf{Optimization Values}
      \\  \hline
     $\delta_{\gamma} $ & 0.0000000012
\\  \hline
$\gamma_0 $ & 1.0001 \\ \hline $K_0$ ($M_{\odot} \,
\mathrm{Kpc}^{-3} \, (\mathrm{km/s})^{2}$)& 7050  \\ \hline
    \end{tabular}
  \end{center}
\end{table}
\begin{table}[h!]
  \begin{center}
    \caption{NFW  Optimization Values}
    \label{NavaroNGC2915}
     \begin{tabular}{|r|r|}
     \hline
      \textbf{Parameter}   & \textbf{Optimization Values}
      \\  \hline
   $\rho_s$   & $0.0027\times 10^9$
\\  \hline
$r_s$&  20
\\  \hline
    \end{tabular}
  \end{center}
\end{table}
\begin{table}[h!]
  \begin{center}
    \caption{Burkert Optimization Values}
    \label{BuckertNGC2915}
     \begin{tabular}{|r|r|}
     \hline
      \textbf{Parameter}   & \textbf{Optimization Values}
      \\  \hline
     $\rho_0^B$  & $0.049\times 10^9$
\\  \hline
$r_0$&  6
\\  \hline
    \end{tabular}
  \end{center}
\end{table}
\begin{table}[h!]
  \begin{center}
    \caption{Einasto Optimization Values}
    \label{EinastoNGC2915}
    \begin{tabular}{|r|r|}
     \hline
      \textbf{Parameter}   & \textbf{Optimization Values}
      \\  \hline
     $\rho_e$  & $0.0018\times 10^9$
\\  \hline
$r_e$ & 10
\\  \hline
$n_e$ & 0.11
\\  \hline
    \end{tabular}
  \end{center}
\end{table}
\begin{table}[h!]
\centering \caption{Physical assessment of collisional DM
parameters (NGC2915).}
\begin{tabular}{lcc}
\hline
Parameter & Value & Physical Verdict \\
\hline
$\gamma_0$ & $1.0001$ & Essentially isothermal  \\
$\delta_\gamma$ & $9.6\times10^{-5}$ & Small \\
$r_\gamma$ & $1.5\ \mathrm{Kpc}$ & Transition radius in inner halo \\
$K_0$ & $2.90\times10^{3}$ & Moderate entropy   \\
$r_c$ & $0.5\ \mathrm{Kpc}$ & Small core scale- plausible for inner region \\
$p$ & $0.01$ & Extremely shallow $K(r)$ slope \\
\hline
Overall &-& Physically consistent but functionally nearly isothermal \\
\hline
\end{tabular}
\label{EVALUATIONNGC2915}
\end{table}

\subsubsection{The Galaxy NGC3769}

\begin{figure}[h!]
\centering
\includegraphics[width=20pc]{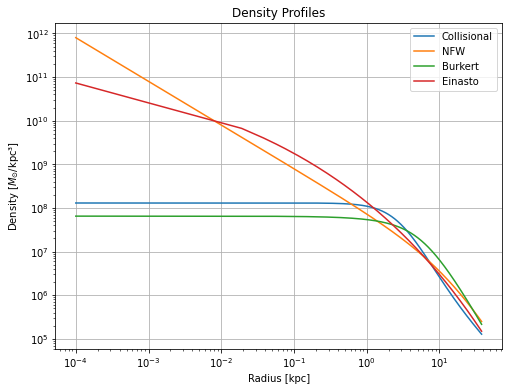}
\caption{The density of the collisional DM model (\ref{tanhmodel})
for the galaxy NGC3769, as a function of the radius.}
\label{NGC3769dens}
\end{figure}

\begin{figure}[h!]
\centering
\includegraphics[width=20pc]{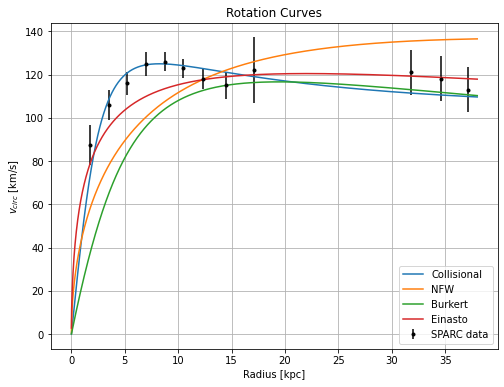}
\caption{The predicted rotation curves after using an optimization
for the collisional DM model (\ref{tanhmodel}), versus the SPARC
observational data for the galaxy NGC3769. We also plotted the
optimized curves for the NFW model, the Burkert model and the
Einasto model.} \label{NGC3769}
\end{figure}

\begin{figure}[h!]
\centering
\includegraphics[width=20pc]{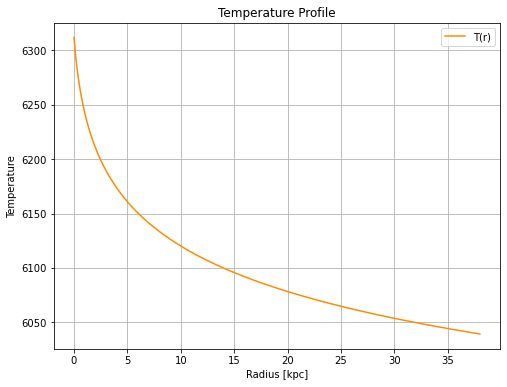}
\caption{The temperature as a function of the radius for the
collisional DM model (\ref{tanhmodel}) for the galaxy NGC3769.}
\label{NGC3769temp}
\end{figure}

For this galaxy, we shall choose $\rho_0=1.3\times
10^8$$M_{\odot}/\mathrm{Kpc}^{3}$. NGC3769 is a barred spiral
galaxy (morphological type often quoted as SB(r)b / SBb) located
in the constellation Ursa Major. Its distance is $D \sim 18.6\
\mathrm{Mpc}$. In Figs. \ref{NGC3769dens}, \ref{NGC3769} and
\ref{NGC3769temp} we present the density of the collisional DM
model, the predicted rotation curves after using an optimization
for the collisional DM model (\ref{tanhmodel}), versus the SPARC
observational data and the temperature parameter as a function of
the radius respectively. As it can be seen, the SIDM model
produces viable rotation curves compatible with the SPARC data.
Also in Tables \ref{collNGC3769}, \ref{NavaroNGC3769},
\ref{BuckertNGC3769} and \ref{EinastoNGC3769} we present the
optimization values for the SIDM model, and the other DM profiles.
Also in Table \ref{EVALUATIONNGC3769} we present the overall
evaluation of the SIDM model for the galaxy at hand. The resulting
phenomenology is viable.
\begin{table}[h!]
  \begin{center}
    \caption{Collisional Dark Matter Optimization Values}
    \label{collNGC3769}
     \begin{tabular}{|r|r|}
     \hline
      \textbf{Parameter}   & \textbf{Optimization Values}
      \\  \hline
     $\delta_{\gamma} $ & 0.0000000012
\\  \hline
$\gamma_0 $ & 1.0001 \\ \hline $K_0$ ($M_{\odot} \,
\mathrm{Kpc}^{-3} \, (\mathrm{km/s})^{2}$)& 6300  \\ \hline
    \end{tabular}
  \end{center}
\end{table}
\begin{table}[h!]
  \begin{center}
    \caption{NFW  Optimization Values}
    \label{NavaroNGC3769}
     \begin{tabular}{|r|r|}
     \hline
      \textbf{Parameter}   & \textbf{Optimization Values}
      \\  \hline
   $\rho_s$   & $0.004\times 10^9$
\\  \hline
$r_s$&  20
\\  \hline
    \end{tabular}
  \end{center}
\end{table}
\begin{table}[h!]
  \begin{center}
    \caption{Burkert Optimization Values}
    \label{BuckertNGC3769}
     \begin{tabular}{|r|r|}
     \hline
      \textbf{Parameter}   & \textbf{Optimization Values}
      \\  \hline
     $\rho_0^B$  & $0.065\times 10^9$
\\  \hline
$r_0$&  6
\\  \hline
    \end{tabular}
  \end{center}
\end{table}
\begin{table}[h!]
  \begin{center}
    \caption{Einasto Optimization Values}
    \label{EinastoNGC3769}
    \begin{tabular}{|r|r|}
     \hline
      \textbf{Parameter}   & \textbf{Optimization Values}
      \\  \hline
     $\rho_e$  & $0.003\times 10^9$
\\  \hline
$r_e$ & 10
\\  \hline
$n_e$ & 0.17
\\  \hline
    \end{tabular}
  \end{center}
\end{table}
\begin{table}[h!]
\centering \caption{Physical assessment of collisional DM
parameters (NGC3769).}
\begin{tabular}{lcc}
\hline
Parameter & Value & Physical Verdict \\
\hline
$\gamma_0$ & $1.0001$ & Essentially isothermal  \\
$\delta_\gamma$ & $1.2\times10^{-9}$ & Negligible   \\
$r_\gamma$ & $1.5\ \mathrm{Kpc}$ & Transition radius in inner halo   \\
$K_0$ & $6.30\times10^{3}$ & Moderate entropy  \\
$r_c$ & $0.5\ \mathrm{Kpc}$ & Small core scale   \\
$p$ & $0.01$ & Extremely shallow $K(r)$ slope \\
\hline
Overall &-& Physically consistent but functionally nearly isothermal \\
\hline
\end{tabular}
\label{EVALUATIONNGC3769}
\end{table}

\subsubsection{The Galaxy NGC3893}

\begin{figure}[h!]
\centering
\includegraphics[width=20pc]{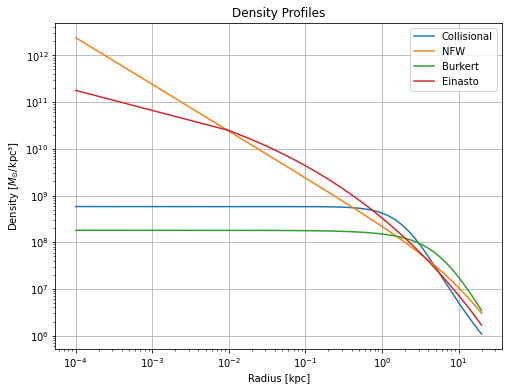}
\caption{The density of the collisional DM model (\ref{tanhmodel})
for the galaxy NGC3893, as a function of the radius.}
\label{NGC3893dens}
\end{figure}
\begin{figure}[h!]
\centering
\includegraphics[width=20pc]{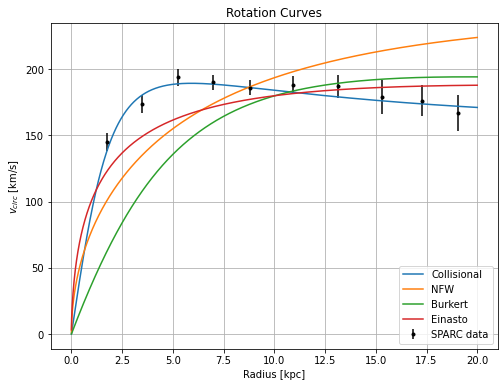}
\caption{The predicted rotation curves after using an optimization
for the collisional DM model (\ref{tanhmodel}), versus the SPARC
observational data for the galaxy NGC3893. We also plotted the
optimized curves for the NFW model, the Burkert model and the
Einasto model.} \label{NGC3893}
\end{figure}
\begin{figure}[h!]
\centering
\includegraphics[width=20pc]{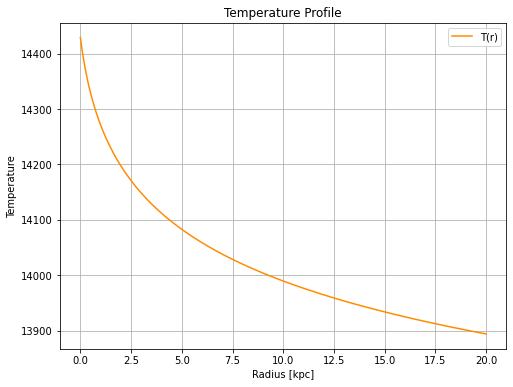}
\caption{The temperature as a function of the radius for the
collisional DM model (\ref{tanhmodel}) for the galaxy NGC3893.}
\label{NGC3893temp}
\end{figure}

For this galaxy, we shall choose $\rho_0=5.8\times
10^8$$M_{\odot}/\mathrm{Kpc}^{3}$. NGC3893 is a grand-design
spiral galaxy located in the constellation Ursa Major. It is
classified morphologically as SAB(rs)c, indicating a weakly barred
spiral with relatively loosely wound arms. It is a member of the
M\,109 group and interacts with its small companion NGC\,3896. The
distance to NGC3893, is $D \sim 14.5\ \mathrm{Mpc}$. In Figs.
\ref{NGC3893dens}, \ref{NGC3893} and \ref{NGC3893temp} we present
the density of the collisional DM model, the predicted rotation
curves after using an optimization for the collisional DM model
(\ref{tanhmodel}), versus the SPARC observational data and the
temperature parameter as a function of the radius respectively. As
it can be seen, the SIDM model produces viable rotation curves
compatible with the SPARC data. Also in Tables \ref{collNGC3893},
\ref{NavaroNGC3893}, \ref{BuckertNGC3893} and \ref{EinastoNGC3893}
we present the optimization values for the SIDM model, and the
other DM profiles. Also in Table \ref{EVALUATIONNGC3893} we
present the overall evaluation of the SIDM model for the galaxy at
hand. The resulting phenomenology is viable.
\begin{table}[h!]
  \begin{center}
    \caption{Collisional Dark Matter Optimization Values}
    \label{collNGC3893}
     \begin{tabular}{|r|r|}
     \hline
      \textbf{Parameter}   & \textbf{Optimization Values}
      \\  \hline
     $\delta_{\gamma} $ & 0.0000000012
\\  \hline
$\gamma_0 $ & 1.0001 \\ \hline $K_0$ ($M_{\odot} \,
\mathrm{Kpc}^{-3} \, (\mathrm{km/s})^{2}$)& 14400 \\ \hline
    \end{tabular}
  \end{center}
\end{table}
\begin{table}[h!]
  \begin{center}
    \caption{NFW  Optimization Values}
    \label{NavaroNGC3893}
     \begin{tabular}{|r|r|}
     \hline
      \textbf{Parameter}   & \textbf{Optimization Values}
      \\  \hline
   $\rho_s$   & $0.012\times 10^9$
\\  \hline
$r_s$&  20
\\  \hline
    \end{tabular}
  \end{center}
\end{table}
\begin{table}[h!]
  \begin{center}
    \caption{Burkert Optimization Values}
    \label{BuckertNGC3893}
     \begin{tabular}{|r|r|}
     \hline
      \textbf{Parameter}   & \textbf{Optimization Values}
      \\  \hline
     $\rho_0^B$  & $0.18\times 10^9$
\\  \hline
$r_0$&  6
\\  \hline
    \end{tabular}
  \end{center}
\end{table}
\begin{table}[h!]
  \begin{center}
    \caption{Einasto Optimization Values}
    \label{EinastoNGC3893}
    \begin{tabular}{|r|r|}
     \hline
      \textbf{Parameter}   & \textbf{Optimization Values}
      \\  \hline
     $\rho_e$  & $0.0073\times 10^9$
\\  \hline
$r_e$ & 10
\\  \hline
$n_e$ & 0.17
\\  \hline
    \end{tabular}
  \end{center}
\end{table}
\begin{table}[h!]
\centering \caption{Physical assessment of collisional DM
parameters (NGC3893).}
\begin{tabular}{lcc}
\hline
Parameter & Value & Physical Verdict \\
\hline
$\gamma_0$ & $1.0001$ & Essentially isothermal  \\
$\delta_\gamma$ & $1.2\times10^{-9}$ & Negligible   \\
$r_\gamma$ & $1.5\ \mathrm{Kpc}$ & Transition radius in inner halo   \\
$K_0$ & $1.44\times10^{4}$ & Moderate-to-large entropy   \\
$r_c$ & $0.5\ \mathrm{Kpc}$ & Small core scale   \\
$p$ & $0.01$ & Extremely shallow $K(r)$ slope \\
\hline
Overall &-& Physically consistent but functionally nearly isothermal  \\
\hline
\end{tabular}
\label{EVALUATIONNGC3893}
\end{table}

\subsubsection{The Galaxy NGC4157}

\begin{figure}[h!]
\centering
\includegraphics[width=20pc]{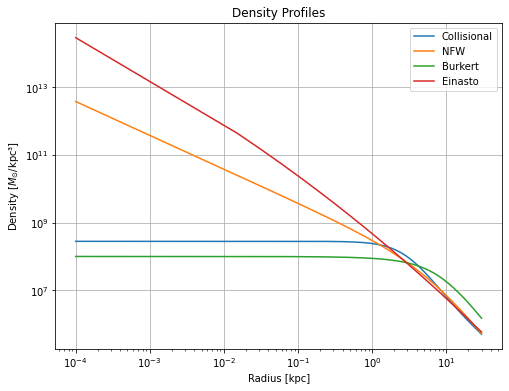}
\caption{The density of the collisional DM model (\ref{tanhmodel})
for the galaxy NGC4157, as a function of the radius.}
\label{NGC4157dens}
\end{figure}
\begin{figure}[h!]
\centering
\includegraphics[width=20pc]{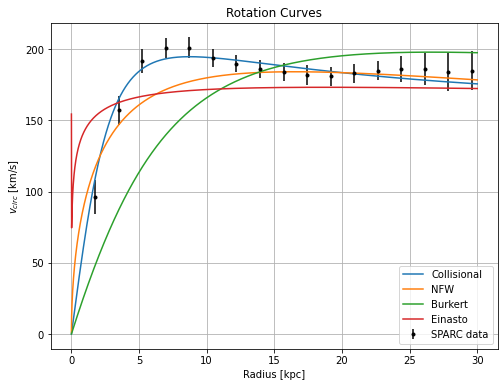}
\caption{The predicted rotation curves after using an optimization
for the collisional DM model (\ref{tanhmodel}), versus the SPARC
observational data for the galaxy NGC4157. We also plotted the
optimized curves for the NFW model, the Burkert model and the
Einasto model.} \label{NGC4157}
\end{figure}
\begin{figure}[h!]
\centering
\includegraphics[width=20pc]{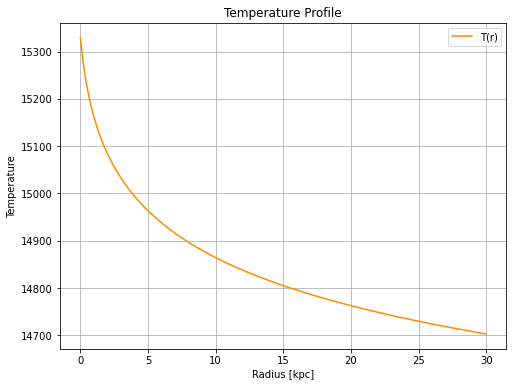}
\caption{The temperature as a function of the radius for the
collisional DM model (\ref{tanhmodel}) for the galaxy NGC4157.}
\label{NGC4157temp}
\end{figure}

For this galaxy, we shall choose $\rho_0=2.8\times
10^8$$M_{\odot}/\mathrm{Kpc}^{3}$. NGC4157 is a nearly edge-on
intermediate spiral galaxy type SAB(s)b in the Ursa Major group at
a distance of about \(D \sim 17.1\pm 3.1\ \mathrm{Mpc}\). In Figs.
\ref{NGC4157dens}, \ref{NGC4157} and \ref{NGC4157temp} we present
the density of the collisional DM model, the predicted rotation
curves after using an optimization for the collisional DM model
(\ref{tanhmodel}), versus the SPARC observational data and the
temperature parameter as a function of the radius respectively. As
it can be seen, the SIDM model produces viable rotation curves
compatible with the SPARC data. Also in Tables \ref{collNGC4157},
\ref{NavaroNGC4157}, \ref{BuckertNGC4157} and \ref{EinastoNGC4157}
we present the optimization values for the SIDM model, and the
other DM profiles. Also in Table \ref{EVALUATIONNGC4157} we
present the overall evaluation of the SIDM model for the galaxy at
hand. The resulting phenomenology is viable.
\begin{table}[h!]
  \begin{center}
    \caption{Collisional Dark Matter Optimization Values}
    \label{collNGC4157}
     \begin{tabular}{|r|r|}
     \hline
      \textbf{Parameter}   & \textbf{Optimization Values}
      \\  \hline
     $\delta_{\gamma} $ & 0.0000000012
\\  \hline
$\gamma_0 $ & 1.0001 \\ \hline $K_0$ ($M_{\odot} \,
\mathrm{Kpc}^{-3} \, (\mathrm{km/s})^{2}$)& 15300  \\ \hline
    \end{tabular}
  \end{center}
\end{table}
\begin{table}[h!]
  \begin{center}
    \caption{NFW  Optimization Values}
    \label{NavaroNGC4157}
     \begin{tabular}{|r|r|}
     \hline
      \textbf{Parameter}   & \textbf{Optimization Values}
      \\  \hline
   $\rho_s$   & $5\times 10^7$
\\  \hline
$r_s$&  7.62
\\  \hline
    \end{tabular}
  \end{center}
\end{table}
\begin{table}[h!]
  \begin{center}
    \caption{Burkert Optimization Values}
    \label{BuckertNGC4157}
     \begin{tabular}{|r|r|}
     \hline
      \textbf{Parameter}   & \textbf{Optimization Values}
      \\  \hline
     $\rho_0^B$  & $1\times 10^8$
\\  \hline
$r_0$&  8.21
\\  \hline
    \end{tabular}
  \end{center}
\end{table}
\begin{table}[h!]
  \begin{center}
    \caption{Einasto Optimization Values}
    \label{EinastoNGC4157}
    \begin{tabular}{|r|r|}
     \hline
      \textbf{Parameter}   & \textbf{Optimization Values}
      \\  \hline
     $\rho_e$  &$1\times 10^7$
\\  \hline
$r_e$ & 7.61
\\  \hline
$n_e$ & 0.05
\\  \hline
    \end{tabular}
  \end{center}
\end{table}
\begin{table}[h!]
\centering \caption{Physical assessment of collisional DM
parameters (NGC4157).}
\begin{tabular}{lcc}
\hline
Parameter & Value & Physical Verdict \\
\hline
$\gamma_0$ & $1.0001$ & Essentially isothermal \\
$\delta_\gamma$ & $1.2\times10^{-9}$ & Effectively zero   \\
$r_\gamma$ & $1.5\ \mathrm{Kpc}$ & Reasonable transition radius  \\
$K_0$ & $1.53\times10^{4}$ & High entropy scale \\
$r_c$ & $0.5\ \mathrm{Kpc}$ & Small core scale \\
$p$ & $0.01$ & Extremely shallow decline of $K(r)$; $K$ nearly constant radially \\
\hline
Overall &-& Physically consistent \\
\hline
\end{tabular}
\label{EVALUATIONNGC4157}
\end{table}

\subsubsection{The Galaxy UGC01281}

\begin{figure}[h!]
\centering
\includegraphics[width=20pc]{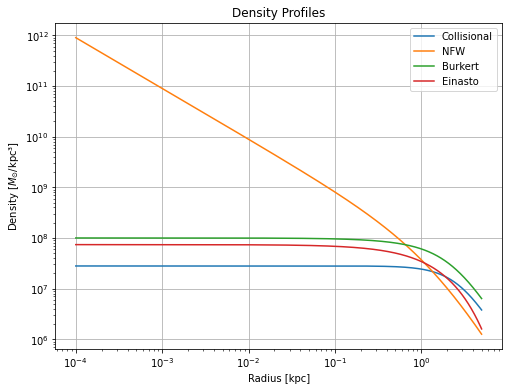}
\caption{The density of the collisional DM model (\ref{tanhmodel})
for the galaxy UGC01281, as a function of the radius.}
\label{UGC01281dens}
\end{figure}

\begin{figure}[h!]
\centering
\includegraphics[width=20pc]{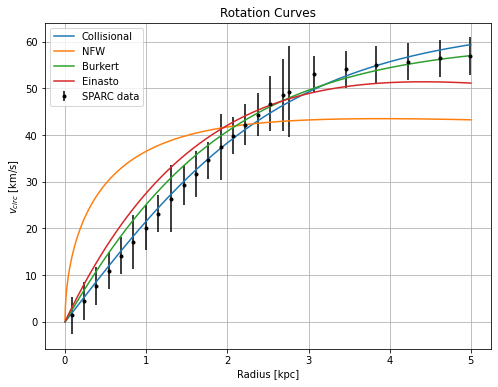}
\caption{The predicted rotation curves after using an optimization
for the collisional DM model (\ref{tanhmodel}), versus the SPARC
observational data for the galaxy UGC01281. We also plotted the
optimized curves for the NFW model, the Burkert model and the
Einasto model.} \label{UGC01281}
\end{figure}

\begin{figure}[h!]
\centering
\includegraphics[width=20pc]{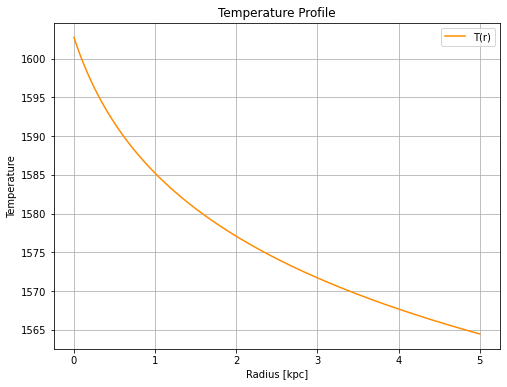}
\caption{The temperature as a function of the radius for the
collisional DM model (\ref{tanhmodel}) for the galaxy UGC01281.}
\label{UGC01281temp}
\end{figure}

For this galaxy, we shall choose $\rho_0=2.8\times
10^7$$M_{\odot}/\mathrm{Kpc}^{3}$. UGC01281 is a nearby, nearly
edge-on dwarf/late-type galaxy in Triangulum with a distance of
order $D\sim 5\text{-}6\ \mathrm{Mpc}$. In Figs.
\ref{UGC01281dens}, \ref{UGC01281} and \ref{UGC01281temp} we
present the density of the collisional DM model, the predicted
rotation curves after using an optimization for the collisional DM
model (\ref{tanhmodel}), versus the SPARC observational data and
the temperature parameter as a function of the radius
respectively. As it can be seen, the SIDM model produces viable
rotation curves compatible with the SPARC data. Also in Tables
\ref{collUGC01281}, \ref{NavaroUGC01281}, \ref{BuckertUGC01281}
and \ref{EinastoUGC01281} we present the optimization values for
the SIDM model, and the other DM profiles. Also in Table
\ref{EVALUATIONUGC01281} we present the overall evaluation of the
SIDM model for the galaxy at hand. The resulting phenomenology is
viable.
\begin{table}[h!]
  \begin{center}
    \caption{Collisional Dark Matter Optimization Values}
    \label{collUGC01281}
     \begin{tabular}{|r|r|}
     \hline
      \textbf{Parameter}   & \textbf{Optimization Values}
      \\  \hline
     $\delta_{\gamma} $ & 0.0000000012
\\  \hline
$\gamma_0 $ & 1.0001  \\ \hline $K_0$ ($M_{\odot} \,
\mathrm{Kpc}^{-3} \, (\mathrm{km/s})^{2}$)& 1600  \\ \hline
    \end{tabular}
  \end{center}
\end{table}
\begin{table}[h!]
  \begin{center}
    \caption{NFW  Optimization Values}
    \label{NavaroUGC01281}
     \begin{tabular}{|r|r|}
     \hline
      \textbf{Parameter}   & \textbf{Optimization Values}
      \\  \hline
   $\rho_s$   & $5\times 10^7$
\\  \hline
$r_s$&  1.80
\\  \hline
    \end{tabular}
  \end{center}
\end{table}
\begin{table}[h!]
  \begin{center}
    \caption{Burkert Optimization Values}
    \label{BuckertUGC01281}
     \begin{tabular}{|r|r|}
     \hline
      \textbf{Parameter}   & \textbf{Optimization Values}
      \\  \hline
     $\rho_0^B$  & $1\times 10^8$
\\  \hline
$r_0$&  2.45
\\  \hline
    \end{tabular}
  \end{center}
\end{table}
\begin{table}[h!]
  \begin{center}
    \caption{Einasto Optimization Values}
    \label{EinastoUGC01281}
    \begin{tabular}{|r|r|}
     \hline
      \textbf{Parameter}   & \textbf{Optimization Values}
      \\  \hline
     $\rho_e$  &$1\times 10^7$
\\  \hline
$r_e$ & 2.61
\\  \hline
$n_e$ & 1
\\  \hline
    \end{tabular}
  \end{center}
\end{table}
\begin{table}[h!]
\centering \caption{Physical assessment of collisional DM
parameters for UGC01281.}
\begin{tabular}{lcc}
\hline
Parameter & Value & Physical Verdict \\
\hline
$\gamma_0$ & 1.0001 & Practically isothermal \\
$\delta_\gamma$ & $1.2\times10^{-9}$ & Negligible variation \\
$r_\gamma$ & 1.5 Kpc & Transition radius \\
$K_0$ & $1.6\times10^3$ & Moderate entropy scale \\
$r_c$ & 0.5 Kpc & Small core scale, physically plausible \\
$p$ & 0.01 & Extremely shallow K(r) decrease, nearly constant across halo \\
\hline
Overall &-& Physically simple and nearly isothermal \\
\hline
\end{tabular}
\label{EVALUATIONUGC01281}
\end{table}

\subsubsection{The Galaxy UGC04483}

\begin{figure}[h!]
\centering
\includegraphics[width=20pc]{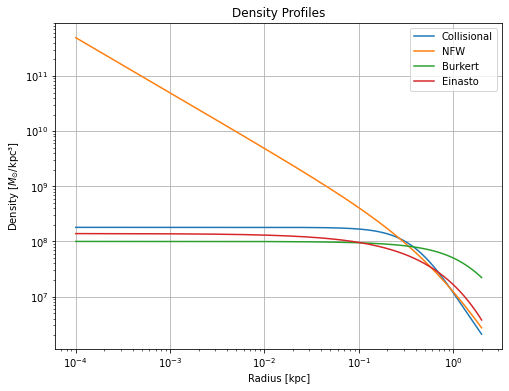}
\caption{The density of the collisional DM model (\ref{tanhmodel})
for the galaxy UGC04483, as a function of the radius.}
\label{UGC04483dens}
\end{figure}
\begin{figure}[h!]
\centering
\includegraphics[width=20pc]{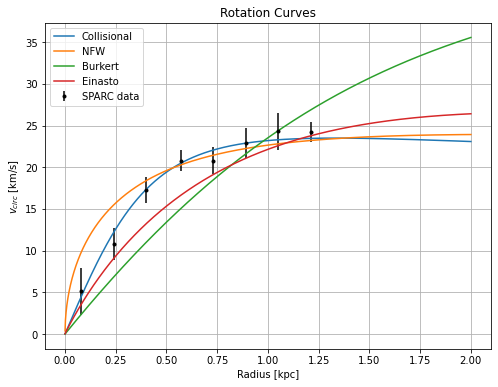}
\caption{The predicted rotation curves after using an optimization
for the collisional DM model (\ref{tanhmodel}), versus the SPARC
observational data for the galaxy UGC04483. We also plotted the
optimized curves for the NFW model, the Burkert model and the
Einasto model.} \label{UGC04483}
\end{figure}
\begin{figure}[h!]
\centering
\includegraphics[width=20pc]{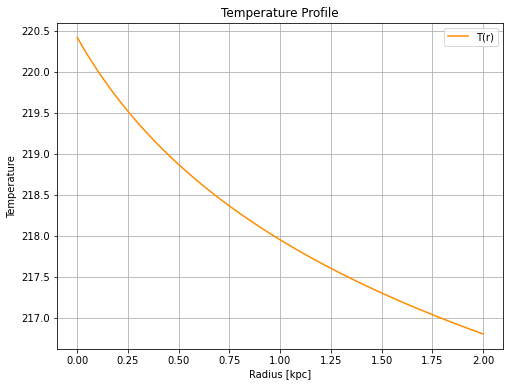}
\caption{The temperature as a function of the radius for the
collisional DM model (\ref{tanhmodel}) for the galaxy UGC04483.}
\label{UGC04483temp}
\end{figure}

For this galaxy, we shall choose $\rho_0=1.6\times
10^8$$M_{\odot}/\mathrm{Kpc}^{3}$. UGC04483 is a nearby
blue-compact/dwarf irregular galaxy in the M81/NGC2403 complex,
characterized by a compact, high-surface-brightness starburst
region and an extended, low-surface-brightness stellar body at a
distance $\sim3.2$--$3.4\,$Mpc. In Figs. \ref{UGC04483dens},
\ref{UGC04483} and \ref{UGC04483temp} we present the density of
the collisional DM model, the predicted rotation curves after
using an optimization for the collisional DM model
(\ref{tanhmodel}), versus the SPARC observational data and the
temperature parameter as a function of the radius respectively. As
it can be seen, the SIDM model produces viable rotation curves
compatible with the SPARC data. Also in Tables \ref{collUGC04483},
\ref{NavaroUGC04483}, \ref{BuckertUGC04483} and
\ref{EinastoUGC04483} we present the optimization values for the
SIDM model, and the other DM profiles. Also in Table
\ref{EVALUATIONUGC04483} we present the overall evaluation of the
SIDM model for the galaxy at hand. The resulting phenomenology is
viable.
\begin{table}[h!]
  \begin{center}
    \caption{Collisional Dark Matter Optimization Values}
    \label{collUGC04483}
     \begin{tabular}{|r|r|}
     \hline
      \textbf{Parameter}   & \textbf{Optimization Values}
      \\  \hline
     $\delta_{\gamma} $ & 0.00000012
\\  \hline
$\gamma_0 $ & 1.0001  \\ \hline $K_0$ ($M_{\odot} \,
\mathrm{Kpc}^{-3} \, (\mathrm{km/s})^{2}$)& 220  \\ \hline
    \end{tabular}
  \end{center}
\end{table}
\begin{table}[h!]
  \begin{center}
    \caption{NFW  Optimization Values}
    \label{NavaroUGC04483}
     \begin{tabular}{|r|r|}
     \hline
      \textbf{Parameter}   & \textbf{Optimization Values}
      \\  \hline
   $\rho_s$   & $5\times 10^7$
\\  \hline
$r_s$&  0.99
\\  \hline
    \end{tabular}
  \end{center}
\end{table}
\begin{table}[h!]
  \begin{center}
    \caption{Burkert Optimization Values}
    \label{BuckertUGC04483}
     \begin{tabular}{|r|r|}
     \hline
      \textbf{Parameter}   & \textbf{Optimization Values}
      \\  \hline
     $\rho_0^B$  & $1\times 10^8$
\\  \hline
$r_0$&  1.85
\\  \hline
    \end{tabular}
  \end{center}
\end{table}
\begin{table}[h!]
  \begin{center}
    \caption{Einasto Optimization Values}
    \label{EinastoUGC04483}
    \begin{tabular}{|r|r|}
     \hline
      \textbf{Parameter}   & \textbf{Optimization Values}
      \\  \hline
     $\rho_e$  &$1\times 10^7$
\\  \hline
$r_e$ & 1.32
\\  \hline
$n_e$ & 0.76
\\  \hline
    \end{tabular}
  \end{center}
\end{table}
\begin{table}[h!]
\centering \caption{Physical assessment of collisional DM
parameters for UGC04483.}
\begin{tabular}{lcc}
\hline
Parameter & Value & Physical Verdict \\
\hline
$\gamma_0$ & $1.0001$ & Almost exactly isothermal \\
$\delta_\gamma$ & $1.2\times10^{-9}$ & Practically zero \\
$r_\gamma$ & $1.5~\mathrm{Kpc}$ & Plausible transition scale \\
$K_0$ & $2.2\times10^{2}$ & Moderately medium pressure support \\
$r_c$ & $0.5~\mathrm{Kpc}$ & Small core scale \\
$p$ & $0.01$ & Extremely shallow decline \\
\hline
Overall & --- & Nearly isothermal \\
\hline
\end{tabular}
\label{EVALUATIONUGC04483}
\end{table}

\subsubsection{The Galaxy UGC05750}

\begin{figure}[h!]
\centering
\includegraphics[width=20pc]{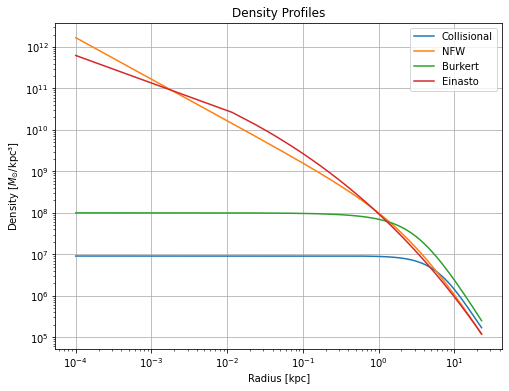}
\caption{The density of the collisional DM model (\ref{tanhmodel})
for the galaxy UGC05750, as a function of the radius.}
\label{UGC05750dens}
\end{figure}
\begin{figure}[h!]
\centering
\includegraphics[width=20pc]{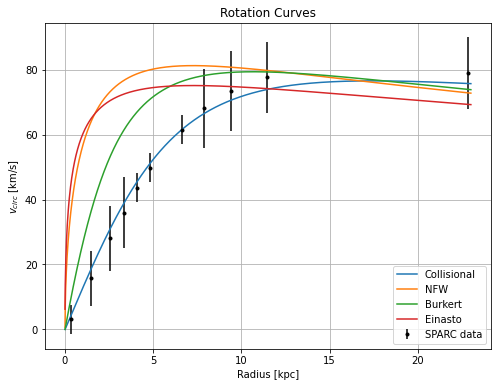}
\caption{The predicted rotation curves after using an optimization
for the collisional DM model (\ref{tanhmodel}), versus the SPARC
observational data for the galaxy UGC05750. We also plotted the
optimized curves for the NFW model, the Burkert model and the
Einasto model.} \label{UGC05750}
\end{figure}
\begin{figure}[h!]
\centering
\includegraphics[width=20pc]{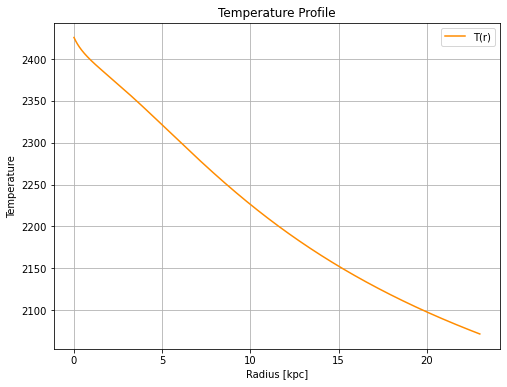}
\caption{The temperature as a function of the radius for the
collisional DM model (\ref{tanhmodel}) for the galaxy UGC05750.}
\label{UGC05750temp}
\end{figure}

For this galaxy, we shall choose $\rho_0=9.6\times
10^6$$M_{\odot}/\mathrm{Kpc}^{3}$. UGC05750 is a late-type,
low-surface-brightness dwarf/irregular disk galaxy with a
slowly-rising rotation curve. Its distance is of the order  \(\sim
2\times10^{1}\) Mpc. In Figs. \ref{UGC05750dens}, \ref{UGC05750}
and \ref{UGC05750temp} we present the density of the collisional
DM model, the predicted rotation curves after using an
optimization for the collisional DM model (\ref{tanhmodel}),
versus the SPARC observational data and the temperature parameter
as a function of the radius respectively. As it can be seen, the
SIDM model produces viable rotation curves compatible with the
SPARC data. Also in Tables \ref{collUGC05750},
\ref{NavaroUGC05750}, \ref{BuckertUGC05750} and
\ref{EinastoUGC05750} we present the optimization values for the
SIDM model, and the other DM profiles. Also in Table
\ref{EVALUATIONUGC05750} we present the overall evaluation of the
SIDM model for the galaxy at hand. The resulting phenomenology is
viable.
\begin{table}[h!]
  \begin{center}
    \caption{Collisional Dark Matter Optimization Values}
    \label{collUGC05750}
     \begin{tabular}{|r|r|}
     \hline
      \textbf{Parameter}   & \textbf{Optimization Values}
      \\  \hline
     $\delta_{\gamma} $ & 0.0000000012
\\  \hline
$\gamma_0 $ & 1.0001 \\ \hline $K_0$ ($M_{\odot} \,
\mathrm{Kpc}^{-3} \, (\mathrm{km/s})^{2}$)& 1800  \\ \hline
    \end{tabular}
  \end{center}
\end{table}
\begin{table}[h!]
  \begin{center}
    \caption{NFW  Optimization Values}
    \label{NavaroUGC05750}
     \begin{tabular}{|r|r|}
     \hline
      \textbf{Parameter}   & \textbf{Optimization Values}
      \\  \hline
   $\rho_s$  &  $5\times 10^7$
\\  \hline
$r_s$&  3.36
\\  \hline
    \end{tabular}
  \end{center}
\end{table}
\begin{table}[h!]
  \begin{center}
    \caption{Burkert Optimization Values}
    \label{BuckertUGC05750}
     \begin{tabular}{|r|r|}
     \hline
      \textbf{Parameter}   & \textbf{Optimization Values}
      \\  \hline
     $\rho_0^B$  & $1\times 10^8$
\\  \hline
$r_0$&  3.29
\\  \hline
    \end{tabular}
  \end{center}
\end{table}
\begin{table}[h!]
  \begin{center}
    \caption{Einasto Optimization Values}
    \label{EinastoUGC05750}
    \begin{tabular}{|r|r|}
     \hline
      \textbf{Parameter}   & \textbf{Optimization Values}
      \\  \hline
     $\rho_e$  &$1\times 10^7$
\\  \hline
$r_e$ & 3.23
\\  \hline
$n_e$ & 0.14
\\  \hline
    \end{tabular}
  \end{center}
\end{table}
\begin{table}[h!]
\centering \caption{Physical assessment of collisional DM
parameters for UGC05750.}
\begin{tabular}{lcc}
\hline
Parameter & Value & Physical Verdict \\
\hline
$\gamma_0$ & $1.0001$ & Slightly above isothermal \\
$\delta_\gamma$ & $1.2\times10^{-9}$ & Negligible variation \\
$r_\gamma$ & $1.5\ \mathrm{Kpc}$ & Transition radius inside inner halo \\
$K_0$ ($M_{\odot}\,\mathrm{Kpc}^{-3}\,(\mathrm{km/s})^{2}$) & $1.8\times10^{3}$ & Moderately high pressure support \\
$r_c$ & $0.5\ \mathrm{Kpc}$ & Small core scale  \\
$p$ & $0.01$ & Very shallow radial decline of $K(r)$ \\
Overall & -- & Model effectively reduces to an almost-isothermal\\
\hline
\end{tabular}
\label{EVALUATIONUGC05750}
\end{table}

\subsubsection{The Galaxy UGC08837 Viable}

\begin{figure}[h!]
\centering
\includegraphics[width=20pc]{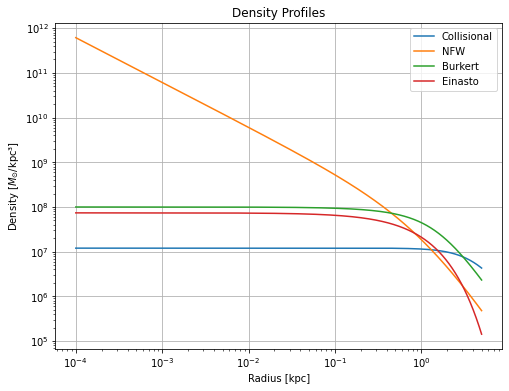}
\caption{The density of the collisional DM model (\ref{tanhmodel})
for the galaxy UGC08837, as a function of the radius.}
\label{UGC08837dens}
\end{figure}

\begin{figure}[h!]
\centering
\includegraphics[width=20pc]{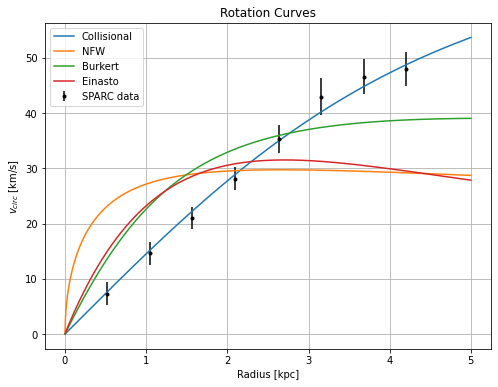}
\caption{The predicted rotation curves after using an optimization
for the collisional DM model (\ref{tanhmodel}), versus the SPARC
observational data for the galaxy UGC08837. We also plotted the
optimized curves for the NFW model, the Burkert model and the
Einasto model.} \label{UGC08837}
\end{figure}

\begin{figure}[h!]
\centering
\includegraphics[width=20pc]{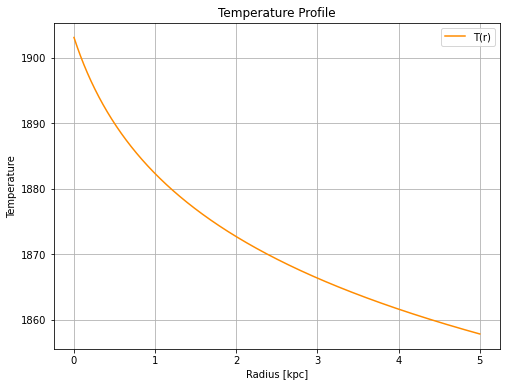}
\caption{The temperature as a function of the radius for the
collisional DM model (\ref{tanhmodel}) for the galaxy UGC08837.}
\label{UGC08837temp}
\end{figure}

For this galaxy, we shall choose $\rho_0=1.2\times
10^7$$M_{\odot}/\mathrm{Kpc}^{3}$. UGC08837, is a dwarf irregular
galaxy (type IB(s)m) located at a distance of about $7.4
\mathrm{Mpc}$. In Figs. \ref{UGC08837dens}, \ref{UGC08837} and
\ref{UGC08837temp} we present the density of the collisional DM
model, the predicted rotation curves after using an optimization
for the collisional DM model (\ref{tanhmodel}), versus the SPARC
observational data and the temperature parameter as a function of
the radius respectively. As it can be seen, the SIDM model
produces viable rotation curves compatible with the SPARC data.
Also in Tables \ref{collUGC08837}, \ref{NavaroUGC08837},
\ref{BuckertUGC08837} and \ref{EinastoUGC08837} we present the
optimization values for the SIDM model, and the other DM profiles.
Also in Table \ref{EVALUATIONUGC08837} we present the overall
evaluation of the SIDM model for the galaxy at hand. The resulting
phenomenology is viable.
\begin{table}[h!]
  \begin{center}
    \caption{Collisional Dark Matter Optimization Values}
    \label{collUGC08837}
     \begin{tabular}{|r|r|}
     \hline
      \textbf{Parameter}   & \textbf{Optimization Values}
      \\  \hline
     $\delta_{\gamma} $ & 0.0000000012
\\  \hline
$\gamma_0 $ & 1.0001 \\ \hline $K_0$ ($M_{\odot} \,
\mathrm{Kpc}^{-3} \, (\mathrm{km/s})^{2}$)& 1900 \\ \hline
    \end{tabular}
  \end{center}
\end{table}
\begin{table}[h!]
  \begin{center}
    \caption{NFW  Optimization Values}
    \label{NavaroUGC08837}
     \begin{tabular}{|r|r|}
     \hline
      \textbf{Parameter}   & \textbf{Optimization Values}
      \\  \hline
   $\rho_s$   & $5\times 10^7$
\\  \hline
$r_s$&  1.23
\\  \hline
    \end{tabular}
  \end{center}
\end{table}
\begin{table}[h!]
  \begin{center}
    \caption{Burkert Optimization Values}
    \label{BuckertUGC08837}
     \begin{tabular}{|r|r|}
     \hline
      \textbf{Parameter}   & \textbf{Optimization Values}
      \\  \hline
     $\rho_0^B$  & $1\times 10^8$
\\  \hline
$r_0$&  1.62
\\  \hline
    \end{tabular}
  \end{center}
\end{table}
\begin{table}[h!]
  \begin{center}
    \caption{Einasto Optimization Values}
    \label{EinastoUGC08837}
    \begin{tabular}{|r|r|}
     \hline
      \textbf{Parameter}   & \textbf{Optimization Values}
      \\  \hline
     $\rho_e$  &$1\times 10^7$
\\  \hline
$r_e$ & 1.60
\\  \hline
$n_e$ & 1
\\  \hline
    \end{tabular}
  \end{center}
\end{table}
\begin{table}[h!]
\centering \caption{Physical assessment of collisional DM
parameters for UGC08837.}
\begin{tabular}{lcc}
\hline
Parameter & Value   & Physical Verdict \\
\hline
$\gamma_0$ & $1.0001$ & Practically isothermal\\
$\delta_\gamma$ & $1.2\times10^{-9}$ & Negligible  \\
$r_\gamma$ & $1.5\ \mathrm{Kpc}$ & Transition radius irrelevant \\
$K_0$ & $1.9\times10^{3}$ & Pressure support acceptable, high  \\
$r_c$ & $0.5\ \mathrm{Kpc}$ & Small core scale  \\
$p$ & $0.01$ & Very shallow decline \\
\hline
Overall &-& Numerically stable and physically plausible for a low-mass/isothermal halo \\
\hline
\end{tabular}
\label{EVALUATIONUGC08837}
\end{table}

\subsection{Analysis and Simulation of SPARC Galaxy Data and Fitting with two-parameter Simple SIDM Model: A Sample of Marginally Viable Galaxies}

\subsubsection{The Galaxy UGC03546 Marginally Late-time Spiral
Large Initial Density}

\begin{figure}[h!]
\centering
\includegraphics[width=20pc]{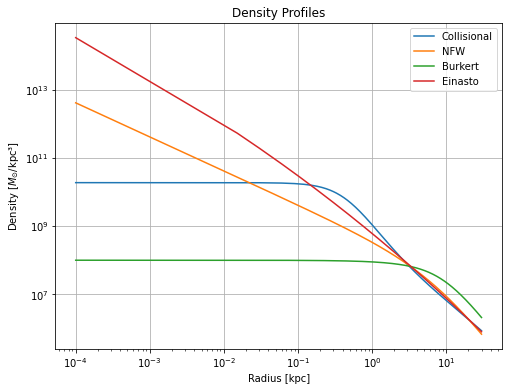}
\caption{The density of the collisional DM model (\ref{tanhmodel})
for the galaxy UGC03546, as a function of the radius.}
\label{UGC03546dens}
\end{figure}

\begin{figure}[h!]
\centering
\includegraphics[width=20pc]{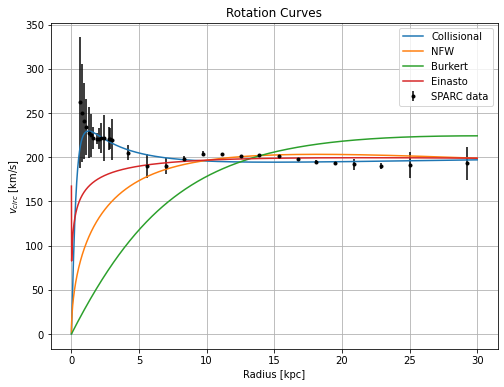}
\caption{The predicted rotation curves after using an optimization
for the collisional DM model (\ref{tanhmodel}), versus the SPARC
observational data for the galaxy UGC03546. We also plotted the
optimized curves for the NFW model, the Burkert model and the
Einasto model.} \label{UGC03546}
\end{figure}

\begin{figure}[h!]
\centering
\includegraphics[width=20pc]{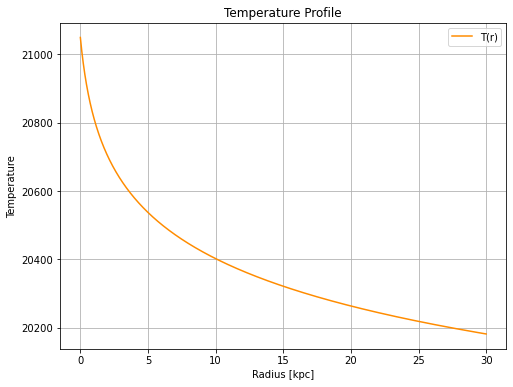}
\caption{The temperature as a function of the radius for the
collisional DM model (\ref{tanhmodel}) for the galaxy UGC03546.}
\label{UGC03546temp}
\end{figure}

\begin{figure}[h!]
\centering
\includegraphics[width=20pc]{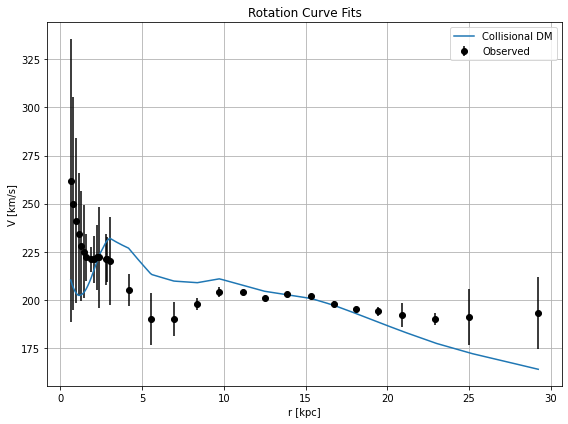}
\caption{The predicted rotation curves after using an optimization
for the collisional DM model (\ref{tanhmodel}), versus the
extended SPARC observational data for the galaxy UGC03546. The
model includes the rotation curves from all the components of the
galaxy, including gas and disk velocities, along with the
collisional DM model.} \label{extendedUGC03546}
\end{figure}

For this galaxy, we shall choose $\rho_0=1.9\times
10^{10}$$M_{\odot}/\mathrm{Kpc}^{3}$. UGC03546 (also known as
NGC2273) is a barred spiral galaxy of type SB(rs)b, located in the
constellation Lynx. It lies at a distance of approximately \(28.5\
\mathrm{Mpc}\). In Figs. \ref{UGC03546dens}, \ref{UGC03546} and
\ref{UGC03546temp} we present the density of the collisional DM
model, the predicted rotation curves after using an optimization
for the collisional DM model (\ref{tanhmodel}), versus the SPARC
observational data and the temperature parameter as a function of
the radius respectively. As it can be seen, the SIDM model
produces marginally viable rotation curves compatible with the
SPARC data. Also in Tables \ref{collUGC03546},
\ref{NavaroUGC03546}, \ref{BuckertUGC03546} and
\ref{EinastoUGC03546} we present the optimization values for the
SIDM model, and the other DM profiles. Also in Table
\ref{EVALUATIONUGC03546} we present the overall evaluation of the
SIDM model for the galaxy at hand. The resulting phenomenology is
marginally viable.
\begin{table}[h!]
  \begin{center}
    \caption{Collisional Dark Matter Optimization Values}
    \label{collUGC03546}
     \begin{tabular}{|r|r|}
     \hline
      \textbf{Parameter}   & \textbf{Optimization Values}
      \\  \hline
     $\delta_{\gamma} $ & 0.0000000012
\\  \hline
$\gamma_0 $ & 1.0001 \\ \hline $K_0$ ($M_{\odot} \,
\mathrm{Kpc}^{-3} \, (\mathrm{km/s})^{2}$)& 21000  \\ \hline
    \end{tabular}
  \end{center}
\end{table}
\begin{table}[h!]
  \begin{center}
    \caption{NFW  Optimization Values}
    \label{NavaroUGC03546}
     \begin{tabular}{|r|r|}
     \hline
      \textbf{Parameter}   & \textbf{Optimization Values}
      \\  \hline
   $\rho_s$   & $5\times 10^7$
\\  \hline
$r_s$&  8.41
\\  \hline
    \end{tabular}
  \end{center}
\end{table}
\begin{table}[h!]
  \begin{center}
    \caption{Burkert Optimization Values}
    \label{BuckertUGC03546}
     \begin{tabular}{|r|r|}
     \hline
      \textbf{Parameter}   & \textbf{Optimization Values}
      \\  \hline
     $\rho_0^B$  & $1\times 10^8$
\\  \hline
$r_0$&  9.30
\\  \hline
    \end{tabular}
  \end{center}
\end{table}
\begin{table}[h!]
  \begin{center}
    \caption{Einasto Optimization Values}
    \label{EinastoUGC03546}
    \begin{tabular}{|r|r|}
     \hline
      \textbf{Parameter}   & \textbf{Optimization Values}
      \\  \hline
     $\rho_e$  &$1\times 10^7$
\\  \hline
$r_e$ & 8.76
\\  \hline
$n_e$ & 0.05
\\  \hline
    \end{tabular}
  \end{center}
\end{table}
\begin{table}[h!]
\centering \caption{Physical assessment of collisional DM
parameters for UGC03546.}
\begin{tabular}{lcc}
\hline
Parameter & Value & Physical Verdict \\
\hline
$\gamma_0$ & $1.0001$ & Almost exactly isothermal  \\
$\delta_\gamma$ & $1.2\times10^{-9}$ & Practically zero   \\
$r_\gamma$ & $1.5~\mathrm{Kpc}$ & Reasonable transition scale \\
$K_0$ & $2.1\times10^{4}$ & Large pressure support \\
$r_c$ & $0.5~\mathrm{Kpc}$ & Small core scale; consistent with near-constant $K(r)$ \\
$p$ & $0.01$ & Extremely shallow decline \\
\hline
Overall &-& Nearly isothermal \\
\hline
\end{tabular}
\label{EVALUATIONUGC03546}
\end{table}
Now the extended picture including the rotation velocity from the
other components of the galaxy, such as the disk and gas, makes
the collisional DM model viable for this galaxy. In Fig.
\ref{extendedUGC03546} we present the combined rotation curves
including the other components of the galaxy along with the
collisional matter. As it can be seen, the extended collisional DM
model is marginally viable.
Also in Table \ref{evaluationextendedUGC03546} we present the
values of the free parameters of the collisional DM model for
which the maximum compatibility with the SPARC data comes for the
galaxy UGC03546.
\begin{table}[h!]
\centering \caption{Physical assessment of Extended collisional DM
parameters for galaxy UGC03546.}
\begin{tabular}{lcc}
\hline
Parameter & Value & Physical Verdict \\
\hline
$\gamma_0$ & 1.05008017 & Slightly above isothermal  \\
$\delta_\gamma$ & 0.0001 & Extremely small variation  across radii \\
$K_0$ & 3000 & Moderate entropy   \\
$ml_{disk}$ & 0.95463856 & High but physically plausible disk M/L \\
$ml_{bulge}$ & 0.5 & Moderately large bulge M/L \\
\hline
Overall &-& Physically plausible \\
\hline
\end{tabular}
\label{evaluationextendedUGC03546}
\end{table}

\subsubsection{The Galaxy NGC1090 Marginally Viable Large Spiral,
Extended Viable}

\begin{figure}[h!]
\centering
\includegraphics[width=20pc]{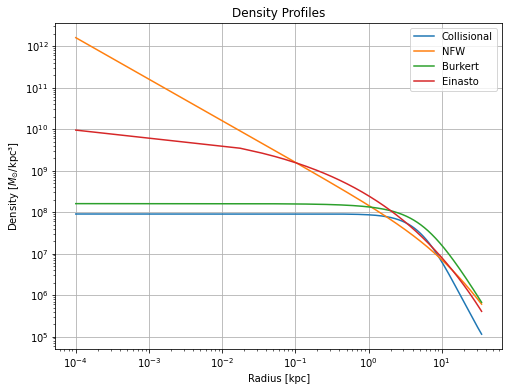}
\caption{The density of the collisional DM model (\ref{tanhmodel})
for the galaxy NGC1090, as a function of the radius.}
\label{NGC1090dens}
\end{figure}

\begin{figure}[h!]
\centering
\includegraphics[width=20pc]{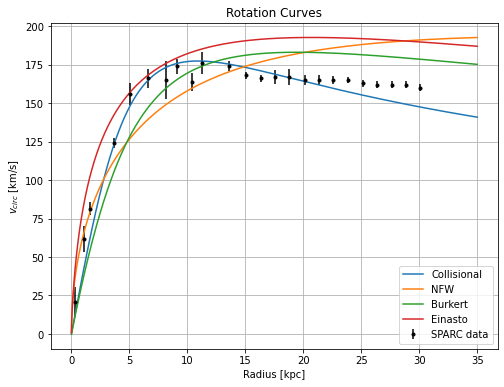}
\caption{The predicted rotation curves after using an optimization
for the collisional DM model (\ref{tanhmodel}), versus the SPARC
observational data for the galaxy NGC1090. We also plotted the
optimized curves for the NFW model, the Burkert model and the
Einasto model.} \label{NGC1090}
\end{figure}

\begin{figure}[h!]
\centering
\includegraphics[width=20pc]{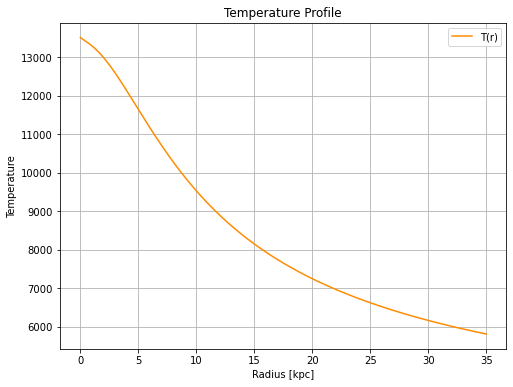}
\caption{The temperature as a function of the radius for the
collisional DM model (\ref{tanhmodel}) for the galaxy NGC1090.}
\label{NGC1090temp}
\end{figure}

\begin{figure}[h!]
\centering
\includegraphics[width=20pc]{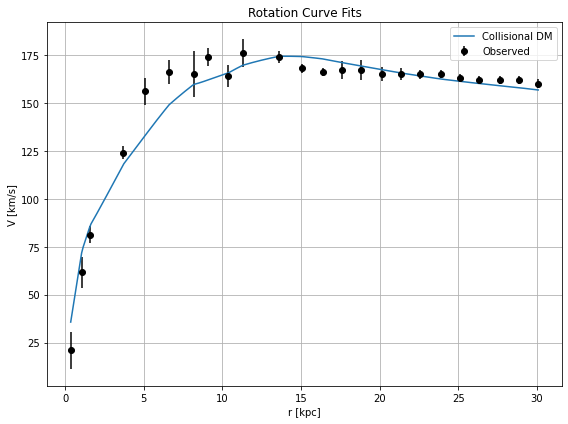}
\caption{The predicted rotation curves after using an optimization
for the collisional DM model (\ref{tanhmodel}), versus the
extended SPARC observational data for the galaxy NGC1090. The
model includes the rotation curves from all the components of the
galaxy, including gas and disk velocities, along with the
collisional DM model.} \label{extendedNGC1090}
\end{figure}

For this galaxy, we shall choose $\rho_0=1.4\times
10^8$$M_{\odot}/\mathrm{Kpc}^{3}$. NGC\,1090 is a barred spiral
galaxy, type $\mathrm{SB(rs)bc}$. Its distance from Milky Way is
about $D \sim 38\;\mathrm{Mpc}$. In Figs. \ref{NGC1090dens},
\ref{NGC1090} and \ref{NGC1090temp} we present the density of the
collisional DM model, the predicted rotation curves after using an
optimization for the collisional DM model (\ref{tanhmodel}),
versus the SPARC observational data and the temperature parameter
as a function of the radius respectively. As it can be seen, the
SIDM model produces viable rotation curves marginally compatible
with the SPARC data. Also in Tables \ref{collNGC1090},
\ref{NavaroNGC1090}, \ref{BuckertNGC1090} and \ref{EinastoNGC1090}
we present the optimization values for the SIDM model, and the
other DM profiles. Also in Table \ref{EVALUATIONNGC1090} we
present the overall evaluation of the SIDM model for the galaxy at
hand. The resulting phenomenology is  marginally viable.
\begin{table}[h!]
  \begin{center}
    \caption{Collisional Dark Matter Optimization Values}
    \label{collNGC1090}
     \begin{tabular}{|r|r|}
     \hline
      \textbf{Parameter}   & \textbf{Optimization Values}
      \\  \hline
     $\delta_{\gamma} $ & 0.0000000012
\\  \hline
$\gamma_0 $ & 1.0001 \\ \hline $K_0$ ($M_{\odot} \,
\mathrm{Kpc}^{-3} \, (\mathrm{km/s})^{2}$)& 12000 \\ \hline
    \end{tabular}
  \end{center}
\end{table}
\begin{table}[h!]
  \begin{center}
    \caption{NFW  Optimization Values}
    \label{NavaroNGC1090}
     \begin{tabular}{|r|r|}
     \hline
      \textbf{Parameter}   & \textbf{Optimization Values}
      \\  \hline
   $\rho_s$   & $0.008\times 10^9$
\\  \hline
$r_s$&  20
\\  \hline
    \end{tabular}
  \end{center}
\end{table}
\begin{table}[h!]
  \begin{center}
    \caption{Burkert Optimization Values}
    \label{BuckertNGC1090}
     \begin{tabular}{|r|r|}
     \hline
      \textbf{Parameter}   & \textbf{Optimization Values}
      \\  \hline
     $\rho_0^B$  & $0.16\times 10^9$
\\  \hline
$r_0$&  6
\\  \hline
    \end{tabular}
  \end{center}
\end{table}
\begin{table}[h!]
  \begin{center}
    \caption{Einasto Optimization Values}
    \label{EinastoNGC1090}
    \begin{tabular}{|r|r|}
     \hline
      \textbf{Parameter}   & \textbf{Optimization Values}
      \\  \hline
     $\rho_e$  & $0.008\times 10^9$
\\  \hline
$r_e$ & 10
\\  \hline
$n_e$ & 0.27
\\  \hline
    \end{tabular}
  \end{center}
\end{table}
\begin{table}[h!]
\centering \caption{Physical assessment of collisional DM
parameters for NGC1090.}
\begin{tabular}{lcc}
\hline
Parameter & Value & Physical Verdict \\
\hline
$\gamma_0$ & $1.0001$ & Effectively isothermal; $P \sim \rho$ in inner halo \\
$\delta_\gamma$ & $1.2\times10^{-9}$ & Negligible; $\gamma(r)$ constant \\
$r_\gamma$ & $1.5\ \mathrm{Kpc}$ & Transition radius irrelevant \\
$K_0$ & $1.2\times10^4$ & Central temperature high \\
$r_c$ & $0.5\ \mathrm{Kpc}$ & Small core; reasonable for inner halo \\
$p$ & $0.01$ & Nearly constant entropy; minimal radial variation \\
\hline
Overall &-& Physically plausible; inner halo nearly isothermal, moderate central density \\
\hline
\end{tabular}
\label{EVALUATIONNGC1090}
\end{table}
Now the extended picture including the rotation velocity from the
other components of the galaxy, such as the disk and gas, makes
the collisional DM model viable for this galaxy. In Fig.
\ref{extendedNGC1090} we present the combined rotation curves
including the other components of the galaxy along with the
collisional matter. As it can be seen, the extended collisional DM
model is viable.
Also in Table \ref{evaluationextendedNGC1090} we present the
values of the free parameters of the collisional DM model for
which the maximum compatibility with the SPARC data comes for the
galaxy NGC1090.
\begin{table}[h!]
\centering \caption{Physical assessment of Extended collisional DM
parameters for galaxy NGC1090.}
\begin{tabular}{lcc}
\hline
Parameter & Value & Physical Verdict \\
\hline
$\gamma_0$ & 1.05586851 & Slightly above isothermal \\
$\delta_\gamma$ & 0.001 & Negligible radial variation \\
$K_0$ & 3000 & Moderate entropy scale \\
$ml_{disk}$ & 0.72364758 & Moderate disk M/L \\
$ml_{bulge}$ & 0.00000000 & No bulge contribution \\
\hline
Overall &-& Physically plausible \\
\hline
\end{tabular}
\label{evaluationextendedNGC1090}
\end{table}

\subsection{Analysis and Simulation of SPARC Galaxy Data and Fitting with two-parameter Simple SIDM Model: A Sample of Non-Viable Galaxies}

\subsubsection{The Galaxy UGC02916 Non-viable Late type Spiral,
Extended non-viable too}

\begin{figure}[h!]
\centering
\includegraphics[width=20pc]{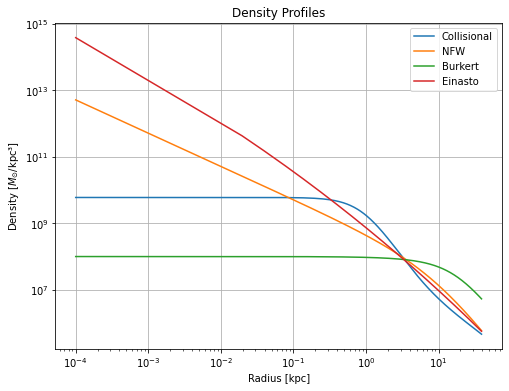}
\caption{The density of the collisional DM model (\ref{tanhmodel})
for the galaxy UGC02916, as a function of the radius.}
\label{UGC02916dens}
\end{figure}
\begin{figure}[h!]
\centering
\includegraphics[width=20pc]{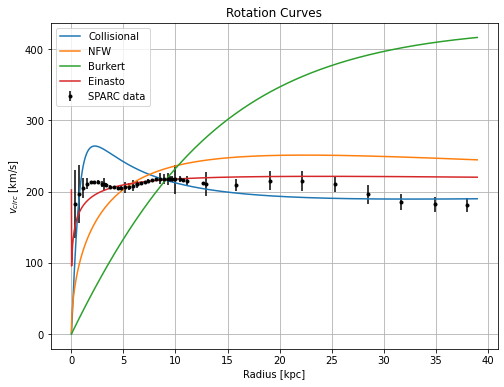}
\caption{The predicted rotation curves after using an optimization
for the collisional DM model (\ref{tanhmodel}), versus the SPARC
observational data for the galaxy UGC02916. We also plotted the
optimized curves for the NFW model, the Burkert model and the
Einasto model.} \label{UGC02916}
\end{figure}
\begin{figure}[h!]
\centering
\includegraphics[width=20pc]{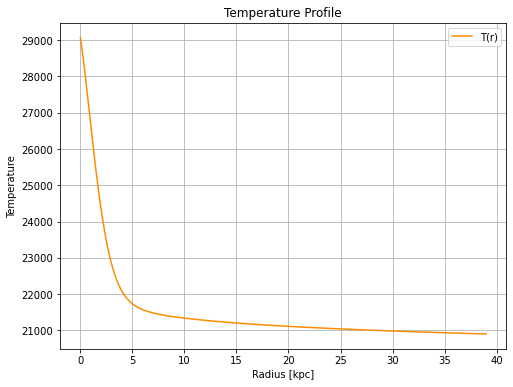}
\caption{The temperature as a function of the radius for the
collisional DM model (\ref{tanhmodel}) for the galaxy UGC02916.}
\label{UGC02916temp}
\end{figure}
\begin{figure}[h!]
\centering
\includegraphics[width=20pc]{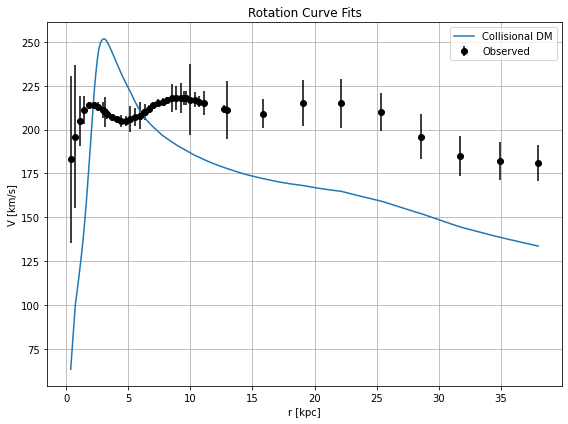}
\caption{The predicted rotation curves after using an optimization
for the collisional DM model (\ref{tanhmodel}), versus the
extended SPARC observational data for the galaxy UGC02916. The
model includes the rotation curves from all the components of the
galaxy, including gas and disk velocities, along with the
collisional DM model.} \label{extendedUGC02916}
\end{figure}

For this galaxy, we shall choose $\rho_0=0.5\times
10^7$$M_{\odot}/\mathrm{Kpc}^{3}$. UGC2916 is catalogued as a
faint Sa-Sb spiral in the constellation Camelopardalis. In Figs.
\ref{UGC02916dens}, \ref{UGC02916} and \ref{UGC02916temp} we
present the density of the collisional DM model, the predicted
rotation curves after using an optimization for the collisional DM
model (\ref{tanhmodel}), versus the SPARC observational data and
the temperature parameter as a function of the radius
respectively. As it can be seen, the SIDM model produces
non-viable rotation curves with the SPARC data. Also in Tables
\ref{collUGC02916}, \ref{NavaroUGC02916}, \ref{BuckertUGC02916}
and \ref{EinastoUGC02916} we present the optimization values for
the SIDM model, and the other DM profiles. Also in Table
\ref{EVALUATIONUGC02916} we present the overall evaluation of the
SIDM model for the galaxy at hand. The resulting phenomenology is
non-viable.
\begin{table}[h!]
  \begin{center}
    \caption{Collisional Dark Matter Optimization Values}
    \label{collUGC02916}
     \begin{tabular}{|r|r|}
     \hline
      \textbf{Parameter}   & \textbf{Optimization Values}
      \\  \hline
     $\delta_{\gamma} $ & 0.0000000012
\\  \hline
$\gamma_0 $ & 1.0001  \\ \hline $K_0$ ($M_{\odot} \,
\mathrm{Kpc}^{-3} \, (\mathrm{km/s})^{2}$)& 21000  \\ \hline
    \end{tabular}
  \end{center}
\end{table}
\begin{table}[h!]
  \begin{center}
    \caption{NFW  Optimization Values}
    \label{NavaroUGC02916}
     \begin{tabular}{|r|r|}
     \hline
      \textbf{Parameter}   & \textbf{Optimization Values}
      \\  \hline
   $\rho_s$   & $5\times 10^7$
\\  \hline
$r_s$&  10.39
\\  \hline
    \end{tabular}
  \end{center}
\end{table}
\begin{table}[h!]
  \begin{center}
    \caption{Burkert Optimization Values}
    \label{BuckertUGC02916}
     \begin{tabular}{|r|r|}
     \hline
      \textbf{Parameter}   & \textbf{Optimization Values}
      \\  \hline
     $\rho_0^B$  & $1\times 10^8$
\\  \hline
$r_0$& 17.69
\\  \hline
    \end{tabular}
  \end{center}
\end{table}
\begin{table}[h!]
  \begin{center}
    \caption{Einasto Optimization Values}
    \label{EinastoUGC02916}
    \begin{tabular}{|r|r|}
     \hline
      \textbf{Parameter}   & \textbf{Optimization Values}
      \\  \hline
     $\rho_e$  &$1\times 10^7$
\\  \hline
$r_e$ & 9.72
\\  \hline
$n_e$ & 0.05
\\  \hline
    \end{tabular}
  \end{center}
\end{table}
\begin{table}[h!]
\centering \caption{Physical assessment of collisional DM
parameters (UGC02916).}
\begin{tabular}{lcc}
\hline
Parameter & Value & Physical Verdict \\
\hline
$\gamma_0$ & $1.0001$ & Nearly isothermal; very soft EoS, favors shallow inner slope \\
$\delta_\gamma$ & $0.0000000012$ & Negligible radial variation  \\
$r_\gamma$ & $1.5\ \mathrm{Kpc}$ & Transition radius inside inner halo, but little effect due to tiny $\delta_\gamma$ \\
$K_0$ & $2.1\times10^{4}$ & Entropy large  \\
$r_c$ & $0.5\ \mathrm{Kpc}$ & Small core scale; reasonable for compact inner core \\
$p$ & $0.01$ & Practically constant $K(r)$; no significant radial entropy decline \\
\hline
Overall &-& Physically plausible as a nearly-isothermal, cored halo \\
\hline
\end{tabular}
\label{EVALUATIONUGC02916}
\end{table}
Now the extended picture including the rotation velocity from the
other components of the galaxy, such as the disk and gas, makes
the collisional DM model viable for this galaxy. In Fig.
\ref{extendedUGC02916} we present the combined rotation curves
including the other components of the galaxy along with the
collisional matter. As it can be seen, the extended collisional DM
model is non-viable.
Also in Table \ref{evaluationextendedUGC02916} we present the
values of the free parameters of the collisional DM model for
which the maximum compatibility with the SPARC data comes for the
galaxy UGC02916.
\begin{table}[h!]
\centering \caption{Physical assessment of Extended collisional DM
parameters (second set) for UGC02916.}
\begin{tabular}{lcc}
\hline
Parameter & Value & Physical Verdict \\
\hline
$\gamma_0$ & 1.29485641 & Moderately above isothermal; higher central pressure/support \\
$\delta_\gamma$ & 0.31096862 & Large variation  \\
$K_0$ & 3000 & Moderate entropy  \\
$ml_{disk}$ & 1.00000000 & Disk-dominated mass-to-light, typical for maximal-disk fits \\
$ml_{bulge}$ & 0.50000000 & Substantial bulge contribution; non-negligible central baryonic potential \\
\hline
Overall &-& Plausible \\
\hline
\end{tabular}
\label{evaluationextendedUGC02916}
\end{table}

\subsubsection{The Galaxy NGC4214, Exceptional Case Non-viable
Dwarf}

\begin{figure}[h!]
\centering
\includegraphics[width=20pc]{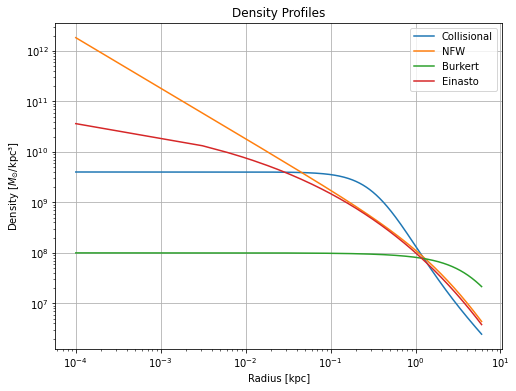}
\caption{The density of the collisional DM model (\ref{tanhmodel})
for the galaxy NGC4214, as a function of the radius.}
\label{NGC4214dens}
\end{figure}
\begin{figure}[h!]
\centering
\includegraphics[width=20pc]{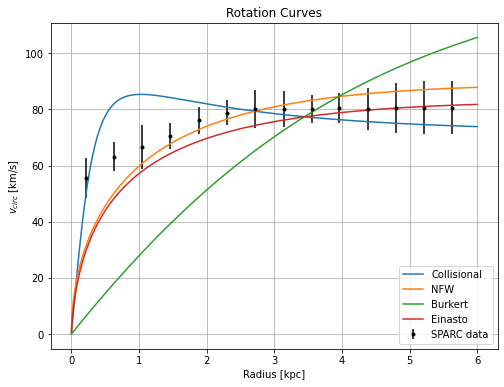}
\caption{The predicted rotation curves after using an optimization
for the collisional DM model (\ref{tanhmodel}), versus the SPARC
observational data for the galaxy NGC4214. We also plotted the
optimized curves for the NFW model, the Burkert model and the
Einasto model.} \label{NGC4214}
\end{figure}
\begin{figure}[h!]
\centering
\includegraphics[width=20pc]{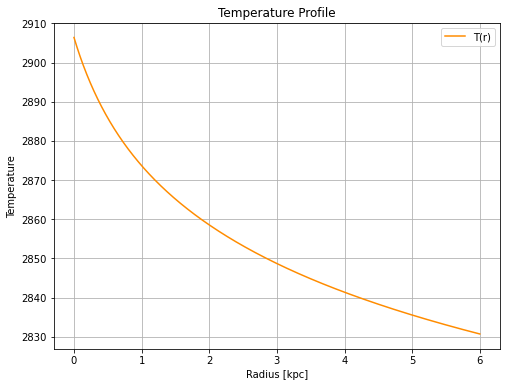}
\caption{The temperature as a function of the radius for the
collisional DM model (\ref{tanhmodel}) for the galaxy NGC4214.}
\label{NGC4214temp}
\end{figure}
\begin{figure}[h!]
\centering
\includegraphics[width=20pc]{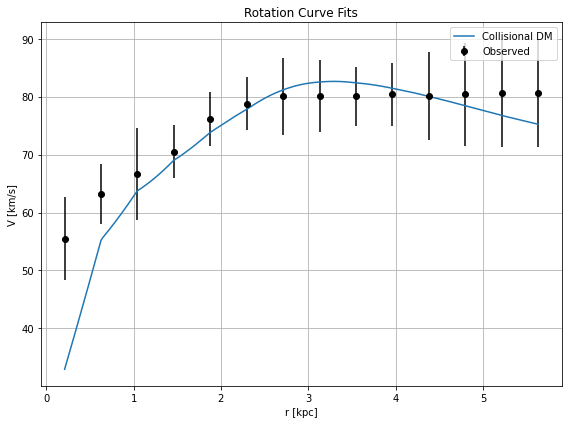}
\caption{The predicted rotation curves after using an optimization
for the collisional DM model (\ref{tanhmodel}), versus the
extended SPARC observational data for the galaxy NGC4214. The
model includes the rotation curves from all the components of the
galaxy, including gas and disk velocities, along with the
collisional DM model.} \label{extendedNGC4214}
\end{figure}

For this galaxy, we shall choose $\rho_0=4\times
10^9$$M_{\odot}/\mathrm{Kpc}^{3}$. NGC4214 is a dwarf irregular
galaxy of type IAB(s)m, located approximately \(2.98 \pm 0.25\)
Mpc from the Milky Way in the constellation Canes Venatici. It is
a member of the M94 Group and exhibits active star formation,
particularly in its central regions, which host super star
clusters rich in Wolf-Rayet stars. In Figs. \ref{NGC4214dens},
\ref{NGC4214} and \ref{NGC4214temp} we present the density of the
collisional DM model, the predicted rotation curves after using an
optimization for the collisional DM model (\ref{tanhmodel}),
versus the SPARC observational data and the temperature parameter
as a function of the radius respectively. As it can be seen, the
SIDM model produces non-viable rotation curves incompatible with
the SPARC data. Also in Tables \ref{collNGC4214},
\ref{NavaroNGC4214}, \ref{BuckertNGC4214} and \ref{EinastoNGC4214}
we present the optimization values for the SIDM model, and the
other DM profiles. Also in Table \ref{EVALUATIONNGC4214} we
present the overall evaluation of the SIDM model for the galaxy at
hand. The resulting phenomenology is non-viable.
\begin{table}[h!]
  \begin{center}
    \caption{Collisional Dark Matter Optimization Values}
    \label{collNGC4214}
     \begin{tabular}{|r|r|}
     \hline
      \textbf{Parameter}   & \textbf{Optimization Values}
      \\  \hline
     $\delta_{\gamma} $ & 0.0000000012
\\  \hline
$\gamma_0 $ &  1.0001 \\ \hline $K_0$ ($M_{\odot} \,
\mathrm{Kpc}^{-3} \, (\mathrm{km/s})^{2}$)& 2900 \\ \hline
    \end{tabular}
  \end{center}
\end{table}
\begin{table}[h!]
  \begin{center}
    \caption{NFW  Optimization Values}
    \label{NavaroNGC4214}
     \begin{tabular}{|r|r|}
     \hline
      \textbf{Parameter}   & \textbf{Optimization Values}
      \\  \hline
   $\rho_s$   & $5\times 10^7$
\\  \hline
$r_s$& 3.66
\\  \hline
    \end{tabular}
  \end{center}
\end{table}
\begin{table}[h!]
  \begin{center}
    \caption{Burkert Optimization Values}
    \label{BuckertNGC4214}
     \begin{tabular}{|r|r|}
     \hline
      \textbf{Parameter}   & \textbf{Optimization Values}
      \\  \hline
     $\rho_0^B$  & $1\times 10^8$
\\  \hline
$r_0$&  5.45
\\  \hline
    \end{tabular}
  \end{center}
\end{table}
\begin{table}[h!]
  \begin{center}
    \caption{Einasto Optimization Values}
    \label{EinastoNGC4214}
    \begin{tabular}{|r|r|}
     \hline
      \textbf{Parameter}   & \textbf{Optimization Values}
      \\  \hline
     $\rho_e$  &$1\times 10^7$
\\  \hline
$r_e$ & 3.79
\\  \hline
$n_e$ & 0.22
\\  \hline
    \end{tabular}
  \end{center}
\end{table}
\begin{table}[h!]
\centering \caption{Physical assessment of collisional DM
parameters (NGC4214).}
\begin{tabular}{lcc}
\hline
Parameter & Value & Physical Verdict \\
\hline
$\gamma_0$ & $1.0001$ & Essentially isothermal \\
$\delta_\gamma$ & $1.2\times10^{-9}$ & Effectively zero   \\
$r_\gamma$ & $1.5\ \mathrm{Kpc}$ & Reasonable transition radius  \\
$K_0$ & $2.90\times10^{3}$ & Modest pressure support \\
$r_c$ & $0.5\ \mathrm{Kpc}$ & Small core scale \\
$p$ & $0.01$ & Extremely shallow decline of $K(r)$ \\
\hline
Overall &-& Physically consistent \\
\hline
\end{tabular}
\label{EVALUATIONNGC4214}
\end{table}
Now the extended picture including the rotation velocity from the
other components of the galaxy, such as the disk and gas, makes
the collisional DM model viable for this galaxy. In Fig.
\ref{extendedNGC4214} we present the combined rotation curves
including the other components of the galaxy along with the
collisional matter. As it can be seen, the extended collisional DM
model is non-viable.
Also in Table \ref{evaluationextendedNGC4214} we present the
values of the free parameters of the collisional DM model for
which the maximum compatibility with the SPARC data comes for the
galaxy NGC4214.
\begin{table}[h!]
\centering \caption{Physical assessment of Extended collisional DM
parameters for NGC4214.}
\begin{tabular}{lcc}
\hline
Parameter & Value & Physical Verdict \\
\hline
$\gamma_0$ & 1.05704654 & Nearly isothermal core\\
$\delta_\gamma$ & 0.11759971 & Moderate radial variation  \\
$K_0$ & 3000 & Moderate entropy  \\
$ml_{\text{disk}}$ & 1.00000000 & At the upper bound  \\
$ml_{\text{bulge}}$ & 0.00000000 & No bulge contribution \\
\hline
Overall &-& Physically plausible \\
\hline
\end{tabular}
\label{evaluationextendedNGC4214}
\end{table}

\subsection{Verdict for Simple SPARC Data Galaxies}

Now let us analyze the results of the previous section and discuss
the physical outcomes of it in some detail. Essentially what we
found for the viable galaxies is that the SIDM model that
reproduces successfully their galactic rotation curves has the
following characteristics: It describes an almost isothermal halo
due to the fact that  \(\gamma(r)\sim 1\) and \(\delta_\gamma \ll
1\). Also, essentially the energy and dispersion scale is set by
\(K_0\) since $K(r)$ varies quite smoothly with increasing radius.
So the model fits better low-mass, gas-rich,
low-surface-brightness and dwarf galaxies with slowly rising
rotation curves and weak central baryonic concentrations, and it
marginally fits or fails to fit completely, galactic systems with
strong central baryonic dominance, steep inner rises, pronounced
non-circular motions, or rotation curves that decline or peak
sharply. So basically many massive, high-surface-brightness
spirals and bulge-dominated or disk-dominated galaxies.

Let us theorize and ponder on the results at this point. We chose
the parameter values as,
\[
\gamma_0 \sim 1.0001,\qquad \delta_\gamma \sim 10^{-9}
\]
so \(\gamma(r)\) is effectively constant and almost equal to unity
across the galactic halo. Thus the EoS is therefore virtually
isothermal, so \cite{Sommer-Larsen:1996lnb,isothermal},
\[
P \sim \rho\, T ,\,\,\,P\sim K\,\rho.
\]
With the function $K(r)$ chosen as,
\[
K(r)=K_0\left(1+\frac{r}{r_c}\right)^{-p},\qquad p=0.01,\
r_c=0.5\,\mathrm{Kpc},
\]
the function \(K(r)\) hardly changes with radius, therefore the
halo has a single characteristic velocity-dispersion scale,
\[
\sigma \sim \sqrt{K_0}.
\]
Hence, the SIDM halo model is morphologically an isothermal-sphere
parameterized essentially by \(\rho_0\) and \(K_0\) solely. Thus,
we have this physical picture: any rotation curve which is
reproducible by a nearly isothermal halo plus a fairly simple
baryonic contribution will be fitted well. On the contrary, any
galactic structure that requires radial variation of the overall
pressure support, or a strong cuspy and contracted inner mass
profile, or even complex large scale kinematics, will not be
fitted well by the model. So the answer to the question why the
model of scale-dependent SIDM mimics very well and reproduces the
SPARC rotation curves of (most of the) dwarfs, low-luminosity and
low-surface-brightness spirals, is simple: dwarfs and
low-surface-brightness spirals typically have slowly rising and
strongly cored rotation curves and in addition very low central
baryonic mass. Thus, an isothermal-like dark halo naturally
reproduces their rotation curves. These galaxies are known to be
DM dominated at all radii (most of them), hence by neglecting the
contribution of the gas and the bulge components has essentially
no impact on the final shape of the rotation curve because the
scale-dependent SIDM model quantifies the dominant dynamical mass,
which is DM basically. Also most of the chosen \(K_0\) values
(hundreds or a few thousands) give \(\sqrt{K_0}\) of order
$\mathcal{O}(10-100)$ which is the right scale for low-mass
systems.

The reasons why the model fails for large galaxies is simple:
massive spirals have a strong central baryonic contribution in
their bulge-if present-. Thus by omitting bulge, disk stars, and
the gaseous component of the galaxy, the SIDM driven hydrostatic
solution cannot reproduce the steep inner rises of the rotation
velocity \(v_c(r)\). This is reasonable, recalling that in large
spirals 15$\%$ of the galaxy is composed by baryons, 10$\%$ in gas
and $5\%$ in luminous baryons. Also large spirals do not have only
the baryon issue, there are additional dynamical structures that
may or may not affect the halo and vice versa, for example bar
excitations, warps and additional gas related phenomena, such as
supernova outflows that may make the core a bit more cusp
pronounced.

In conclusion, the physical picture is crystal clear:
scale-dependent SIDM with a nearly isothermal EoS, produces
nearly-isothermal halos  which in turn are consistent with the
thermalization of the inner halo. Thus, the heat conduction and
thermal Bremsstrahlung scattering among the SIDM particles
flattens the cusps centers to cores, which naturally helps explain
dwarfs, low-luminosity and low-surface-brightness spiral cores.
The natural explanation of the cusp-core problem by the
scale-dependent EoS SIDM is one of the major outcomes of this
work. This is also seen in the figures of the densities for the
galaxies presented, even for the non-viable ones.

\section{$K_0$--$V_{\mathrm{max}}$ Correlation and Its Physical Significance in Viable SIDM Galaxies}

A key relation which serves as a diagnostic of scale-dependent
SIDM models with a variable polytropic constant \(K(r)\) is the
relation between the central normalization parameter \(K_0\) which
appears in the equation of state,
\[
P = K(r)\,\left(\frac{\rho(r)}{\rho_{\star}}\right)^{\gamma(r)},
\]
and the maximum rotation velocity \(V_{\mathrm{max}}\) of the
galaxy under investigation. From now on we shall consider only the
galaxies that are perfectly fitted by the scale-dependent SIDM, so
basically only the viable galaxies of a previous section which are
100 galaxies from the SPARC data \cite{Lelli:2016zqa}. The
parameter \(K_0\) encapsulates the entropy and the temperature
scale of the scale-dependent polytropic DM fluid, while the
maximum rotation velocity \(V_{\mathrm{max}}\) is a direct measure
the depth of the gravitational potential well created by the SIDM
DM in galaxies dominated by DM (we omit the baryon contribution
argument for the moment and we will consider this later on, where
the flat velocity must be used to measure the depth of the DM
gravitational potential). Hence, their correlation directly probes
the thermodynamic equilibrium structure of SIDM halos in a direct
way.

In the inner halo where we have approximately \(K(r)\sim K_0\) and
basically since for all radii we have \(\gamma(r)\sim \gamma_0\),
which indicates a nearly isothermal EoS $P=\mathcal{B}\rho T$
\cite{Sommer-Larsen:1996lnb,isothermal}, we get the following
general relation,
\begin{equation}\label{relation1K0}
K_0 \simeq \frac{\mathcal{B}
T_0}{\rho_0^{\,\gamma_0-1}}\rho_{\star}^{\gamma_0},
\end{equation}
where $\mathcal{B}$ is a general thermodynamics related constant
corresponding to the isothermal EoS and depends on the Boltzmann
constant and the DM particle mass. Hence \(K_0\) directly captures
the entropy normalization of the SIDM fluid. A large \(K_0\)
implies higher temperature or equivalently lower central density,
corresponding to more extended and less compact cores.

On the other hand, the maximum rotation velocity indicates the
depth of the gravitational potential of the SIDM fluid, in
galaxies dominated by SIDM, like the viable galaxies we shall
consider\footnote{It would be more appropriate to disentangle the
effect of baryons and DM, thus the flat rotation velocity would be
a more appropriate measure of the DM gravitational potential
depth. We shall consider this later on.}. So we have,
\[
V_{\mathrm{max}}^2 \sim \frac{G\,M(r_{\max})}{r_{\max}}\, ,
\]
so in hydrodynamic equilibrium, we get,
\[
\frac{dP}{dr} = -\rho\,\frac{G\,M(r)}{r^2},
\]
hence halos with deeper potentials, or equivalently larger
\(V_{\mathrm{max}}\), require proportionally higher central
pressure, and therefore larger \(K_0\). Thus a monotonic
correlation between \(K_0\) and \(V_{\mathrm{max}}\) naturally
emerges.

Now let us see how to construct the theoretical prediction for the
\(K_0\)-\(V_{\mathrm{max}}\) relation and we will verify from the
data we found, our predictions for \(K_0\) and also the
corresponding \(V_{\mathrm{max}}\) from the SPARC data, whether
such a relation is verified from the data. We will assume a
virialized SIDM halo, so the following relations hold true,
\[
T_0 \sim V_{\mathrm{max}}^2, \qquad \rho_0 \sim
V_{\mathrm{max}}^{-2},
\]
thus using Eq. (\ref{relation1K0}) we obtain the following
relation,
\begin{equation}\label{finalequationk0vmax}
K_0 \sim \frac{T_0}{\rho_0^{\,\gamma_0 -
1}}\rho_{\star}^{\gamma_0} \sim V_{\mathrm{max}}^{\,2 + 2(\gamma_0
- 1)}\, .
\end{equation}
For nearly isothermal SIDM which is our case, since we have
\(\gamma_0 \simeq 1.0001\) and $\delta_{\gamma}\sim 10^{-9}$, the
exponent in Eq. (\ref{finalequationk0vmax}) is essentially a
perfect square, thus we have the following theoretical prediction
for the \(K_0\)-\(V_{\mathrm{max}}\) relation,
\begin{equation}\label{theoreticalpredictionK0Vmaxrelation}
K_0 \sim V_{\mathrm{max}}^2.
\end{equation}
This quadratic scaling of Eq.
(\ref{theoreticalpredictionK0Vmaxrelation}) represents the
thermodynamic equilibrium expectation for self-gravitating and
thermalized SIDM halos.

Now relation (\ref{theoreticalpredictionK0Vmaxrelation}) is
extracted based entirely on the assumption of having thermalized
and virialized SIDM in thermal equilibrium and hydrodynamic
equilibrium. It is a theoretical prediction for a nearly
isothermal SIDM halo. Thus our aim now is to verify from the data
whether this relation also holds true. Thus for the 100 viable
galaxies of a previous section, most of which are dwarfs,
low-surface-brightness spirals and low-luminosity galaxies, we
will use the corresponding parameter $K_0$ for these, and also the
corresponding $\rho_0$ and $\gamma_0$ (recall that
$\rho_{\star}=1\,M_{\odot}/\mathrm{Kpc}^3$) and use the
corresponding $V_{\mathrm{max}}$ from the SPARC data and we shall
verify whether the theoretical relation
(\ref{theoreticalpredictionK0Vmaxrelation}) holds indeed true.
Plotting the parameter \(K_0\) versus \(V_{\mathrm{max}}\) tests
directly whether the fitted halos obey the theoretically expected
thermodynamic scaling between the entropy normalization
(quantified by $K_0$) and the SIDM halo potential depth
(quantified by $V_{\mathrm{max}}$). From that
$K_0$-$V_{\mathrm{max}}$ plot, a nearly quadratic relation would
imply a self-regulated virialized SIDM halo in thermodynamic
equilibrium, while a large scatter would indicate a
non-universality or non-thermal effects. Specifically, deviations
from the  \(K_0 \sim V_{\mathrm{max}}^2\) behavior, of the form
\(K_0\sim V_{\mathrm{max}}^n\) are physically informative and
would indicate a systematic variation of the effective polytropic
index or the self-interaction regime. Or even possibly that the
SIDM halo is not sufficiently virialized.

Now to proceed, we will use the data coming from SPARC, for the
100 viable galaxies of a previous section, which are mostly DM
dominated, so mostly low-luminosity, dwarfs and
low-surface-brightness spirals, namely the maximum rotation
velocity from the SPARC data for each corresponding galaxy and the
$K_0$ parameter we found for each one of the 100 viable galaxies
we found, and we shall plot the \(K_0\)-\(V_{\mathrm{max}}\).
Specifically, we will fit the relation
\[
K_0 = A \left(V_{\mathrm{max}}^2\,\rho_0^{\,1 -
\gamma_0}\rho_{\star}^{\gamma_0}\right)^n
\]
which allows us to extract the physical interpretation of both the
parameter $A$ and $n$. From a theoretical point of view, taking
into account the hydrostatic and virial thermal equilibrium, the
relation is,
\[
K_0 \sim V_{\mathrm{max}}^2 \rho_0^{\,1 - \gamma_0}
\rho_{\star}^{\gamma_0}\quad \rightarrow \quad n_{\text{theory}}
\simeq 1,
\]
for a isothermal halo. Hence, if we find $n \simeq 1$ from the
fitting, it will prove that the halo follows the universal SIDM
thermodynamic scaling of Eq.
(\ref{theoreticalpredictionK0Vmaxrelation}). Significant deviation
from  $n= 1$ would signal different relaxation efficiency,
variations in $\gamma(r)$ (unlikely in our case though), or
non-thermal behavior, which is the most probable explanation for a
deviation from the thermodynamic equilibrium of the halo.

\begin{figure}[h!]
\centering
\includegraphics[width=20pc]{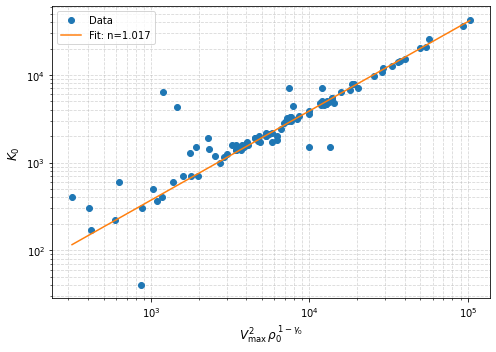}
\caption{The plot $K_0$-$V_{\mathrm{max}}$ for the 100 viable
galaxies we found from the SPARC data, for which the
scale-dependent SIDM model fits well their galactic rotation
curves. We used the $K_0$ we found for each galaxy and the
$V_{\mathrm{max}}$ from the SPARC data for the corresponding
galaxy. We also fitted the relation $K_0 = A
\left(V_{\mathrm{max}}^2\,\rho_0^{\,1 -
\gamma_0}\rho_{\star}^{\gamma_0}\right)^n$ with the orange line
and we found $A = 3.315\times 10^{-1}$ and $n = 1.017\pm 0.019$
which indicates that the isothermal SIDM model prediction for the
scaling $K_0\sim V_{\mathrm{max}}^2$ in Eq.
(\ref{theoreticalpredictionK0Vmaxrelation}) is verified by the
data.} \label{K0vmaxscatter}
\end{figure}

Now let us proceed to the results. In Fig. \ref{K0vmaxscatter} we
present the plot $K_0$-$V_{\mathrm{max}}$ and the fit of the
relation $K_0 = A \left(V_{\mathrm{max}}^2\,\rho_0^{\,1 -
\gamma_0}\rho_{\star}^{\gamma_0}\right)^n$ and now let us analyze
the results.
We analyzed the numerical data obtained from the SIDM halo models,
examining the dependence of the effective entropy constant \( K_0
\) on the composite dynamical quantity \( V_{\mathrm{max}}^2
\rho_0^{\,1-\gamma_0}\rho_{\star}^{\gamma_0} \). The empirical
relation was modelled as a power-law of the form,
\begin{equation}
    K_0 = A \left( V_{\mathrm{max}}^2 \rho_0^{\,1-\gamma_0} \rho_{\star}^{\gamma_0}\right)^{n},
    \label{eq:powerlaw_fit}
\end{equation}
where \( A \) and \( n \) are free parameters determined by a
nonlinear least-squares fit in log-log space. The results appear
in Fig. \ref{K0vmaxscatter}. From the fit, we obtained the
following best-fit parameters:
\begin{align}
    A &= (3.315 \pm 0.050) \times 10^{-1}, \\
    n &= 1.017 \pm 0.019.
\end{align}
Thus the resulting relation is,
\begin{equation}
    K_0 \simeq 0.33 \, \left( V_{\mathrm{max}}^2 \rho_0^{\,1-\gamma_0} \rho_{\star}^{\gamma_0}\right)^{1.017}.
\end{equation}
The fitted exponent \( n = 1.017 \pm 0.019 \) is remarkably close
to unity, hence implying a nearly linear scaling between \( K_0 \)
and \( V_{\mathrm{max}}^2
\rho_0^{\,1-\gamma_0}\rho_{\star}^{\gamma_0} \). This result
provides direct empirical support of the theoretical prediction of
Eq. (\ref{theoreticalpredictionK0Vmaxrelation}). Hence, the
near-unity slope directly indicates a quantitative confirmation
that the entropy normalization $K_0$ for each galaxy encapsulates
the dynamical temperature of the halo, thus linking microphysical
SIDM properties to macroscopic galactic observables. Also it
directly proves that the scale-dependent SIDM halo is virialized
and in thermal equilibrium.

Now it is tempting to go a bit further and give an estimate of the
DM particle $m_{\chi}$, since we have available the parameter $A$
in Eq. (\ref{eq:powerlaw_fit}). Using simple numerical arguments
since theoretically and ideally, $A\sim 1/m_{\chi}$, we
reluctantly point out that the DM particle mass is of the order
$m_{\chi}=\mathcal{O}(23.4)$MeV. But this is neither a proof nor a
prediction, since we made crucial assumptions for this derivation.
So we briefly mention it and we hope to formally address this in
the future.

Now it is tempting to investigate whether the present framework
can lead to the theoretical explanation of the Tully-Fisher
relation, since $L\sim V_{\mathrm{max}}^4$. We address this issue
in the following section.

\section{From Halo Physics to the Canonical Tully-Fisher Law: A Semi-Theoretical Proof}

In this section we will show that the Tully-Fisher law is obeyed
by the scale-dependent SIDM model for the 100 low-luminosity,
dwarfs and low-surface-brightness spirals which the model fits
well. We will demonstrate that the Tully-Fisher law is proved
semi-theoretically and semi-empirically. In the previous section
we showed that $K_0\sim V_{\mathrm{max}}^2$ and we proved this
theoretically by assuming a nearly isothermal DM halo and verified
the relation by the data. In this section by using the data we
will show that $L\sim K_0^2$ thus the model reproduces empirically
$L\sim V_{\mathrm{max}}^4$. Hence the Tully-Fisher law is
reproduced by the model using a semi-theoretical and
semi-empirical approach.

Thus, we shall consider the theoretically derived scaling $K_0\sim
V_{\mathrm{max}}^2$, which was derived on the basis of an
isothermal halo in hydrostatic and thermal equilibrium, and by
using an empirical fit of the luminosity with $K_0$,
\(\displaystyle L = A\,K_0^{\,n}\), with the measured parameters
\(A\) and \(n\), we will combine theory and the new data to obtain
the implied Tully-Fisher relation \(L\sim
V_{\mathrm{max}}^\alpha\). We will also take into account the
numerical propagation of uncertainties, and also take the results
for the fits of the previous section into account.

\begin{figure}[h!]
\centering
\includegraphics[width=20pc]{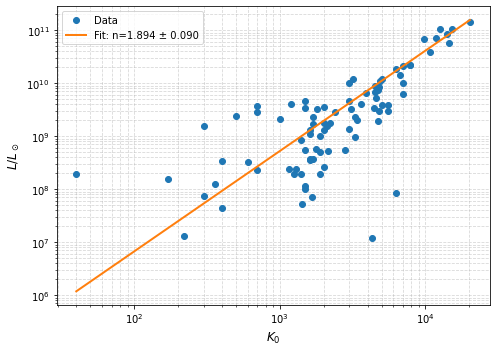}
\caption{The plot $L-K_0$ for the 100 viable galaxies we found
from the SPARC data, for which the scale-dependent SIDM model fits
well their galactic rotation curves. We used the $K_0$ we found
for each galaxy and the luminosity $L$ at 3.6 $\mathrm{\mu m}$ at
infrared from the SPARC data for the corresponding galaxy. We also
fitted the relation $L= A\,K_0^{\,n}$ with the orange line and we
found $A = \left(1.093 \pm 0.9497\right)\times 10^{3}$ and $n =
1.894 \pm 0.090$ which indicates that the isothermal SIDM model
satisfies the Tully-Fisher law, which is reproduced
semi-theoretically and empirically by the model.} \label{lumplot}
\end{figure}

We start with the theoretically derived result from halo
thermodynamics:
\begin{equation}
K_0 \;\sim\; V_{\mathrm{max}}^{\,2}. \label{eq:K0V2}
\end{equation}
We will use the SPARC data and we will fit the luminosity $L$ with
$K_0$, assuming initially a power law of the form,
\begin{equation}
L \;=\; A\,K_0^{\,n}\, .\label{eq:LofK}
\end{equation}
Using the SPARC data for the luminosities at 3.6 $\mathrm{\mu m}$
at infrared of the 100 viable galaxies, and the corresponding
values of $K_0$, in Fig. \ref{lumplot} we present the results of
the fitting.
From our data fit analysis we found
\begin{align}
A &= \left(1.093 \pm 0.9497\right)\times 10^{3}, \label{eq:Aval}\\
n &= 1.894 \pm 0.090. \label{eq:nvalnew}
\end{align}
From the results in Eq. (\ref{eq:nvalnew}) we need to note that
the normalization $A$ carries a large fractional uncertainty
(\(\sim 87\%\)) which affects the absolute scaling but it does not
directly change the slope analysis that follows below. The slope
\(n\) is the quantity of our main physical interest.

Let us combine Eqs. \eqref{eq:K0V2} and \eqref{eq:LofK}.  Firstly
we rewrite \eqref{eq:K0V2} as follows,
\[
K_0 \;\sim\; (V_{\mathrm{max}}^2 \rho_0^{\,1-\gamma_0})^{\,n_1},
\]
and more generally, we keep the following algebraic form,
\[
K_0 \;=\; A_1 \left(V_{\mathrm{max}}^2
\rho_0^{\,1-\gamma_0}\right)^{n_1}.
\]
Substituting the above in (\ref{eq:LofK}) gives,
\begin{equation}
L \;=\; A_2 \left(V_{\mathrm{max}}^2
\rho_0^{\,1-\gamma_0}\right)^{n_1 n_2}, \label{eq:Lcombined}
\end{equation}
where for clarity we denoted \(n_2 \equiv n\) (the measured
exponent in (\ref{eq:LofK})) and also \(A_2\) is the combined
normalization. Then we have,
\begin{equation}
L \;\sim\; V_{\mathrm{max}}^{\,2 n_1 n_2}. \label{eq:LvsV}
\end{equation}
We define,
\begin{equation}
p \equiv n_1 n_2, \qquad \alpha \equiv 2p = 2 n_1 n_2,
\end{equation}
thus the fractional uncertainty on the product \(p=n_1 n_2\) is,
\begin{equation}
\frac{\sigma_p}{p} \;=\;
\sqrt{\left(\frac{\sigma_{n_1}}{n_1}\right)^2 +
\left(\frac{\sigma_{n_2}}{n_2}\right)^2 }\, , \label{eq:prop}
\end{equation}
hence \(\sigma_\alpha = 2\sigma_p\). If we adopt the derived from
the data values for $n_1$ and $n_2$, we get,
\[
n_1 = 1.017 \pm 0.019,\qquad n_2 = 1.894 \pm 0.090.
\]
Thus the product and the propagated uncertainty are,
\begin{align*}
p &= n_1 n_2 = 1.92620\pm 0.09835,\\
\alpha &= 2p = 3.85240\pm 0.19670.
\end{align*}
Hence, the combined $L-V_{\mathrm{max}}$ relation is
\[
L \;\sim\; V_{\mathrm{max}}^{\,3.85 \pm 0.20}.
\]
A comparison to the canonical Tully-Fisher slope
\(\alpha_{\mathrm{TF}}\simeq 4.00\) has a difference,
\[
\Delta \equiv 4 - \alpha \sim 0.15
\]
which corresponds to a nearly \(0.8\sigma\) deviation which is
statistically insignificant.

In conclusion, by taking the theoretically derived (and data
confirmed) scaling relation \(K_0\sim V_{\mathrm{max}}^2\) as
given, and combining it with the measured relation \(L =
(1.09\pm0.95)\times 10^3\;K_0^{\,1.894\pm0.090}\), yields the
following $L-V_{\mathrm{max}}$ scaling relation,
\[
L \sim V_{\mathrm{max}}^{\,3.85\pm0.20},
\]
which is consistent with the canonical Tully-Fisher relation
\(L\sim V_{\mathrm{max}}^4\) within the combined uncertainties and
errors. The slope of the fit is robust, but we have large
uncertainties in the normalization. Hence, under the assumptions
of the halo thermodynamics model and for the present 100 viable
galaxy sample, the Tully-Fisher scaling emerges naturally as a
theoretical-empirical consequence.

Now a natural question emerges, if the results derived in this
section comply with the baryonic Tully-Fisher relation. This is
the subject of the next section.

\section{Thermodynamic Origin of the Baryonic Tully-Fisher Relation:
A Theoretical-Empirical Derivation}

Our focus now is devoted to the baryonic Tully-Fisher relation and
how this universal galactic law is reproduced in a
semi-theoretical and semi-empirical way from the scale-dependent
EoS SIDM model. We shall discuss the necessity for studying the
baryonic Tully-Fisher law, since it provides a more robust result
compared to the canonical Tully-Fisher law presented in the
previous section. Then we shall show that the baryonic
Tully-Fisher law emerges naturally from the scale-dependent EoS
SIDM model.

Let us show what we will prove in this section in order to have a
mental compass of what we aim to achieve.  We will show that the
scale-dependent EoS SIDM model yields theoretically,
\[
K_0 \sim V_{\rm flat}^2
\]
assuming an  isothermal EoS, which is also verified by the data.
Then, from the data of the 100 viable galaxies we will show that,
\[
K_0 \sim M_b^{1/2},
\]
which results to the baryonic Tully-Fisher scaling law $M_b\sim
V_{\rm flat}^4$. This will indicate that the internal halo
structure, quantified by $K_0$, is linked directly to the
gravitational potential depth quantified by $V_{\mathrm{flat}}^2$,
and it serves as an empirical and semi-theoretical relation
between $K_0$ and the total baryonic content. Our result will
establish a semi-theoretical bridge between dark halo physics and
the observed baryonic scaling laws, offering a novel
theoretical-empirical derivation of the Tully-Fisher relation.

Now the question is why \(V_{\mathrm{flat}}\) may provide a better
description of the DM gravitational potential depth, compared with
\(V_{\mathrm{max}}\), and why the baryonic Tully-Fisher relation
might be more robust when it comes to DM physics. Let us answer
these questions in the following, and then we address the baryonic
Tully-Fisher relation more concretely. The quantities
\(V_{\mathrm{max}}\) and \(V_{\mathrm{flat}}\) characterize
different dynamical regimes of the galactic rotation curves. The
quantity \(V_{\mathrm{max}}\) stands for the maximum observed
circular velocity, which corresponds to the peak of the
gravitational potential \(V(r)\). In contrast, the flat rotation
velocity \(V_{\mathrm{flat}}\) denotes the asymptotic velocity
reached in the outer skirts, approximately flat portion of the
rotation curve, if such a region is observed for a galaxy given.
The gravitational potential depth of a galaxy's dark halo is most
accurately traced by the velocity at radii where the halo
dominates the mass distribution, which typically is well beyond
the stellar disk. Thus the flat velocity curve
\(V_{\mathrm{flat}}\) is more appropriate for DM physics. In
high-surface-brightness spirals, the luminous stellar disk
dominates the inner mass distribution and the rotation curve in
these systems can rise steeply, until it reaches a peak, and then
decline slightly, before flattening (in most cases). Consequently,
\(V_{\mathrm{max}}\) often occurs within the baryon-dominated
region of the galactic structure, and this reflects the combined
disk-halo structure, and is not representative of the halo
potential by itself. On the other hand, at large galactic radii,
where gas (HI) and DM dominates, the baryonic contribution becomes
negligible, and therefore the rotation curve approaches a flat
regime controlled mostly by the dark matter halo and the HI gas.

In this region of the galaxy, the flat rotation curve
\(V_{\mathrm{flat}}\) measures directly  the circular velocity of
the DM halo, so the depth of its gravitational potential well.

In dwarfs and, low luminosity and low-surface-brightness galaxies,
which is exactly the case for most of the 100 viable galaxies that
our model best fits their rotation curves, the situation becomes
more critical. These systems are DM dominated even near their
centers, yet their rotation curves rise gradually and often do not
reach a flat plateau within the observed range. As a result,
\(V_{\mathrm{max}}\) is measured before the true asymptotic
velocity is attained, while \(V_{\mathrm{flat}}\) may not be
observable at all. For such galaxies, \(V_{\mathrm{max}}\)
systematically underestimates the true circular velocity and the
potential depth of the DM halo. This may lead to a shallower
\(L\)-\(V_{\mathrm{max}}\) relation or even a
\(M_b\)-\(V_{\mathrm{max}}\) relation, a behavior that naturally
explains the deviations observed in faint and diffuse galaxies.
Thus for our analysis we shall use the flat rotation velocity
\(V_{\mathrm{flat}}\) for the 100 viable galaxies, and when it is
not available we shall use the maximum rotation velocity
\(V_{\mathrm{max}}\).

Now regarding the other question, why the baryonic Tully-Fisher
relation is more robust compared to the canonical luminosity-based
(3.6~$\mu$m) Tully-Fisher, especially for dwarf and
low-surface-brightness and low-luminosity spiral galaxies, the
reason is easily understood if we consider the problems regarding
luminosity and velocity relations. For faint,
low-surface-brightness, or dwarf galaxies, the stellar luminosity
is not a reliable tracer of the total baryonic mass, for several
reasons. In bright spiral galaxies, the near-infrared light (at
3.6~$\mu$m) characterizes mostly the old population II stars
fairly well, but in dwarfs and low-surface-brightness systems, the
stellar populations are typically younger population I stars. In
effect, the mass-to-light ratio may vary by large factors. Thus,
the infrared luminosity $L_{3.6\mu{\rm m}}$ ceases to trace
concretely the total stellar mass. Also in these low-luminosity
galactic systems, the neutral gas mass often exceeds the star mass
by an order of magnitude, but the $L_{3.6\mu{\rm m}}$ luminosity
traces only the stellar component. Hence, using only the
luminosity based Tully-Fisher relation leads to underestimation of
the actual baryonic mass. The baryonic Tully-Fisher relation fixes
these problems because the mass includes the mass of the gas and
all the stellar populations (I and II), and the total DM, gas and
baryon gravitational potential is now traced by $V_{\rm flat}$
(whenever is present or exists). Now let us demonstrate how the
baryonic Tully-Fisher relation emerges from our theoretical
framework in a semi-theoretical and semi-empirical way.

In the scale-dependent EoS SIDM framework, the EoS is
\begin{equation}
    P(r) = K(r)\,\left(\rho(r)/\rho_{\star}\right)^{\gamma(r)},
\end{equation}
and since we found that the model fits perfectly 100 galaxies with
$\gamma_0\sim 1$, this is an isothermal EoS, so we have,
\begin{equation}
    P = \mathcal{B}\rho T,
\end{equation}
Equating the two expressions for $P$ at $r=r_0$ where the density
is $\rho_0$, yields the exact identity,
\begin{equation}
    K_0 = \mathcal{B} T_0\rho_0^{\,1-\gamma_0}\rho_{\star}^{\gamma_0},
    \label{eq:K0_definition}
\end{equation}
which links the macroscopic polytropic constant $K_0$ to the
microscopic kinetic temperature.

In a quasi-virialized and thermalized halo, the temperature is
expected to characterize the depth of the gravitational potential
so we may write,
\begin{equation}
    T_0 \simeq\,V_{\mathrm{flat}}^{\,2},
\end{equation}
thus we have,
\begin{equation}
    K_0 \simeq \mathcal{B}\,V_{\mathrm{flat}}^{\,2}\,\rho_0^{\,1-\gamma_0}\rho_{\star}^{\gamma_0},
    \label{eq:K0_virial}
\end{equation}
which expresses the polytropic constant in terms of observable
quantities. Since $\gamma_0\simeq 1$ we get theoretically the
$K_0-V_{\mathrm{flat}}$ scaling law,
\begin{equation}
    K_0 \sim V_{\mathrm{flat}}^{\,2}\, ,
    \label{eq:K0_Vflat2}
\end{equation}
which is derived on a theoretical ground. Now, to connect the DM
thermodynamics to the baryonic observables, we introduce an
empirical scaling law to be checked by using the data,
\begin{equation}
    K_0 \sim M_b^{\,q},
    \label{eq:K0_Mb}
\end{equation}
with $M_b$ being the total baryonic mass (stars plus gas
components) of the galaxy. Combining Eqs.~(\ref{eq:K0_virial}) and
(\ref{eq:K0_Mb}) eliminates $K_0$ and we get,
\begin{equation}
    M_b^{\,q} \sim V_{\mathrm{flat}}^{\,2}\,\rho_0^{\,1-\gamma_0}\rho_{\star}^{\gamma_0}
    \quad \Rightarrow \quad
    M_b \sim V_{\mathrm{flat}}^{\,2/q}\,\rho_0^{\,(1-\gamma_0)/q}\rho_{\star}^{\gamma_0/q}.
    \label{eq:BTFR_derived}
\end{equation}
For $\gamma_0\simeq1$, the density term is essentially constant
and the predicted baryonic Tully--Fisher relation becomes,
\begin{equation}
    M_b \sim V_{\mathrm{flat}}^{\,\alpha}, \qquad
    \alpha = \frac{2}{q}.
    \label{eq:BTFR_slope}
\end{equation}
The canonical observed baryonic Tully-Fisher slope $\alpha\simeq4$
is thus recovered if $q$ is approximately $q\simeq0.5$. Therefore,
determining the $K_0$-$M_b$ for the scale-dependent EoS DM model,
provides a direct test of the theoretical link between SIDM
thermodynamics and the baryonic Tully-Fisher relation.

\begin{figure}[h!]
\centering
\includegraphics[width=20pc]{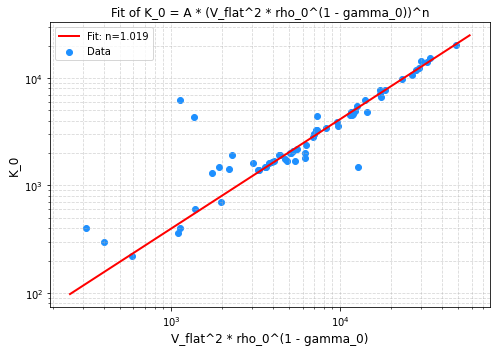}
\caption{The plot $K_0-V_{\mathrm{flat}}$ for the 100 viable
galaxies we found from the SPARC data, for which the
scale-dependent SIDM model fits well their galactic rotation
curves. We used the $K_0$ we found for each galaxy and the flat
velocity (or the maximum velocity if the flat velocity is not
available) from the SPARC data for the corresponding galaxy. We
also fitted the relation $K_0 = A
\left(V_{\mathrm{flat}}^{\,2}\,\rho_0^{\,1-\gamma_0}\rho_{\star}^{\gamma_0}\right)^n$
with the orange line and we found $A = (3.4724 \pm 1.4233)\times
10^{-1}$ and $n = 1.0188 \pm 0.0404$ which indicates that $K_0
\sim V_{\mathrm{flat}}^{\,2}$. This verifies the theoretical
expectation for a SIDM isothermal halo. } \label{K0vflat}
\end{figure}

Let us proceed to the analysis of the 100 viable galaxies and the
data for the baryonic mass for these galaxies coming from the
SPARC data. We performed the following regression in logarithmic
space: the $K_0$-$V_{\mathrm{flat}}$ relation by fitting the
functional form,
\begin{equation}
    K_0 = A
    \left(V_{\mathrm{flat}}^{\,2}\,\rho_0^{\,1-\gamma_0}\rho_{\star}^{\gamma_0}\right)^n\,
    .
\end{equation}
The result appears in Fig. \ref{K0vflat}.
The fit of our analysis indicates that,
\[
    A = (3.4724 \pm 1.4233)\times
10^{-1}, \qquad n = 1.0188 \pm 0.0404,
\]
indicating an almost perfect linear scaling $K_0\sim
V_{\mathrm{flat}}^{\,2}$, in excellent agreement with
Eq.~\ref{eq:K0_Vflat2}.

\begin{figure}[h!]
\centering
\includegraphics[width=20pc]{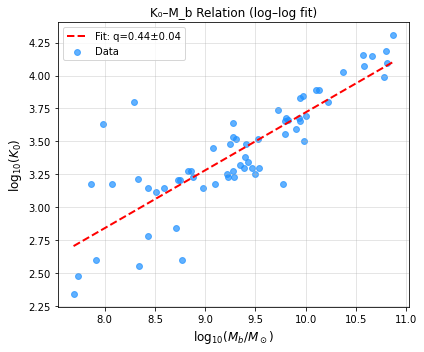}
\caption{The plot $Log(K_0)-Log(M_b/M_{\odot})$ for the 100 viable
galaxies we found from the SPARC data, for which the
scale-dependent SIDM model fits well their galactic rotation
curves. We used the $K_0$ we found for each galaxy and the
baryonic mass $M_b$ from the SPARC data for the corresponding
galaxy. We also fitted the relation $\log K_0 = a + q\,\log M_b$
with the orange line and we found $a =-0.6761 \pm 0.103$ and $q =
0.4397 \pm 0.0367$ which indicates that we have approximately $K_0
\sim M_b^{1/2}$.} \label{K0mb}
\end{figure}

\begin{figure}[h!]
\centering
\includegraphics[width=30pc]{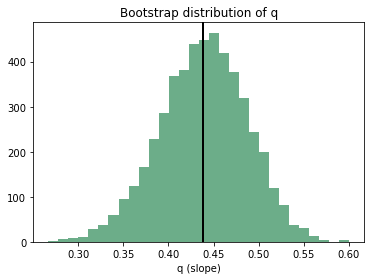}
\caption{Bootstrap sampling of the parameter $q$, the result is $q
= 0.4377 \pm 0.0485$.} \label{boot}
\end{figure}

Now upon using the baryonic masses from the SPARC data, we perform
a linear fit in logarithmic space, for the relation,
\begin{equation}
    \log K_0 = a + q\,\log M_b,
\end{equation}
and the result appears in Fig. \ref{K0mb}.
The result of the fit is,
\[
    a = -0.6761 \pm 0.103, \qquad q =
0.4397 \pm 0.0367
\]
which was independently confirmed by bootstrap sampling ($q =
0.4377 \pm 0.0485$), see Fig. \ref{boot}.
With the measured $q$ and upon substituting in
Eq.~\ref{eq:BTFR_slope} we get the baryonic Tully-Fisher exponent,
\[
    \alpha = \frac{2}{q}=4.55\pm 0.38,
\]
which agrees very well with the empirical value
$\alpha_{\mathrm{obs}} = 4.0 \pm 0.3$ obtained by the SPARC sample
itself.

Hence, the measured slope $q\simeq 0.45$ indicates that the
effective entropy scale $K_0$ of the scale-dependent EoS SIDM halo
increases in a sub-linear way with baryonic mass $K_0\sim
M_b^{0.45}$, so more massive galaxies possess deeper potential
wells and higher effective DM halo temperatures, but the
dependence is moderate. Upon combining this with the theoretically
derived and data-proved virial relation between $K_0$ and
$V_{\mathrm{flat}}$, for the 100 viable galaxies, naturally
produces the phenomenological baryonic Tully-Fisher relation
without any parameter tuning. This result provides another strong
evidence that the SIDM polytropic framework encapsulates the
correct thermodynamic coupling between the baryonic and DM
components. The baryonic Tully-Fisher relation hence emerges
naturally from the thermal equilibrium and hydrodynamic
equilibrium of the SIDM of a self-interacting halo whose central
entropy is regulated by the gravitational potential depth, instead
from fine-tuned feedback processes.

\section{Qualitative Comparison with MOND-like Phenomenology}

MOND is not a theory per se, but it is founded on certain axioms
that seem to provide desirable characteristic galactic
phenomenology. Scale-dependent SIDM seems to reproduce these
desirable characteristics without using an unjustified axiomatic
approach on modifying gravity in a non-fundamental way but rather
ad hoc.

MOND is based on a breakdown of Newton's second law at low
accelerations, introducing an interpolation function $\mu(x)$ such
that,
\begin{equation}
a\,\mu\!\left(\frac{a}{a_0}\right) = \frac{GM_b(r)}{r^2},
\label{eq:mond}
\end{equation}
where $a_0$ is the characteristic MOND acceleration scale and
$M_b(r)$ is the enclosed baryonic mass. In the deep-MOND limit
($a\ll a_0$), Eq.~(\ref{eq:mond}) yields,
\begin{equation}
a = \sqrt{a_0\,\frac{GM_b(r)}{r^2}} \;\;\Rightarrow\;\; v_\phi^4 =
G\,M_b\,a_0, \label{eq:btfr_mond}
\end{equation}
which reproduces the empirical baryonic Tully-Fisher relation,
\begin{equation}
M_b \sim v_\phi^4,
\end{equation}
and also it naturally explains the flattening of rotation curves
without invoking dark matter.

In contrast, the scale-dependent SIDM model we introduced in this
work, retains standard Newtonian gravity but it modifies the DM
thermodynamics through a spatially dependent EoS,
\begin{equation}
P(r) = K(r)\,\rho(r)^{\gamma(r)}, \label{eq:eos_variable}
\end{equation}
where $\gamma(r)$ and $K(r)$ describe local variations in the
effective compressibility and the entropy of the collisional DM
fluid. The system satisfies hydrostatic equilibrium,
\begin{equation}
\frac{dP}{dr} = -\rho(r)\,\frac{G\,M(r)}{r^2}, \label{eq:hydro_eq}
\end{equation}
where $M(r)$ is the total (mostly DM) enclosed mass. The circular
velocity follows the usual definition,
\begin{equation}
v_\phi^2(r) = \frac{G\,M(r)}{r}. \label{eq:v_circ}
\end{equation}
In the specific class of models analyzed in this article, the
polytropic index is parameterized as,
\begin{equation}
\gamma(r) = \gamma_0 - \delta_\gamma\,\tanh\!\left(\frac{r -
r_\gamma}{2.0\mathrm{Kpc}}\right),
\end{equation}
where the best-fit parameters for galaxies exhibiting good fitted
and flat rotation curves and compatible with the SPARC data, are
found to be,
\[
\gamma_0 = 1.0001, \qquad \delta_\gamma = 1.2\times10^{-9}.
\]
This implies a nearly isothermal EoS in the core and inner halo
regions ($\gamma \simeq 1$), corresponding to a locally
thermalized SIDM with efficient heat conduction.

Let us consider the effects of the isothermal limit and the
emergent flat rotation curves. For an isothermal fluid ($\gamma
\rightarrow 1$), Eq.~(\ref{eq:eos_variable}) reduces to,
\begin{equation}
P = K_0\,\rho,
\end{equation}
with $K_0$ representing the effective entropy constant. Combining
Eqs.~(\ref{eq:hydro_eq}) and (\ref{eq:eos_variable}) in this limit
we get,
\begin{equation}
\frac{1}{\rho}\frac{d\rho}{dr} = -\frac{G\,M(r)}{K_0\,r^2}.
\end{equation}
For a quasi-isothermal sphere, we get approximatelly $\rho(r)\sim
r^{-2}$, one thus obtains,
\begin{equation}
M(r)\sim r, \quad \rightarrow \quad v_\phi^2(r)=\frac{GM(r)}{r} =
\text{constant}.
\end{equation}
Thus, the flat rotation curves emerge naturally from the
thermodynamic equilibrium condition of a nearly isothermal SIDM
halo.

\section{Conclusions and Phenomenological Discussion}

In this work we introduced the concept of scale-dependent SIDM in
the context of which the DM is collisional and its EoS is
radius-dependent and has the form
$P(r)=K(r)\left(\frac{\rho(r)}{\rho_{\star}}\right)^{\gamma(r)}$.
This theoretical assumption is motivated by mirror DM which is
composed by atoms and elementary particles along with radiation,
and thus its EoS may vary in an environment-dependent way. With
this assumption for the DM EoS we studied 174 galaxies from the
SPARC data and confronted the observational derived galactic
rotation curves with the predictions of scale-dependent SIDM and
we found that 100 galaxies can be perfectly fitted, 27 marginally
fitted and the rest cannot be fitted well. The scale-dependent
SIDM proves to fit well dwarfs galaxies, low-surface-brightness
galaxies and low-luminosity galaxies, which are known to be DM
dominated. On the other hand, larger galaxies in which baryons
contribute to the rotation curves in an indistinguishable way, the
scale-dependent SIDM model provides marginal fits. The
scale-dependent SIDM solves in a natural way the cusp-core
problems of dwarfs galaxies, low-surface-brightness galaxies and
low-luminosity galaxies, since it generates a cored center of the
galaxy by construction, and we demonstrated this explicitly. More
importantly, the structure of the scale-dependent SIDM model
solves semi-theoretically and semi-empirically the canonical
Tully-Fisher and the baryonic Tully-Fisher relations when these
100 viable dwarfs, low-surface-brightness and low-luminosity
galaxies are taken into account. Specifically the perfect fit for
the 100 viable dwarfs, low-surface-brightness and low-luminosity
galaxies is achieved for $\gamma(r)\sim 1$, so for a nearly
isothermal EoS
$P(r)=K(r)\left(\frac{\rho(r)}{\rho_{\star}}\right)$, with
$K(r)=K_0\times\left(1+\frac{r}{r_c} \right)^{-p}$. We proved on a
theoretical basis that the entropy parameter $K_0$ should be
related to the maximum rotation velocity $V_{\mathrm{max}}$ as
$K_0\sim V_{\mathrm{max}}^2$, based on the assumption of a nearly
isothermal SIDM halo. It proved that indeed the entropy parameter
$K_0$ is strongly correlated to the observational maximum rotation
velocity of each of the 100 viable dwarfs, low-surface-brightness
and low-luminosity galaxies to which the scale-dependent SIDM
provides a perfect fit. This was a theoretical prediction of the
scale-dependent SIDM model which was also confirmed empirically
from the rotation curve fits. Apart from that, we showed
empirically from the data that for these 100 viable dwarfs,
low-surface-brightness and low-luminosity galaxies to which the
scale-dependent SIDM provides a perfect fit, the infrared
$3.6\mu$m luminosity scales as $L\sim K_0^2$, thus the canonical
Tully-Fisher relation is reproduced in a semi-theoretical and
semi-empirical manner. The empirically confirmed theoretical
scaling $K_0\sim V_{\mathrm{max}}^2$, directly proves that the
scale-dependent SIDM halo is virialized and is in thermal
equilibrium. The parameter \(K_0\) directly captures the entropy
normalization of the SIDM fluid and a large \(K_0\) implies higher
temperature or equivalently lower central density, corresponding
to more extended and less compact cores. The parameter \(K_0\)
encapsulates the entropy and the temperature scale of the
scale-dependent polytropic DM fluid, while the maximum rotation
velocity \(V_{\mathrm{max}}\) is a direct measure the depth of the
gravitational potential which is created by the SIDM DM in
galaxies dominated by DM. Hence, their correlation directly probes
the thermodynamic equilibrium structure of SIDM halos in a direct
way. Now, taking into account that \(V_{\mathrm{max}}\) may be
also be affected by baryons, we did the same procedure taking into
account the flat rotation velocity, where this was available, for
these 100 viable galaxies. The gravitational potential depth of a
galactic dark halo is mostly traced by the velocity at radii where
the dark halo dominates the mass distribution, which typically is
well beyond the stellar disk. Thus the flat velocity curve
\(V_{\mathrm{flat}}\) is more appropriate for DM-dominated
galaxies. In high-surface-brightness spirals, the luminous stellar
disk dominates the inner mass distribution and the rotation curve
in these systems can rise steeply, until it reaches a peak, and
then decline slightly, before flattening. Consequently,
\(V_{\mathrm{max}}\) often occurs within the baryon-dominated
region of the galactic structure, and this reflects the combined
disk-halo structure, and is not representative of the halo
potential by itself. Thus we did the same procedure replacing the
maximum velocity with the flat rotation curve (where available),
in which case the theoretical prediction of a virialized nearly
isothermal halo gives $K_0\sim V_{\mathrm{flat}}^2$. We confirmed
the theoretical prediction empirically for the 100 DM dominated
viable galaxies. In addition, we found an empirical correlation of
$K_0$ with the baryonic mass of each of these 100 viable galaxies
and proved that we approximately have the scaling relation
$K_0\sim M_b^{0.5}$, hence the baryonic Tully-Fisher law naturally
emerges in semi-theoretical and semi-empirical manner.

There is still much work to be done to fully understand the
phenomenological consequences of scale-dependent SIDM. For example
what are the fundamental procedures that alter the value of the
entropy parameter $K_0$ from galaxy to galaxy. Another example is
the issue of dark disk formation, along with the baryonic disk in
spirals, which is a consequence of the dissipative nature of
scale-dependent SIDM, see Refs.
\cite{Foot:2014uba,Fan:2013yva,McCullough:2013jma} for the dark
disk formation issue. Another issue that deserves proper study is
the gravothermal catastrophe issue which we did not address in
this introductory work. Also how would this scale-dependent SIDM
would affect the inner baryonic structures of the galaxy, like the
supermassive black hole at the center and the bars in massive
spirals should also be addressed appropriately. In addition, if DM
is dissipative and dark structures form, like dark stars and
supernovae and consequently central dark gas inflow/outflow would
occur, this feedback could affect significantly the EoS OF SIDM.
There are also various other galactic scale phenomenological
aspects of scale-dependent SIDM which we did not address in this
introductory work, but we aim to address in the near future. In
addition, in our future plans we include the cosmological scale
implications of a radius-dependent SIDM. At cosmological scales,
the polytropic EoS might be a limiting case of the radius
dependent EoS, since $K$ and the polytropic index might be
constant at various evolutionary epochs of our Universe, thus the
radius dependence applies only at galactic scales or even at
supercluster scales. In this case one must explain what would
$K_0$ be and how a universal value of it should be interpreted. We
aim to address these issues soon. But the $K_0$ issue is also
important for single galaxies, why does it vary from galaxy to
galaxy, what microphysical galactic processes modify its value
from galaxy to galaxy. These are questions of fundamental
importance that need to be addressed properly.

\section*{Acknowledgments}

This research has been is funded by the Committee of Science of
the Ministry of Education and Science of the Republic of
Kazakhstan (Grant No. AP26194585) (V.K. Oikonomou). I am grateful
to my colleague in science and good friend Porfyrios for inspiring
discussions on the role of scale-dependent self-interacting dark
matter in the large scale structure.

\appendix
\section*{Appendix: Complete List of Galaxy Simulations Using SIDM with scale-dependent EoS}

In this appendix we present the complete confrontation of the SIDM
model we introduced with a scale-dependent EoS, with the SPARC
galaxies.

\subsection{The Galaxy CamB}

\begin{figure}[h!]
\centering
\includegraphics[width=20pc]{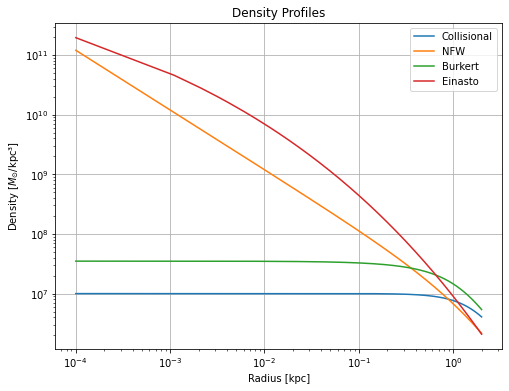}
\caption{The density of the collisional DM model (\ref{tanhmodel})
for the galaxy CamB, as a function of the radius.}
\label{CamBdens}
\end{figure}

\begin{figure}[h!]
\centering
\includegraphics[width=20pc]{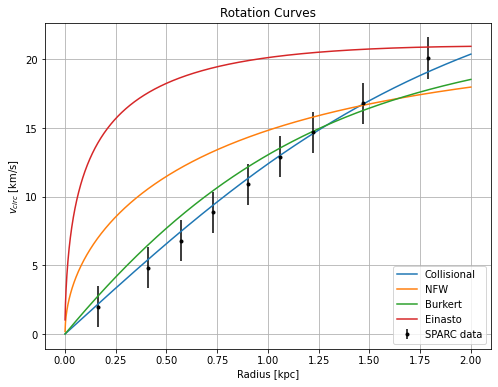}
\caption{The predicted rotation curves after using an optimization
for the collisional DM model (\ref{tanhmodel}), versus the SPARC
observational data for the galaxy CamB. We also plotted the
optimized curves for the NFW model, the Burkert model and the
Einasto model.} \label{CamB}
\end{figure}

\begin{figure}[h!]
\centering
\includegraphics[width=20pc]{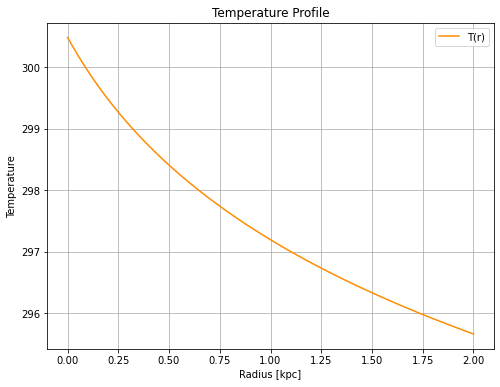}
\caption{The temperature as a function of the radius for the
collisional DM model (\ref{tanhmodel}) for the galaxy CamB.}
\label{CamBtemp}
\end{figure}

For this galaxy, we shall choose $\rho_0=1\times
10^7$$M_{\odot}/\mathrm{Kpc}^{3}$. The galaxy Camelopardalis B
(Cam B) is a dwarf irregular extremely faint, low-luminosity,
gas-rich system. Typical of faint dwarf irregulars, it has an
irregular shape, an extended HI disk, and very
low-surface-brightness. Its optical radius is roughly 0.7-1 Kpc.
In Figs. \ref{CamBdens}, \ref{CamB} and \ref{CamBtemp} we present
the density of the collisional DM model, the for predicted
rotation curves after using an optimization for the collisional DM
model (\ref{tanhmodel}), versus the SPARC observational data and
the temperature parameter as a function of the radius
respectively. As it can be seen, the SIDM model produces viable
rotation curves compatible with the SPARC data. Also in Tables
\ref{collCamB}, \ref{NavaroCamB}, \ref{BuckertCamB} and
\ref{EinastoCamB} we present the optimization values for the SIDM
model, and the other DM profiles. Also in Table
\ref{EVALUATIONCamB} we present the overall evaluation of the SIDM
model for the galaxy at hand. The resulting phenomenology is
viable.
\begin{table}[h!]
  \begin{center}
    \caption{Collisional Dark Matter Optimization Values}
    \label{collCamB}
     \begin{tabular}{|r|r|}
     \hline
      \textbf{Parameter}   & \textbf{Optimization Values}
      \\  \hline
     $\delta_{\gamma} $ & 0.0000000012
\\  \hline
$\gamma_0 $ & 1.0001 \\  \hline $K_0$ ($M_{\odot} \,
\mathrm{Kpc}^{-3} \, (\mathrm{km/s})^{2}$)& 300
\\  \hline
    \end{tabular}
  \end{center}
\end{table}
\begin{table}[h!]
  \begin{center}
    \caption{NFW  Optimization Values}
    \label{NavaroCamB}
     \begin{tabular}{|r|r|}
     \hline
      \textbf{Parameter}   & \textbf{Optimization Values}
      \\  \hline
   $\rho_s$   & $0.004\times 10^9$
\\  \hline
$r_s$&  3
\\  \hline
    \end{tabular}
  \end{center}
\end{table}
\begin{table}[h!]
  \begin{center}
    \caption{Burkert Optimization Values}
    \label{BuckertCamB}
     \begin{tabular}{|r|r|}
     \hline
      \textbf{Parameter}   & \textbf{Optimization Values}
      \\  \hline
     $\rho_0^B$  & $0.035\times 10^9$
\\  \hline
$r_0$&  1.5
\\  \hline
    \end{tabular}
  \end{center}
\end{table}
\begin{table}[h!]
  \begin{center}
    \caption{Einasto Optimization Values}
    \label{EinastoCamB}
    \begin{tabular}{|r|r|}
     \hline
      \textbf{Parameter}   & \textbf{Optimization Values}
      \\  \hline
     $\rho_e$  & $0.009\times 10^9$
\\  \hline
$r_e$ & 1
\\  \hline
$n_e$ & 0.15
\\  \hline
    \end{tabular}
  \end{center}
\end{table}
\begin{table}[h!]
\centering \caption{Physical assessment of collisional DM
parameters (Camb).}
\begin{tabular}{lcc}
\hline
Parameter & Value & Physical Verdict \\
\hline
$\gamma_0$ & $1.0001$ & Essentially isothermal  \\
$\delta_\gamma$ & $1.2\times10^{-9}$ & Vanishingly small variation \\
$r_\gamma$ & $1.5\ \mathrm{Kpc}$ & Reasonable inner-halo scale \\
$K_0$ & $3\times10^{2}$ & Moderate entropy/pressure scale provides support. \\
$r_c$ & $0.5\ \mathrm{Kpc}$ & Small core radius \\
$p$ & $0.01$ & Almost flat $K(r)$; negligible radial change in entropy. \\
\hline
Overall & - & Physically consistent and numerically stable \\
\hline
\end{tabular}
\label{EVALUATIONCamB}
\end{table}


\subsection{The Galaxy D512-2}

\begin{figure}[h!]
\centering
\includegraphics[width=20pc]{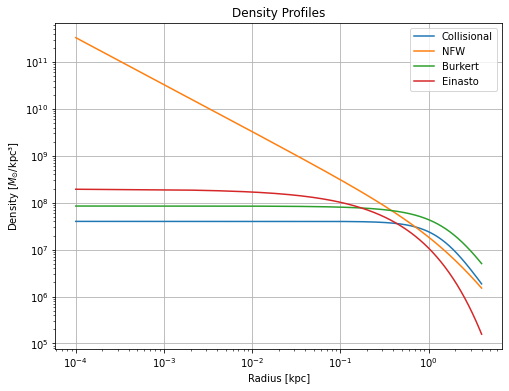}
\caption{The density of the collisional DM model (\ref{tanhmodel})
for the galaxy D512-2, as a function of the radius.}
\label{D512-2dens}
\end{figure}

\begin{figure}[h!]
\centering
\includegraphics[width=20pc]{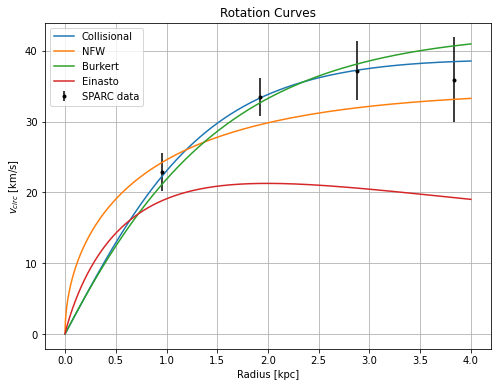}
\caption{The predicted rotation curves after using an optimization
for the collisional DM model (\ref{tanhmodel}), versus the SPARC
observational data for the galaxy D512-2. We also plotted the
optimized curves for the NFW model, the Burkert model and the
Einasto model.} \label{D512-2}
\end{figure}

\begin{figure}[h!]
\centering
\includegraphics[width=20pc]{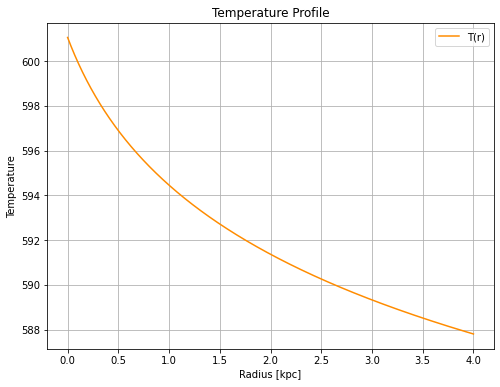}
\caption{The temperature as a function of the radius for the
collisional DM model (\ref{tanhmodel}) for the galaxy D512-2.}
\label{D512-2temp}
\end{figure}

For this galaxy, we shall choose $\rho_0=4\times
10^7$$M_{\odot}/\mathrm{Kpc}^{3}$. D512-2 is a small Magellanic/Sm
dwarf irregular galaxy (late-type dwarf). The assumed distance
used in the H\,I kinematics is $14.1 \pm 2.2 \ \mathrm{Mpc}$. In
Figs. \ref{D512-2dens}, \ref{D512-2} and \ref{D512-2temp} we
present the density of the collisional DM model, the for predicted
rotation curves after using an optimization for the collisional DM
model (\ref{tanhmodel}), versus the SPARC observational data and
the temperature parameter as a function of the radius
respectively. As it can be seen, the SIDM model produces viable
rotation curves compatible with the SPARC data. Also in Tables
\ref{collD512-2}, \ref{NavaroD512-2}, \ref{BuckertD512-2} and
\ref{EinastoD512-2} we present the optimization values for the
SIDM model, and the other DM profiles. Also in Table
\ref{EVALUATIOND512-2} we present the overall evaluation of the
SIDM model for the galaxy at hand. The resulting phenomenology is
viable.
\begin{table}[h!]
  \begin{center}
    \caption{Collisional Dark Matter Optimization Values}
    \label{collD512-2}
     \begin{tabular}{|r|r|}
     \hline
      \textbf{Parameter}   & \textbf{Optimization Values}
      \\  \hline
     $\delta_{\gamma} $ & 0.0000000012
\\  \hline
$\gamma_0 $ & 1.0001 \\  \hline $K_0$ ($M_{\odot} \,
\mathrm{Kpc}^{-3} \, (\mathrm{km/s})^{2}$)& 600
\\  \hline
    \end{tabular}
  \end{center}
\end{table}
\begin{table}[h!]
  \begin{center}
    \caption{NFW  Optimization Values}
    \label{NavaroD512-2}
     \begin{tabular}{|r|r|}
     \hline
      \textbf{Parameter}   & \textbf{Optimization Values}
      \\  \hline
   $\rho_s$   & $0.011\times 10^9$
\\  \hline
$r_s$&  3
\\  \hline
    \end{tabular}
  \end{center}
\end{table}
\begin{table}[h!]
  \begin{center}
    \caption{Burkert Optimization Values}
    \label{BuckertD512-2}
     \begin{tabular}{|r|r|}
     \hline
      \textbf{Parameter}   & \textbf{Optimization Values}
      \\  \hline
     $\rho_0^B$  & $0.085\times 10^9$
\\  \hline
$r_0$&  1.5
\\  \hline
    \end{tabular}
  \end{center}
\end{table}
\begin{table}[h!]
  \begin{center}
    \caption{Einasto Optimization Values}
    \label{EinastoD512-2}
    \begin{tabular}{|r|r|}
     \hline
      \textbf{Parameter}   & \textbf{Optimization Values}
      \\  \hline
     $\rho_e$  & $0.009\times 10^9$
\\  \hline
$r_e$ & 1.1
\\  \hline
$n_e$ & 0.65
\\  \hline
    \end{tabular}
  \end{center}
\end{table}
\begin{table}[h!]
\centering \caption{Physical assessment of collisional DM
parameters (D512-2).}
\begin{tabular}{lcc}
\hline
Parameter & Value  & Physical Verdict \\
\hline
$\gamma_0$ & $1.0001$ & Essentially isothermal  \\
$\delta_\gamma$ & $1.2\times10^{-9}$ & Vanishingly small variation   \\
$r_\gamma$ & $1.5\ \mathrm{Kpc}$ & Reasonable inner-halo scale  \\
$K_0$ & $6\times10^{2}$ & Moderate entropy/pressure scale. \\
$r_c$ & $0.5\ \mathrm{Kpc}$ & Small core radius \\
$p$ & $0.01$ & Almost flat $K(r)$; negligible radial change in entropy. \\
\hline Overall & - & Physically consistent and numerically
stable.\\ \hline
\end{tabular}
\label{EVALUATIOND512-2}
\end{table}


\subsection{The Galaxy D564-8}

\begin{figure}[h!]
\centering
\includegraphics[width=20pc]{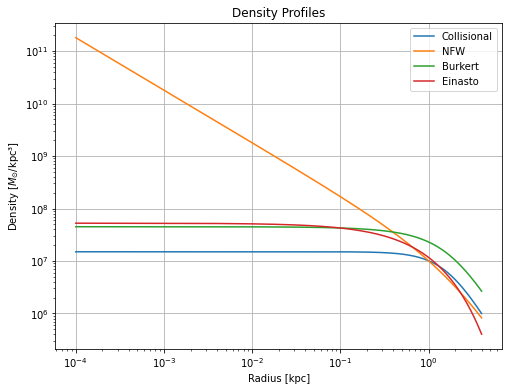}
\caption{The density of the collisional DM model (\ref{tanhmodel})
for the galaxy D564-8, as a function of the radius.}
\label{D564-8dens}
\end{figure}

\begin{figure}[h!]
\centering
\includegraphics[width=20pc]{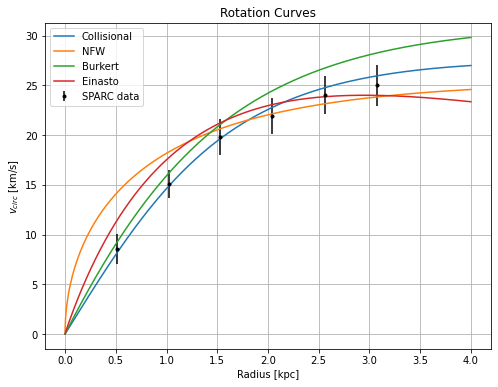}
\caption{The predicted rotation curves after using an optimization
for the collisional DM model (\ref{tanhmodel}), versus the SPARC
observational data for the galaxy D564-8. We also plotted the
optimized curves for the NFW model, the Burkert model and the
Einasto model.} \label{D564-8}
\end{figure}

\begin{figure}[h!]
\centering
\includegraphics[width=20pc]{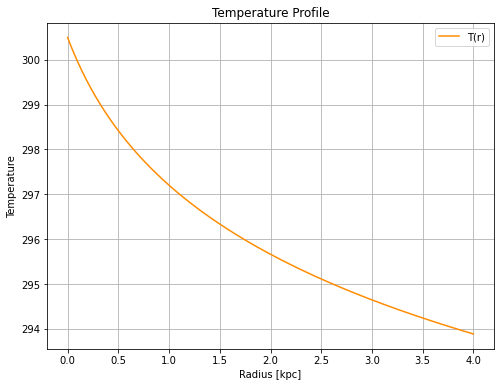}
\caption{The temperature as a function of the radius for the
collisional DM model (\ref{tanhmodel}) for the galaxy D564-8.}
\label{D564-8temp}
\end{figure}

For this galaxy, we shall choose $\rho_0=0.6\times
10^7$$M_{\odot}/\mathrm{Kpc}^{3}$. The galaxy D564-8 is classified
as a dwarf irregular galaxy. It was assumed to lie at a distance
of about 6.5 Mpc. It's a low mass, gas-rich late-type galaxy. In
Figs. \ref{D564-8dens}, \ref{D564-8} and \ref{D564-8temp} we
present the density of the collisional DM model, the for predicted
rotation curves after using an optimization for the collisional DM
model (\ref{tanhmodel}), versus the SPARC observational data and
the temperature parameter as a function of the radius
respectively. As it can be seen, the SIDM model produces viable
rotation curves compatible with the SPARC data. Also in Tables
\ref{collD564-8}, \ref{NavaroD564-8}, \ref{BuckertD564-8} and
\ref{EinastoD564-8} we present the optimization values for the
SIDM model, and the other DM profiles. Also in Table
\ref{EVALUATIOND564-8} we present the overall evaluation of the
SIDM model for the galaxy at hand. The resulting phenomenology is
viable.
\begin{table}[h!]
  \begin{center}
    \caption{Collisional Dark Matter Optimization Values}
    \label{collD564-8}
     \begin{tabular}{|r|r|}
     \hline
      \textbf{Parameter}   & \textbf{Optimization Values}
      \\  \hline
     $\delta_{\gamma} $ & 0.0000000012
\\  \hline
$\gamma_0 $ & 1.0001 \\ \hline $K_0$ ($M_{\odot} \,
\mathrm{Kpc}^{-3} \, (\mathrm{km/s})^{2}$)& 300
\\  \hline
    \end{tabular}
  \end{center}
\end{table}
\begin{table}[h!]
  \begin{center}
    \caption{NFW  Optimization Values}
    \label{NavaroD564-8}
     \begin{tabular}{|r|r|}
     \hline
      \textbf{Parameter}   & \textbf{Optimization Values}
      \\  \hline
   $\rho_s$   & $0.006\times 10^9$
\\  \hline
$r_s$&  3
\\  \hline
    \end{tabular}
  \end{center}
\end{table}
\begin{table}[h!]
  \begin{center}
    \caption{Burkert Optimization Values}
    \label{BuckertD564-8}
     \begin{tabular}{|r|r|}
     \hline
      \textbf{Parameter}   & \textbf{Optimization Values}
      \\  \hline
     $\rho_0^B$  & $0.045\times 10^9$
\\  \hline
$r_0$&  1.5
\\  \hline
    \end{tabular}
  \end{center}
\end{table}
\begin{table}[h!]
  \begin{center}
    \caption{Einasto Optimization Values}
    \label{EinastoD564-8}
    \begin{tabular}{|r|r|}
     \hline
      \textbf{Parameter}   & \textbf{Optimization Values}
      \\  \hline
     $\rho_e$  & $0.005\times 10^9$
\\  \hline
$r_e$ & 1.7
\\  \hline
$n_e$ & 0.85
\\  \hline
    \end{tabular}
  \end{center}
\end{table}
\begin{table}[h!]
\centering \caption{Physical assessment of collisional DM
parameters (D564-8).}
\begin{tabular}{lcc}
\hline
Parameter & Value  & Physical Verdict \\
\hline
$\gamma_0$ & $1.0001$ & Essentially isothermal  \\
$\delta_\gamma$ & $1.2\times10^{-9}$ & Vanishingly small variation   \\
$r_\gamma$ & $1.5\ \mathrm{Kpc}$ & Reasonable inner-halo scale  \\
$K_0$ & $3\times10^{2}$ & Moderate entropy/pressure scale. \\
$r_c$ & $0.5\ \mathrm{Kpc}$ & Small core radius  \\
$p$ & $0.01$ & Almost flat $K(r)$; negligible radial change in entropy. \\
\hline
Overall & - & Physically consistent and numerically stable.\\
\hline
\end{tabular}
\label{EVALUATIOND564-8}
\end{table}

\subsection{The Galaxy DDO168 Marginally Viable, the extended 3 Parameter Model Viable Density.}

\begin{figure}[h!]
\centering
\includegraphics[width=20pc]{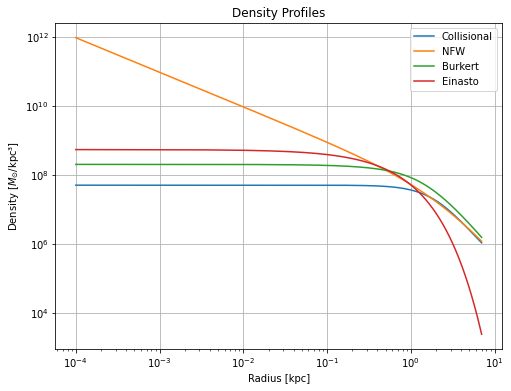}
\caption{The density of the collisional DM model (\ref{tanhmodel})
for the galaxy DDO168, as a function of the radius.}
\label{DDO168dens}
\end{figure}
\begin{figure}[h!]
\centering
\includegraphics[width=20pc]{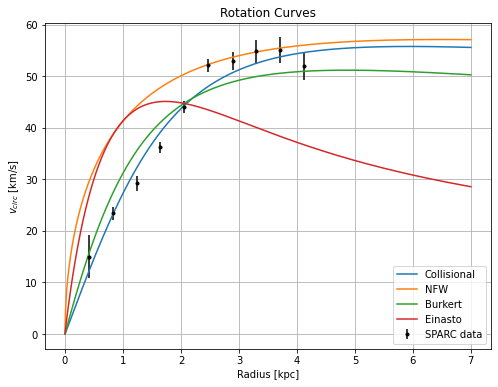}
\caption{The predicted rotation curves after using an optimization
for the collisional DM model (\ref{tanhmodel}), versus the SPARC
observational data for the galaxy DDO168. We also plotted the
optimized curves for the NFW model, the Burkert model and the
Einasto model.} \label{DDO168}
\end{figure}
\begin{figure}[h!]
\centering
\includegraphics[width=20pc]{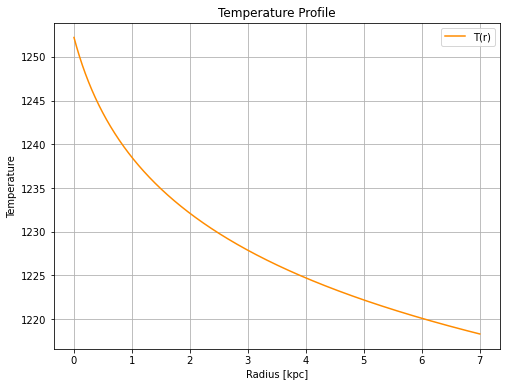}
\caption{The temperature as a function of the radius for the
collisional DM model (\ref{tanhmodel}) for the galaxy DDO168.}
\label{DDO168temp}
\end{figure}

For this galaxy, we shall choose $\rho_0=5\times
10^7$$M_{\odot}/\mathrm{Kpc}^{3}$. DDO\,168 is observed as a dwarf
irregular galaxy located at approximately $4.3\,\mathrm{Mpc}$. In
Figs. \ref{DDO168dens}, \ref{DDO168} and \ref{DDO168temp} we
present the density of the collisional DM model, the for predicted
rotation curves after using an optimization for the collisional DM
model (\ref{tanhmodel}), versus the SPARC observational data and
the temperature parameter as a function of the radius
respectively. As it can be seen, the SIDM model produces
marginally viable rotation curves compatible with the SPARC data.
Also in Tables \ref{collDDO168}, \ref{NavaroDDO168},
\ref{BuckertDDO168} and \ref{EinastoDDO168} we present the
optimization values for the SIDM model, and the other DM profiles.
Also in Table \ref{EVALUATIONDDO168} we present the overall
evaluation of the SIDM model for the galaxy at hand. The resulting
phenomenology is marginally viable.
\begin{table}[h!]
  \begin{center}
    \caption{Collisional Dark Matter Optimization Values}
    \label{collDDO168}
     \begin{tabular}{|r|r|}
     \hline
      \textbf{Parameter}   & \textbf{Optimization Values}
      \\  \hline
     $\delta_{\gamma} $ & 0.0000000012
\\  \hline
$\gamma_0 $ & 1.0001 \\  \hline $K_0$ ($M_{\odot} \,
\mathrm{Kpc}^{-3} \, (\mathrm{km/s})^{2}$)& 1250
\\  \hline
    \end{tabular}
  \end{center}
\end{table}
\begin{table}[h!]
  \begin{center}
    \caption{NFW  Optimization Values}
    \label{NavaroDDO168}
     \begin{tabular}{|r|r|}
     \hline
      \textbf{Parameter}   & \textbf{Optimization Values}
      \\  \hline
   $\rho_s$   & $0.031\times 10^9$
\\  \hline
$r_s$&  3
\\  \hline
    \end{tabular}
  \end{center}
\end{table}
\begin{table}[h!]
  \begin{center}
    \caption{Burkert Optimization Values}
    \label{BuckertDDO168}
     \begin{tabular}{|r|r|}
     \hline
      \textbf{Parameter}   & \textbf{Optimization Values}
      \\  \hline
     $\rho_0^B$  & $0.2\times 10^9$
\\  \hline
$r_0$&  1.5
\\  \hline
    \end{tabular}
  \end{center}
\end{table}

\begin{figure}[h!]
\centering
\includegraphics[width=20pc]{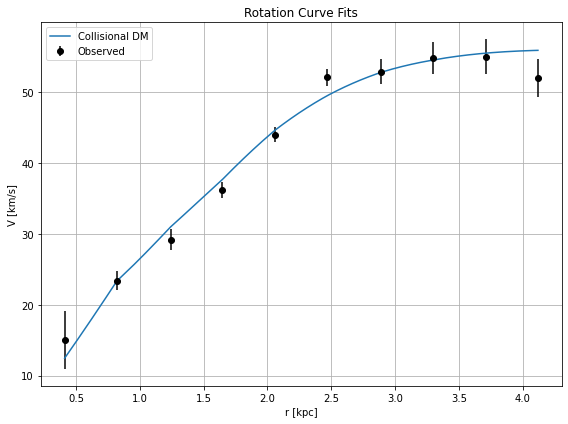}
\caption{The predicted rotation curves after using an optimization
for the collisional DM model (\ref{tanhmodel}), versus the
extended SPARC observational data for the galaxy DDO168. The model
includes the rotation curves from all the components of the
galaxy, including gas and disk velocities, along with the
collisional DM model.} \label{extendedDDO168}
\end{figure}

\begin{table}[h!]
  \begin{center}
    \caption{Einasto Optimization Values}
    \label{EinastoDDO168}
    \begin{tabular}{|r|r|}
     \hline
      \textbf{Parameter}   & \textbf{Optimization Values}
      \\  \hline
     $\rho_e$  & $0.051\times 10^9$
\\  \hline
$r_e$ & 1
\\  \hline
$n_e$ & 0.85
\\  \hline
    \end{tabular}
  \end{center}
\end{table}
\begin{table}[h!]
\centering \caption{Physical assessment of collisional DM
parameters (DDO-168).}
\begin{tabular}{lcc}
\hline
Parameter & Value  & Physical Verdict \\
\hline
$\gamma_0$ & $1.0001$ & Essentially isothermal  \\
$\delta_\gamma$ & $1.2\times10^{-9}$ & Vanishingly small  \\
$r_\gamma$ & $1.5\ \mathrm{Kpc}$ & Reasonable transition scale in principle  \\
$K_0$ & $1.25\times10^{3}$ & Moderate-to-large entropy/pressure scale. \\
$r_c$ & $0.5\ \mathrm{Kpc}$ & Small core radius  \\
$p$ & $0.01$ & Almost flat $K(r)$; negligible radial change in entropy. \\
Overall & - & Numerically stable and physically consistent. \\
\hline
\end{tabular}
\label{EVALUATIONDDO168}
\end{table}
Now the extended picture including the rotation velocity from the
other components of the galaxy, such as the disk and gas, makes
the collisional DM model viable for this galaxy. In Fig.
\ref{extendedDDO168} we present the combined rotation curves
including the other components of the galaxy along with the
collisional matter. As it can be seen, the extended collisional DM
model is viable.
Also in Table \ref{evaluationextendedDDO168} we present the values
of the free parameters of the collisional DM model for which the
maximum compatibility with the SPARC data comes for the galaxy
ESO116-G012.
\begin{table}[h!]
\centering \caption{Physical assessment of Extended collisional DM
parameters for DDO168.}
\begin{tabular}{lcc}
\hline
Parameter & Value & Physical Verdict \\
\hline
$\gamma_0$ & 0.98442775 & Moderately below isothermal, slightly stiffer core \\
$\delta_\gamma$ & 0.06937176 & Noticeable radial variation; $\gamma(r)$ increases outward \\
$K_0$ & 50 & Very low entropy \\
$ml_{\text{disk}}$ & 0.00000249 & Low disk contribution \\
$ml_{\text{bulge}}$ & 0.0000 & No bulge component; disk-dominated morphology consistent with a dwarf irregular galaxy \\
\hline
Overall & - & Physically plausible for a dwarf galaxy \\
\hline
\end{tabular}
\label{evaluationextendedDDO168}
\end{table}




\subsection{The Galaxy D631-7}

\begin{figure}[h!]
\centering
\includegraphics[width=20pc]{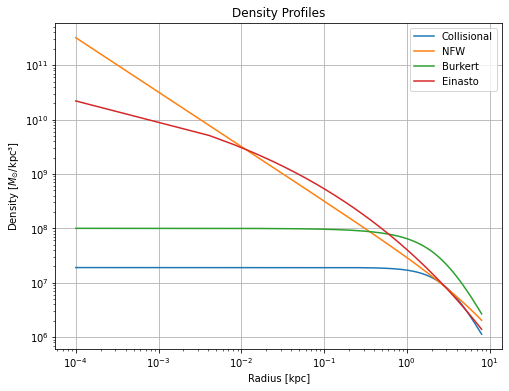}
\caption{The density of the collisional DM model (\ref{tanhmodel})
for the galaxy D631-7, as a function of the radius.}
\label{D631-7dens}
\end{figure}
\begin{figure}[h!]
\centering
\includegraphics[width=20pc]{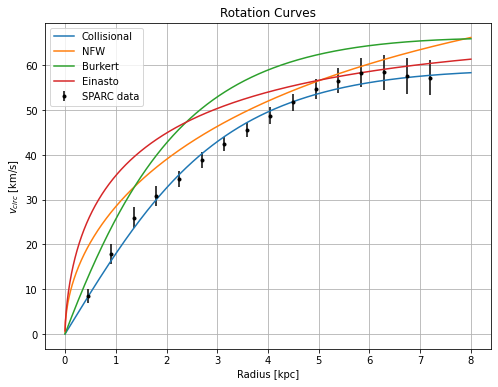}
\caption{The predicted rotation curves after using an optimization
for the collisional DM model (\ref{tanhmodel}), versus the SPARC
observational data for the galaxy D631-7. We also plotted the
optimized curves for the NFW model, the Burkert model and the
Einasto model.} \label{D631-7}
\end{figure}
\begin{figure}[h!]
\centering
\includegraphics[width=20pc]{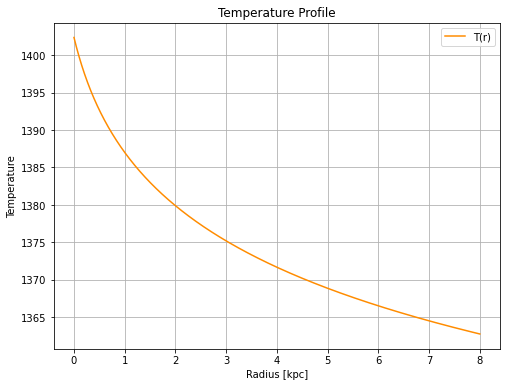}
\caption{The temperature as a function of the radius for the
collisional DM model (\ref{tanhmodel}) for the galaxy D631-7.}
\label{D631-7temp}
\end{figure}

For this galaxy, we shall choose $\rho_0=1.9\times
10^7$$M_{\odot}/\mathrm{Kpc}^{3}$. D631-7, also known as UGC 4115,
is classified as a dwarf irregular galaxy. Its distance is about
$5.5 \pm 0.6$ Mpc. The optical radius ($R_{25}$) is approximately
$0.8$ Kpc. In Figs. \ref{D631-7dens}, \ref{D631-7} and
\ref{D631-7temp} we present the density of the collisional DM
model, the for predicted rotation curves after using an
optimization for the collisional DM model (\ref{tanhmodel}),
versus the SPARC observational data and the temperature parameter
as a function of the radius respectively. As it can be seen, the
SIDM model produces viable rotation curves compatible with the
SPARC data. Also in Tables \ref{collD631-7}, \ref{NavaroD631-7},
\ref{BuckertD631-7} and \ref{EinastoD631-7} we present the
optimization values for the SIDM model, and the other DM profiles.
Also in Table \ref{EVALUATIOND631-7} we present the overall
evaluation of the SIDM model for the galaxy at hand. The resulting
phenomenology is viable.
\begin{table}[h!]
  \begin{center}
    \caption{Collisional Dark Matter Optimization Values}
    \label{collD631-7}
     \begin{tabular}{|r|r|}
     \hline
      \textbf{Parameter}   & \textbf{Optimization Values}
      \\  \hline
     $\delta_{\gamma} $ & 0.0000000012
\\  \hline
$\gamma_0 $ & 1.0001 \\ \hline $K_0$ ($M_{\odot} \,
\mathrm{Kpc}^{-3} \, (\mathrm{km/s})^{2}$)& 1400
\\  \hline
    \end{tabular}
  \end{center}
\end{table}
\begin{table}[h!]
  \begin{center}
    \caption{NFW  Optimization Values}
    \label{NavaroD631-7}
     \begin{tabular}{|r|r|}
     \hline
      \textbf{Parameter}   & \textbf{Optimization Values}
      \\  \hline
   $\rho_s$   & $0.0016\times 10^9$
\\  \hline
$r_s$&  20
\\  \hline
    \end{tabular}
  \end{center}
\end{table}
\begin{table}[h!]
  \begin{center}
    \caption{Burkert Optimization Values}
    \label{BuckertD631-7}
     \begin{tabular}{|r|r|}
     \hline
      \textbf{Parameter}   & \textbf{Optimization Values}
      \\  \hline
     $\rho_0^B$  & $0.1\times 10^9$
\\  \hline
$r_0$&  2.74
\\  \hline
    \end{tabular}
  \end{center}
\end{table}
\begin{table}[h!]
  \begin{center}
    \caption{Einasto Optimization Values}
    \label{EinastoD631-7}
    \begin{tabular}{|r|r|}
     \hline
      \textbf{Parameter}   & \textbf{Optimization Values}
      \\  \hline
     $\rho_e$  & $0.0009\times 10^9$
\\  \hline
$r_e$ & 10
\\  \hline
$n_e$ & 0.17
\\  \hline
    \end{tabular}
  \end{center}
\end{table}
\begin{table}[h!]
\centering \caption{Physical assessment of collisional DM
parameters (D631-7).}
\begin{tabular}{lcc}
\hline
Parameter & Value   & Physical Verdict \\
\hline
$\gamma_0$ & $1.0001$ & Essentially isothermal  \\
$\delta_\gamma$ & $1.2\times10^{-9}$ & Vanishingly small variation \\
$r_\gamma$ & $1.5\ \mathrm{Kpc}$ & Reasonable scale in principle  \\
$K_0$ & $3\times10^{2}$ & Moderate entropy/pressure scale. \\
$r_c$ & $0.5\ \mathrm{Kpc}$ & Small core radius  \\
$p$ & $0.01$ & Almost flat $K(r)$; negligible radial change in entropy. \\
\hline Overall & - & Numerically stable and physically
consistent.\\ \hline
\end{tabular}
\label{EVALUATIOND631-7}
\end{table}


\subsection{The Galaxy DDO064}

\begin{figure}[h!]
\centering
\includegraphics[width=20pc]{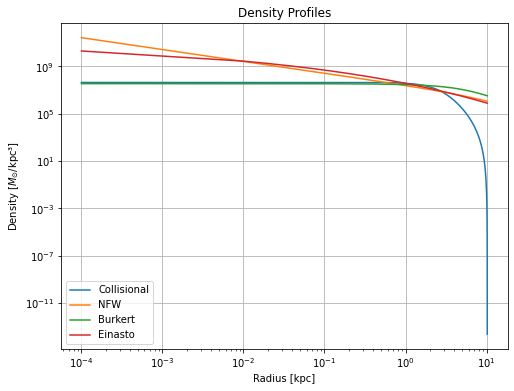}
\caption{The density of the collisional DM model (\ref{tanhmodel})
for the galaxy DDO064, as a function of the radius.}
\label{DDO064dens}
\end{figure}
\begin{figure}[h!]
\centering
\includegraphics[width=20pc]{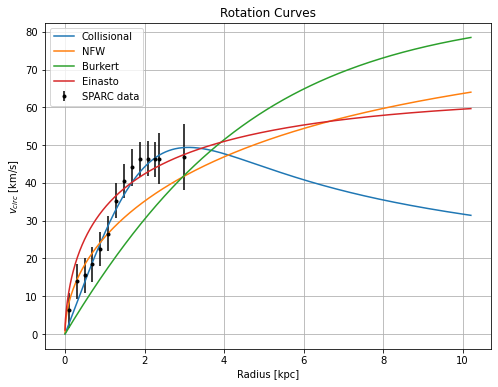}
\caption{The predicted rotation curves after using an optimization
for the collisional DM model (\ref{tanhmodel}), versus the SPARC
observational data for the galaxy DDO064. We also plotted the
optimized curves for the NFW model, the Burkert model and the
Einasto model.} \label{DDO064}
\end{figure}
\begin{figure}[h!]
\centering
\includegraphics[width=20pc]{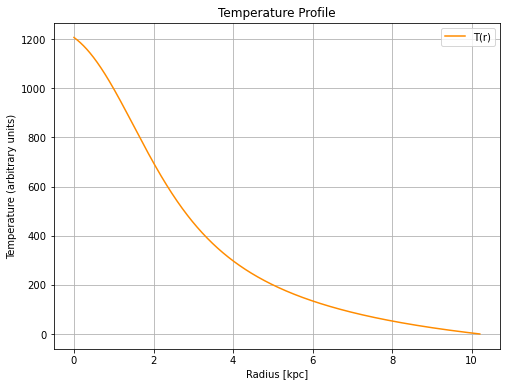}
\caption{The temperature as a function of the radius for the
collisional DM model (\ref{tanhmodel}) for the galaxy DDO064.}
\label{DDO064temp}
\end{figure}

For this galaxy, we shall choose $\rho_0=4.3\times
10^7$$M_{\odot}/\mathrm{Kpc}^{3}$ DDO064 is a dwarf irregular
galaxy, a low-luminosity, gas-rich system with an irregular
optical appearance and an extended HI disk. Its distance from the
Milky Way is about 7.5 Mpc. In Figs. \ref{DDO064dens},
\ref{DDO064} and \ref{DDO064temp} we present the density of the
collisional DM model, the for predicted rotation curves after
using an optimization for the collisional DM model
(\ref{tanhmodel}), versus the SPARC observational data and the
temperature parameter as a function of the radius respectively. As
it can be seen, the SIDM model produces viable rotation curves
compatible with the SPARC data. Also in Tables \ref{collDDO064},
\ref{NavaroDDO064}, \ref{BuckertDDO064} and \ref{EinastoDDO064} we
present the optimization values for the SIDM model, and the other
DM profiles. Also in Table \ref{EVALUATIONDDO064} we present the
overall evaluation of the SIDM model for the galaxy at hand. The
resulting phenomenology is viable.
Optimization values:
\begin{table}[h!]
  \begin{center}
    \caption{Collisional Dark Matter Optimization Values}
    \label{collDDO064}
     \begin{tabular}{|r|r|}
     \hline
      \textbf{Parameter}   & \textbf{Optimization Values}
      \\  \hline
     $\delta_{\gamma} $ & 0.02
\\  \hline
$\gamma_0 $ & 1.26
\\  \hline
    \end{tabular}
  \end{center}
\end{table}
\begin{table}[h!]
  \begin{center}
    \caption{NFW  Optimization Values}
    \label{NavaroDDO064}
     \begin{tabular}{|r|r|}
     \hline
      \textbf{Parameter}   & \textbf{Optimization Values}
      \\  \hline
   $\rho_s$   & $0.0013\times 10^9$
\\  \hline
$r_s$&  20
\\  \hline
    \end{tabular}
  \end{center}
\end{table}
\begin{table}[h!]
  \begin{center}
    \caption{Burkert Optimization Values}
    \label{BuckertDDO064}
     \begin{tabular}{|r|r|}
     \hline
      \textbf{Parameter}   & \textbf{Optimization Values}
      \\  \hline
     $\rho_0^B$  & $0.034\times 10^9$
\\  \hline
$r_0$&  6
\\  \hline
    \end{tabular}
  \end{center}
\end{table}
\begin{table}[h!]
  \begin{center}
    \caption{Einasto Optimization Values}
    \label{EinastoDDO064}
    \begin{tabular}{|r|r|}
     \hline
      \textbf{Parameter}   & \textbf{Optimization Values}
      \\  \hline
     $\rho_e$  & $0.0008\times 10^9$
\\  \hline
$r_e$ & 10
\\  \hline
$n_e$ & 0.17
\\  \hline
    \end{tabular}
  \end{center}
\end{table}
\begin{table}[h!]
\centering \caption{Physical assessment of collisional-DM
parameters for DDO064.}
\begin{tabular}{lcc}
\hline
Parameter & Value & Physical verdict \\
\hline
$\gamma_0$ & $1.26$ & Mildly stiffer than isothermal ($\gamma=1$) \\
$\delta_\gamma$ & $0.02$ & Very small radial variation  \\
$r_\gamma$ & $1.5\ \mathrm{Kpc}$ & Transition radius in the observable inner halo \\
$K_0$ & $10$ & Moderate entropy  \\
$r_c$ & $0.5\ \mathrm{Kpc}$ & Small core scale in $K(r)$ \\
$p$ & $0.01$ & Nearly flat $K(r)$ - EoS nearly spatially uniform \\
\hline Overall & -- & Physically plausible for a dwarf\\ \hline
\end{tabular}
\label{EVALUATIONDDO064}
\end{table}


\subsection{The Galaxy DDO154}

\begin{figure}[h!]
\centering
\includegraphics[width=20pc]{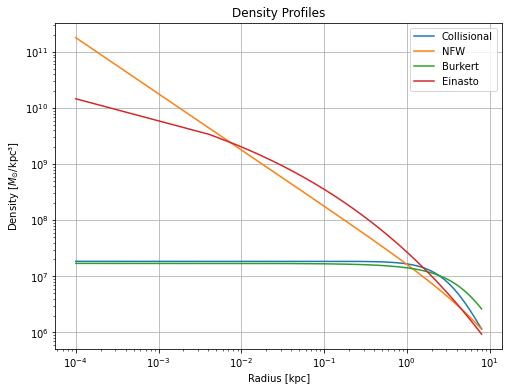}
\caption{The density of the collisional DM model (\ref{tanhmodel})
for the galaxy DDO154, as a function of the radius.}
\label{DDO154dens}
\end{figure}
\begin{figure}[h!]
\centering
\includegraphics[width=20pc]{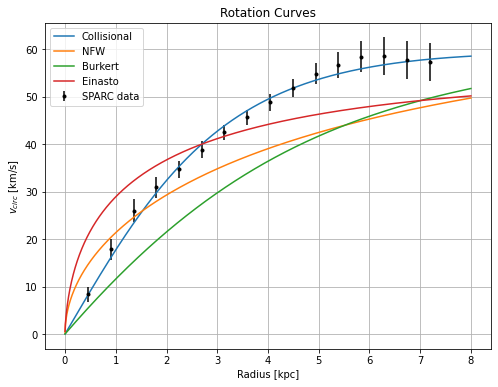}
\caption{The predicted rotation curves after using an optimization
for the collisional DM model (\ref{tanhmodel}), versus the SPARC
observational data for the galaxy DDO154. We also plotted the
optimized curves for the NFW model, the Burkert model and the
Einasto model.} \label{DDO154}
\end{figure}
\begin{figure}[h!]
\centering
\includegraphics[width=20pc]{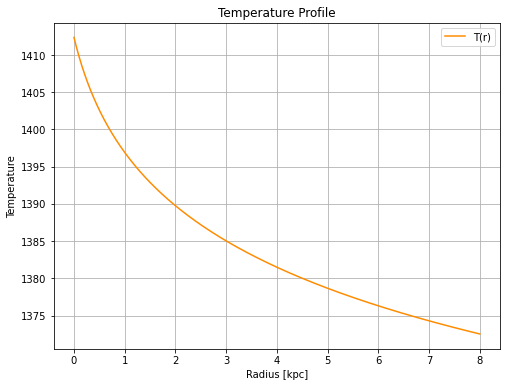}
\caption{The temperature as a function of the radius for the
collisional DM model (\ref{tanhmodel}) for the galaxy DDO154.}
\label{DDO154temp}
\end{figure}

For this galaxy, we shall choose $\rho_0=1.85\times
10^7$$M_{\odot}/\mathrm{Kpc}^{3}$. DDO\,154 is a well-studied
nearby dwarf irregular galaxy, and is considered one of the
archetypal gas-rich systems with a slowly rising rotation curve
dominated by dark matter.  Its distance is about $4.3$ Mpc,
placing it in the local volume of galaxies beyond the Local Group.
The morphological type is dwarf irregular. In Figs.
\ref{DDO154dens}, \ref{DDO154} and \ref{DDO154temp} we present the
density of the collisional DM model, the for predicted rotation
curves after using an optimization for the collisional DM model
(\ref{tanhmodel}), versus the SPARC observational data and the
temperature parameter as a function of the radius respectively. As
it can be seen, the SIDM model produces viable rotation curves
compatible with the SPARC data. Also in Tables \ref{collDDO154},
\ref{NavaroDDO154}, \ref{BuckertDDO154} and \ref{EinastoDDO154} we
present the optimization values for the SIDM model, and the other
DM profiles. Also in Table \ref{EVALUATIONDDO154} we present the
overall evaluation of the SIDM model for the galaxy at hand. The
resulting phenomenology is viable.
\begin{table}[h!]
  \begin{center}
    \caption{Collisional Dark Matter Optimization Values}
    \label{collDDO154}
     \begin{tabular}{|r|r|}
     \hline
      \textbf{Parameter}   & \textbf{Optimization Values}
      \\  \hline
     $\delta_{\gamma} $ & 0.0000000012
\\  \hline
$\gamma_0 $ & 1.0001 \\ \hline $K_0$ ($M_{\odot} \,
\mathrm{Kpc}^{-3} \, (\mathrm{km/s})^{2}$)& 1410
\\  \hline
    \end{tabular}
  \end{center}
\end{table}
\begin{table}[h!]
  \begin{center}
    \caption{NFW  Optimization Values}
    \label{NavaroDDO154}
     \begin{tabular}{|r|r|}
     \hline
      \textbf{Parameter}   & \textbf{Optimization Values}
      \\  \hline
   $\rho_s$   & $0.0009\times 10^9$
\\  \hline
$r_s$&  20
\\  \hline
    \end{tabular}
  \end{center}
\end{table}
\begin{table}[h!]
  \begin{center}
    \caption{Burkert Optimization Values}
    \label{BuckertDDO154}
     \begin{tabular}{|r|r|}
     \hline
      \textbf{Parameter}   & \textbf{Optimization Values}
      \\  \hline
     $\rho_0^B$  & $0.017\times 10^9$
\\  \hline
$r_0$&  6
\\  \hline
    \end{tabular}
  \end{center}
\end{table}
\begin{table}[h!]
  \begin{center}
    \caption{Einasto Optimization Values}
    \label{EinastoDDO154}
    \begin{tabular}{|r|r|}
     \hline
      \textbf{Parameter}   & \textbf{Optimization Values}
      \\  \hline
     $\rho_e$  & $0.0006\times 10^9$
\\  \hline
$r_e$ & 10
\\  \hline
$n_e$ & 0.17
\\  \hline
    \end{tabular}
  \end{center}
\end{table}
\begin{table}[h!]
\centering \caption{Physical assessment of collisional DM
parameters for DDO154.}
\begin{tabular}{lcc}
\hline
Parameter & Value & Physical verdict \\
\hline
$\gamma_0$ & $1.271$ & Moderately above isothermal \\
$\delta_{\gamma}$ & $1\times10^{-5}$ & Practically zero \\
$r_{\gamma}$ & $1.5\ \mathrm{Kpc}$ & Transition radius nominally inside inner halo \\
$K_0$ & $\simeq 10$ & Moderate entropy/temperature scale \\
$r_c$ & $0.5\ \mathrm{Kpc}$ & Small core radius for $K(r)$  \\
$p$ & $0.01$ & Very shallow radial decline of $K(r)$ \\
\hline Overall & - & Physically plausible for a cored dwarf halo
\\ \hline
\end{tabular}
\label{EVALUATIONDDO154}
\end{table}


\subsection{The Galaxy DDO161 Marginally- Non-viable Dwarf, Extended Marginally no improvement}

\begin{figure}[h!]
\centering
\includegraphics[width=20pc]{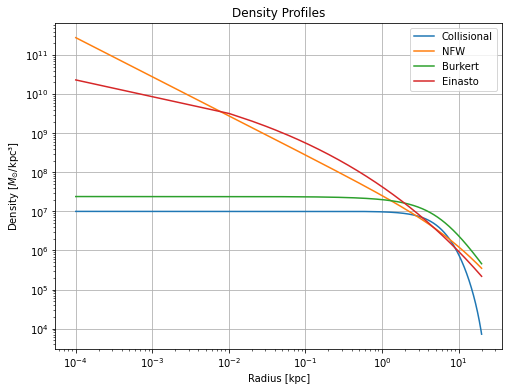}
\caption{The density of the collisional DM model (\ref{tanhmodel})
for the galaxy DDO161, as a function of the radius.}
\label{DDO161dens}
\end{figure}
\begin{figure}[h!]
\centering
\includegraphics[width=20pc]{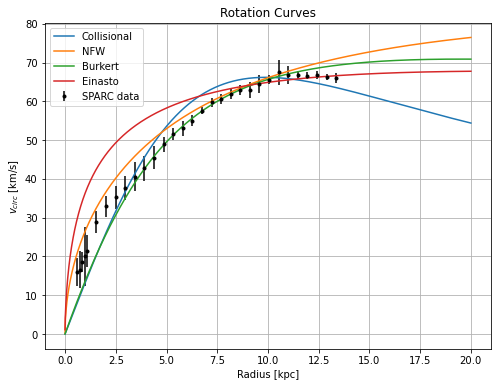}
\caption{The predicted rotation curves after using an optimization
for the collisional DM model (\ref{tanhmodel}), versus the SPARC
observational data for the galaxy DDO161. We also plotted the
optimized curves for the NFW model, the Burkert model and the
Einasto model.} \label{DDO161}
\end{figure}
\begin{figure}[h!]
\centering
\includegraphics[width=20pc]{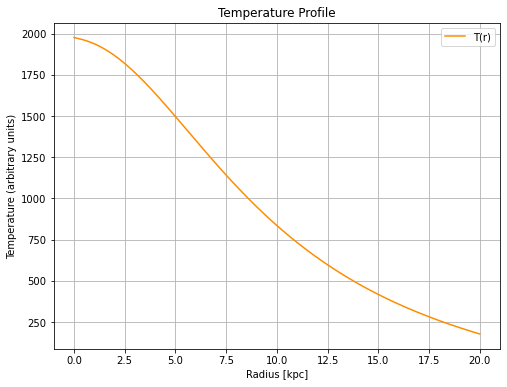}
\caption{The temperature as a function of the radius for the
collisional DM model (\ref{tanhmodel}) for the galaxy DDO161.}
\label{DDO161temp}
\end{figure}
\begin{figure}[h!]
\centering
\includegraphics[width=20pc]{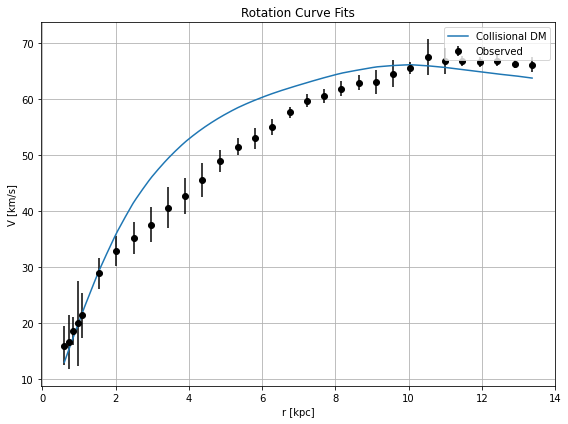}
\caption{The predicted rotation curves after using an optimization
for the collisional DM model (\ref{tanhmodel}), versus the
extended SPARC observational data for the galaxy DDO161. The model
includes the rotation curves from all the components of the
galaxy, including gas and disk velocities, along with the
collisional DM model.} \label{extendedDDO161}
\end{figure}

For this galaxy, we shall choose
$\rho_0=10^7$$M_{\odot}/\mathrm{Kpc}^{3}$. DDO\,161 is a gas-rich
dwarf galaxy of morphological type Sm, often classified as a dwarf
irregular. It lies at a distance of $D =
6.03^{+0.29}_{-0.21}\,\mathrm{Mpc}$. The system forms an isolated
pair together with UGCA\,319 and is characterized by active star
formation and a substantial neutral hydrogen content. In Figs.
\ref{DDO161dens}, \ref{DDO161} and \ref{DDO161temp} we present the
density of the collisional DM model, the for predicted rotation
curves after using an optimization for the collisional DM model
(\ref{tanhmodel}), versus the SPARC observational data and the
temperature parameter as a function of the radius respectively. As
it can be seen, the SIDM model produces marginally viable rotation
curves compared to the SPARC data. Also in Tables
\ref{collDDO161}, \ref{NavaroDDO161}, \ref{BuckertDDO161} and
\ref{EinastoDDO161} we present the optimization values for the
SIDM model, and the other DM profiles. Also in Table
\ref{EVALUATIONDDO161} we present the overall evaluation of the
SIDM model for the galaxy at hand. The resulting phenomenology is
marginally viable.
\begin{table}[h!]
  \begin{center}
    \caption{Collisional Dark Matter Optimization Values}
    \label{collDDO161}
     \begin{tabular}{|r|r|}
     \hline
      \textbf{Parameter}   & \textbf{Optimization Values}
      \\  \hline
     $\delta_{\gamma} $ & 0.0000000001
\\  \hline
$\gamma_0 $ & 1.328
\\  \hline
    \end{tabular}
  \end{center}
\end{table}
\begin{table}[h!]
  \begin{center}
    \caption{NFW  Optimization Values}
    \label{NavaroDDO161}
     \begin{tabular}{|r|r|}
     \hline
      \textbf{Parameter}   & \textbf{Optimization Values}
      \\  \hline
   $\rho_s$   & $0.0014\times 10^9$
\\  \hline
$r_s$&  20
\\  \hline
    \end{tabular}
  \end{center}
\end{table}
\begin{table}[h!]
  \begin{center}
    \caption{Burkert Optimization Values}
    \label{BuckertDDO161}
     \begin{tabular}{|r|r|}
     \hline
      \textbf{Parameter}   & \textbf{Optimization Values}
      \\  \hline
     $\rho_0^B$  & $0.024\times 10^9$
\\  \hline
$r_0$&  6
\\  \hline
    \end{tabular}
  \end{center}
\end{table}
\begin{table}[h!]
  \begin{center}
    \caption{Einasto Optimization Values}
    \label{EinastoDDO161}
    \begin{tabular}{|r|r|}
     \hline
      \textbf{Parameter}   & \textbf{Optimization Values}
      \\  \hline
     $\rho_e$  & $0.00095\times 10^9$
\\  \hline
$r_e$ & 10
\\  \hline
$n_e$ & 0.17
\\  \hline
    \end{tabular}
  \end{center}
\end{table}
\begin{table}[h!]
\centering \caption{Physical assessment of collisional DM
parameters (DDO161).}
\begin{tabular}{lcc}
\hline
Parameter & Value & Physical Verdict \\
\hline
$\gamma_0$ & $1.328$ & Reasonable (between isothermal and adiabatic) \\
$\delta_\gamma$ & $1\times10^{-10}$ & Practically zero  \\
$r_\gamma$ & $1.5\ \mathrm{Kpc}$ & Transition radius chosen inside inner halo; acceptable \\
$K_0$ & $10$ & Numerical entropy scale \\
$r_c$ & $0.5\ \mathrm{Kpc}$ & Small core scale, plausible for dwarf halo modelling \\
$p$ & $0.01$ & Nearly constant $K(r)$ \\
\hline
Overall & --- & Physically plausible baseline \\
\hline
\end{tabular}
\label{EVALUATIONDDO161}
\end{table}
Now the extended picture including the rotation velocity from the
other components of the galaxy, such as the disk and gas, makes
the collisional DM model viable for this galaxy. In Fig.
\ref{extendedDDO161} we present the combined rotation curves
including the other components of the galaxy along with the
collisional matter. As it can be seen, the extended collisional DM
model is marginally viable.
Also in Table \ref{evaluationextendedDDO161} we present the values
of the free parameters of the collisional DM model for which the
marginal  maximum compatibility with the SPARC data comes for the
galaxy DDO161.
\begin{table}[h!]
\centering \caption{Physical assessment of Extended collisional DM
parameters for galaxy DDO161.}
\begin{tabular}{lcc}
\hline
Parameter & Value & Physical Verdict \\
\hline
$\gamma_0$ & 0.94913086 & Slightly below isothermal; suggests softer core, lower central pressure \\
$\delta_\gamma$ & 0.0001 & Negligible variation  with radius \\
$K_0$ & 3000 & Moderate entropy scale; consistent with low-mass dwarf dark matter halos \\
$ml_{disk}$ & 0.5 & Subdominant stellar disk; DM dominates the rotation curve \\
$ml_{bulge}$ & 0.00000000 & No bulge contribution; typical for irregular dwarf systems \\
\hline
Overall & - & Physically viable; nearly isothermal halo with low baryonic influence \\
\hline
\end{tabular}
\label{evaluationextendedDDO161}
\end{table}


\subsection{The Galaxy DDO170}

\begin{figure}[h!]
\centering
\includegraphics[width=20pc]{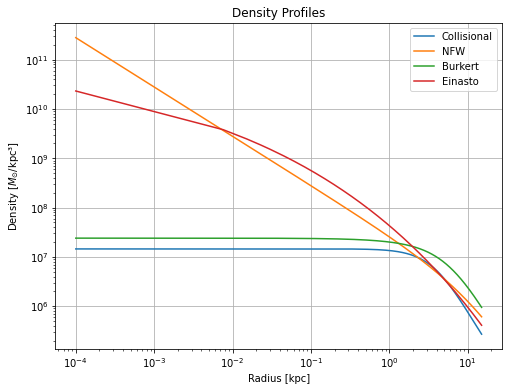}
\caption{The density of the collisional DM model (\ref{tanhmodel})
for the galaxy DDO170, as a function of the radius.}
\label{DDO170dens}
\end{figure}

\begin{figure}[h!]
\centering
\includegraphics[width=20pc]{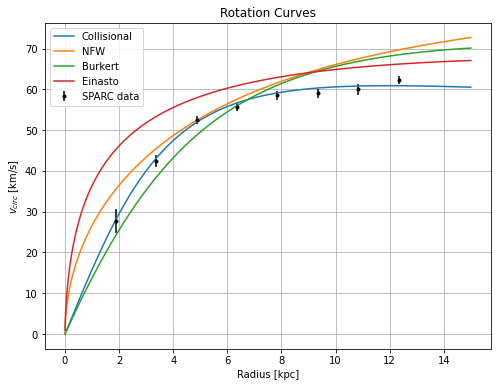}
\caption{The predicted rotation curves after using an optimization
for the collisional DM model (\ref{tanhmodel}), versus the SPARC
observational data for the galaxy DDO170. We also plotted the
optimized curves for the NFW model, the Burkert model and the
Einasto model.} \label{DDO170}
\end{figure}

\begin{figure}[h!]
\centering
\includegraphics[width=20pc]{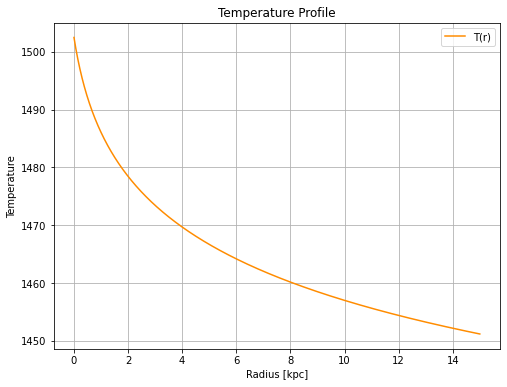}
\caption{The temperature as a function of the radius for the
collisional DM model (\ref{tanhmodel}) for the galaxy DDO170.}
\label{DDO170temp}
\end{figure}

For this galaxy, we shall choose $\rho_0=1.45\times
10^7$$M_{\odot}/\mathrm{Kpc}^{3}$. DDO\,170 is a gas-rich dwarf
irregular galaxy, often described as a faint late-type system with
a strong dark matter dominance. In Figs. \ref{DDO170dens},
\ref{DDO170} and \ref{DDO170temp} we present the density of the
collisional DM model, the for predicted rotation curves after
using an optimization for the collisional DM model
(\ref{tanhmodel}), versus the SPARC observational data and the
temperature parameter as a function of the radius respectively. As
it can be seen, the SIDM model produces viable rotation curves
compatible with the SPARC data. Also in Tables \ref{collDDO170},
\ref{NavaroDDO170}, \ref{BuckertDDO170} and \ref{EinastoDDO170} we
present the optimization values for the SIDM model, and the other
DM profiles. Also in Table \ref{EVALUATIONDDO170} we present the
overall evaluation of the SIDM model for the galaxy at hand. The
resulting phenomenology is viable.
\begin{table}[h!]
  \begin{center}
    \caption{Collisional Dark Matter Optimization Values}
    \label{collDDO170}
     \begin{tabular}{|r|r|}
     \hline
      \textbf{Parameter}   & \textbf{Optimization Values}
      \\  \hline
     $\delta_{\gamma} $ & 0.0000000012
\\  \hline
$\gamma_0 $ & 1.0001 \\ \hline $K_0$ ($M_{\odot} \,
\mathrm{Kpc}^{-3} \, (\mathrm{km/s})^{2}$)& 1500
\\  \hline
    \end{tabular}
  \end{center}
\end{table}
\begin{table}[h!]
  \begin{center}
    \caption{NFW  Optimization Values}
    \label{NavaroDDO170}
     \begin{tabular}{|r|r|}
     \hline
      \textbf{Parameter}   & \textbf{Optimization Values}
      \\  \hline
   $\rho_s$   & $0.0014\times 10^9$
\\  \hline
$r_s$&  20
\\  \hline
    \end{tabular}
  \end{center}
\end{table}
\begin{table}[h!]
  \begin{center}
    \caption{Burkert Optimization Values}
    \label{BuckertDDO170}
     \begin{tabular}{|r|r|}
     \hline
      \textbf{Parameter}   & \textbf{Optimization Values}
      \\  \hline
     $\rho_0^B$  & $0.024\times 10^9$
\\  \hline
$r_0$&  6
\\  \hline
    \end{tabular}
  \end{center}
\end{table}
\begin{table}[h!]
  \begin{center}
    \caption{Einasto Optimization Values}
    \label{EinastoDDO170}
    \begin{tabular}{|r|r|}
     \hline
      \textbf{Parameter}   & \textbf{Optimization Values}
      \\  \hline
     $\rho_e$  & $0.00095\times 10^9$
\\  \hline
$r_e$ & 10
\\  \hline
$n_e$ & 0.17
\\  \hline
    \end{tabular}
  \end{center}
\end{table}
\begin{table}[h!]
\centering \caption{Physical assessment of collisional DM
parameters (DDO170).}
\begin{tabular}{lcc}
\hline
Parameter & Value & Physical Verdict \\
\hline
$\gamma_0$ & $1.31$ & Acceptable: between isothermal and adiabatic \\
$\delta_\gamma$ & $1\times10^{-14}$ & Practically zero \\
$r_\gamma$ & $1.5\ \mathrm{Kpc}$ & Transition radius placed inside inner halo \\
$K_0$ & $10$ & Numerical entropy/pressure scale \\
$r_c$ & $0.5\ \mathrm{Kpc}$ & Small core scale consistent with dwarf-halo core modelling \\
$p$ & $0.01$ & Nearly zero: $K(r)$ effectively constant, no radial entropy gradient \\
\hline
Overall & --- & Baseline single-polytrope is physically plausible \\
\hline
\end{tabular}
\label{EVALUATIONDDO170}
\end{table}


\subsection{The Galaxy ESO079-G014 Non-viable}

\begin{figure}[h!]
\centering
\includegraphics[width=20pc]{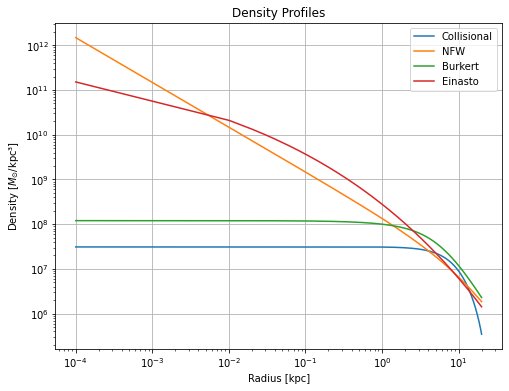}
\caption{The density of the collisional DM model (\ref{tanhmodel})
for the galaxy ESO079-G014, as a function of the radius.}
\label{ESO079-G014dens}
\end{figure}
\begin{figure}[h!]
\centering
\includegraphics[width=20pc]{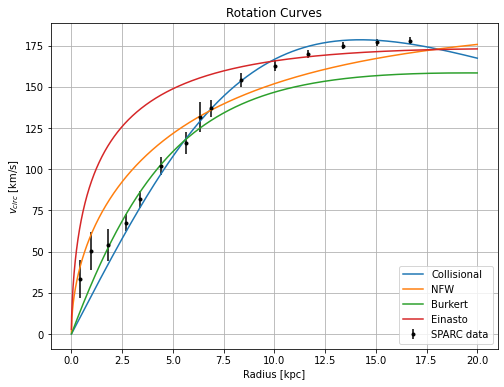}
\caption{The predicted rotation curves after using an optimization
for the collisional DM model (\ref{tanhmodel}), versus the SPARC
observational data for the galaxy ESO079-G014. We also plotted the
optimized curves for the NFW model, the Burkert model and the
Einasto model.} \label{ESO079-G014}
\end{figure}
\begin{figure}[h!]
\centering
\includegraphics[width=20pc]{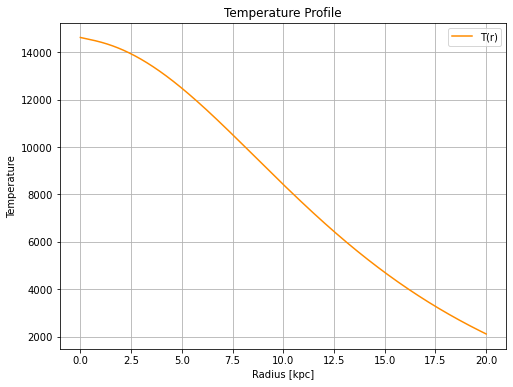}
\caption{The temperature as a function of the radius for the
collisional DM model (\ref{tanhmodel}) for the galaxy
ESO079-G014.} \label{ESO079-G014temp}
\end{figure}
\begin{figure}[h!]
\centering
\includegraphics[width=20pc]{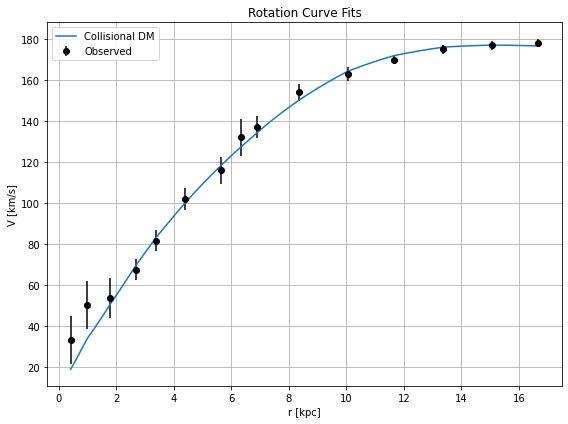}
\caption{The predicted rotation curves after using an optimization
for the collisional DM model (\ref{tanhmodel}), versus the
extended SPARC observational data for the galaxy ESO079-G014. The
model includes the rotation curves from all the components of the
galaxy, including gas and disk velocities, along with the
collisional DM model.} \label{extendedESO079-G014}
\end{figure}

For this galaxy, we shall choose $\rho_0=3.1\times
10^7$$M_{\odot}/\mathrm{Kpc}^{3}$. ESO079-G014 is a spiral galaxy
included in the SPARC sample, at a distance of about
\(31.6\pm3.0\) Mpc. It exhibits a rotation-dominated disk and
extended H I (so it is a late-type spiral). In Figs.
\ref{ESO079-G014dens}, \ref{ESO079-G014} and \ref{ESO079-G014temp}
we present the density of the collisional DM model, the for
predicted rotation curves after using an optimization for the
collisional DM model (\ref{tanhmodel}), versus the SPARC
observational data and the temperature parameter as a function of
the radius respectively. As it can be seen, the SIDM model
produces non-viable rotation curves overall incompatible with the
SPARC data. Also in Tables \ref{collESO079-G014},
\ref{NavaroESO079-G014}, \ref{BuckertESO079-G014} and
\ref{EinastoESO079-G014} we present the optimization values for
the SIDM model, and the other DM profiles. Also in Table
\ref{EVALUATIONESO079-G014} we present the overall evaluation of
the SIDM model for the galaxy at hand. The resulting phenomenology
is non-viable.
\begin{table}[h!]
  \begin{center}
    \caption{Collisional Dark Matter Optimization Values}
    \label{collESO079-G014}
     \begin{tabular}{|r|r|}
     \hline
      \textbf{Parameter}   & \textbf{Optimization Values}
      \\  \hline
     $\delta_{\gamma} $ & 0.0000000002
\\  \hline
$\gamma_0 $ & 1.4225
\\  \hline
    \end{tabular}
  \end{center}
\end{table}
\begin{table}[h!]
  \begin{center}
    \caption{NFW  Optimization Values}
    \label{NavaroESO079-G014}
     \begin{tabular}{|r|r|}
     \hline
      \textbf{Parameter}   & \textbf{Optimization Values}
      \\  \hline
   $\rho_s$   & $0.0074\times 10^9$
\\  \hline
$r_s$&  20
\\  \hline
    \end{tabular}
  \end{center}
\end{table}
\begin{table}[h!]
  \begin{center}
    \caption{Burkert Optimization Values}
    \label{BuckertESO079-G014}
     \begin{tabular}{|r|r|}
     \hline
      \textbf{Parameter}   & \textbf{Optimization Values}
      \\  \hline
     $\rho_0^B$  & $0.12\times 10^9$
\\  \hline
$r_0$&  6
\\  \hline
    \end{tabular}
  \end{center}
\end{table}
\begin{table}[h!]
  \begin{center}
    \caption{Einasto Optimization Values}
    \label{EinastoESO079-G014}
    \begin{tabular}{|r|r|}
     \hline
      \textbf{Parameter}   & \textbf{Optimization Values}
      \\  \hline
     $\rho_e$  & $0.0062\times 10^9$
\\  \hline
$r_e$ & 10
\\  \hline
$n_e$ & 0.17
\\  \hline
    \end{tabular}
  \end{center}
\end{table}
\begin{table}[h!]
\centering \caption{Physical assessment of collisional DM
parameters.}
\begin{tabular}{lcc}
\hline
Parameter & Value & Physical verdict \\
\hline
$\gamma_0$ & $1.4225$ & Mildly polytropic \\
$\delta_\gamma$ & $2\times10^{-10}$ & Practically zero \\
$r_\gamma$ & $1.5\ \mathrm{Kpc}$ & Transition radius irrelevant  \\
$K_0$ & $10$ & Moderate entropy  \\
$r_c$ & $0.5\ \mathrm{Kpc}$ & Small core scale  \\
$p$ & $0.01$ & Extremely shallow decline of $K(r)$ \\
\hline
Overall & --- & Physically plausible as a near-isothermal \\
\hline
\end{tabular}
\label{EVALUATIONESO079-G014}
\end{table}
Now the extended picture including the rotation velocity from the
other components of the galaxy, such as the disk and gas, makes
the collisional DM model non-viable for this galaxy. In Fig.
\ref{extendedESO079-G014} we present the combined rotation curves
including the other components of the galaxy along with the
collisional matter. As it can be seen, the extended collisional DM
model is non-viable.
Also in Table \ref{evaluationextendedESO079-G014} we present the
values of the free parameters of the collisional DM model for
which the maximum compatibility with the SPARC data comes for the
galaxy ESO079-G014.
\begin{table}[h!]
\centering \caption{Physical assessment of Extended collisional DM
parameters for ESO079-G014.}
\begin{tabular}{lcc}
\hline
Parameter & Value & Physical Verdict \\
\hline
$\gamma_0$ & 1.1137 & Nearly isothermal core; stable inner region \\
$\delta_\gamma$ & 0.04196 & Mild variation; $\gamma(r)$ rises slowly outward \\
$K_0$ & 3000 & Moderate entropy  \\
$ml_{\text{disk}}$ & 0.6132 & Reasonable stellar mass-to-light ratio for late-type spiral \\
$ml_{\text{bulge}}$ & 0.0000 & No bulge component, consistent with disk-dominated morphology \\
\hline
Overall & - & Physically viable; halo close to isothermal, balanced baryonic contribution \\
\hline
\end{tabular}
\label{evaluationextendedESO079-G014}
\end{table}


\subsection{The Galaxy ESO116-G012 Non-viable}

\begin{figure}[h!]
\centering
\includegraphics[width=20pc]{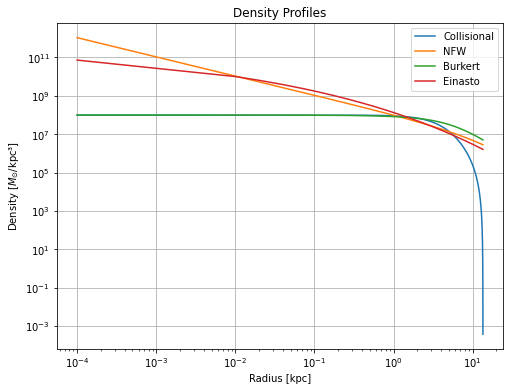}
\caption{The density of the collisional DM model (\ref{tanhmodel})
for the galaxy ESO116-G012, as a function of the radius.}
\label{ESO116-G012dens}
\end{figure}
\begin{figure}[h!]
\centering
\includegraphics[width=20pc]{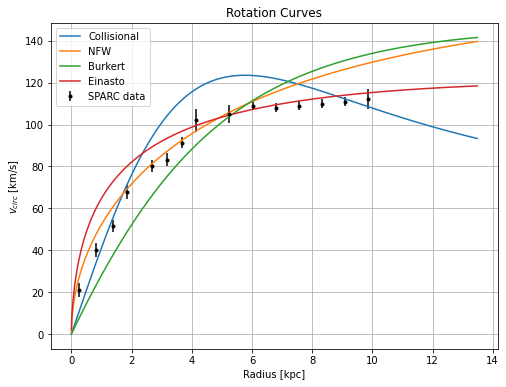}
\caption{The predicted rotation curves after using an optimization
for the collisional DM model (\ref{tanhmodel}), versus the SPARC
observational data for the galaxy ESO116-G012. We also plotted the
optimized curves for the NFW model, the Burkert model and the
Einasto model.} \label{ESO116-G012}
\end{figure}
\begin{figure}[h!]
\centering
\includegraphics[width=20pc]{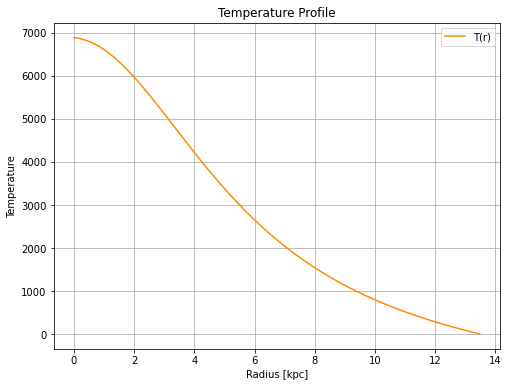}
\caption{The temperature as a function of the radius for the
collisional DM model (\ref{tanhmodel}) for the galaxy
ESO116-G012.} \label{ESO116-G012temp}
\end{figure}
\begin{figure}[h!]
\centering
\includegraphics[width=20pc]{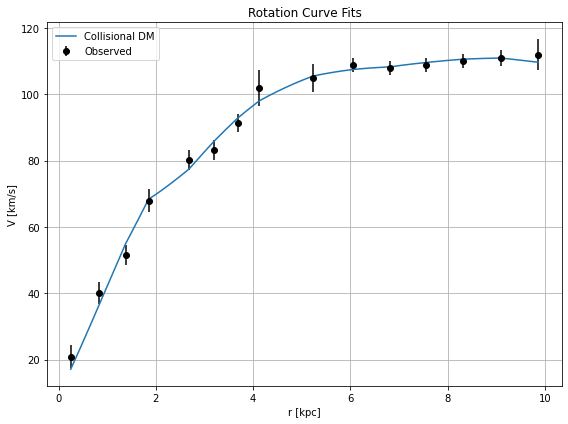}
\caption{The predicted rotation curves after using an optimization
for the collisional DM model (\ref{tanhmodel}), versus the
extended SPARC observational data for the galaxy ESO079-G014. The
model includes the rotation curves from all the components of the
galaxy, including gas and disk velocities, along with the
collisional DM model.} \label{extendedESO116-G012}
\end{figure}

For this galaxy, we shall choose $\rho_0=9.9\times
10^7$$M_{\odot}/\mathrm{Kpc}^{3}$. ESO\,116-G\,012 (PGC\,11984) is
an edge-on barred spiral galaxy of type SBc. In Figs.
\ref{ESO116-G012dens}, \ref{ESO116-G012} and \ref{ESO116-G012temp}
we present the density of the collisional DM model, the for
predicted rotation curves after using an optimization for the
collisional DM model (\ref{tanhmodel}), versus the SPARC
observational data and the temperature parameter as a function of
the radius respectively. As it can be seen, the SIDM model
produces non-viable rotation curves compatible with the SPARC
data. Also in Tables \ref{collESO116-G012},
\ref{NavaroESO116-G012}, \ref{BuckertESO116-G012} and
\ref{EinastoESO116-G012} we present the optimization values for
the SIDM model, and the other DM profiles. Also in Table
\ref{EVALUATIONESO116-G012} we present the overall evaluation of
the SIDM model for the galaxy at hand. The resulting phenomenology
is non-viable.
\begin{table}[h!]
  \begin{center}
    \caption{Collisional Dark Matter Optimization Values}
    \label{collESO116-G012}
     \begin{tabular}{|r|r|}
     \hline
      \textbf{Parameter}   & \textbf{Optimization Values}
      \\  \hline
     $\delta_{\gamma} $ & 0.0000000002
\\  \hline
$\gamma_0 $ & 1.355
\\  \hline
    \end{tabular}
  \end{center}
\end{table}
\begin{table}[h!]
  \begin{center}
    \caption{NFW  Optimization Values}
    \label{NavaroESO116-G012}
     \begin{tabular}{|r|r|}
     \hline
      \textbf{Parameter}   & \textbf{Optimization Values}
      \\  \hline
   $\rho_s$   & $0.0054\times 10^9$
\\  \hline
$r_s$&  20
\\  \hline
    \end{tabular}
  \end{center}
\end{table}
\begin{table}[h!]
  \begin{center}
    \caption{Burkert Optimization Values}
    \label{BuckertESO116-G012}
     \begin{tabular}{|r|r|}
     \hline
      \textbf{Parameter}   & \textbf{Optimization Values}
      \\  \hline
     $\rho_0^B$  & $0.1\times 10^9$
\\  \hline
$r_0$&  6
\\  \hline
    \end{tabular}
  \end{center}
\end{table}
\begin{table}[h!]
  \begin{center}
    \caption{Einasto Optimization Values}
    \label{EinastoESO116-G012}
    \begin{tabular}{|r|r|}
     \hline
      \textbf{Parameter}   & \textbf{Optimization Values}
      \\  \hline
     $\rho_e$  & $0.003\times 10^9$
\\  \hline
$r_e$ & 0.1
\\  \hline
$n_e$ & 0.17
\\  \hline
    \end{tabular}
  \end{center}
\end{table}
\begin{table}[h!]
\centering \caption{Physical assessment of collisional DM
parameters.}
\begin{tabular}{lcc}
\hline
Parameter & Value   & Physical verdict \\
\hline
$\gamma_0$ & $1.355$ & Mild polytropic index \\
$\delta_\gamma$ & $2\times10^{-10}$ & Practically zero \\
$r_\gamma$ & $1.5\ \mathrm{Kpc}$ & Transition radius inside inner halo \\
$K_0$ & $10$ & Moderate numeric value. \\
$r_c$ & $0.5\ \mathrm{Kpc}$ & Small core scale  \\
$p$ & $0.01$ & Extremely shallow decline \\
\hline
Overall & --- & Model behaves like constant-$\gamma$, constant-$K$ polytrope. \\
\hline
\end{tabular}
\label{EVALUATIONESO116-G012}
\end{table}
Now the extended picture including the rotation velocity from the
other components of the galaxy, such as the disk and gas, makes
the collisional DM model non-viable for this galaxy. In Fig.
\ref{extendedESO116-G012} we present the combined rotation curves
including the other components of the galaxy along with the
collisional matter. As it can be seen, the extended collisional DM
model is non-viable.
Also in Table \ref{evaluationextendedESO116-G012} we present the
values of the free parameters of the collisional DM model for
which the maximum compatibility with the SPARC data comes for the
galaxy ESO116-G012.
\begin{table}[h!]
\centering \caption{Physical assessment of Extended collisional DM
parameters for ESO116-G012.}
\begin{tabular}{lcc}
\hline
Parameter & Value & Physical Verdict \\
\hline
$\gamma_0$ & 1.0516 & Near-isothermal core; slightly above isothermal, low central pressure \\
$\delta_\gamma$ & 0.04245 & Small radial variation; $\gamma(r)$ rises slowly with radius \\
$K_0$ & 3000 & Moderate entropy  \\
$ml_{\text{disk}}$ & 0.9579 & Relatively high mass-to-light ratio \\
$ml_{\text{bulge}}$ & 0.0000 & No bulge component; disk-dominated morphology \\
\hline
Overall & - & Physically plausible; inner halo close to isothermal, baryons dominated by the disk, limited EoS flexibility \\
\hline
\end{tabular}
\label{evaluationextendedESO116-G012}
\end{table}


\subsection{The Galaxy ESO444-G084}

\begin{figure}[h!]
\centering
\includegraphics[width=20pc]{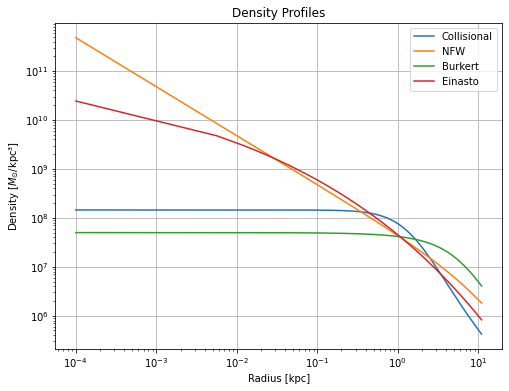}
\caption{The density of the collisional DM model (\ref{tanhmodel})
for the galaxy ESO444-G084, as a function of the radius.}
\label{ESO444-G084dens}
\end{figure}

\begin{figure}[h!]
\centering
\includegraphics[width=20pc]{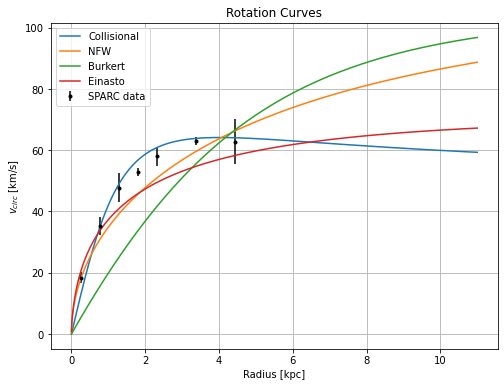}
\caption{The predicted rotation curves after using an optimization
for the collisional DM model (\ref{tanhmodel}), versus the SPARC
observational data for the galaxy ESO444-G084. We also plotted the
optimized curves for the NFW model, the Burkert model and the
Einasto model.} \label{ESO444-G084}
\end{figure}

\begin{figure}[h!]
\centering
\includegraphics[width=20pc]{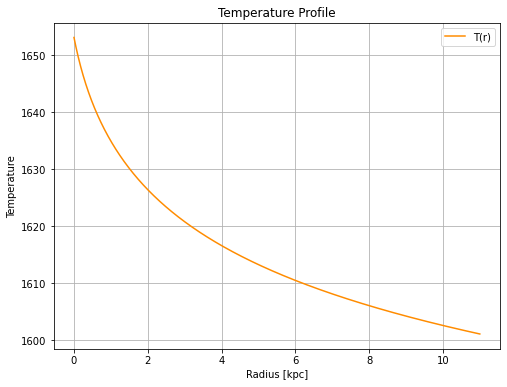}
\caption{The temperature as a function of the radius for the
collisional DM model (\ref{tanhmodel}) for the galaxy
ESO444-G084.} \label{ESO444-G084temp}
\end{figure}

For this galaxy, we shall choose $\rho_0=1.45\times
10^8$$M_{\odot}/\mathrm{Kpc}^{3}$. ESO 444-G084 is a gas-rich
dwarf irregular galaxy located in the Centaurus A group. Its
distance, determined via the tip of the red giant branch, is
approximately 4.6 Mpc. The galaxy exhibits extremely
low-surface-brightness and is characterized by an extended,
centrally concentrated HI disk. Its properties make it a
particularly valuable laboratory for investigating dark matter
distributions and halo structure in low-mass galaxies. As we now
show, the SIDM model provides a natural explanation of the
rotation  curves for this galaxy. In Figs. \ref{ESO444-G084dens},
\ref{ESO444-G084} and \ref{ESO444-G084temp} we present the density
of the collisional DM model, the for predicted rotation curves
after using an optimization for the collisional DM model
(\ref{tanhmodel}), versus the SPARC observational data and the
temperature parameter as a function of the radius respectively. As
it can be seen, the SIDM model produces viable rotation curves
compatible with the SPARC data. Also in Tables
\ref{collESO444-G084}, \ref{NavaroESO444-G084},
\ref{BuckertESO444-G084} and \ref{EinastoESO444-G084} we present
the optimization values for the SIDM model, and the other DM
profiles. Also in Table \ref{EVALUATIONESO444-G084} we present the
overall evaluation of the SIDM model for the galaxy at hand. The
resulting phenomenology is viable.
\begin{table}[h!]
  \begin{center}
    \caption{Collisional Dark Matter Optimization Values}
    \label{collESO444-G084}
     \begin{tabular}{|r|r|}
     \hline
      \textbf{Parameter}   & \textbf{Optimization Values}
      \\  \hline
     $\delta_{\gamma} $ & 0.0000000012
\\  \hline
$\gamma_0 $ & 1.0001 \\ \hline $K_0$ ($M_{\odot} \,
\mathrm{Kpc}^{-3} \, (\mathrm{km/s})^{2}$)& 1650
\\  \hline
    \end{tabular}
  \end{center}
\end{table}
\begin{table}[h!]
  \begin{center}
    \caption{NFW  Optimization Values}
    \label{NavaroESO444-G084}
     \begin{tabular}{|r|r|}
     \hline
      \textbf{Parameter}   & \textbf{Optimization Values}
      \\  \hline
   $\rho_s$   & $0.0024\times 10^9$
\\  \hline
$r_s$&  20
\\  \hline
    \end{tabular}
  \end{center}
\end{table}
\begin{table}[h!]
  \begin{center}
    \caption{Burkert Optimization Values}
    \label{BuckertESO444-G084}
     \begin{tabular}{|r|r|}
     \hline
      \textbf{Parameter}   & \textbf{Optimization Values}
      \\  \hline
     $\rho_0^B$  & $0.05\times 10^9$
\\  \hline
$r_0$&  6
\\  \hline
    \end{tabular}
  \end{center}
\end{table}
\begin{table}[h!]
  \begin{center}
    \caption{Einasto Optimization Values}
    \label{EinastoESO444-G084}
    \begin{tabular}{|r|r|}
     \hline
      \textbf{Parameter}   & \textbf{Optimization Values}
      \\  \hline
     $\rho_e$  & $0.001\times 10^9$
\\  \hline
$r_e$ & 10
\\  \hline
$n_e$ & 0.17
\\  \hline
    \end{tabular}
  \end{center}
\end{table}
\begin{table}[h!]
\centering \caption{Physical assessment of collisional DM
parameters (ESO444-G084).}
\begin{tabular}{lcc}
\hline
Parameter & Value & Physical Verdict \\
\hline
$\gamma_0$ & $1.28$ & Mildly super-isothermal; plausible for thermalized DM \\
$\delta_\gamma$ & $0.02$ & Extremely small radial variation  \\
$r_\gamma$ & $1.5\ \mathrm{Kpc}$ & Inner-halo transition radius \\
$K_0$ & $0.1\times10^{2}=10$ & Moderate numeric scale \\
$r_c$ & $0.5\ \mathrm{Kpc}$ & Small core scale; consistent with compact inner core \\
$p$ & $0.01$ & Nearly constant $K(r)$; very weak radial entropy gradient \\
\hline
Overall & - & Physically plausible  \\
\hline
\end{tabular}
\label{EVALUATIONESO444-G084}
\end{table}


\subsection{The Galaxy ESO\,563-G021 Marginally Viable}

\begin{figure}[h!]
\centering
\includegraphics[width=20pc]{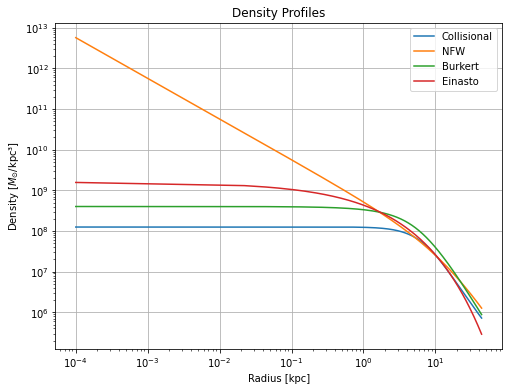}
\caption{The density of the collisional DM model (\ref{tanhmodel})
for the galaxy ESO563-G021, as a function of the radius.}
\label{ESO563-G021dens}
\end{figure}
\begin{figure}[h!]
\centering
\includegraphics[width=20pc]{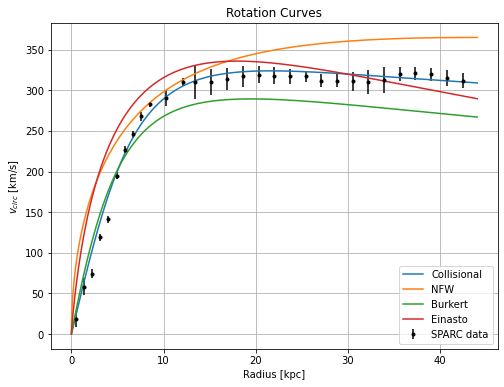}
\caption{The predicted rotation curves after using an optimization
for the collisional DM model (\ref{tanhmodel}), versus the SPARC
observational data for the galaxy ESO563-G021. We also plotted the
optimized curves for the NFW model, the Burkert model and the
Einasto model.} \label{ESO563-G021}
\end{figure}
\begin{figure}[h!]
\centering
\includegraphics[width=20pc]{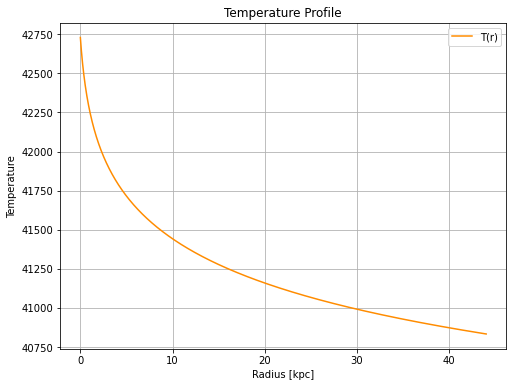}
\caption{The temperature as a function of the radius for the
collisional DM model (\ref{tanhmodel}) for the galaxy
ESO563-G021.} \label{ESO563-G021temp}
\end{figure}

For this galaxy, we shall choose $\rho_0=1.25\times
10^8$$M_{\odot}/\mathrm{Kpc}^{3}$. ESO\,563-G021 is a disk galaxy
included in the SPARC database of well-measured rotation-curve
systems. According to detailed rotation curve fitting analysis
that marginalize over inclination, stellar mass-to-light ratio,
and distance, its distance is determined to be $D = 88.1 \pm 4.9\
\mathrm{Mpc}$. In Figs. \ref{ESO563-G021dens}, \ref{ESO563-G021}
and \ref{ESO563-G021temp} we present the density of the
collisional DM model, the for predicted rotation curves after
using an optimization for the collisional DM model
(\ref{tanhmodel}), versus the SPARC observational data and the
temperature parameter as a function of the radius respectively. As
it can be seen, the SIDM model produces marginally viable rotation
curves compatible with the SPARC data. Also in Tables
\ref{collESO563-G021}, \ref{NavaroESO563-G021},
\ref{BuckertESO563-G021} and \ref{EinastoESO563-G021} we present
the optimization values for the SIDM model, and the other DM
profiles. Also in Table \ref{EVALUATIONESO563G021} we present the
overall evaluation of the SIDM model for the galaxy at hand. The
resulting phenomenology is marginally viable.
\begin{table}[h!]
  \begin{center}
    \caption{Collisional Dark Matter Optimization Values}
    \label{collESO563-G021}
     \begin{tabular}{|r|r|}
     \hline
      \textbf{Parameter}   & \textbf{Optimization Values}
      \\  \hline
     $\delta_{\gamma} $ & 0.0000000012
\\  \hline
$\gamma_0 $ & 1.0001 \\ \hline $K_0$ ($M_{\odot} \,
\mathrm{Kpc}^{-3} \, (\mathrm{km/s})^{2}$)& 42650
\\  \hline
    \end{tabular}
  \end{center}
\end{table}
\begin{table}[h!]
  \begin{center}
    \caption{NFW  Optimization Values}
    \label{NavaroESO563-G021}
     \begin{tabular}{|r|r|}
     \hline
      \textbf{Parameter}   & \textbf{Optimization Values}
      \\  \hline
   $\rho_s$   & $0.0285\times 10^9$
\\  \hline
$r_s$&  20
\\  \hline
    \end{tabular}
  \end{center}
\end{table}
\begin{table}[h!]
  \begin{center}
    \caption{Burkert Optimization Values}
    \label{BuckertESO563-G021}
     \begin{tabular}{|r|r|}
     \hline
      \textbf{Parameter}   & \textbf{Optimization Values}
      \\  \hline
     $\rho_0^B$  & $0.4\times 10^9$
\\  \hline
$r_0$&  6
\\  \hline
    \end{tabular}
  \end{center}
\end{table}
\begin{table}[h!]
  \begin{center}
    \caption{Einasto Optimization Values}
    \label{EinastoESO563-G021}
    \begin{tabular}{|r|r|}
     \hline
      \textbf{Parameter}   & \textbf{Optimization Values}
      \\  \hline
     $\rho_e$  & $0.029\times 10^9$
\\  \hline
$r_e$ & 9.5
\\  \hline
$n_e$ & 0.5
\\  \hline
    \end{tabular}
  \end{center}
\end{table}
\begin{table}[h!]
\centering \caption{Physical assessment of collisional DM
parameters (ESO563-G021).}
\begin{tabular}{lcc}
\hline
Parameter & Value & Physical Verdict \\
\hline
$\gamma_0$ & $1.0001$ & Essentially isothermal  \\
$\delta_\gamma$ & $1.2\times10^{-9}$ & Negligible  \\
$r_\gamma$ & $1.5\ \mathrm{Kpc}$ & Reasonable inner transition radius  \\
$K_0$ & $4.265\times10^{4}$ & Large entropy/temperature scale\\
$r_c$ & $0.5\ \mathrm{Kpc}$ & Small core scale for $K(r)$; physically plausible \\
$p$ & $0.01$ & Very shallow decline \\
\hline Overall & - & Halo behaves like an isothermal sphere;
physically consistent \\
\hline
\end{tabular}
\label{EVALUATIONESO563G021}
\end{table}


\subsection{The Galaxy F561-1}

\begin{figure}[h!]
\centering
\includegraphics[width=20pc]{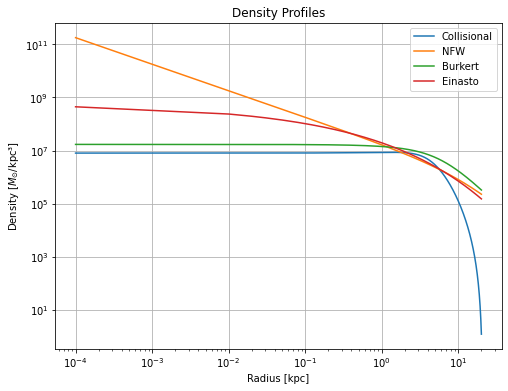}
\caption{The density of the collisional DM model (\ref{tanhmodel})
for the galaxy F561-1, as a function of the radius.}
\label{F561-1dens}
\end{figure}
\begin{figure}[h!]
\centering
\includegraphics[width=20pc]{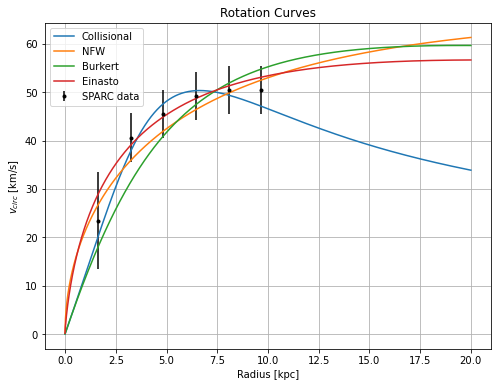}
\caption{The predicted rotation curves after using an optimization
for the collisional DM model (\ref{tanhmodel}), versus the SPARC
observational data for the galaxy F561-1. We also plotted the
optimized curves for the NFW model, the Burkert model and the
Einasto model.} \label{F561-1}
\end{figure}
\begin{figure}[h!]
\centering
\includegraphics[width=20pc]{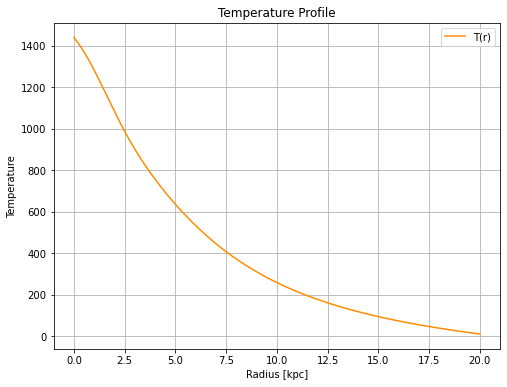}
\caption{The temperature as a function of the radius for the
collisional DM model (\ref{tanhmodel}) for the galaxy F561-1.}
\label{F561-1temp}
\end{figure}

For this galaxy, we shall choose $\rho_0=1\times
10^7$$M_{\odot}/\mathrm{Kpc}^{3}$. F561-1 is categorized as a
low-surface-brightness disk galaxy, characterized by a faint
central brightness and dominating dark matter content. Its
estimated distance, is on the order of \(D\sim
30\text{-}50\)\,Mpc. In Figs. \ref{F561-1dens}, \ref{F561-1} and
\ref{F561-1temp} we present the density of the collisional DM
model, the for predicted rotation curves after using an
optimization for the collisional DM model (\ref{tanhmodel}),
versus the SPARC observational data and the temperature parameter
as a function of the radius respectively. As it can be seen, the
SIDM model produces viable rotation curves compatible with the
SPARC data. Also in Tables \ref{collF561-1}, \ref{NavaroF561-1},
\ref{BuckertF561-1} and \ref{EinastoF561-1} we present the
optimization values for the SIDM model, and the other DM profiles.
Also in Table \ref{EVALUATIONF561-1} we present the overall
evaluation of the SIDM model for the galaxy at hand. The resulting
phenomenology is viable.
\begin{table}[h!]
  \begin{center}
    \caption{Collisional Dark Matter Optimization Values}
    \label{collF561-1}
     \begin{tabular}{|r|r|}
     \hline
      \textbf{Parameter}   & \textbf{Optimization Values}
      \\  \hline
     $\delta_{\gamma} $ & 0.0000000012
\\  \hline
$\gamma_0 $ & 1.0001 \\ \hline $K_0$ ($M_{\odot} \,
\mathrm{Kpc}^{-3} \, (\mathrm{km/s})^{2}$)& 1200
\\  \hline
    \end{tabular}
  \end{center}
\end{table}
\begin{table}[h!]
  \begin{center}
    \caption{NFW  Optimization Values}
    \label{NavaroF561-1}
     \begin{tabular}{|r|r|}
     \hline
      \textbf{Parameter}   & \textbf{Optimization Values}
      \\  \hline
   $\rho_s$   & $0.0009\times 10^9$
\\  \hline
$r_s$&  20
\\  \hline
    \end{tabular}
  \end{center}
\end{table}
\begin{table}[h!]
  \begin{center}
    \caption{Burkert Optimization Values}
    \label{BuckertF561-1}
     \begin{tabular}{|r|r|}
     \hline
      \textbf{Parameter}   & \textbf{Optimization Values}
      \\  \hline
     $\rho_0^B$  & $0.017\times 10^9$
\\  \hline
$r_0$&  6
\\  \hline
    \end{tabular}
  \end{center}
\end{table}
\begin{table}[h!]
  \begin{center}
    \caption{Einasto Optimization Values}
    \label{EinastoF561-1}
    \begin{tabular}{|r|r|}
     \hline
      \textbf{Parameter}   & \textbf{Optimization Values}
      \\  \hline
     $\rho_e$  & $0.0007\times 10^9$
\\  \hline
$r_e$ & 10
\\  \hline
$n_e$ & 0.3
\\  \hline
    \end{tabular}
  \end{center}
\end{table}
\begin{table}[h!]
\centering \caption{Physical assessment of collisional DM
parameters (F561-1).}
\begin{tabular}{lcc}
\hline
Parameter & Value & Physical Verdict \\
\hline
$\gamma_0$ & $1.0001$ & Practically isothermal \\
$\delta_\gamma$ & $1.2\times10^{-9}$ & Negligible  \\
$r_\gamma$ & $1.5\ \mathrm{Kpc}$ & Reasonable transition radius   \\
$K_0$ & $1.2\times10^{3}$ & Sets nearly-constant temperature \\
$r_c$ & $0.5\ \mathrm{Kpc}$ & Small core scale for $K(r)$; physically plausible \\
$p$ & $0.01$ & Very shallow decline  \\
\hline
Overall & - & Model  isothermal halo: physically consistent. \\
\hline
\end{tabular}
\label{EVALUATIONF561-1}
\end{table}


\subsection{The Galaxy F563-1}

\begin{figure}[h!]
\centering
\includegraphics[width=20pc]{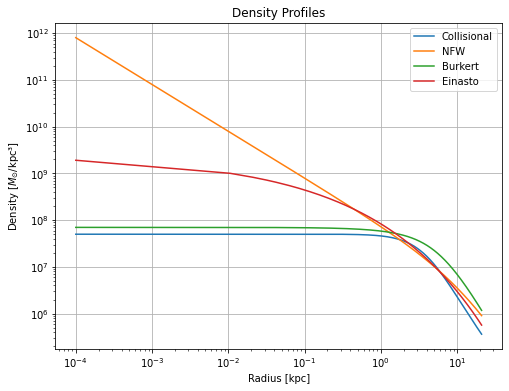}
\caption{The density of the collisional DM model (\ref{tanhmodel})
for the galaxy F563-V1, as a function of the radius.}
\label{F563-1dens}
\end{figure}
\begin{figure}[h!]
\centering
\includegraphics[width=20pc]{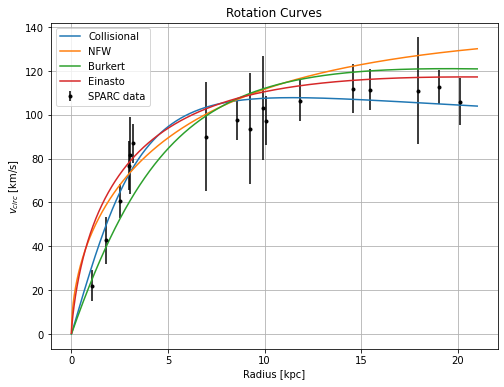}
\caption{The predicted rotation curves after using an optimization
for the collisional DM model (\ref{tanhmodel}), versus the SPARC
observational data for the galaxy F563-1. We also plotted the
optimized curves for the NFW model, the Burkert model and the
Einasto model.} \label{F563-1}
\end{figure}
\begin{figure}[h!]
\centering
\includegraphics[width=20pc]{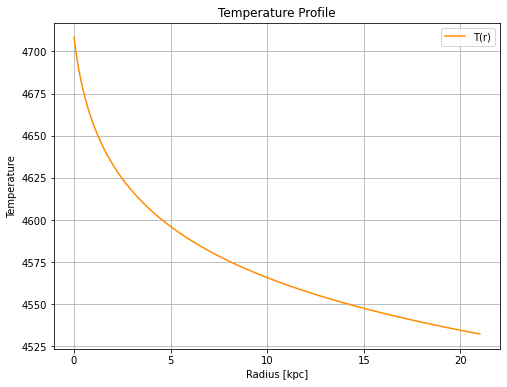}
\caption{The temperature as a function of the radius for the
collisional DM model (\ref{tanhmodel}) for the galaxy F563-1.}
\label{F563-1temp}
\end{figure}

For this galaxy, we shall choose $\rho_0=5\times
10^7$$M_{\odot}/\mathrm{Kpc}^{3}$. F563\!-\!1 is a
low-surface-brightness disc galaxy, often classified as a dwarf or
diffuse spiral (not a giant normal spiral) dominated by dark
matter. Its distance is about \(46.8\) Mpc. In Figs.
\ref{F563-1dens}, \ref{F563-1} and \ref{F563-1temp} we present the
density of the collisional DM model, the for predicted rotation
curves after using an optimization for the collisional DM model
(\ref{tanhmodel}), versus the SPARC observational data and the
temperature parameter as a function of the radius respectively. As
it can be seen, the SIDM model produces viable rotation curves
compatible with the SPARC data. Also in Tables \ref{collF563-1},
\ref{NavaroF563-1}, \ref{BuckertF563-1} and \ref{EinastoF563-1} we
present the optimization values for the SIDM model, and the other
DM profiles. Also in Table \ref{EVALUATIONF563-1} we present the
overall evaluation of the SIDM model for the galaxy at hand. The
resulting phenomenology is viable.
\begin{table}[h!]
  \begin{center}
    \caption{Collisional Dark Matter Optimization Values}
    \label{collF563-1}
     \begin{tabular}{|r|r|}
     \hline
      \textbf{Parameter}   & \textbf{Optimization Values}
      \\  \hline
     $\delta_{\gamma} $ & 0.0000000012
\\  \hline
$\gamma_0 $ & 1.0001 \\ \hline $K_0$ ($M_{\odot} \,
\mathrm{Kpc}^{-3} \, (\mathrm{km/s})^{2}$)&  4700
\\  \hline
    \end{tabular}
  \end{center}
\end{table}
\begin{table}[h!]
  \begin{center}
    \caption{NFW  Optimization Values}
    \label{NavaroF563-1}
     \begin{tabular}{|r|r|}
     \hline
      \textbf{Parameter}   & \textbf{Optimization Values}
      \\  \hline
   $\rho_s$   & $0.00035\times 10^9$
\\  \hline
$r_s$&  20
\\  \hline
    \end{tabular}
  \end{center}
\end{table}
\begin{table}[h!]
  \begin{center}
    \caption{Burkert Optimization Values}
    \label{BuckertF563-1}
     \begin{tabular}{|r|r|}
     \hline
      \textbf{Parameter}   & \textbf{Optimization Values}
      \\  \hline
     $\rho_0^B$  & $0.006\times 10^9$
\\  \hline
$r_0$&  6
\\  \hline
    \end{tabular}
  \end{center}
\end{table}
\begin{table}[h!]
  \begin{center}
    \caption{Einasto Optimization Values}
    \label{EinastoF563-1}
    \begin{tabular}{|r|r|}
     \hline
      \textbf{Parameter}   & \textbf{Optimization Values}
      \\  \hline
     $\rho_e$  & $0.0003\times 10^9$
\\  \hline
$r_e$ & 10
\\  \hline
$n_e$ & 0.5
\\  \hline
    \end{tabular}
  \end{center}
\end{table}
\begin{table}[h!]
\centering \caption{Physical assessment of collisional DM
parameters (F563-1).}
\begin{tabular}{lcc}
\hline
Parameter & Value & Physical Verdict \\
\hline
$\gamma_0$ & $1.0001$ & Practically isothermal \\
$\delta_\gamma$ & $1.2\times10^{-9}$ & Negligible  \\
$r_\gamma$ & $1.5\ \mathrm{Kpc}$ & Reasonable transition radius   \\
$K_0$ & $4.7\times10^{3}$ & Sets nearly-constant temperature \\
$r_c$ & $0.5\ \mathrm{Kpc}$ & Small core scale for $K(r)$; physically plausible \\
$p$ & $0.01$ & Very shallow decline  \\
\hline
Overall & - & Models a nearly isothermal halo. \\
\hline
\end{tabular}
\label{EVALUATIONF563-1}
\end{table}


\subsection{The Galaxy F563-V1}

\begin{figure}[h!]
\centering
\includegraphics[width=20pc]{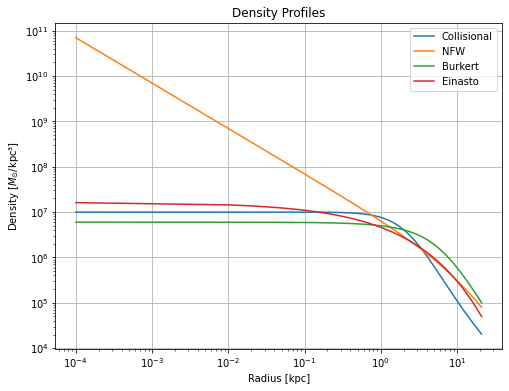}
\caption{The density of the collisional DM model (\ref{tanhmodel})
for the galaxy F563-V1, as a function of the radius.}
\label{F563-V1dens}
\end{figure}

\begin{figure}[h!]
\centering
\includegraphics[width=20pc]{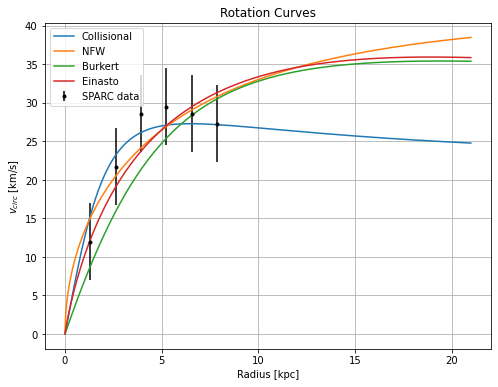}
\caption{The predicted rotation curves after using an optimization
for the collisional DM model (\ref{tanhmodel}), versus the SPARC
observational data for the galaxy F563-V1. We also plotted the
optimized curves for the NFW model, the Burkert model and the
Einasto model.} \label{F563-V1}
\end{figure}

\begin{figure}[h!]
\centering
\includegraphics[width=20pc]{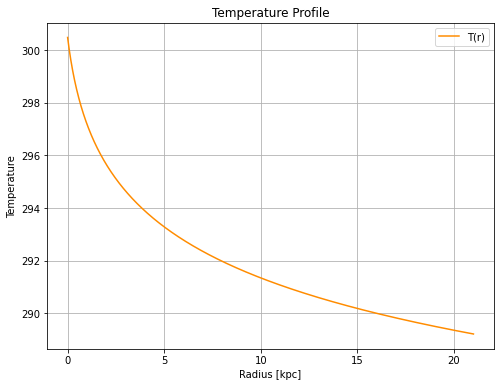}
\caption{The temperature as a function of the radius for the
collisional DM model (\ref{tanhmodel}) for the galaxy F563-V1.}
\label{F563-V1temp}
\end{figure}

For this galaxy, we shall choose $\rho_0=1\times
10^7$$M_{\odot}/\mathrm{Kpc}^{3}$. F563-V1 is a late-type,
low-surface-brightness  disk galaxy. It features modest luminosity
and significant gas content, characteristic of
low-surface-brightness disks that are dark-matter dominated and
slowly evolving. It is  a faint, extended disk system. In Figs.
\ref{F563-V1dens}, \ref{F563-V1} and \ref{F563-V1temp} we present
the density of the collisional DM model, the for predicted
rotation curves after using an optimization for the collisional DM
model (\ref{tanhmodel}), versus the SPARC observational data and
the temperature parameter as a function of the radius
respectively. As it can be seen, the SIDM model produces viable
rotation curves compatible with the SPARC data. Also in Tables
\ref{collF563-V1}, \ref{NavaroF563-V1}, \ref{BuckertF563-V1} and
\ref{EinastoF563-V1} we present the optimization values for the
SIDM model, and the other DM profiles. Also in Table
\ref{EVALUATIONF563-V1} we present the overall evaluation of the
SIDM model for the galaxy at hand. The resulting phenomenology is
viable.
\begin{table}[h!]
  \begin{center}
    \caption{Collisional Dark Matter Optimization Values}
    \label{collF563-V1}
     \begin{tabular}{|r|r|}
     \hline
      \textbf{Parameter}   & \textbf{Optimization Values}
      \\  \hline
     $\delta_{\gamma} $ & 0.0000000012
\\  \hline
$\gamma_0 $ & 1.0001  \\ \hline $K_0$ ($M_{\odot} \,
\mathrm{Kpc}^{-3} \, (\mathrm{km/s})^{2}$)&  300
\\  \hline
    \end{tabular}
  \end{center}
\end{table}
\begin{table}[h!]
  \begin{center}
    \caption{NFW  Optimization Values}
    \label{NavaroF563-V1}
     \begin{tabular}{|r|r|}
     \hline
      \textbf{Parameter}   & \textbf{Optimization Values}
      \\  \hline
   $\rho_s$   & $0.00035\times 10^9$
\\  \hline
$r_s$&  20
\\  \hline
    \end{tabular}
  \end{center}
\end{table}
\begin{table}[h!]
  \begin{center}
    \caption{Burkert Optimization Values}
    \label{BuckertF563-V1}
     \begin{tabular}{|r|r|}
     \hline
      \textbf{Parameter}   & \textbf{Optimization Values}
      \\  \hline
     $\rho_0^B$  & $0.006\times 10^9$
\\  \hline
$r_0$&  6
\\  \hline
    \end{tabular}
  \end{center}
\end{table}
\begin{table}[h!]
  \begin{center}
    \caption{Einasto Optimization Values}
    \label{EinastoF563-V1}
    \begin{tabular}{|r|r|}
     \hline
      \textbf{Parameter}   & \textbf{Optimization Values}
      \\  \hline
     $\rho_e$  & $0.0003\times 10^9$
\\  \hline
$r_e$ & 10
\\  \hline
$n_e$ & 0.5
\\  \hline
    \end{tabular}
  \end{center}
\end{table}
\begin{table}[h!]
\centering \caption{Physical assessment of collisional DM
parameters (F563-V1).}
\begin{tabular}{lcc}
\hline
Parameter & Value & Physical Verdict \\
\hline
$\gamma_0$ & $1.0001$ & Practically isothermal \\
$\delta_\gamma$ & $1.2\times10^{-9}$ & Negligible  \\
$r_\gamma$ & $1.5\ \mathrm{Kpc}$ & Reasonable transition radius   \\
$K_0$ & $3.0\times10^{2}$ & Sets nearly-constant temperature \\
$r_c$ & $0.5\ \mathrm{Kpc}$ & Small core scale for $K(r)$; physically plausible \\
$p$ & $0.01$ & Very shallow decline  \\
\hline
Overall &-& Models isothermal halo: physically consistent. \\
\hline
\end{tabular}
\label{EVALUATIONF563-V1}
\end{table}


\subsection{The Galaxy F563-V2: An interesting Galaxy for Doing Physics, Large and Small $K_0$ Scenarios}

\begin{figure}[h!]
\centering
\includegraphics[width=20pc]{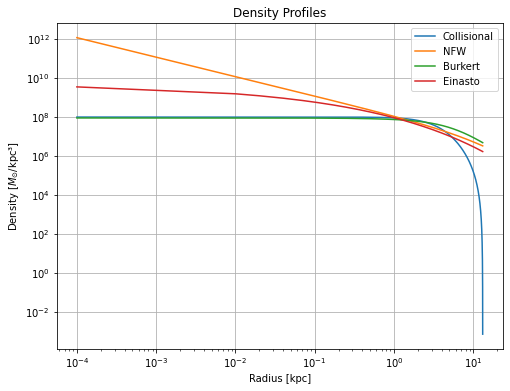}
\caption{The density of the collisional DM model (\ref{tanhmodel})
for the galaxy F563-V2, as a function of the radius.}
\label{F563-V2dens}
\end{figure}

\begin{figure}[h!]
\centering
\includegraphics[width=20pc]{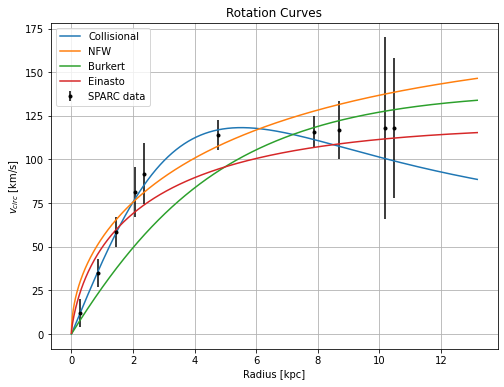}
\caption{The predicted rotation curves after using an optimization
for the collisional DM model (\ref{tanhmodel}), versus the SPARC
observational data for the galaxy F563-V2. We also plotted the
optimized curves for the NFW model, the Burkert model and the
Einasto model.} \label{F563-V2}
\end{figure}

\begin{figure}[h!]
\centering
\includegraphics[width=20pc]{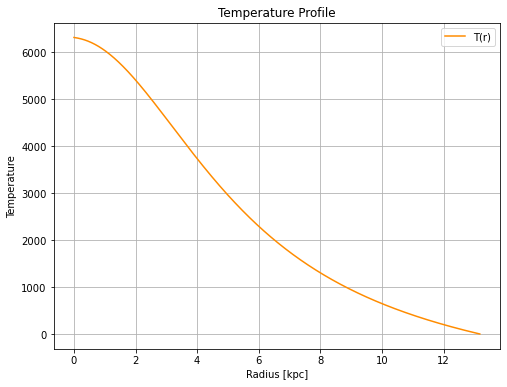}
\caption{The temperature as a function of the radius for the
collisional DM model (\ref{tanhmodel}) for the galaxy F563-V2.}
\label{F563-V2temp}
\end{figure}

\begin{figure}[h!]
\centering
\includegraphics[width=20pc]{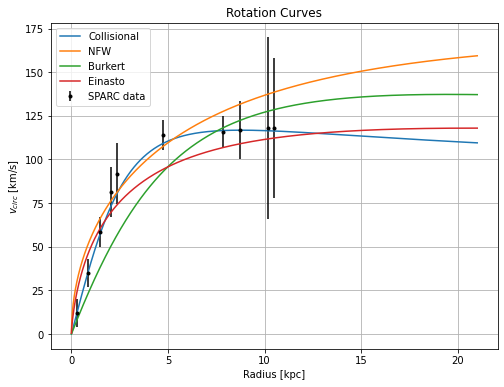}
\caption{The predicted rotation curves after using an optimization
for the collisional DM model (\ref{tanhmodel}), versus the SPARC
observational data for the galaxy F563-V2. We also plotted the
optimized curves for the NFW model, the Burkert model and the
Einasto model.} \label{F563-V2large}
\end{figure}

\begin{figure}[h!]
\centering
\includegraphics[width=20pc]{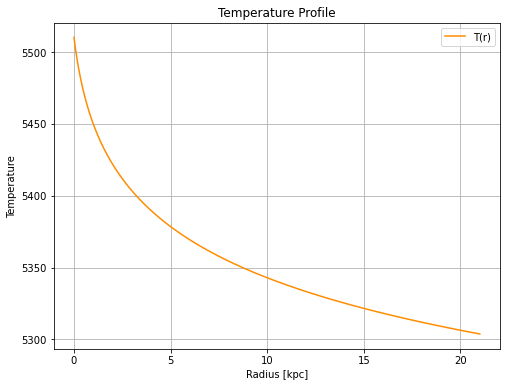}
\caption{The temperature as a function of the radius for the
collisional DM model (\ref{tanhmodel}) for the galaxy F563-V2.}
\label{F563-V2largetemp}
\end{figure}

First we shall deal with a small $K_0$ Case. For this galaxy, we
shall choose $\rho_0=10^8$$M_{\odot}/\mathrm{Kpc}^{3}$. F563-V2 is
a diffuse, late-type low-surface-brightness disk galaxy with a
faint stellar disk, strong gas dominance, and low stellar surface
density, making it highly dark-matter dominated. It lies at a
distance of order tens of megaparsecs. It hosts a slowly rotating
central bar, and its optical disk extends a few kiloparsecs, with
the HI disk expected to reach significantly further, by several
kiloparsecs. In Figs. \ref{F563-V2dens}, \ref{F563-V2} and
\ref{F563-V2temp} we present the density of the collisional DM
model, the for predicted rotation curves after using an
optimization for the collisional DM model (\ref{tanhmodel}),
versus the SPARC observational data and the temperature parameter
as a function of the radius respectively. As it can be seen, the
SIDM model produces viable rotation curves compatible with the
SPARC data. Also in Tables \ref{collF563-V2}, \ref{NavaroF563-V2},
\ref{BuckertF563-V2} and \ref{EinastoF563-V2} we present the
optimization values for the SIDM model, and the other DM profiles.
Also in Table \ref{EVALUATIONF563-V2} we present the overall
evaluation of the SIDM model for the galaxy at hand. The resulting
phenomenology is viable.
\begin{table}[h!]
  \begin{center}
    \caption{Collisional Dark Matter Optimization Values}
    \label{collF563-V2}
     \begin{tabular}{|r|r|}
     \hline
      \textbf{Parameter}   & \textbf{Optimization Values}
      \\  \hline
     $\delta_{\gamma} $ & 0.0000000002
\\  \hline
$\gamma_0 $ & 1.35
\\  \hline
    \end{tabular}
  \end{center}
\end{table}
\begin{table}[h!]
  \begin{center}
    \caption{NFW  Optimization Values}
    \label{NavaroF563-V2}
     \begin{tabular}{|r|r|}
     \hline
      \textbf{Parameter}   & \textbf{Optimization Values}
      \\  \hline
   $\rho_s$   & $0.006\times 10^9$
\\  \hline
$r_s$&  20
\\  \hline
    \end{tabular}
  \end{center}
\end{table}
\begin{table}[h!]
  \begin{center}
    \caption{Burkert Optimization Values}
    \label{BuckertF563-V2}
     \begin{tabular}{|r|r|}
     \hline
      \textbf{Parameter}   & \textbf{Optimization Values}
      \\  \hline
     $\rho_0^B$  & $0.09\times 10^9$
\\  \hline
$r_0$&  6
\\  \hline
    \end{tabular}
  \end{center}
\end{table}
\begin{table}[h!]
  \begin{center}
    \caption{Einasto Optimization Values}
    \label{EinastoF563-V2}
    \begin{tabular}{|r|r|}
     \hline
      \textbf{Parameter}   & \textbf{Optimization Values}
      \\  \hline
     $\rho_e$  & $0.003\times 10^9$
\\  \hline
$r_e$ & 10
\\  \hline
$n_e$ & 0.27
\\  \hline
    \end{tabular}
  \end{center}
\end{table}
\begin{table}[h!]
\centering \caption{Physical assessment of collisional DM
parameters (F563-V2 set).}
\begin{tabular}{lcc}
\toprule
Parameter & Value & Physical verdict \\
$\gamma_0$ & $1.35$ & Mildly super-isothermal; provides moderate pressure support \\
$\delta_\gamma$ & $2\times10^{-10}$ & Essentially zero  \\
$r_\gamma$ & $1.5\ \mathrm{Kpc}$ & Reasonable transition scale   \\
$K_0$ & $10$ & Moderate entropy \\
$r_c$ & $0.5\ \mathrm{Kpc}$ & Very compact scale; concentrates any $K$-variation to inner Kpc \\
$p$ & $0.01$ & Nearly flat $K(r)$; negligible radial decline \\
Overall &- & Physically consistent \\
\end{tabular}
\label{EVALUATIONF563-V2}
\end{table}
Now the large $K_0$ case. Again we choose,
$\rho_0=10^8$$M_{\odot}/\mathrm{Kpc}^{3}$. In Fig.
 \ref{F563-V2large} we present the predicted
rotation curves after using an optimization for the collisional DM
model (\ref{tanhmodel}), versus the SPARC observational data. The
resulting phenomenology is viable. Also in Table
\ref{collF563-V2large}, we present the optimization values for the
SIDM model. In Fig. \ref{F563-V2largetemp} we present the
temperature parameter as a function of the radius respectively.
Also in Table \ref{EVALUATIONF563-V2large} we present the overall
evaluation of the SIDM model for the galaxy at hand. The resulting
phenomenology is viable.
\begin{table}[h!]
  \begin{center}
    \caption{Collisional Dark Matter Optimization Values}
    \label{collF563-V2large}
     \begin{tabular}{|r|r|}
     \hline
      \textbf{Parameter}   & \textbf{Optimization Values}
      \\  \hline
     $\delta_{\gamma} $ & 0.0000000012
\\  \hline
$\gamma_0 $ & 1.0001
\\  \hline
$K_0 $ & 5500
\\  \hline
    \end{tabular}
  \end{center}
\end{table}
\begin{table}[h!]
\centering \caption{Physical assessment of collisional DM
parameters (F563-V2).}
\begin{tabular}{lcc}
\hline
Parameter & Value & Physical Verdict \\
\hline
$\gamma_0$ & $1.0001$ & Practically isothermal   \\
$\delta_\gamma$ & $1.2\times10^{-9}$ & Negligible  \\
$r_\gamma$ & $1.5\ \mathrm{Kpc}$ & Reasonable transition radius   \\
$K_0$ & $5.5\times10^{3}$ & Sets nearly-constant temperature \\
$r_c$ & $0.5\ \mathrm{Kpc}$ & Small core scale for $K(r)$; physically plausible \\
$p$ & $0.01$ & Very shallow decline  \\
\hline
Overall & - & Models a nearly isothermal halo\\
\hline
\end{tabular}
\label{EVALUATIONF563-V2large}
\end{table}
Let us note that in the  $K_0$ large case, the predicted rotation
curves for the SIDM appear flatter rotation curves at large radii,
compared to the small $K_0$ case. This is an interesting scenario.


\subsection{The Galaxy F565-V2}

\begin{figure}[h!]
\centering
\includegraphics[width=20pc]{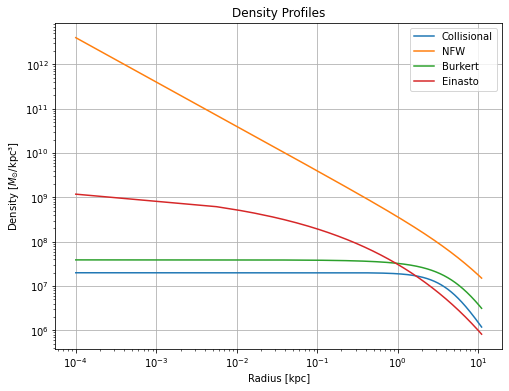}
\caption{The density of the collisional DM model (\ref{tanhmodel})
for the galaxy F565-V2, as a function of the radius.}
\label{F565-V2dens}
\end{figure}

\begin{figure}[h!]
\centering
\includegraphics[width=20pc]{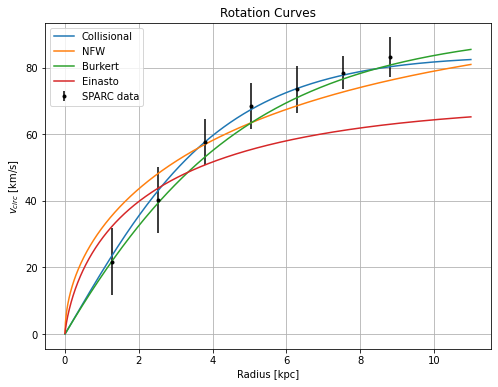}
\caption{The predicted rotation curves after using an optimization
for the collisional DM model (\ref{tanhmodel}), versus the SPARC
observational data for the galaxy F565-V2. We also plotted the
optimized curves for the NFW model, the Burkert model and the
Einasto model.} \label{F565-V2}
\end{figure}

\begin{figure}[h!]
\centering
\includegraphics[width=20pc]{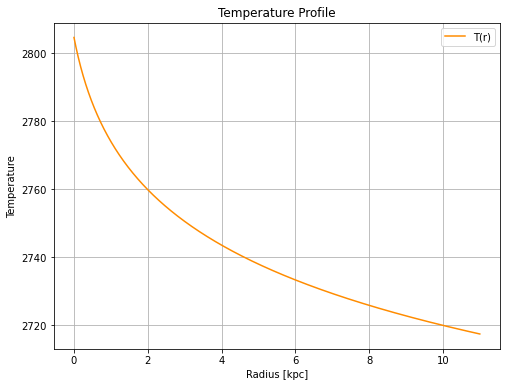}
\caption{The temperature as a function of the radius for the
collisional DM model (\ref{tanhmodel}) for the galaxy F565-V2.}
\label{F565-V2temp}
\end{figure}

For this galaxy, we shall choose $\rho_0=2\times
10^7$$M_{\odot}/\mathrm{Kpc}^{3}$. F563-V2 is a faint late-type
low-surface-brightness spiral galaxy at a distance of several tens
of Mpc, with optical radius of a few Kpc, an extended H\,I disk
out to \(\sim 10\)-\(15\ \mathrm{Kpc}\), and a dark halo extending
to at least several tens of Kpc. In Figs. \ref{F565-V2dens},
\ref{F565-V2} and \ref{F565-V2temp} we present the density of the
collisional DM model, the predicted rotation curves after using an
optimization for the collisional DM model (\ref{tanhmodel}),
versus the SPARC observational data and the temperature parameter
as a function of the radius respectively. As it can be seen, the
SIDM model produces viable rotation curves compatible with the
SPARC data. Also in Tables \ref{collF565-V2}, \ref{NavaroF565-V2},
\ref{BuckertF565-V2} and \ref{EinastoF565-V2} we present the
optimization values for the SIDM model, and the other DM profiles.
Also in Table \ref{EVALUATIONF565-V2} we present the overall
evaluation of the SIDM model for the galaxy at hand. The resulting
phenomenology is viable.
\begin{table}[h!]
  \begin{center}
    \caption{Collisional Dark Matter Optimization Values}
    \label{collF565-V2}
     \begin{tabular}{|r|r|}
     \hline
      \textbf{Parameter}   & \textbf{Optimization Values}
      \\  \hline
     $\delta_{\gamma} $ & 0.0000000012
\\  \hline
$\gamma_0 $ & 1.0001 \\ \hline $K_0$ ($M_{\odot} \,
\mathrm{Kpc}^{-3} \, (\mathrm{km/s})^{2}$)& 2800  \\ \hline
    \end{tabular}
  \end{center}
\end{table}
\begin{table}[h!]
  \begin{center}
    \caption{NFW  Optimization Values}
    \label{NavaroF565-V2}
     \begin{tabular}{|r|r|}
     \hline
      \textbf{Parameter}   & \textbf{Optimization Values}
      \\  \hline
   $\rho_s$   & $0.002\times 10^9$
\\  \hline
$r_s$&  20
\\  \hline
    \end{tabular}
  \end{center}
\end{table}
\begin{table}[h!]
  \begin{center}
    \caption{Burkert Optimization Values}
    \label{BuckertF565-V2}
     \begin{tabular}{|r|r|}
     \hline
      \textbf{Parameter}   & \textbf{Optimization Values}
      \\  \hline
     $\rho_0^B$  & $0.039\times 10^9$
\\  \hline
$r_0$&  6
\\  \hline
    \end{tabular}
  \end{center}
\end{table}
\begin{table}[h!]
  \begin{center}
    \caption{Einasto Optimization Values}
    \label{EinastoF565-V2}
    \begin{tabular}{|r|r|}
     \hline
      \textbf{Parameter}   & \textbf{Optimization Values}
      \\  \hline
     $\rho_e$  & $0.001\times 10^9$
\\  \hline
$r_e$ & 10
\\  \hline
$n_e$ & 0.27
\\  \hline
    \end{tabular}
  \end{center}
\end{table}
\begin{table}[h!]
\centering \caption{Physical assessment of collisional DM
parameters (F565-V2).}
\begin{tabular}{lcc}
\hline
Parameter & Value & Physical Verdict \\
\hline
$\gamma_0$ & $1.0001$ & Practically isothermal   \\
$\delta_\gamma$ & $1.2\times10^{-9}$ & Negligible  \\
$r_\gamma$ & $1.5\ \mathrm{Kpc}$ & Reasonable transition radius   \\
$K_0$ & $2.8\times10^{3}$ & Sets nearly-constant temperature \\
$r_c$ & $0.5\ \mathrm{Kpc}$ & Small core scale for $K(r)$; physically plausible \\
$p$ & $0.01$ & Very shallow decline \\
\hline
Overall & - & Model nearly isothermal halo. \\
\hline
\end{tabular}
\label{EVALUATIONF565-V2}
\end{table}


\subsection{The Galaxy F567-2}

\begin{figure}[h!]
\centering
\includegraphics[width=20pc]{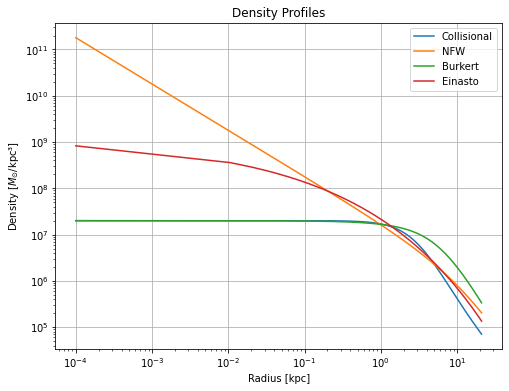}
\caption{The density of the collisional DM model (\ref{tanhmodel})
for the galaxy F567-2, as a function of the radius.}
\label{F567-2dens}
\end{figure}
\begin{figure}[h!]
\centering
\includegraphics[width=20pc]{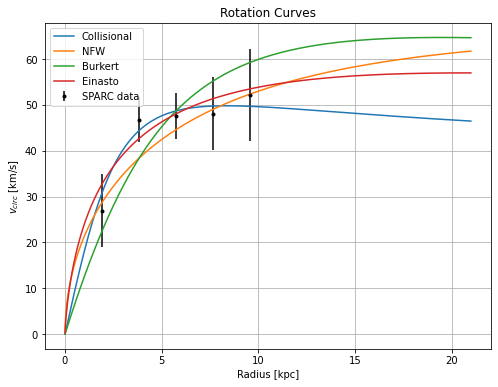}
\caption{The predicted rotation curves after using an optimization
for the collisional DM model (\ref{tanhmodel}), versus the SPARC
observational data for the galaxy F567-2. We also plotted the
optimized curves for the NFW model, the Burkert model and the
Einasto model.} \label{F567-2}
\end{figure}
\begin{figure}[h!]
\centering
\includegraphics[width=20pc]{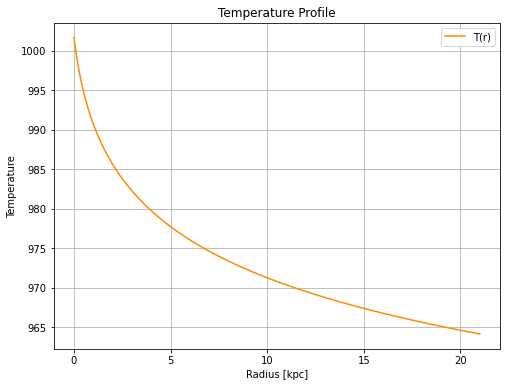}
\caption{The temperature as a function of the radius for the
collisional DM model (\ref{tanhmodel}) for the galaxy F567-2.}
\label{F567-2temp}
\end{figure}

For this galaxy, we shall choose $\rho_0=1.05\times
10^7$$M_{\odot}/\mathrm{Kpc}^{3}$. F567-2's is classified as a
low-surface-brightness galaxy. In Figs. \ref{F567-2dens},
\ref{F567-2} and \ref{F567-2temp} we present the density of the
collisional DM model, the predicted rotation curves after using an
optimization for the collisional DM model (\ref{tanhmodel}),
versus the SPARC observational data and the temperature parameter
as a function of the radius respectively. As it can be seen, the
SIDM model produces viable rotation curves compatible with the
SPARC data. Also in Tables \ref{collF567-2}, \ref{NavaroF567-2},
\ref{BuckertF567-2} and \ref{EinastoF567-2} we present the
optimization values for the SIDM model, and the other DM profiles.
Also in Table \ref{EVALUATIONF567-2} we present the overall
evaluation of the SIDM model for the galaxy at hand. The resulting
phenomenology is viable.
\begin{table}[h!]
  \begin{center}
    \caption{Collisional Dark Matter Optimization Values}
    \label{collF567-2}
     \begin{tabular}{|r|r|}
     \hline
      \textbf{Parameter}   & \textbf{Optimization Values}
      \\  \hline
     $\delta_{\gamma} $ & 0.0000000012
\\  \hline
$\gamma_0 $ & 1.0001 \\ \hline $K_0$ ($M_{\odot} \,
\mathrm{Kpc}^{-3} \, (\mathrm{km/s})^{2}$)& 1000  \\ \hline
    \end{tabular}
  \end{center}
\end{table}
\begin{table}[h!]
  \begin{center}
    \caption{NFW  Optimization Values}
    \label{NavaroF567-2}
     \begin{tabular}{|r|r|}
     \hline
      \textbf{Parameter}   & \textbf{Optimization Values}
      \\  \hline
   $\rho_s$   & $0.0009\times 10^9$
\\  \hline
$r_s$&  20
\\  \hline
    \end{tabular}
  \end{center}
\end{table}
\begin{table}[h!]
  \begin{center}
    \caption{Burkert Optimization Values}
    \label{BuckertF567-2}
     \begin{tabular}{|r|r|}
     \hline
      \textbf{Parameter}   & \textbf{Optimization Values}
      \\  \hline
     $\rho_0^B$  & $0.02\times 10^9$
\\  \hline
$r_0$&  6
\\  \hline
    \end{tabular}
  \end{center}
\end{table}
\begin{table}[h!]
  \begin{center}
    \caption{Einasto Optimization Values}
    \label{EinastoF567-2}
    \begin{tabular}{|r|r|}
     \hline
      \textbf{Parameter}   & \textbf{Optimization Values}
      \\  \hline
     $\rho_e$  & $0.0007\times 10^9$
\\  \hline
$r_e$ & 10
\\  \hline
$n_e$ & 0.27
\\  \hline
    \end{tabular}
  \end{center}
\end{table}
\begin{table}[h!]
\centering \caption{Physical assessment of collisional DM
parameters (F567-2).}
\begin{tabular}{lcc}
\hline
Parameter & Value & Physical Verdict \\
\hline
$\gamma_0$ & $1.0001$ & Practically isothermal   \\
$\delta_\gamma$ & $1.2\times10^{-9}$ & Negligible  \\
$r_\gamma$ & $1.5\ \mathrm{Kpc}$ & Reasonable transition radius   \\
$K_0$ & $1.0\times10^{3}$ & Sets nearly-constant temperature \\
$r_c$ & $0.5\ \mathrm{Kpc}$ & Small core scale for $K(r)$; physically plausible \\
$p$ & $0.01$ & Very shallow decline  \\
\hline
Overall & -& Models a nearly isothermal halo \\
\hline
\end{tabular}
\label{EVALUATIONF567-2}
\end{table}


\subsection{The Galaxy F568-1}

\begin{figure}[h!]
\centering
\includegraphics[width=20pc]{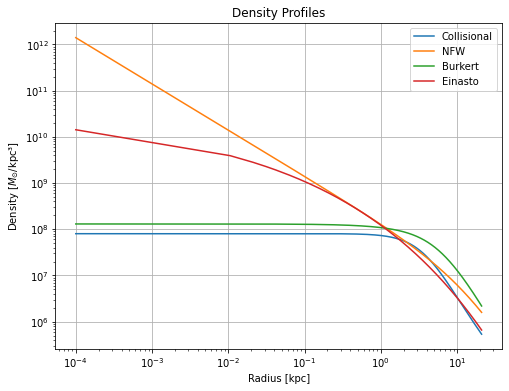}
\caption{The density of the collisional DM model (\ref{tanhmodel})
for the galaxy F568-1, as a function of the radius.}
\label{F568-1dens}
\end{figure}
\begin{figure}[h!]
\centering
\includegraphics[width=20pc]{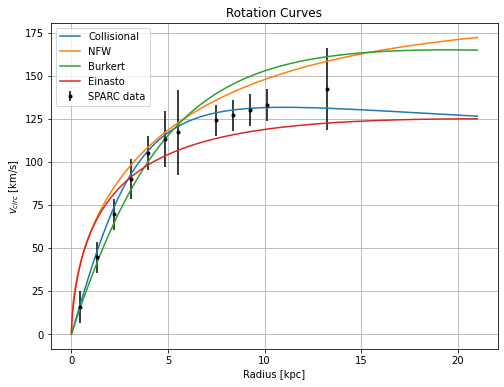}
\caption{The predicted rotation curves after using an optimization
for the collisional DM model (\ref{tanhmodel}), versus the SPARC
observational data for the galaxy F568-1. We also plotted the
optimized curves for the NFW model, the Burkert model and the
Einasto model.} \label{F568-1}
\end{figure}
\begin{figure}[h!]
\centering
\includegraphics[width=20pc]{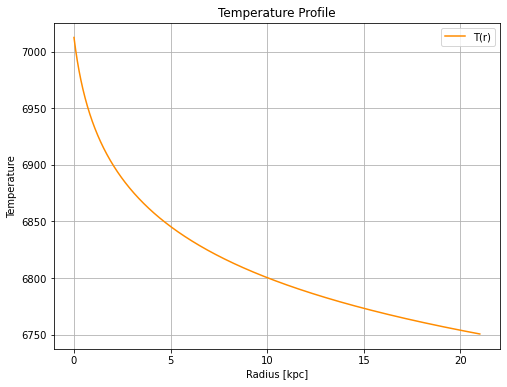}
\caption{The temperature as a function of the radius for the
collisional DM model (\ref{tanhmodel}) for the galaxy F568-1.}
\label{F568-1temp}
\end{figure}

For this galaxy, we shall choose $\rho_0=8\times
10^7$$M_{\odot}/\mathrm{Kpc}^{3}$. Galaxy F568-1 is a
low-surface-brightness   spiral galaxy, classified as type
SA(s)cd, indicating an unbarred spiral structure with loosely
wound arms and a late-type morphology. It is located at a distance
of approximately 2.54 Mpc from our Galaxy, placing it within the
realm of nearby galaxies suitable for detailed study. In terms of
its classification, F568-1 is considered a dwarf spiral galaxy due
to its low luminosity and mass compared to typical spiral
galaxies. Its dark matter halo is relatively extended, with a core
radius indicative of a shallow central density profile, aligning
with the characteristics observed in other low-surface-brightness
galaxies. In Figs. \ref{F568-1dens}, \ref{F568-1} and
\ref{F568-1temp} we present the density of the collisional DM
model, the predicted rotation curves after using an optimization
for the collisional DM model (\ref{tanhmodel}), versus the SPARC
observational data and the temperature parameter as a function of
the radius respectively. As it can be seen, the SIDM model
produces viable rotation curves compatible with the SPARC data.
Also in Tables \ref{collF568-1}, \ref{NavaroF568-1},
\ref{BuckertF568-1} and \ref{EinastoF568-1} we present the
optimization values for the SIDM model, and the other DM profiles.
Also in Table \ref{EVALUATIONF568-1} we present the overall
evaluation of the SIDM model for the galaxy at hand. The resulting
phenomenology is viable.
\begin{table}[h!]
  \begin{center}
    \caption{Collisional Dark Matter Optimization Values}
    \label{collF568-1}
     \begin{tabular}{|r|r|}
     \hline
      \textbf{Parameter}   & \textbf{Optimization Values}
      \\  \hline
     $\delta_{\gamma} $ & 0.0000000012
\\  \hline
$\gamma_0 $ & 1.0001 \\ \hline $K_0$ ($M_{\odot} \,
\mathrm{Kpc}^{-3} \, (\mathrm{km/s})^{2}$)& 7000  \\ \hline
    \end{tabular}
  \end{center}
\end{table}
\begin{table}[h!]
  \begin{center}
    \caption{NFW  Optimization Values}
    \label{NavaroF568-1}
     \begin{tabular}{|r|r|}
     \hline
      \textbf{Parameter}   & \textbf{Optimization Values}
      \\  \hline
   $\rho_s$   & $0.007\times 10^9$
\\  \hline
$r_s$&  20
\\  \hline
    \end{tabular}
  \end{center}
\end{table}
\begin{table}[h!]
  \begin{center}
    \caption{Burkert Optimization Values}
    \label{BuckertF568-1}
     \begin{tabular}{|r|r|}
     \hline
      \textbf{Parameter}   & \textbf{Optimization Values}
      \\  \hline
     $\rho_0^B$  & $0.13\times 10^9$
\\  \hline
$r_0$&  6
\\  \hline
    \end{tabular}
  \end{center}
\end{table}
\begin{table}[h!]
  \begin{center}
    \caption{Einasto Optimization Values}
    \label{EinastoF568-1}
    \begin{tabular}{|r|r|}
     \hline
      \textbf{Parameter}   & \textbf{Optimization Values}
      \\  \hline
     $\rho_e$  & $0.0033\times 10^9$
\\  \hline
$r_e$ & 10
\\  \hline
$n_e$ & 0.22
\\  \hline
    \end{tabular}
  \end{center}
\end{table}
\begin{table}[h!]
\centering \caption{Physical assessment of collisional DM
parameters (F568-1).}
\begin{tabular}{lcc}
\hline
Parameter & Value & Physical Verdict \\
\hline
$\gamma_0$ & $1.0001$ & Practically isothermal   \\
$\delta_\gamma$ & $1.2\times10^{-9}$ & Negligible  \\
$r_\gamma$ & $1.5\ \mathrm{Kpc}$ & Reasonable transition radius   \\
$K_0$ & $7.0\times10^{3}$ & Sets nearly-constant temperature \\
$r_c$ & $0.5\ \mathrm{Kpc}$ & Small core scale for $K(r)$; physically plausible \\
$p$ & $0.01$ & Very shallow decline  \\
\hline
Overall & -& Models a nearly isothermal halo \\
\hline
\end{tabular}
\label{EVALUATIONF568-1}
\end{table}


\subsection{The Galaxy F568-3}

\begin{figure}[h!]
\centering
\includegraphics[width=20pc]{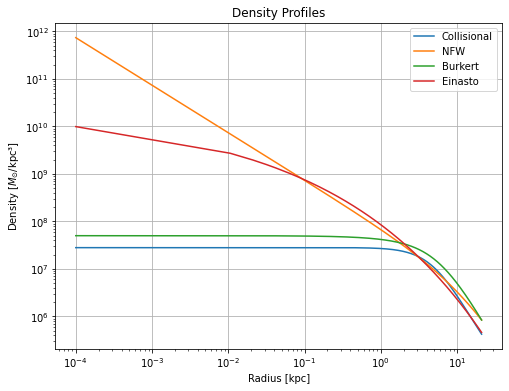}
\caption{The density of the collisional DM model (\ref{tanhmodel})
for the galaxy F568-3, as a function of the radius.}
\label{F568-3dens}
\end{figure}
\begin{figure}[h!]
\centering
\includegraphics[width=20pc]{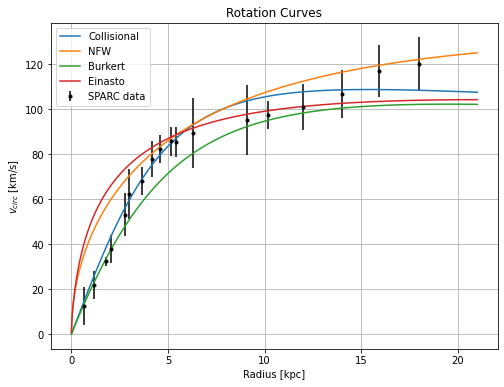}
\caption{The predicted rotation curves after using an optimization
for the collisional DM model (\ref{tanhmodel}), versus the SPARC
observational data for the galaxy F568-3. We also plotted the
optimized curves for the NFW model, the Burkert model and the
Einasto model.} \label{F568-3}
\end{figure}
\begin{figure}[h!]
\centering
\includegraphics[width=20pc]{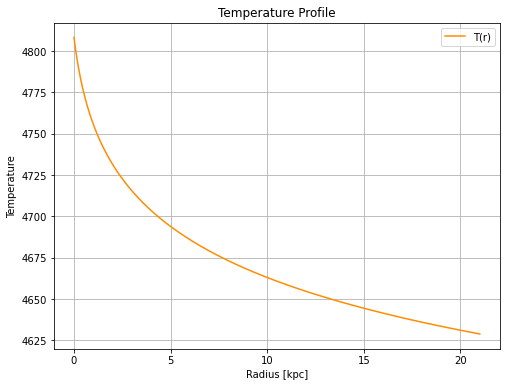}
\caption{The temperature as a function of the radius for the
collisional DM model (\ref{tanhmodel}) for the galaxy F568-3.}
\label{F568-3temp}
\end{figure}

For this galaxy, we shall choose $\rho_0=2.8\times
10^7$$M_{\odot}/\mathrm{Kpc}^{3}$. F568-3 is catalogued as a
low-surface-brightness, late--type/spiral galaxy; high-resolution
rotation-curve studies place it among the prototypical
low-surface-brightness systems used to study dark matter cores
versus cusps. In Figs. \ref{F568-3dens}, \ref{F568-3} and
\ref{F568-3temp} we present the density of the collisional DM
model, the predicted rotation curves after using an optimization
for the collisional DM model (\ref{tanhmodel}), versus the SPARC
observational data and the temperature parameter as a function of
the radius respectively. As it can be seen, the SIDM model
produces viable rotation curves compatible with the SPARC data.
Also in Tables \ref{collF568-3}, \ref{NavaroF568-3},
\ref{BuckertF568-3} and \ref{EinastoF568-3} we present the
optimization values for the SIDM model, and the other DM profiles.
Also in Table \ref{EVALUATIONF568-3} we present the overall
evaluation of the SIDM model for the galaxy at hand. The resulting
phenomenology is viable.
\begin{table}[h!]
  \begin{center}
    \caption{Collisional Dark Matter Optimization Values}
    \label{collF568-3}
     \begin{tabular}{|r|r|}
     \hline
      \textbf{Parameter}   & \textbf{Optimization Values}
      \\  \hline
     $\delta_{\gamma} $ & 0.0000000012
\\  \hline
$\gamma_0 $ & 1.0001 \\ \hline $K_0$ ($M_{\odot} \,
\mathrm{Kpc}^{-3} \, (\mathrm{km/s})^{2}$)& 4800  \\ \hline
    \end{tabular}
  \end{center}
\end{table}
\begin{table}[h!]
  \begin{center}
    \caption{NFW  Optimization Values}
    \label{NavaroF568-3}
     \begin{tabular}{|r|r|}
     \hline
      \textbf{Parameter}   & \textbf{Optimization Values}
      \\  \hline
   $\rho_s$   & $0.0037\times 10^9$
\\  \hline
$r_s$&  20
\\  \hline
    \end{tabular}
  \end{center}
\end{table}
\begin{table}[h!]
  \begin{center}
    \caption{Burkert Optimization Values}
    \label{BuckertF568-3}
     \begin{tabular}{|r|r|}
     \hline
      \textbf{Parameter}   & \textbf{Optimization Values}
      \\  \hline
     $\rho_0^B$  & $0.05\times 10^9$
\\  \hline
$r_0$&  6
\\  \hline
    \end{tabular}
  \end{center}
\end{table}
\begin{table}[h!]
  \begin{center}
    \caption{Einasto Optimization Values}
    \label{EinastoF568-3}
    \begin{tabular}{|r|r|}
     \hline
      \textbf{Parameter}   & \textbf{Optimization Values}
      \\  \hline
     $\rho_e$  & $0.0023\times 10^9$
\\  \hline
$r_e$ & 10
\\  \hline
$n_e$ & 0.22
\\  \hline
    \end{tabular}
  \end{center}
\end{table}
\begin{table}[h!]
\centering \caption{Physical assessment of collisional DM
parameters (F568-3).}
\begin{tabular}{lcc}
\hline
Parameter & Value & Physical Verdict \\
\hline
$\gamma_0$ & $1.0001$ & Practically isothermal   \\
$\delta_\gamma$ & $1.2\times10^{-9}$ & Negligible  \\
$r_\gamma$ & $1.5\ \mathrm{Kpc}$ & Reasonable transition radius   \\
$K_0$ & $4.8\times10^{3}$ & Sets nearly-constant temperature \\
$r_c$ & $0.5\ \mathrm{Kpc}$ & Small core scale for $K(r)$; physically plausible \\
$p$ & $0.01$ & Very shallow decline  \\
\hline
Overall &-& Model behaves like an isothermal halo \\
\hline
\end{tabular}
\label{EVALUATIONF568-3}
\end{table}


\subsection{The Galaxy F568-V1}

\begin{figure}[h!]
\centering
\includegraphics[width=20pc]{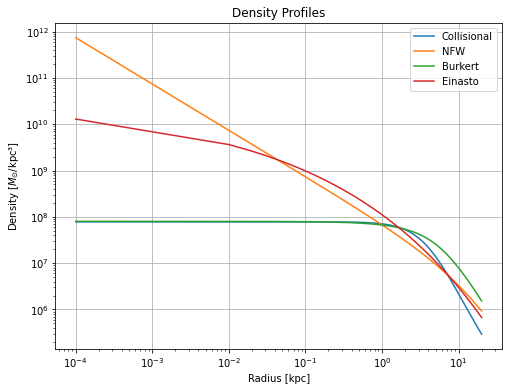}
\caption{The density of the collisional DM model (\ref{tanhmodel})
for the galaxy F568-V1, as a function of the radius.}
\label{F568-V1dens}
\end{figure}
\begin{figure}[h!]
\centering
\includegraphics[width=20pc]{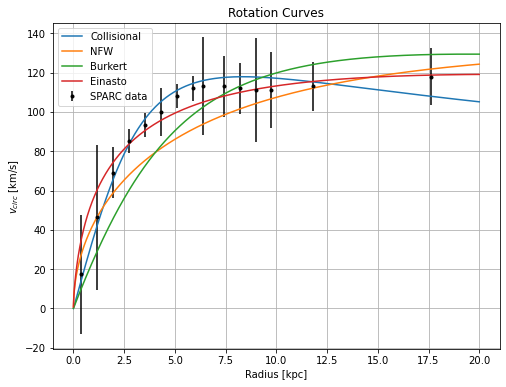}
\caption{The predicted rotation curves after using an optimization
for the collisional DM model (\ref{tanhmodel}), versus the SPARC
observational data for the galaxy F568-V1. We also plotted the
optimized curves for the NFW model, the Burkert model and the
Einasto model.} \label{F568-V1}
\end{figure}
\begin{figure}[h!]
\centering
\includegraphics[width=20pc]{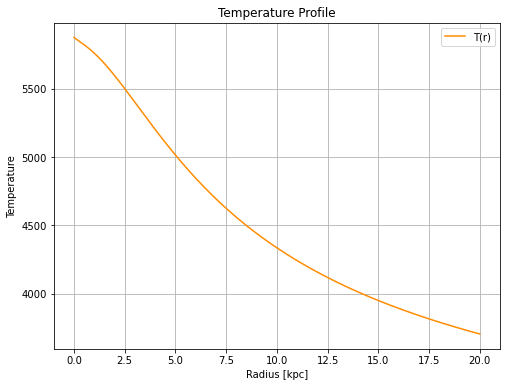}
\caption{The temperature as a function of the radius for the
collisional DM model (\ref{tanhmodel}) for the galaxy F568-V1.}
\label{F568-V1temp}
\end{figure}

For this galaxy, we shall choose $\rho_0=7.8\times
10^7$$M_{\odot}/\mathrm{Kpc}^{3}$. F568-V1 is a faint late-type
low-surface-brightness disk galaxy at a distance of approximately
\(60\;\mathrm{Mpc}\). In Figs. \ref{F568-V1dens}, \ref{F568-V1}
and \ref{F568-V1temp} we present the density of the collisional DM
model, the predicted rotation curves after using an optimization
for the collisional DM model (\ref{tanhmodel}), versus the SPARC
observational data and the temperature parameter as a function of
the radius respectively. As it can be seen, the SIDM model
produces viable rotation curves compatible with the SPARC data.
Also in Tables \ref{collF568-V1}, \ref{NavaroF568-V1},
\ref{BuckertF568-V1} and \ref{EinastoF568-V1} we present the
optimization values for the SIDM model, and the other DM profiles.
Also in Table \ref{EVALUATIONF568-V1} we present the overall
evaluation of the SIDM model for the galaxy at hand. The resulting
phenomenology is viable.
\begin{table}[h!]
  \begin{center}
    \caption{Collisional Dark Matter Optimization Values}
    \label{collF568-V1}
     \begin{tabular}{|r|r|}
     \hline
      \textbf{Parameter}   & \textbf{Optimization Values}
      \\  \hline
     $\delta_{\gamma} $ & 0.0000000012
\\  \hline
$\gamma_0 $ & 1.0001 \\ \hline $K_0$ ($M_{\odot} \,
\mathrm{Kpc}^{-3} \, (\mathrm{km/s})^{2}$)& 5500  \\ \hline
    \end{tabular}
  \end{center}
\end{table}
\begin{table}[h!]
  \begin{center}
    \caption{NFW  Optimization Values}
    \label{NavaroF568-V1}
     \begin{tabular}{|r|r|}
     \hline
      \textbf{Parameter}   & \textbf{Optimization Values}
      \\  \hline
   $\rho_s$   & $0.0037\times 10^9$
\\  \hline
$r_s$&  20
\\  \hline
    \end{tabular}
  \end{center}
\end{table}
\begin{table}[h!]
  \begin{center}
    \caption{Burkert Optimization Values}
    \label{BuckertF568-V1}
     \begin{tabular}{|r|r|}
     \hline
      \textbf{Parameter}   & \textbf{Optimization Values}
      \\  \hline
     $\rho_0^B$  & $0.08\times 10^9$
\\  \hline
$r_0$&  6
\\  \hline
    \end{tabular}
  \end{center}
\end{table}
\begin{table}[h!]
  \begin{center}
    \caption{Einasto Optimization Values}
    \label{EinastoF568-V1}
    \begin{tabular}{|r|r|}
     \hline
      \textbf{Parameter}   & \textbf{Optimization Values}
      \\  \hline
     $\rho_e$  & $0.003\times 10^9$
\\  \hline
$r_e$ & 10
\\  \hline
$n_e$ & 0.15
\\  \hline
    \end{tabular}
  \end{center}
\end{table}
\begin{table}[h!]
\centering \caption{Physical assessment of collisional DM
parameters for F568-V1.}
\begin{tabular}{lcc}
\hline
Parameter & Value & Physical Verdict \\
\hline
$\gamma_0$ & $1.0001$ & Practically isothermal \\
$\delta_\gamma$ & $1.2\times10^{-9}$ & Essentially zero \\
$r_\gamma$ & $1.5\ \mathrm{Kpc}$ & Transition radius chosen inside outer halo \\
$K_0$ & $5.5\times10^{3}$ & Sets temperature/entropy scale \\
$r_c$ & $0.5\ \mathrm{Kpc}$ & Small core scale; yields compact central core \\
$p$ & $0.01$ & Very shallow decline of $K(r)$  \\
\hline
Overall &-& Physically consistent but nearly isothermal and almost spatially uniform \\
\hline
\end{tabular}
\label{EVALUATIONF568-V1}
\end{table}


\subsection{The Galaxy F571-8 Non-viable Low-Surface-Brightness Galaxy}

\begin{figure}[h!]
\centering
\includegraphics[width=20pc]{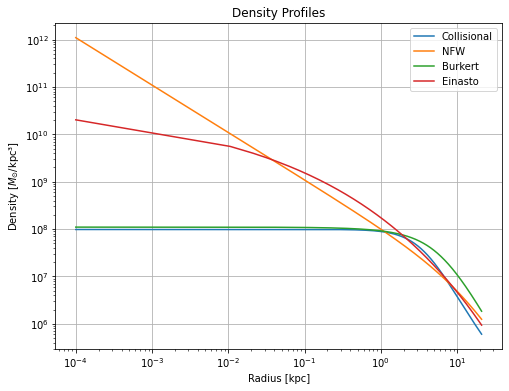}
\caption{The density of the collisional DM model (\ref{tanhmodel})
for the galaxy F571-8, as a function of the radius.}
\label{F571-8dens}
\end{figure}
\begin{figure}[h!]
\centering
\includegraphics[width=20pc]{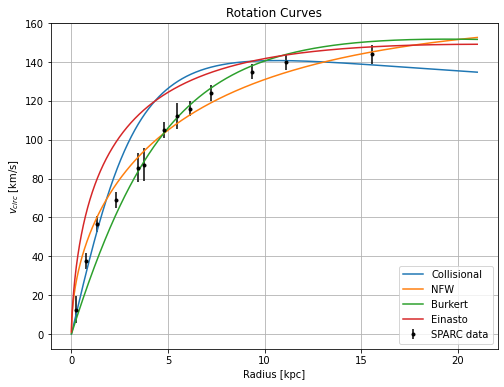}
\caption{The predicted rotation curves after using an optimization
for the collisional DM model (\ref{tanhmodel}), versus the SPARC
observational data for the galaxy F571-8. We also plotted the
optimized curves for the NFW model, the Burkert model and the
Einasto model.} \label{F571-8}
\end{figure}
\begin{figure}[h!]
\centering
\includegraphics[width=20pc]{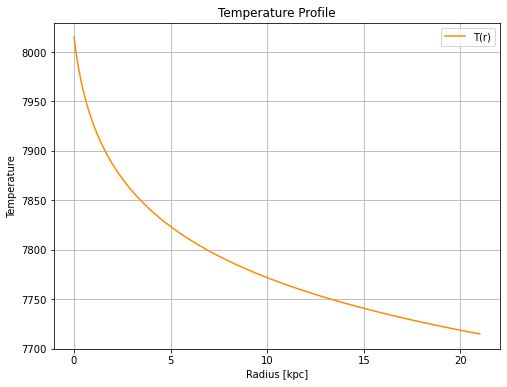}
\caption{The temperature as a function of the radius for the
collisional DM model (\ref{tanhmodel}) for the galaxy F571-8.}
\label{F571-8temp}
\end{figure}
\begin{figure}[h!]
\centering
\includegraphics[width=20pc]{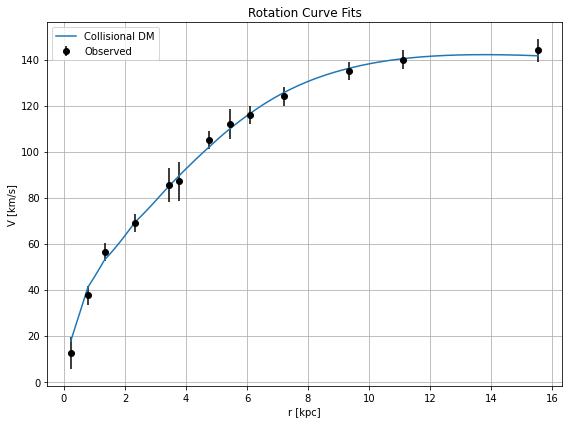}
\caption{The predicted rotation curves after using an optimization
for the collisional DM model (\ref{tanhmodel}), versus the
extended SPARC observational data for the galaxy F571-8. The model
includes the rotation curves from all the components of the
galaxy, including gas and disk velocities, along with the
collisional DM model.} \label{extendedF571-8}
\end{figure}

For this galaxy, we shall choose $\rho_0=9.8\times
10^7$$M_{\odot}/\mathrm{Kpc}^{3}$. F571-8 is an edge-on
low-surface-brightness spiral with a large bulge. Its distance is
at $\sim {48}{Mpc}$. In Figs. \ref{F571-8dens}, \ref{F571-8} and
\ref{F571-8temp} we present the density of the collisional DM
model, the predicted rotation curves after using an optimization
for the collisional DM model (\ref{tanhmodel}), versus the SPARC
observational data and the temperature parameter as a function of
the radius respectively. As it can be seen, the SIDM model
produces non-viable rotation curves compatible with the SPARC
data. Also in Tables \ref{collF571-8}, \ref{NavaroF571-8},
\ref{BuckertF571-8} and \ref{EinastoF571-8} we present the
optimization values for the SIDM model, and the other DM profiles.
Also in Table \ref{EVALUATIONF571-8} we present the overall
evaluation of the SIDM model for the galaxy at hand. The resulting
phenomenology is non-viable.
\begin{table}[h!]
  \begin{center}
    \caption{Collisional Dark Matter Optimization Values}
    \label{collF571-8}
     \begin{tabular}{|r|r|}
     \hline
      \textbf{Parameter}   & \textbf{Optimization Values}
      \\  \hline
     $\delta_{\gamma} $ & 0.0000000012
\\  \hline
$\gamma_0 $ &  1.0001 \\ \hline $K_0$ ($M_{\odot} \,
\mathrm{Kpc}^{-3} \, (\mathrm{km/s})^{2}$)& 8000 \\ \hline
    \end{tabular}
  \end{center}
\end{table}
\begin{table}[h!]
  \begin{center}
    \caption{NFW  Optimization Values}
    \label{NavaroF571-8}
     \begin{tabular}{|r|r|}
     \hline
      \textbf{Parameter}   & \textbf{Optimization Values}
      \\  \hline
   $\rho_s$   & $0.0055\times 10^9$
\\  \hline
$r_s$&  20
\\  \hline
    \end{tabular}
  \end{center}
\end{table}
\begin{table}[h!]
  \begin{center}
    \caption{Burkert Optimization Values}
    \label{BuckertF571-8}
     \begin{tabular}{|r|r|}
     \hline
      \textbf{Parameter}   & \textbf{Optimization Values}
      \\  \hline
     $\rho_0^B$  & $0.11\times 10^9$
\\  \hline
$r_0$&  6
\\  \hline
    \end{tabular}
  \end{center}
\end{table}
\begin{table}[h!]
  \begin{center}
    \caption{Einasto Optimization Values}
    \label{EinastoF571-8}
    \begin{tabular}{|r|r|}
     \hline
      \textbf{Parameter}   & \textbf{Optimization Values}
      \\  \hline
     $\rho_e$  & $0.0047\times 10^9$
\\  \hline
$r_e$ & 10
\\  \hline
$n_e$ & 0.22
\\  \hline
    \end{tabular}
  \end{center}
\end{table}
\begin{table}[h!]
\centering \caption{Physical assessment of collisional DM
parameters for F571-8.}
\begin{tabular}{lcc}
\hline
Parameter & Value & Physical Verdict \\
\hline
$\gamma_0$ & $1.0001$ & Practically isothermal \\
$\delta_\gamma$ & $1.2\times10^{-9}$ & Essentially zero\\
$r_\gamma$ & $1.5\ \mathrm{Kpc}$ & Reasonable transition radius   \\
$K_0$ & $8.0\times10^{3}$ & Sets temperature/entropy scale \\
$r_c$ & $0.5\ \mathrm{Kpc}$ & Small core scale; yields a compact central core \\
$p$ & $0.01$ & Very shallow decline of $K(r)$  \\
\hline
Overall &-& Physically consistent but effectively isothermal and spatially uniform \\
\hline
\end{tabular}
\label{EVALUATIONF571-8}
\end{table}
Now the extended picture including the rotation velocity from the
other components of the galaxy, such as the disk and gas, makes
the collisional DM model viable for this galaxy. In Fig.
\ref{extendedF571-8} we present the combined rotation curves
including the other components of the galaxy along with the
collisional matter. As it can be seen, the extended collisional DM
model is non-viable.
Also in Table \ref{evaluationextendedF571-8} we present the values
of the free parameters of the collisional DM model for which the
maximum compatibility with the SPARC data comes for the galaxy
ESO116-G012.
\begin{table}[h!]
\centering \caption{Physical assessment of Extended collisional DM
parameters for F571-8.}
\begin{tabular}{lcc}
\hline
Parameter & Value & Physical Verdict \\
\hline
$\gamma_0$ & 1.1076 & Nearly isothermal core \\
$\delta_\gamma$ & 0.04999 & Small but noticeable radial variation \\
$K_0$ & 3000 & Moderate entropy \\
$ml_{\text{disk}}$ & 0.4512 & Reasonable low mass-to-light ratio \\
$ml_{\text{bulge}}$ & 0.0000 & No bulge component \\
\hline
Overall &-& Physically plausible \\
\hline
\end{tabular}
\label{evaluationextendedF571-8}
\end{table}


\subsection{The Galaxy F571-V1}

\begin{figure}[h!]
\centering
\includegraphics[width=20pc]{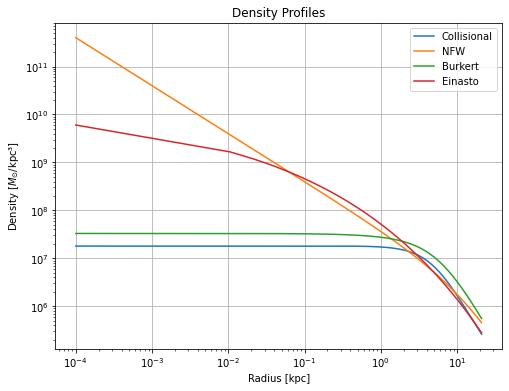}
\caption{The density of the collisional DM model (\ref{tanhmodel})
for the galaxy F571-V1, as a function of the radius.}
\label{F571-V1dens}
\end{figure}
\begin{figure}[h!]
\centering
\includegraphics[width=20pc]{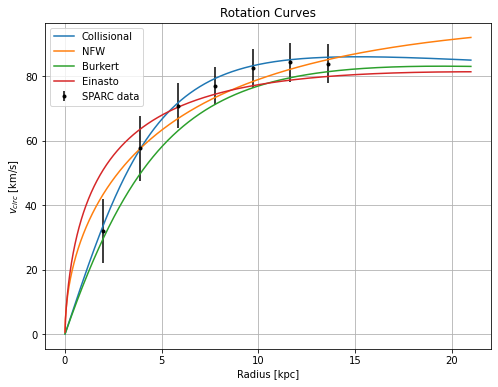}
\caption{The predicted rotation curves after using an optimization
for the collisional DM model (\ref{tanhmodel}), versus the SPARC
observational data for the galaxy F571-V1. We also plotted the
optimized curves for the NFW model, the Burkert model and the
Einasto model.} \label{F571-V1}
\end{figure}
\begin{figure}[h!]
\centering
\includegraphics[width=20pc]{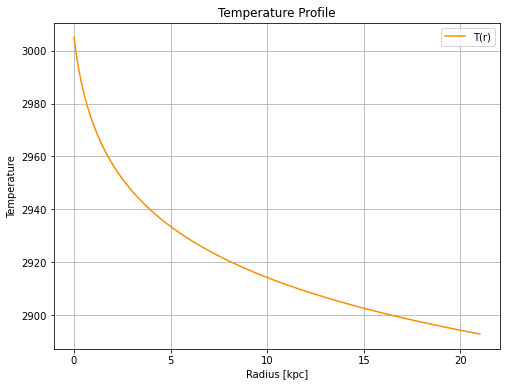}
\caption{The temperature as a function of the radius for the
collisional DM model (\ref{tanhmodel}) for the galaxy F571-V1.}
\label{F571-V1temp}
\end{figure}

For this galaxy, we shall choose $\rho_0=1.8\times
10^7$$M_{\odot}/\mathrm{Kpc}^{3}$. F571-V1 is a diffuse, late-type
low-surface-brightness spiral at $51$Mpc, with a large optical
disk radius ($\sim {25}{Kpc}$) and an extended HI disk ($\sim
{40}{Kpc}$). Its morphology and dynamics place it firmly among the
prototypical dark-matter-dominated low-surface-brightness
galaxies. In Figs. \ref{F571-V1dens}, \ref{F571-V1} and
\ref{F571-V1temp} we present the density of the collisional DM
model, the predicted rotation curves after using an optimization
for the collisional DM model (\ref{tanhmodel}), versus the SPARC
observational data and the temperature parameter as a function of
the radius respectively. As it can be seen, the SIDM model
produces viable rotation curves compatible with the SPARC data.
Also in Tables \ref{collF571-V1}, \ref{NavaroF571-V1},
\ref{BuckertF571-V1} and \ref{EinastoF571-V1} we present the
optimization values for the SIDM model, and the other DM profiles.
Also in Table \ref{EVALUATIONF571-V1} we present the overall
evaluation of the SIDM model for the galaxy at hand. The resulting
phenomenology is viable.
\begin{table}[h!]
  \begin{center}
    \caption{Collisional Dark Matter Optimization Values}
    \label{collF571-V1}
     \begin{tabular}{|r|r|}
     \hline
      \textbf{Parameter}   & \textbf{Optimization Values}
      \\  \hline
     $\delta_{\gamma} $ & 0.0000000012
\\  \hline
$\gamma_0 $ & 1.0001 \\ \hline $K_0$ ($M_{\odot} \,
\mathrm{Kpc}^{-3} \, (\mathrm{km/s})^{2}$)& 3000  \\ \hline
    \end{tabular}
  \end{center}
\end{table}
\begin{table}[h!]
  \begin{center}
    \caption{NFW  Optimization Values}
    \label{NavaroF571-V1}
     \begin{tabular}{|r|r|}
     \hline
      \textbf{Parameter}   & \textbf{Optimization Values}
      \\  \hline
   $\rho_s$   & $0.002\times 10^9$
\\  \hline
$r_s$&  20
\\  \hline
    \end{tabular}
  \end{center}
\end{table}
\begin{table}[h!]
  \begin{center}
    \caption{Burkert Optimization Values}
    \label{BuckertF571-V1}
     \begin{tabular}{|r|r|}
     \hline
      \textbf{Parameter}   & \textbf{Optimization Values}
      \\  \hline
     $\rho_0^B$  & $0.033\times 10^9$
\\  \hline
$r_0$&  6
\\  \hline
    \end{tabular}
  \end{center}
\end{table}
\begin{table}[h!]
  \begin{center}
    \caption{Einasto Optimization Values}
    \label{EinastoF571-V1}
    \begin{tabular}{|r|r|}
     \hline
      \textbf{Parameter}   & \textbf{Optimization Values}
      \\  \hline
     $\rho_e$  & $0.0014\times 10^9$
\\  \hline
$r_e$ & 10
\\  \hline
$n_e$ & 0.22
\\  \hline
    \end{tabular}
  \end{center}
\end{table}
\begin{table}[h!]
\centering \caption{Physical assessment of collisional DM
parameters for F571-V1.}
\begin{tabular}{lcc}
\hline
Parameter & Value & Physical Verdict \\
\hline
$\gamma_0$ & $1.0001$ & Practically isothermal  \\
$\delta_\gamma$ & $1.2\times10^{-9}$ & Essentially zero   \\
$r_\gamma$ & $1.5\ \mathrm{Kpc}$ & Reasonable transition radius   \\
$K_0$ & $3.0\times10^{3}$ & Sets temperature/entropy scale \\
$r_c$ & $0.5\ \mathrm{Kpc}$ & Small core scale; yields a compact central core \\
$p$ & $0.01$ & Very shallow decline of $K(r)$  \\
\hline
Overall &-& Physically consistent but effectively isothermal and spatially uniform \\
\hline
\end{tabular}
\label{EVALUATIONF571-V1}
\end{table}


\subsection{The Galaxy F574-1}

\begin{figure}[h!]
\centering
\includegraphics[width=20pc]{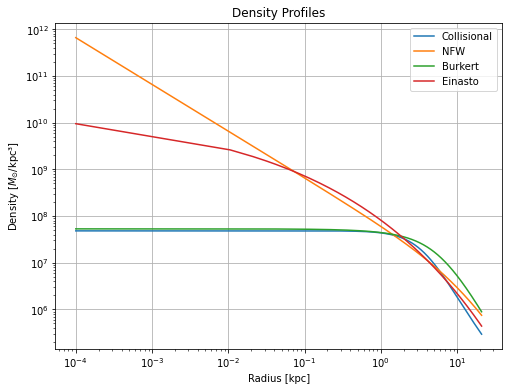}
\caption{The density of the collisional DM model (\ref{tanhmodel})
for the galaxy F574-1, as a function of the radius.}
\label{F574-1dens}
\end{figure}
\begin{figure}[h!]
\centering
\includegraphics[width=20pc]{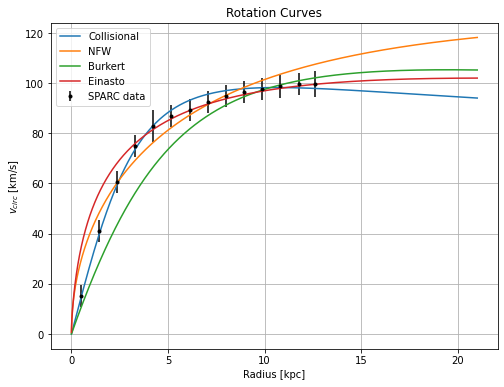}
\caption{The predicted rotation curves after using an optimization
for the collisional DM model (\ref{tanhmodel}), versus the SPARC
observational data for the galaxy F574-1. We also plotted the
optimized curves for the NFW model, the Burkert model and the
Einasto model.} \label{F574-1}
\end{figure}
\begin{figure}[h!]
\centering
\includegraphics[width=20pc]{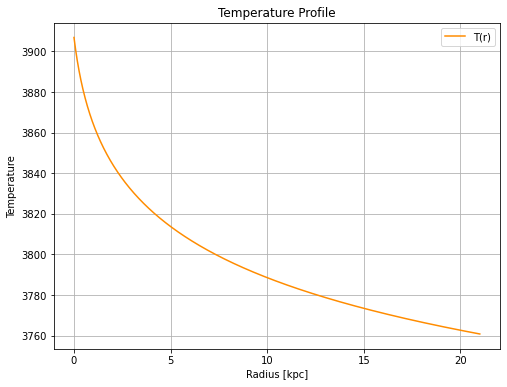}
\caption{The temperature as a function of the radius for the
collisional DM model (\ref{tanhmodel}) for the galaxy F574-1.}
\label{F574-1temp}
\end{figure}

For this galaxy, we shall choose $\rho_0=4.8\times
10^7$$M_{\odot}/\mathrm{Kpc}^{3}$. F574-1 is classified in the
literature as a late--type, low-surface-brightness disk galaxy
that is strongly dark--matter dominated in its outer parts. Its
distance is $D = 95.5\ \mathrm{Mpc}$. In Figs. \ref{F574-1dens},
\ref{F574-1} and \ref{F574-1temp} we present the density of the
collisional DM model, the predicted rotation curves after using an
optimization for the collisional DM model (\ref{tanhmodel}),
versus the SPARC observational data and the temperature parameter
as a function of the radius respectively. As it can be seen, the
SIDM model produces viable rotation curves compatible with the
SPARC data. Also in Tables \ref{collF574-1}, \ref{NavaroF574-1},
\ref{BuckertF574-1} and \ref{EinastoF574-1} we present the
optimization values for the SIDM model, and the other DM profiles.
Also in Table \ref{EVALUATIONF574-1} we present the overall
evaluation of the SIDM model for the galaxy at hand. The resulting
phenomenology is viable.
\begin{table}[h!]
  \begin{center}
    \caption{Collisional Dark Matter Optimization Values}
    \label{collF574-1}
     \begin{tabular}{|r|r|}
     \hline
      \textbf{Parameter}   & \textbf{Optimization Values}
      \\  \hline
     $\delta_{\gamma} $ & 0.0000000012
\\  \hline
$\gamma_0 $ & 1.0001 \\ \hline $K_0$ ($M_{\odot} \,
\mathrm{Kpc}^{-3} \, (\mathrm{km/s})^{2}$)& 3900  \\ \hline
    \end{tabular}
  \end{center}
\end{table}
\begin{table}[h!]
  \begin{center}
    \caption{NFW  Optimization Values}
    \label{NavaroF574-1}
     \begin{tabular}{|r|r|}
     \hline
      \textbf{Parameter}   & \textbf{Optimization Values}
      \\  \hline
   $\rho_s$   & $0.0033\times 10^9$
\\  \hline
$r_s$&  20
\\  \hline
    \end{tabular}
  \end{center}
\end{table}
\begin{table}[h!]
  \begin{center}
    \caption{Burkert Optimization Values}
    \label{BuckertF574-1}
     \begin{tabular}{|r|r|}
     \hline
      \textbf{Parameter}   & \textbf{Optimization Values}
      \\  \hline
     $\rho_0^B$  & $0.053\times 10^9$
\\  \hline
$r_0$&  6
\\  \hline
    \end{tabular}
  \end{center}
\end{table}
\begin{table}[h!]
  \begin{center}
    \caption{Einasto Optimization Values}
    \label{EinastoF574-1}
    \begin{tabular}{|r|r|}
     \hline
      \textbf{Parameter}   & \textbf{Optimization Values}
      \\  \hline
     $\rho_e$  & $0.0022\times 10^9$
\\  \hline
$r_e$ & 10
\\  \hline
$n_e$ & 0.22
\\  \hline
    \end{tabular}
  \end{center}
\end{table}
\begin{table}[h!]
\centering \caption{Physical assessment of collisional DM
parameters for F574-1.}
\begin{tabular}{lcc}
\hline
Parameter & Value & Physical Verdict \\
\hline
$\gamma_0$ & $1.0001$ & Practically isothermal  \\
$\delta_\gamma$ & $1.2\times10^{-9}$ & Essentially zero   \\
$r_\gamma$ & $1.5\ \mathrm{Kpc}$ & Reasonable transition radius   \\
$K_0$ & $3.9\times10^{3}$ & Sets temperature/entropy scale \\
$r_c$ & $0.5\ \mathrm{Kpc}$ & Small core scale; yields a compact central core \\
$p$ & $0.01$ & Very shallow decline of $K(r)$  \\
\hline
Overall &-& Physically consistent but effectively isothermal and spatially uniform\\
\hline
\end{tabular}
\label{EVALUATIONF574-1}
\end{table}


\subsection{The Galaxy F574-2}

\begin{figure}[h!]
\centering
\includegraphics[width=20pc]{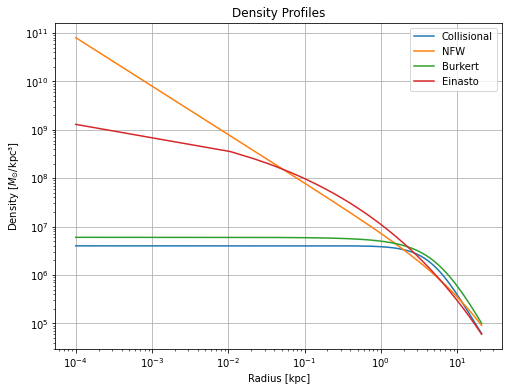}
\caption{The density of the collisional DM model (\ref{tanhmodel})
for the galaxy F574-2, as a function of the radius.}
\label{F574-2dens}
\end{figure}
\begin{figure}[h!]
\centering
\includegraphics[width=20pc]{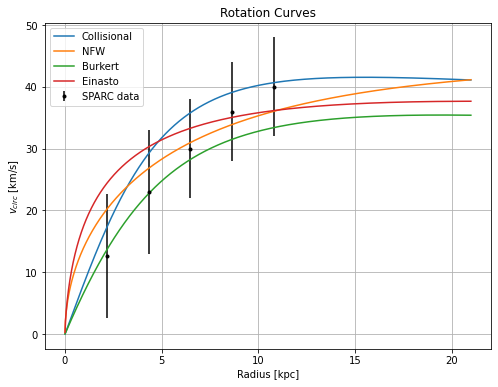}
\caption{The predicted rotation curves after using an optimization
for the collisional DM model (\ref{tanhmodel}), versus the SPARC
observational data for the galaxy F574-2. We also plotted the
optimized curves for the NFW model, the Burkert model and the
Einasto model.} \label{F574-2}
\end{figure}
\begin{figure}[h!]
\centering
\includegraphics[width=20pc]{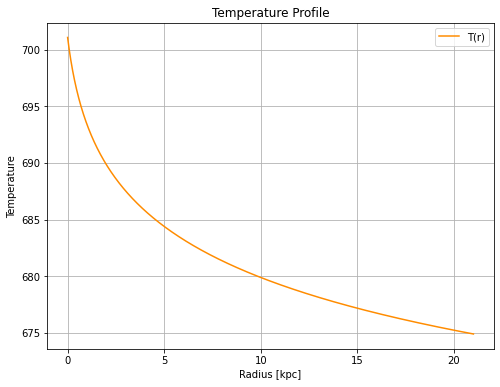}
\caption{The temperature as a function of the radius for the
collisional DM model (\ref{tanhmodel}) for the galaxy F574-2.}
\label{F574-2temp}
\end{figure}

For this galaxy, we shall choose $\rho_0=4\times
10^6$$M_{\odot}/\mathrm{Kpc}^{3}$. F574-2 is a late-type,
low-surface-brightness   spiral galaxy included in the SPARC. It
is characterized by a significant dark matter component, which
becomes evident in its extended rotation curve that deviates from
the predictions based solely on luminous matter. The best
published distance estimate for F574-2 is
\[
D = 95.5 \pm 8.7 \ \mathrm{Mpc}.
\]
In Figs. \ref{F574-2dens}, \ref{F574-2} and \ref{F574-2temp} we
present the density of the collisional DM model, the predicted
rotation curves after using an optimization for the collisional DM
model (\ref{tanhmodel}), versus the SPARC observational data and
the temperature parameter as a function of the radius
respectively. As it can be seen, the SIDM model produces viable
rotation curves compatible with the SPARC data. Also in Tables
\ref{collF574-2}, \ref{NavaroF574-2}, \ref{BuckertF574-2} and
\ref{EinastoF574-2} we present the optimization values for the
SIDM model, and the other DM profiles. Also in Table
\ref{EVALUATIONF574-2} we present the overall evaluation of the
SIDM model for the galaxy at hand. The resulting phenomenology is
viable.
\begin{table}[h!]
  \begin{center}
    \caption{Collisional Dark Matter Optimization Values}
    \label{collF574-2}
     \begin{tabular}{|r|r|}
     \hline
      \textbf{Parameter}   & \textbf{Optimization Values}
      \\  \hline
     $\delta_{\gamma} $ & 0.0000000012
\\  \hline
$\gamma_0 $ & 1.0001 \\ \hline $K_0$ ($M_{\odot} \,
\mathrm{Kpc}^{-3} \, (\mathrm{km/s})^{2}$)& 700  \\ \hline
    \end{tabular}
  \end{center}
\end{table}
\begin{table}[h!]
  \begin{center}
    \caption{NFW  Optimization Values}
    \label{NavaroF574-2}
     \begin{tabular}{|r|r|}
     \hline
      \textbf{Parameter}   & \textbf{Optimization Values}
      \\  \hline
   $\rho_s$   & $0.0004\times 10^9$
\\  \hline
$r_s$&  20
\\  \hline
    \end{tabular}
  \end{center}
\end{table}
\begin{table}[h!]
  \begin{center}
    \caption{Burkert Optimization Values}
    \label{BuckertF574-2}
     \begin{tabular}{|r|r|}
     \hline
      \textbf{Parameter}   & \textbf{Optimization Values}
      \\  \hline
     $\rho_0^B$  & $0.006\times 10^9$
\\  \hline
$r_0$&  6
\\  \hline
    \end{tabular}
  \end{center}
\end{table}
\begin{table}[h!]
  \begin{center}
    \caption{Einasto Optimization Values}
    \label{EinastoF574-2}
    \begin{tabular}{|r|r|}
     \hline
      \textbf{Parameter}   & \textbf{Optimization Values}
      \\  \hline
     $\rho_e$  & $0.0003\times 10^9$
\\  \hline
$r_e$ & 10
\\  \hline
$n_e$ & 0.22
\\  \hline
    \end{tabular}
  \end{center}
\end{table}
\begin{table}[h!]
\centering \caption{Physical assessment of collisional DM
parameters for F574-2.}
\begin{tabular}{lcc}
\hline
Parameter & Value & Physical Verdict \\
\hline
$\gamma_0$ & $1.0001$ & Practically isothermal  \\
$\delta_\gamma$ & $1.2\times10^{-9}$ & Essentially zero   \\
$r_\gamma$ & $1.5\ \mathrm{Kpc}$ & Transition radius chosen   \\
$K_0$ & $7.0\times10^{2}$ & Sets temperature/entropy scale \\
$r_c$ & $0.5\ \mathrm{Kpc}$ & Small core scale; yields compact central core \\
$p$ & $0.01$ & Very shallow decline of $K(r)$  \\
\hline
Overall &-& Physically consistent but effectively isothermal and spatially uniform \\
\hline
\end{tabular}
\label{EVALUATIONF574-2}
\end{table}


\subsection{The Galaxy F579-V1}

\begin{figure}[h!]
\centering
\includegraphics[width=20pc]{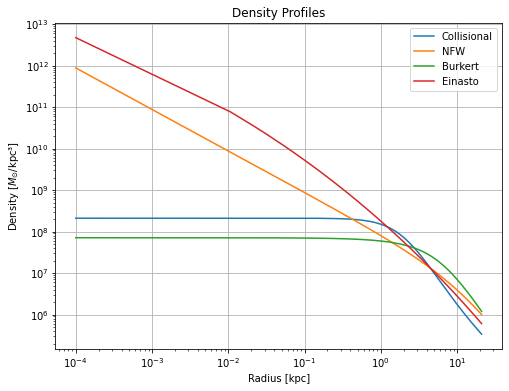}
\caption{The density of the collisional DM model (\ref{tanhmodel})
for the galaxy F579-V1, as a function of the radius.}
\label{F579-V1dens}
\end{figure}
\begin{figure}[h!]
\centering
\includegraphics[width=20pc]{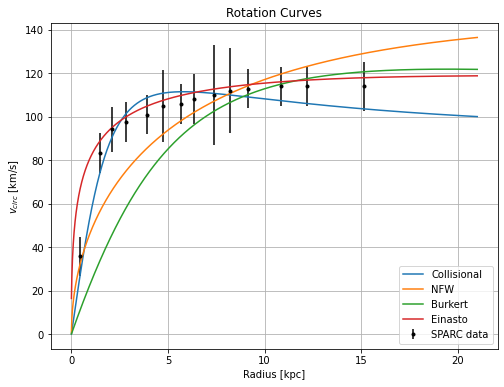}
\caption{The predicted rotation curves after using an optimization
for the collisional DM model (\ref{tanhmodel}), versus the SPARC
observational data for the galaxy F579-V1. We also plotted the
optimized curves for the NFW model, the Burkert model and the
Einasto model.} \label{F579-V1}
\end{figure}
\begin{figure}[h!]
\centering
\includegraphics[width=20pc]{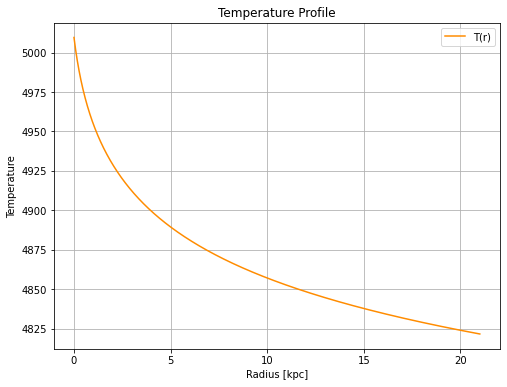}
\caption{The temperature as a function of the radius for the
collisional DM model (\ref{tanhmodel}) for the galaxy F579-V1.}
\label{F579-V1temp}
\end{figure}

For this galaxy, we shall choose $\rho_0=2.1\times
10^8$$M_{\odot}/\mathrm{Kpc}^{3}$. F579-V1 is catalogued in the
literature as a low--surface--brightness  , late--type disk
galaxy. Its distance from the Milky Way is approximately 90Mpc. In
Figs. \ref{F579-V1dens}, \ref{F579-V1} and \ref{F579-V1temp} we
present the density of the collisional DM model, the predicted
rotation curves after using an optimization for the collisional DM
model (\ref{tanhmodel}), versus the SPARC observational data and
the temperature parameter as a function of the radius
respectively. As it can be seen, the SIDM model produces viable
rotation curves compatible with the SPARC data. Also in Tables
\ref{collF579-V1}, \ref{NavaroF579-V1}, \ref{BuckertF579-V1} and
\ref{EinastoF579-V1} we present the optimization values for the
SIDM model, and the other DM profiles. Also in Table
\ref{EVALUATIONF579-V1} we present the overall evaluation of the
SIDM model for the galaxy at hand. The resulting phenomenology is
viable.
\begin{table}[h!]
  \begin{center}
    \caption{Collisional Dark Matter Optimization Values}
    \label{collF579-V1}
     \begin{tabular}{|r|r|}
     \hline
      \textbf{Parameter}   & \textbf{Optimization Values}
      \\  \hline
     $\delta_{\gamma} $ & 0.0000000012
\\  \hline
$\gamma_0 $ & 1.0001 \\ \hline $K_0$ ($M_{\odot} \,
\mathrm{Kpc}^{-3} \, (\mathrm{km/s})^{2}$)& 5000 \\ \hline
    \end{tabular}
  \end{center}
\end{table}
\begin{table}[h!]
  \begin{center}
    \caption{NFW  Optimization Values}
    \label{NavaroF579-V1}
     \begin{tabular}{|r|r|}
     \hline
      \textbf{Parameter}   & \textbf{Optimization Values}
      \\  \hline
   $\rho_s$   & $0.0044\times 10^9$
\\  \hline
$r_s$&  20
\\  \hline
    \end{tabular}
  \end{center}
\end{table}
\begin{table}[h!]
  \begin{center}
    \caption{Burkert Optimization Values}
    \label{BuckertF579-V1}
     \begin{tabular}{|r|r|}
     \hline
      \textbf{Parameter}   & \textbf{Optimization Values}
      \\  \hline
     $\rho_0^B$  & $0.071\times 10^9$
\\  \hline
$r_0$&  6
\\  \hline
    \end{tabular}
  \end{center}
\end{table}
\begin{table}[h!]
  \begin{center}
    \caption{Einasto Optimization Values}
    \label{EinastoF579-V1}
    \begin{tabular}{|r|r|}
     \hline
      \textbf{Parameter}   & \textbf{Optimization Values}
      \\  \hline
     $\rho_e$  & $0.0028\times 10^9$
\\  \hline
$r_e$ & 10
\\  \hline
$n_e$ & 0.09
\\  \hline
    \end{tabular}
  \end{center}
\end{table}
\begin{table}[h!]
\centering \caption{Physical assessment of collisional DM
parameters for F579-V1.}
\begin{tabular}{lcc}
\hline
Parameter & Value & Physical Verdict \\
\hline
$\gamma_0$ & $1.0001$ & Practically isothermal  \\
$\delta_\gamma$ & $1.2\times10^{-9}$ & Essentially zero   \\
$r_\gamma$ & $1.5\ \mathrm{Kpc}$ & Reasonable transition radius   \\
$K_0$ & $5.0\times10^{3}$ & Sets temperature/entropy scale\\
$r_c$ & $0.5\ \mathrm{Kpc}$ & Small core scale; yields a compact central core \\
$p$ & $0.01$ & Very shallow decline of $K(r)$  \\
\hline
Overall &-& Physically consistent but effectively isothermal and spatially uniform \\
\hline
\end{tabular}
\label{EVALUATIONF579-V1}
\end{table}


\subsection{The Galaxy F583-1}

\begin{figure}[h!]
\centering
\includegraphics[width=20pc]{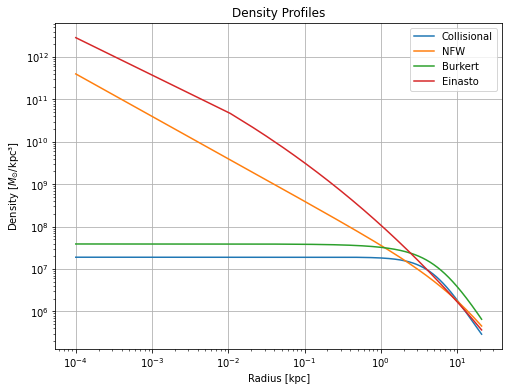}
\caption{The density of the collisional DM model (\ref{tanhmodel})
for the galaxy F583-1, as a function of the radius.}
\label{F583-1dens}
\end{figure}

\begin{figure}[h!]
\centering
\includegraphics[width=20pc]{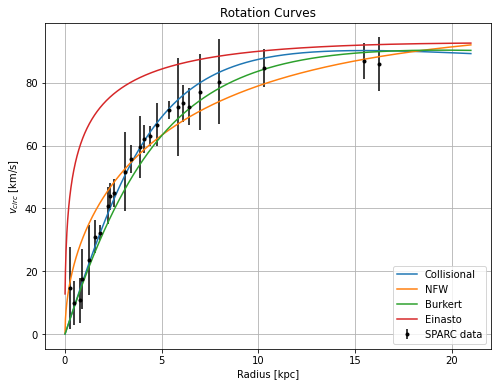}
\caption{The predicted rotation curves after using an optimization
for the collisional DM model (\ref{tanhmodel}), versus the SPARC
observational data for the galaxy F583-1. We also plotted the
optimized curves for the NFW model, the Burkert model and the
Einasto model.} \label{F583-1}
\end{figure}

\begin{figure}[h!]
\centering
\includegraphics[width=20pc]{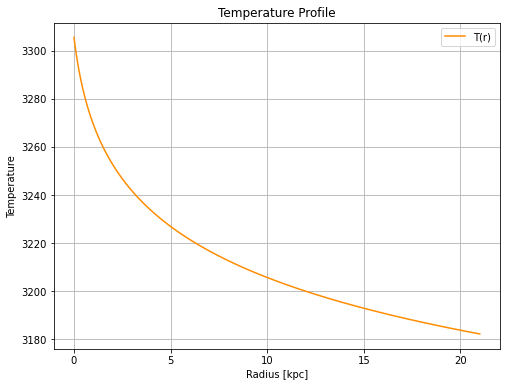}
\caption{The temperature as a function of the radius for the
collisional DM model (\ref{tanhmodel}) for the galaxy F583-1.}
\label{F583-1temp}
\end{figure}

For this galaxy, we shall choose $\rho_0=1.9\times
10^7$$M_{\odot}/\mathrm{Kpc}^{3}$. F583-1 is a well--studied
low--surface--brightness  , late--type spiral galaxy belonging to
the F583 pair. It is commonly used in rotation--curve studies as a
prototypical low-surface-brightness system. Its morphology is that
of a diffuse, gas--rich disk galaxy rather than a compact dwarf.
Its distance is 49 Mpc. In Figs. \ref{F583-1dens}, \ref{F583-1}
and \ref{F583-1temp} we present the density of the collisional DM
model, the predicted rotation curves after using an optimization
for the collisional DM model (\ref{tanhmodel}), versus the SPARC
observational data and the temperature parameter as a function of
the radius respectively. As it can be seen, the SIDM model
produces viable rotation curves compatible with the SPARC data.
Also in Tables \ref{collF583-1}, \ref{NavaroF583-1},
\ref{BuckertF583-1} and \ref{EinastoF583-1} we present the
optimization values for the SIDM model, and the other DM profiles.
Also in Table \ref{EVALUATIONF583-1} we present the overall
evaluation of the SIDM model for the galaxy at hand. The resulting
phenomenology is viable.
\begin{table}[h!]
  \begin{center}
    \caption{Collisional Dark Matter Optimization Values}
    \label{collF583-1}
     \begin{tabular}{|r|r|}
     \hline
      \textbf{Parameter}   & \textbf{Optimization Values}
      \\  \hline
     $\delta_{\gamma} $ & 0.0000000012
\\  \hline
$\gamma_0 $ & 1.0001  \\ \hline $K_0$ ($M_{\odot} \,
\mathrm{Kpc}^{-3} \, (\mathrm{km/s})^{2}$)& 3300  \\ \hline
    \end{tabular}
  \end{center}
\end{table}
\begin{table}[h!]
  \begin{center}
    \caption{NFW  Optimization Values}
    \label{NavaroF583-1}
     \begin{tabular}{|r|r|}
     \hline
      \textbf{Parameter}   & \textbf{Optimization Values}
      \\  \hline
   $\rho_s$   & $0.002\times 10^9$
\\  \hline
$r_s$&  20
\\  \hline
    \end{tabular}
  \end{center}
\end{table}
\begin{table}[h!]
  \begin{center}
    \caption{Burkert Optimization Values}
    \label{BuckertF583-1}
     \begin{tabular}{|r|r|}
     \hline
      \textbf{Parameter}   & \textbf{Optimization Values}
      \\  \hline
     $\rho_0^B$  & $0.039\times 10^9$
\\  \hline
$r_0$&  6
\\  \hline
    \end{tabular}
  \end{center}
\end{table}
\begin{table}[h!]
  \begin{center}
    \caption{Einasto Optimization Values}
    \label{EinastoF583-1}
    \begin{tabular}{|r|r|}
     \hline
      \textbf{Parameter}   & \textbf{Optimization Values}
      \\  \hline
     $\rho_e$  & $0.0017\times 10^9$
\\  \hline
$r_e$ & 10
\\  \hline
$n_e$ & 0.09
\\  \hline
    \end{tabular}
  \end{center}
\end{table}
\begin{table}[h!]
\centering \caption{Physical assessment of collisional DM
parameters for F583-1.}
\begin{tabular}{lcc}
\hline
Parameter & Value & Physical Verdict \\
\hline
$\gamma_0$ & $1.0001$ & Practically isothermal  \\
$\delta_\gamma$ & $1.2\times10^{-9}$ & Essentially zero   \\
$r_\gamma$ & $1.5\ \mathrm{Kpc}$ & Reasonable transition radius   \\
$K_0$ & $3.3\times10^{3}$ & Sets temperature/entropy scale \\
$r_c$ & $0.5\ \mathrm{Kpc}$ & Small core scale; yields a compact central core \\
$p$ & $0.01$ & Very shallow decline of $K(r)$  \\
\hline
Overall &-& Physically consistent but effectively isothermal and spatially uniform \\
\hline
\end{tabular}
\label{EVALUATIONF583-1}
\end{table}


\subsection{The Galaxy F583-4 Curious Model Viable}

\begin{figure}[h!]
\centering
\includegraphics[width=20pc]{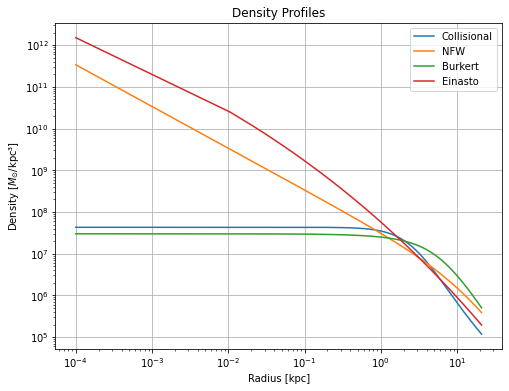}
\caption{The density of the collisional DM model (\ref{tanhmodel})
for the galaxy F583-4, as a function of the radius.}
\label{F583-4dens}
\end{figure}

\begin{figure}[h!]
\centering
\includegraphics[width=20pc]{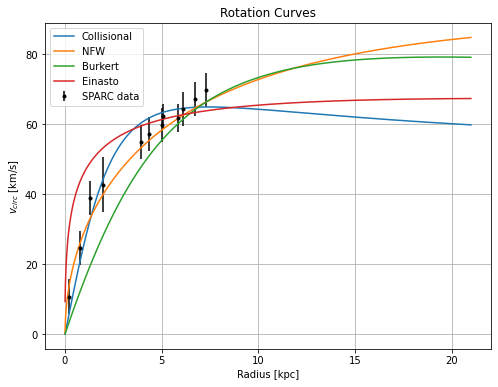}
\caption{The predicted rotation curves after using an optimization
for the collisional DM model (\ref{tanhmodel}), versus the SPARC
observational data for the galaxy F583-4. We also plotted the
optimized curves for the NFW model, the Burkert model and the
Einasto model.} \label{F583-4}
\end{figure}

\begin{figure}[h!]
\centering
\includegraphics[width=20pc]{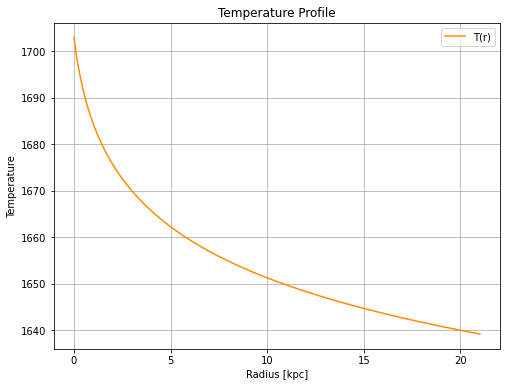}
\caption{The temperature as a function of the radius for the
collisional DM model (\ref{tanhmodel}) for the galaxy F583-4.}
\label{F583-4temp}
\end{figure}

For this galaxy, we shall choose $\rho_0=4.3\times
10^7$$M_{\odot}/\mathrm{Kpc}^{3}$. F583-4 is a
low-surface-brightness  , late-type disc galaxy, gas-rich, dark
matter dominated, but a diffuse, late-spiral with low stellar
surface brightness. Its distance is $D \simeq 50.1\ \mathrm{Mpc}$.
In Figs. \ref{F583-4dens}, \ref{F583-4} and \ref{F583-4temp} we
present the density of the collisional DM model, the predicted
rotation curves after using an optimization for the collisional DM
model (\ref{tanhmodel}), versus the SPARC observational data and
the temperature parameter as a function of the radius
respectively. As it can be seen, the SIDM model produces viable
rotation curves compatible with the SPARC data. Also in Tables
\ref{collF583-4}, \ref{NavaroF583-4}, \ref{BuckertF583-4} and
\ref{EinastoF583-4} we present the optimization values for the
SIDM model, and the other DM profiles. Also in Table
\ref{EVALUATIONF583-4} we present the overall evaluation of the
SIDM model for the galaxy at hand. The resulting phenomenology is
viable.
\begin{table}[h!]
  \begin{center}
    \caption{Collisional Dark Matter Optimization Values}
    \label{collF583-4}
     \begin{tabular}{|r|r|}
     \hline
      \textbf{Parameter}   & \textbf{Optimization Values}
      \\  \hline
     $\delta_{\gamma} $ & 0.0000000012
\\  \hline
$\gamma_0 $ &  1.0001 \\ \hline $K_0$ ($M_{\odot} \,
\mathrm{Kpc}^{-3} \, (\mathrm{km/s})^{2}$)& 1700  \\ \hline
    \end{tabular}
  \end{center}
\end{table}
\begin{table}[h!]
  \begin{center}
    \caption{NFW  Optimization Values}
    \label{NavaroF583-4}
     \begin{tabular}{|r|r|}
     \hline
      \textbf{Parameter}   & \textbf{Optimization Values}
      \\  \hline
   $\rho_s$   & $0.0017\times 10^9$
\\  \hline
$r_s$&  20
\\  \hline
    \end{tabular}
  \end{center}
\end{table}
\begin{table}[h!]
  \begin{center}
    \caption{Burkert Optimization Values}
    \label{BuckertF583-4}
     \begin{tabular}{|r|r|}
     \hline
      \textbf{Parameter}   & \textbf{Optimization Values}
      \\  \hline
     $\rho_0^B$  & $0.03\times 10^9$
\\  \hline
$r_0$&  6
\\  \hline
    \end{tabular}
  \end{center}
\end{table}
\begin{table}[h!]
  \begin{center}
    \caption{Einasto Optimization Values}
    \label{EinastoF583-4}
    \begin{tabular}{|r|r|}
     \hline
      \textbf{Parameter}   & \textbf{Optimization Values}
      \\  \hline
     $\rho_e$  & $0.0009\times 10^9$
\\  \hline
$r_e$ & 10
\\  \hline
$n_e$ & 0.09
\\  \hline
    \end{tabular}
  \end{center}
\end{table}
\begin{table}[h!]
\centering \caption{Physical assessment of collisional DM
parameters for F583-4.}
\begin{tabular}{lcc}
\hline
Parameter & Value & Physical Verdict \\
\hline
$\gamma_0$ & $1.0001$ & Practically isothermal  \\
$\delta_\gamma$ & $1.2\times10^{-9}$ & Essentially zero   \\
$r_\gamma$ & $1.5\ \mathrm{Kpc}$ & Transition radius chosen   \\
$K_0$ & $1.7\times10^{3}$ & Sets temperature/entropy scale \\
$r_c$ & $0.5\ \mathrm{Kpc}$ & Small core scale; yields a compact central core \\
$p$ & $0.01$ & Very shallow decline of $K(r)$  \\
\hline
Overall &-& Physically consistent but effectively isothermal and spatially uniform \\
\hline
\end{tabular}
\label{EVALUATIONF583-4}
\end{table}


\subsection{The Galaxy IC2574}

\begin{figure}[h!]
\centering
\includegraphics[width=20pc]{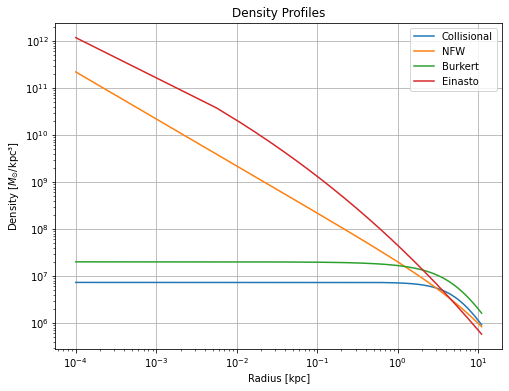}
\caption{The density of the collisional DM model (\ref{tanhmodel})
for the galaxy IC2574, as a function of the radius.}
\label{IC2574dens}
\end{figure}
\begin{figure}[h!]
\centering
\includegraphics[width=20pc]{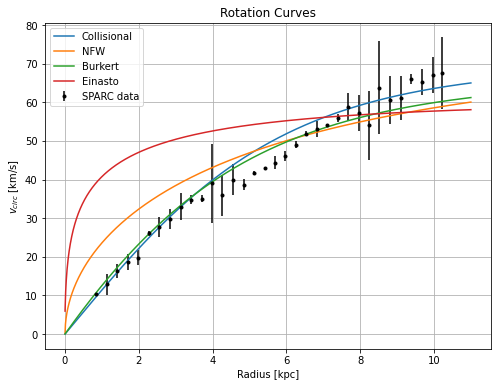}
\caption{The predicted rotation curves after using an optimization
for the collisional DM model (\ref{tanhmodel}), versus the SPARC
observational data for the galaxy IC2574. We also plotted the
optimized curves for the NFW model, the Burkert model and the
Einasto model.} \label{IC2574}
\end{figure}
\begin{figure}[h!]
\centering
\includegraphics[width=20pc]{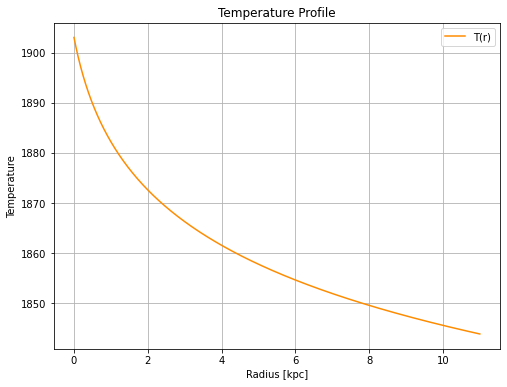}
\caption{The temperature as a function of the radius for the
collisional DM model (\ref{tanhmodel}) for the galaxy IC2574.}
\label{IC2574temp}
\end{figure}

For this galaxy, we shall choose $\rho_0=7.3\times
10^6$$M_{\odot}/\mathrm{Kpc}^{3}$. IC\,2574 (also known as
Coddington's Nebula) is a gas--rich, low--surface--brightness
  dwarf irregular galaxy belonging to the M81 Group. It is
classified morphologically as a dwarf spiral/irregular system
rather than an ordinary high--surface--brightness spiral. Its
stellar component is diffuse and irregular, while its interstellar
medium is dominated by extended HI structures and star--forming
regions. The galaxy lies at a distance of $D \simeq 4.0\
\text{Mpc}$. In Figs. \ref{IC2574dens}, \ref{IC2574} and
\ref{IC2574temp} we present the density of the collisional DM
model, the predicted rotation curves after using an optimization
for the collisional DM model (\ref{tanhmodel}), versus the SPARC
observational data and the temperature parameter as a function of
the radius respectively. As it can be seen, the SIDM model
produces viable rotation curves compatible with the SPARC data.
Also in Tables \ref{collIC2574}, \ref{NavaroIC2574},
\ref{BuckertIC2574} and \ref{EinastoIC2574} we present the
optimization values for the SIDM model, and the other DM profiles.
Also in Table \ref{EVALUATIONIC2574} we present the overall
evaluation of the SIDM model for the galaxy at hand. The resulting
phenomenology is viable.
\begin{table}[h!]
  \begin{center}
    \caption{Collisional Dark Matter Optimization Values}
    \label{collIC2574}
     \begin{tabular}{|r|r|}
     \hline
      \textbf{Parameter}   & \textbf{Optimization Values}
      \\  \hline
     $\delta_{\gamma} $ & 0.0000000012
\\  \hline
$\gamma_0 $ &  1.0001 \\ \hline $K_0$ ($M_{\odot} \,
\mathrm{Kpc}^{-3} \, (\mathrm{km/s})^{2}$)& 1900  \\ \hline
    \end{tabular}
  \end{center}
\end{table}
\begin{table}[h!]
  \begin{center}
    \caption{NFW  Optimization Values}
    \label{NavaroIC2574}
     \begin{tabular}{|r|r|}
     \hline
      \textbf{Parameter}   & \textbf{Optimization Values}
      \\  \hline
   $\rho_s$   & $0.0013\times 10^9$
\\  \hline
$r_s$&  20
\\  \hline
    \end{tabular}
  \end{center}
\end{table}
\begin{table}[h!]
  \begin{center}
    \caption{Burkert Optimization Values}
    \label{BuckertIC2574}
     \begin{tabular}{|r|r|}
     \hline
      \textbf{Parameter}   & \textbf{Optimization Values}
      \\  \hline
     $\rho_0^B$  & $0.023\times 10^9$
\\  \hline
$r_0$&  6
\\  \hline
    \end{tabular}
  \end{center}
\end{table}
\begin{table}[h!]
  \begin{center}
    \caption{Einasto Optimization Values}
    \label{EinastoIC2574}
    \begin{tabular}{|r|r|}
     \hline
      \textbf{Parameter}   & \textbf{Optimization Values}
      \\  \hline
     $\rho_e$  & $0.0009\times 10^9$
\\  \hline
$r_e$ & 10
\\  \hline
$n_e$ & 0.09
\\  \hline
    \end{tabular}
  \end{center}
\end{table}
\begin{table}[h!]
\centering \caption{Physical assessment of collisional DM
parameters for IC2574.}
\begin{tabular}{lcc}
\hline
Parameter & Value & Physical Verdict \\
\hline
$\gamma_0$ & $1.0001$ & Practically isothermal  \\
$\delta_\gamma$ & $1.2\times10^{-9}$ & Essentially zero   \\
$r_\gamma$ & $1.5\ \mathrm{Kpc}$ & Transition radius chosen   \\
$K_0$ & $1.9\times10^{3}$ & Sets temperature/entropy scale \\
$r_c$ & $0.5\ \mathrm{Kpc}$ & Small core scale; yields a compact central core \\
$p$ & $0.01$ & Very shallow decline of $K(r)$  \\
\hline
Overall &-& Physically consistent but effectively isothermal and spatially uniform \\
\hline
\end{tabular}
\label{EVALUATIONIC2574}
\end{table}


\subsection{The Galaxy IC4202 Marginally, Extended Marginally too}

\begin{figure}[h!]
\centering
\includegraphics[width=20pc]{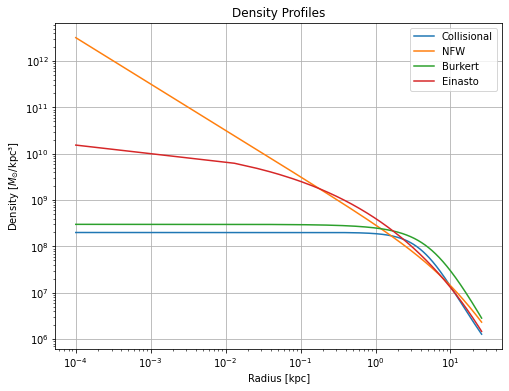}
\caption{The density of the collisional DM model (\ref{tanhmodel})
for the galaxy IC4202, as a function of the radius.}
\label{IC4202dens}
\end{figure}
\begin{figure}[h!]
\centering
\includegraphics[width=20pc]{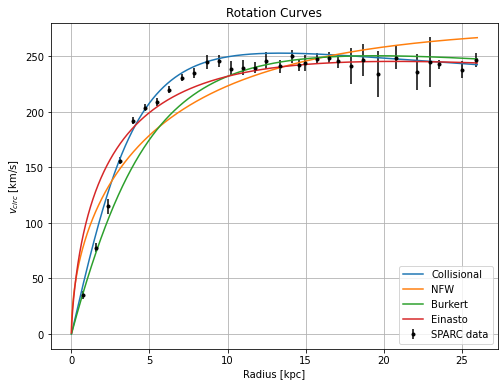}
\caption{The predicted rotation curves after using an optimization
for the collisional DM model (\ref{tanhmodel}), versus the SPARC
observational data for the galaxy IC4202. We also plotted the
optimized curves for the NFW model, the Burkert model and the
Einasto model.} \label{IC4202}
\end{figure}
\begin{figure}[h!]
\centering
\includegraphics[width=20pc]{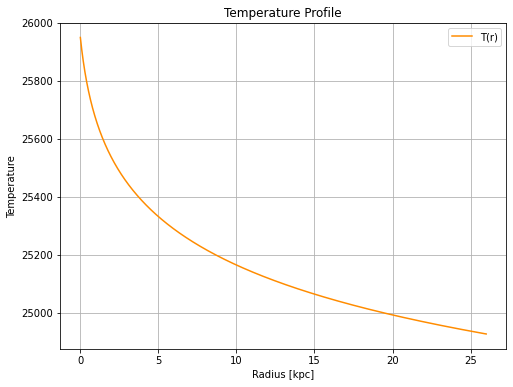}
\caption{The temperature as a function of the radius for the
collisional DM model (\ref{tanhmodel}) for the galaxy IC4202.}
\label{IC4202temp}
\end{figure}
\begin{figure}[h!]
\centering
\includegraphics[width=20pc]{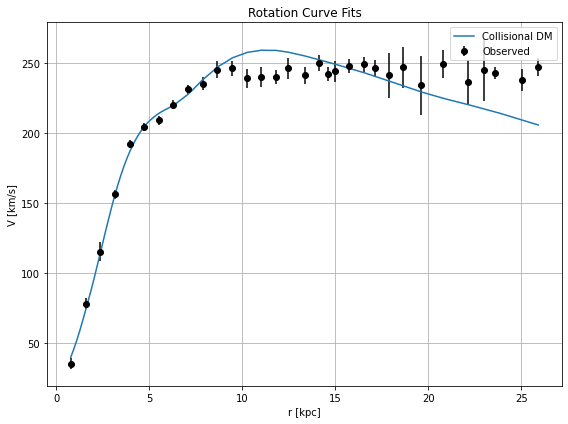}
\caption{The predicted rotation curves after using an optimization
for the collisional DM model (\ref{tanhmodel}), versus the
extended SPARC observational data for the galaxy IC4202. The model
includes the rotation curves from all the components of the
galaxy, including gas and disk velocities, along with the
collisional DM model.} \label{extendedIC4202}
\end{figure}

For this galaxy, we shall choose $\rho_0=2\times
10^8$$M_{\odot}/\mathrm{Kpc}^{3}$. IC\,4202 is classified in
catalogs as a spiral galaxy of type Sbc. It is a relatively large,
ordinary spiral with significant stellar disk structure. The
galaxy lies at a distance of about $D \simeq 109.0\ \mathrm{Mpc}$.
In Figs. \ref{IC4202dens}, \ref{IC4202} and \ref{IC4202temp} we
present the density of the collisional DM model, the predicted
rotation curves after using an optimization for the collisional DM
model (\ref{tanhmodel}), versus the SPARC observational data and
the temperature parameter as a function of the radius
respectively. As it can be seen, the SIDM model produces
marginally viable rotation curves compatible with the SPARC data.
Also in Tables \ref{collIC4202}, \ref{NavaroIC4202},
\ref{BuckertIC4202} and \ref{EinastoIC4202} we present the
optimization values for the SIDM model, and the other DM profiles.
Also in Table \ref{EVALUATIONIC4202} we present the overall
evaluation of the SIDM model for the galaxy at hand. The resulting
phenomenology is marginally viable.
\begin{table}[h!]
  \begin{center}
    \caption{Collisional Dark Matter Optimization Values}
    \label{collIC4202}
     \begin{tabular}{|r|r|}
     \hline
      \textbf{Parameter}   & \textbf{Optimization Values}
      \\  \hline
     $\delta_{\gamma} $ & 0.0000000012
\\  \hline
$\gamma_0 $ & 1.0001 \\ \hline $K_0$ ($M_{\odot} \,
\mathrm{Kpc}^{-3} \, (\mathrm{km/s})^{2}$)& 25900  \\ \hline
    \end{tabular}
  \end{center}
\end{table}
\begin{table}[h!]
  \begin{center}
    \caption{NFW  Optimization Values}
    \label{NavaroIC4202}
     \begin{tabular}{|r|r|}
     \hline
      \textbf{Parameter}   & \textbf{Optimization Values}
      \\  \hline
   $\rho_s$   & $0.016\times 10^9$
\\  \hline
$r_s$&  20
\\  \hline
    \end{tabular}
  \end{center}
\end{table}
\begin{table}[h!]
  \begin{center}
    \caption{Burkert Optimization Values}
    \label{BuckertIC4202}
     \begin{tabular}{|r|r|}
     \hline
      \textbf{Parameter}   & \textbf{Optimization Values}
      \\  \hline
     $\rho_0^B$  & $0.3\times 10^9$
\\  \hline
$r_0$&  6
\\  \hline
    \end{tabular}
  \end{center}
\end{table}
\begin{table}[h!]
  \begin{center}
    \caption{Einasto Optimization Values}
    \label{EinastoIC4202}
    \begin{tabular}{|r|r|}
     \hline
      \textbf{Parameter}   & \textbf{Optimization Values}
      \\  \hline
     $\rho_e$  & $0.013\times 10^9$
\\  \hline
$r_e$ & 10
\\  \hline
$n_e$ & 0.27
\\  \hline
    \end{tabular}
  \end{center}
\end{table}
\begin{table}[h!]
\centering \caption{Physical assessment of collisional DM
parameters for IC4202.}
\begin{tabular}{lcc}
\hline
Parameter & Value & Physical Verdict \\
\hline
$\gamma_0$ & $1.0001$ & Practically isothermal  \\
$\delta_\gamma$ & $1.2\times10^{-9}$ & Essentially zero   \\
$r_\gamma$ & $1.5\ \mathrm{Kpc}$ & Transition radius chosen   \\
$K_0$ & $2.59\times10^{4}$ & High entropy/temperature scale;\\
$r_c$ & $0.5\ \mathrm{Kpc}$ & Small core scale; compact central core \\
$p$ & $0.01$ & Very shallow decline of $K(r)$  \\
\hline
Overall &-& Physically consistent but effectively isothermal \\
\hline
\end{tabular}
\label{EVALUATIONIC4202}
\end{table}
Now the extended picture including the rotation velocity from the
other components of the galaxy, such as the disk and gas, makes
the collisional DM model viable for this galaxy. In Fig.
\ref{extendedIC4202} we present the combined rotation curves
including the other components of the galaxy along with the
collisional matter. As it can be seen, the extended collisional DM
model is marginally viable.
Also in Table \ref{evaluationextendedIC4202} we present the values
of the free parameters of the collisional DM model for which the
maximum compatibility with the SPARC data comes for the galaxy
IC4202.
\begin{table}[h!]
\centering \caption{Physical assessment of Extended collisional DM
parameters for galaxy IC4202.}
\begin{tabular}{lcc}
\hline
Parameter & Value & Physical Verdict \\
\hline
$\gamma_0$ & 1.22111920 & Slightly above isothermal \\
$\delta_\gamma$ & 0.17612945 & Moderate radial variation\\
$K_0$ & 3000 & Moderate entropy scale\\
$ml_{disk}$ & 1.00000000 & Relatively high stellar M/L for the disk \\
$ml_{bulge}$ & 0.21735196 & Small-to-moderate bulge M/L \\
\hline
Overall &-& Physically plausible \\
\hline
\end{tabular}
\label{evaluationextendedIC4202}
\end{table}


\subsection{The Galaxy KK98-251}

\begin{figure}[h!]
\centering
\includegraphics[width=20pc]{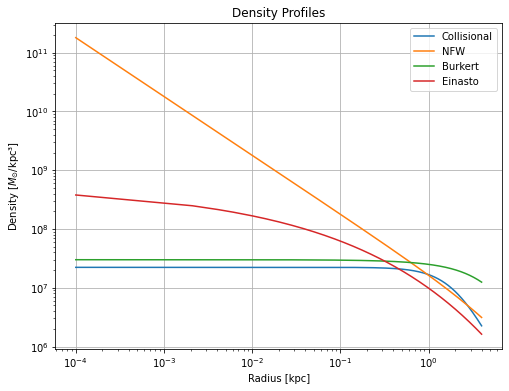}
\caption{The density of the collisional DM model (\ref{tanhmodel})
for the galaxy KK98-251, as a function of the radius.}
\label{KK98-251dens}
\end{figure}
\begin{figure}[h!]
\centering
\includegraphics[width=20pc]{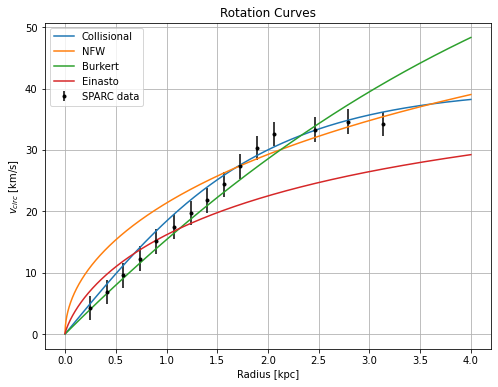}
\caption{The predicted rotation curves after using an optimization
for the collisional DM model (\ref{tanhmodel}), versus the SPARC
observational data for the galaxy KK98-251. We also plotted the
optimized curves for the NFW model, the Burkert model and the
Einasto model.} \label{KK98-251}
\end{figure}
\begin{figure}[h!]
\centering
\includegraphics[width=20pc]{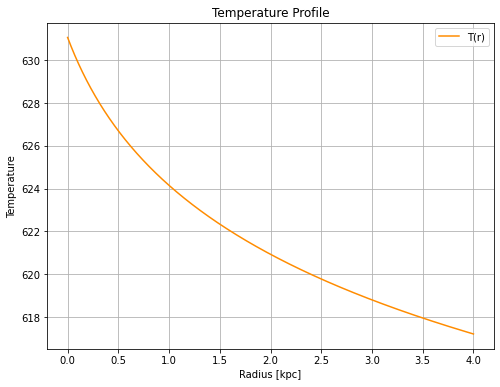}
\caption{The temperature as a function of the radius for the
collisional DM model (\ref{tanhmodel}) for the galaxy KK98-251.}
\label{KK98-251temp}
\end{figure}

For this galaxy, we shall choose $\rho_0=2.225\times
10^7$$M_{\odot}/\mathrm{Kpc}^{3}$. The galaxy KK98-251 is a dwarf
galaxy located in the NGC 6946 group. Its distance from Earth is
estimated to be approximately 5.6 Mpc, based on measurements from
the brightest stars in eight members of the group. KK98-251 is
classified as a dwarf irregular galaxy, characterized by its
low-surface-brightness and irregular shape. KK98-251 thus is a
faint, irregular dwarf galaxy with a significant HI mass and a
dark matter halo that is typical for galaxies of its type. In
Figs. \ref{KK98-251dens}, \ref{KK98-251} and \ref{KK98-251temp} we
present the density of the collisional DM model, the predicted
rotation curves after using an optimization for the collisional DM
model (\ref{tanhmodel}), versus the SPARC observational data and
the temperature parameter as a function of the radius
respectively. As it can be seen, the SIDM model produces viable
rotation curves compatible with the SPARC data. Also in Tables
\ref{collKK98-251}, \ref{NavaroKK98-251}, \ref{BuckertKK98-251}
and \ref{EinastoKK98-251} we present the optimization values for
the SIDM model, and the other DM profiles. Also in Table
\ref{EVALUATIONKK98-251} we present the overall evaluation of the
SIDM model for the galaxy at hand. The resulting phenomenology is
viable.
\begin{table}[h!]
  \begin{center}
    \caption{Collisional Dark Matter Optimization Values}
    \label{collKK98-251}
     \begin{tabular}{|r|r|}
     \hline
      \textbf{Parameter}   & \textbf{Optimization Values}
      \\  \hline
     $\delta_{\gamma} $ &  0.0000000012
\\  \hline
$\gamma_0 $ & 1.0001 \\ \hline $K_0$ ($M_{\odot} \,
\mathrm{Kpc}^{-3} \, (\mathrm{km/s})^{2}$)& 6300  \\ \hline
    \end{tabular}
  \end{center}
\end{table}
\begin{table}[h!]
  \begin{center}
    \caption{NFW  Optimization Values}
    \label{NavaroKK98-251}
     \begin{tabular}{|r|r|}
     \hline
      \textbf{Parameter}   & \textbf{Optimization Values}
      \\  \hline
   $\rho_s$   & $0.0009\times 10^9$
\\  \hline
$r_s$&  20
\\  \hline
    \end{tabular}
  \end{center}
\end{table}
\begin{table}[h!]
  \begin{center}
    \caption{Burkert Optimization Values}
    \label{BuckertKK98-251}
     \begin{tabular}{|r|r|}
     \hline
      \textbf{Parameter}   & \textbf{Optimization Values}
      \\  \hline
     $\rho_0^B$  & $0.3\times 10^9$
\\  \hline
$r_0$&  6
\\  \hline
    \end{tabular}
  \end{center}
\end{table}
\begin{table}[h!]
  \begin{center}
    \caption{Einasto Optimization Values}
    \label{EinastoKK98-251}
    \begin{tabular}{|r|r|}
     \hline
      \textbf{Parameter}   & \textbf{Optimization Values}
      \\  \hline
     $\rho_e$  & $0.00032\times 10^9$
\\  \hline
$r_e$ & 10
\\  \hline
$n_e$ & 0.09
\\  \hline
    \end{tabular}
  \end{center}
\end{table}
\begin{table}[h!]
\centering \caption{Physical assessment of collisional DM
parameters for KK98-251.}
\begin{tabular}{lcc}
\hline
Parameter & Value & Physical Verdict \\
\hline
$\gamma_0$ & $1.0001$ & Effectively isothermal  \\
$\delta_\gamma$ & $1.2\times10^{-9}$ & Essentially zero - $\gamma(r)$ constant \\
$r_\gamma$ & $1.5\ \mathrm{Kpc}$ & Transition radius irrelevant due to tiny $\delta_\gamma$ \\
$K_0$ & $6.3\times10^2$ & Moderate entropy\\
$r_c$ & $0.5\ \mathrm{Kpc}$ & Small core; reasonable for dwarf halo \\
$p$ & $0.01$ & Nearly constant entropy; minimal radial variation \\
\hline
Overall &-& Physically plausible for dwarf galaxy; inner halo nearly isothermal and uniform \\
\hline
\end{tabular}
\label{EVALUATIONKK98-251}
\end{table}


\subsection{The Galaxy NGC0024}

\begin{figure}[h!]
\centering
\includegraphics[width=20pc]{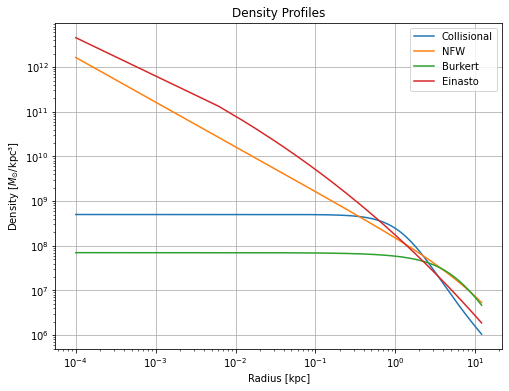}
\caption{The density of the collisional DM model (\ref{tanhmodel})
for the galaxy NGC0024, as a function of the radius.}
\label{NGC0024dens}
\end{figure}

\begin{figure}[h!]
\centering
\includegraphics[width=20pc]{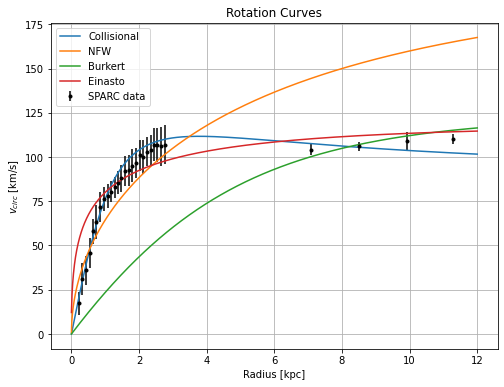}
\caption{The predicted rotation curves after using an optimization
for the collisional DM model (\ref{tanhmodel}), versus the SPARC
observational data for the galaxy NGC0024. We also plotted the
optimized curves for the NFW model, the Burkert model and the
Einasto model.} \label{NGC0024}
\end{figure}

\begin{figure}[h!]
\centering
\includegraphics[width=20pc]{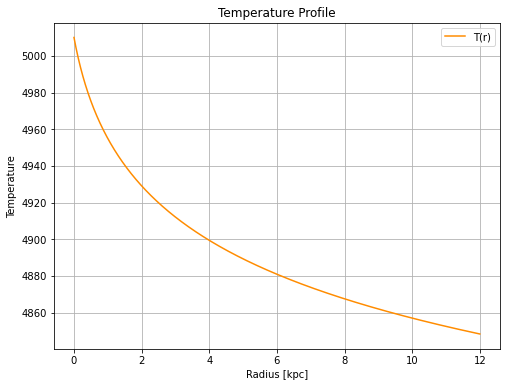}
\caption{The temperature as a function of the radius for the
collisional DM model (\ref{tanhmodel}) for the galaxy NGC0024.}
\label{NGC0024temp}
\end{figure}

For this galaxy, we shall choose $\rho_0=5\times
10^8$$M_{\odot}/\mathrm{Kpc}^{3}$. NGC\,24 is an unbarred spiral
galaxy of morphological type $\mathrm{SA(s)c}$ hence a late-type
ordinary spiral. It lies in the direction of the Sculptor
constellation and has been treated in the literature as a
background galaxy relative to the Sculptor Group. It is
approximately 6.8 Mpc away from the Milky Way. In Figs.
\ref{NGC0024dens}, \ref{NGC0024} and \ref{NGC0024temp} we present
the density of the collisional DM model, the predicted rotation
curves after using an optimization for the collisional DM model
(\ref{tanhmodel}), versus the SPARC observational data and the
temperature parameter as a function of the radius respectively. As
it can be seen, the SIDM model produces viable rotation curves
compatible with the SPARC data. Also in Tables \ref{collNGC0024},
\ref{NavaroNGC0024}, \ref{BuckertNGC0024} and \ref{EinastoNGC0024}
we present the optimization values for the SIDM model, and the
other DM profiles. Also in Table \ref{EVALUATIONNGC0024} we
present the overall evaluation of the SIDM model for the galaxy at
hand. The resulting phenomenology is viable.
\begin{table}[h!]
  \begin{center}
    \caption{Collisional Dark Matter Optimization Values}
    \label{collNGC0024}
     \begin{tabular}{|r|r|}
     \hline
      \textbf{Parameter}   & \textbf{Optimization Values}
      \\  \hline
     $\delta_{\gamma} $ & 0.0000000012
\\  \hline
$\gamma_0 $ & 1.0001 \\ \hline $K_0$ ($M_{\odot} \,
\mathrm{Kpc}^{-3} \, (\mathrm{km/s})^{2}$)& 5000 \\ \hline
    \end{tabular}
  \end{center}
\end{table}
\begin{table}[h!]
  \begin{center}
    \caption{NFW  Optimization Values}
    \label{NavaroNGC0024}
     \begin{tabular}{|r|r|}
     \hline
      \textbf{Parameter}   & \textbf{Optimization Values}
      \\  \hline
   $\rho_s$   & $0.0082\times 10^9$
\\  \hline
$r_s$&  20
\\  \hline
    \end{tabular}
  \end{center}
\end{table}
\begin{table}[h!]
  \begin{center}
    \caption{Burkert Optimization Values}
    \label{BuckertNGC0024}
     \begin{tabular}{|r|r|}
     \hline
      \textbf{Parameter}   & \textbf{Optimization Values}
      \\  \hline
     $\rho_0^B$  & $0.07\times 10^9$
\\  \hline
$r_0$&  6
\\  \hline
    \end{tabular}
  \end{center}
\end{table}
\begin{table}[h!]
  \begin{center}
    \caption{Einasto Optimization Values}
    \label{EinastoNGC0024}
    \begin{tabular}{|r|r|}
     \hline
      \textbf{Parameter}   & \textbf{Optimization Values}
      \\  \hline
     $\rho_e$  & $0.0027\times 10^9$
\\  \hline
$r_e$ & 10
\\  \hline
$n_e$ & 0.09
\\  \hline
    \end{tabular}
  \end{center}
\end{table}
\begin{table}[h!]
\centering \caption{Physical assessment of collisional DM
parameters for NGC0024.}
\begin{tabular}{lcc}
\hline
Parameter & Value & Physical Verdict \\
\hline
$\gamma_0$ & $1.0001$ & Effectively isothermal  \\
$\delta_\gamma$ & $1.2\times10^{-9}$ & Essentially zero - $\gamma(r)$ constant \\
$r_\gamma$ & $1.5\ \mathrm{Kpc}$ & Transition radius irrelevant due to tiny $\delta_\gamma$ \\
$K_0$ & $5.0\times10^3$ & Moderately high pressure support \\
$r_c$ & $0.5\ \mathrm{Kpc}$ & Small core; reasonable for inner halo \\
$p$ & $0.01$ & Nearly constant entropy; minimal radial variation \\
\hline
Overall &-& Physically plausible; inner halo nearly isothermal and uniform \\
\hline
\end{tabular}
\label{EVALUATIONNGC0024}
\end{table}


\subsection{The Galaxy NGC0055}

\begin{figure}[h!]
\centering
\includegraphics[width=20pc]{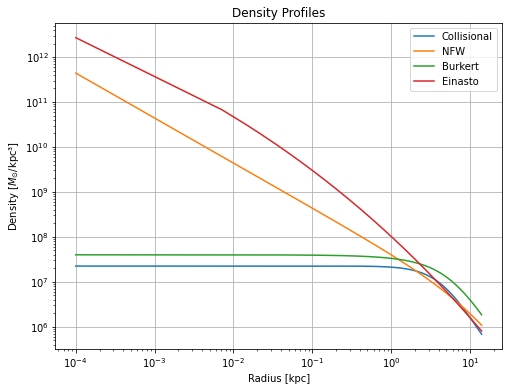}
\caption{The density of the collisional DM model (\ref{tanhmodel})
for the galaxy NGC0055, as a function of the radius.}
\label{NGC0055dens}
\end{figure}

\begin{figure}[h!]
\centering
\includegraphics[width=20pc]{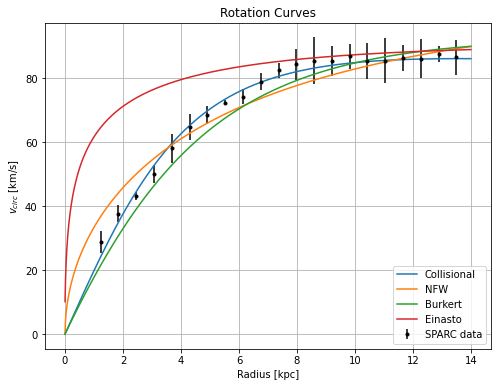}
\caption{The predicted rotation curves after using an optimization
for the collisional DM model (\ref{tanhmodel}), versus the SPARC
observational data for the galaxy NGC0055. We also plotted the
optimized curves for the NFW model, the Burkert model and the
Einasto model.} \label{NGC0055}
\end{figure}

\begin{figure}[h!]
\centering
\includegraphics[width=20pc]{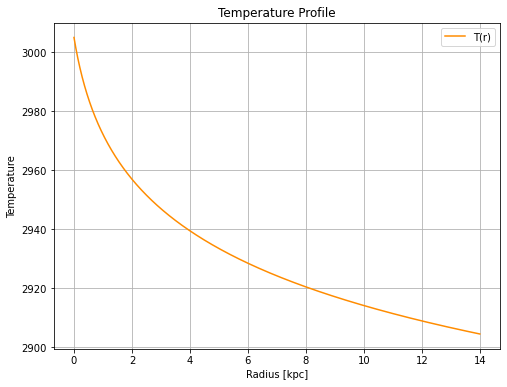}
\caption{The temperature as a function of the radius for the
collisional DM model (\ref{tanhmodel}) for the galaxy NGC0055.}
\label{NGC0055temp}
\end{figure}

For this galaxy, we shall choose $\rho_0=2.25\times
10^7$$M_{\odot}/\mathrm{Kpc}^{3}$. NGC\,55 is a Magellanic-type
barred spiral galaxy, often classified as $\mathrm{SB(s)m}$,
viewed nearly edge-on; it is a low-mass spiral/irregular analogue.
It is one of the closer galaxies to the Local Group and lies in or
near the Sculptor Group. From Cepheid, planetary nebula, and other
indicators its distance is well constrained to be $D \sim 2.0 \pm
0.2\ \mathrm{Mpc}$. In Figs. \ref{NGC0055dens}, \ref{NGC0055} and
\ref{NGC0055temp} we present the density of the collisional DM
model, the predicted rotation curves after using an optimization
for the collisional DM model (\ref{tanhmodel}), versus the SPARC
observational data and the temperature parameter as a function of
the radius respectively. As it can be seen, the SIDM model
produces viable rotation curves compatible with the SPARC data.
Also in Tables \ref{collNGC0055}, \ref{NavaroNGC0055},
\ref{BuckertNGC0055} and \ref{EinastoNGC0055} we present the
optimization values for the SIDM model, and the other DM profiles.
Also in Table \ref{EVALUATIONNGC0055} we present the overall
evaluation of the SIDM model for the galaxy at hand. The resulting
phenomenology is viable.
\begin{table}[h!]
  \begin{center}
    \caption{Collisional Dark Matter Optimization Values}
    \label{collNGC0055}
     \begin{tabular}{|r|r|}
     \hline
      \textbf{Parameter}   & \textbf{Optimization Values}
      \\  \hline
     $\delta_{\gamma} $ & 0.0000000012
\\  \hline
$\gamma_0 $ & 1.0001 \\ \hline $K_0$ ($M_{\odot} \,
\mathrm{Kpc}^{-3} \, (\mathrm{km/s})^{2}$)& 3000  \\ \hline
    \end{tabular}
  \end{center}
\end{table}
\begin{table}[h!]
  \begin{center}
    \caption{NFW  Optimization Values}
    \label{NavaroNGC0055}
     \begin{tabular}{|r|r|}
     \hline
      \textbf{Parameter}   & \textbf{Optimization Values}
      \\  \hline
   $\rho_s$   & $0.0022\times 10^9$
\\  \hline
$r_s$&  20
\\  \hline
    \end{tabular}
  \end{center}
\end{table}
\begin{table}[h!]
  \begin{center}
    \caption{Burkert Optimization Values}
    \label{BuckertNGC0055}
     \begin{tabular}{|r|r|}
     \hline
      \textbf{Parameter}   & \textbf{Optimization Values}
      \\  \hline
     $\rho_0^B$  & $0.04\times 10^9$
\\  \hline
$r_0$&  6
\\  \hline
    \end{tabular}
  \end{center}
\end{table}
\begin{table}[h!]
  \begin{center}
    \caption{Einasto Optimization Values}
    \label{EinastoNGC0055}
    \begin{tabular}{|r|r|}
     \hline
      \textbf{Parameter}   & \textbf{Optimization Values}
      \\  \hline
     $\rho_e$  & $0.0016\times 10^9$
\\  \hline
$r_e$ & 10
\\  \hline
$n_e$ & 0.09
\\  \hline
    \end{tabular}
  \end{center}
\end{table}
\begin{table}[h!]
\centering \caption{Physical assessment of collisional DM
parameters for NGC0055.}
\begin{tabular}{lcc}
\hline
Parameter & Value & Physical Verdict \\
\hline
$\gamma_0$ & $1.0001$ & Effectively isothermal  \\
$\delta_\gamma$ & $1.2\times10^{-9}$ & Essentially zero - $\gamma(r)$ constant \\
$r_\gamma$ & $1.5\ \mathrm{Kpc}$ & Transition radius irrelevant due to tiny $\delta_\gamma$ \\
$K_0$ & $3.0\times10^3$ & Moderately high pressure support \\
$r_c$ & $0.5\ \mathrm{Kpc}$ & Small core; reasonable for inner halo \\
$p$ & $0.01$ & Nearly constant entropy; minimal radial variation \\
\hline
Overall &-& Physically plausible; inner halo nearly isothermal and uniform \\
\hline
\end{tabular}
\label{EVALUATIONNGC0055}
\end{table}


\subsection{The Galaxy NGC0100}

\begin{figure}[h!]
\centering
\includegraphics[width=20pc]{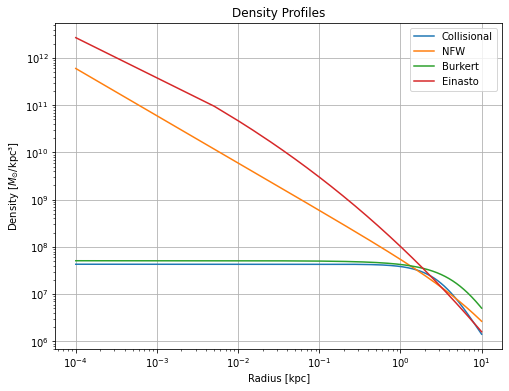}
\caption{The density of the collisional DM model (\ref{tanhmodel})
for the galaxy NGC0100, as a function of the radius.}
\label{NGC0100dens}
\end{figure}

\begin{figure}[h!]
\centering
\includegraphics[width=20pc]{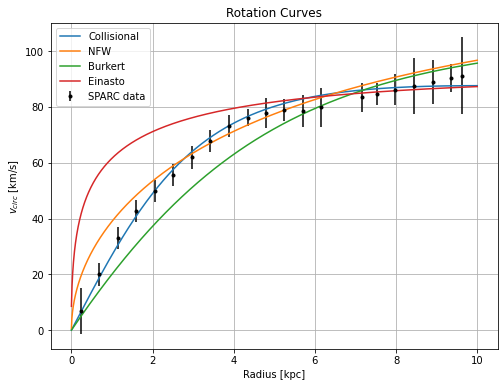}
\caption{The predicted rotation curves after using an optimization
for the collisional DM model (\ref{tanhmodel}), versus the SPARC
observational data for the galaxy NGC0100. We also plotted the
optimized curves for the NFW model, the Burkert model and the
Einasto model.} \label{NGC0100}
\end{figure}

\begin{figure}[h!]
\centering
\includegraphics[width=20pc]{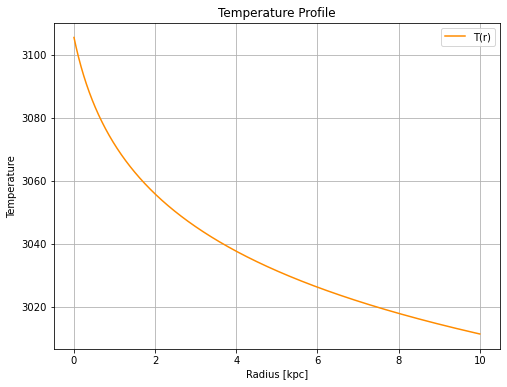}
\caption{The temperature as a function of the radius for the
collisional DM model (\ref{tanhmodel}) for the galaxy NGC0100.}
\label{NGC0100temp}
\end{figure}

For this galaxy, we shall choose $\rho_0=4.3\times
10^7$$M_{\odot}/\mathrm{Kpc}^{3}$. NGC\,100 is a spiral galaxy of
morphological type approximately $\mathrm{Scd}$ in the
constellation Pisces.  It is an ordinary (though somewhat
elongated / super-thin) spiral. From redshift and
redshift-independent methods its distance is $D \sim 18.45 \pm
0.20\;\mathrm{Mpc}$. In Figs. \ref{NGC0100dens}, \ref{NGC0100} and
\ref{NGC0100temp} we present the density of the collisional DM
model, the predicted rotation curves after using an optimization
for the collisional DM model (\ref{tanhmodel}), versus the SPARC
observational data and the temperature parameter as a function of
the radius respectively. As it can be seen, the SIDM model
produces viable rotation curves compatible with the SPARC data.
Also in Tables \ref{collNGC0100}, \ref{NavaroNGC0100},
\ref{BuckertNGC0100} and \ref{EinastoNGC0100} we present the
optimization values for the SIDM model, and the other DM profiles.
Also in Table \ref{EVALUATIONNGC0100} we present the overall
evaluation of the SIDM model for the galaxy at hand. The resulting
phenomenology is viable.
\begin{table}[h!]
  \begin{center}
    \caption{Collisional Dark Matter Optimization Values}
    \label{collNGC0100}
     \begin{tabular}{|r|r|}
     \hline
      \textbf{Parameter}   & \textbf{Optimization Values}
      \\  \hline
     $\delta_{\gamma} $ & 0.0000000012
\\  \hline
$\gamma_0 $ &1.0001  \\ \hline $K_0$ ($M_{\odot} \,
\mathrm{Kpc}^{-3} \, (\mathrm{km/s})^{2}$)& 3100  \\ \hline
    \end{tabular}
  \end{center}
\end{table}
\begin{table}[h!]
  \begin{center}
    \caption{NFW  Optimization Values}
    \label{NavaroNGC0100}
     \begin{tabular}{|r|r|}
     \hline
      \textbf{Parameter}   & \textbf{Optimization Values}
      \\  \hline
   $\rho_s$   & $0.003\times 10^9$
\\  \hline
$r_s$&  20
\\  \hline
    \end{tabular}
  \end{center}
\end{table}
\begin{table}[h!]
  \begin{center}
    \caption{Burkert Optimization Values}
    \label{BuckertNGC0100}
     \begin{tabular}{|r|r|}
     \hline
      \textbf{Parameter}   & \textbf{Optimization Values}
      \\  \hline
     $\rho_0^B$  & $0.051\times 10^9$
\\  \hline
$r_0$&  6
\\  \hline
    \end{tabular}
  \end{center}
\end{table}
\begin{table}[h!]
  \begin{center}
    \caption{Einasto Optimization Values}
    \label{EinastoNGC0100}
    \begin{tabular}{|r|r|}
     \hline
      \textbf{Parameter}   & \textbf{Optimization Values}
      \\  \hline
     $\rho_e$  & $0.0016\times 10^9$
\\  \hline
$r_e$ & 10
\\  \hline
$n_e$ & 0.09
\\  \hline
    \end{tabular}
  \end{center}
\end{table}
\begin{table}[h!]
\centering \caption{Physical assessment of collisional DM
parameters for NGC0100.}
\begin{tabular}{lcc}
\hline
Parameter & Value & Physical Verdict \\
\hline
$\gamma_0$ & $1.0001$ & Effectively isothermal  \\
$\delta_\gamma$ & $1.2\times10^{-9}$ & Essentially zero - $\gamma(r)$ constant \\
$r_\gamma$ & $1.5\ \mathrm{Kpc}$ & Transition radius irrelevant due to tiny $\delta_\gamma$ \\
$K_0$ & $3.1\times10^3$ & Moderately high pressure support \\
$r_c$ & $0.5\ \mathrm{Kpc}$ & Small core; reasonable for inner halo \\
$p$ & $0.01$ & Nearly constant entropy; minimal radial variation \\
\hline
Overall &-& Physically plausible; inner halo nearly isothermal and uniform \\
\hline
\end{tabular}
\label{EVALUATIONNGC0100}
\end{table}


\subsection{The Galaxy NGC0247 Non-viable}

\begin{figure}[h!]
\centering
\includegraphics[width=20pc]{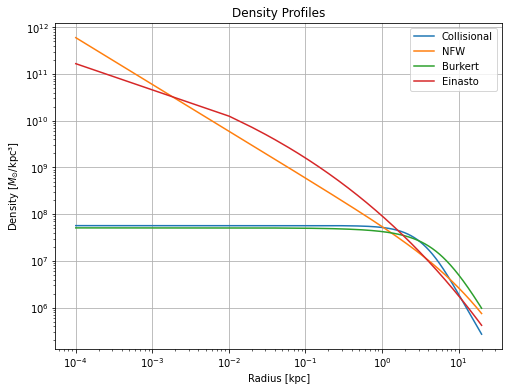}
\caption{The density of the collisional DM model (\ref{tanhmodel})
for the galaxy NGC0247, as a function of the radius.}
\label{NGC0247dens}
\end{figure}

\begin{figure}[h!]
\centering
\includegraphics[width=20pc]{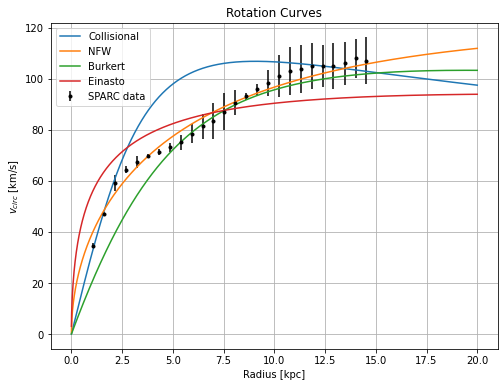}
\caption{The predicted rotation curves after using an optimization
for the collisional DM model (\ref{tanhmodel}), versus the SPARC
observational data for the galaxy NGC0247. We also plotted the
optimized curves for the NFW model, the Burkert model and the
Einasto model.} \label{NGC0247}
\end{figure}

\begin{figure}[h!]
\centering
\includegraphics[width=20pc]{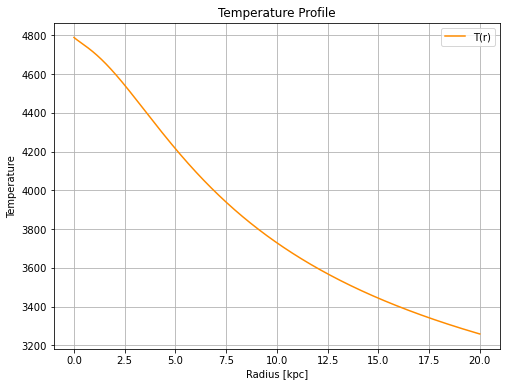}
\caption{The temperature as a function of the radius for the
collisional DM model (\ref{tanhmodel}) for the galaxy NGC0247.}
\label{NGC0247temp}
\end{figure}

\begin{figure}[h!]
\centering
\includegraphics[width=20pc]{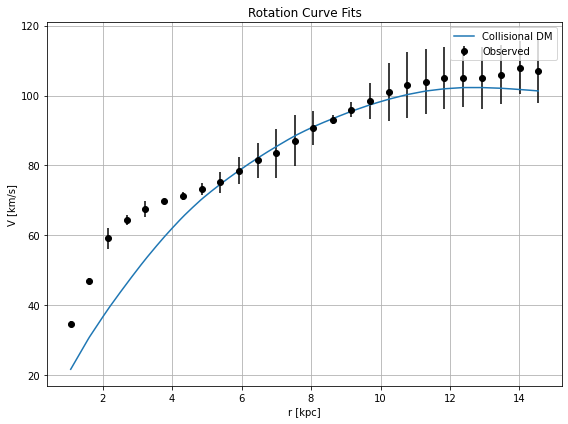}
\caption{The predicted rotation curves after using an optimization
for the collisional DM model (\ref{tanhmodel}), versus the
extended SPARC observational data for the galaxy NGC0247. The
model includes the rotation curves from all the components of the
galaxy, including gas and disk velocities, along with the
collisional DM model.} \label{extendedNGC0247}
\end{figure}

For this galaxy, we shall choose $\rho_0=5.3\times
10^7$$M_{\odot}/\mathrm{Kpc}^{3}$. NGC~247 is an
\(\mathrm{SAB(s)d}\) late-type spiral galaxy (weakly barred,
loosely wound arms) in the Sculptor Group, at a distance of about
\(3.4 \,\mathrm{Mpc}\). In Figs. \ref{NGC0247dens}, \ref{NGC0247}
and \ref{NGC0247temp} we present the density of the collisional DM
model, the predicted rotation curves after using an optimization
for the collisional DM model (\ref{tanhmodel}), versus the SPARC
observational data and the temperature parameter as a function of
the radius respectively. As it can be seen, the SIDM model
produces non-viable rotation curves incompatible with the SPARC
data. Also in Tables \ref{collNGC0247}, \ref{NavaroNGC0247},
\ref{BuckertNGC0247} and \ref{EinastoNGC0247} we present the
optimization values for the SIDM model, and the other DM profiles.
Also in Table \ref{EVALUATIONNGC0247} we present the overall
evaluation of the SIDM model for the galaxy at hand. The resulting
phenomenology is non-viable.
\begin{table}[h!]
  \begin{center}
    \caption{Collisional Dark Matter Optimization Values}
    \label{collNGC0247}
     \begin{tabular}{|r|r|}
     \hline
      \textbf{Parameter}   & \textbf{Optimization Values}
      \\  \hline
     $\delta_{\gamma} $ & 0.0000000012
\\  \hline
$\gamma_0 $ & 1.0001 \\ \hline $K_0$ ($M_{\odot} \,
\mathrm{Kpc}^{-3} \, (\mathrm{km/s})^{2}$)& 4200  \\ \hline
    \end{tabular}
  \end{center}
\end{table}
\begin{table}[h!]
  \begin{center}
    \caption{NFW  Optimization Values}
    \label{NavaroNGC0247}
     \begin{tabular}{|r|r|}
     \hline
      \textbf{Parameter}   & \textbf{Optimization Values}
      \\  \hline
   $\rho_s$   & $0.003\times 10^9$
\\  \hline
$r_s$&  20
\\  \hline
    \end{tabular}
  \end{center}
\end{table}
\begin{table}[h!]
  \begin{center}
    \caption{Burkert Optimization Values}
    \label{BuckertNGC0247}
     \begin{tabular}{|r|r|}
     \hline
      \textbf{Parameter}   & \textbf{Optimization Values}
      \\  \hline
     $\rho_0^B$  & $0.051\times 10^9$
\\  \hline
$r_0$&  6
\\  \hline
    \end{tabular}
  \end{center}
\end{table}
\begin{table}[h!]
  \begin{center}
    \caption{Einasto Optimization Values}
    \label{EinastoNGC0247}
    \begin{tabular}{|r|r|}
     \hline
      \textbf{Parameter}   & \textbf{Optimization Values}
      \\  \hline
     $\rho_e$  & $0.0018\times 10^9$
\\  \hline
$r_e$ & 10
\\  \hline
$n_e$ & 0.14
\\  \hline
    \end{tabular}
  \end{center}
\end{table}
\begin{table}[h!]
\centering \caption{Physical assessment of collisional DM
parameters for NGC0247.}
\begin{tabular}{lcc}
\hline
Parameter & Value & Physical Verdict \\
\hline
$\gamma_0$ & $1.0001$ & Effectively isothermal  \\
$\delta_\gamma$ & $1.2\times10^{-9}$ & Essentially zero - $\gamma(r)$ constant \\
$r_\gamma$ & $1.5\ \mathrm{Kpc}$ & Transition radius irrelevant due to tiny $\delta_\gamma$ \\
$K_0$ & $4.2\times10^3$ & Moderately high pressure support \\
$r_c$ & $0.5\ \mathrm{Kpc}$ & Small core; reasonable for inner halo \\
$p$ & $0.01$ & Nearly constant entropy; minimal radial variation \\
\hline
Overall &-& Physically plausible; inner halo nearly isothermal and uniform \\
\hline
\end{tabular}
\label{EVALUATIONNGC0247}
\end{table}
Now the extended picture including the rotation velocity from the
other components of the galaxy, such as the disk and gas, makes
the collisional DM model viable for this galaxy. In Fig.
\ref{extendedNGC0247} we present the combined rotation curves
including the other components of the galaxy along with the
collisional matter. As it can be seen, the extended collisional DM
model is non-viable.
Also in Table \ref{evaluationextendedNGC0247} we present the
values of the free parameters of the collisional DM model for
which the maximum compatibility with the SPARC data comes for the
galaxy NGC0247.
\begin{table}[h!]
\centering \caption{Physical assessment of Extended collisional DM
parameters for NGC0247.}
\begin{tabular}{lcc}
\hline
Parameter & Value & Physical Verdict \\
\hline
$\gamma_0$ & 1.0100 & Very close to isothermal \\
$\delta_\gamma$ & 0.0010 & Negligible radial variation \\
$K_0$ & 3000 & Moderate entropy \\
$ml_{\text{disk}}$ & 0.6184 & Reasonable stellar $M/L$ for a late-type galaxy; physically plausible \\
$ml_{\text{bulge}}$ & 0.0000 & No bulge component; disk-dominated system \\
\hline
Overall &-& Physically plausible \\
\hline
\end{tabular}
\label{evaluationextendedNGC0247}
\end{table}
Overall, the model is marginally viable even when the extended
features of the galaxy are added.


\subsection{The Galaxy NGC0289 Non-viable}

\begin{figure}[h!]
\centering
\includegraphics[width=20pc]{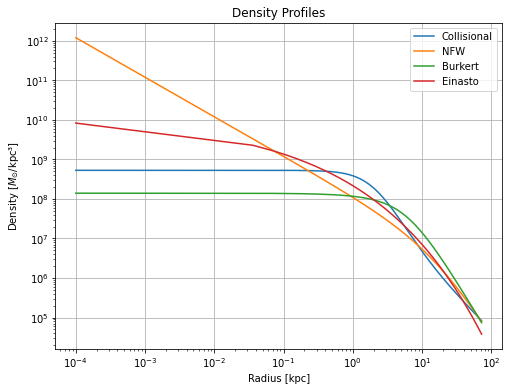}
\caption{The density of the collisional DM model (\ref{tanhmodel})
for the galaxy NGC0289, as a function of the radius.}
\label{NGC0289dens}
\end{figure}
\begin{figure}[h!]
\centering
\includegraphics[width=20pc]{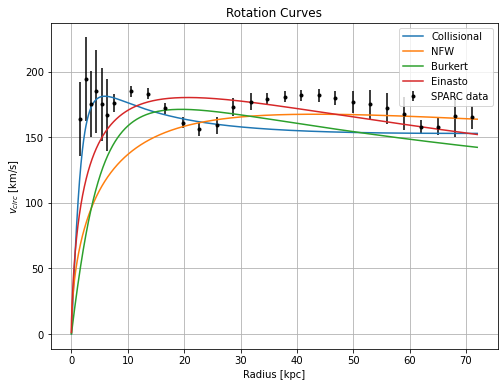}
\caption{The predicted rotation curves after using an optimization
for the collisional DM model (\ref{tanhmodel}), versus the SPARC
observational data for the galaxy NGC0289. We also plotted the
optimized curves for the NFW model, the Burkert model and the
Einasto model.} \label{NGC0289}
\end{figure}
\begin{figure}[h!]
\centering
\includegraphics[width=20pc]{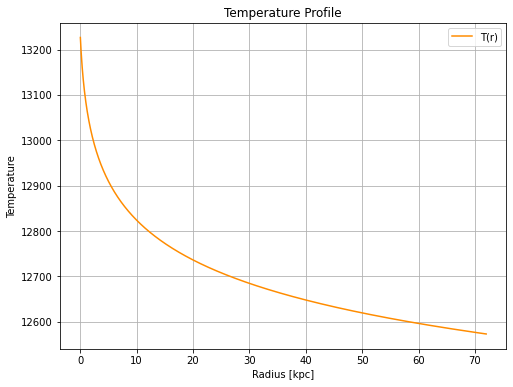}
\caption{The temperature as a function of the radius for the
collisional DM model (\ref{tanhmodel}) for the galaxy NGC0289.}
\label{NGC0289temp}
\end{figure}

For this galaxy, we shall choose $\rho_0=5.3\times
10^8$$M_{\odot}/\mathrm{Kpc}^{3}$. NGC\,289 is a barred spiral
galaxy, of Hubble type \(\mathrm{SB(rs)bc}\), located in the
constellation Sculptor. It is a large, gas-rich,
low-surface-brightness spiral with a small central bar and a weak
bulge, and displays Seyfert II nuclear activity. Its distance is
approximately $D \sim 23.3\;\mathrm{Mpc}$. In Figs.
\ref{NGC0289dens}, \ref{NGC0289} and \ref{NGC0289temp} we present
the density of the collisional DM model, the predicted rotation
curves after using an optimization for the collisional DM model
(\ref{tanhmodel}), versus the SPARC observational data and the
temperature parameter as a function of the radius respectively. As
it can be seen, the SIDM model produces non-viable rotation curves
incompatible with the SPARC data. Also in Tables
\ref{collNGC0289}, \ref{NavaroNGC0289}, \ref{BuckertNGC0289} and
\ref{EinastoNGC0289} we present the optimization values for the
SIDM model, and the other DM profiles. Also in Table
\ref{EVALUATIONNGC0289} we present the overall evaluation of the
SIDM model for the galaxy at hand. The resulting phenomenology is
non-viable.
\begin{table}[h!]
  \begin{center}
    \caption{Collisional Dark Matter Optimization Values}
    \label{collNGC0289}
     \begin{tabular}{|r|r|}
     \hline
      \textbf{Parameter}   & \textbf{Optimization Values}
      \\  \hline
     $\delta_{\gamma} $ & 0.0000000012
\\  \hline
$\gamma_0 $ & 1.0001 \\ \hline $K_0$ ($M_{\odot} \,
\mathrm{Kpc}^{-3} \, (\mathrm{km/s})^{2}$)& 13200  \\ \hline
    \end{tabular}
  \end{center}
\end{table}

\begin{figure}[h!]
\centering
\includegraphics[width=20pc]{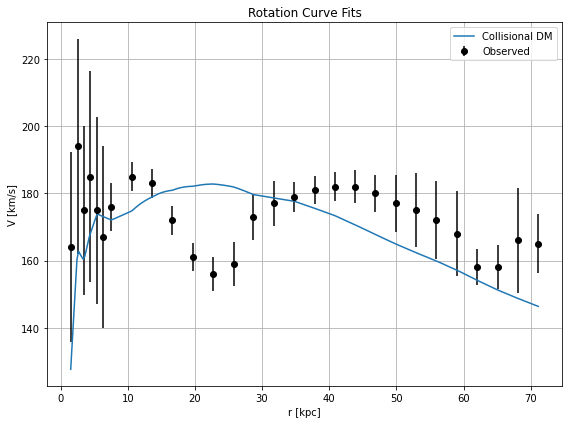}
\caption{The predicted rotation curves after using an optimization
for the collisional DM model (\ref{tanhmodel}), versus the
extended SPARC observational data for the galaxy NGC0289. The
model includes the rotation curves from all the components of the
galaxy, including gas and disk velocities, along with the
collisional DM model.} \label{extendedNGC0289}
\end{figure}

\begin{table}[h!]
  \begin{center}
    \caption{NFW  Optimization Values}
    \label{NavaroNGC0289}
     \begin{tabular}{|r|r|}
     \hline
      \textbf{Parameter}   & \textbf{Optimization Values}
      \\  \hline
   $\rho_s$   & $0.006\times 10^9$
\\  \hline
$r_s$&  20
\\  \hline
    \end{tabular}
  \end{center}
\end{table}
\begin{table}[h!]
  \begin{center}
    \caption{Burkert Optimization Values}
    \label{BuckertNGC0289}
     \begin{tabular}{|r|r|}
     \hline
      \textbf{Parameter}   & \textbf{Optimization Values}
      \\  \hline
     $\rho_0^B$  & $0.14\times 10^9$
\\  \hline
$r_0$&  6
\\  \hline
    \end{tabular}
  \end{center}
\end{table}
\begin{table}[h!]
  \begin{center}
    \caption{Einasto Optimization Values}
    \label{EinastoNGC0289}
    \begin{tabular}{|r|r|}
     \hline
      \textbf{Parameter}   & \textbf{Optimization Values}
      \\  \hline
     $\rho_e$  & $0.007\times 10^9$
\\  \hline
$r_e$ & 10
\\  \hline
$n_e$ & 0.27
\\  \hline
    \end{tabular}
  \end{center}
\end{table}
\begin{table}[h!]
\centering \caption{Physical assessment of collisional DM
parameters for NGC0289.}
\begin{tabular}{lcc}
\hline
Parameter & Value & Physical Verdict \\
\hline
$\gamma_0$ & $1.0001$ & Effectively isothermal  \\
$\delta_\gamma$ & $1.2\times10^{-9}$ & Essentially zero - $\gamma(r)$ constant \\
$r_\gamma$ & $1.5\ \mathrm{Kpc}$ & Transition radius irrelevant due to tiny $\delta_\gamma$ \\
$K_0$ & $1.32\times10^4$ & High entropy scale \\
$r_c$ & $0.5\ \mathrm{Kpc}$ & Small core; reasonable for inner halo \\
$p$ & $0.01$ & Nearly constant entropy; minimal radial variation \\
\hline
Overall &-& Physically plausible; inner halo nearly isothermal, high central density \\
\hline
\end{tabular}
\label{EVALUATIONNGC0289}
\end{table}
Now the extended picture including the rotation velocity from the
other components of the galaxy, such as the disk and gas, makes
the collisional DM model viable for this galaxy. In Fig.
\ref{extendedNGC0289} we present the combined rotation curves
including the other components of the galaxy along with the
collisional matter. As it can be seen, the extended collisional DM
model is marginally viable.
Also in Table \ref{evaluationextendedNGC0289} we present the
values of the free parameters of the collisional DM model for
which the maximum compatibility with the SPARC data comes for the
galaxy NGC0289.
\begin{table}[h!]
\centering \caption{Physical assessment of Extended collisional DM
parameters for NGC0247.}
\begin{tabular}{lcc}
\hline
Parameter & Value & Physical Verdict \\
\hline
$\gamma_0$ & 1.0781 & Mildly above isothermal \\
$\delta_\gamma$ & 0.0000 & No radial variation \\
$K_0$ & 3000 & Moderate entropy  \\
$ml_{\text{disk}}$ & 0.7009 & Moderately high stellar $M/L$ \\
$ml_{\text{bulge}}$ & 0.0000 & No bulge component \\
\hline
Overall &-& Physically viable \\
\hline
\end{tabular}
\label{evaluationextendedNGC0289}
\end{table}

\subsection{The Galaxy NGC0300 Marginally, Extended Viable}

\begin{figure}[h!]
\centering
\includegraphics[width=20pc]{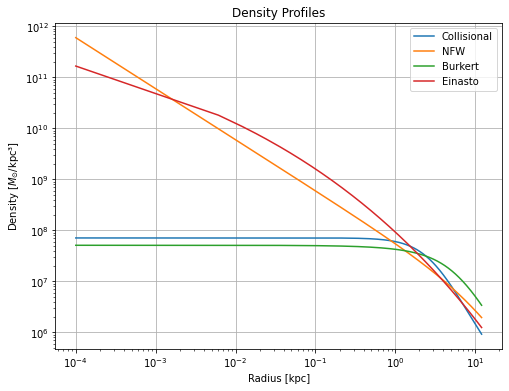}
\caption{The density of the collisional DM model (\ref{tanhmodel})
for the galaxy NGC0300, as a function of the radius.}
\label{NGC0300dens}
\end{figure}
\begin{figure}[h!]
\centering
\includegraphics[width=20pc]{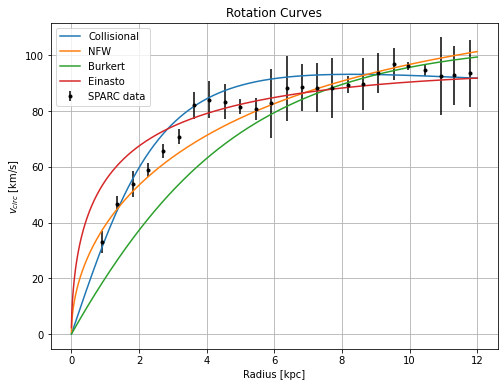}
\caption{The predicted rotation curves after using an optimization
for the collisional DM model (\ref{tanhmodel}), versus the SPARC
observational data for the galaxy NGC0300. We also plotted the
optimized curves for the NFW model, the Burkert model and the
Einasto model.} \label{NGC0300}
\end{figure}
\begin{figure}[h!]
\centering
\includegraphics[width=20pc]{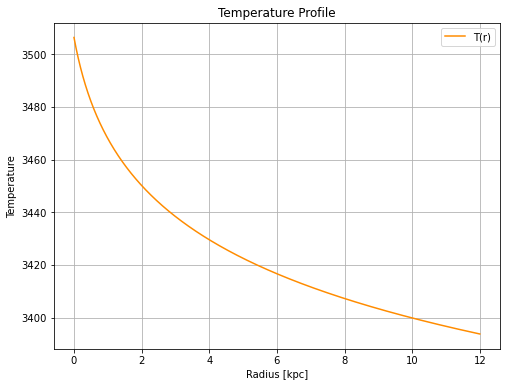}
\caption{The temperature as a function of the radius for the
collisional DM model (\ref{tanhmodel}) for the galaxy NGC0300.}
\label{NGC0300temp}
\end{figure}
\begin{figure}[h!]
\centering
\includegraphics[width=20pc]{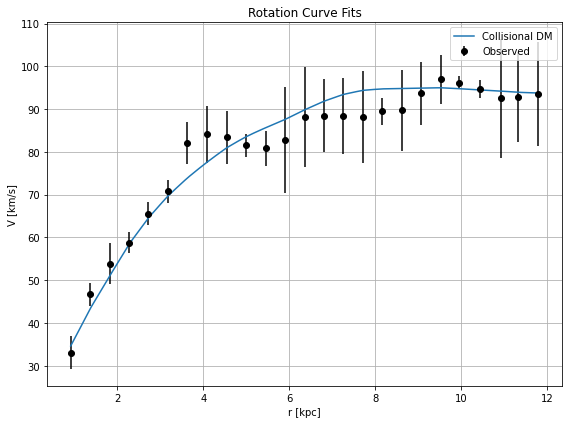}
\caption{The predicted rotation curves after using an optimization
for the collisional DM model (\ref{tanhmodel}), versus the
extended SPARC observational data for the galaxy NGC0300. The
model includes the rotation curves from all the components of the
galaxy, including gas and disk velocities, along with the
collisional DM model.} \label{extendedNGC0300}
\end{figure}

For this galaxy, we shall choose $\rho_0=7.1\times
10^7$$M_{\odot}/\mathrm{Kpc}^{3}$. NGC\,300 is a nearby late-type
spiral galaxy, of Hubble type \(\mathrm{SA(s)d}\), in the Sculptor
region, with little or no strong bulge and a relatively low mass
and surface brightness disk. Its distance is approximately $D \sim
1.86\;\mathrm{Mpc}$. In Figs. \ref{NGC0300dens}, \ref{NGC0300} and
\ref{NGC0300temp} we present the density of the collisional DM
model, the predicted rotation curves after using an optimization
for the collisional DM model (\ref{tanhmodel}), versus the SPARC
observational data and the temperature parameter as a function of
the radius respectively. As it can be seen, the SIDM model
produces viable rotation curves marginally compatible with the
SPARC data. Also in Tables \ref{collNGC0300}, \ref{NavaroNGC0300},
\ref{BuckertNGC0300} and \ref{EinastoNGC0300} we present the
optimization values for the SIDM model, and the other DM profiles.
Also in Table \ref{EVALUATIONNGC0300} we present the overall
evaluation of the SIDM model for the galaxy at hand. The resulting
phenomenology is marginally viable.
\begin{table}[h!]
  \begin{center}
    \caption{Collisional Dark Matter Optimization Values}
    \label{collNGC0300}
     \begin{tabular}{|r|r|}
     \hline
      \textbf{Parameter}   & \textbf{Optimization Values}
      \\  \hline
     $\delta_{\gamma} $ &  0.0000000012
\\  \hline
$\gamma_0 $ & 1.0001 \\ \hline $K_0$ ($M_{\odot} \,
\mathrm{Kpc}^{-3} \, (\mathrm{km/s})^{2}$)& 1500  \\ \hline
    \end{tabular}
  \end{center}
\end{table}
\begin{table}[h!]
  \begin{center}
    \caption{NFW  Optimization Values}
    \label{NavaroNGC0300}
     \begin{tabular}{|r|r|}
     \hline
      \textbf{Parameter}   & \textbf{Optimization Values}
      \\  \hline
   $\rho_s$   & $0.003\times 10^9$
\\  \hline
$r_s$&  20
\\  \hline
    \end{tabular}
  \end{center}
\end{table}
\begin{table}[h!]
  \begin{center}
    \caption{Burkert Optimization Values}
    \label{BuckertNGC0300}
     \begin{tabular}{|r|r|}
     \hline
      \textbf{Parameter}   & \textbf{Optimization Values}
      \\  \hline
     $\rho_0^B$  & $0.051\times 10^9$
\\  \hline
$r_0$&  6
\\  \hline
    \end{tabular}
  \end{center}
\end{table}
\begin{table}[h!]
  \begin{center}
    \caption{Einasto Optimization Values}
    \label{EinastoNGC0300}
    \begin{tabular}{|r|r|}
     \hline
      \textbf{Parameter}   & \textbf{Optimization Values}
      \\  \hline
     $\rho_e$  & $0.0018\times 10^9$
\\  \hline
$r_e$ & 10
\\  \hline
$n_e$ & 0.14
\\  \hline
    \end{tabular}
  \end{center}
\end{table}
\begin{table}[h!]
\centering \caption{Physical assessment of collisional DM
parameters for NGC0300.}
\begin{tabular}{lcc}
\hline
Parameter & Value & Physical Verdict \\
\hline
$\gamma_0$ & $1.0001$ & Effectively isothermal  \\
$\delta_\gamma$ & $1.2\times10^{-9}$ & Essentially zero - $\gamma(r)$ constant \\
$r_\gamma$ & $1.5\ \mathrm{Kpc}$ & Transition radius irrelevant due to tiny $\delta_\gamma$ \\
$K_0$ & $3.5\times10^3$ & Enough pressure support \\
$r_c$ & $0.5\ \mathrm{Kpc}$ & Small core; reasonable for inner halo \\
$p$ & $0.01$ & Nearly constant entropy; minimal radial variation \\
\hline
Overall &-& Physically plausible; inner halo nearly isothermal, moderate central density \\
\hline
\end{tabular}
\label{EVALUATIONNGC0300}
\end{table}
Now the extended picture including the rotation velocity from the
other components of the galaxy, such as the disk and gas, makes
the collisional DM model viable for this galaxy. In Fig.
\ref{extendedNGC0300} we present the combined rotation curves
including the other components of the galaxy along with the
collisional matter. As it can be seen, the extended collisional DM
model is viable.
Also in Table \ref{evaluationextendedNGC0300} we present the
values of the free parameters of the collisional DM model for
which the maximum compatibility with the SPARC data comes for the
galaxy NGC0300.
\begin{table}[h!]
\centering \caption{Physical assessment of Extended collisional DM
parameters for galaxy NGC0300.}
\begin{tabular}{lcc}
\hline
Parameter & Value & Physical Verdict \\
\hline
$\gamma_0$ & 1.01256201 & Near-isothermal (slightly above 1.0) \\
$\delta_\gamma$ & 0.01362152 & Small radial variation \\
$K_0$ & 3000 & Moderate entropy scale \\
$ml_{disk}$ & 1.00000000 & Relatively high disk M/L \\
$ml_{bulge}$ & 0.00000000 & No bulge contribution \\
\hline
Overall &-& Physically plausible; inner halo is nearly isothermal \\
\hline
\end{tabular}
\label{evaluationextendedNGC0300}
\end{table}

\subsection{The Galaxy NGC0801, Exceptional Case, Marginally viable}

\begin{figure}[h!]
\centering
\includegraphics[width=20pc]{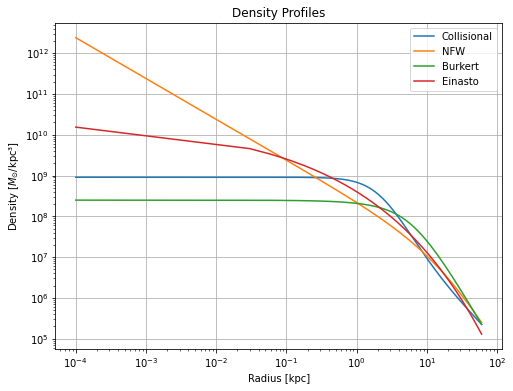}
\caption{The density of the collisional DM model (\ref{tanhmodel})
for the galaxy NGC0801, as a function of the radius.}
\label{NGC0801dens}
\end{figure}

\begin{figure}[h!]
\centering
\includegraphics[width=20pc]{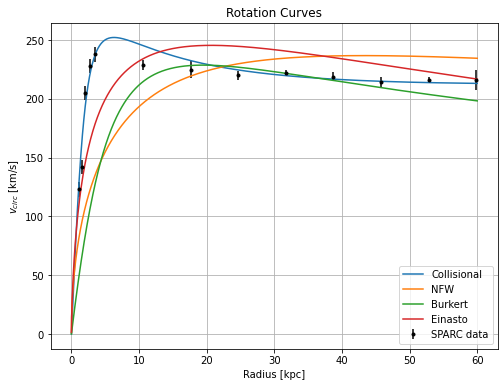}
\caption{The predicted rotation curves after using an optimization
for the collisional DM model (\ref{tanhmodel}), versus the SPARC
observational data for the galaxy NGC0801. We also plotted the
optimized curves for the NFW model, the Burkert model and the
Einasto model.} \label{NGC0801}
\end{figure}

\begin{figure}[h!]
\centering
\includegraphics[width=20pc]{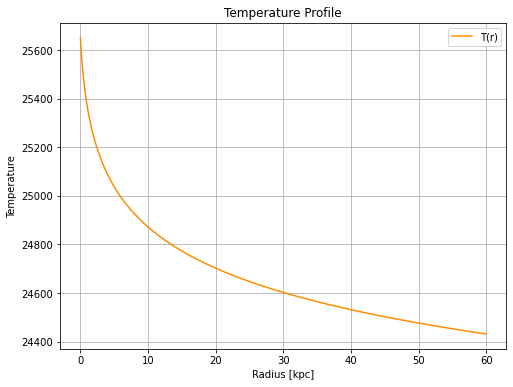}
\caption{The temperature as a function of the radius for the
collisional DM model (\ref{tanhmodel}) for the galaxy NGC0801.}
\label{NGC0801temp}
\end{figure}

For this galaxy, we shall choose $\rho_0=9.1\times
10^8$$M_{\odot}/\mathrm{Kpc}^{3}$. NGC\,801 is a large spiral
galaxy of Hubble type \(\mathrm{Sc}\), located in the
constellation Andromeda. Its distance is approximately $D \sim
80\;\mathrm{Mpc}$. In Figs. \ref{NGC0801dens}, \ref{NGC0801} and
\ref{NGC0801temp} we present the density of the collisional DM
model, the predicted rotation curves after using an optimization
for the collisional DM model (\ref{tanhmodel}), versus the SPARC
observational data and the temperature parameter as a function of
the radius respectively. As it can be seen, the SIDM model
produces marginally viable rotation curves compatible with the
SPARC data. Also in Tables \ref{collNGC0801}, \ref{NavaroNGC0801},
\ref{BuckertNGC0801} and \ref{EinastoNGC0801} we present the
optimization values for the SIDM model, and the other DM profiles.
Also in Table \ref{EVALUATIONNGC0801} we present the overall
evaluation of the SIDM model for the galaxy at hand. The resulting
phenomenology is marginally viable.
\begin{table}[h!]
  \begin{center}
    \caption{Collisional Dark Matter Optimization Values}
    \label{collNGC0801}
     \begin{tabular}{|r|r|}
     \hline
      \textbf{Parameter}   & \textbf{Optimization Values}
      \\  \hline
     $\delta_{\gamma} $ & 0.0000000012
\\  \hline
$\gamma_0 $ & 1.0001 \\ \hline $K_0$ ($M_{\odot} \,
\mathrm{Kpc}^{-3} \, (\mathrm{km/s})^{2}$)& 25600  \\ \hline
    \end{tabular}
  \end{center}
\end{table}
\begin{table}[h!]
  \begin{center}
    \caption{NFW  Optimization Values}
    \label{NavaroNGC0801}
     \begin{tabular}{|r|r|}
     \hline
      \textbf{Parameter}   & \textbf{Optimization Values}
      \\  \hline
   $\rho_s$   & $0.012\times 10^9$
\\  \hline
$r_s$&  20
\\  \hline
    \end{tabular}
  \end{center}
\end{table}
\begin{table}[h!]
  \begin{center}
    \caption{Burkert Optimization Values}
    \label{BuckertNGC0801}
     \begin{tabular}{|r|r|}
     \hline
      \textbf{Parameter}   & \textbf{Optimization Values}
      \\  \hline
     $\rho_0^B$  & $0.25\times 10^9$
\\  \hline
$r_0$&  6
\\  \hline
    \end{tabular}
  \end{center}
\end{table}
\begin{table}[h!]
  \begin{center}
    \caption{Einasto Optimization Values}
    \label{EinastoNGC0801}
    \begin{tabular}{|r|r|}
     \hline
      \textbf{Parameter}   & \textbf{Optimization Values}
      \\  \hline
     $\rho_e$  & $0.013\times 10^9$
\\  \hline
$r_e$ & 10
\\  \hline
$n_e$ & 0.27
\\  \hline
    \end{tabular}
  \end{center}
\end{table}
\begin{table}[h!]
\centering \caption{Physical assessment of collisional DM
parameters for NGC0801.}
\begin{tabular}{lcc}
\hline
Parameter & Value & Physical Verdict \\
\hline
$\gamma_0$ & $1.0001$ & Effectively isothermal; inner halo behaves as $P \sim \rho$ \\
$\delta_\gamma$ & $1.2\times10^{-9}$ & Essentially zero - $\gamma(r)$ constant \\
$r_\gamma$ & $1.5\ \mathrm{Kpc}$ & Transition radius irrelevant due to tiny $\delta_\gamma$ \\
$K_0$ & $2.56\times10^4$ & High entropy scale \\
$r_c$ & $0.5\ \mathrm{Kpc}$ & Small core; reasonable for inner halo \\
$p$ & $0.01$ & Nearly constant entropy; minimal radial variation \\
\hline
Overall &-& Physically plausible; inner halo nearly isothermal, moderate-to-high central density \\
\hline
\end{tabular}
\label{EVALUATIONNGC0801}
\end{table}


\subsection{The Galaxy NGC0891  Marginally Viable Large galaxy}

\begin{figure}[h!]
\centering
\includegraphics[width=20pc]{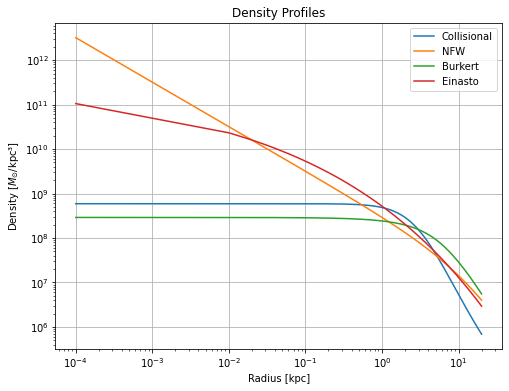}
\caption{The density of the collisional DM model (\ref{tanhmodel})
for the galaxy NGC0891, as a function of the radius.}
\label{NGC0891dens}
\end{figure}

\begin{figure}[h!]
\centering
\includegraphics[width=20pc]{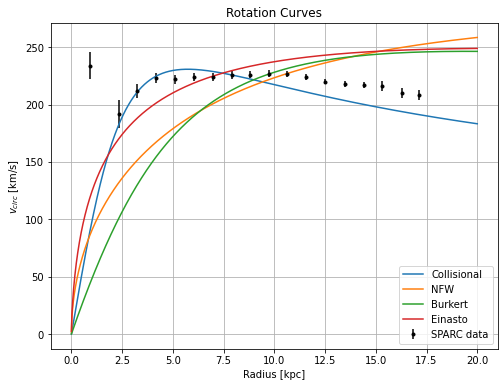}
\caption{The predicted rotation curves after using an optimization
for the collisional DM model (\ref{tanhmodel}), versus the SPARC
observational data for the galaxy NGC0891. We also plotted the
optimized curves for the NFW model, the Burkert model and the
Einasto model.} \label{NGC0891}
\end{figure}

\begin{figure}[h!]
\centering
\includegraphics[width=20pc]{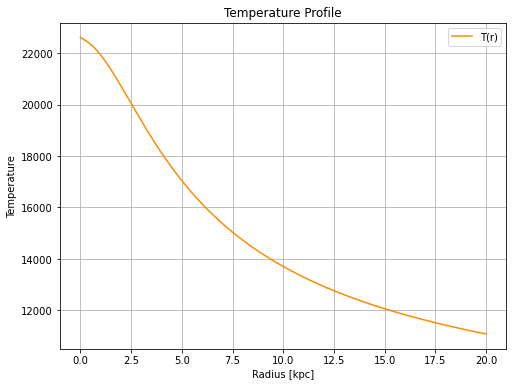}
\caption{The temperature as a function of the radius for the
collisional DM model (\ref{tanhmodel}) for the galaxy NGC0891.}
\label{NGC0891temp}
\end{figure}

For this galaxy, we shall choose $\rho_0=7.1\times
10^8$$M_{\odot}/\mathrm{Kpc}^{3}$. NGC0891 is an edge-on spiral
galaxy located in the constellation Andromeda. It is classified as
type SA(s)b and is approximately $8.4 \pm 0.5$ Mpc away from the
Milky Way. In Figs. \ref{NGC0891dens}, \ref{NGC0891} and
\ref{NGC0891temp} we present the density of the collisional DM
model, the predicted rotation curves after using an optimization
for the collisional DM model (\ref{tanhmodel}), versus the SPARC
observational data and the temperature parameter as a function of
the radius respectively. As it can be seen, the SIDM model
produces marginally viable rotation curves compatible with the
SPARC data. Also in Tables \ref{collNGC0891}, \ref{NavaroNGC0891},
\ref{BuckertNGC0891} and \ref{EinastoNGC0891} we present the
optimization values for the SIDM model, and the other DM profiles.
Also in Table \ref{EVALUATIONNGC0891} we present the overall
evaluation of the SIDM model for the galaxy at hand. The resulting
phenomenology is marginally viable.
\begin{table}[h!]
  \begin{center}
    \caption{Collisional Dark Matter Optimization Values}
    \label{collNGC0891}
     \begin{tabular}{|r|r|}
     \hline
      \textbf{Parameter}   & \textbf{Optimization Values}
      \\  \hline
     $\delta_{\gamma} $ & 0.0000000012
\\  \hline
$\gamma_0 $ & 1.0001 \\ \hline $K_0$ ($M_{\odot} \,
\mathrm{Kpc}^{-3} \, (\mathrm{km/s})^{2}$)& 21000 \\ \hline
    \end{tabular}
  \end{center}
\end{table}
\begin{table}[h!]
  \begin{center}
    \caption{NFW  Optimization Values}
    \label{NavaroNGC0891}
     \begin{tabular}{|r|r|}
     \hline
      \textbf{Parameter}   & \textbf{Optimization Values}
      \\  \hline
   $\rho_s$   & $0.016\times 10^9$
\\  \hline
$r_s$&  20
\\  \hline
    \end{tabular}
  \end{center}
\end{table}
\begin{table}[h!]
  \begin{center}
    \caption{Burkert Optimization Values}
    \label{BuckertNGC0891}
     \begin{tabular}{|r|r|}
     \hline
      \textbf{Parameter}   & \textbf{Optimization Values}
      \\  \hline
     $\rho_0^B$  & $0.29\times 10^9$
\\  \hline
$r_0$&  6
\\  \hline
    \end{tabular}
  \end{center}
\end{table}
\begin{table}[h!]
  \begin{center}
    \caption{Einasto Optimization Values}
    \label{EinastoNGC0891}
    \begin{tabular}{|r|r|}
     \hline
      \textbf{Parameter}   & \textbf{Optimization Values}
      \\  \hline
     $\rho_e$  & $0.013\times 10^9$
\\  \hline
$r_e$ & 10
\\  \hline
$n_e$ & 0.2
\\  \hline
    \end{tabular}
  \end{center}
\end{table}
\begin{table}[h!]
\centering \caption{Physical assessment of collisional DM
parameters for NGC0891.}
\begin{tabular}{lcc}
\hline
Parameter & Value & Physical Verdict \\
\hline
$\gamma_0$ & $1.0001$ & Effectively isothermal; inner halo behaves as $P \sim \rho$ \\
$\delta_\gamma$ & $1.2\times10^{-9}$ & Essentially zero - $\gamma(r)$ constant \\
$r_\gamma$ & $1.5\ \mathrm{Kpc}$ & Transition radius irrelevant due to tiny $\delta_\gamma$ \\
$K_0$ & $2.10\times10^4$ & High entropy scale \\
$r_c$ & $0.5\ \mathrm{Kpc}$ & Small core; reasonable for inner halo \\
$p$ & $0.01$ & Nearly constant entropy; minimal radial variation \\
\hline
Overall &-& Physically plausible; inner halo nearly isothermal, moderate central density \\
\hline
\end{tabular}
\label{EVALUATIONNGC0891}
\end{table}


\subsection{The Galaxy NGC1003 Non-viable}

\begin{figure}[h!]
\centering
\includegraphics[width=20pc]{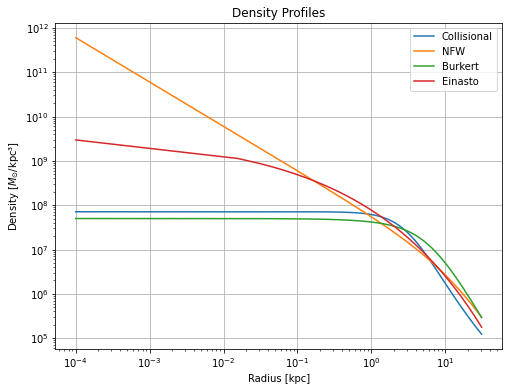}
\caption{The density of the collisional DM model (\ref{tanhmodel})
for the galaxy NGC1003, as a function of the radius.}
\label{NGC1003dens}
\end{figure}
\begin{figure}[h!]
\centering
\includegraphics[width=20pc]{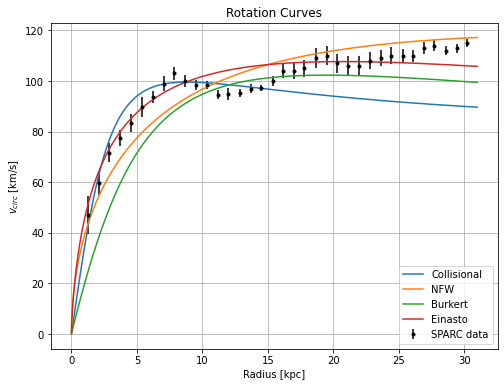}
\caption{The predicted rotation curves after using an optimization
for the collisional DM model (\ref{tanhmodel}), versus the SPARC
observational data for the galaxy NGC1003. We also plotted the
optimized curves for the NFW model, the Burkert model and the
Einasto model.} \label{NGC1003}
\end{figure}
\begin{figure}[h!]
\centering
\includegraphics[width=20pc]{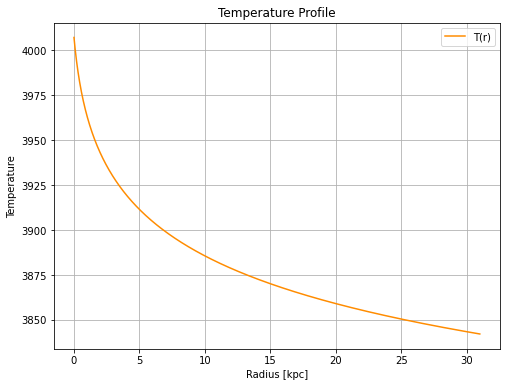}
\caption{The temperature as a function of the radius for the
collisional DM model (\ref{tanhmodel}) for the galaxy NGC1003.}
\label{NGC1003temp}
\end{figure}
\begin{figure}[h!]
\centering
\includegraphics[width=20pc]{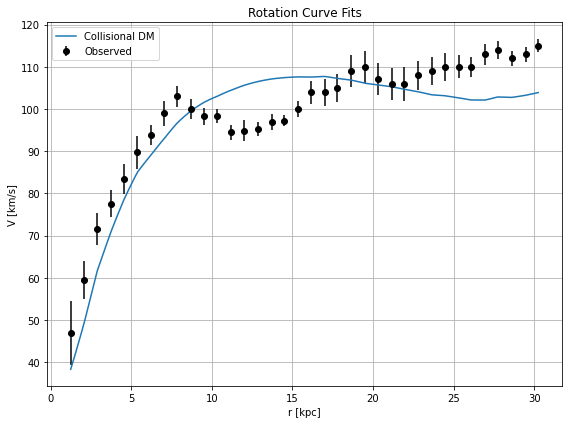}
\caption{The predicted rotation curves after using an optimization
for the collisional DM model (\ref{tanhmodel}), versus the
extended SPARC observational data for the galaxy NGC1003. The
model includes the rotation curves from all the components of the
galaxy, including gas and disk velocities, along with the
collisional DM model.} \label{extendedNGC1003}
\end{figure}

For this galaxy, we shall choose $\rho_0=7.1\times
10^7$$M_{\odot}/\mathrm{Kpc}^{3}$. NGC1003 is a spiral galaxy
located in the constellation Perseus. It is classified as type
SAcd, indicating an unbarred spiral galaxy with loosely wound
arms. The galaxy is situated at a distance of approximately $9.5$
Mpc from Milky Way. In Figs. \ref{NGC1003dens}, \ref{NGC1003} and
\ref{NGC1003temp} we present the density of the collisional DM
model, the predicted rotation curves after using an optimization
for the collisional DM model (\ref{tanhmodel}), versus the SPARC
observational data and the temperature parameter as a function of
the radius respectively. As it can be seen, the SIDM model
produces non-viable rotation curves incompatible with the SPARC
data. Also in Tables \ref{collNGC1003}, \ref{NavaroNGC1003},
\ref{BuckertNGC1003} and \ref{EinastoNGC1003} we present the
optimization values for the SIDM model, and the other DM profiles.
Also in Table \ref{EVALUATIONNGC1003} we present the overall
evaluation of the SIDM model for the galaxy at hand. The resulting
phenomenology is non-viable.
\begin{table}[h!]
  \begin{center}
    \caption{Collisional Dark Matter Optimization Values}
    \label{collNGC1003}
     \begin{tabular}{|r|r|}
     \hline
      \textbf{Parameter}   & \textbf{Optimization Values}
      \\  \hline
     $\delta_{\gamma} $ & 0.0000000012
\\  \hline
$\gamma_0 $ & 1.0001 \\ \hline $K_0$ ($M_{\odot} \,
\mathrm{Kpc}^{-3} \, (\mathrm{km/s})^{2}$)& 4000  \\ \hline
    \end{tabular}
  \end{center}
\end{table}
\begin{table}[h!]
  \begin{center}
    \caption{NFW  Optimization Values}
    \label{NavaroNGC1003}
     \begin{tabular}{|r|r|}
     \hline
      \textbf{Parameter}   & \textbf{Optimization Values}
      \\  \hline
   $\rho_s$   & $0.003\times 10^9$
\\  \hline
$r_s$&  20
\\  \hline
    \end{tabular}
  \end{center}
\end{table}
\begin{table}[h!]
  \begin{center}
    \caption{Burkert Optimization Values}
    \label{BuckertNGC1003}
     \begin{tabular}{|r|r|}
     \hline
      \textbf{Parameter}   & \textbf{Optimization Values}
      \\  \hline
     $\rho_0^B$  & $0.05\times 10^9$
\\  \hline
$r_0$&  6
\\  \hline
    \end{tabular}
  \end{center}
\end{table}
\begin{table}[h!]
  \begin{center}
    \caption{Einasto Optimization Values}
    \label{EinastoNGC1003}
    \begin{tabular}{|r|r|}
     \hline
      \textbf{Parameter}   & \textbf{Optimization Values}
      \\  \hline
     $\rho_e$  & $0.0025\times 10^9$
\\  \hline
$r_e$ & 10
\\  \hline
$n_e$ & 0.27
\\  \hline
    \end{tabular}
  \end{center}
\end{table}
\begin{table}[h!]
\centering \caption{Physical assessment of collisional DM
parameters for NGC1003.}
\begin{tabular}{lcc}
\hline
Parameter & Value & Physical Verdict \\
\hline
$\gamma_0$ & $1.0001$ & Effectively isothermal; inner halo behaves as $P \sim \rho$ \\
$\delta_\gamma$ & $1.2\times10^{-9}$ & Essentially zero - $\gamma(r)$ constant \\
$r_\gamma$ & $1.5\ \mathrm{Kpc}$ & Transition radius irrelevant due to tiny $\delta_\gamma$ \\
$K_0$ & $4.0\times10^3$ & Moderate entropy scale \\
$r_c$ & $0.5\ \mathrm{Kpc}$ & Small core; reasonable for inner halo \\
$p$ & $0.01$ & Nearly constant entropy; minimal radial variation \\
\hline
Overall &-& Physically plausible; inner halo nearly isothermal, low central density \\
\hline
\end{tabular}
\label{EVALUATIONNGC1003}
\end{table}
Now the extended picture including the rotation velocity from the
other components of the galaxy, such as the disk and gas, makes
the collisional DM model viable for this galaxy. In Fig.
\ref{extendedNGC1003} we present the combined rotation curves
including the other components of the galaxy along with the
collisional matter. As it can be seen, the extended collisional DM
model is not viable.
Also in Table \ref{evaluationextendedNGC1003} we present the
values of the free parameters of the collisional DM model for
which the maximum compatibility with the SPARC data comes for the
galaxy NGC1003.
\begin{table}[h!]
\centering \caption{Physical assessment of Extended collisional DM
parameters for NGC1003.}
\begin{tabular}{lcc}
\hline
Parameter & Value & Physical Verdict \\
\hline
$\gamma_0$ & 1.0212 & Very close to isothermal; low central pressure, stable inner halo \\
$\delta_\gamma$ & 0.0000 & No radial variation \\
$K_0$ & 3000 & Moderate entropy  \\
$ml_{\text{disk}}$ & 0.5831 & Reasonable stellar $M/L$ for a late-type disk; physically plausible \\
$ml_{\text{bulge}}$ & 0.0000 & No bulge component; disk-dominated morphology \\
\hline
Overall &-& Physically viable; halo effectively polytropic with fixed $\gamma$ \\
\hline
\end{tabular}
\label{evaluationextendedNGC1003}
\end{table}



\subsection{The Galaxy NGC1705 Remarkably Viable}

\begin{figure}[h!]
\centering
\includegraphics[width=20pc]{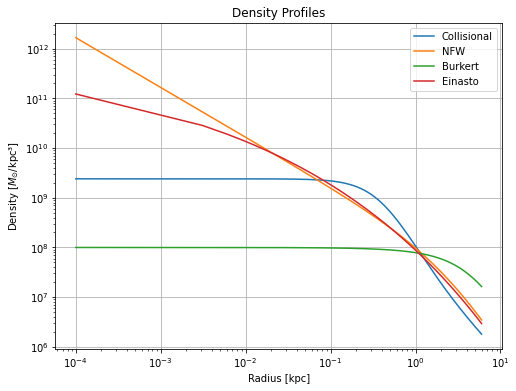}
\caption{The density of the collisional DM model (\ref{tanhmodel})
for the galaxy NGC1705, as a function of the radius.}
\label{NGC1705dens}
\end{figure}

\begin{figure}[h!]
\centering
\includegraphics[width=20pc]{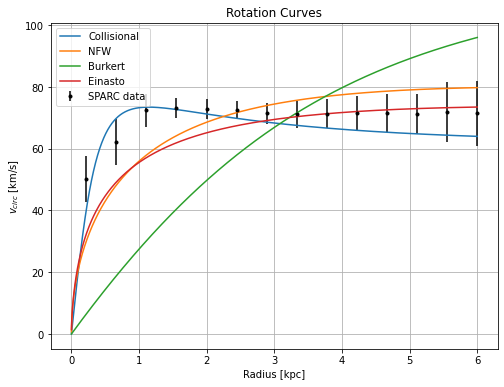}
\caption{The predicted rotation curves after using an optimization
for the collisional DM model (\ref{tanhmodel}), versus the SPARC
observational data for the galaxy NGC1705. We also plotted the
optimized curves for the NFW model, the Burkert model and the
Einasto model.} \label{NGC1705}
\end{figure}

\begin{figure}[h!]
\centering
\includegraphics[width=20pc]{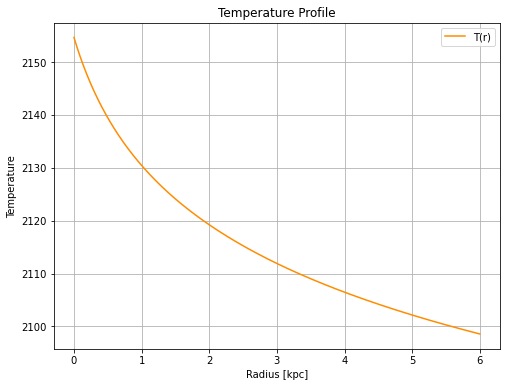}
\caption{The temperature as a function of the radius for the
collisional DM model (\ref{tanhmodel}) for the galaxy NGC1705.}
\label{NGC1705temp}
\end{figure}

For this galaxy, we shall choose $\rho_0=2.4\times
10^9$$M_{\odot}/\mathrm{Kpc}^{3}$. NGC\,1705 is a blue compact
dwarf galaxy (also classified as peculiar lenticular/SA0)
undergoing a starburst. Its distance from Earth is about $D \sim
5.1 \pm 0.6\;\mathrm{Mpc}$. It is a member of the Dorado Group. In
Figs. \ref{NGC1705dens}, \ref{NGC1705} and \ref{NGC1705temp} we
present the density of the collisional DM model, the predicted
rotation curves after using an optimization for the collisional DM
model (\ref{tanhmodel}), versus the SPARC observational data and
the temperature parameter as a function of the radius
respectively. As it can be seen, the SIDM model produces viable
rotation curves compatible with the SPARC data. Also in Tables
\ref{collNGC1705}, \ref{NavaroNGC1705}, \ref{BuckertNGC1705} and
\ref{EinastoNGC1705} we present the optimization values for the
SIDM model, and the other DM profiles. Also in Table
\ref{EVALUATIONNGC1705} we present the overall evaluation of the
SIDM model for the galaxy at hand. The resulting phenomenology is
viable.
\begin{table}[h!]
  \begin{center}
    \caption{Collisional Dark Matter Optimization Values}
    \label{collNGC1705}
     \begin{tabular}{|r|r|}
     \hline
      \textbf{Parameter}   & \textbf{Optimization Values}
      \\  \hline
     $\delta_{\gamma} $ & 0.0000000012
\\  \hline
$\gamma_0 $ & 1.0001  \\ \hline $K_0$ ($M_{\odot} \,
\mathrm{Kpc}^{-3} \, (\mathrm{km/s})^{2}$)& 2150  \\ \hline
    \end{tabular}
  \end{center}
\end{table}
\begin{table}[h!]
  \begin{center}
    \caption{NFW  Optimization Values}
    \label{NavaroNGC1705}
     \begin{tabular}{|r|r|}
     \hline
      \textbf{Parameter}   & \textbf{Optimization Values}
      \\  \hline
   $\rho_s$   & $0.004\times 10^9$
\\  \hline
$r_s$&  20
\\  \hline
    \end{tabular}
  \end{center}
\end{table}
\begin{table}[h!]
  \begin{center}
    \caption{Burkert Optimization Values}
    \label{BuckertNGC1705}
     \begin{tabular}{|r|r|}
     \hline
      \textbf{Parameter}   & \textbf{Optimization Values}
      \\  \hline
     $\rho_0^B$  & $0.035\times 10^9$
\\  \hline
$r_0$&  6
\\  \hline
    \end{tabular}
  \end{center}
\end{table}
\begin{table}[h!]
  \begin{center}
    \caption{Einasto Optimization Values}
    \label{EinastoNGC1705}
    \begin{tabular}{|r|r|}
     \hline
      \textbf{Parameter}   & \textbf{Optimization Values}
      \\  \hline
     $\rho_e$  & $0.009\times 10^9$
\\  \hline
$r_e$ & 10
\\  \hline
$n_e$ & 0.15
\\  \hline
    \end{tabular}
  \end{center}
\end{table}
\begin{table}[h!]
\centering \caption{Physical assessment of collisional DM
parameters for NGC1705.}
\begin{tabular}{lcc}
\hline
Parameter & Value & Physical Verdict \\
\hline
$\gamma_0$ & $1.0001$ & Effectively isothermal; $P \sim \rho$ in inner halo \\
$\delta_\gamma$ & $1.2\times10^{-9}$ & Negligible; $\gamma(r)$ constant \\
$r_\gamma$ & $1.5\ \mathrm{Kpc}$ & Transition radius irrelevant \\
$K_0$ & $2.15\times10^3$ & Acceptable Central Core Pressure Support  \\
$r_c$ & $0.5\ \mathrm{Kpc}$ & Small core; reasonable for dwarf halo \\
$p$ & $0.01$ & Nearly constant entropy; minimal radial variation \\
\hline
Overall &-& Physically plausible; inner halo nearly isothermal \\
\hline
\end{tabular}
\label{EVALUATIONNGC1705}
\end{table}


\subsection{The Galaxy NGC2366}

\begin{figure}[h!]
\centering
\includegraphics[width=20pc]{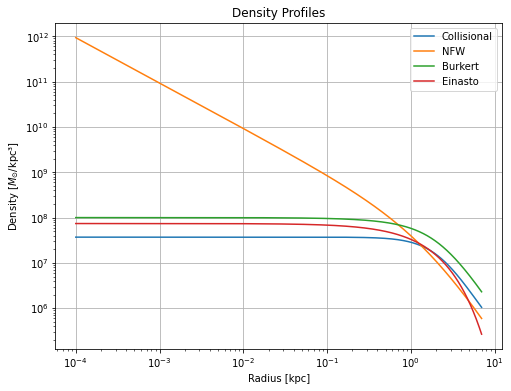}
\caption{The density of the collisional DM model (\ref{tanhmodel})
for the galaxy NGC2366, as a function of the radius.}
\label{NGC2366dens}
\end{figure}
\begin{figure}[h!]
\centering
\includegraphics[width=20pc]{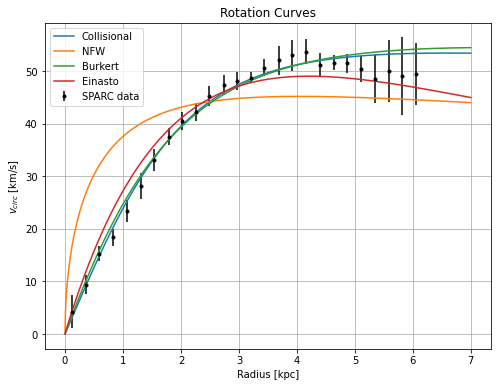}
\caption{The predicted rotation curves after using an optimization
for the collisional DM model (\ref{tanhmodel}), versus the SPARC
observational data for the galaxy NGC2366. We also plotted the
optimized curves for the NFW model, the Burkert model and the
Einasto model.} \label{NGC2366}
\end{figure}
\begin{figure}[h!]
\centering
\includegraphics[width=20pc]{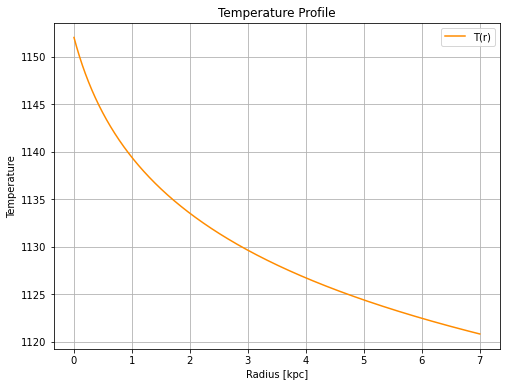}
\caption{The temperature as a function of the radius for the
collisional DM model (\ref{tanhmodel}) for the galaxy NGC2366.}
\label{NGC2366temp}
\end{figure}

For this galaxy, we shall choose $\rho_0=3.7\times
10^7$$M_{\odot}/\mathrm{Kpc}^{3}$. NGC\,2366 is a Magellanic-type
barred irregular (IB(s)m) dwarf galaxy. It lies at a distance $D
\sim 3.4\ \mathrm{Mpc}$. In Figs. \ref{NGC2366dens}, \ref{NGC2366}
and \ref{NGC2366temp} we present the density of the collisional DM
model, the predicted rotation curves after using an optimization
for the collisional DM model (\ref{tanhmodel}), versus the SPARC
observational data and the temperature parameter as a function of
the radius respectively. As it can be seen, the SIDM model
produces viable rotation curves compatible with the SPARC data.
Also in Tables \ref{collNGC2366}, \ref{NavaroNGC2366},
\ref{BuckertNGC2366} and \ref{EinastoNGC2366} we present the
optimization values for the SIDM model, and the other DM profiles.
Also in Table \ref{EVALUATIONNGC2366} we present the overall
evaluation of the SIDM model for the galaxy at hand. The resulting
phenomenology is viable.
\begin{table}[h!]
  \begin{center}
    \caption{Collisional Dark Matter Optimization Values}
    \label{collNGC2366}
     \begin{tabular}{|r|r|}
     \hline
      \textbf{Parameter}   & \textbf{Optimization Values}
      \\  \hline
     $\delta_{\gamma} $ & 0.0000000012
\\  \hline
$\gamma_0 $ & 1.0001 \\ \hline $K_0$ ($M_{\odot} \,
\mathrm{Kpc}^{-3} \, (\mathrm{km/s})^{2}$)& 1150  \\ \hline
    \end{tabular}
  \end{center}
\end{table}
\begin{table}[h!]
  \begin{center}
    \caption{NFW  Optimization Values}
    \label{NavaroNGC2366}
     \begin{tabular}{|r|r|}
     \hline
      \textbf{Parameter}   & \textbf{Optimization Values}
      \\  \hline
   $\rho_s$   & $5\times 10^7$
\\  \hline
$r_s$&  1.87
\\  \hline
    \end{tabular}
  \end{center}
\end{table}
\begin{table}[h!]
  \begin{center}
    \caption{Burkert Optimization Values}
    \label{BuckertNGC2366}
     \begin{tabular}{|r|r|}
     \hline
      \textbf{Parameter}   & \textbf{Optimization Values}
      \\  \hline
     $\rho_0^B$  & $1\times 10^8$
\\  \hline
$r_0$&  2.26
\\  \hline
    \end{tabular}
  \end{center}
\end{table}
\begin{table}[h!]
  \begin{center}
    \caption{Einasto Optimization Values}
    \label{EinastoNGC2366}
    \begin{tabular}{|r|r|}
     \hline
      \textbf{Parameter}   & \textbf{Optimization Values}
      \\  \hline
     $\rho_e$  & $  1 \times 10^7$
\\  \hline
$r_e$ & 2.49
\\  \hline
$n_e$ & 1
\\  \hline
    \end{tabular}
  \end{center}
\end{table}
\begin{table}[h!]
\centering \caption{Physical assessment of collisional DM
parameters (NGC2366).}
\begin{tabular}{lcc}
\hline
Parameter & Value & Physical Verdict \\
\hline
$\gamma_0$ & $1.0001$ & Essentially isothermal  \\
$\delta_\gamma$ & $1.2\times10^{-9}$ & Negligible   \\
$r_\gamma$ & $1.5\ \mathrm{Kpc}$ & Transition radius set in inner halo \\
$K_0$ & $1.15\times10^{3}$ & Moderate entropy scale \\
$r_c$ & $0.5\ \mathrm{Kpc}$ & Small core scale - plausible for inner halo concentration \\
$p$ & $0.01$ & Extremely shallow $K(r)$ slope; $K$ practically constant with radius \\
\hline
Overall &-& Physically consistent but functionally nearly isothermal \\
\hline
\end{tabular}
\label{EVALUATIONNGC2366}
\end{table}


\subsection{The Galaxy NGC2403 Non-viable}

\begin{figure}[h!]
\centering
\includegraphics[width=20pc]{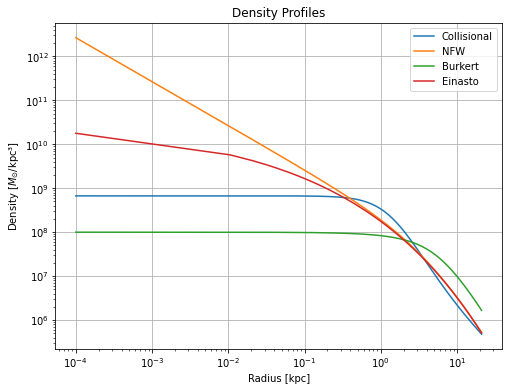}
\caption{The density of the collisional DM model (\ref{tanhmodel})
for the galaxy NGC2403, as a function of the radius.}
\label{NGC2403dens}
\end{figure}
\begin{figure}[h!]
\centering
\includegraphics[width=20pc]{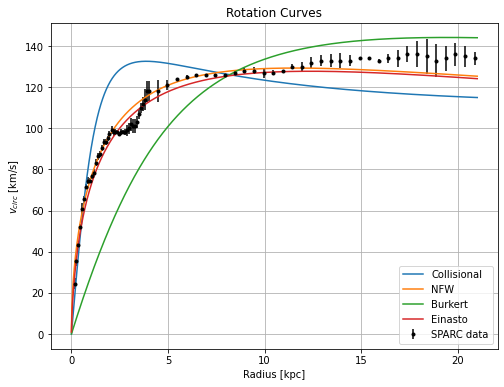}
\caption{The predicted rotation curves after using an optimization
for the collisional DM model (\ref{tanhmodel}), versus the SPARC
observational data for the galaxy NGC2403. We also plotted the
optimized curves for the NFW model, the Burkert model and the
Einasto model.} \label{NGC2403}
\end{figure}
\begin{figure}[h!]
\centering
\includegraphics[width=20pc]{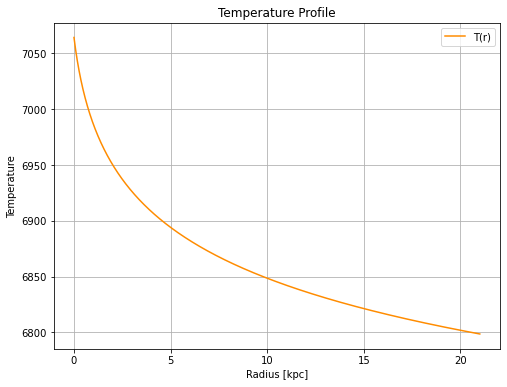}
\caption{The temperature as a function of the radius for the
collisional DM model (\ref{tanhmodel}) for the galaxy NGC2403.}
\label{NGC2403temp}
\end{figure}
\begin{figure}[h!]
\centering
\includegraphics[width=20pc]{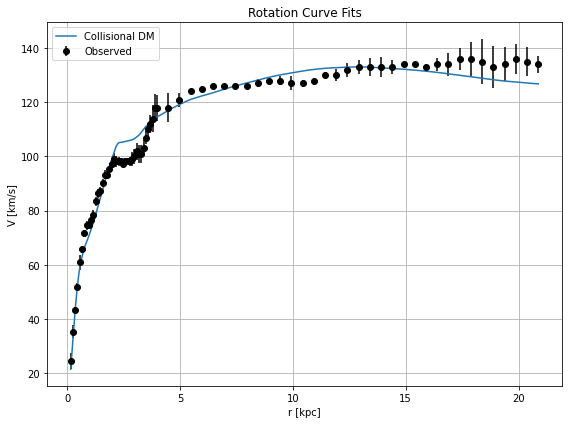}
\caption{The predicted rotation curves after using an optimization
for the collisional DM model (\ref{tanhmodel}), versus the
extended SPARC observational data for the galaxy NGC2403. The
model includes the rotation curves from all the components of the
galaxy, including gas and disk velocities, along with the
collisional DM model.} \label{extendedNGC2403}
\end{figure}

For this galaxy, we shall choose $\rho_0=6.7\times
10^8$$M_{\odot}/\mathrm{Kpc}^{3}$. NGC\,2403 is an intermediate
spiral galaxy of type SAB(s)cd, located in the constellation
Camelopardalis. It is a member of the M81 Group and is often
considered a scaled-down analogue of M33. The galaxy lies at a
distance of about $D \sim 3.2\ \mathrm{Mpc}$. In Figs.
\ref{NGC2403dens}, \ref{NGC2403} and \ref{NGC2403temp} we present
the density of the collisional DM model, the predicted rotation
curves after using an optimization for the collisional DM model
(\ref{tanhmodel}), versus the SPARC observational data and the
temperature parameter as a function of the radius respectively. As
it can be seen, the SIDM model produces non-viable rotation curves
incompatible with the SPARC data. Also in Tables
\ref{collNGC2403}, \ref{NavaroNGC2403}, \ref{BuckertNGC2403} and
\ref{EinastoNGC2403} we present the optimization values for the
SIDM model, and the other DM profiles. Also in Table
\ref{EVALUATIONNGC2403} we present the overall evaluation of the
SIDM model for the galaxy at hand. The resulting phenomenology is
non-viable.
\begin{table}[h!]
  \begin{center}
    \caption{Collisional Dark Matter Optimization Values}
    \label{collNGC2403}
     \begin{tabular}{|r|r|}
     \hline
      \textbf{Parameter}   & \textbf{Optimization Values}
      \\  \hline
     $\delta_{\gamma} $ & 0.0000000012
\\  \hline
$\gamma_0 $ &  1.0001 \\ \hline $K_0$ ($M_{\odot} \,
\mathrm{Kpc}^{-3} \, (\mathrm{km/s})^{2}$)& 7050  \\ \hline
    \end{tabular}
  \end{center}
\end{table}
\begin{table}[h!]
  \begin{center}
    \caption{NFW  Optimization Values}
    \label{NavaroNGC2403}
     \begin{tabular}{|r|r|}
     \hline
      \textbf{Parameter}   & \textbf{Optimization Values}
      \\  \hline
   $\rho_s$   & $5\times 10^7$
\\  \hline
$r_s$&  5.35
\\  \hline
    \end{tabular}
  \end{center}
\end{table}
\begin{table}[h!]
  \begin{center}
    \caption{Burkert Optimization Values}
    \label{BuckertNGC2403}
     \begin{tabular}{|r|r|}
     \hline
      \textbf{Parameter}   & \textbf{Optimization Values}
      \\  \hline
     $\rho_0^B$  & $1\times 10^8$
\\  \hline
$r_0$&  5.98
\\  \hline
    \end{tabular}
  \end{center}
\end{table}
\begin{table}[h!]
  \begin{center}
    \caption{Einasto Optimization Values}
    \label{EinastoNGC2403}
    \begin{tabular}{|r|r|}
     \hline
      \textbf{Parameter}   & \textbf{Optimization Values}
      \\  \hline
     $\rho_e$  & $1\times 10^7$
\\  \hline
$r_e$ & 5.91
\\  \hline
$n_e$ & 0.25
\\  \hline
    \end{tabular}
  \end{center}
\end{table}
\begin{table}[h!]
\centering \caption{Physical assessment of collisional DM
parameters (NGC2403).}
\begin{tabular}{lcc}
\hline
Parameter & Value & Physical Verdict \\
\hline
$\gamma_0$ & $1.0001$ & Essentially isothermal  \\
$\delta_\gamma$ & $1.2\times10^{-9}$ & Negligible   \\
$r_\gamma$ & $1.5\ \mathrm{Kpc}$ & Transition radius in inner halo  \\
$K_0$ & $7.05\times10^{3}$ & Large entropy  \\
$r_c$ & $0.5\ \mathrm{Kpc}$ & Small core scale - plausible but on the compact side for a large spiral \\
$p$ & $0.01$ & Extremely shallow $K(r)$ slope; $K$ practically constant with radius \\
\hline
Overall &-& Physically consistent but functionally nearly isothermal \\
\hline
\end{tabular}
\label{EVALUATIONNGC2403}
\end{table}
Now the extended picture including the rotation velocity from the
other components of the galaxy, such as the disk and gas, makes
the collisional DM model viable for this galaxy. In Fig.
\ref{extendedNGC2403} we present the combined rotation curves
including the other components of the galaxy along with the
collisional matter. As it can be seen, the extended collisional DM
model is non-viable.
Also in Table \ref{evaluationextendedNGC2403} we present the
values of the free parameters of the collisional DM model for
which the maximum compatibility with the SPARC data comes for the
galaxy NGC2403.
\begin{table}[h!]
\centering \caption{Physical assessment of Extended collisional DM
parameters for NGC2403.}
\begin{tabular}{lcc}
\hline
Parameter & Value & Physical Verdict \\
\hline
$\gamma_0$ & 1.0516 & Near-isothermal core \\
$\delta_\gamma$ & 0.01467 & Small radial variation; \\
$K_0$ & 3000 & Moderate entropy \\
$ml_{\text{disk}}$ & 0.91896 & Relatively high stellar $M/L$ \\
$ml_{\text{bulge}}$ & 0.0000 & No bulge component \\
\hline
Overall &-& Physically plausible \\
\hline
\end{tabular}
\label{evaluationextendedNGC2403}
\end{table}


\subsection{The Galaxy NGC2683 Non-viable}

\begin{figure}[h!]
\centering
\includegraphics[width=20pc]{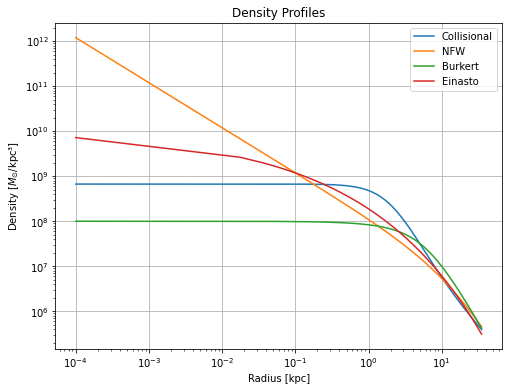}
\caption{The density of the collisional DM model (\ref{tanhmodel})
for the galaxy NGC2683, as a function of the radius.}
\label{NGC2683dens}
\end{figure}
\begin{figure}[h!]
\centering
\includegraphics[width=20pc]{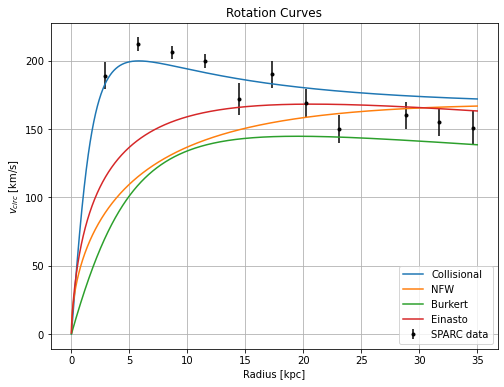}
\caption{The predicted rotation curves after using an optimization
for the collisional DM model (\ref{tanhmodel}), versus the SPARC
observational data for the galaxy NGC2683. We also plotted the
optimized curves for the NFW model, the Burkert model and the
Einasto model.} \label{NGC2683}
\end{figure}
\begin{figure}[h!]
\centering
\includegraphics[width=20pc]{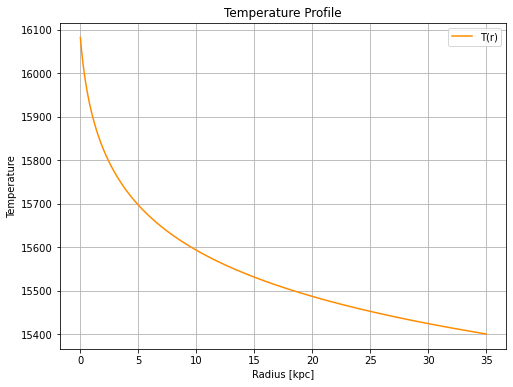}
\caption{The temperature as a function of the radius for the
collisional DM model (\ref{tanhmodel}) for the galaxy NGC2683.}
\label{NGC2683temp}
\end{figure}

For this galaxy, we shall choose $\rho_0=6.7\times
10^8$$M_{\odot}/\mathrm{Kpc}^{3}$. NGC\,2683 is a barred spiral
galaxy of type SA(rs)b, located in the northern constellation of
Lynx. It is nicknamed the ''UFO Galaxy'' due to its edge-on
orientation, which makes it resemble a flying saucer. The galaxy
lies at a distance of approximately $D \sim 7.5\ \mathrm{Mpc}$. In
Figs. \ref{NGC2683dens}, \ref{NGC2683} and \ref{NGC2683temp} we
present the density of the collisional DM model, the predicted
rotation curves after using an optimization for the collisional DM
model (\ref{tanhmodel}), versus the SPARC observational data and
the temperature parameter as a function of the radius
respectively. As it can be seen, the SIDM model produces
non-viable rotation curves incompatible with the SPARC data. Also
in Tables \ref{collNGC2683}, \ref{NavaroNGC2683},
\ref{BuckertNGC2683} and \ref{EinastoNGC2683} we present the
optimization values for the SIDM model, and the other DM profiles.
Also in Table \ref{EVALUATIONNGC2683} we present the overall
evaluation of the SIDM model for the galaxy at hand. The resulting
phenomenology is non-viable.
\begin{table}[h!]
  \begin{center}
    \caption{Collisional Dark Matter Optimization Values}
    \label{collNGC2683}
     \begin{tabular}{|r|r|}
     \hline
      \textbf{Parameter}   & \textbf{Optimization Values}
      \\  \hline
     $\delta_{\gamma} $ & 0.0000000012
\\  \hline
$\gamma_0 $ & 1.0001\\ \hline $K_0$ ($M_{\odot} \,
\mathrm{Kpc}^{-3} \, (\mathrm{km/s})^{2}$)& 16050  \\ \hline
    \end{tabular}
  \end{center}
\end{table}
\begin{table}[h!]
  \begin{center}
    \caption{NFW  Optimization Values}
    \label{NavaroNGC2683}
     \begin{tabular}{|r|r|}
     \hline
      \textbf{Parameter}   & \textbf{Optimization Values}
      \\  \hline
   $\rho_s$   & $0.006\times 10^9$
\\  \hline
$r_s$&  20
\\  \hline
    \end{tabular}
  \end{center}
\end{table}
\begin{table}[h!]
  \begin{center}
    \caption{Burkert Optimization Values}
    \label{BuckertNGC2683}
     \begin{tabular}{|r|r|}
     \hline
      \textbf{Parameter}   & \textbf{Optimization Values}
      \\  \hline
     $\rho_0^B$  & $0.1\times 10^9$
\\  \hline
$r_0$&  6
\\  \hline
    \end{tabular}
  \end{center}
\end{table}
\begin{table}[h!]
  \begin{center}
    \caption{Einasto Optimization Values}
    \label{EinastoNGC2683}
    \begin{tabular}{|r|r|}
     \hline
      \textbf{Parameter}   & \textbf{Optimization Values}
      \\  \hline
     $\rho_e$  & $0.0061\times 10^9$
\\  \hline
$r_e$ & 10
\\  \hline
$n_e$ & 0.27
\\  \hline
    \end{tabular}
  \end{center}
\end{table}
\begin{table}[h!]
\centering \caption{Physical assessment of collisional DM
parameters (NGC2683).}
\begin{tabular}{lcc}
\hline
Parameter & Value & Physical Verdict \\
\hline
$\gamma_0$ & $1.0001$ & Essentially isothermal  \\
$\delta_\gamma$ & $1.2\times10^{-9}$ & Negligible   \\
$r_\gamma$ & $1.5\ \mathrm{Kpc}$ & Transition radius in inner halo  \\
$K_0$ & $1.605\times10^{4}$ & Large entropy \\
$r_c$ & $0.5\ \mathrm{Kpc}$ & Small core scale- plausible for the inner region but compact for an extended halo \\
$p$ & $0.01$ & Extremely shallow $K(r)$ slope\\
\hline
Overall &-& Physically consistent but functionally nearly isothermal \\
\hline
\end{tabular}
\label{EVALUATIONNGC2683}
\end{table}


\subsection{The Galaxy NGC2841 Non-viable}

\begin{figure}[h!]
\centering
\includegraphics[width=20pc]{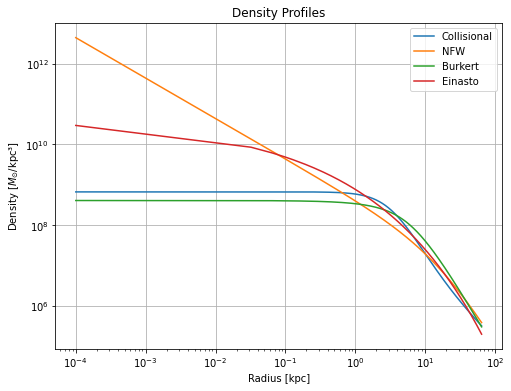}
\caption{The density of the collisional DM model (\ref{tanhmodel})
for the galaxy NGC2841, as a function of the radius.}
\label{NGC2841dens}
\end{figure}
\begin{figure}[h!]
\centering
\includegraphics[width=20pc]{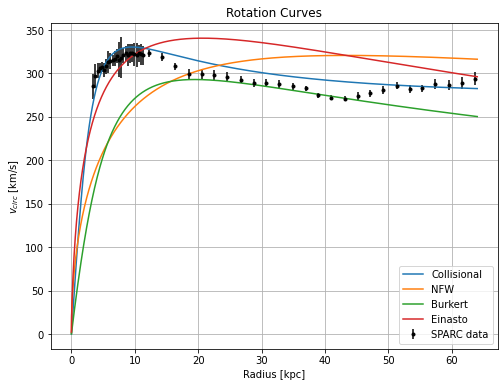}
\caption{The predicted rotation curves after using an optimization
for the collisional DM model (\ref{tanhmodel}), versus the SPARC
observational data for the galaxy NGC2841. We also plotted the
optimized curves for the NFW model, the Burkert model and the
Einasto model.} \label{NGC2841}
\end{figure}
\begin{figure}[h!]
\centering
\includegraphics[width=20pc]{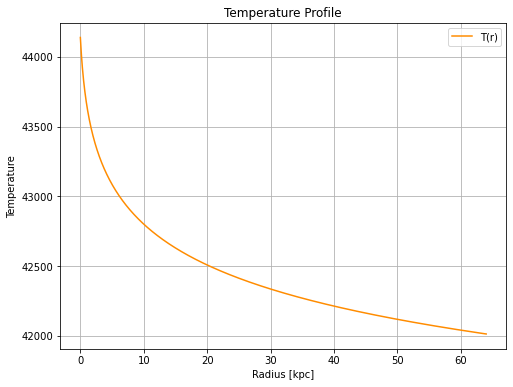}
\caption{The temperature as a function of the radius for the
collisional DM model (\ref{tanhmodel}) for the galaxy NGC2841.}
\label{NGC2841temp}
\end{figure}
\begin{figure}[h!]
\centering
\includegraphics[width=20pc]{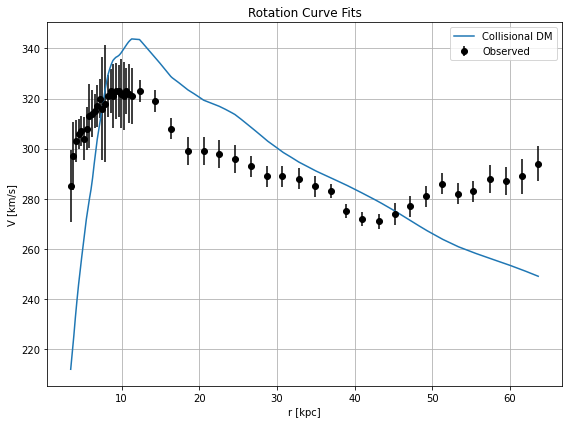}
\caption{The predicted rotation curves after using an optimization
for the collisional DM model (\ref{tanhmodel}), versus the
extended SPARC observational data for the galaxy NGC2841. The
model includes the rotation curves from all the components of the
galaxy, including gas and disk velocities, along with the
collisional DM model.} \label{extendedNGC2841}
\end{figure}

For this galaxy, we shall choose $\rho_0=6.7\times
10^8$$M_{\odot}/\mathrm{Kpc}^{3}$. Galaxy NGC\,2841 is a massive
unbarred spiral galaxy (type SA(r)b / SAa), with tightly
wound/flocculent arms and a prominent nuclear bulge. Its distance
is $D = 14.1 \pm 1.5 \,\mathrm{Mpc}$. In Figs. \ref{NGC2841dens},
\ref{NGC2841} and \ref{NGC2841temp} we present the density of the
collisional DM model, the predicted rotation curves after using an
optimization for the collisional DM model (\ref{tanhmodel}),
versus the SPARC observational data and the temperature parameter
as a function of the radius respectively. As it can be seen, the
SIDM model produces non-viable rotation curves incompatible with
the SPARC data. Also in Tables \ref{collNGC2841},
\ref{NavaroNGC2841}, \ref{BuckertNGC2841} and \ref{EinastoNGC2841}
we present the optimization values for the SIDM model, and the
other DM profiles. Also in Table \ref{EVALUATIONNGC2841} we
present the overall evaluation of the SIDM model for the galaxy at
hand. The resulting phenomenology is non-viable.
\begin{table}[h!]
  \begin{center}
    \caption{Collisional Dark Matter Optimization Values}
    \label{collNGC2841}
     \begin{tabular}{|r|r|}
     \hline
      \textbf{Parameter}   & \textbf{Optimization Values}
      \\  \hline
     $\delta_{\gamma} $ &0.0000000012
\\  \hline
$\gamma_0 $ & 1.0001 \\ \hline $K_0$ ($M_{\odot} \,
\mathrm{Kpc}^{-3} \, (\mathrm{km/s})^{2}$)& 44050  \\ \hline
    \end{tabular}
  \end{center}
\end{table}
\begin{table}[h!]
  \begin{center}
    \caption{NFW  Optimization Values}
    \label{NavaroNGC2841}
     \begin{tabular}{|r|r|}
     \hline
      \textbf{Parameter}   & \textbf{Optimization Values}
      \\  \hline
   $\rho_s$   & $0.022\times 10^9$
\\  \hline
$r_s$&  20
\\  \hline
    \end{tabular}
  \end{center}
\end{table}
\begin{table}[h!]
  \begin{center}
    \caption{Burkert Optimization Values}
    \label{BuckertNGC2841}
     \begin{tabular}{|r|r|}
     \hline
      \textbf{Parameter}   & \textbf{Optimization Values}
      \\  \hline
     $\rho_0^B$  & $0.41\times 10^9$
\\  \hline
$r_0$&  6
\\  \hline
    \end{tabular}
  \end{center}
\end{table}
\begin{table}[h!]
  \begin{center}
    \caption{Einasto Optimization Values}
    \label{EinastoNGC2841}
    \begin{tabular}{|r|r|}
     \hline
      \textbf{Parameter}   & \textbf{Optimization Values}
      \\  \hline
     $\rho_e$  & $0.025\times 10^9$
\\  \hline
$r_e$ & 10
\\  \hline
$n_e$ & 0.27
\\  \hline
    \end{tabular}
  \end{center}
\end{table}
\begin{table}[h!]
\centering \caption{Physical assessment of collisional DM
parameters (NGC2841).}
\begin{tabular}{lcc}
\hline
Parameter & Value & Physical Verdict \\
\hline
$\gamma_0$ & $1.0001$ & Essentially isothermal  \\
$\delta_\gamma$ & $1.2\times10^{-9}$ & Negligible   \\
$r_\gamma$ & $1.5\ \mathrm{Kpc}$ & Transition radius in inner halo  \\
$K_0$ & $4.405\times10^{4}$ & Very large entropy  \\
$r_c$ & $0.5\ \mathrm{Kpc}$ & Small core scale   \\
$p$ & $0.01$ & Extremely shallow $K(r)$ slope\\
\hline
Overall &-& Physically consistent \\
\hline
\end{tabular}
\label{EVALUATIONNGC2841}
\end{table}
Now the extended picture including the rotation velocity from the
other components of the galaxy, such as the disk and gas, makes
the collisional DM model viable for this galaxy. In Fig.
\ref{extendedNGC2841} we present the combined rotation curves
including the other components of the galaxy along with the
collisional matter. As it can be seen, the extended collisional DM
model is marginally viable.
Also in Table \ref{evaluationextendedNGC2841} we present the
values of the free parameters of the collisional DM model for
which the maximum compatibility with the SPARC data comes for the
galaxy NGC2841.
\begin{table}[h!]
\centering \caption{Physical assessment of Extended collisional DM
parameters for NGC2841.}
\begin{tabular}{lcc}
\hline
Parameter & Value & Physical Verdict \\
\hline
$\gamma_0$ & 1.1270 & Mildly above isothermal; low central pressure, stable inner halo \\
$\delta_\gamma$ & 0.0000 & No radial variation; $\gamma(r)$ is constant (pure polytropic EoS) \\
$K_0$ & 3000 & Moderate entropy  \\
$ml_{\text{disk}}$ & 1.2568 & Relatively high stellar $M/L$  \\
$ml_{\text{bulge}}$ & 0.0000 & No bulge component in the fit; disk-dominated assumption \\
\hline
Overall &-& Physically plausible \\
\hline
\end{tabular}
\label{evaluationextendedNGC2841}
\end{table}

\subsection{The Galaxy NGC2903 Marginally viable by only two}

\begin{figure}[h!]
\centering
\includegraphics[width=20pc]{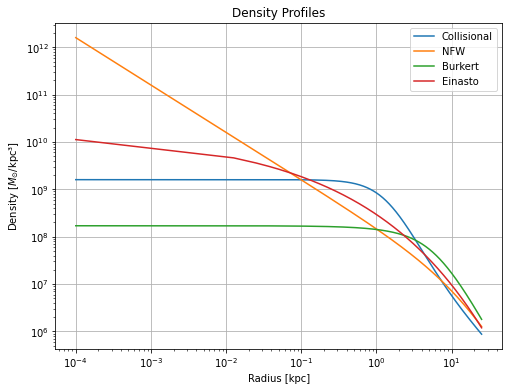}
\caption{The density of the collisional DM model (\ref{tanhmodel})
for the galaxy NGC2903, as a function of the radius.}
\label{NGC2903dens}
\end{figure}

\begin{figure}[h!]
\centering
\includegraphics[width=20pc]{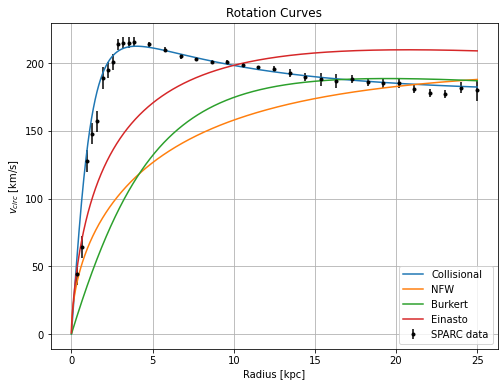}
\caption{The predicted rotation curves after using an optimization
for the collisional DM model (\ref{tanhmodel}), versus the SPARC
observational data for the galaxy NGC2903. We also plotted the
optimized curves for the NFW model, the Burkert model and the
Einasto model.} \label{NGC2903}
\end{figure}

\begin{figure}[h!]
\centering
\includegraphics[width=20pc]{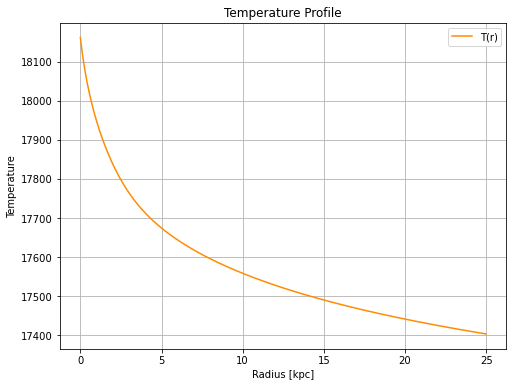}
\caption{The temperature as a function of the radius for the
collisional DM model (\ref{tanhmodel}) for the galaxy NGC2903.}
\label{NGC2903temp}
\end{figure}

\begin{figure}[h!]
\centering
\includegraphics[width=20pc]{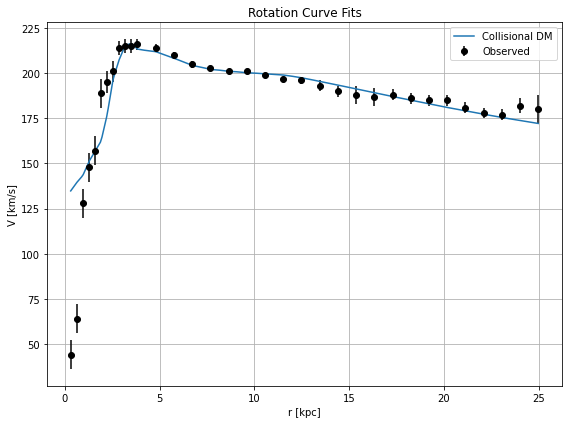}
\caption{The predicted rotation curves after using an optimization
for the collisional DM model (\ref{tanhmodel}), versus the
extended SPARC observational data for the galaxy NGC2903. The
model includes the rotation curves from all the components of the
galaxy, including gas and disk velocities, along with the
collisional DM model.} \label{extendedNGC2903}
\end{figure}

For this galaxy, we shall choose $\rho_0=1.6\times
10^9$$M_{\odot}/\mathrm{Kpc}^{3}$. Galaxy  NGC\,2903 is a barred
spiral galaxy (type SBbc or SAB(rs)bc), a fairly massive field
spiral with a bar, ring-like features, and active star formation
especially in its central region. Its distance is $D \sim 8.9$ to
$9.3 \;\mathrm{Mpc}$. In Figs. \ref{NGC2903dens}, \ref{NGC2903}
and \ref{NGC2903temp} we present the density of the collisional DM
model, the predicted rotation curves after using an optimization
for the collisional DM model (\ref{tanhmodel}), versus the SPARC
observational data and the temperature parameter as a function of
the radius respectively. As it can be seen, the SIDM model
produces viable rotation curves compatible with the SPARC data.
Also in Tables \ref{collNGC2903}, \ref{NavaroNGC2903},
\ref{BuckertNGC2903} and \ref{EinastoNGC2903} we present the
optimization values for the SIDM model, and the other DM profiles.
Also in Table \ref{EVALUATIONNGC2903} we present the overall
evaluation of the SIDM model for the galaxy at hand. The resulting
phenomenology is viable.
\begin{table}[h!]
  \begin{center}
    \caption{Collisional Dark Matter Optimization Values}
    \label{collNGC2903}
     \begin{tabular}{|r|r|}
     \hline
      \textbf{Parameter}   & \textbf{Optimization Values}
      \\  \hline
     $\delta_{\gamma} $ & 0.0000000012
\\  \hline
$\gamma_0 $ & 1.0001 \\ \hline $K_0$ ($M_{\odot} \,
\mathrm{Kpc}^{-3} \, (\mathrm{km/s})^{2}$)& 18100  \\ \hline
    \end{tabular}
  \end{center}
\end{table}
\begin{table}[h!]
  \begin{center}
    \caption{NFW  Optimization Values}
    \label{NavaroNGC2903}
     \begin{tabular}{|r|r|}
     \hline
      \textbf{Parameter}   & \textbf{Optimization Values}
      \\  \hline
   $\rho_s$   & $0.008\times 10^9$
\\  \hline
$r_s$&  20
\\  \hline
    \end{tabular}
  \end{center}
\end{table}
\begin{table}[h!]
  \begin{center}
    \caption{Burkert Optimization Values}
    \label{BuckertNGC2903}
     \begin{tabular}{|r|r|}
     \hline
      \textbf{Parameter}   & \textbf{Optimization Values}
      \\  \hline
     $\rho_0^B$  & $0.17\times 10^9$
\\  \hline
$r_0$&  6
\\  \hline
    \end{tabular}
  \end{center}
\end{table}
\begin{table}[h!]
  \begin{center}
    \caption{Einasto Optimization Values}
    \label{EinastoNGC2903}
    \begin{tabular}{|r|r|}
     \hline
      \textbf{Parameter}   & \textbf{Optimization Values}
      \\  \hline
     $\rho_e$  & $0.0095\times 10^9$
\\  \hline
$r_e$ & 10
\\  \hline
$n_e$ & 0.27
\\  \hline
    \end{tabular}
  \end{center}
\end{table}
\begin{table}[h!]
\centering \caption{Physical assessment of collisional DM
parameters (NGC2903).}
\begin{tabular}{lcc}
\hline
Parameter & Value & Physical Verdict \\
\hline
$\gamma_0$ & $1.0001$ & Essentially isothermal  \\
$\delta_\gamma$ & $12\times10^{-9}$ & Negligible \\
$r_\gamma$ & $1.5\ \mathrm{Kpc}$ & Transition radius in inner halo \\
$K_0$ & $1.81\times10^{4}$ & Moderate-to-large entropy   \\
$r_c$ & $0.5\ \mathrm{Kpc}$ & Small core scale \\
$p$ & $0.01$ & Extremely shallow $K(r)$ slope; $K$ practically constant across 0-25 Kpc \\
\hline
Overall &-& Physically consistent but functionally nearly isothermal \\
\hline
\end{tabular}
\label{EVALUATIONNGC2903}
\end{table}
Now the extended picture including the rotation velocity from the
other components of the galaxy, such as the disk and gas, makes
the collisional DM model viable for this galaxy. In Fig.
\ref{extendedNGC2903} we present the combined rotation curves
including the other components of the galaxy along with the
collisional matter. As it can be seen, the extended collisional DM
model is viable.
Also in Table \ref{evaluationextendedNGC2903} we present the
values of the free parameters of the collisional DM model for
which the maximum compatibility with the SPARC data comes for the
galaxy NGC2903.
\begin{table}[h!]
\centering \caption{Physical assessment of Extended collisional DM
parameters for galaxy NGC2903.}
\begin{tabular}{lcc}
\hline
Parameter & Value & Physical Verdict \\
\hline
$\gamma_0$ & 1.12295295 & Slightly above isothermal \\
$\delta_\gamma$ & 0.05153307 & Small-to-moderate radial variation \\
$K_0$ & 3000 & Moderate entropy scale; consistent with intermediate-mass spiral halos \\
$ml_{disk}$ & 0.68098091 & Moderate disk M/L \\
$ml_{bulge}$ & 0.00000000 & No bulge contribution \\
\hline
Overall &-& Physically plausible \\
\hline
\end{tabular}
\label{evaluationextendedNGC2903}
\end{table}



\subsection{The Galaxy NGC2955 Non-viable, Extended Non-viable}

\begin{figure}[h!]
\centering
\includegraphics[width=20pc]{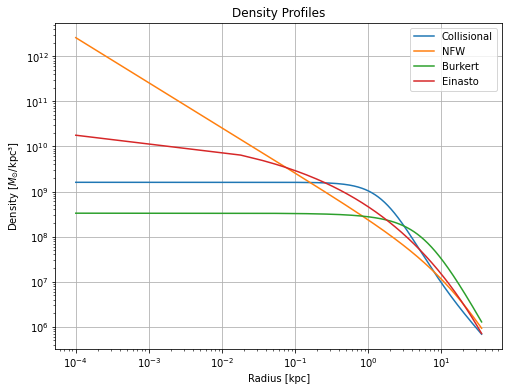}
\caption{The density of the collisional DM model (\ref{tanhmodel})
for the galaxy NGC2955, as a function of the radius.}
\label{NGC2955dens}
\end{figure}
\begin{figure}[h!]
\centering
\includegraphics[width=20pc]{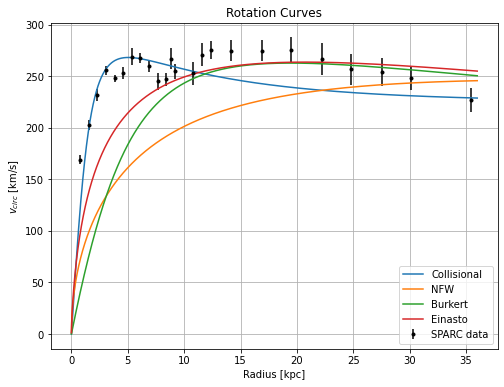}
\caption{The predicted rotation curves after using an optimization
for the collisional DM model (\ref{tanhmodel}), versus the SPARC
observational data for the galaxy NGC2955. We also plotted the
optimized curves for the NFW model, the Burkert model and the
Einasto model.} \label{NGC2955}
\end{figure}
\begin{figure}[h!]
\centering
\includegraphics[width=20pc]{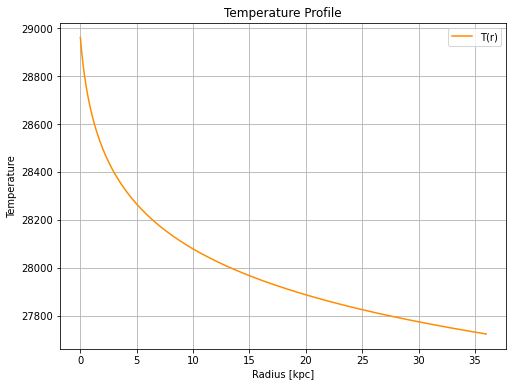}
\caption{The temperature as a function of the radius for the
collisional DM model (\ref{tanhmodel}) for the galaxy NGC2955.}
\label{NGC2955temp}
\end{figure}
\begin{figure}[h!]
\centering
\includegraphics[width=20pc]{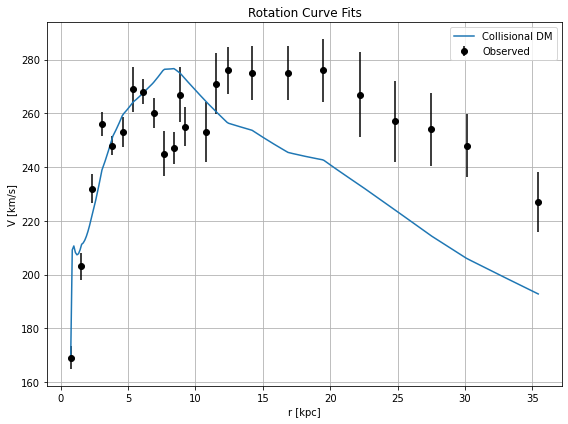}
\caption{The predicted rotation curves after using an optimization
for the collisional DM model (\ref{tanhmodel}), versus the
extended SPARC observational data for the galaxy NGC2955. The
model includes the rotation curves from all the components of the
galaxy, including gas and disk velocities, along with the
collisional DM model.} \label{extendedNGC2955}
\end{figure}

For this galaxy, we shall choose $\rho_0=1.6\times
10^9$$M_{\odot}/\mathrm{Kpc}^{3}$. The galaxy NGC\,2955 is a field
spiral galaxy of type Sb in the constellation Leo Minor. It is a
mid-sized ordinary spiral, with a ring/outer pseudo-ring structure
in its morphology. Its distance is $D \sim 95.7\ \mathrm{Mpc}$. In
Figs. \ref{NGC2955dens}, \ref{NGC2955} and \ref{NGC2955temp} we
present the density of the collisional DM model, the predicted
rotation curves after using an optimization for the collisional DM
model (\ref{tanhmodel}), versus the SPARC observational data and
the temperature parameter as a function of the radius
respectively. As it can be seen, the SIDM model produces
non-viable rotation curves incompatible with the SPARC data. Also
in Tables \ref{collNGC2955}, \ref{NavaroNGC2955},
\ref{BuckertNGC2955} and \ref{EinastoNGC2955} we present the
optimization values for the SIDM model, and the other DM profiles.
Also in Table \ref{EVALUATIONNGC2955} we present the overall
evaluation of the SIDM model for the galaxy at hand. The resulting
phenomenology is non-viable.
\begin{table}[h!]
  \begin{center}
    \caption{Collisional Dark Matter Optimization Values}
    \label{collNGC2955}
     \begin{tabular}{|r|r|}
     \hline
      \textbf{Parameter}   & \textbf{Optimization Values}
      \\  \hline
     $\delta_{\gamma} $ & 0.0000000012
\\  \hline
$\gamma_0 $ &  1.0001 \\ \hline $K_0$ ($M_{\odot} \,
\mathrm{Kpc}^{-3} \, (\mathrm{km/s})^{2}$)& 28950  \\ \hline
    \end{tabular}
  \end{center}
\end{table}
\begin{table}[h!]
  \begin{center}
    \caption{NFW  Optimization Values}
    \label{NavaroNGC2955}
     \begin{tabular}{|r|r|}
     \hline
      \textbf{Parameter}   & \textbf{Optimization Values}
      \\  \hline
   $\rho_s$   & $0.013\times 10^9$
\\  \hline
$r_s$&  20
\\  \hline
    \end{tabular}
  \end{center}
\end{table}
\begin{table}[h!]
  \begin{center}
    \caption{Burkert Optimization Values}
    \label{BuckertNGC2955}
     \begin{tabular}{|r|r|}
     \hline
      \textbf{Parameter}   & \textbf{Optimization Values}
      \\  \hline
     $\rho_0^B$  & $0.33\times 10^9$
\\  \hline
$r_0$&  6
\\  \hline
    \end{tabular}
  \end{center}
\end{table}
\begin{table}[h!]
  \begin{center}
    \caption{Einasto Optimization Values}
    \label{EinastoNGC2955}
    \begin{tabular}{|r|r|}
     \hline
      \textbf{Parameter}   & \textbf{Optimization Values}
      \\  \hline
     $\rho_e$  & $0.015\times 10^9$
\\  \hline
$r_e$ & 10
\\  \hline
$n_e$ & 0.27
\\  \hline
    \end{tabular}
  \end{center}
\end{table}
\begin{table}[h!]
\centering \caption{Physical assessment of collisional DM
parameters (NGC2955).}
\begin{tabular}{lcc}
\hline
Parameter & Value & Physical Verdict \\
\hline
$\gamma_0$ & $1.0001$ & Essentially isothermal  \\
$\delta_\gamma$ & $1.2\times10^{-9}$ & Negligible   \\
$r_\gamma$ & $1.5\ \mathrm{Kpc}$ & Transition radius in inner halo \\
$K_0$ & $2.89\times10^{4}$ & Large entropy   \\
$r_c$ & $0.5\ \mathrm{Kpc}$ & Small core scale \\
$p$ & $0.01$ & Extremely shallow $K(r)$ slope \\
\hline
Overall &-& Physically consistent but functionally nearly isothermal \\
\hline
\end{tabular}
\label{EVALUATIONNGC2955}
\end{table}
Now the extended picture including the rotation velocity from the
other components of the galaxy, such as the disk and gas, makes
the collisional DM model viable for this galaxy. In Fig.
\ref{extendedNGC2955} we present the combined rotation curves
including the other components of the galaxy along with the
collisional matter. As it can be seen, the extended collisional DM
model is non-viable.
Also in Table \ref{evaluationextendedNGC2955} we present the
values of the free parameters of the collisional DM model for
which the maximum compatibility with the SPARC data comes for the
galaxy NGC2955.
\begin{table}[h!]
\centering \caption{Physical assessment of Extended collisional DM
parameters for NGC2955.}
\begin{tabular}{lcc}
\hline
Parameter & Value & Physical Verdict \\
\hline
$\gamma_0$ & 1.37211021 & Significantly stiffer than isothermal \\
$\delta_\gamma$ & 1.04158659 & Very large radial variation\\
$K_0$ & 3000 & Moderate entropy   \\
$ml_{\text{disk}}$ & 0.92868947 & High but plausible disk $M/L$  \\
$ml_{\text{bulge}}$ & 0.63432131 & Reasonable bulge $M/L$ for a classical bulge \\
\hline
Overall &-& Marginal physical maximum compatibility \\
\hline
\end{tabular}
\label{evaluationextendedNGC2955}
\end{table}


\subsection{The Galaxy NGC2976}

\begin{figure}[h!]
\centering
\includegraphics[width=20pc]{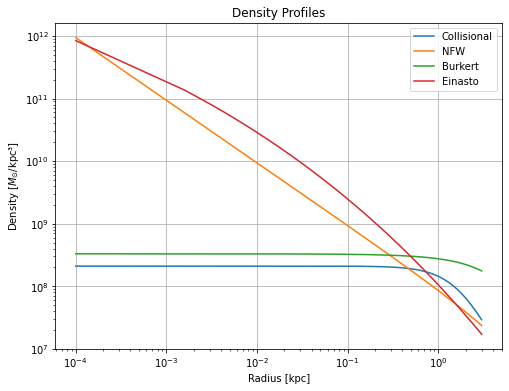}
\caption{The density of the collisional DM model (\ref{tanhmodel})
for the galaxy NGC2976, as a function of the radius.}
\label{NGC2976dens}
\end{figure}

\begin{figure}[h!]
\centering
\includegraphics[width=20pc]{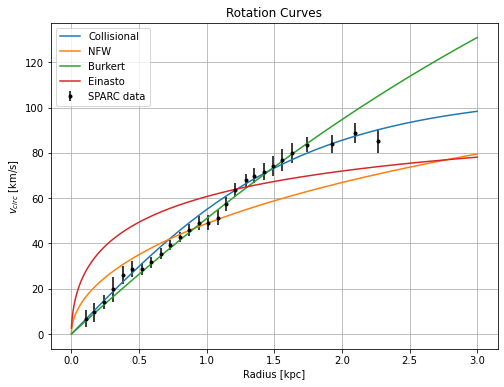}
\caption{The predicted rotation curves after using an optimization
for the collisional DM model (\ref{tanhmodel}), versus the SPARC
observational data for the galaxy NGC2976. We also plotted the
optimized curves for the NFW model, the Burkert model and the
Einasto model.} \label{NGC2976}
\end{figure}

\begin{figure}[h!]
\centering
\includegraphics[width=20pc]{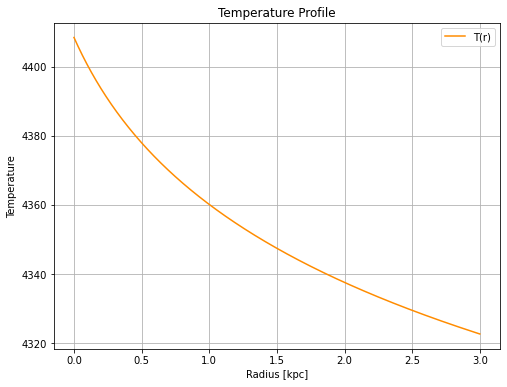}
\caption{The temperature as a function of the radius for the
collisional DM model (\ref{tanhmodel}) for the galaxy NGC2976.}
\label{NGC2976temp}
\end{figure}

For this galaxy, we shall choose $\rho_0=2.1\times
10^8$$M_{\odot}/\mathrm{Kpc}^{3}$. NGC 2976 is a peculiar dwarf
spiral galaxy classified as SAa, indicating an unbarred spiral
galaxy with tightly wound arms. It resides in the M81 Group,
approximately 3.4 Mpc from the Milky Way. In Figs.
\ref{NGC2976dens}, \ref{NGC2976} and \ref{NGC2976temp} we present
the density of the collisional DM model, the predicted rotation
curves after using an optimization for the collisional DM model
(\ref{tanhmodel}), versus the SPARC observational data and the
temperature parameter as a function of the radius respectively. As
it can be seen, the SIDM model produces viable rotation curves
compatible with the SPARC data. Also in Tables \ref{collNGC2976},
\ref{NavaroNGC2976}, \ref{BuckertNGC2976} and \ref{EinastoNGC2976}
we present the optimization values for the SIDM model, and the
other DM profiles. Also in Table \ref{EVALUATIONNGC2976} we
present the overall evaluation of the SIDM model for the galaxy at
hand. The resulting phenomenology is viable.
\begin{table}[h!]
  \begin{center}
    \caption{Collisional Dark Matter Optimization Values}
    \label{collNGC2976}
     \begin{tabular}{|r|r|}
     \hline
      \textbf{Parameter}   & \textbf{Optimization Values}
      \\  \hline
     $\delta_{\gamma} $ & 0.0000000012
\\  \hline
$\gamma_0 $ & 1.0001 \\ \hline $K_0$ ($M_{\odot} \,
\mathrm{Kpc}^{-3} \, (\mathrm{km/s})^{2}$)& 4400  \\ \hline
    \end{tabular}
  \end{center}
\end{table}
\begin{table}[h!]
  \begin{center}
    \caption{NFW  Optimization Values}
    \label{NavaroNGC2976}
     \begin{tabular}{|r|r|}
     \hline
      \textbf{Parameter}   & \textbf{Optimization Values}
      \\  \hline
   $\rho_s$   & $0.0047\times 10^9$
\\  \hline
$r_s$&  20
\\  \hline
    \end{tabular}
  \end{center}
\end{table}
\begin{table}[h!]
  \begin{center}
    \caption{Burkert Optimization Values}
    \label{BuckertNGC2976}
     \begin{tabular}{|r|r|}
     \hline
      \textbf{Parameter}   & \textbf{Optimization Values}
      \\  \hline
     $\rho_0^B$  & $0.33\times 10^9$
\\  \hline
$r_0$&  6
\\  \hline
    \end{tabular}
  \end{center}
\end{table}
\begin{table}[h!]
  \begin{center}
    \caption{Einasto Optimization Values}
    \label{EinastoNGC2976}
    \begin{tabular}{|r|r|}
     \hline
      \textbf{Parameter}   & \textbf{Optimization Values}
      \\  \hline
     $\rho_e$  & $0.0018\times 10^9$
\\  \hline
$r_e$ & 10
\\  \hline
$n_e$ & 0.11
\\  \hline
    \end{tabular}
  \end{center}
\end{table}
\begin{table}[h!]
\centering \caption{Physical assessment of collisional DM
parameters (NGC2976).}
\begin{tabular}{lcc}
\hline
Parameter & Value & Physical Verdict \\
\hline
$\gamma_0$ & $1.0001$ & Essentially isothermal  \\
$\delta_\gamma$ & $1.2\times10^{-9}$ & Negligible   \\
$r_\gamma$ & $1.5\ \mathrm{Kpc}$ & Transition radius in inner halo \\
$K_0$ & $4.40\times10^{3}$ & Moderate entropy   \\
$r_c$ & $0.5\ \mathrm{Kpc}$ & Small core/entropy radius \\
$p$ & $0.01$ & Extremely shallow $K(r)$ slope; $K$ effectively constant across domain \\
\hline
Overall &-& Physically consistent \\
\hline
\end{tabular}
\label{EVALUATIONNGC2976}
\end{table}


\subsection{The Galaxy NGC2998 Non-viable, Extended Marginally Viable}

\begin{figure}[h!]
\centering
\includegraphics[width=20pc]{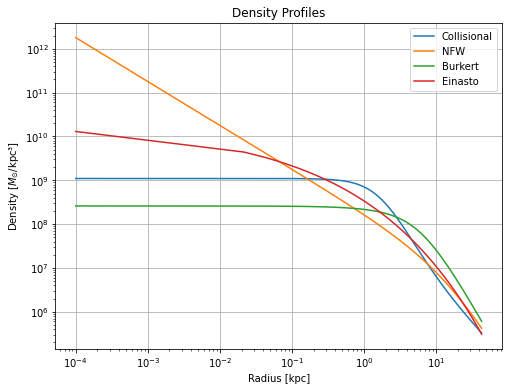}
\caption{The density of the collisional DM model (\ref{tanhmodel})
for the galaxy NGC2998, as a function of the radius.}
\label{NGC2998dens}
\end{figure}
\begin{figure}[h!]
\centering
\includegraphics[width=20pc]{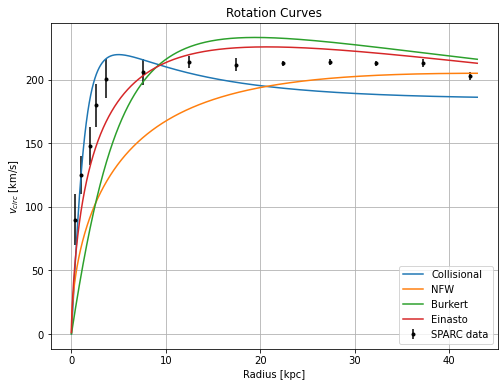}
\caption{The predicted rotation curves after using an optimization
for the collisional DM model (\ref{tanhmodel}), versus the SPARC
observational data for the galaxy NGC2998. We also plotted the
optimized curves for the NFW model, the Burkert model and the
Einasto model.} \label{NGC2998}
\end{figure}
\begin{figure}[h!]
\centering
\includegraphics[width=20pc]{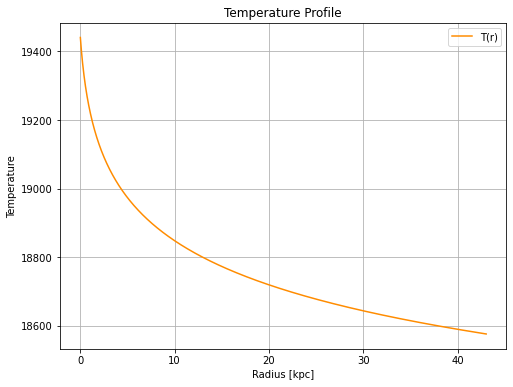}
\caption{The temperature as a function of the radius for the
collisional DM model (\ref{tanhmodel}) for the galaxy NGC2998.}
\label{NGC2998temp}
\end{figure}
\begin{figure}[h!]
\centering
\includegraphics[width=20pc]{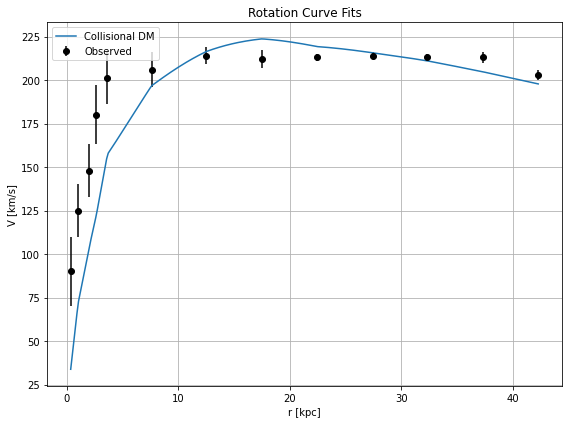}
\caption{The predicted rotation curves after using an optimization
for the collisional DM model (\ref{tanhmodel}), versus the
extended SPARC observational data for the galaxy NGC2998. The
model includes the rotation curves from all the components of the
galaxy, including gas and disk velocities, along with the
collisional DM model.} \label{extendedNGC2998}
\end{figure}

For this galaxy, we shall choose $\rho_0=1.1\times
10^9$$M_{\odot}/\mathrm{Kpc}^{3}$. NGC 2998 is a barred spiral
galaxy classified as SAB(rs)c, indicating an intermediate-type
spiral galaxy with a bar and loosely wound spiral arms. It is
situated in the constellation Ursa Major and lies approximately
67.4 Mpc from Milky Way. In Figs. \ref{NGC2998dens}, \ref{NGC2998}
and \ref{NGC2998temp} we present the density of the collisional DM
model, the predicted rotation curves after using an optimization
for the collisional DM model (\ref{tanhmodel}), versus the SPARC
observational data and the temperature parameter as a function of
the radius respectively. As it can be seen, the SIDM model
produces non-viable rotation curves incompatible with the SPARC
data. Also in Tables \ref{collNGC2998}, \ref{NavaroNGC2998},
\ref{BuckertNGC2998} and \ref{EinastoNGC2998} we present the
optimization values for the SIDM model, and the other DM profiles.
Also in Table \ref{EVALUATIONNGC2998} we present the overall
evaluation of the SIDM model for the galaxy at hand. The resulting
phenomenology is non-viable.
\begin{table}[h!]
  \begin{center}
    \caption{Collisional Dark Matter Optimization Values}
    \label{collNGC2998}
     \begin{tabular}{|r|r|}
     \hline
      \textbf{Parameter}   & \textbf{Optimization Values}
      \\  \hline
     $\delta_{\gamma} $ & 0.0000000012
\\  \hline
$\gamma_0 $ & 1.0001 \\ \hline $K_0$ ($M_{\odot} \,
\mathrm{Kpc}^{-3} \, (\mathrm{km/s})^{2}$)& 19400 \\ \hline
    \end{tabular}
  \end{center}
\end{table}
\begin{table}[h!]
  \begin{center}
    \caption{NFW  Optimization Values}
    \label{NavaroNGC2998}
     \begin{tabular}{|r|r|}
     \hline
      \textbf{Parameter}   & \textbf{Optimization Values}
      \\  \hline
   $\rho_s$   & $0.009\times 10^9$
\\  \hline
$r_s$&  20
\\  \hline
    \end{tabular}
  \end{center}
\end{table}
\begin{table}[h!]
  \begin{center}
    \caption{Burkert Optimization Values}
    \label{BuckertNGC2998}
     \begin{tabular}{|r|r|}
     \hline
      \textbf{Parameter}   & \textbf{Optimization Values}
      \\  \hline
     $\rho_0^B$  & $0.26\times 10^9$
\\  \hline
$r_0$&  6
\\  \hline
    \end{tabular}
  \end{center}
\end{table}
\begin{table}[h!]
  \begin{center}
    \caption{Einasto Optimization Values}
    \label{EinastoNGC2998}
    \begin{tabular}{|r|r|}
     \hline
      \textbf{Parameter}   & \textbf{Optimization Values}
      \\  \hline
     $\rho_e$  & $0.011\times 10^9$
\\  \hline
$r_e$ & 10
\\  \hline
$n_e$ & 0.27
\\  \hline
    \end{tabular}
  \end{center}
\end{table}
\begin{table}[h!]
\centering \caption{Physical assessment of collisional DM
parameters (NGC2998).}
\begin{tabular}{lcc}
\hline
Parameter & Value & Physical Verdict \\
\hline
$\gamma_0$ & $1.0001$ & Essentially isothermal  \\
$\delta_\gamma$ & $1.2\times10^{-9}$ & Negligible   \\
$r_\gamma$ & $1.5\ \mathrm{Kpc}$ & Transition radius in inner halo   \\
$K_0$ & $1.94\times10^{4}$ & Moderate-to-large entropy   \\
$r_c$ & $0.5\ \mathrm{Kpc}$ & Small core scale  \\
$p$ & $0.01$ & Extremely shallow $K(r)$ slope\\
\hline
Overall &-& Physically consistent but functionally nearly isothermal \\
\hline
\end{tabular}
\label{EVALUATIONNGC2998}
\end{table}
Now the extended picture including the rotation velocity from the
other components of the galaxy, such as the disk and gas, makes
the collisional DM model viable for this galaxy. In Fig.
\ref{extendedNGC2998} we present the combined rotation curves
including the other components of the galaxy along with the
collisional matter. As it can be seen, the extended collisional DM
model is marginally viable.
Also in Table \ref{evaluationextendedNGC2998} we present the
values of the free parameters of the collisional DM model for
which the maximum compatibility with the SPARC data comes for the
galaxy NGC2998.
\begin{table}[h!]
\centering \caption{Physical assessment of Extended collisional DM
parameters for NGC2998.}
\begin{tabular}{lcc}
\hline
Parameter & Value & Physical Verdict \\
\hline
$\gamma_0$ & 1.0885 & Near-isothermal core \\
$\delta_\gamma$ & 0.0000 & No radial variation \\
$K_0$ & 3000 & Moderate entropy   \\
$ml_{\text{disk}}$ & 0.7493 & Moderately high stellar $M/L$  \\
$ml_{\text{bulge}}$ & 0.0000 & No bulge component \\
\hline
Overall &-& Physically viable \\
\hline
\end{tabular}
\label{evaluationextendedNGC2998}
\end{table}


\subsection{The Galaxy NGC3109}

\begin{figure}[h!]
\centering
\includegraphics[width=20pc]{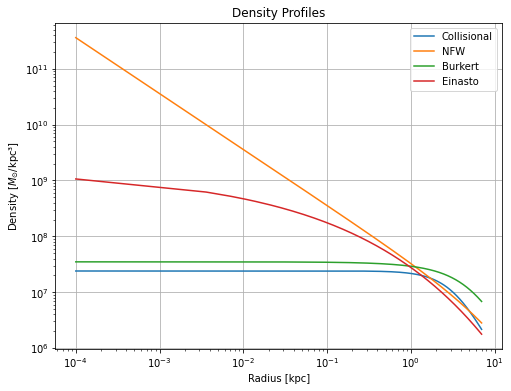}
\caption{The density of the collisional DM model (\ref{tanhmodel})
for the galaxy NGC3109, as a function of the radius.}
\label{NGC3109dens}
\end{figure}

\begin{figure}[h!]
\centering
\includegraphics[width=20pc]{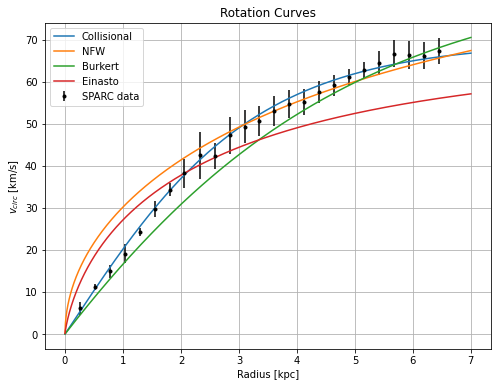}
\caption{The predicted rotation curves after using an optimization
for the collisional DM model (\ref{tanhmodel}), versus the SPARC
observational data for the galaxy NGC3109. We also plotted the
optimized curves for the NFW model, the Burkert model and the
Einasto model.} \label{NGC3109}
\end{figure}

\begin{figure}[h!]
\centering
\includegraphics[width=20pc]{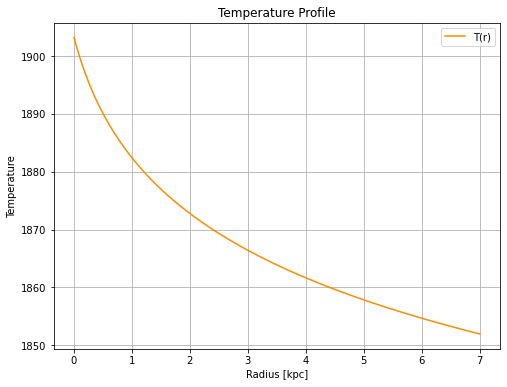}
\caption{The temperature as a function of the radius for the
collisional DM model (\ref{tanhmodel}) for the galaxy NGC3109.}
\label{NGC3109temp}
\end{figure}

For this galaxy, we shall choose $\rho_0=2.4\times
10^7$$M_{\odot}/\mathrm{Kpc}^{3}$. NGC3109 is a barred
Magellanic-type spiral galaxy, classified as SB(s)m, situated in
the constellation Hydra. It is located approximately 1.33 Mpc from
the Milky Way, placing it at the outskirts of the Local Group. In
Figs. \ref{NGC3109dens}, \ref{NGC3109} and \ref{NGC3109temp} we
present the density of the collisional DM model, the predicted
rotation curves after using an optimization for the collisional DM
model (\ref{tanhmodel}), versus the SPARC observational data and
the temperature parameter as a function of the radius
respectively. As it can be seen, the SIDM model produces viable
rotation curves compatible with the SPARC data. Also in Tables
\ref{collNGC3109}, \ref{NavaroNGC3109}, \ref{BuckertNGC3109} and
\ref{EinastoNGC3109} we present the optimization values for the
SIDM model, and the other DM profiles. Also in Table
\ref{EVALUATIONNGC3109} we present the overall evaluation of the
SIDM model for the galaxy at hand. The resulting phenomenology is
viable.
\begin{table}[h!]
  \begin{center}
    \caption{Collisional Dark Matter Optimization Values}
    \label{collNGC3109}
     \begin{tabular}{|r|r|}
     \hline
      \textbf{Parameter}   & \textbf{Optimization Values}
      \\  \hline
     $\delta_{\gamma} $ & 0.0000000012
\\  \hline
$\gamma_0 $ & 1.0001 \\ \hline $K_0$ ($M_{\odot} \,
\mathrm{Kpc}^{-3} \, (\mathrm{km/s})^{2}$)& 1900 \\ \hline
    \end{tabular}
  \end{center}
\end{table}
\begin{table}[h!]
  \begin{center}
    \caption{NFW  Optimization Values}
    \label{NavaroNGC3109}
     \begin{tabular}{|r|r|}
     \hline
      \textbf{Parameter}   & \textbf{Optimization Values}
      \\  \hline
   $\rho_s$   & $0.0018\times 10^9$
\\  \hline
$r_s$&  20
\\  \hline
    \end{tabular}
  \end{center}
\end{table}
\begin{table}[h!]
  \begin{center}
    \caption{Burkert Optimization Values}
    \label{BuckertNGC3109}
     \begin{tabular}{|r|r|}
     \hline
      \textbf{Parameter}   & \textbf{Optimization Values}
      \\  \hline
     $\rho_0^B$  & $0.035\times 10^9$
\\  \hline
$r_0$&  6
\\  \hline
    \end{tabular}
  \end{center}
\end{table}
\begin{table}[h!]
  \begin{center}
    \caption{Einasto Optimization Values}
    \label{EinastoNGC3109}
    \begin{tabular}{|r|r|}
     \hline
      \textbf{Parameter}   & \textbf{Optimization Values}
      \\  \hline
     $\rho_e$  & $0.0009\times 10^9$
\\  \hline
$r_e$ & 10
\\  \hline
$n_e$ & 0.11
\\  \hline
    \end{tabular}
  \end{center}
\end{table}
\begin{table}[h!]
\centering \caption{Physical assessment of collisional DM
parameters (NGC3109).}
\begin{tabular}{lcc}
\hline
Parameter & Value & Physical Verdict \\
\hline
$\gamma_0$ & $1.0001$ & Essentially isothermal  \\
$\delta_\gamma$ & $1.2\times10^{-9}$ & Negligible   \\
$r_\gamma$ & $1.5\ \mathrm{Kpc}$ & Transition radius in inner halo \\
$K_0$ & $1.90\times10^{3}$ & Modest entropy   \\
$r_c$ & $0.5\ \mathrm{Kpc}$ & Small core/entropy radius \\
$p$ & $0.01$ & Extremely shallow $K(r)$ slope\\
\hline
Overall &-& Physically consistent but functionally nearly isothermal \\
\hline
\end{tabular}
\label{EVALUATIONNGC3109}
\end{table}


\subsection{The Galaxy NGC3198 Marginally Viable}

\begin{figure}[h!]
\centering
\includegraphics[width=20pc]{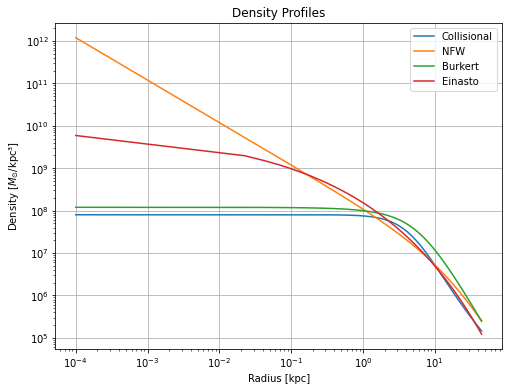}
\caption{The density of the collisional DM model (\ref{tanhmodel})
for the galaxy NGC3198, as a function of the radius.}
\label{NGC3198dens}
\end{figure}
\begin{figure}[h!]
\centering
\includegraphics[width=20pc]{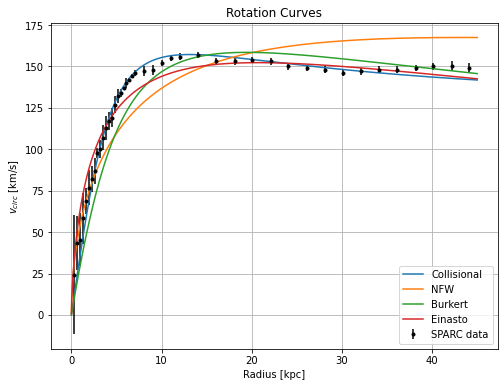}
\caption{The predicted rotation curves after using an optimization
for the collisional DM model (\ref{tanhmodel}), versus the SPARC
observational data for the galaxy NGC3198. We also plotted the
optimized curves for the NFW model, the Burkert model and the
Einasto model.} \label{NGC3198}
\end{figure}
\begin{figure}[h!]
\centering
\includegraphics[width=20pc]{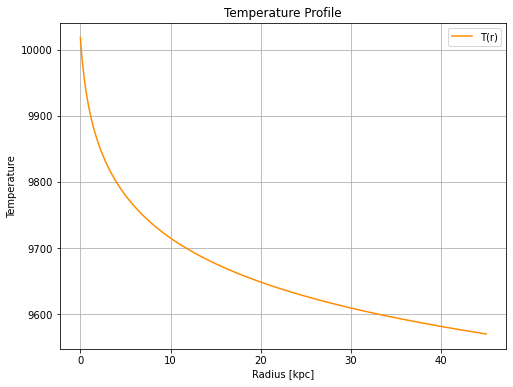}
\caption{The temperature as a function of the radius for the
collisional DM model (\ref{tanhmodel}) for the galaxy NGC3198.}
\label{NGC3198temp}
\end{figure}
\begin{figure}[h!]
\centering
\includegraphics[width=20pc]{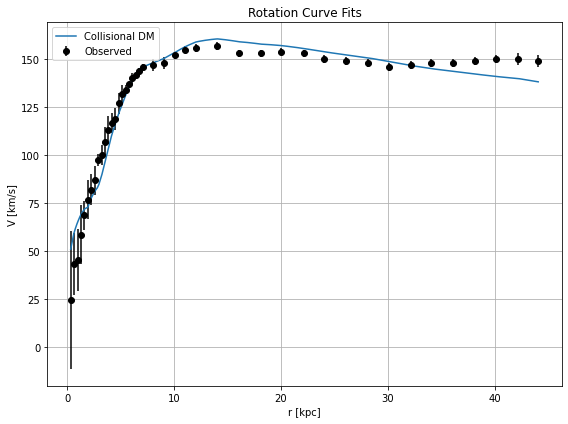}
\caption{The predicted rotation curves after using an optimization
for the collisional DM model (\ref{tanhmodel}), versus the
extended SPARC observational data for the galaxy NGC3198. The
model includes the rotation curves from all the components of the
galaxy, including gas and disk velocities, along with the
collisional DM model.} \label{extendedNGC3198}
\end{figure}

For this galaxy, we shall choose $\rho_0=8\times
10^7$$M_{\odot}/\mathrm{Kpc}^{3}$. NGC3198 is a barred spiral
galaxy classified as SB(rs)c, situated in the constellation Ursa
Major. It lies approximately 14.5 Mpc from the Milky Way. In Figs.
\ref{NGC3198dens}, \ref{NGC3198} and \ref{NGC3198temp} we present
the density of the collisional DM model, the predicted rotation
curves after using an optimization for the collisional DM model
(\ref{tanhmodel}), versus the SPARC observational data and the
temperature parameter as a function of the radius respectively. As
it can be seen, the SIDM model produces marginally viable rotation
curves compatible with the SPARC data. Also in Tables
\ref{collNGC3198}, \ref{NavaroNGC3198}, \ref{BuckertNGC3198} and
\ref{EinastoNGC3198} we present the optimization values for the
SIDM model, and the other DM profiles. Also in Table
\ref{EVALUATIONNGC3198} we present the overall evaluation of the
SIDM model for the galaxy at hand. The resulting phenomenology is
marginally viable.
\begin{table}[h!]
  \begin{center}
    \caption{Collisional Dark Matter Optimization Values}
    \label{collNGC3198}
     \begin{tabular}{|r|r|}
     \hline
      \textbf{Parameter}   & \textbf{Optimization Values}
      \\  \hline
     $\delta_{\gamma} $ & 0.0000000012
\\  \hline
$\gamma_0 $ & 1.0001\\ \hline $K_0$ ($M_{\odot} \,
\mathrm{Kpc}^{-3} \, (\mathrm{km/s})^{2}$)& 10000  \\ \hline
    \end{tabular}
  \end{center}
\end{table}
\begin{table}[h!]
  \begin{center}
    \caption{NFW  Optimization Values}
    \label{NavaroNGC3198}
     \begin{tabular}{|r|r|}
     \hline
      \textbf{Parameter}   & \textbf{Optimization Values}
      \\  \hline
   $\rho_s$   & $0.006\times 10^9$
\\  \hline
$r_s$&  20
\\  \hline
    \end{tabular}
  \end{center}
\end{table}
\begin{table}[h!]
  \begin{center}
    \caption{Burkert Optimization Values}
    \label{BuckertNGC3198}
     \begin{tabular}{|r|r|}
     \hline
      \textbf{Parameter}   & \textbf{Optimization Values}
      \\  \hline
     $\rho_0^B$  & $0.12\times 10^9$
\\  \hline
$r_0$&  6
\\  \hline
    \end{tabular}
  \end{center}
\end{table}
\begin{table}[h!]
  \begin{center}
    \caption{Einasto Optimization Values}
    \label{EinastoNGC3198}
    \begin{tabular}{|r|r|}
     \hline
      \textbf{Parameter}   & \textbf{Optimization Values}
      \\  \hline
     $\rho_e$  & $0.005\times 10^9$
\\  \hline
$r_e$ & 10
\\  \hline
$n_e$ & 0.27
\\  \hline
    \end{tabular}
  \end{center}
\end{table}
\begin{table}[h!]
\centering \caption{Physical assessment of collisional DM
parameters (NGC3198).}
\begin{tabular}{lcc}
\hline
Parameter & Value & Physical Verdict \\
\hline
$\gamma_0$ & $1.0001$ & Essentially isothermal  \\
$\delta_\gamma$ & $1.2\times10^{-9}$ & Negligible   \\
$r_\gamma$ & $1.5\ \mathrm{Kpc}$ & Transition radius in inner halo   \\
$K_0$ & $1.00\times10^{4}$ & Large entropy   \\
$r_c$ & $0.5\ \mathrm{Kpc}$ & Small core/entropy radius  \\
$p$ & $0.01$ & Extremely shallow $K(r)$ slope \\
\hline
Overall &-& Physically consistent but functionally nearly isothermal \\
\hline
\end{tabular}
\label{EVALUATIONNGC3198}
\end{table}
Now the extended picture including the rotation velocity from the
other components of the galaxy, such as the disk and gas, makes
the collisional DM model viable for this galaxy. In Fig.
\ref{extendedNGC3198} we present the combined rotation curves
including the other components of the galaxy along with the
collisional matter. As it can be seen, the extended collisional DM
model is viable.
Also in Table \ref{evaluationextendedNGC3198} we present the
values of the free parameters of the collisional DM model for
which the maximum compatibility with the SPARC data comes for the
galaxy NGC3198.
\begin{table}[h!]
\centering \caption{Physical assessment of Extended collisional DM
parameters for galaxy NGC3198.}
\begin{tabular}{lcc}
\hline
Parameter & Value & Physical Verdict \\
\hline
$\gamma_0$ & 1.05656567 & Slightly above isothermal \\
$\delta_\gamma$ & 0.00001 & No radial variation  \\
$K_0$ & 3000 & Moderate entropy scale \\
$ml_{disk}$ & 0.80314979 & Moderate-to-high disk M/L \\
$ml_{bulge}$ & 0.00000000 & No bulge contribution  \\
\hline
Overall &-& Physically plausible \\
\hline
\end{tabular}
\label{evaluationextendedNGC3198}
\end{table}

\subsection{The Galaxy NGC3521 Marginally Viable}

\begin{figure}[h!]
\centering
\includegraphics[width=20pc]{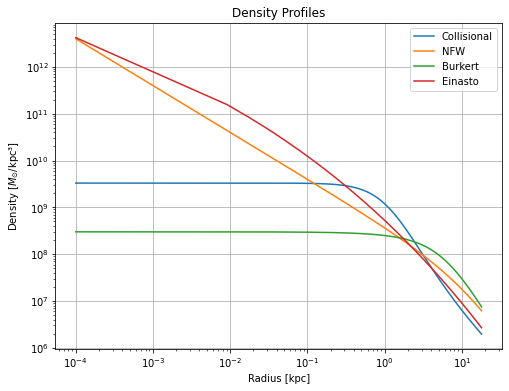}
\caption{The density of the collisional DM model (\ref{tanhmodel})
for the galaxy NGC3521, as a function of the radius.}
\label{NGC3521dens}
\end{figure}
\begin{figure}[h!]
\centering
\includegraphics[width=20pc]{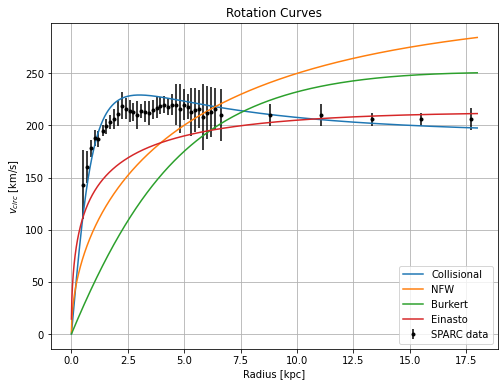}
\caption{The predicted rotation curves after using an optimization
for the collisional DM model (\ref{tanhmodel}), versus the SPARC
observational data for the galaxy NGC3521. We also plotted the
optimized curves for the NFW model, the Burkert model and the
Einasto model.} \label{NGC3521}
\end{figure}
\begin{figure}[h!]
\centering
\includegraphics[width=20pc]{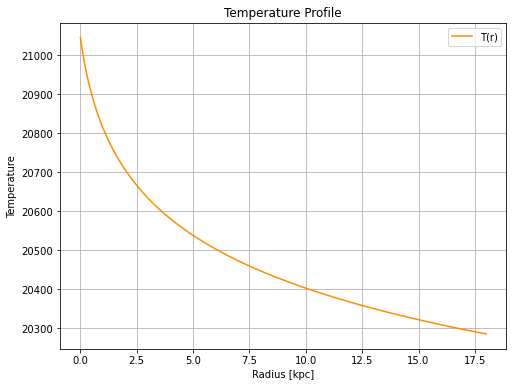}
\caption{The temperature as a function of the radius for the
collisional DM model (\ref{tanhmodel}) for the galaxy NGC3521.}
\label{NGC3521temp}
\end{figure}
\begin{figure}[h!]
\centering
\includegraphics[width=20pc]{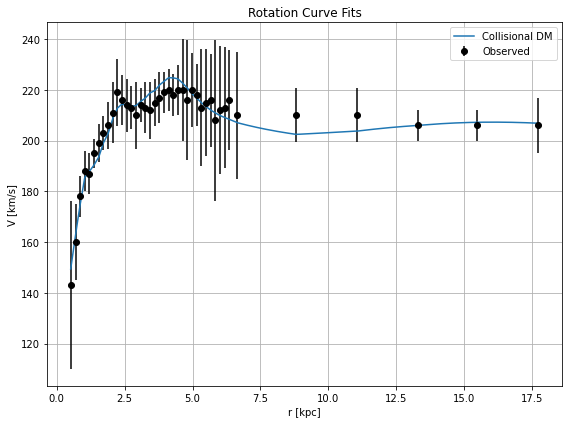}
\caption{The predicted rotation curves after using an optimization
for the collisional DM model (\ref{tanhmodel}), versus the
extended SPARC observational data for the galaxy NGC3521. The
model includes the rotation curves from all the components of the
galaxy, including gas and disk velocities, along with the
collisional DM model.} \label{extendedNGC3521}
\end{figure}

For this galaxy, we shall choose $\rho_0=3.3\times
10^9$$M_{\odot}/\mathrm{Kpc}^{3}$. NGC3521 is an intermediate
spiral galaxy of morphological type SAB(rs)bc. It is flocculent in
structure. Its distance is $D \sim 10.7\ \mathrm{Mpc}$. In Figs.
\ref{NGC3521dens}, \ref{NGC3521} and \ref{NGC3521temp} we present
the density of the collisional DM model, the predicted rotation
curves after using an optimization for the collisional DM model
(\ref{tanhmodel}), versus the SPARC observational data and the
temperature parameter as a function of the radius respectively. As
it can be seen, the SIDM model produces marginally viable rotation
curves compatible with the SPARC data. Also in Tables
\ref{collNGC3521}, \ref{NavaroNGC3521}, \ref{BuckertNGC3521} and
\ref{EinastoNGC3521} we present the optimization values for the
SIDM model, and the other DM profiles. Also in Table
\ref{EVALUATIONNGC3521} we present the overall evaluation of the
SIDM model for the galaxy at hand. The resulting phenomenology is
marginally viable.
Optimization values:
\begin{table}[h!]
  \begin{center}
    \caption{Collisional Dark Matter Optimization Values}
    \label{collNGC3521}
     \begin{tabular}{|r|r|}
     \hline
      \textbf{Parameter}   & \textbf{Optimization Values}
      \\  \hline
     $\delta_{\gamma} $ & 0.0000000012
\\  \hline
$\gamma_0 $ & 1.0001  \\ \hline $K_0$ ($M_{\odot} \,
\mathrm{Kpc}^{-3} \, (\mathrm{km/s})^{2}$)& 21000 \\ \hline
    \end{tabular}
  \end{center}
\end{table}
\begin{table}[h!]
  \begin{center}
    \caption{NFW  Optimization Values}
    \label{NavaroNGC3521}
     \begin{tabular}{|r|r|}
     \hline
      \textbf{Parameter}   & \textbf{Optimization Values}
      \\  \hline
   $\rho_s$   & $0.02\times 10^9$
\\  \hline
$r_s$&  20
\\  \hline
    \end{tabular}
  \end{center}
\end{table}
\begin{table}[h!]
  \begin{center}
    \caption{Burkert Optimization Values}
    \label{BuckertNGC3521}
     \begin{tabular}{|r|r|}
     \hline
      \textbf{Parameter}   & \textbf{Optimization Values}
      \\  \hline
     $\rho_0^B$  & $0.3\times 10^9$
\\  \hline
$r_0$&  6
\\  \hline
    \end{tabular}
  \end{center}
\end{table}
\begin{table}[h!]
  \begin{center}
    \caption{Einasto Optimization Values}
    \label{EinastoNGC3521}
    \begin{tabular}{|r|r|}
     \hline
      \textbf{Parameter}   & \textbf{Optimization Values}
      \\  \hline
     $\rho_e$  & $0.009\times 10^9$
\\  \hline
$r_e$ & 10
\\  \hline
$n_e$ & 0.11
\\  \hline
    \end{tabular}
  \end{center}
\end{table}
\begin{table}[h!]
\centering \caption{Physical assessment of collisional DM
parameters (NGC3521).}
\begin{tabular}{lcc}
\hline
Parameter & Value & Physical Verdict \\
\hline
$\gamma_0$ & $1.0001$ & Essentially isothermal  \\
$\delta_\gamma$ & $1.2\times10^{-9}$ & Negligible   \\
$r_\gamma$ & $1.5\ \mathrm{Kpc}$ & Transition radius set in inner halo   \\
$K_0$ & $2.10\times10^{4}$ & Large entropy  \\
$r_c$ & $0.5\ \mathrm{Kpc}$ & Small core scale   \\
$p$ & $0.01$ & Extremely shallow $K(r)$ slope; $K$ practically constant across 0-18 Kpc \\
\hline
Overall &-& Physically consistent but functionally nearly isothermal \\
\hline
\end{tabular}
\label{EVALUATIONNGC3521}
\end{table}
Now the extended picture including the rotation velocity from the
other components of the galaxy, such as the disk and gas, makes
the collisional DM model viable for this galaxy. In Fig.
\ref{extendedNGC3521} we present the combined rotation curves
including the other components of the galaxy along with the
collisional matter. As it can be seen, the extended collisional DM
model is viable.
Also in Table \ref{evaluationextendedNGC3521} we present the
values of the free parameters of the collisional DM model for
which the maximum compatibility with the SPARC data comes for the
galaxy NGC3521.
\begin{table}[h!]
\centering \caption{Physical assessment of Extended collisional DM
parameters for galaxy NGC3521.}
\begin{tabular}{lcc}
\hline
Parameter & Value & Physical Verdict \\
\hline
$\gamma_0$ & 1.09663580 & Slightly above isothermal \\
$\delta_\gamma$ & 0.00168535 & Very small radial variation \\
$K_0$ & 3000 & Moderate entropy scale; consistent with intermediate-mass spiral halos \\
$ml_{disk}$ & 0.76246054 & Moderate disk M/L \\
$ml_{bulge}$ & 0.00000000 & No bulge contribution \\
\hline
Overall &-& Physically plausible \\
\hline
\end{tabular}
\label{evaluationextendedNGC3521}
\end{table}

\subsection{The Galaxy NGC3726 Marginally Viable, Extended Marginally Too}

\begin{figure}[h!]
\centering
\includegraphics[width=20pc]{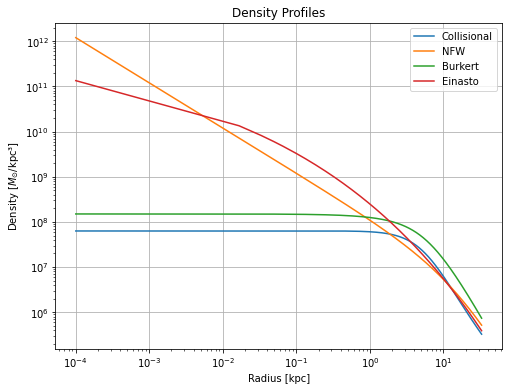}
\caption{The density of the collisional DM model (\ref{tanhmodel})
for the galaxy NGC3726, as a function of the radius.}
\label{NGC3726dens}
\end{figure}
\begin{figure}[h!]
\centering
\includegraphics[width=20pc]{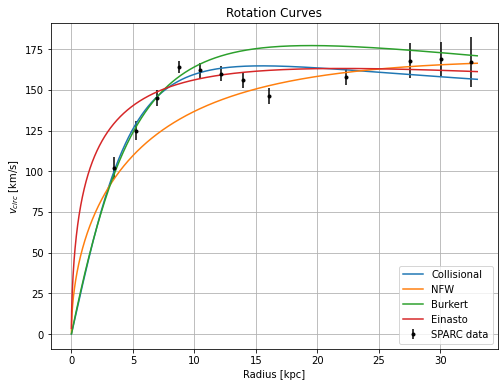}
\caption{The predicted rotation curves after using an optimization
for the collisional DM model (\ref{tanhmodel}), versus the SPARC
observational data for the galaxy NGC3726. We also plotted the
optimized curves for the NFW model, the Burkert model and the
Einasto model.} \label{NGC3726}
\end{figure}
\begin{figure}[h!]
\centering
\includegraphics[width=20pc]{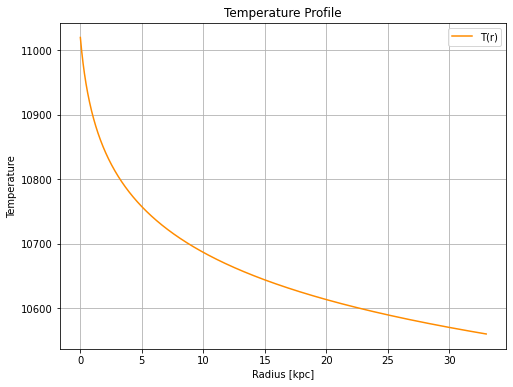}
\caption{The temperature as a function of the radius for the
collisional DM model (\ref{tanhmodel}) for the galaxy NGC3726.}
\label{NGC3726temp}
\end{figure}
\begin{figure}[h!]
\centering
\includegraphics[width=20pc]{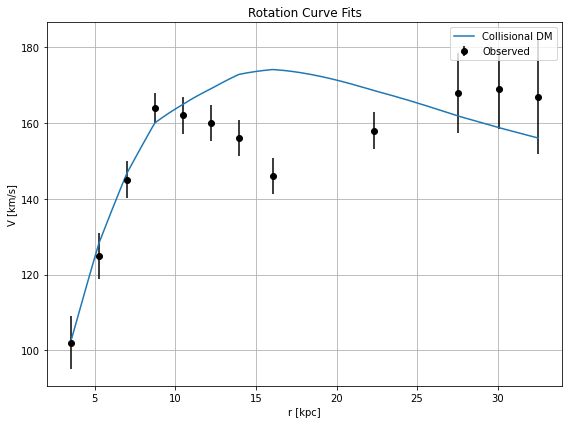}
\caption{The predicted rotation curves after using an optimization
for the collisional DM model (\ref{tanhmodel}), versus the
extended SPARC observational data for the galaxy NGC3726. The
model includes the rotation curves from all the components of the
galaxy, including gas and disk velocities, along with the
collisional DM model.} \label{extendedNGC3726}
\end{figure}

For this galaxy, we shall choose $\rho_0=6.3\times
10^7$$M_{\odot}/\mathrm{Kpc}^{3}$. NGC\,3726 is a barred spiral
galaxy of morphological type SAB(r)c, with a small central bar and
inner ring structure, medium inclination, with well-defined spiral
arms and a massive dark matter halo. Its distance is $D \sim 14.3
\pm 3.4\ \mathrm{Mpc}$. In Figs. \ref{NGC3726dens}, \ref{NGC3726}
and \ref{NGC3726temp} we present the density of the collisional DM
model, the predicted rotation curves after using an optimization
for the collisional DM model (\ref{tanhmodel}), versus the SPARC
observational data and the temperature parameter as a function of
the radius respectively. As it can be seen, the SIDM model
produces marginally viable rotation curves compatible with the
SPARC data. Also in Tables \ref{collNGC3726}, \ref{NavaroNGC3726},
\ref{BuckertNGC3726} and \ref{EinastoNGC3726} we present the
optimization values for the SIDM model, and the other DM profiles.
Also in Table \ref{EVALUATIONNGC3726} we present the overall
evaluation of the SIDM model for the galaxy at hand. The resulting
phenomenology is marginally viable.
\begin{table}[h!]
  \begin{center}
    \caption{Collisional Dark Matter Optimization Values}
    \label{collNGC3726}
     \begin{tabular}{|r|r|}
     \hline
      \textbf{Parameter}   & \textbf{Optimization Values}
      \\  \hline
     $\delta_{\gamma} $ & 0.0000000012
\\  \hline
$\gamma_0 $ &  1.0001 \\ \hline $K_0$ ($M_{\odot} \,
\mathrm{Kpc}^{-3} \, (\mathrm{km/s})^{2}$)& 11000  \\ \hline
    \end{tabular}
  \end{center}
\end{table}
\begin{table}[h!]
  \begin{center}
    \caption{NFW  Optimization Values}
    \label{NavaroNGC3726}
     \begin{tabular}{|r|r|}
     \hline
      \textbf{Parameter}   & \textbf{Optimization Values}
      \\  \hline
   $\rho_s$   & $0.006\times 10^9$
\\  \hline
$r_s$&  20
\\  \hline
    \end{tabular}
  \end{center}
\end{table}
\begin{table}[h!]
  \begin{center}
    \caption{Burkert Optimization Values}
    \label{BuckertNGC3726}
     \begin{tabular}{|r|r|}
     \hline
      \textbf{Parameter}   & \textbf{Optimization Values}
      \\  \hline
     $\rho_0^B$  & $0.15\times 10^9$
\\  \hline
$r_0$&  6
\\  \hline
    \end{tabular}
  \end{center}
\end{table}
\begin{table}[h!]
  \begin{center}
    \caption{Einasto Optimization Values}
    \label{EinastoNGC3726}
    \begin{tabular}{|r|r|}
     \hline
      \textbf{Parameter}   & \textbf{Optimization Values}
      \\  \hline
     $\rho_e$  & $0.0055\times 10^9$
\\  \hline
$r_e$ & 10
\\  \hline
$n_e$ & 0.17
\\  \hline
    \end{tabular}
  \end{center}
\end{table}
\begin{table}[h!]
\centering \caption{Physical assessment of collisional DM
parameters (NGC3726).}
\begin{tabular}{lcc}
\hline
Parameter & Value & Physical Verdict \\
\hline
$\gamma_0$ & $1.0001$ & Essentially isothermal  \\
$\delta_\gamma$ & $1.2\times10^{-9}$ & Negligible   \\
$r_\gamma$ & $1.5\ \mathrm{Kpc}$ & Transition radius set in inner halo   \\
$K_0$ & $1.1\times10^{4}$ & Moderate entropy \\
$r_c$ & $0.5\ \mathrm{Kpc}$ & Small core scale \\
$p$ & $0.01$ & Extremely shallow $K(r)$ slope \\
\hline
Overall &-& Physically plausible; inner halo nearly isothermal \\
\hline
\end{tabular}
\label{EVALUATIONNGC3726}
\end{table}
Now the extended picture including the rotation velocity from the
other components of the galaxy, such as the disk and gas, makes
the collisional DM model viable for this galaxy. In Fig.
\ref{extendedNGC3726} we present the combined rotation curves
including the other components of the galaxy along with the
collisional matter. As it can be seen, the extended collisional DM
model is viable.
Also in Table \ref{evaluationextendedNGC3726} we present the
values of the free parameters of the collisional DM model for
which the maximum compatibility with the SPARC data comes for the
galaxy NGC3726.
\begin{table}[h!]
\centering \caption{Physical assessment of Extended collisional DM
parameters for galaxy NGC3726.}
\begin{tabular}{lcc}
\hline
Parameter & Value & Physical Verdict \\
\hline
$\gamma_0$ & 1.06600000 & Slightly above isothermal \\
$\delta_\gamma$ & 0.0000144775 & Negligible radial variation  \\
$K_0$ & 3000 & Moderate entropy scale; consistent with intermediate-mass spiral halos \\
$ml_{disk}$ & 0.63620167 & Moderate disk M/L; disk contributes noticeably to inner rotation curve but DM still important \\
$ml_{bulge}$ & 0.00000000 & No bulge contribution  \\
\hline
Overall &-& Physically plausible \\
\hline
\end{tabular}
\label{evaluationextendedNGC3726}
\end{table}

\subsection{The Galaxy NGC3741 Non-viable Dwarf-One of the Few Cases}

\begin{figure}[h!]
\centering
\includegraphics[width=20pc]{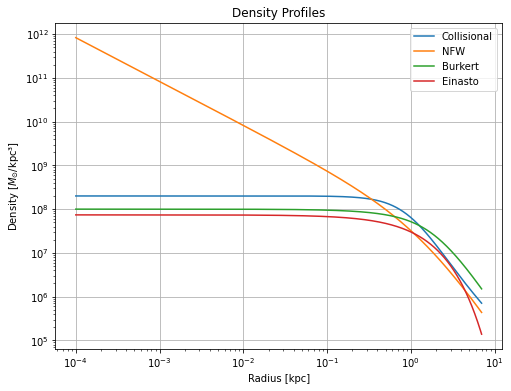}
\caption{The density of the collisional DM model (\ref{tanhmodel})
for the galaxy NGC3741, as a function of the radius.}
\label{NGC3741dens}
\end{figure}

\begin{figure}[h!]
\centering
\includegraphics[width=20pc]{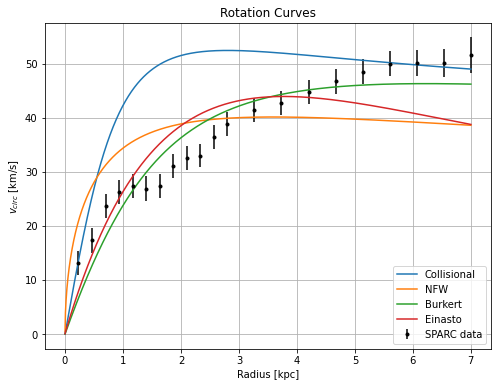}
\caption{The predicted rotation curves after using an optimization
for the collisional DM model (\ref{tanhmodel}), versus the SPARC
observational data for the galaxy NGC3741. We also plotted the
optimized curves for the NFW model, the Burkert model and the
Einasto model.} \label{NGC3741}
\end{figure}

\begin{figure}[h!]
\centering
\includegraphics[width=20pc]{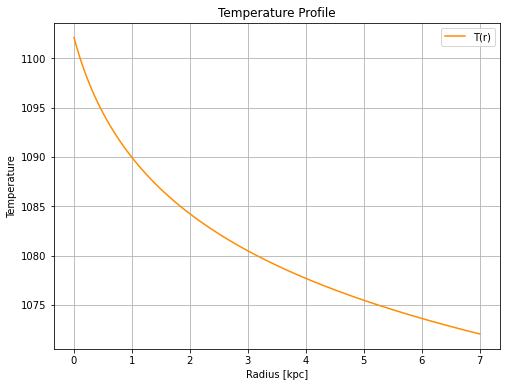}
\caption{The temperature as a function of the radius for the
collisional DM model (\ref{tanhmodel}) for the galaxy NGC3741.}
\label{NGC3741temp}
\end{figure}

\begin{figure}[h!]
\centering
\includegraphics[width=20pc]{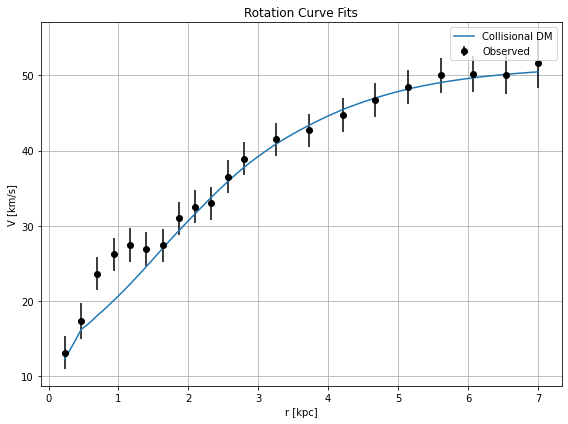}
\caption{The predicted rotation curves after using an optimization
for the collisional DM model (\ref{tanhmodel}), versus the
extended SPARC observational data for the galaxy NGC3741. The
model includes the rotation curves from all the components of the
galaxy, including gas and disk velocities, along with the
collisional DM model.} \label{extendedNGC3741}
\end{figure}

For this galaxy, we shall choose $\rho_0=1\times
10^8$$M_{\odot}/\mathrm{Kpc}^{3}$. NGC3741 is a dwarf irregular
galaxy of type ImIII/BCD, located in the constellation Ursa Major.
The galaxy is characterized by a compact stellar component and an
exceptionally extended neutral hydrogen HI disk. The HI disk is
strongly warped but symmetrically distributed, extending
approximately 7 Kpc from the center. This makes NGC3741 one of the
most gas-rich and dark matter-dominated galaxies known. In Figs.
\ref{NGC3741dens}, \ref{NGC3741} and \ref{NGC3741temp} we present
the density of the collisional DM model, the predicted rotation
curves after using an optimization for the collisional DM model
(\ref{tanhmodel}), versus the SPARC observational data and the
temperature parameter as a function of the radius respectively. As
it can be seen, the SIDM model produces non-viable rotation curves
incompatible with the SPARC data. Also in Tables
\ref{collNGC3741}, \ref{NavaroNGC3741}, \ref{BuckertNGC3741} and
\ref{EinastoNGC3741} we present the optimization values for the
SIDM model, and the other DM profiles. Also in Table
\ref{EVALUATIONNGC3741} we present the overall evaluation of the
SIDM model for the galaxy at hand. The resulting phenomenology is
non-viable.
\begin{table}[h!]
  \begin{center}
    \caption{Collisional Dark Matter Optimization Values}
    \label{collNGC3741}
     \begin{tabular}{|r|r|}
     \hline
      \textbf{Parameter}   & \textbf{Optimization Values}
      \\  \hline
     $\delta_{\gamma} $ & 0.0000000012
\\  \hline
$\gamma_0 $ & 1.0001 \\ \hline $K_0$ ($M_{\odot} \,
\mathrm{Kpc}^{-3} \, (\mathrm{km/s})^{2}$)& 1100  \\ \hline
    \end{tabular}
  \end{center}
\end{table}
\begin{table}[h!]
  \begin{center}
    \caption{NFW  Optimization Values}
    \label{NavaroNGC3741}
     \begin{tabular}{|r|r|}
     \hline
      \textbf{Parameter}   & \textbf{Optimization Values}
      \\  \hline
   $\rho_s$   & $0.001\times 10^9$
\\  \hline
$r_s$&  20
\\  \hline
    \end{tabular}
  \end{center}
\end{table}
\begin{table}[h!]
  \begin{center}
    \caption{Burkert Optimization Values}
    \label{BuckertNGC3741}
     \begin{tabular}{|r|r|}
     \hline
      \textbf{Parameter}   & \textbf{Optimization Values}
      \\  \hline
     $\rho_0^B$  & $0.025\times 10^9$
\\  \hline
$r_0$&  6
\\  \hline
    \end{tabular}
  \end{center}
\end{table}
\begin{table}[h!]
  \begin{center}
    \caption{Einasto Optimization Values}
    \label{EinastoNGC3741}
    \begin{tabular}{|r|r|}
     \hline
      \textbf{Parameter}   & \textbf{Optimization Values}
      \\  \hline
     $\rho_e$  & $0.0055\times 10^9$
\\  \hline
$r_e$ & 10
\\  \hline
$n_e$ & 0.17
\\  \hline
    \end{tabular}
  \end{center}
\end{table}
\begin{table}[h!]
\centering \caption{Physical assessment of collisional DM
parameters (NGC3741).}
\begin{tabular}{lcc}
\hline
Parameter & Value & Physical Verdict \\
\hline
$\gamma_0$ & $1.0001$ & Essentially isothermal  \\
$\delta_\gamma$ & $1.2\times10^{-9}$ & Negligible   \\
$r_\gamma$ & $1.5\ \mathrm{Kpc}$ & Transition radius in inner halo \\
$K_0$ & $1.10\times10^{3}$ & Moderate entropy   \\
$r_c$ & $0.5\ \mathrm{Kpc}$ & Small core scale  \\
$p$ & $0.01$ & Extremely shallow $K(r)$ slope \\
\hline
Overall &-& Physically consistent but functionally nearly isothermal \\
\hline
\end{tabular}
\label{EVALUATIONNGC3741}
\end{table}
Now the extended picture including the rotation velocity from the
other components of the galaxy, such as the disk and gas, makes
the collisional DM model viable for this galaxy. In Fig.
\ref{extendedNGC3741} we present the combined rotation curves
including the other components of the galaxy along with the
collisional matter. As it can be seen, the extended collisional DM
model is non-viable.
Also in Table \ref{evaluationextendedNGC3741} we present the
values of the free parameters of the collisional DM model for
which the approximate maximum compatibility with the SPARC data
comes for the galaxy NGC3741.
\begin{table}[h!]
\centering \caption{Physical assessment of Extended collisional DM
parameters for NGC3741.}
\begin{tabular}{lcc}
\hline
Parameter & Value & Physical Verdict \\
\hline
$\gamma_0$ & 0.9378 & Slightly below isothermal  \\
$\delta_\gamma$ & 0.00763 & Very small radial variation \\
$K_0$ & 3000 & Moderate entropy  \\
$ml_{\text{disk}}$ & 0.9909 & Moderate-to-high stellar $M/L$ \\
$ml_{\text{bulge}}$ & 0.0000 & No bulge component; disk-dominated morphology \\
\hline
Overall &-& Physically plausible \\
\hline
\end{tabular}
\label{evaluationextendedNGC3741}
\end{table}

\subsection{The Galaxy NGC3877}

\begin{figure}[h!]
\centering
\includegraphics[width=20pc]{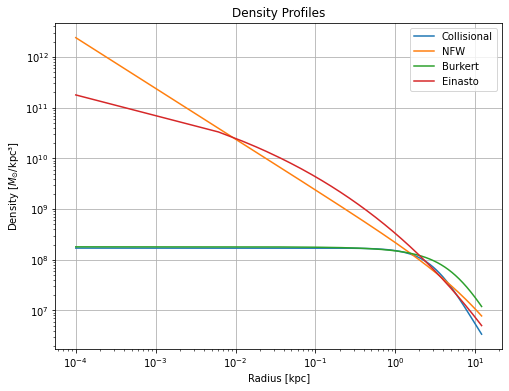}
\caption{The density of the collisional DM model (\ref{tanhmodel})
for the galaxy NGC3877, as a function of the radius.}
\label{NGC3877dens}
\end{figure}
\begin{figure}[h!]
\centering
\includegraphics[width=20pc]{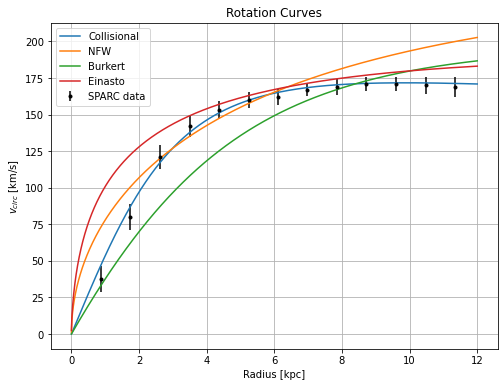}
\caption{The predicted rotation curves after using an optimization
for the collisional DM model (\ref{tanhmodel}), versus the SPARC
observational data for the galaxy NGC3877. We also plotted the
optimized curves for the NFW model, the Burkert model and the
Einasto model.} \label{NGC3877}
\end{figure}
\begin{figure}[h!]
\centering
\includegraphics[width=20pc]{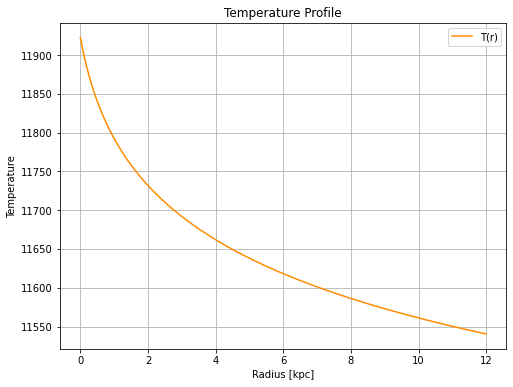}
\caption{The temperature as a function of the radius for the
collisional DM model (\ref{tanhmodel}) for the galaxy NGC3877.}
\label{NGC3877temp}
\end{figure}

For this galaxy, we shall choose $\rho_0=1.7\times
10^8$$M_{\odot}/\mathrm{Kpc}^{3}$. NGC3877 is an edge-on spiral
galaxy located in the constellation Ursa Major. It is classified
morphologically as an Sc-type spiral, seen nearly edge-on. It is a
member of the Ursa Major Cluster. Its distance is $D \sim 14.5\
\mathrm{Mpc}$. In Figs. \ref{NGC3877dens}, \ref{NGC3877} and
\ref{NGC3877temp} we present the density of the collisional DM
model, the predicted rotation curves after using an optimization
for the collisional DM model (\ref{tanhmodel}), versus the SPARC
observational data and the temperature parameter as a function of
the radius respectively. As it can be seen, the SIDM model
produces viable rotation curves compatible with the SPARC data.
Also in Tables \ref{collNGC3877}, \ref{NavaroNGC3877},
\ref{BuckertNGC3877} and \ref{EinastoNGC3877} we present the
optimization values for the SIDM model, and the other DM profiles.
Also in Table \ref{EVALUATIONNGC3877} we present the overall
evaluation of the SIDM model for the galaxy at hand. The resulting
phenomenology is viable.
\begin{table}[h!]
  \begin{center}
    \caption{Collisional Dark Matter Optimization Values}
    \label{collNGC3877}
     \begin{tabular}{|r|r|}
     \hline
      \textbf{Parameter}   & \textbf{Optimization Values}
      \\  \hline
     $\delta_{\gamma} $ & 0.0000000012
\\  \hline
$\gamma_0 $ & 1.0001 \\ \hline $K_0$ ($M_{\odot} \,
\mathrm{Kpc}^{-3} \, (\mathrm{km/s})^{2}$)& 11900  \\ \hline
    \end{tabular}
  \end{center}
\end{table}
\begin{table}[h!]
  \begin{center}
    \caption{NFW  Optimization Values}
    \label{NavaroNGC3877}
     \begin{tabular}{|r|r|}
     \hline
      \textbf{Parameter}   & \textbf{Optimization Values}
      \\  \hline
   $\rho_s$   & $0.012\times 10^9$
\\  \hline
$r_s$&  20
\\  \hline
    \end{tabular}
  \end{center}
\end{table}
\begin{table}[h!]
  \begin{center}
    \caption{Burkert Optimization Values}
    \label{BuckertNGC3877}
     \begin{tabular}{|r|r|}
     \hline
      \textbf{Parameter}   & \textbf{Optimization Values}
      \\  \hline
     $\rho_0^B$  & $0.18\times 10^9$
\\  \hline
$r_0$&  6
\\  \hline
    \end{tabular}
  \end{center}
\end{table}
\begin{table}[h!]
  \begin{center}
    \caption{Einasto Optimization Values}
    \label{EinastoNGC3877}
    \begin{tabular}{|r|r|}
     \hline
      \textbf{Parameter}   & \textbf{Optimization Values}
      \\  \hline
     $\rho_e$  & $0.0073\times 10^9$
\\  \hline
$r_e$ & 10
\\  \hline
$n_e$ & 0.17
\\  \hline
    \end{tabular}
  \end{center}
\end{table}
\begin{table}[h!]
\centering \caption{Physical assessment of collisional DM
parameters (NGC3877).}
\begin{tabular}{lcc}
\hline
Parameter & Value & Physical Verdict \\
\hline
$\gamma_0$ & $1.0001$ & Essentially isothermal  \\
$\delta_\gamma$ & $1.2\times10^{-9}$ & Negligible   \\
$r_\gamma$ & $1.5\ \mathrm{Kpc}$ & Transition radius in inner halo   \\
$K_0$ & $1.19\times10^{4}$ & Moderate entropy   \\
$r_c$ & $0.5\ \mathrm{Kpc}$ & Small core/entropy radius   \\
$p$ & $0.01$ & Extremely shallow $K(r)$ slope \\
\hline
Overall &-& Physically consistent but functionally nearly isothermal \\
\hline
\end{tabular}
\label{EVALUATIONNGC3877}
\end{table}



\subsection{The Galaxy NGC3917}

\begin{figure}[h!]
\centering
\includegraphics[width=20pc]{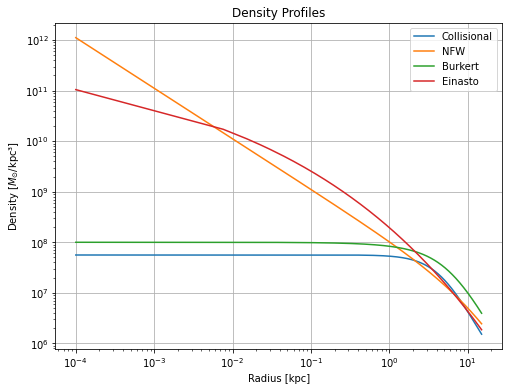}
\caption{The density of the collisional DM model (\ref{tanhmodel})
for the galaxy NGC3917, as a function of the radius.}
\label{NGC3917dens}
\end{figure}
\begin{figure}[h!]
\centering
\includegraphics[width=20pc]{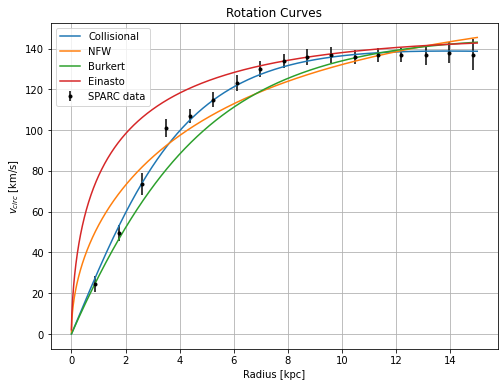}
\caption{The predicted rotation curves after using an optimization
for the collisional DM model (\ref{tanhmodel}), versus the SPARC
observational data for the galaxy NGC3917. We also plotted the
optimized curves for the NFW model, the Burkert model and the
Einasto model.} \label{NGC3917}
\end{figure}
\begin{figure}[h!]
\centering
\includegraphics[width=20pc]{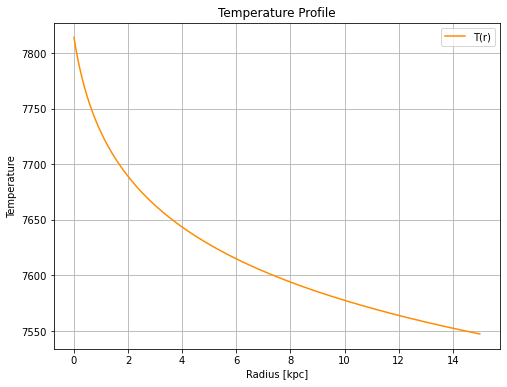}
\caption{The temperature as a function of the radius for the
collisional DM model (\ref{tanhmodel}) for the galaxy NGC3917.}
\label{NGC3917temp}
\end{figure}

For this galaxy, we shall choose $\rho_0=5.6\times
10^7$$M_{\odot}/\mathrm{Kpc}^{3}$. NGC3917 is a spiral galaxy of
Hubble type SAc/Scd (late-type spiral). The distance is $D \sim
13.6\,\mathrm{Mpc}$. In Figs. \ref{NGC3917dens}, \ref{NGC3917} and
\ref{NGC3917temp} we present the density of the collisional DM
model, the predicted rotation curves after using an optimization
for the collisional DM model (\ref{tanhmodel}), versus the SPARC
observational data and the temperature parameter as a function of
the radius respectively. As it can be seen, the SIDM model
produces viable rotation curves compatible with the SPARC data.
Also in Tables \ref{collNGC3917}, \ref{NavaroNGC3917},
\ref{BuckertNGC3917} and \ref{EinastoNGC3917} we present the
optimization values for the SIDM model, and the other DM profiles.
Also in Table \ref{EVALUATIONNGC3917} we present the overall
evaluation of the SIDM model for the galaxy at hand. The resulting
phenomenology is viable.
\begin{table}[h!]
  \begin{center}
    \caption{Collisional Dark Matter Optimization Values}
    \label{collNGC3917}
     \begin{tabular}{|r|r|}
     \hline
      \textbf{Parameter}   & \textbf{Optimization Values}
      \\  \hline
     $\delta_{\gamma} $ & 0.0000000012
\\  \hline
$\gamma_0 $ & 1.0001 \\ \hline $K_0$ ($M_{\odot} \,
\mathrm{Kpc}^{-3} \, (\mathrm{km/s})^{2}$)& 7800  \\ \hline
    \end{tabular}
  \end{center}
\end{table}
\begin{table}[h!]
  \begin{center}
    \caption{NFW  Optimization Values}
    \label{NavaroNGC3917}
     \begin{tabular}{|r|r|}
     \hline
      \textbf{Parameter}   & \textbf{Optimization Values}
      \\  \hline
   $\rho_s$   & $0.0056\times 10^9$
\\  \hline
$r_s$&  20
\\  \hline
    \end{tabular}
  \end{center}
\end{table}
\begin{table}[h!]
  \begin{center}
    \caption{Burkert Optimization Values}
    \label{BuckertNGC3917}
     \begin{tabular}{|r|r|}
     \hline
      \textbf{Parameter}   & \textbf{Optimization Values}
      \\  \hline
     $\rho_0^B$  & $0.1\times 10^9$
\\  \hline
$r_0$&  6
\\  \hline
    \end{tabular}
  \end{center}
\end{table}
\begin{table}[h!]
  \begin{center}
    \caption{Einasto Optimization Values}
    \label{EinastoNGC3917}
    \begin{tabular}{|r|r|}
     \hline
      \textbf{Parameter}   & \textbf{Optimization Values}
      \\  \hline
     $\rho_e$  & $0.0043\times 10^9$
\\  \hline
$r_e$ & 10
\\  \hline
$n_e$ & 0.17
\\  \hline
    \end{tabular}
  \end{center}
\end{table}
\begin{table}[h!]
\centering \caption{Physical assessment of collisional DM
parameters (NGC3917).}
\begin{tabular}{lcc}
\hline
Parameter & Value & Physical Verdict \\
\hline
$\gamma_0$ & $1.0001$ & Essentially isothermal  \\
$\delta_\gamma$ & $1.2\times10^{-9}$ & Negligible   \\
$r_\gamma$ & $1.5\ \mathrm{Kpc}$ & Transition radius in inner halo   \\
$K_0$ & $7.8\times10^{3}$ & Moderate entropy   \\
$r_c$ & $0.5\ \mathrm{Kpc}$ & Small core scale   \\
$p$ & $0.01$ & Extremely shallow $K(r)$ slope \\
\hline
Overall &-& Physically consistent but functionally nearly isothermal  \\
\hline
\end{tabular}
\label{EVALUATIONNGC3917}
\end{table}


\subsection{The Galaxy NGC3949}

\begin{figure}[h!]
\centering
\includegraphics[width=20pc]{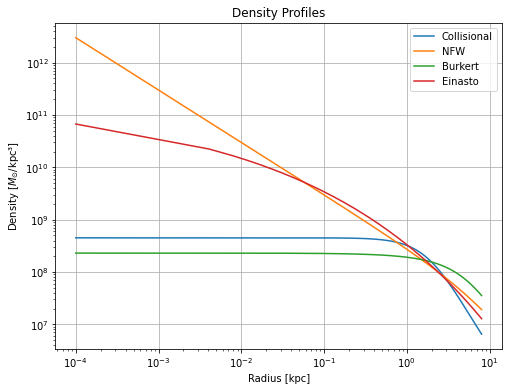}
\caption{The density of the collisional DM model (\ref{tanhmodel})
for the galaxy NGC3949, as a function of the radius.}
\label{NGC3949dens}
\end{figure}
\begin{figure}[h!]
\centering
\includegraphics[width=20pc]{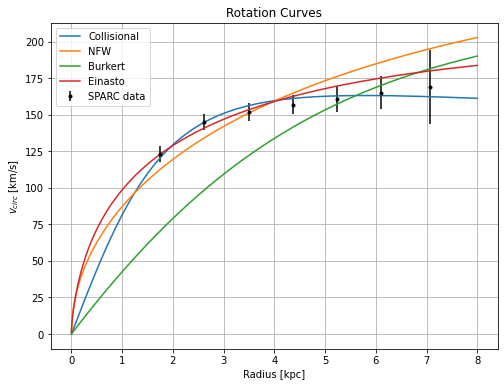}
\caption{The predicted rotation curves after using an optimization
for the collisional DM model (\ref{tanhmodel}), versus the SPARC
observational data for the galaxy NGC3949. We also plotted the
optimized curves for the NFW model, the Burkert model and the
Einasto model.} \label{NGC3949}
\end{figure}
\begin{figure}[h!]
\centering
\includegraphics[width=20pc]{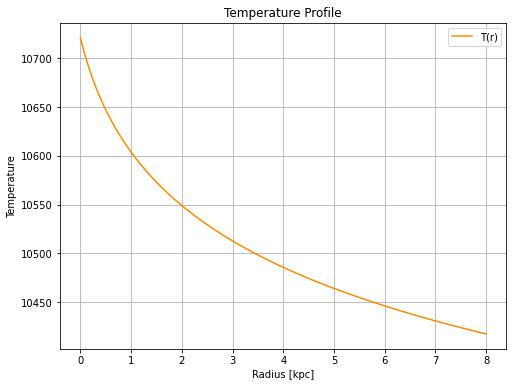}
\caption{The temperature as a function of the radius for the
collisional DM model (\ref{tanhmodel}) for the galaxy NGC3949.}
\label{NGC3949temp}
\end{figure}

For this galaxy, we shall choose $\rho_0=4.5\times
10^8$$M_{\odot}/\mathrm{Kpc}^{3}$. NGC3949 is an unbarred spiral
galaxy of Hubble type SA(s)bc. Its distance is $D\sim
14.9\;\mathrm{Mpc}$. In Figs. \ref{NGC3949dens}, \ref{NGC3949} and
\ref{NGC3949temp} we present the density of the collisional DM
model, the predicted rotation curves after using an optimization
for the collisional DM model (\ref{tanhmodel}), versus the SPARC
observational data and the temperature parameter as a function of
the radius respectively. As it can be seen, the SIDM model
produces viable rotation curves compatible with the SPARC data.
Also in Tables \ref{collNGC3949}, \ref{NavaroNGC3949},
\ref{BuckertNGC3949} and \ref{EinastoNGC3949} we present the
optimization values for the SIDM model, and the other DM profiles.
Also in Table \ref{EVALUATIONNGC3949} we present the overall
evaluation of the SIDM model for the galaxy at hand. The resulting
phenomenology is viable.
\begin{table}[h!]
  \begin{center}
    \caption{Collisional Dark Matter Optimization Values}
    \label{collNGC3949}
     \begin{tabular}{|r|r|}
     \hline
      \textbf{Parameter}   & \textbf{Optimization Values}
      \\  \hline
     $\delta_{\gamma} $ & 0.0000000012
\\  \hline
$\gamma_0 $ & 1.0001 \\ \hline $K_0$ ($M_{\odot} \,
\mathrm{Kpc}^{-3} \, (\mathrm{km/s})^{2}$)& 10700  \\ \hline
    \end{tabular}
  \end{center}
\end{table}
\begin{table}[h!]
  \begin{center}
    \caption{NFW  Optimization Values}
    \label{NavaroNGC3949}
     \begin{tabular}{|r|r|}
     \hline
      \textbf{Parameter}   & \textbf{Optimization Values}
      \\  \hline
   $\rho_s$   & $0.015\times 10^9$
\\  \hline
$r_s$&  9
\\  \hline
    \end{tabular}
  \end{center}
\end{table}
\begin{table}[h!]
  \begin{center}
    \caption{Burkert Optimization Values}
    \label{BuckertNGC3949}
     \begin{tabular}{|r|r|}
     \hline
      \textbf{Parameter}   & \textbf{Optimization Values}
      \\  \hline
     $\rho_0^B$  & $0.23\times 10^9$
\\  \hline
$r_0$&  2
\\  \hline
    \end{tabular}
  \end{center}
\end{table}
\begin{table}[h!]
  \begin{center}
    \caption{Einasto Optimization Values}
    \label{EinastoNGC3949}
    \begin{tabular}{|r|r|}
     \hline
      \textbf{Parameter}   & \textbf{Optimization Values}
      \\  \hline
     $\rho_e$  & $0.0083\times 10^9$
\\  \hline
$r_e$ & 10
\\  \hline
$n_e$ & 0.2
\\  \hline
    \end{tabular}
  \end{center}
\end{table}
\begin{table}[h!]
\centering \caption{Physical assessment of collisional DM
parameters (NGC3949).}
\begin{tabular}{lcc}
\hline
Parameter & Value & Physical Verdict \\
\hline
$\gamma_0$ & $1.0001$ & Essentially isothermal  \\
$\delta_\gamma$ & $1.2\times10^{-9}$ & Negligible variation \\
$r_\gamma$ & $1.5\ \mathrm{Kpc}$ & Reasonable transition radius   \\
$K_0$ & $1.07\times10^{4}$ & Enough central pressure support \\
$r_c$ & $0.5\ \mathrm{Kpc}$ & Small core scale - acceptable for inner halo structure \\
$p$ & $0.01$ & Very shallow decline of $K(r)$, practically constant entropy \\
\hline
Overall &-& Physically consistent \\
\hline
\end{tabular}
\label{EVALUATIONNGC3949}
\end{table}


\subsection{The Galaxy NGC3953}

\begin{figure}[h!]
\centering
\includegraphics[width=20pc]{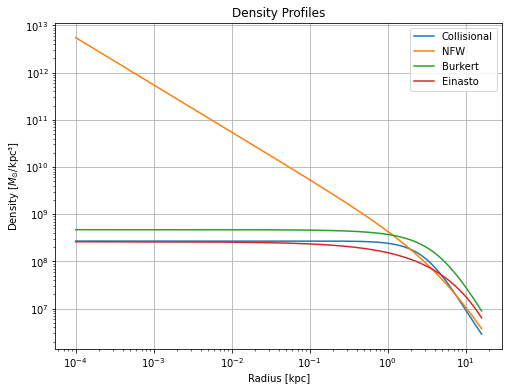}
\caption{The density of the collisional DM model (\ref{tanhmodel})
for the galaxy NGC3953, as a function of the radius.}
\label{NGC3953dens}
\end{figure}

\begin{figure}[h!]
\centering
\includegraphics[width=20pc]{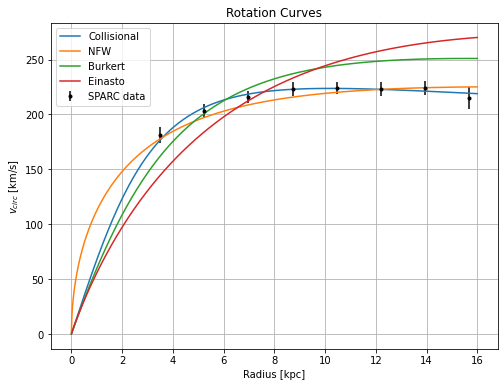}
\caption{The predicted rotation curves after using an optimization
for the collisional DM model (\ref{tanhmodel}), versus the SPARC
observational data for the galaxy NGC3953. We also plotted the
optimized curves for the NFW model, the Burkert model and the
Einasto model.} \label{NGC3953}
\end{figure}

\begin{figure}[h!]
\centering
\includegraphics[width=20pc]{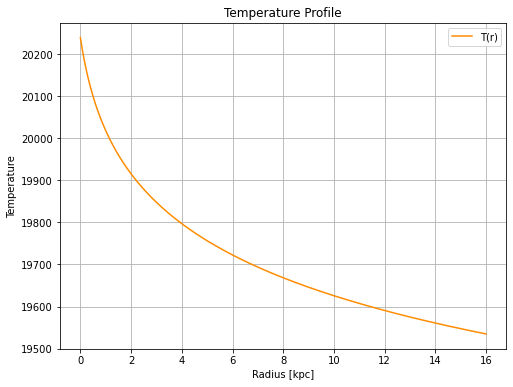}
\caption{The temperature as a function of the radius for the
collisional DM model (\ref{tanhmodel}) for the galaxy NGC3953.}
\label{NGC3953temp}
\end{figure}

For this galaxy, we shall choose $\rho_0=2.7\times
10^8$$M_{\odot}/\mathrm{Kpc}^{3}$. NGC3953 is a barred spiral
galaxy of Hubble type SB(r)bc, with an inner ring structure and a
moderately bright nucleus. Its distance is estimated at
approximately $D \sim 15.4\;\mathrm{Mpc}$. In Figs.
\ref{NGC3953dens}, \ref{NGC3953} and \ref{NGC3953temp} we present
the density of the collisional DM model, the predicted rotation
curves after using an optimization for the collisional DM model
(\ref{tanhmodel}), versus the SPARC observational data and the
temperature parameter as a function of the radius respectively. As
it can be seen, the SIDM model produces viable rotation curves
compatible with the SPARC data. Also in Tables \ref{collNGC3953},
\ref{NavaroNGC3953}, \ref{BuckertNGC3953} and \ref{EinastoNGC3953}
we present the optimization values for the SIDM model, and the
other DM profiles. Also in Table \ref{EVALUATIONNGC3953} we
present the overall evaluation of the SIDM model for the galaxy at
hand. The resulting phenomenology is viable.
\begin{table}[h!]
  \begin{center}
    \caption{Collisional Dark Matter Optimization Values}
    \label{collNGC3953}
     \begin{tabular}{|r|r|}
     \hline
      \textbf{Parameter}   & \textbf{Optimization Values}
      \\  \hline
     $\delta_{\gamma} $ & 0.0000000012
\\  \hline
$\gamma_0 $ & 1.0001 \\ \hline $K_0$ ($M_{\odot} \,
\mathrm{Kpc}^{-3} \, (\mathrm{km/s})^{2}$)& 20200  \\ \hline
    \end{tabular}
  \end{center}
\end{table}
\begin{table}[h!]
  \begin{center}
    \caption{NFW  Optimization Values}
    \label{NavaroNGC3953}
     \begin{tabular}{|r|r|}
     \hline
      \textbf{Parameter}   & \textbf{Optimization Values}
      \\  \hline
   $\rho_s$   & $0.0695\times 10^9$
\\  \hline
$r_s$&  7.9
\\  \hline
    \end{tabular}
  \end{center}
\end{table}

\begin{table}[h!]
  \begin{center}
    \caption{Burkert Optimization Values}
    \label{BuckertNGC3953}
     \begin{tabular}{|r|r|}
     \hline
      \textbf{Parameter}   & \textbf{Optimization Values}
      \\  \hline
     $\rho_0^B$  & $0.47\times 10^9$
\\  \hline
$r_0$&  4.8
\\  \hline
    \end{tabular}
  \end{center}
\end{table}
\begin{table}[h!]
  \begin{center}
    \caption{Einasto Optimization Values}
    \label{EinastoNGC3953}
    \begin{tabular}{|r|r|}
     \hline
      \textbf{Parameter}   & \textbf{Optimization Values}
      \\  \hline
     $\rho_e$  & $0.015\times 10^9$
\\  \hline
$r_e$ & 11
\\  \hline
$n_e$ & 0.7
\\  \hline
    \end{tabular}
  \end{center}
\end{table}
\begin{table}[h!]
\centering \caption{Physical assessment of collisional DM
parameters (NGC3953).}
\begin{tabular}{lcc}
\hline
Parameter & Value & Physical Verdict \\
\hline
$\gamma_0$ & $1.0001$ & Essentially isothermal  \\
$\delta_\gamma$ & $1.2\times10^{-9}$ & Negligible variation  \\
$r_\gamma$ & $1.5\ \mathrm{Kpc}$ & Reasonable transition radius   \\
$K_0$ & $2.02\times10^{4}$ & Significant pressure support at the center \\
$r_c$ & $0.5\ \mathrm{Kpc}$ & Small core scale - acceptable for inner halo structure \\
$p$ & $0.01$ & Very shallow decline of $K(r)$ \\
\hline
Overall &-& Physically consistent and almost-isothermal \\
\hline
\end{tabular}
\label{EVALUATIONNGC3953}
\end{table}

\subsection{The Galaxy NGC3972}

\begin{figure}[h!]
\centering
\includegraphics[width=20pc]{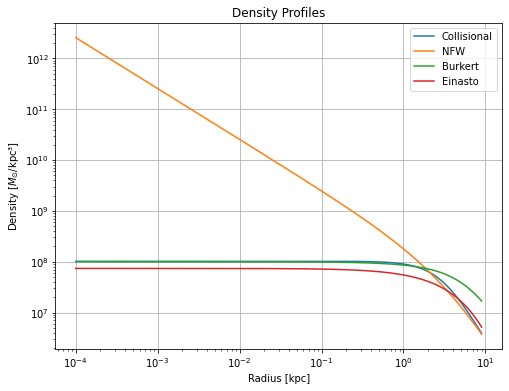}
\caption{The density of the collisional DM model (\ref{tanhmodel})
for the galaxy NGC3972, as a function of the radius.}
\label{NGC3972dens}
\end{figure}

\begin{figure}[h!]
\centering
\includegraphics[width=20pc]{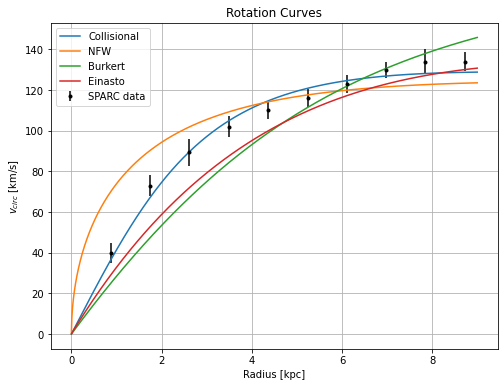}
\caption{The predicted rotation curves after using an optimization
for the collisional DM model (\ref{tanhmodel}), versus the SPARC
observational data for the galaxy NGC3972. We also plotted the
optimized curves for the NFW model, the Burkert model and the
Einasto model.} \label{NGC3972}
\end{figure}

\begin{figure}[h!]
\centering
\includegraphics[width=20pc]{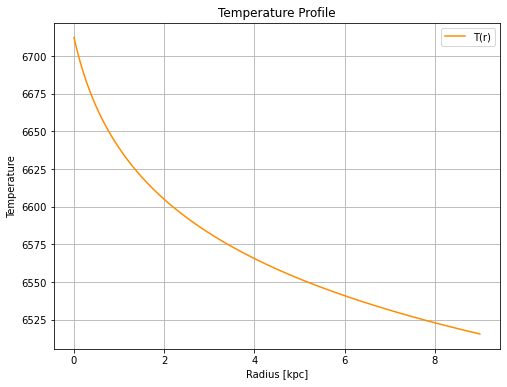}
\caption{The temperature as a function of the radius for the
collisional DM model (\ref{tanhmodel}) for the galaxy NGC3972.}
\label{NGC3972temp}
\end{figure}

For this galaxy, we shall choose $\rho_0=1.02\times
10^8$$M_{\odot}/\mathrm{Kpc}^{3}$. NGC3972 is a spiral galaxy in
the constellation Ursa Major, classified in catalogs as an
unbarred to weakly barred intermediate spiral (SA(s)bc / SABb).
The galaxy lies at a distance of about $20$ Mpc, based on Cepheid
and other distance estimates. In Figs. \ref{NGC3972dens},
\ref{NGC3972} and \ref{NGC3972temp} we present the density of the
collisional DM model, the predicted rotation curves after using an
optimization for the collisional DM model (\ref{tanhmodel}),
versus the SPARC observational data and the temperature parameter
as a function of the radius respectively. As it can be seen, the
SIDM model produces viable rotation curves compatible with the
SPARC data. Also in Tables \ref{collNGC3972}, \ref{NavaroNGC3972},
\ref{BuckertNGC3972} and \ref{EinastoNGC3972} we present the
optimization values for the SIDM model, and the other DM profiles.
Also in Table \ref{EVALUATIONNGC3972} we present the overall
evaluation of the SIDM model for the galaxy at hand. The resulting
phenomenology is viable.
\begin{table}[h!]
  \begin{center}
    \caption{Collisional Dark Matter Optimization Values}
    \label{collNGC3972}
     \begin{tabular}{|r|r|}
     \hline
      \textbf{Parameter}   & \textbf{Optimization Values}
      \\  \hline
     $\delta_{\gamma} $ & 0.0000000012
\\  \hline
$\gamma_0 $ & 1.0001 \\ \hline $K_0$ ($M_{\odot} \,
\mathrm{Kpc}^{-3} \, (\mathrm{km/s})^{2}$)& 6700  \\ \hline
    \end{tabular}
  \end{center}
\end{table}
\begin{table}[h!]
  \begin{center}
    \caption{NFW  Optimization Values}
    \label{NavaroNGC3972}
     \begin{tabular}{|r|r|}
     \hline
      \textbf{Parameter}   & \textbf{Optimization Values}
      \\  \hline
   $\rho_s$   & $5\times 10^7$
\\  \hline
$r_s$&  5.13
\\  \hline
    \end{tabular}
  \end{center}
\end{table}
\begin{table}[h!]
  \begin{center}
    \caption{Burkert Optimization Values}
    \label{BuckertNGC3972}
     \begin{tabular}{|r|r|}
     \hline
      \textbf{Parameter}   & \textbf{Optimization Values}
      \\  \hline
     $\rho_0^B$  & $ 10^8$
\\  \hline
$r_0$&  7.11
\\  \hline
    \end{tabular}
  \end{center}
\end{table}
\begin{table}[h!]
  \begin{center}
    \caption{Einasto Optimization Values}
    \label{EinastoNGC3972}
    \begin{tabular}{|r|r|}
     \hline
      \textbf{Parameter}   & \textbf{Optimization Values}
      \\  \hline
     $\rho_e$  & $1\times 10^7$
\\  \hline
$r_e$ & 6.77
\\  \hline
$n_e$ & 1
\\  \hline
    \end{tabular}
  \end{center}
\end{table}
\begin{table}[h!]
\centering \caption{Physical assessment of collisional DM
parameters (NGC3972).}
\begin{tabular}{lcc}
\hline
Parameter & Value & Physical Verdict \\
\hline
$\gamma_0$ & $1.0001$ & Essentially isothermal  \\
$\delta_\gamma$ & $1.2\times10^{-9}$ & Negligible variation  \\
$r_\gamma$ & $1.5\ \mathrm{Kpc}$ & Reasonable transition radius   \\
$K_0$ & $6.70\times10^{3}$ & Enough central pressure support \\
$r_c$ & $0.5\ \mathrm{Kpc}$ & Small core scale  \\
$p$ & $0.01$ & Very shallow $K(r)$ decline; practically constant entropy \\
\hline
Overall &-& Physically consistent but almost-isothermal \\
\hline
\end{tabular}
\label{EVALUATIONNGC3972}
\end{table}


\subsection{The Galaxy NGC3992 Marginally Viable, Extended Viable}

\begin{figure}[h!]
\centering
\includegraphics[width=20pc]{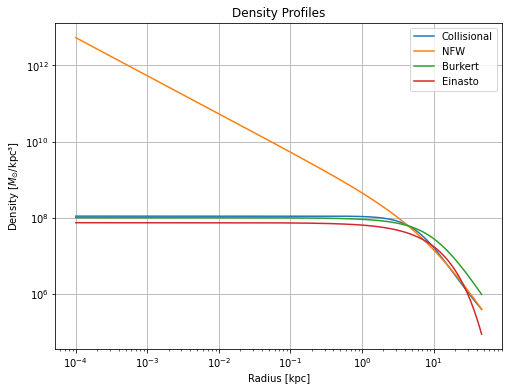}
\caption{The density of the collisional DM model (\ref{tanhmodel})
for the galaxy NGC3992, as a function of the radius.}
\label{NGC3992dens}
\end{figure}
\begin{figure}[h!]
\centering
\includegraphics[width=20pc]{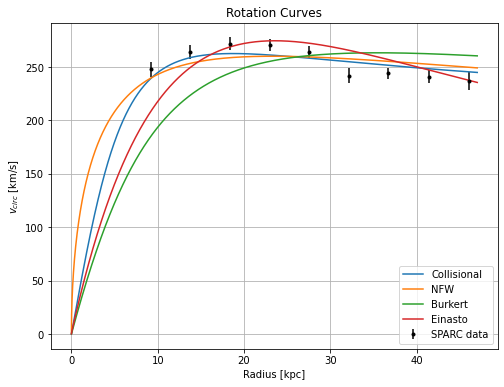}
\caption{The predicted rotation curves after using an optimization
for the collisional DM model (\ref{tanhmodel}), versus the SPARC
observational data for the galaxy NGC3992. We also plotted the
optimized curves for the NFW model, the Burkert model and the
Einasto model.} \label{NGC3992}
\end{figure}
\begin{figure}[h!]
\centering
\includegraphics[width=20pc]{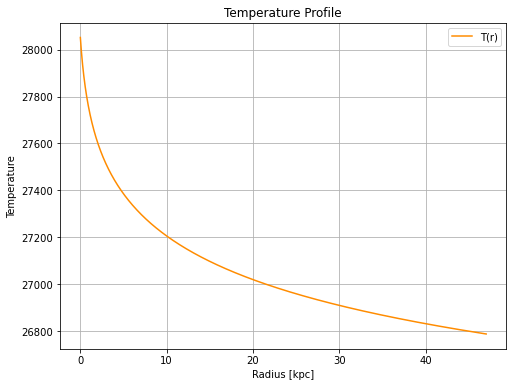}
\caption{The temperature as a function of the radius for the
collisional DM model (\ref{tanhmodel}) for the galaxy NGC3992.}
\label{NGC3992temp}
\end{figure}
\begin{figure}[h!]
\centering
\includegraphics[width=20pc]{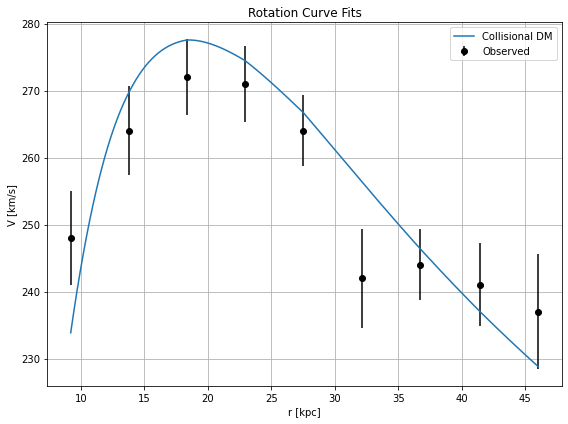}
\caption{The predicted rotation curves after using an optimization
for the collisional DM model (\ref{tanhmodel}), versus the
extended SPARC observational data for the galaxy NGC3992. The
model includes the rotation curves from all the components of the
galaxy, including gas and disk velocities, along with the
collisional DM model.} \label{extendedNGC3992}
\end{figure}

For this galaxy, we shall choose $\rho_0=1.1\times
10^8$$M_{\odot}/\mathrm{Kpc}^{3}$. NGC3992 is a barred spiral
galaxy (type SBbc / SB(rs)bc) at a distance $D \sim 18.6
\,\mathrm{Mpc}$. In Figs. \ref{NGC3992dens}, \ref{NGC3992} and
\ref{NGC3992temp} we present the density of the collisional DM
model, the predicted rotation curves after using an optimization
for the collisional DM model (\ref{tanhmodel}), versus the SPARC
observational data and the temperature parameter as a function of
the radius respectively. As it can be seen, the SIDM model
produces viable rotation curves marginally compatible with the
SPARC data. Also in Tables \ref{collNGC3992}, \ref{NavaroNGC3992},
\ref{BuckertNGC3992} and \ref{EinastoNGC3992} we present the
optimization values for the SIDM model, and the other DM profiles.
Also in Table \ref{EVALUATIONNGC3992} we present the overall
evaluation of the SIDM model for the galaxy at hand. The resulting
phenomenology is marginally viable.
\begin{table}[h!]
  \begin{center}
    \caption{Collisional Dark Matter Optimization Values}
    \label{collNGC3992}
     \begin{tabular}{|r|r|}
     \hline
      \textbf{Parameter}   & \textbf{Optimization Values}
      \\  \hline
     $\delta_{\gamma} $ & 0.0000000012
\\  \hline
$\gamma_0 $ & 1.0001 \\ \hline $K_0$ ($M_{\odot} \,
\mathrm{Kpc}^{-3} \, (\mathrm{km/s})^{2}$)& 28000  \\ \hline
    \end{tabular}
  \end{center}
\end{table}
\begin{table}[h!]
  \begin{center}
    \caption{NFW  Optimization Values}
    \label{NavaroNGC3992}
     \begin{tabular}{|r|r|}
     \hline
      \textbf{Parameter}   & \textbf{Optimization Values}
      \\  \hline
   $\rho_s$   & $5\times 10^7$
\\  \hline
$r_s$&  10.76
\\  \hline
    \end{tabular}
  \end{center}
\end{table}
\begin{table}[h!]
  \begin{center}
    \caption{Burkert Optimization Values}
    \label{BuckertNGC3992}
     \begin{tabular}{|r|r|}
     \hline
      \textbf{Parameter}   & \textbf{Optimization Values}
      \\  \hline
     $\rho_0^B$  & $ 10^8$
\\  \hline
$r_0$&  10.92
\\  \hline
    \end{tabular}
  \end{center}
\end{table}
\begin{table}[h!]
  \begin{center}
    \caption{Einasto Optimization Values}
    \label{EinastoNGC3992}
    \begin{tabular}{|r|r|}
     \hline
      \textbf{Parameter}   & \textbf{Optimization Values}
      \\  \hline
     $\rho_e$  & $10^7$
\\  \hline
$r_e$ & 13.94
\\  \hline
$n_e$ & 1
\\  \hline
    \end{tabular}
  \end{center}
\end{table}
\begin{table}[h!]
\centering \caption{Physical assessment of collisional DM
parameters (NGC3992).}
\begin{tabular}{lcc}
\hline
Parameter & Value & Physical Verdict \\
\hline
$\gamma_0$ & $1.0001$ & Essentially isothermal  \\
$\delta_\gamma$ & $1.2\times10^{-9}$ & Negligible variation  \\
$r_\gamma$ & $1.5\ \mathrm{Kpc}$ & Reasonable transition radius   \\
$K_0$ & $2.80\times10^{4}$ & Enough central pressure support \\
$r_c$ & $0.5\ \mathrm{Kpc}$ & Small core scale \\
$p$ & $0.01$ & Very shallow $K(r)$ decline; practically constant entropy \\
\hline
Overall &-& Physically consistent nearly isothermal \\
\hline
\end{tabular}
\label{EVALUATIONNGC3992}
\end{table}
Now the extended picture including the rotation velocity from the
other components of the galaxy, such as the disk and gas, makes
the collisional DM model viable for this galaxy. In Fig.
\ref{extendedNGC3992} we present the combined rotation curves
including the other components of the galaxy along with the
collisional matter. As it can be seen, the extended collisional DM
model is viable.
Also in Table \ref{evaluationextendedNGC3992} we present the
values of the free parameters of the collisional DM model for
which the maximum compatibility with the SPARC data comes for the
galaxy NGC3992.
\begin{table}[h!]
\centering \caption{Physical assessment of Extended collisional DM
parameters for galaxy NGC3992.}
\begin{tabular}{lcc}
\hline
Parameter & Value & Physical Verdict \\
\hline
$\gamma_0$ & 1.24100000 & Above isothermal \\
$\delta_\gamma$ & 0.10750000 & Moderate radial variation\\
$K_0$ & 3000 & Moderate entropy scale \\
$ml_{disk}$ & 0.00001916 & Essentially zero stellar M/L \\
$ml_{bulge}$ & 0.00000000 & No bulge contribution \\
\hline
Overall &-& Mixed plausibility \\
\hline
\end{tabular}
\label{evaluationextendedNGC3992}
\end{table}


\subsection{The Galaxy NGC4010 Non-viable}

\begin{figure}[h!]
\centering
\includegraphics[width=20pc]{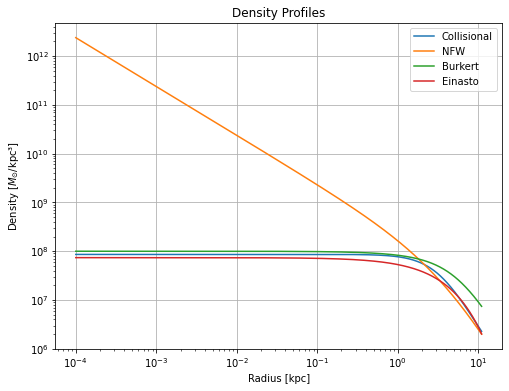}
\caption{The density of the collisional DM model (\ref{tanhmodel})
for the galaxy NGC4010, as a function of the radius.}
\label{NGC4010dens}
\end{figure}
\begin{figure}[h!]
\centering
\includegraphics[width=20pc]{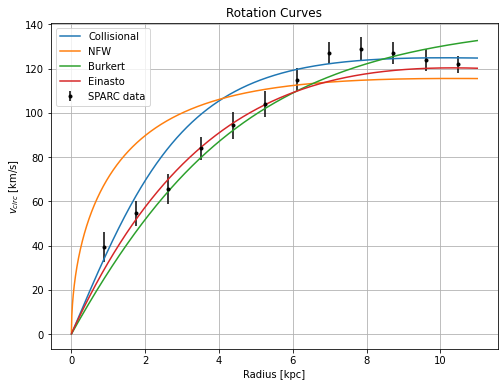}
\caption{The predicted rotation curves after using an optimization
for the collisional DM model (\ref{tanhmodel}), versus the SPARC
observational data for the galaxy NGC4010. We also plotted the
optimized curves for the NFW model, the Burkert model and the
Einasto model.} \label{NGC4010}
\end{figure}
\begin{figure}[h!]
\centering
\includegraphics[width=20pc]{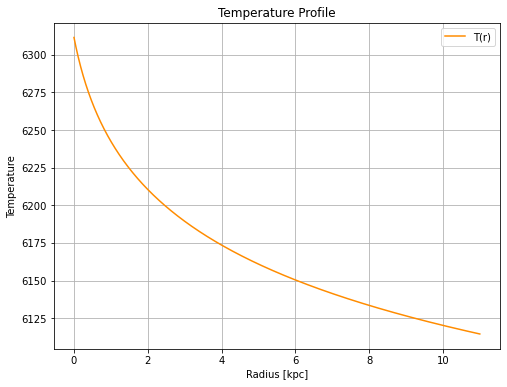}
\caption{The temperature as a function of the radius for the
collisional DM model (\ref{tanhmodel}) for the galaxy NGC4010.}
\label{NGC4010temp}
\end{figure}
\begin{figure}[h!]
\centering
\includegraphics[width=20pc]{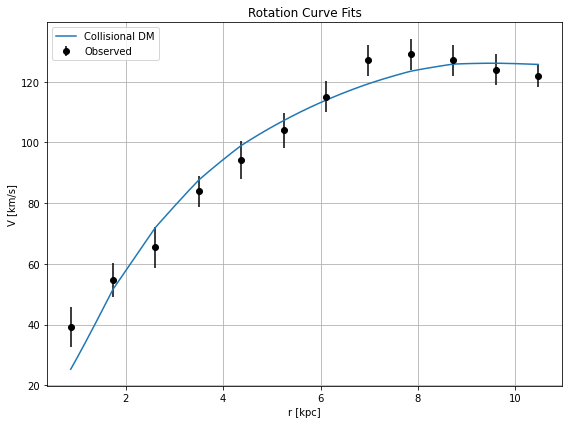}
\caption{The predicted rotation curves after using an optimization
for the collisional DM model (\ref{tanhmodel}), versus the
extended SPARC observational data for the galaxy NGC4010. The
model includes the rotation curves from all the components of the
galaxy, including gas and disk velocities, along with the
collisional DM model.} \label{extendedNGC4010}
\end{figure}

For this galaxy, we shall choose $\rho_0=8.6\times
10^7$$M_{\odot}/\mathrm{Kpc}^{3}$. NGC4010 is a late-type barred
spiral galaxy at a distance $D \sim 13.1\;\mathrm{Mpc}$. In Figs.
\ref{NGC4010dens}, \ref{NGC4010} and \ref{NGC4010temp} we present
the density of the collisional DM model, the predicted rotation
curves after using an optimization for the collisional DM model
(\ref{tanhmodel}), versus the SPARC observational data and the
temperature parameter as a function of the radius respectively. As
it can be seen, the SIDM model produces non-viable rotation curves
incompatible with the SPARC data. Also in Tables
\ref{collNGC4010}, \ref{NavaroNGC4010}, \ref{BuckertNGC4010} and
\ref{EinastoNGC4010} we present the optimization values for the
SIDM model, and the other DM profiles. Also in Table
\ref{EVALUATIONNGC4010} we present the overall evaluation of the
SIDM model for the galaxy at hand. The resulting phenomenology is
non-viable.
\begin{table}[h!]
  \begin{center}
    \caption{Collisional Dark Matter Optimization Values}
    \label{collNGC4010}
     \begin{tabular}{|r|r|}
     \hline
      \textbf{Parameter}   & \textbf{Optimization Values}
      \\  \hline
     $\delta_{\gamma} $ & 0.0000000012
\\  \hline
$\gamma_0 $ & 1.0001 \\ \hline $K_0$ ($M_{\odot} \,
\mathrm{Kpc}^{-3} \, (\mathrm{km/s})^{2}$)& 6300  \\ \hline
    \end{tabular}
  \end{center}
\end{table}
\begin{table}[h!]
  \begin{center}
    \caption{NFW  Optimization Values}
    \label{NavaroNGC4010}
     \begin{tabular}{|r|r|}
     \hline
      \textbf{Parameter}   & \textbf{Optimization Values}
      \\  \hline
   $\rho_s$   & $5\times 10^7$
\\  \hline
$r_s$&  4.78
\\  \hline
    \end{tabular}
  \end{center}
\end{table}
\begin{table}[h!]
  \begin{center}
    \caption{Burkert Optimization Values}
    \label{BuckertNGC4010}
     \begin{tabular}{|r|r|}
     \hline
      \textbf{Parameter}   & \textbf{Optimization Values}
      \\  \hline
     $\rho_0^B$  & $ 10^8$
\\  \hline
$r_0$&  5.77
\\  \hline
    \end{tabular}
  \end{center}
\end{table}
\begin{table}[h!]
  \begin{center}
    \caption{Einasto Optimization Values}
    \label{EinastoNGC4010}
    \begin{tabular}{|r|r|}
     \hline
      \textbf{Parameter}   & \textbf{Optimization Values}
      \\  \hline
     $\rho_e$  & $10^7$
\\  \hline
$r_e$ & 6.11
\\  \hline
$n_e$ & 1
\\  \hline
    \end{tabular}
  \end{center}
\end{table}
\begin{table}[h!]
\centering \caption{Physical assessment of collisional DM
parameters (NGC4010).}
\begin{tabular}{lcc}
\hline
Parameter & Value & Physical Verdict \\
\hline
$\gamma_0$ & $1.0001$ & Essentially isothermal  \\
$\delta_\gamma$ & $1.2\times10^{-9}$ & Negligible variation  \\
$r_\gamma$ & $1.5\ \mathrm{Kpc}$ & Reasonable transition radius   \\
$K_0$ & $6.30\times10^{3}$ & Enough pressure support \\
$r_c$ & $0.5\ \mathrm{Kpc}$ & Small core scale   \\
$p$ & $0.01$ & Very shallow $K(r)$ decline; practically constant entropy \\
\hline
Overall &-& Physically consistent nearly isothermal \\
\hline
\end{tabular}
\label{EVALUATIONNGC4010}
\end{table}
Now the extended picture including the rotation velocity from the
other components of the galaxy, such as the disk and gas, makes
the collisional DM model viable for this galaxy. In Fig.
\ref{extendedNGC4010} we present the combined rotation curves
including the other components of the galaxy along with the
collisional matter. As it can be seen, the extended collisional DM
model is non-viable.
Also in Table \ref{evaluationextendedNGC4010} we present the
values of the free parameters of the collisional DM model for
which the maximum compatibility with the SPARC data comes for the
galaxy NGC4010.
\begin{table}[h!]
\centering \caption{Physical assessment of Extended collisional DM
parameters for NGC4010.}
\begin{tabular}{lcc}
\hline
Parameter & Value & Physical Verdict \\
\hline
$\gamma_0$ & 1.0636 & Near-isothermal core \\
$\delta_\gamma$ & 0.03615 & Small radial variation \\
$K_0$ & 3000 & Moderate entropy \\
$ml_{\text{disk}}$ & 0.6310 & Reasonable stellar $M/L$ \\
$ml_{\text{bulge}}$ & 0.0000 & No bulge component\\
\hline
Overall &-& Physically plausible \\
\hline
\end{tabular}
\label{evaluationextendedNGC4010}
\end{table}


\subsection{The Galaxy NGC4013 Non-viable}

\begin{figure}[h!]
\centering
\includegraphics[width=20pc]{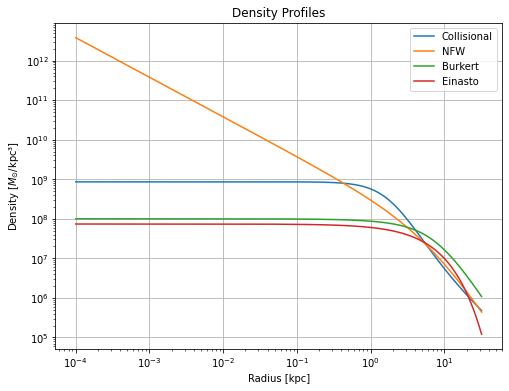}
\caption{The density of the collisional DM model (\ref{tanhmodel})
for the galaxy NGC4013, as a function of the radius.}
\label{NGC4013dens}
\end{figure}
\begin{figure}[h!]
\centering
\includegraphics[width=20pc]{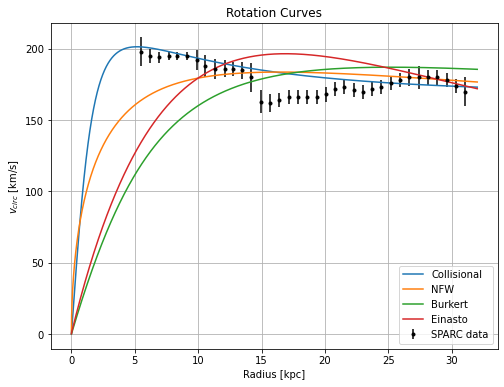}
\caption{The predicted rotation curves after using an optimization
for the collisional DM model (\ref{tanhmodel}), versus the SPARC
observational data for the galaxy NGC4013. We also plotted the
optimized curves for the NFW model, the Burkert model and the
Einasto model.} \label{NGC4013}
\end{figure}
\begin{figure}[h!]
\centering
\includegraphics[width=20pc]{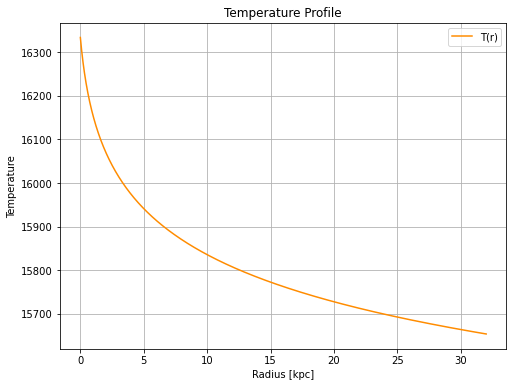}
\caption{The temperature as a function of the radius for the
collisional DM model (\ref{tanhmodel}) for the galaxy NGC4013.}
\label{NGC4013temp}
\end{figure}
\begin{figure}[h!]
\centering
\includegraphics[width=20pc]{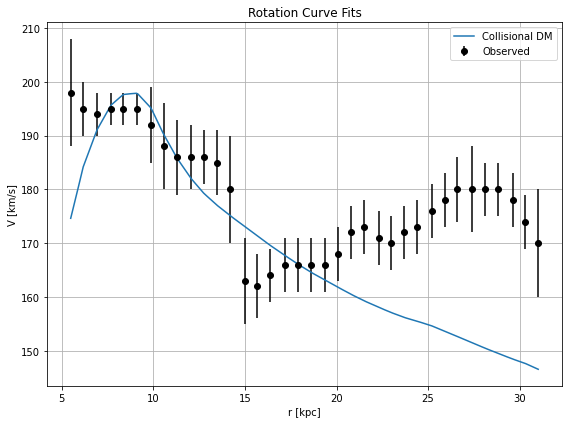}
\caption{The predicted rotation curves after using an optimization
for the collisional DM model (\ref{tanhmodel}), versus the
extended SPARC observational data for the galaxy NGC4013. The
model includes the rotation curves from all the components of the
galaxy, including gas and disk velocities, along with the
collisional DM model.} \label{extendedNGC4013}
\end{figure}

For this galaxy, we shall choose $\rho_0=8.6\times
10^8$$M_{\odot}/\mathrm{Kpc}^{3}$. NGC4013 is an edge-on barred
spiral (SBa/Sb-like) at $D\sim18.6\ \mathrm{Mpc}$. In Figs.
\ref{NGC4013dens}, \ref{NGC4013} and \ref{NGC4013temp} we present
the density of the collisional DM model, the predicted rotation
curves after using an optimization for the collisional DM model
(\ref{tanhmodel}), versus the SPARC observational data and the
temperature parameter as a function of the radius respectively. As
it can be seen, the SIDM model produces non-viable rotation curves
incompatible with the SPARC data. Also in Tables
\ref{collNGC4013}, \ref{NavaroNGC4013}, \ref{BuckertNGC4013} and
\ref{EinastoNGC4013} we present the optimization values for the
SIDM model, and the other DM profiles. Also in Table
\ref{EVALUATIONNGC4013} we present the overall evaluation of the
SIDM model for the galaxy at hand. The resulting phenomenology is
non-viable.
\begin{table}[h!]
  \begin{center}
    \caption{Collisional Dark Matter Optimization Values}
    \label{collNGC4013}
     \begin{tabular}{|r|r|}
     \hline
      \textbf{Parameter}   & \textbf{Optimization Values}
      \\  \hline
     $\delta_{\gamma} $ & 0.0000000012
\\  \hline
$\gamma_0 $ & 1.0001 \\ \hline $K_0$ ($M_{\odot} \,
\mathrm{Kpc}^{-3} \, (\mathrm{km/s})^{2}$)& 16300 \\ \hline
    \end{tabular}
  \end{center}
\end{table}
\begin{table}[h!]
  \begin{center}
    \caption{NFW  Optimization Values}
    \label{NavaroNGC4013}
     \begin{tabular}{|r|r|}
     \hline
      \textbf{Parameter}   & \textbf{Optimization Values}
      \\  \hline
   $\rho_s$   & $5\times 10^7$
\\  \hline
$r_s$&  7.6
\\  \hline
    \end{tabular}
  \end{center}
\end{table}
\begin{table}[h!]
  \begin{center}
    \caption{Burkert Optimization Values}
    \label{BuckertNGC4013}
     \begin{tabular}{|r|r|}
     \hline
      \textbf{Parameter}   & \textbf{Optimization Values}
      \\  \hline
     $\rho_0^B$  & $10^8$
\\  \hline
$r_0$&  7.76
\\  \hline
    \end{tabular}
  \end{center}
\end{table}
\begin{table}[h!]
  \begin{center}
    \caption{Einasto Optimization Values}
    \label{EinastoNGC4013}
    \begin{tabular}{|r|r|}
     \hline
      \textbf{Parameter}   & \textbf{Optimization Values}
      \\  \hline
     $\rho_e$  & $10^7$
\\  \hline
$r_e$ & 9.98
\\  \hline
$n_e$ & 1
\\  \hline
    \end{tabular}
  \end{center}
\end{table}
\begin{table}[h!]
\centering \caption{Physical assessment of collisional DM
parameters (NGC4013).}
\begin{tabular}{lcc}
\hline
Parameter & Value & Physical Verdict \\
\hline
$\gamma_0$ & $1.0001$ & Essentially isothermal  \\
$\delta_\gamma$ & $1.2\times10^{-9}$ & Negligible variation  \\
$r_\gamma$ & $1.5\ \mathrm{Kpc}$ & Reasonable transition radius   \\
$K_0$ & $1.63\times10^{4}$ & Enough pressure support \\
$r_c$ & $0.5\ \mathrm{Kpc}$ & Small core scale   \\
$p$ & $0.01$ & Very shallow $K(r)$ decline; practically constant entropy \\
\hline
Overall &-& Physically consistent nearly isothermal \\
\hline
\end{tabular}
\label{EVALUATIONNGC4013}
\end{table}
Now the extended picture including the rotation velocity from the
other components of the galaxy, such as the disk and gas, makes
the collisional DM model viable for this galaxy. In Fig.
\ref{extendedNGC4013} we present the combined rotation curves
including the other components of the galaxy along with the
collisional matter. As it can be seen, the extended collisional DM
model is not viable.
Also in Table \ref{evaluationextendedNGC4013} we present the
values of the free parameters of the collisional DM model for
which the maximum compatibility with the SPARC data comes for the
galaxy NGC4013.
\begin{table}[h!]
\centering \caption{Physical assessment of Extended collisional DM
parameters for NGC4013.}
\begin{tabular}{lcc}
\hline
Parameter & Value & Physical Verdict \\
\hline
$\gamma_0$ & 1.07017382 & Nearly isothermal core \\
$\delta_\gamma$ & 0.00000001 & No variation \\
$K_0$ & 3000 & Moderate entropy  \\
$ml_{\text{disk}}$ & 0.79303305 & Reasonable stellar mass-to-light ratio \\
$ml_{\text{bulge}}$ & 0.00000000 & Absence of bulge component \\
\hline
Overall &-& Physically plausible \\
\hline
\end{tabular}
\label{evaluationextendedNGC4013}
\end{table}


\subsection{The Galaxy NGC4051 Marginally Viable, Extended Viable}

\begin{figure}[h!]
\centering
\includegraphics[width=20pc]{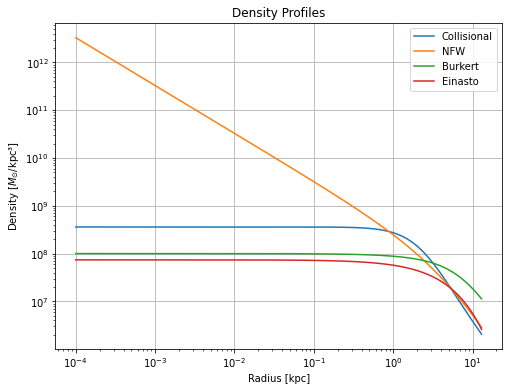}
\caption{The density of the collisional DM model (\ref{tanhmodel})
for the galaxy NGC4051, as a function of the radius.}
\label{NGC4051dens}
\end{figure}

\begin{figure}[h!]
\centering
\includegraphics[width=20pc]{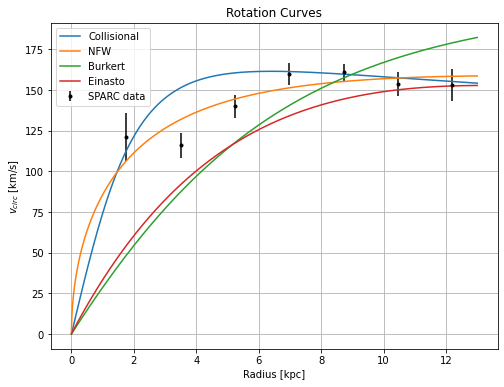}
\caption{The predicted rotation curves after using an optimization
for the collisional DM model (\ref{tanhmodel}), versus the SPARC
observational data for the galaxy NGC4051. We also plotted the
optimized curves for the NFW model, the Burkert model and the
Einasto model.} \label{NGC4051}
\end{figure}

\begin{figure}[h!]
\centering
\includegraphics[width=20pc]{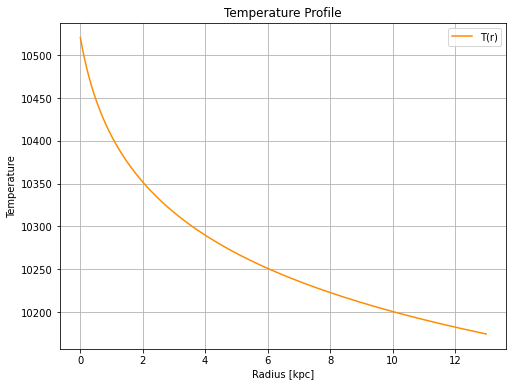}
\caption{The temperature as a function of the radius for the
collisional DM model (\ref{tanhmodel}) for the galaxy NGC4051.}
\label{NGC4051temp}
\end{figure}

\begin{figure}[h!]
\centering
\includegraphics[width=20pc]{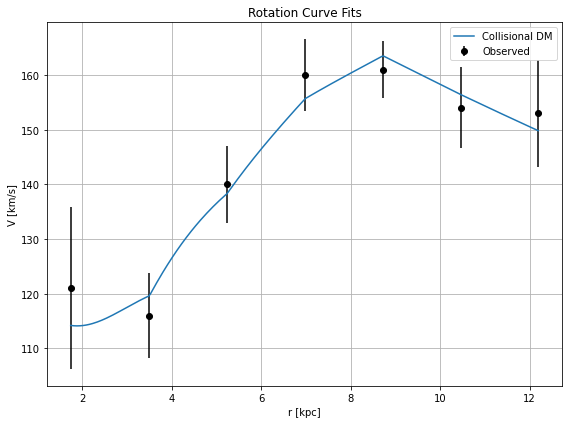}
\caption{The predicted rotation curves after using an optimization
for the collisional DM model (\ref{tanhmodel}), versus the
extended SPARC observational data for the galaxy NGC4051. The
model includes the rotation curves from all the components of the
galaxy, including gas and disk velocities, along with the
collisional DM model.} \label{extendedNGC4051}
\end{figure}

For this galaxy, we shall choose $\rho_0=3.6\times
10^8$$M_{\odot}/\mathrm{Kpc}^{3}$. NGC4051 is an intermediate
barred spiral galaxy located in the constellation Ursa Major. The
galaxy is a member of the Ursa Major Cluster and exhibits active
galactic nucleus activity. The distance to NGC4051 has been
determined using Cepheid variable stars, yielding a value of $D =
16.6 \pm 0.3 \, \mathrm{Mpc}$. In Figs. \ref{NGC4051dens},
\ref{NGC4051} and \ref{NGC4051temp} we present the density of the
collisional DM model, the predicted rotation curves after using an
optimization for the collisional DM model (\ref{tanhmodel}),
versus the SPARC observational data and the temperature parameter
as a function of the radius respectively. As it can be seen, the
SIDM model produces marginally viable rotation curves compatible
with the SPARC data. Also in Tables \ref{collNGC4051},
\ref{NavaroNGC4051}, \ref{BuckertNGC4051} and \ref{EinastoNGC4051}
we present the optimization values for the SIDM model, and the
other DM profiles. Also in Table \ref{EVALUATIONNGC4051} we
present the overall evaluation of the SIDM model for the galaxy at
hand. The resulting phenomenology is marginally viable.
\begin{table}[h!]
  \begin{center}
    \caption{Collisional Dark Matter Optimization Values}
    \label{collNGC4051}
     \begin{tabular}{|r|r|}
     \hline
      \textbf{Parameter}   & \textbf{Optimization Values}
      \\  \hline
     $\delta_{\gamma} $ & 0.0000000012
\\  \hline
$\gamma_0 $ & 1.0001 \\ \hline $K_0$ ($M_{\odot} \,
\mathrm{Kpc}^{-3} \, (\mathrm{km/s})^{2}$)& 10500  \\ \hline
    \end{tabular}
  \end{center}
\end{table}
\begin{table}[h!]
  \begin{center}
    \caption{NFW  Optimization Values}
    \label{NavaroNGC4051}
     \begin{tabular}{|r|r|}
     \hline
      \textbf{Parameter}   & \textbf{Optimization Values}
      \\  \hline
   $\rho_s$   & $5\times 10^7$
\\  \hline
$r_s$&  6.57
\\  \hline
    \end{tabular}
  \end{center}
\end{table}
\begin{table}[h!]
  \begin{center}
    \caption{Burkert Optimization Values}
    \label{BuckertNGC4051}
     \begin{tabular}{|r|r|}
     \hline
      \textbf{Parameter}   & \textbf{Optimization Values}
      \\  \hline
     $\rho_0^B$  & $1\times 10^8$
\\  \hline
$r_0$&  8.3
\\  \hline
    \end{tabular}
  \end{center}
\end{table}
\begin{table}[h!]
  \begin{center}
    \caption{Einasto Optimization Values}
    \label{EinastoNGC4051}
    \begin{tabular}{|r|r|}
     \hline
      \textbf{Parameter}   & \textbf{Optimization Values}
      \\  \hline
     $\rho_e$  &$1\times 10^7$
\\  \hline
$r_e$ & 7.76
\\  \hline
$n_e$ & 1
\\  \hline
    \end{tabular}
  \end{center}
\end{table}
\begin{table}[h!]
\centering \caption{Physical assessment of collisional DM
parameters (NGC4051).}
\begin{tabular}{lcc}
\hline
Parameter & Value & Physical Verdict \\
\hline
$\gamma_0$ & $1.0001$ & Essentially isothermal  \\
$\delta_\gamma$ & $1.2\times10^{-9}$ & Negligible variation  \\
$r_\gamma$ & $1.5\ \mathrm{Kpc}$ & Reasonable transition radius   \\
$K_0$ & $1.05\times10^{4}$ & Enough central pressure support \\
$r_c$ & $0.5\ \mathrm{Kpc}$ & Small core scale   \\
$p$ & $0.01$ & Very shallow $K(r)$ decline; practically constant entropy \\
\hline
Overall &-& Physically consistent nearly isothermal \\
\hline
\end{tabular}
\label{EVALUATIONNGC4051}
\end{table}
Now the extended picture including the rotation velocity from the
other components of the galaxy, such as the disk and gas, makes
the collisional DM model viable for this galaxy. In Fig.
\ref{extendedNGC4051} we present the combined rotation curves
including the other components of the galaxy along with the
collisional matter. As it can be seen, the extended collisional DM
model is viable.
Also in Table \ref{evaluationextendedNGC4051} we present the
values of the free parameters of the collisional DM model for
which the maximum compatibility with the SPARC data comes for the
galaxy NGC4051.
\begin{table}[h!]
\centering \caption{Physical assessment of Extended collisional DM
parameters for galaxy NGC4051.}
\begin{tabular}{lcc}
\hline
Parameter & Value & Physical Verdict \\
\hline
$\gamma_0$ & 1.09399536 & Slightly above isothermal \\
$\delta_\gamma$ & 0.08512064 & Moderate radial variation \\
$K_0$ & 3000 & Moderate entropy scale; compatible with intermediate-mass spiral halos \\
$ml_{disk}$ & 0.63653782 & Moderate disk M/L \\
$ml_{bulge}$ & 0.00000000 & No bulge contribution \\
\hline
Overall &-& Physically plausible \\
\hline
\end{tabular}
\label{evaluationextendedNGC4051}
\end{table}


\subsection{The Galaxy NGC4068}

\begin{figure}[h!]
\centering
\includegraphics[width=20pc]{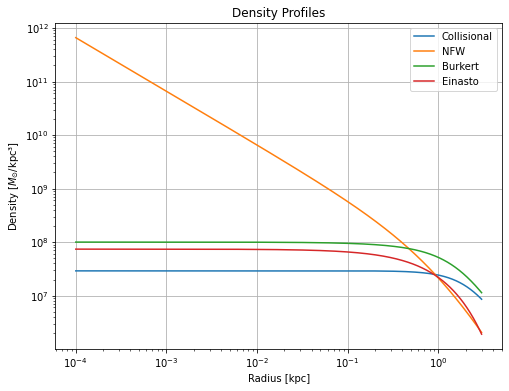}
\caption{The density of the collisional DM model (\ref{tanhmodel})
for the galaxy NGC4068, as a function of the radius.}
\label{NGC4068dens}
\end{figure}
\begin{figure}[h!]
\centering
\includegraphics[width=20pc]{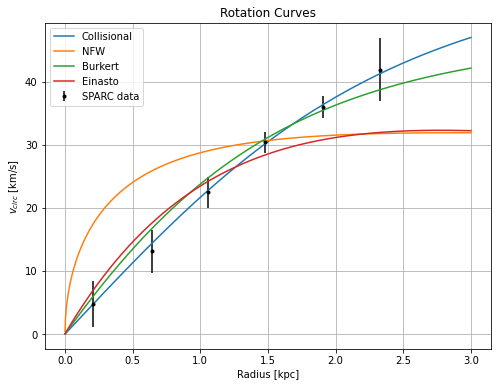}
\caption{The predicted rotation curves after using an optimization
for the collisional DM model (\ref{tanhmodel}), versus the SPARC
observational data for the galaxy NGC4068. We also plotted the
optimized curves for the NFW model, the Burkert model and the
Einasto model.} \label{NGC4068}
\end{figure}
\begin{figure}[h!]
\centering
\includegraphics[width=20pc]{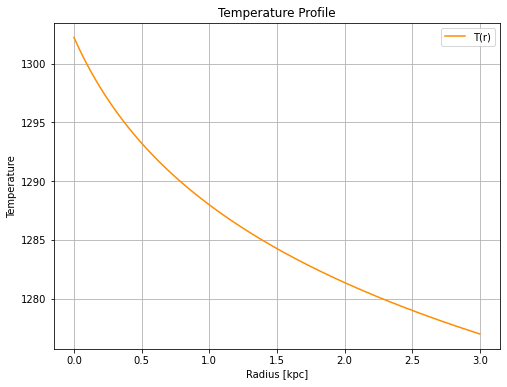}
\caption{The temperature as a function of the radius for the
collisional DM model (\ref{tanhmodel}) for the galaxy NGC4068.}
\label{NGC4068temp}
\end{figure}

For this galaxy, we shall choose $\rho_0=2.9\times
10^7$$M_{\odot}/\mathrm{Kpc}^{3}$. NGC4068 is a dwarf irregular
galaxy located approximately 4.36 Mpc away in the Ursa Major
constellation. In Figs. \ref{NGC4068dens}, \ref{NGC4068} and
\ref{NGC4068temp} we present the density of the collisional DM
model, the predicted rotation curves after using an optimization
for the collisional DM model (\ref{tanhmodel}), versus the SPARC
observational data and the temperature parameter as a function of
the radius respectively. As it can be seen, the SIDM model
produces viable rotation curves compatible with the SPARC data.
Also in Tables \ref{collNGC4068}, \ref{NavaroNGC4068},
\ref{BuckertNGC4068} and \ref{EinastoNGC4068} we present the
optimization values for the SIDM model, and the other DM profiles.
Also in Table \ref{EVALUATIONNGC4068} we present the overall
evaluation of the SIDM model for the galaxy at hand. The resulting
phenomenology is viable.
\begin{table}[h!]
  \begin{center}
    \caption{Collisional Dark Matter Optimization Values}
    \label{collNGC4068}
     \begin{tabular}{|r|r|}
     \hline
      \textbf{Parameter}   & \textbf{Optimization Values}
      \\  \hline
     $\delta_{\gamma} $ & 0.0000000012
\\  \hline
$\gamma_0 $ & 1.0001 \\ \hline $K_0$ ($M_{\odot} \,
\mathrm{Kpc}^{-3} \, (\mathrm{km/s})^{2}$)& 1300 \\ \hline
    \end{tabular}
  \end{center}
\end{table}
\begin{table}[h!]
  \begin{center}
    \caption{NFW  Optimization Values}
    \label{NavaroNGC4068}
     \begin{tabular}{|r|r|}
     \hline
      \textbf{Parameter}   & \textbf{Optimization Values}
      \\  \hline
   $\rho_s$   & $5\times 10^7$
\\  \hline
$r_s$&  1.32
\\  \hline
    \end{tabular}
  \end{center}
\end{table}
\begin{table}[h!]
  \begin{center}
    \caption{Burkert Optimization Values}
    \label{BuckertNGC4068}
     \begin{tabular}{|r|r|}
     \hline
      \textbf{Parameter}   & \textbf{Optimization Values}
      \\  \hline
     $\rho_0^B$  & $1\times 10^8$
\\  \hline
$r_0$&  1.92
\\  \hline
    \end{tabular}
  \end{center}
\end{table}
\begin{table}[h!]
  \begin{center}
    \caption{Einasto Optimization Values}
    \label{EinastoNGC4068}
    \begin{tabular}{|r|r|}
     \hline
      \textbf{Parameter}   & \textbf{Optimization Values}
      \\  \hline
     $\rho_e$  &$1\times 10^7$
\\  \hline
$r_e$ & 1.64
\\  \hline
$n_e$ & 1
\\  \hline
    \end{tabular}
  \end{center}
\end{table}
\begin{table}[h!]
\centering \caption{Physical assessment of collisional DM
parameters (NGC4068).}
\begin{tabular}{lcc}
\hline
Parameter & Value & Physical Verdict \\
\hline
$\gamma_0$ & $1.0001$ & Essentially isothermal  \\
$\delta_\gamma$ & $1.2\times10^{-9}$ & Negligible variation  \\
$r_\gamma$ & $1.5\ \mathrm{Kpc}$ & Reasonable transition radius   \\
$K_0$ & $1.30\times10^{3}$ & Enough central pressure support \\
$r_c$ & $0.5\ \mathrm{Kpc}$ & Small core scale   \\
$p$ & $0.01$ & Very shallow $K(r)$ decline; practically constant entropy \\
\hline
Overall &-& Physically consistent for a low-mass system nearly isothermal \\
\hline
\end{tabular}
\label{EVALUATIONNGC4068}
\end{table}


\subsection{The Galaxy NGC4085}

\begin{figure}[h!]
\centering
\includegraphics[width=20pc]{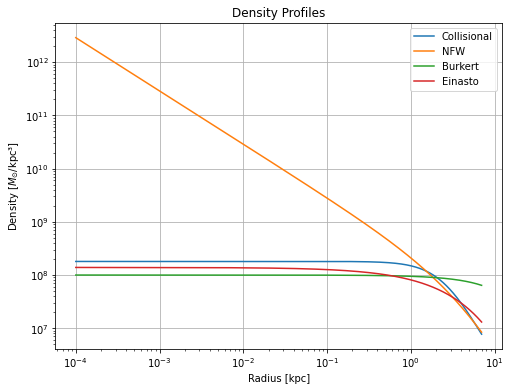}
\caption{The density of the collisional DM model (\ref{tanhmodel})
for the galaxy NGC4085, as a function of the radius.}
\label{NGC4085dens}
\end{figure}
\begin{figure}[h!]
\centering
\includegraphics[width=20pc]{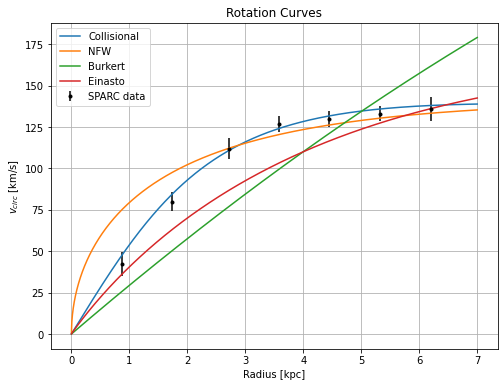}
\caption{The predicted rotation curves after using an optimization
for the collisional DM model (\ref{tanhmodel}), versus the SPARC
observational data for the galaxy NGC4085. We also plotted the
optimized curves for the NFW model, the Burkert model and the
Einasto model.} \label{NGC4085}
\end{figure}
\begin{figure}[h!]
\centering
\includegraphics[width=20pc]{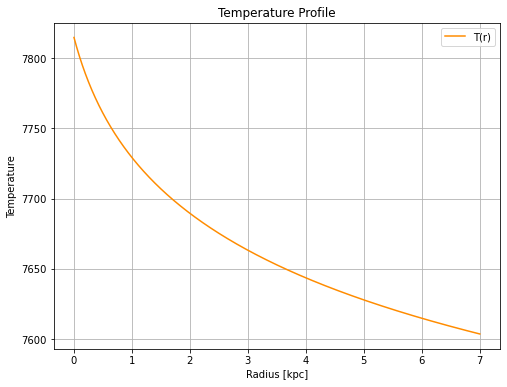}
\caption{The temperature as a function of the radius for the
collisional DM model (\ref{tanhmodel}) for the galaxy NGC4085.}
\label{NGC4085temp}
\end{figure}

For this galaxy, we shall choose $\rho_0=1.8\times
10^8$$M_{\odot}/\mathrm{Kpc}^{3}$. NGC4085 is an intermediate
spiral galaxy located approximately 16.6 Mpc away in the
constellation Ursa Major. In Figs. \ref{NGC4085dens},
\ref{NGC4085} and \ref{NGC4085temp} we present the density of the
collisional DM model, the predicted rotation curves after using an
optimization for the collisional DM model (\ref{tanhmodel}),
versus the SPARC observational data and the temperature parameter
as a function of the radius respectively. As it can be seen, the
SIDM model produces viable rotation curves compatible with the
SPARC data. Also in Tables \ref{collNGC4085}, \ref{NavaroNGC4085},
\ref{BuckertNGC4085} and \ref{EinastoNGC4085} we present the
optimization values for the SIDM model, and the other DM profiles.
Also in Table \ref{EVALUATIONNGC4085} we present the overall
evaluation of the SIDM model for the galaxy at hand. The resulting
phenomenology is viable.
\begin{table}[h!]
  \begin{center}
    \caption{Collisional Dark Matter Optimization Values}
    \label{collNGC4085}
     \begin{tabular}{|r|r|}
     \hline
      \textbf{Parameter}   & \textbf{Optimization Values}
      \\  \hline
     $\delta_{\gamma} $ & 0.0000000012
\\  \hline
$\gamma_0 $ & 1.0001 \\ \hline $K_0$ ($M_{\odot} \,
\mathrm{Kpc}^{-3} \, (\mathrm{km/s})^{2}$)& 7800  \\ \hline
    \end{tabular}
  \end{center}
\end{table}
\begin{table}[h!]
  \begin{center}
    \caption{NFW  Optimization Values}
    \label{NavaroNGC4085}
     \begin{tabular}{|r|r|}
     \hline
      \textbf{Parameter}   & \textbf{Optimization Values}
      \\  \hline
   $\rho_s$   & $5\times 10^7$
\\  \hline
$r_s$&  5.78
\\  \hline
    \end{tabular}
  \end{center}
\end{table}
\begin{table}[h!]
  \begin{center}
    \caption{Burkert Optimization Values}
    \label{BuckertNGC4085}
     \begin{tabular}{|r|r|}
     \hline
      \textbf{Parameter}   & \textbf{Optimization Values}
      \\  \hline
     $\rho_0^B$  & $1\times 10^8$
\\  \hline
$r_0$&  18.84
\\  \hline
    \end{tabular}
  \end{center}
\end{table}
\begin{table}[h!]
  \begin{center}
    \caption{Einasto Optimization Values}
    \label{EinastoNGC4085}
    \begin{tabular}{|r|r|}
     \hline
      \textbf{Parameter}   & \textbf{Optimization Values}
      \\  \hline
     $\rho_e$  &$1\times 10^7$
\\  \hline
$r_e$ & 8.07
\\  \hline
$n_e$ & 0.76
\\  \hline
    \end{tabular}
  \end{center}
\end{table}
\begin{table}[h!]
\centering \caption{Physical assessment of collisional DM
parameters (NGC4085).}
\begin{tabular}{lcc}
\hline
Parameter & Value & Physical Verdict \\
\hline
$\gamma_0$ & $1.0001$ & Essentially isothermal  \\
$\delta_\gamma$ & $1.2\times10^{-9}$ & Negligible variation  \\
$r_\gamma$ & $1.5\ \mathrm{Kpc}$ & Reasonable transition radius   \\
$K_0$ & $7.80\times10^{3}$ & Enough pressure support \\
$r_c$ & $0.5\ \mathrm{Kpc}$ & Small core scale   \\
$p$ & $0.01$ & Very shallow $K(r)$ decline; practically constant entropy \\
\hline
Overall &-& Physically consistent nearly isothermal \\
\hline
\end{tabular}
\label{EVALUATIONNGC4085}
\end{table}

\subsection{The Galaxy NGC4088}

\begin{figure}[h!]
\centering
\includegraphics[width=20pc]{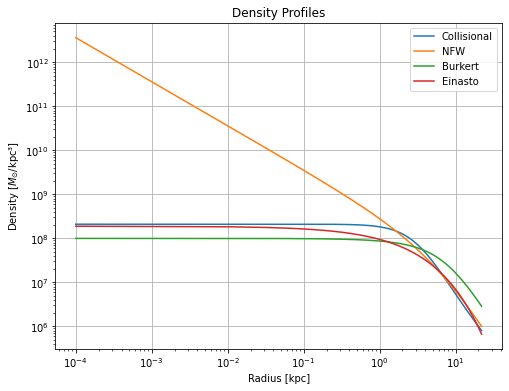}
\caption{The density of the collisional DM model (\ref{tanhmodel})
for the galaxy NGC4088, as a function of the radius.}
\label{NGC4088dens}
\end{figure}
\begin{figure}[h!]
\centering
\includegraphics[width=20pc]{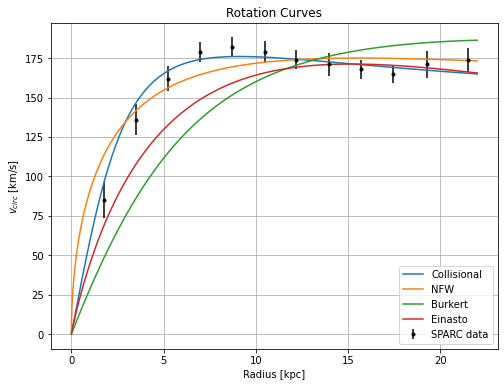}
\caption{The predicted rotation curves after using an optimization
for the collisional DM model (\ref{tanhmodel}), versus the SPARC
observational data for the galaxy NGC4088. We also plotted the
optimized curves for the NFW model, the Burkert model and the
Einasto model.} \label{NGC4088}
\end{figure}
\begin{figure}[h!]
\centering
\includegraphics[width=20pc]{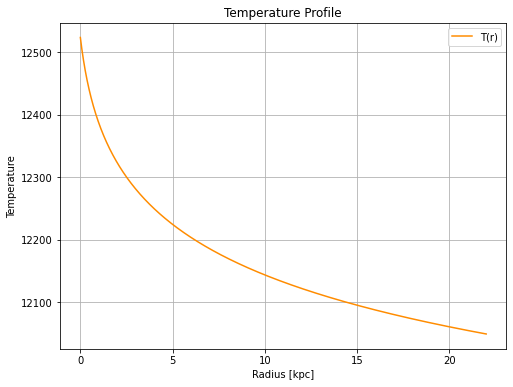}
\caption{The temperature as a function of the radius for the
collisional DM model (\ref{tanhmodel}) for the galaxy NGC4088.}
\label{NGC4088temp}
\end{figure}

For this galaxy, we shall choose $\rho_0=2.1\times
10^8$$M_{\odot}/\mathrm{Kpc}^{3}$. NGC4088 is an intermediate
spiral galaxy in the Ursa Major group, at a distance of about $D
\sim 15-17\ \mathrm{Mpc}$. In Figs. \ref{NGC4088dens},
\ref{NGC4088} and \ref{NGC4088temp} we present the density of the
collisional DM model, the predicted rotation curves after using an
optimization for the collisional DM model (\ref{tanhmodel}),
versus the SPARC observational data and the temperature parameter
as a function of the radius respectively. As it can be seen, the
SIDM model produces viable rotation curves compatible with the
SPARC data. Also in Tables \ref{collNGC4088}, \ref{NavaroNGC4088},
\ref{BuckertNGC4088} and \ref{EinastoNGC4088} we present the
optimization values for the SIDM model, and the other DM profiles.
Also in Table \ref{EVALUATIONNGC4088} we present the overall
evaluation of the SIDM model for the galaxy at hand. The resulting
phenomenology is viable.
\begin{table}[h!]
  \begin{center}
    \caption{Collisional Dark Matter Optimization Values}
    \label{collNGC4088}
     \begin{tabular}{|r|r|}
     \hline
      \textbf{Parameter}   & \textbf{Optimization Values}
      \\  \hline
     $\delta_{\gamma} $ & 0.0000000012
\\  \hline
$\gamma_0 $ & 1.0001 \\ \hline $K_0$ ($M_{\odot} \,
\mathrm{Kpc}^{-3} \, (\mathrm{km/s})^{2}$)& 12500  \\ \hline
    \end{tabular}
  \end{center}
\end{table}
\begin{table}[h!]
  \begin{center}
    \caption{NFW  Optimization Values}
    \label{NavaroNGC4088}
     \begin{tabular}{|r|r|}
     \hline
      \textbf{Parameter}   & \textbf{Optimization Values}
      \\  \hline
   $\rho_s$   & $5\times 10^7$
\\  \hline
$r_s$& 7.24
\\  \hline
    \end{tabular}
  \end{center}
\end{table}
\begin{table}[h!]
  \begin{center}
    \caption{Burkert Optimization Values}
    \label{BuckertNGC4088}
     \begin{tabular}{|r|r|}
     \hline
      \textbf{Parameter}   & \textbf{Optimization Values}
      \\  \hline
     $\rho_0^B$  & $1\times 10^8$
\\  \hline
$r_0$&  7.75
\\  \hline
    \end{tabular}
  \end{center}
\end{table}
\begin{table}[h!]
  \begin{center}
    \caption{Einasto Optimization Values}
    \label{EinastoNGC4088}
    \begin{tabular}{|r|r|}
     \hline
      \textbf{Parameter}   & \textbf{Optimization Values}
      \\  \hline
     $\rho_e$  &$1\times 10^7$
\\  \hline
$r_e$ & 8.43
\\  \hline
$n_e$ & 0.68
\\  \hline
    \end{tabular}
  \end{center}
\end{table}
\begin{table}[h!]
\centering \caption{Physical assessment of collisional DM
parameters (NGC4088).}
\begin{tabular}{lcc}
\hline
Parameter & Value & Physical Verdict \\
\hline
$\gamma_0$ & $1.0001$ & Essentially isothermal \\
$\delta_\gamma$ & $1.2\times10^{-9}$ & Effectively zero \\
$r_\gamma$ & $1.5\ \mathrm{Kpc}$ & Reasonable transition radius \\
$K_0$ & $1.25\times10^{4}$ & High entropy scale \\
$r_c$ & $0.5\ \mathrm{Kpc}$ & Small core scale -plausible for inner halo flattening \\
$p$ & $0.01$ & Extremely shallow decline of $K(r)$, $K$ nearly constant radially \\
\hline
Overall &-& Physically consistent but nearly spatially constant EoS \\
\hline
\end{tabular}
\label{EVALUATIONNGC4088}
\end{table}

\subsection{The Galaxy NGC4100 Non viable}

\begin{figure}[h!]
\centering
\includegraphics[width=20pc]{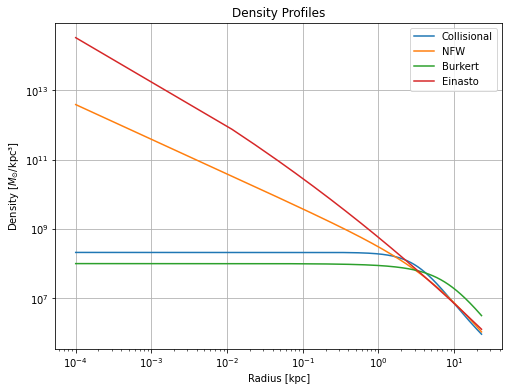}
\caption{The density of the collisional DM model (\ref{tanhmodel})
for the galaxy NGC4100, as a function of the radius.}
\label{NGC4100dens}
\end{figure}
\begin{figure}[h!]
\centering
\includegraphics[width=20pc]{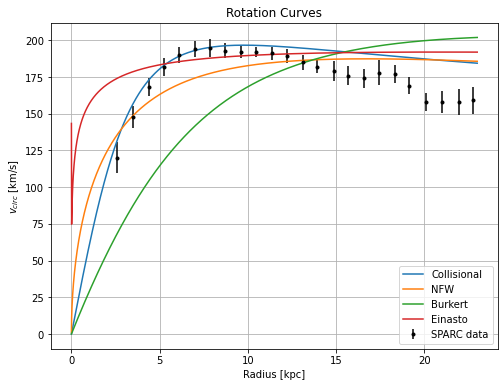}
\caption{The predicted rotation curves after using an optimization
for the collisional DM model (\ref{tanhmodel}), versus the SPARC
observational data for the galaxy NGC4100. We also plotted the
optimized curves for the NFW model, the Burkert model and the
Einasto model.} \label{NGC4100}
\end{figure}
\begin{figure}[h!]
\centering
\includegraphics[width=20pc]{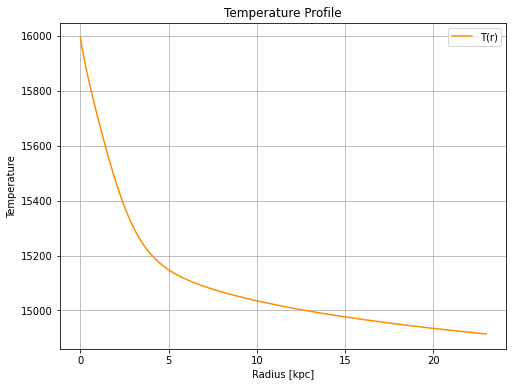}
\caption{The temperature as a function of the radius for the
collisional DM model (\ref{tanhmodel}) for the galaxy NGC4100.}
\label{NGC4100temp}
\end{figure}
\begin{figure}[h!]
\centering
\includegraphics[width=20pc]{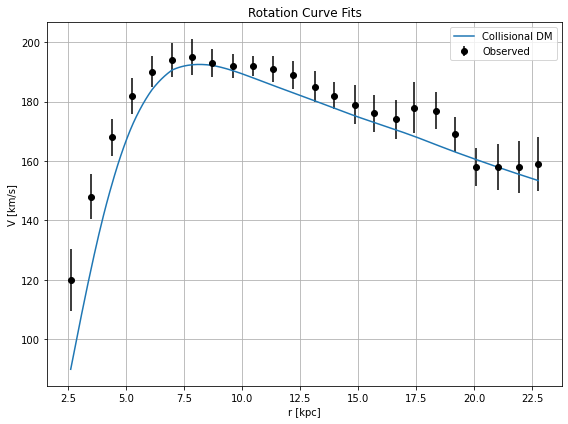}
\caption{The predicted rotation curves after using an optimization
for the collisional DM model (\ref{tanhmodel}), versus the
extended SPARC observational data for the galaxy NGC4100. The
model includes the rotation curves from all the components of the
galaxy, including gas and disk velocities, along with the
collisional DM model.} \label{extendedNGC4100}
\end{figure}

For this galaxy, we shall choose $\rho_0=2.1\times
10^8$$M_{\odot}/\mathrm{Kpc}^{3}$. NGC4100 is a fairly typical
spiral galaxy of type SAbc in the constellation Ursa Major,
belonging to the NGC 3992 (M109) group. In Figs.
\ref{NGC4100dens}, \ref{NGC4100} and \ref{NGC4100temp} we present
the density of the collisional DM model, the predicted rotation
curves after using an optimization for the collisional DM model
(\ref{tanhmodel}), versus the SPARC observational data and the
temperature parameter as a function of the radius respectively. As
it can be seen, the SIDM model produces non-viable rotation curves
incompatible with the SPARC data. Also in Tables
\ref{collNGC4100}, \ref{NavaroNGC4100}, \ref{BuckertNGC4100} and
\ref{EinastoNGC4100} we present the optimization values for the
SIDM model, and the other DM profiles. Also in Table
\ref{EVALUATIONNGC4100} we present the overall evaluation of the
SIDM model for the galaxy at hand. The resulting phenomenology is
non-viable.
\begin{table}[h!]
  \begin{center}
    \caption{Collisional Dark Matter Optimization Values}
    \label{collNGC4100}
     \begin{tabular}{|r|r|}
     \hline
      \textbf{Parameter}   & \textbf{Optimization Values}
      \\  \hline
     $\delta_{\gamma} $ & 0.0000000012
\\  \hline
$\gamma_0 $ & 1.0001 \\ \hline $K_0$ ($M_{\odot} \,
\mathrm{Kpc}^{-3} \, (\mathrm{km/s})^{2}$)& 15500  \\ \hline
    \end{tabular}
  \end{center}
\end{table}
\begin{table}[h!]
  \begin{center}
    \caption{NFW  Optimization Values}
    \label{NavaroNGC4100}
     \begin{tabular}{|r|r|}
     \hline
      \textbf{Parameter}   & \textbf{Optimization Values}
      \\  \hline
   $\rho_s$   & $5\times 10^7$
\\  \hline
$r_s$&  7.75
\\  \hline
    \end{tabular}
  \end{center}
\end{table}
\begin{table}[h!]
  \begin{center}
    \caption{Burkert Optimization Values}
    \label{BuckertNGC4100}
     \begin{tabular}{|r|r|}
     \hline
      \textbf{Parameter}   & \textbf{Optimization Values}
      \\  \hline
     $\rho_0^B$  & $1\times 10^8$
\\  \hline
$r_0$&  8.41
\\  \hline
    \end{tabular}
  \end{center}
\end{table}
\begin{table}[h!]
  \begin{center}
    \caption{Einasto Optimization Values}
    \label{EinastoNGC4100}
    \begin{tabular}{|r|r|}
     \hline
      \textbf{Parameter}   & \textbf{Optimization Values}
      \\  \hline
     $\rho_e$  &$1\times 10^7$
\\  \hline
$r_e$ & 8.43
\\  \hline
$n_e$ & 0.05
\\  \hline
    \end{tabular}
  \end{center}
\end{table}
\begin{table}[h!]
\centering \caption{Physical assessment of collisional DM
parameters (NGC4100).}
\begin{tabular}{lcc}
\hline
Parameter & Value & Physical Verdict \\
\hline
$\gamma_0$ & $1.0001$ & Essentially isothermal \\
$\delta_\gamma$ & $1.2\times10^{-9}$ & Effectively zero \\
$r_\gamma$ & $1.5\ \mathrm{Kpc}$ & Reasonable transition radius  \\
$K_0$ & $1.55\times10^{4}$ & Sets temperature/entropy scale \\
$r_c$ & $0.5\ \mathrm{Kpc}$ & Small core scale \\
$p$ & $0.01$ & Extremely shallow decline of $K(r)$; $K$ nearly constant radially \\
\hline
Overall &-& Physically consistent  \\
\hline
\end{tabular}
\label{EVALUATIONNGC4100}
\end{table}
Now the extended picture including the rotation velocity from the
other components of the galaxy, such as the disk and gas, makes
the collisional DM model viable for this galaxy. In Fig.
\ref{extendedNGC4100} we present the combined rotation curves
including the other components of the galaxy along with the
collisional matter. As it can be seen, the extended collisional DM
model is non-viable.
Also in Table \ref{evaluationextendedNGC4100} we present the
values of the free parameters of the collisional DM model for
which the optimized compatibility with the SPARC data comes for
the galaxy NGC4100.
\begin{table}[h!]
\centering \caption{Physical assessment of Extended collisional DM
parameters for NGC4100.}
\begin{tabular}{lcc}
\hline
Parameter & Value & Physical Verdict \\
\hline
$\gamma_0$ & 1.20600000 & Slightly above isothermal\\
$\delta_\gamma$ & 0.13400000 & Moderate variation\\
$K_0$ & 3000 & High entropy  \\
$ml_{\text{disk}}$ & 0.28693574 & Low-to-moderate stellar $M/L$  \\
$ml_{\text{bulge}}$ & 0.00000000 & No bulge contribution \\
\hline
Overall &-& Physically plausible \\
\hline
\end{tabular}
\label{evaluationextendedNGC4100}
\end{table}


\subsection{The Galaxy NGC4138 Marginally, Extended Viable}

\begin{figure}[h!]
\centering
\includegraphics[width=20pc]{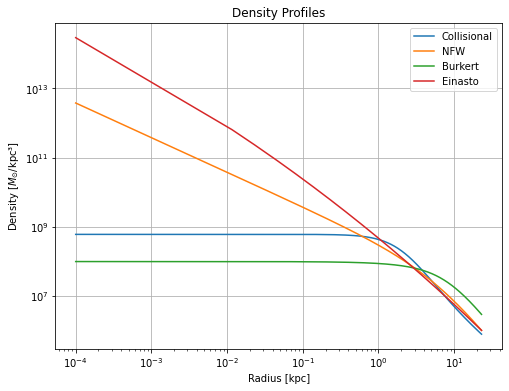}
\caption{The density of the collisional DM model (\ref{tanhmodel})
for the galaxy NGC4138, as a function of the radius.}
\label{NGC4138dens}
\end{figure}
\begin{figure}[h!]
\centering
\includegraphics[width=20pc]{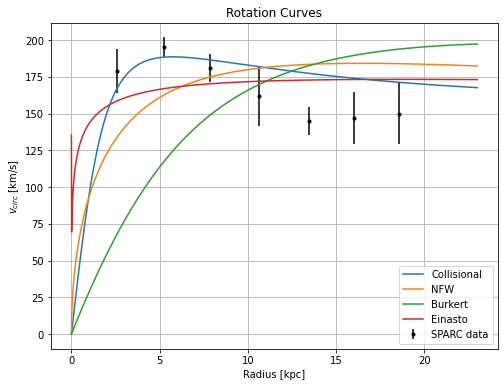}
\caption{The predicted rotation curves after using an optimization
for the collisional DM model (\ref{tanhmodel}), versus the SPARC
observational data for the galaxy NGC4138. We also plotted the
optimized curves for the NFW model, the Burkert model and the
Einasto model.} \label{NGC4138}
\end{figure}
\begin{figure}[h!]
\centering
\includegraphics[width=20pc]{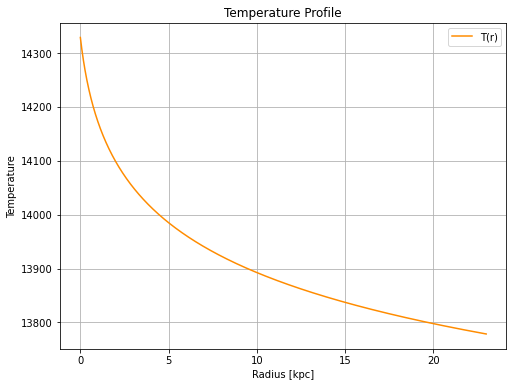}
\caption{The temperature as a function of the radius for the
collisional DM model (\ref{tanhmodel}) for the galaxy NGC4138.}
\label{NGC4138temp}
\end{figure}
\begin{figure}[h!]
\centering
\includegraphics[width=20pc]{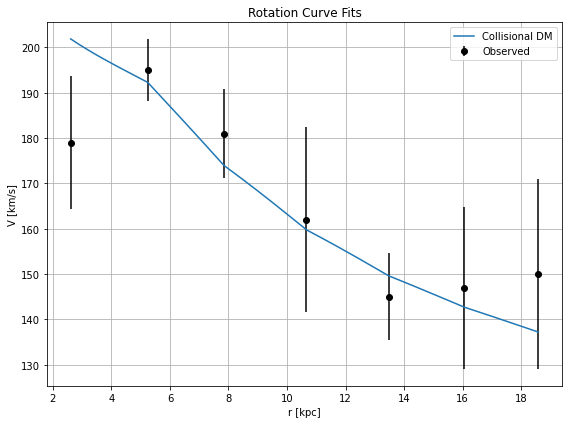}
\caption{The predicted rotation curves after using an optimization
for the collisional DM model (\ref{tanhmodel}), versus the
extended SPARC observational data for the galaxy NGC4138. The
model includes the rotation curves from all the components of the
galaxy, including gas and disk velocities, along with the
collisional DM model.} \label{extendedNGC4138}
\end{figure}

For this galaxy, we shall choose $\rho_0=6.1\times
10^8$$M_{\odot}/\mathrm{Kpc}^{3}$. NGC4138 is a
lenticular/early-spiral galaxy at a distance of about
$D\sim16\;\mathrm{Mpc}$. In Figs. \ref{NGC4138dens}, \ref{NGC4138}
and \ref{NGC4138temp} we present the density of the collisional DM
model, the predicted rotation curves after using an optimization
for the collisional DM model (\ref{tanhmodel}), versus the SPARC
observational data and the temperature parameter as a function of
the radius respectively. As it can be seen, the SIDM model
produces viable rotation curves marginally compatible with the
SPARC data. Also in Tables \ref{collNGC4138}, \ref{NavaroNGC4138},
\ref{BuckertNGC4138} and \ref{EinastoNGC4138} we present the
optimization values for the SIDM model, and the other DM profiles.
Also in Table \ref{EVALUATIONNGC4138} we present the overall
evaluation of the SIDM model for the galaxy at hand. The resulting
phenomenology is marginally viable.
\begin{table}[h!]
  \begin{center}
    \caption{Collisional Dark Matter Optimization Values}
    \label{collNGC4138}
     \begin{tabular}{|r|r|}
     \hline
      \textbf{Parameter}   & \textbf{Optimization Values}
      \\  \hline
     $\delta_{\gamma} $ & 0.0000000012
\\  \hline
$\gamma_0 $ & 1.0001 \\ \hline $K_0$ ($M_{\odot} \,
\mathrm{Kpc}^{-3} \, (\mathrm{km/s})^{2}$)& 14300  \\ \hline
    \end{tabular}
  \end{center}
\end{table}
\begin{table}[h!]
  \begin{center}
    \caption{NFW  Optimization Values}
    \label{NavaroNGC4138}
     \begin{tabular}{|r|r|}
     \hline
      \textbf{Parameter}   & \textbf{Optimization Values}
      \\  \hline
   $\rho_s$   & $5\times 10^7$
\\  \hline
$r_s$&  7.62
\\  \hline
    \end{tabular}
  \end{center}
\end{table}
\begin{table}[h!]
  \begin{center}
    \caption{Burkert Optimization Values}
    \label{BuckertNGC4138}
     \begin{tabular}{|r|r|}
     \hline
      \textbf{Parameter}   & \textbf{Optimization Values}
      \\  \hline
     $\rho_0^B$  & $1\times 10^8$
\\  \hline
$r_0$& 8.21
\\  \hline
    \end{tabular}
  \end{center}
\end{table}
\begin{table}[h!]
  \begin{center}
    \caption{Einasto Optimization Values}
    \label{EinastoNGC4138}
    \begin{tabular}{|r|r|}
     \hline
      \textbf{Parameter}   & \textbf{Optimization Values}
      \\  \hline
     $\rho_e$  &$1\times 10^7$
\\  \hline
$r_e$ & 7.61
\\  \hline
$n_e$ & 0.05
\\  \hline
    \end{tabular}
  \end{center}
\end{table}
\begin{table}[h!]
\centering \caption{Physical assessment of collisional DM
parameters (NGC4138).}
\begin{tabular}{lcc}
\hline
Parameter & Value & Physical Verdict \\
\hline
$\gamma_0$ & $1.0001$ & Essentially isothermal\\
$\delta_\gamma$ & $1.2\times10^{-9}$ & Effectively zero \\
$r_\gamma$ & $1.5\ \mathrm{Kpc}$ & Reasonable transition radius  \\
$K_0$ & $1.43\times10^{4}$ & High entropy scale \\
$r_c$ & $0.5\ \mathrm{Kpc}$ & Small core scale\\
$p$ & $0.01$ & Extremely shallow decline of $K(r)$; $K$ nearly constant radially \\
\hline
Overall &-& Physically consistent \\
\hline
\end{tabular}
\label{EVALUATIONNGC4138}
\end{table}
Now the extended picture including the rotation velocity from the
other components of the galaxy, such as the disk and gas, makes
the collisional DM model viable for this galaxy. In Fig.
\ref{extendedNGC4138} we present the combined rotation curves
including the other components of the galaxy along with the
collisional matter. As it can be seen, the extended collisional DM
model is viable.
Also in Table \ref{evaluationextendedNGC4138} we present the
values of the free parameters of the collisional DM model for
which the maximum compatibility with the SPARC data comes for the
galaxy NGC4138.
\begin{table}[h!]
\centering \caption{Physical assessment of Extended collisional DM
parameters for galaxy NGC4138.}
\begin{tabular}{lcc}
\hline
Parameter & Value & Physical Verdict \\
\hline
$\gamma_0$ & 1.03 & Near-isothermal; mild central pressure gradient, inner profile close to isothermal \\
$\delta_\gamma$ & 0.01 & Very small radial variation   \\
$K_0$ & 3000 & Moderate entropy scale; consistent with low-to-intermediate mass spiral halos \\
$ml_{disk}$ & 1.00000000 & High disk M/L \\
$ml_{bulge}$ & 0.00000000 & No bulge contribution \\
\hline
Overall &-& Physically plausible \\
\hline
\end{tabular}
\label{evaluationextendedNGC4138}
\end{table}

\subsection{The Galaxy NGC4183}

\begin{figure}[h!]
\centering
\includegraphics[width=20pc]{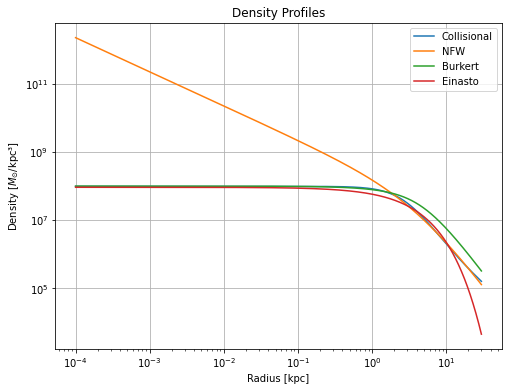}
\caption{The density of the collisional DM model (\ref{tanhmodel})
for the galaxy NGC4183, as a function of the radius.}
\label{NGC4183dens}
\end{figure}
\begin{figure}[h!]
\centering
\includegraphics[width=20pc]{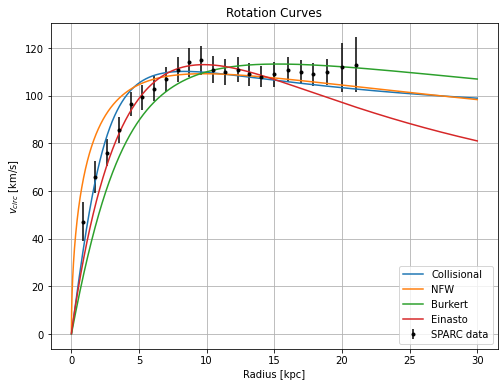}
\caption{The predicted rotation curves after using an optimization
for the collisional DM model (\ref{tanhmodel}), versus the SPARC
observational data for the galaxy NGC4183. We also plotted the
optimized curves for the NFW model, the Burkert model and the
Einasto model.} \label{NGC4183}
\end{figure}
\begin{figure}[h!]
\centering
\includegraphics[width=20pc]{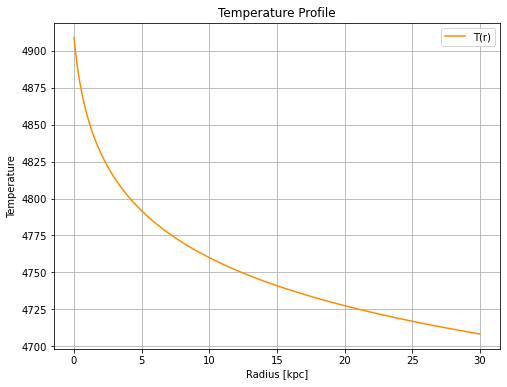}
\caption{The temperature as a function of the radius for the
collisional DM model (\ref{tanhmodel}) for the galaxy NGC4183.}
\label{NGC4183temp}
\end{figure}

For this galaxy, we shall choose $\rho_0=9.8\times
10^7$$M_{\odot}/\mathrm{Kpc}^{3}$. NGC4183 is a normal Sc-type
spiral galaxy in the Ursa Major region, at a distance of about
(\(\sim 17\) Mpc). In Figs. \ref{NGC4183dens}, \ref{NGC4183} and
\ref{NGC4183temp} we present the density of the collisional DM
model, the predicted rotation curves after using an optimization
for the collisional DM model (\ref{tanhmodel}), versus the SPARC
observational data and the temperature parameter as a function of
the radius respectively. As it can be seen, the SIDM model
produces viable rotation curves compatible with the SPARC data.
Also in Tables \ref{collNGC4183}, \ref{NavaroNGC4183},
\ref{BuckertNGC4183} and \ref{EinastoNGC4183} we present the
optimization values for the SIDM model, and the other DM profiles.
Also in Table \ref{EVALUATIONNGC4183} we present the overall
evaluation of the SIDM model for the galaxy at hand. The resulting
phenomenology is viable.
\begin{table}[h!]
  \begin{center}
    \caption{Collisional Dark Matter Optimization Values}
    \label{collNGC4183}
     \begin{tabular}{|r|r|}
     \hline
      \textbf{Parameter}   & \textbf{Optimization Values}
      \\  \hline
     $\delta_{\gamma} $ & 0.0000000012
\\  \hline
$\gamma_0 $ & 1.0001 \\ \hline $K_0$ ($M_{\odot} \,
\mathrm{Kpc}^{-3} \, (\mathrm{km/s})^{2}$)& 4900  \\ \hline
    \end{tabular}
  \end{center}
\end{table}
\begin{table}[h!]
  \begin{center}
    \caption{NFW  Optimization Values}
    \label{NavaroNGC4183}
     \begin{tabular}{|r|r|}
     \hline
      \textbf{Parameter}   & \textbf{Optimization Values}
      \\  \hline
   $\rho_s$   & $5\times 10^7$
\\  \hline
$r_s$&  4.52
\\  \hline
    \end{tabular}
  \end{center}
\end{table}
\begin{table}[h!]
  \begin{center}
    \caption{Burkert Optimization Values}
    \label{BuckertNGC4183}
     \begin{tabular}{|r|r|}
     \hline
      \textbf{Parameter}   & \textbf{Optimization Values}
      \\  \hline
     $\rho_0^B$  & $1\times 10^8$
\\  \hline
$r_0$&  4.70
\\  \hline
    \end{tabular}
  \end{center}
\end{table}
\begin{table}[h!]
  \begin{center}
    \caption{Einasto Optimization Values}
    \label{EinastoNGC4183}
    \begin{tabular}{|r|r|}
     \hline
      \textbf{Parameter}   & \textbf{Optimization Values}
      \\  \hline
     $\rho_e$  &$1\times 10^7$
\\  \hline
$r_e$ & 5.69
\\  \hline
$n_e$ & 0.9
\\  \hline
    \end{tabular}
  \end{center}
\end{table}
\begin{table}[h!]
\centering \caption{Physical assessment of collisional DM
parameters (NGC4183).}
\begin{tabular}{lcc}
\hline
Parameter & Value & Physical Verdict \\
\hline
$\gamma_0$ & $1.0001$ & Essentially isothermal\\
$\delta_\gamma$ & $1.2\times10^{-9}$ & Effectively zero   \\
$r_\gamma$ & $1.5\ \mathrm{Kpc}$ & Reasonable transition radius \\
$K_0$ & $4.90\times10^{3}$ &Pressure support sufficient \\
$r_c$ & $0.5\ \mathrm{Kpc}$ & Small core scale \\
$p$ & $0.01$ & Extremely shallow decline of $K(r)$ \\
\hline
Overall &-& Physically consistent \\
\hline
\end{tabular}
\label{EVALUATIONNGC4183}
\end{table}


\subsection{The Galaxy NGC4217}

\begin{figure}[h!]
\centering
\includegraphics[width=20pc]{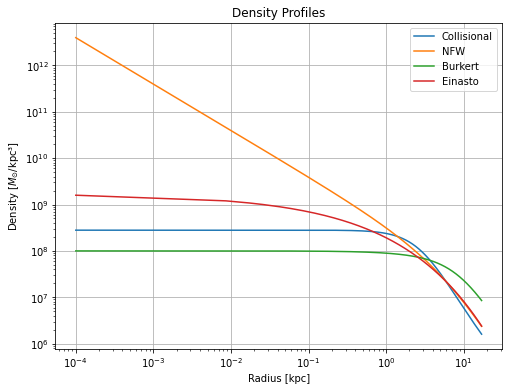}
\caption{The density of the collisional DM model (\ref{tanhmodel})
for the galaxy NGC4217, as a function of the radius.}
\label{NGC4217dens}
\end{figure}
\begin{figure}[h!]
\centering
\includegraphics[width=20pc]{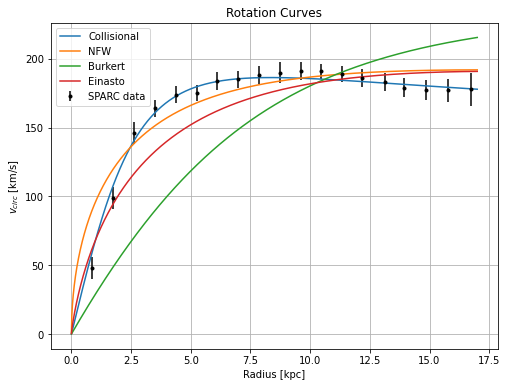}
\caption{The predicted rotation curves after using an optimization
for the collisional DM model (\ref{tanhmodel}), versus the SPARC
observational data for the galaxy NGC4217. We also plotted the
optimized curves for the NFW model, the Burkert model and the
Einasto model.} \label{NGC4217}
\end{figure}
\begin{figure}[h!]
\centering
\includegraphics[width=20pc]{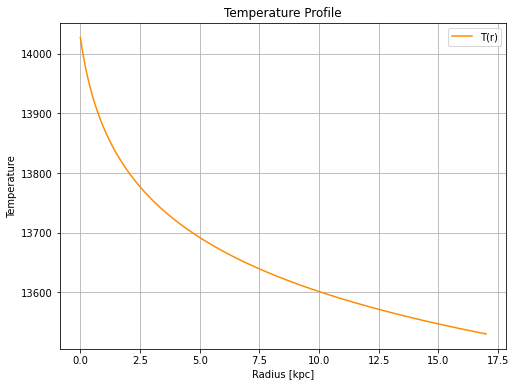}
\caption{The temperature as a function of the radius for the
collisional DM model (\ref{tanhmodel}) for the galaxy NGC4217.}
\label{NGC4217temp}
\end{figure}

For this galaxy, we shall choose $\rho_0=2.5\times
10^8$$M_{\odot}/\mathrm{Kpc}^{3}$. NGC4217 is an edge-on spiral
galaxy of type Sb, located approximately \(D \sim 18\) Mpc from
the Milky Way in the constellation Canes Venatici. In Figs.
\ref{NGC4217dens}, \ref{NGC4217} and \ref{NGC4217temp} we present
the density of the collisional DM model, the predicted rotation
curves after using an optimization for the collisional DM model
(\ref{tanhmodel}), versus the SPARC observational data and the
temperature parameter as a function of the radius respectively. As
it can be seen, the SIDM model produces viable rotation curves
compatible with the SPARC data. Also in Tables \ref{collNGC4217},
\ref{NavaroNGC4217}, \ref{BuckertNGC4217} and \ref{EinastoNGC4217}
we present the optimization values for the SIDM model, and the
other DM profiles. Also in Table \ref{EVALUATIONNGC4217} we
present the overall evaluation of the SIDM model for the galaxy at
hand. The resulting phenomenology is viable.
\begin{table}[h!]
  \begin{center}
    \caption{Collisional Dark Matter Optimization Values}
    \label{collNGC4217}
     \begin{tabular}{|r|r|}
     \hline
      \textbf{Parameter}   & \textbf{Optimization Values}
      \\  \hline
     $\delta_{\gamma} $ & 0.0000000012
\\  \hline
$\gamma_0 $ & 1.0001 \\ \hline $K_0$ ($M_{\odot} \,
\mathrm{Kpc}^{-3} \, (\mathrm{km/s})^{2}$)& 14000  \\ \hline
    \end{tabular}
  \end{center}
\end{table}
\begin{table}[h!]
  \begin{center}
    \caption{NFW  Optimization Values}
    \label{NavaroNGC4217}
     \begin{tabular}{|r|r|}
     \hline
      \textbf{Parameter}   & \textbf{Optimization Values}
      \\  \hline
   $\rho_s$   & $5\times 10^7$
\\  \hline
$r_s$&  7.94
\\  \hline
    \end{tabular}
  \end{center}
\end{table}
\begin{table}[h!]
  \begin{center}
    \caption{Burkert Optimization Values}
    \label{BuckertNGC4217}
     \begin{tabular}{|r|r|}
     \hline
      \textbf{Parameter}   & \textbf{Optimization Values}
      \\  \hline
     $\rho_0^B$  & $1\times 10^8$
\\  \hline
$r_0$&  9.49
\\  \hline
    \end{tabular}
  \end{center}
\end{table}
\begin{table}[h!]
  \begin{center}
    \caption{Einasto Optimization Values}
    \label{EinastoNGC4217}
    \begin{tabular}{|r|r|}
     \hline
      \textbf{Parameter}   & \textbf{Optimization Values}
      \\  \hline
     $\rho_e$  &$1\times 10^7$
\\  \hline
$r_e$ & 9.05
\\  \hline
$n_e$ & 0.39
\\  \hline
    \end{tabular}
  \end{center}
\end{table}
\begin{table}[h!]
\centering \caption{Physical assessment of collisional DM
parameters (NGC4217).}
\begin{tabular}{lcc}
\hline
Parameter & Value & Physical Verdict \\
\hline
$\gamma_0$ & $1.0001$ & Essentially isothermal \\
$\delta_\gamma$ & $1.2\times10^{-9}$ & Effectively zero   \\
$r_\gamma$ & $1.5\ \mathrm{Kpc}$ & Reasonable transition radius\\
$K_0$ & $1.40\times10^{4}$ & Enough cental pressure support \\
$r_c$ & $0.5\ \mathrm{Kpc}$ & Small core scale - plausible for inner halo flattening \\
$p$ & $0.01$ & Extremely shallow decline of $K(r)$; $K$ nearly constant radially \\
\hline
Overall &-& Physically consistent \\
\hline
\end{tabular}
\label{EVALUATIONNGC4217}
\end{table}


\subsection{The Galaxy NGC4389}

\begin{figure}[h!]
\centering
\includegraphics[width=20pc]{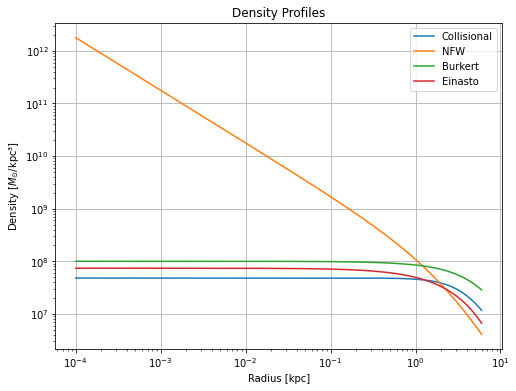}
\caption{The density of the collisional DM model (\ref{tanhmodel})
for the galaxy NGC4389, as a function of the radius.}
\label{NGC4389dens}
\end{figure}

\begin{figure}[h!]
\centering
\includegraphics[width=20pc]{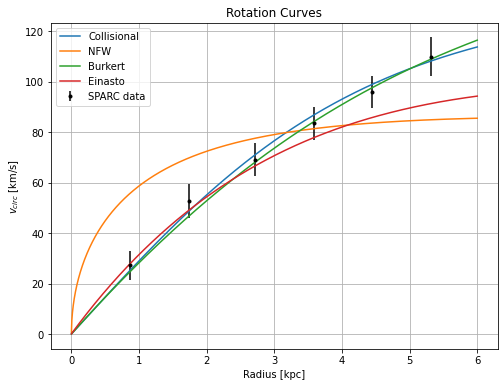}
\caption{The predicted rotation curves after using an optimization
for the collisional DM model (\ref{tanhmodel}), versus the SPARC
observational data for the galaxy NGC4389. We also plotted the
optimized curves for the NFW model, the Burkert model and the
Einasto model.} \label{NGC4389}
\end{figure}

\begin{figure}[h!]
\centering
\includegraphics[width=20pc]{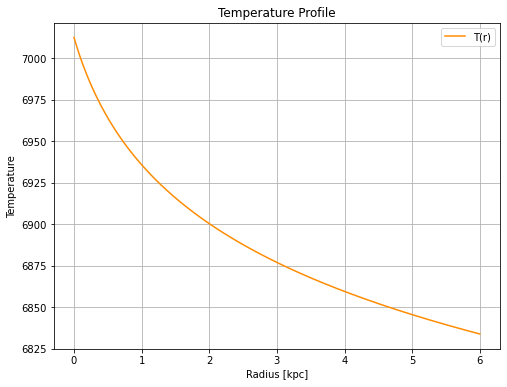}
\caption{The temperature as a function of the radius for the
collisional DM model (\ref{tanhmodel}) for the galaxy NGC4389.}
\label{NGC4389temp}
\end{figure}

For this galaxy, we shall choose $\rho_0=4.8\times
10^7$$M_{\odot}/\mathrm{Kpc}^{3}$. NGC4389 is a barred spiral
galaxy of type SB(rs)cd, located approximately \( D \sim 18.5 \)
Mpc from the Milky Way in the constellation Coma Berenices. In
Figs. \ref{NGC4389dens}, \ref{NGC4389} and \ref{NGC4389temp} we
present the density of the collisional DM model, the predicted
rotation curves after using an optimization for the collisional DM
model (\ref{tanhmodel}), versus the SPARC observational data and
the temperature parameter as a function of the radius
respectively. As it can be seen, the SIDM model produces viable
rotation curves compatible with the SPARC data. Also in Tables
\ref{collNGC4389}, \ref{NavaroNGC4389}, \ref{BuckertNGC4389} and
\ref{EinastoNGC4389} we present the optimization values for the
SIDM model, and the other DM profiles. Also in Table
\ref{EVALUATIONNGC4389} we present the overall evaluation of the
SIDM model for the galaxy at hand. The resulting phenomenology is
viable.
\begin{table}[h!]
  \begin{center}
    \caption{Collisional Dark Matter Optimization Values}
    \label{collNGC4389}
     \begin{tabular}{|r|r|}
     \hline
      \textbf{Parameter}   & \textbf{Optimization Values}
      \\  \hline
     $\delta_{\gamma} $ & 0.0000000012
\\  \hline
$\gamma_0 $ & 1.0001 \\ \hline $K_0$ ($M_{\odot} \,
\mathrm{Kpc}^{-3} \, (\mathrm{km/s})^{2}$)& 7000 \\ \hline
    \end{tabular}
  \end{center}
\end{table}
\begin{table}[h!]
  \begin{center}
    \caption{NFW  Optimization Values}
    \label{NavaroNGC4389}
     \begin{tabular}{|r|r|}
     \hline
      \textbf{Parameter}   & \textbf{Optimization Values}
      \\  \hline
   $\rho_s$   & $5\times 10^7$
\\  \hline
$r_s$&  3.56
\\  \hline
    \end{tabular}
  \end{center}
\end{table}
\begin{table}[h!]
  \begin{center}
    \caption{Burkert Optimization Values}
    \label{BuckertNGC4389}
     \begin{tabular}{|r|r|}
     \hline
      \textbf{Parameter}   & \textbf{Optimization Values}
      \\  \hline
     $\rho_0^B$  & $5\times 10^7$
\\  \hline
$r_0$&  6.59
\\  \hline
    \end{tabular}
  \end{center}
\end{table}
\begin{table}[h!]
  \begin{center}
    \caption{Einasto Optimization Values}
    \label{EinastoNGC4389}
    \begin{tabular}{|r|r|}
     \hline
      \textbf{Parameter}   & \textbf{Optimization Values}
      \\  \hline
     $\rho_e$  &$1.4\times 10^7$
\\  \hline
$r_e$ & 4.98
\\  \hline
$n_e$ & 1
\\  \hline
    \end{tabular}
  \end{center}
\end{table}
\begin{table}[h!]
\centering \caption{Physical assessment of collisional DM
parameters (NGC4389).}
\begin{tabular}{lcc}
\hline
Parameter & Value & Physical Verdict \\
\hline
$\gamma_0$ & $1.0001$ & Essentially isothermal \\
$\delta_\gamma$ & $1.2\times10^{-9}$ & Effectively zero   \\
$r_\gamma$ & $1.5\ \mathrm{Kpc}$ & Reasonable transition radius \\
$K_0$ & $7.0\times10^{3}$ & Modest pressure support \\
$r_c$ & $0.5\ \mathrm{Kpc}$ & Small core scale \\
$p$ & $0.01$ & Extremely shallow decline of $K(r)$ \\
\hline
Overall &-& Physically consistent  \\
\hline
\end{tabular}
\label{EVALUATIONNGC4389}
\end{table}


\subsection{The Galaxy  NGC4559 Non-viable}

\begin{figure}[h!]
\centering
\includegraphics[width=20pc]{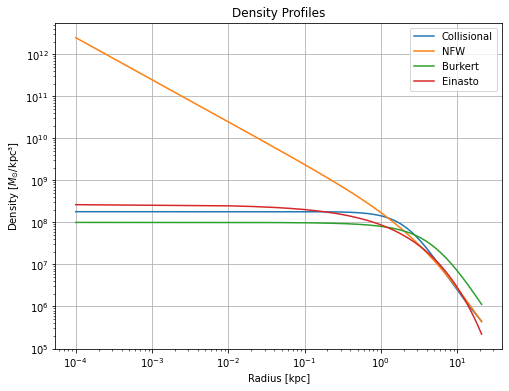}
\caption{The density of the collisional DM model (\ref{tanhmodel})
for the galaxy NGC4559, as a function of the radius.}
\label{NGC4559dens}
\end{figure}
\begin{figure}[h!]
\centering
\includegraphics[width=20pc]{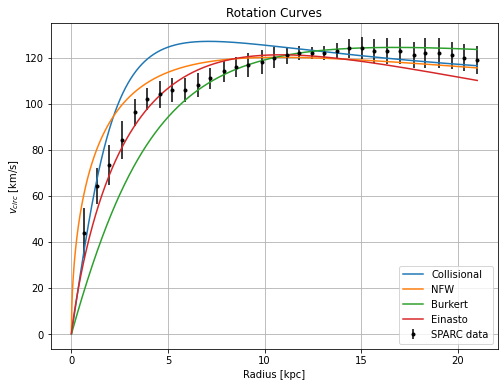}
\caption{The predicted rotation curves after using an optimization
for the collisional DM model (\ref{tanhmodel}), versus the SPARC
observational data for the galaxy NGC4559. We also plotted the
optimized curves for the NFW model, the Burkert model and the
Einasto model.} \label{NGC4559}
\end{figure}
\begin{figure}[h!]
\centering
\includegraphics[width=20pc]{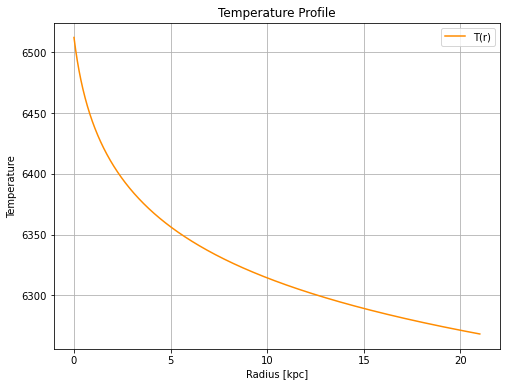}
\caption{The temperature as a function of the radius for the
collisional DM model (\ref{tanhmodel}) for the galaxy NGC4559.}
\label{NGC4559temp}
\end{figure}
\begin{figure}[h!]
\centering
\includegraphics[width=20pc]{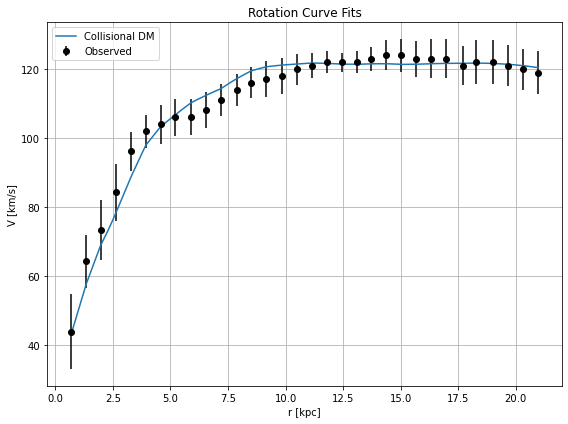}
\caption{The predicted rotation curves after using an optimization
for the collisional DM model (\ref{tanhmodel}), versus the
extended SPARC observational data for the galaxy NGC4559. The
model includes the rotation curves from all the components of the
galaxy, including gas and disk velocities, along with the
collisional DM model.} \label{extendedNGC4559}
\end{figure}

For this galaxy, we shall choose $\rho_0=1.8\times
10^8$$M_{\odot}/\mathrm{Kpc}^{3}$. NGC4559 is a spiral galaxy of
type Scd II (ordinary late-type) at a distance of about
\(9.7\;\mathrm{Mpc}\). In Figs. \ref{NGC4559dens}, \ref{NGC4559}
and \ref{NGC4559temp} we present the density of the collisional DM
model, the predicted rotation curves after using an optimization
for the collisional DM model (\ref{tanhmodel}), versus the SPARC
observational data and the temperature parameter as a function of
the radius respectively. As it can be seen, the SIDM model
produces non-viable rotation curves incompatible with the SPARC
data. Also in Tables \ref{collNGC4559}, \ref{NavaroNGC4559},
\ref{BuckertNGC4559} and \ref{EinastoNGC4559} we present the
optimization values for the SIDM model, and the other DM profiles.
Also in Table \ref{EVALUATIONNGC4559} we present the overall
evaluation of the SIDM model for the galaxy at hand. The resulting
phenomenology is non-viable.
\begin{table}[h!]
  \begin{center}
    \caption{Collisional Dark Matter Optimization Values}
    \label{collNGC4559}
     \begin{tabular}{|r|r|}
     \hline
      \textbf{Parameter}   & \textbf{Optimization Values}
      \\  \hline
     $\delta_{\gamma} $ & 0.0000000012
\\  \hline
$\gamma_0 $ & 1.0001 \\ \hline $K_0$ ($M_{\odot} \,
\mathrm{Kpc}^{-3} \, (\mathrm{km/s})^{2}$)& 6500  \\ \hline
    \end{tabular}
  \end{center}
\end{table}
\begin{table}[h!]
  \begin{center}
    \caption{NFW  Optimization Values}
    \label{NavaroNGC4559}
     \begin{tabular}{|r|r|}
     \hline
      \textbf{Parameter}   & \textbf{Optimization Values}
      \\  \hline
   $\rho_s$   & $5\times 10^7$
\\  \hline
$r_s$&  4.97
\\  \hline
    \end{tabular}
  \end{center}
\end{table}
\begin{table}[h!]
  \begin{center}
    \caption{Burkert Optimization Values}
    \label{BuckertNGC4559}
     \begin{tabular}{|r|r|}
     \hline
      \textbf{Parameter}   & \textbf{Optimization Values}
      \\  \hline
     $\rho_0^B$  & $1\times 10^8$
\\  \hline
$r_0$&  5.16
\\  \hline
    \end{tabular}
  \end{center}
\end{table}
\begin{table}[h!]
  \begin{center}
    \caption{Einasto Optimization Values}
    \label{EinastoNGC4559}
    \begin{tabular}{|r|r|}
     \hline
      \textbf{Parameter}   & \textbf{Optimization Values}
      \\  \hline
     $\rho_e$  &$1\times 10^7$
\\  \hline
$r_e$ & 5.92
\\  \hline
$n_e$ & 0.61
\\  \hline
    \end{tabular}
  \end{center}
\end{table}
\begin{table}[h!]
\centering \caption{Physical assessment of collisional DM
parameters (NGC4559).}
\begin{tabular}{lcc}
\hline
Parameter & Value & Physical Verdict \\
\hline
$\gamma_0$ & $1.0001$ & Essentially isothermal  \\
$\delta_\gamma$ & $1.2\times10^{-9}$ & Effectively zero   \\
$r_\gamma$ & $1.5\ \mathrm{Kpc}$ & Reasonable transition radius \\
$K_0$ & $6.5\times10^{3}$ & Moderate pressure support \\
$r_c$ & $0.5\ \mathrm{Kpc}$ & Small core scale \\
$p$ & $0.01$ & Extremely shallow decline of $K(r)$ \\
\hline
Overall &-& Physically consistent \\
\hline
\end{tabular}
\label{EVALUATIONNGC4559}
\end{table}
Now the extended picture including the rotation velocity from the
other components of the galaxy, such as the disk and gas, makes
the collisional DM model viable for this galaxy. In Fig.
\ref{extendedNGC4559} we present the combined rotation curves
including the other components of the galaxy along with the
collisional matter. As it can be seen, the extended collisional DM
model is non-viable.
Also in Table \ref{evaluationextendedNGC4559} we present the
values of the free parameters of the collisional DM model for
which the maximum compatibility with the SPARC data comes for the
galaxy NGC4559.
\begin{table}[h!]
\centering \caption{Physical assessment of Extended collisional DM
parameters for NGC4559.}
\begin{tabular}{lcc}
\hline
Parameter & Value & Physical Verdict \\
\hline
$\gamma_0$ & 1.02107194 & Very close to isothermal \\
$\delta_\gamma$ & 0.000000012 & No radial variation \\
$K_0$ & 3000 & Moderate entropy  \\
$ml_{\text{disk}}$ & 0.70014448 & At the upper bound \\
$ml_{\text{bulge}}$ & 0.00000000 & No bulge component \\
\hline
Overall &-& Physically plausible\\
\hline
\end{tabular}
\label{evaluationextendedNGC4559}
\end{table}

\subsection{The Galaxy NGC5005 Non-Viable}

\begin{figure}[h!]
\centering
\includegraphics[width=20pc]{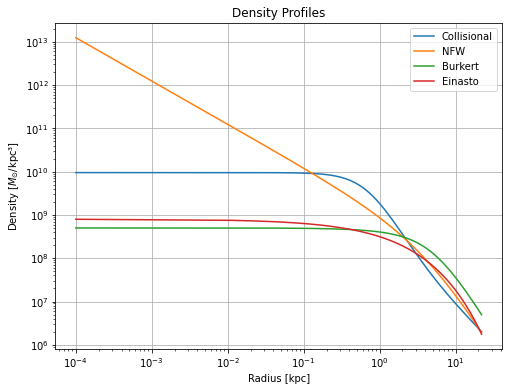}
\caption{The density of the collisional DM model (\ref{tanhmodel})
for the galaxy NGC5005, as a function of the radius.}
\label{NGC5005dens}
\end{figure}
\begin{figure}[h!]
\centering
\includegraphics[width=20pc]{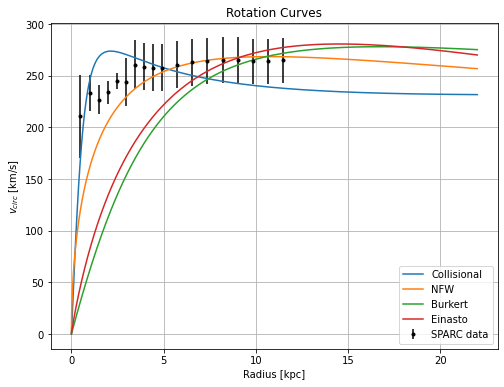}
\caption{The predicted rotation curves after using an optimization
for the collisional DM model (\ref{tanhmodel}), versus the SPARC
observational data for the galaxy NGC5005. We also plotted the
optimized curves for the NFW model, the Burkert model and the
Einasto model.} \label{NGC5005}
\end{figure}
\begin{figure}[h!]
\centering
\includegraphics[width=20pc]{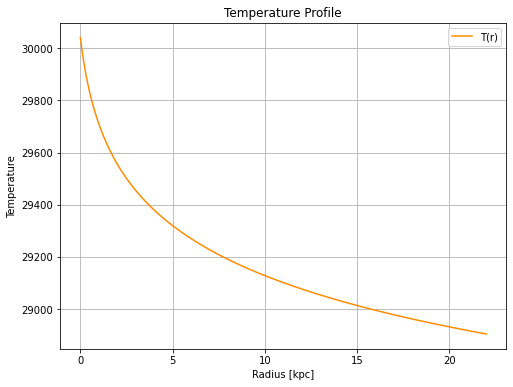}
\caption{The temperature as a function of the radius for the
collisional DM model (\ref{tanhmodel}) for the galaxy NGC5005.}
\label{NGC5005temp}
\end{figure}
\begin{figure}[h!]
\centering
\includegraphics[width=20pc]{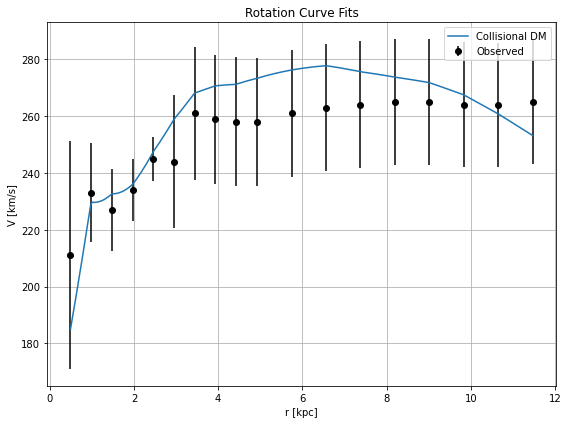}
\caption{The predicted rotation curves after using an optimization
for the collisional DM model (\ref{tanhmodel}), versus the
extended SPARC observational data for the galaxy NGC5005. The
model includes the rotation curves from all the components of the
galaxy, including gas and disk velocities, along with the
collisional DM model.} \label{extendedNGC5005}
\end{figure}

For this galaxy, we shall choose $\rho_0=9.5\times
10^9$$M_{\odot}/\mathrm{Kpc}^{3}$. NGC5005 is an inclined
intermediate barred spiral galaxy located in the constellation
Canes Venatici at a distance of order \(20\ \mathrm{Mpc}\). In
Figs. \ref{NGC5005dens}, \ref{NGC5005} and \ref{NGC5005temp} we
present the density of the collisional DM model, the predicted
rotation curves after using an optimization for the collisional DM
model (\ref{tanhmodel}), versus the SPARC observational data and
the temperature parameter as a function of the radius
respectively. As it can be seen, the SIDM model produces
non-viable rotation curves incompatible with the SPARC data. Also
in Tables \ref{collNGC5005}, \ref{NavaroNGC5005},
\ref{BuckertNGC5005} and \ref{EinastoNGC5005} we present the
optimization values for the SIDM model, and the other DM profiles.
Also in Table \ref{EVALUATIONNGC5005} we present the overall
evaluation of the SIDM model for the galaxy at hand. The resulting
phenomenology is non-viable.
\begin{table}[h!]
  \begin{center}
    \caption{Collisional Dark Matter Optimization Values}
    \label{collNGC5005}
     \begin{tabular}{|r|r|}
     \hline
      \textbf{Parameter}   & \textbf{Optimization Values}
      \\  \hline
     $\delta_{\gamma} $ & 0.0000000012
\\  \hline
$\gamma_0 $ & 1.0001 \\ \hline $K_0$ ($M_{\odot} \,
\mathrm{Kpc}^{-3} \, (\mathrm{km/s})^{2}$)& 30000  \\ \hline
    \end{tabular}
  \end{center}
\end{table}
\begin{table}[h!]
  \begin{center}
    \caption{NFW  Optimization Values}
    \label{NavaroNGC5005}
     \begin{tabular}{|r|r|}
     \hline
      \textbf{Parameter}   & \textbf{Optimization Values}
      \\  \hline
   $\rho_s$   & $52.5\times 10^8$
\\  \hline
$r_s$&  4.97
\\  \hline
    \end{tabular}
  \end{center}
\end{table}
\begin{table}[h!]
  \begin{center}
    \caption{Burkert Optimization Values}
    \label{BuckertNGC5005}
     \begin{tabular}{|r|r|}
     \hline
      \textbf{Parameter}   & \textbf{Optimization Values}
      \\  \hline
     $\rho_0^B$  & $5\times 10^8$
\\  \hline
$r_0$&  5.16
\\  \hline
    \end{tabular}
  \end{center}
\end{table}
\begin{table}[h!]
  \begin{center}
    \caption{Einasto Optimization Values}
    \label{EinastoNGC5005}
    \begin{tabular}{|r|r|}
     \hline
      \textbf{Parameter}   & \textbf{Optimization Values}
      \\  \hline
     $\rho_e$  &$3\times 10^7$
\\  \hline
$r_e$ & 7.92
\\  \hline
$n_e$ & 0.61
\\  \hline
    \end{tabular}
  \end{center}
\end{table}
\begin{table}[h!]
\centering \caption{Physical assessment of collisional DM
parameters for NGC5005.}
\begin{tabular}{lcc}
\hline
Parameter & Value & Physical Verdict \\
\hline
$\gamma_0$ & $1.00006$ & Practically isothermal \\
$\delta_\gamma$ & $6.1\times10^{-9}$ & Essentially zero  \\
$r_\gamma$ & $1.5\ \mathrm{Kpc}$ & Transition inside inner disc \\
$K_0$ & $6.5\times10^{3}$ & Enough pressure support \\
$r_c$ & $0.5\ \mathrm{Kpc}$ & Very small core scale \\
$p$ & $0.01$ & Nearly constant $K(r)$  \\
\hline
Overall &-& Numerically stable and almost-isothermal \\
\hline
\end{tabular}
\label{EVALUATIONNGC5005}
\end{table}
Now the extended picture including the rotation velocity from the
other components of the galaxy, such as the disk and gas, makes
the collisional DM model viable for this galaxy. In Fig.
\ref{extendedNGC5005} we present the combined rotation curves
including the other components of the galaxy along with the
collisional matter. As it can be seen, the extended collisional DM
model is non-viable.
Also in Table \ref{evaluationextendedNGC5005} we present the
values of the free parameters of the collisional DM model for
which the maximum compatibility with the SPARC data comes for the
galaxy NGC5005.
\begin{table}[h!]
\centering \caption{Physical assessment of Extended collisional DM
parameters for NGC5005.}
\begin{tabular}{lcc}
\hline
Parameter & Value & Physical Verdict \\
\hline
$\gamma_0$ & 1.14640231 & Slightly above isothermal \\
$\delta_\gamma$ & 0.10000000 & Moderate radial variation  \\
$K_0$ & 3000 & High entropy   \\
$ml_{\text{disk}}$ & 0.77388584 & Moderate-to-high stellar $M/L$; disk contributes substantially but is not fully maximal \\
$ml_{\text{bulge}}$ & 0.70000000 & Significant bulge mass-to-light ratio \\
\hline
Overall &-& Physically plausible\\
\hline
\end{tabular}
\label{evaluationextendedNGC5005}
\end{table}

\subsection{The Galaxy NGC5033 Non-viable Extended Marginally Viable}

\begin{figure}[h!]
\centering
\includegraphics[width=20pc]{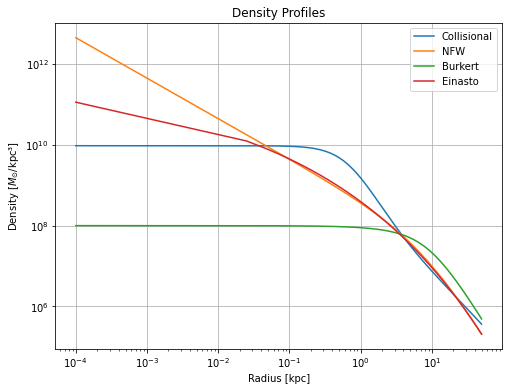}
\caption{The density of the collisional DM model (\ref{tanhmodel})
for the galaxy NGC5033, as a function of the radius.}
\label{NGC5033dens}
\end{figure}

\begin{figure}[h!]
\centering
\includegraphics[width=20pc]{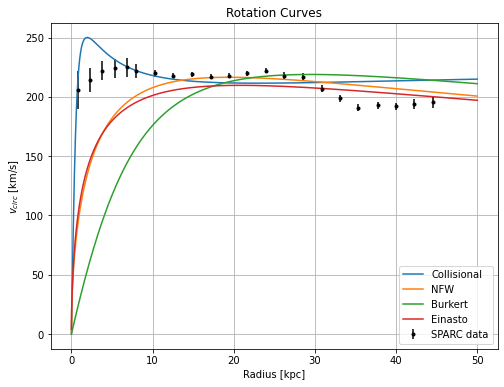}
\caption{The predicted rotation curves after using an optimization
for the collisional DM model (\ref{tanhmodel}), versus the SPARC
observational data for the galaxy NGC5033. We also plotted the
optimized curves for the NFW model, the Burkert model and the
Einasto model.} \label{NGC5033}
\end{figure}

\begin{figure}[h!]
\centering
\includegraphics[width=20pc]{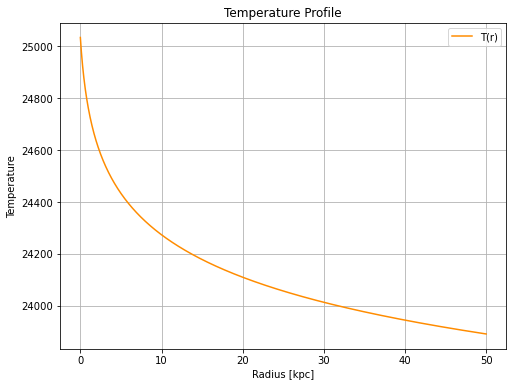}
\caption{The temperature as a function of the radius for the
collisional DM model (\ref{tanhmodel}) for the galaxy NGC5033.}
\label{NGC5033temp}
\end{figure}

\begin{figure}[h!]
\centering
\includegraphics[width=20pc]{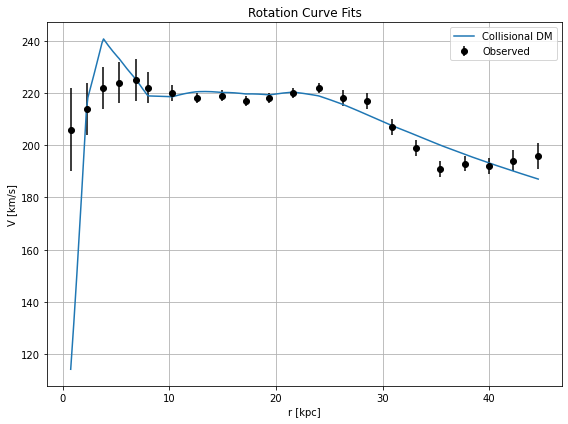}
\caption{The predicted rotation curves after using an optimization
for the collisional DM model (\ref{tanhmodel}), versus the
extended SPARC observational data for the galaxy NGC5033. The
model includes the rotation curves from all the components of the
galaxy, including gas and disk velocities, along with the
collisional DM model.} \label{extendedNGC5033}
\end{figure}

For this galaxy, we shall choose $\rho_0=9.5\times
10^9$$M_{\odot}/\mathrm{Kpc}^{3}$. NGC5033 is a large unbarred
spiral galaxy of type SA(s)c, located in the constellation Canes
Venatici at a distance of about \(18\ \mathrm{Mpc}\). In Figs.
\ref{NGC5033dens}, \ref{NGC5033} and \ref{NGC5033temp} we present
the density of the collisional DM model, the predicted rotation
curves after using an optimization for the collisional DM model
(\ref{tanhmodel}), versus the SPARC observational data and the
temperature parameter as a function of the radius respectively. As
it can be seen, the SIDM model produces non-viable rotation curves
incompatible with the SPARC data. Also in Tables
\ref{collNGC5033}, \ref{NavaroNGC5033}, \ref{BuckertNGC5033} and
\ref{EinastoNGC5033} we present the optimization values for the
SIDM model, and the other DM profiles. Also in Table
\ref{EVALUATIONNGC5033} we present the overall evaluation of the
SIDM model for the galaxy at hand. The resulting phenomenology is
non-viable.
\begin{table}[h!]
  \begin{center}
    \caption{Collisional Dark Matter Optimization Values}
    \label{collNGC5033}
     \begin{tabular}{|r|r|}
     \hline
      \textbf{Parameter}   & \textbf{Optimization Values}
      \\  \hline
     $\delta_{\gamma} $ &  0.0000000012
\\  \hline
$\gamma_0 $ & 1.0001 \\ \hline $K_0$ ($M_{\odot} \,
\mathrm{Kpc}^{-3} \, (\mathrm{km/s})^{2}$)& 25000  \\ \hline
    \end{tabular}
  \end{center}
\end{table}
\begin{table}[h!]
  \begin{center}
    \caption{NFW  Optimization Values}
    \label{NavaroNGC5033}
     \begin{tabular}{|r|r|}
     \hline
      \textbf{Parameter}   & \textbf{Optimization Values}
      \\  \hline
   $\rho_s$   & $5\times 10^7$
\\  \hline
$r_s$&  8.96
\\  \hline
    \end{tabular}
  \end{center}
\end{table}
\begin{table}[h!]
  \begin{center}
    \caption{Burkert Optimization Values}
    \label{BuckertNGC5033}
     \begin{tabular}{|r|r|}
     \hline
      \textbf{Parameter}   & \textbf{Optimization Values}
      \\  \hline
     $\rho_0^B$  & $1\times 10^8$
\\  \hline
$r_0$&  9.08
\\  \hline
    \end{tabular}
  \end{center}
\end{table}
\begin{table}[h!]
  \begin{center}
    \caption{Einasto Optimization Values}
    \label{EinastoNGC5033}
    \begin{tabular}{|r|r|}
     \hline
      \textbf{Parameter}   & \textbf{Optimization Values}
      \\  \hline
     $\rho_e$  &$1\times 10^7$
\\  \hline
$r_e$ & 9.58
\\  \hline
$n_e$ & 0.19
\\  \hline
    \end{tabular}
  \end{center}
\end{table}
\begin{table}[h!]
\centering \caption{Physical assessment of collisional DM
parameters (NGC5033).}
\begin{tabular}{lcc}
\hline
Parameter & Value & Physical Verdict \\
\hline
$\gamma_0$ & 1.00006 & Essentially isothermal \\
$\delta_\gamma$ & $6.1\times 10^{-9}$ & Negligible variation \\
$r_\gamma$ & 1.5 Kpc & Transition suppressed \\
$K_0$ & $2.5\times 10^{4}$ & Very large entropy scale \\
$r_c$ & 0.5 Kpc & Compact entropy core radius \\
$p$ & 0.01 & $K(r)$ nearly flat, minimal radial dependence \\
\hline
Overall &-& Model reduces to pure isothermal \\
\hline
\end{tabular}
\label{EVALUATIONNGC5033}
\end{table}
Now the extended picture including the rotation velocity from the
other components of the galaxy, such as the disk and gas, makes
the collisional DM model viable for this galaxy. In Fig.
\ref{extendedNGC5033} we present the combined rotation curves
including the other components of the galaxy along with the
collisional matter. As it can be seen, the extended collisional DM
model is marginally viable.
Also in Table \ref{evaluationextendedNGC5033} we present the
values of the free parameters of the collisional DM model for
which the maximum compatibility with the SPARC data comes for the
galaxy NGC5033.
\begin{table}[h!]
\centering \caption{Physical assessment of Extended collisional DM
parameters for NGC5005.}
\begin{tabular}{lcc}
\hline
Parameter & Value & Physical Verdict \\
\hline
$\gamma_0$ & 1.09025825 & Nearly isothermal core, stable central pressure \\
$\delta_\gamma$ & 0.00164083 & Negligible variation, effectively constant $\gamma(r)$ \\
$K_0$ & 3000 & Moderate entropy  \\
$M/L_{\text{disk}}$ & 0.96977186 & Realistic disk mass-to-light ratio \\
$M/L_{\text{bulge}}$ & 0.2376 & Low bulge contribution \\
\hline
Overall &-& Physically viable \\
\hline
\end{tabular}
\label{evaluationextendedNGC5033}
\end{table}

\subsection{The Galaxy NGC5055 Non-viable, Extended Model Marginally Viable}

\begin{figure}[h!]
\centering
\includegraphics[width=20pc]{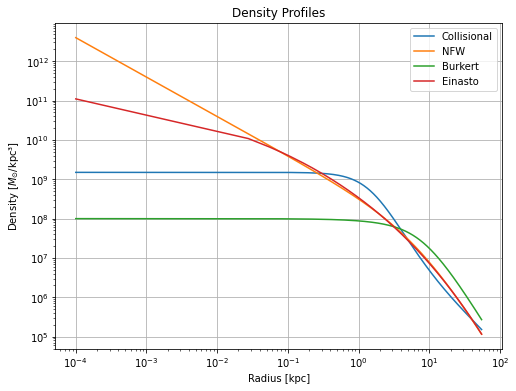}
\caption{The density of the collisional DM model (\ref{tanhmodel})
for the galaxy NGC5055, as a function of the radius.}
\label{NGC5055dens}
\end{figure}

\begin{figure}[h!]
\centering
\includegraphics[width=20pc]{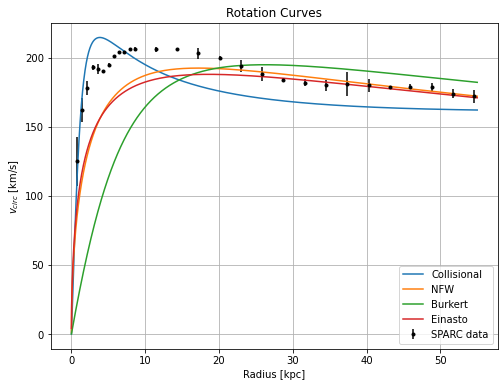}
\caption{The predicted rotation curves after using an optimization
for the collisional DM model (\ref{tanhmodel}), versus the SPARC
observational data for the galaxy NGC5055. We also plotted the
optimized curves for the NFW model, the Burkert model and the
Einasto model.} \label{NGC5055}
\end{figure}

\begin{figure}[h!]
\centering
\includegraphics[width=20pc]{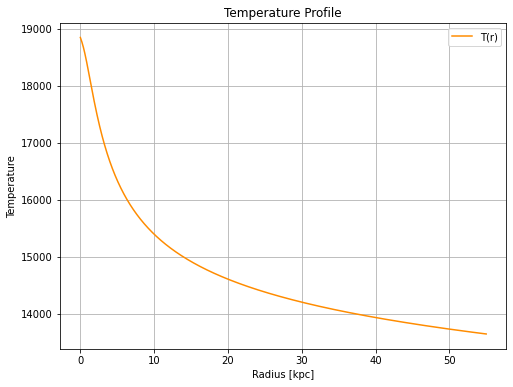}
\caption{The temperature as a function of the radius for the
collisional DM model (\ref{tanhmodel}) for the galaxy NGC5055.}
\label{NGC5055temp}
\end{figure}

For this galaxy, we shall choose $\rho_0=8.5\times
10^8$$M_{\odot}/\mathrm{Kpc}^{3}$. NGC5055, also known as the
''Sunflower Galaxy'' is a massive, nearly face-on spiral galaxy of
type SAbc, classified as an ordinary spiral, located in the
constellation Canes Venatici at a distance of approximately \(9.0\
\mathrm{Mpc}\). In Figs. \ref{NGC5055dens}, \ref{NGC5055} and
\ref{NGC5055temp} we present the density of the collisional DM
model, the predicted rotation curves after using an optimization
for the collisional DM model (\ref{tanhmodel}), versus the SPARC
observational data and the temperature parameter as a function of
the radius respectively. As it can be seen, the SIDM model
produces non-viable rotation curves incompatible with the SPARC
data. Also in Tables \ref{collNGC5055}, \ref{NavaroNGC5055},
\ref{BuckertNGC5055} and \ref{EinastoNGC5055} we present the
optimization values for the SIDM model, and the other DM profiles.
Also in Table \ref{EVALUATIONNGC5055} we present the overall
evaluation of the SIDM model for the galaxy at hand. The resulting
phenomenology is non-viable.
\begin{table}[h!]
  \begin{center}
    \caption{Collisional Dark Matter Optimization Values}
    \label{collNGC5055}
     \begin{tabular}{|r|r|}
     \hline
      \textbf{Parameter}   & \textbf{Optimization Values}
      \\  \hline
     $\delta_{\gamma} $ & 0.0000000012
\\  \hline
$\gamma_0 $ & 1.0001 \\ \hline $K_0$ ($M_{\odot} \,
\mathrm{Kpc}^{-3} \, (\mathrm{km/s})^{2}$)& 18000  \\ \hline
    \end{tabular}
  \end{center}
\end{table}
\begin{table}[h!]
  \begin{center}
    \caption{NFW  Optimization Values}
    \label{NavaroNGC5055}
     \begin{tabular}{|r|r|}
     \hline
      \textbf{Parameter}   & \textbf{Optimization Values}
      \\  \hline
   $\rho_s$   & $5\times 10^7$
\\  \hline
$r_s$&  7.96
\\  \hline
    \end{tabular}
  \end{center}
\end{table}
\begin{table}[h!]
  \begin{center}
    \caption{Burkert Optimization Values}
    \label{BuckertNGC5055}
     \begin{tabular}{|r|r|}
     \hline
      \textbf{Parameter}   & \textbf{Optimization Values}
      \\  \hline
     $\rho_0^B$  & $1\times 10^8$
\\  \hline
$r_0$&  8.08
\\  \hline
    \end{tabular}
  \end{center}
\end{table}
\begin{table}[h!]
  \begin{center}
    \caption{Einasto Optimization Values}
    \label{EinastoNGC5055}
    \begin{tabular}{|r|r|}
     \hline
      \textbf{Parameter}   & \textbf{Optimization Values}
      \\  \hline
     $\rho_e$  &$1\times 10^7$
\\  \hline
$r_e$ & 8.58
\\  \hline
$n_e$ & 0.19
\\  \hline
    \end{tabular}
  \end{center}
\end{table}

\begin{figure}[h!]
\centering
\includegraphics[width=20pc]{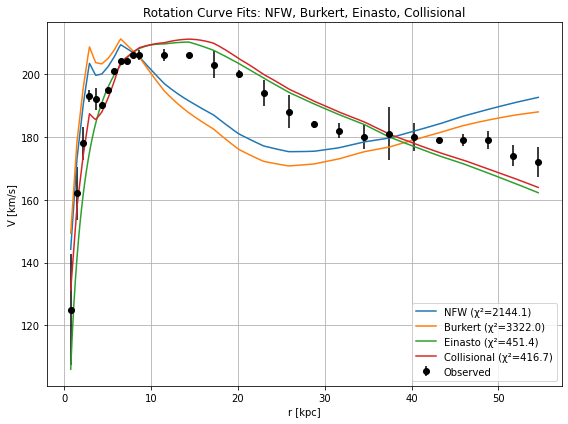}
\caption{The predicted rotation curves after using an optimization
for the collisional DM model (\ref{tanhmodel}), versus the
extended SPARC observational data for the galaxy NGC5055. The
model includes the rotation curves from all the components of the
galaxy, including gas and disk velocities, along with the
collisional DM model.} \label{extendedNGC5055}
\end{figure}

\begin{table}[h!]
\centering \caption{Physical assessment of collisional DM
parameters for NGC 5055.} \label{EVALUATIONNGC5055}
\begin{tabular}{lcc}
\hline
\textbf{Parameter} & \textbf{Value} & \textbf{Physical Verdict} \\
\hline
$\gamma_0$ & 1.0001 & Slightly above isothermal \\
$\delta_\gamma$ & $6.1\times10^{-12}$ & Essentially constant $\gamma(r)$ \\
$r_\gamma$ & 1.5 Kpc & Transition radius inside inner halo \\
$K_0$ & 18000 & High entropy scale \\
$r_c$ & 0.5 Kpc & Small core scale, physically acceptable \\
$p$ & 0.01 & Very shallow $K(r)$ decrease, nearly constant \\
\textbf{Overall} & - & Model is physically plausible in EoS behavior \\
\hline
\end{tabular}
\end{table}
Now the extended picture including the rotation velocity from the
other components of the galaxy, such as the disk and gas, makes
the collisional DM model viable for this galaxy. In Fig.
\ref{extendedNGC5055} we present the combined rotation curves
including the other components of the galaxy along with the
collisional matter. As it can be seen, the extended collisional DM
model is marginally viable.
Also in Table \ref{evaluationextendedNGC5055} we present the
values of the free parameters of the collisional DM model for
which the maximum compatibility with the SPARC data comes for the
galaxy NGC5055.
\begin{table}[h!]
\centering \caption{Physical assessment of Extended collisional DM
parameters for NGC5055.}
\begin{tabular}{lcc}
\hline
Parameter & Value & Physical Verdict \\
\hline
$\gamma_0$ & 1.08161758 & Slightly above isothermal \\
$\delta_\gamma$ & 0.001 & Nearly no radial variation \\
$K_0$ & 3000 & Moderate entropy   \\
$M/L_{\text{disk}}$ & 0.64895803 & Reasonable disk mass-to-light ratio, compatible with stellar populations \\
$M/L_{\text{bulge}}$ & 0.0000000 & No bulge contribution assumed; disk-dominated morphology \\
\hline
Overall &-& Physically plausible \\
\hline
\end{tabular}
\label{evaluationextendedNGC5055}
\end{table}

\subsection{The Galaxy NGC5371 Non-viable, Extended non-viable too}

\begin{figure}[h!]
\centering
\includegraphics[width=20pc]{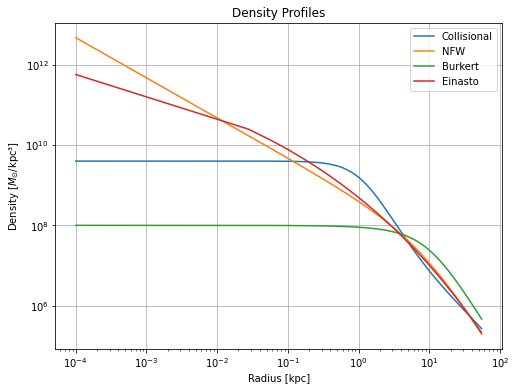}
\caption{The density of the collisional DM model (\ref{tanhmodel})
for the galaxy NGC5371, as a function of the radius.}
\label{NGC5371dens}
\end{figure}

\begin{figure}[h!]
\centering
\includegraphics[width=20pc]{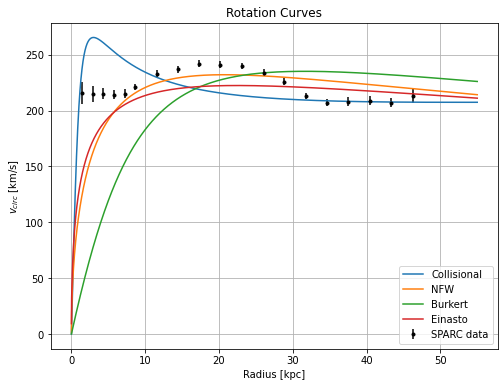}
\caption{The predicted rotation curves after using an optimization
for the collisional DM model (\ref{tanhmodel}), versus the SPARC
observational data for the galaxy NGC5371. We also plotted the
optimized curves for the NFW model, the Burkert model and the
Einasto model.} \label{NGC5371}
\end{figure}

\begin{figure}[h!]
\centering
\includegraphics[width=20pc]{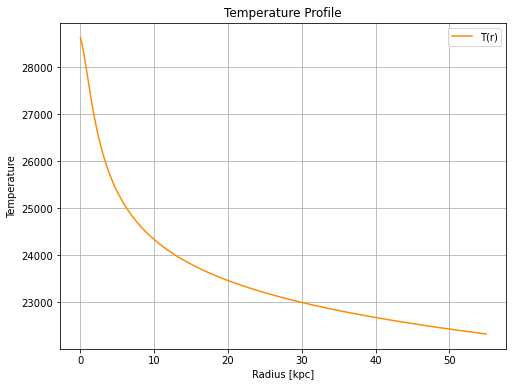}
\caption{The temperature as a function of the radius for the
collisional DM model (\ref{tanhmodel}) for the galaxy NGC5371.}
\label{NGC5371temp}
\end{figure}

\begin{figure}[h!]
\centering
\includegraphics[width=20pc]{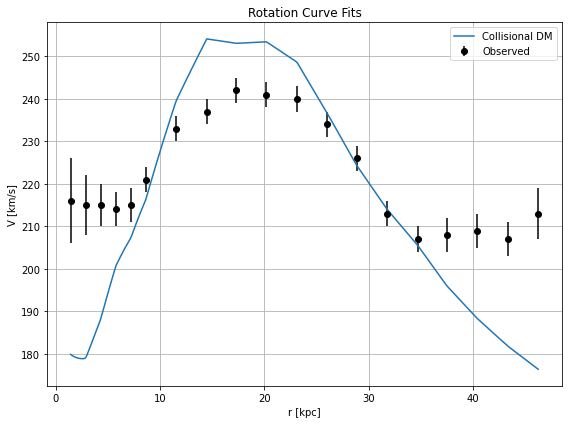}
\caption{The predicted rotation curves after using an optimization
for the collisional DM model (\ref{tanhmodel}), versus the
extended SPARC observational data for the galaxy NGC5371. The
model includes the rotation curves from all the components of the
galaxy, including gas and disk velocities, along with the
collisional DM model.} \label{extendedNGC5371}
\end{figure}

For this galaxy, we shall choose $\rho_0=4.2\times
10^9$$M_{\odot}/\mathrm{Kpc}^{3}$. NGC5371 is an ordinary, weakly
barred spiral galaxy of morphological type SAB(rs)bc, located in
the constellation Canes Venatici at a distance of about \(D \sim
39.7\pm9.9\;\mathrm{Mpc}\). In Figs. \ref{NGC5371dens},
\ref{NGC5371} and \ref{NGC5371temp} we present the density of the
collisional DM model, the predicted rotation curves after using an
optimization for the collisional DM model (\ref{tanhmodel}),
versus the SPARC observational data and the temperature parameter
as a function of the radius respectively. As it can be seen, the
SIDM model produces non-viable rotation curves incompatible with
the SPARC data. Also in Tables \ref{collNGC5371},
\ref{NavaroNGC5371}, \ref{BuckertNGC5371} and \ref{EinastoNGC5371}
we present the optimization values for the SIDM model, and the
other DM profiles. Also in Table \ref{EVALUATIONNGC5371} we
present the overall evaluation of the SIDM model for the galaxy at
hand. The resulting phenomenology is non-viable.
\begin{table}[h!]
  \begin{center}
    \caption{Collisional Dark Matter Optimization Values}
    \label{collNGC5371}
     \begin{tabular}{|r|r|}
     \hline
      \textbf{Parameter}   & \textbf{Optimization Values}
      \\  \hline
     $\delta_{\gamma} $ &  0.0000000012
\\  \hline
$\gamma_0 $ & 1.0001 \\ \hline $K_0$ ($M_{\odot} \,
\mathrm{Kpc}^{-3} \, (\mathrm{km/s})^{2}$)& 18000  \\ \hline
    \end{tabular}
  \end{center}
\end{table}
\begin{table}[h!]
  \begin{center}
    \caption{NFW  Optimization Values}
    \label{NavaroNGC5371}
     \begin{tabular}{|r|r|}
     \hline
      \textbf{Parameter}   & \textbf{Optimization Values}
      \\  \hline
   $\rho_s$   & $5\times 10^7$
\\  \hline
$r_s$&  9.60
\\  \hline
    \end{tabular}
  \end{center}
\end{table}
\begin{table}[h!]
  \begin{center}
    \caption{Burkert Optimization Values}
    \label{BuckertNGC5371}
     \begin{tabular}{|r|r|}
     \hline
      \textbf{Parameter}   & \textbf{Optimization Values}
      \\  \hline
     $\rho_0^B$  & $1\times 10^8$
\\  \hline
$r_0$& 9.75
\\  \hline
    \end{tabular}
  \end{center}
\end{table}
\begin{table}[h!]
  \begin{center}
    \caption{Einasto Optimization Values}
    \label{EinastoNGC5371}
    \begin{tabular}{|r|r|}
     \hline
      \textbf{Parameter}   & \textbf{Optimization Values}
      \\  \hline
     $\rho_e$  &$1\times 10^7$
\\  \hline
$r_e$ & 10.06
\\  \hline
$n_e$ & 0.15
\\  \hline
    \end{tabular}
  \end{center}
\end{table}
\begin{table}[h!]
\centering \caption{Physical assessment of collisional DM
parameters for NGC5371.}
\begin{tabular}{lcc}
\hline
Parameter & Value & Physical verdict \\
\hline
$\gamma_0$ & $1.0001$ & Nearly isothermal; EoS gives $T\sim K$ \\
$\delta_\gamma$ & $9\times10^{-12}$ & Negligible variation  \\
$r_\gamma$ & $1.5\ \mathrm{Kpc}$ & Plausible inner-halo transition radius  \\
$K_0$ & $1.8\times10^{4}$ & Enough central pressure support \\
$r_c$ & $0.5\ \mathrm{Kpc}$ & Small core radius \\
$p$ & $0.01$ & Nearly flat $K(r)$ \\
\hline
Overall &-& Physically consistent numerics\\
\hline
\end{tabular}
\label{EVALUATIONNGC5371}
\end{table}
Now the extended picture including the rotation velocity from the
other components of the galaxy, such as the disk and gas, makes
the collisional DM model viable for this galaxy. In Fig.
\ref{extendedNGC5371} we present the combined rotation curves
including the other components of the galaxy along with the
collisional matter. As it can be seen, the extended collisional DM
model is non-viable.
Also in Table \ref{evaluationextendedNGC5371} we present the
values of the free parameters of the collisional DM model for
which the maximum compatibility with the SPARC data comes for the
galaxy NGC5371.
\begin{table}[h!]
\centering \caption{Physical assessment of Extended collisional DM
parameters for NGC5371.}
\begin{tabular}{lcc}
\hline
Parameter & Value & Physical Verdict \\
\hline
$\gamma_0$ & 1.06739778 & Very close to isothermal \\
$\delta_\gamma$ & 0.02338704 & Small but non-negligible radial variation \\
$K_0$ & 3000 & Moderate entropy   \\
$M/L_{\text{disk}}$ & 0.73997452 & Reasonable disk mass-to-light ratio \\
$M/L_{\text{bulge}}$ & 0.00000000 & No bulge contribution assumed; disk-dominated system \\
\hline
Overall &-& Physically plausible\\
\hline
\end{tabular}
\label{evaluationextendedNGC5371}
\end{table}

\subsection{The Galaxy NGC5585 Non-viable}

\begin{figure}[h!]
\centering
\includegraphics[width=20pc]{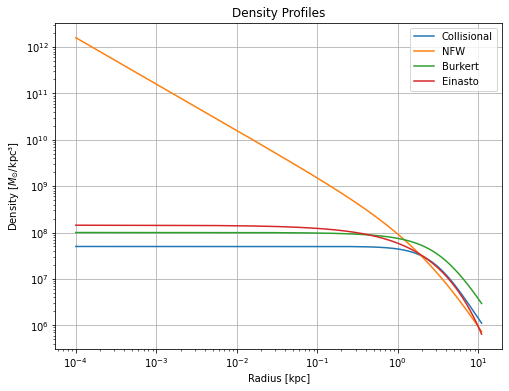}
\caption{The density of the collisional DM model (\ref{tanhmodel})
for the galaxy NGC5585, as a function of the radius.}
\label{NGC5585dens}
\end{figure}

\begin{figure}[h!]
\centering
\includegraphics[width=20pc]{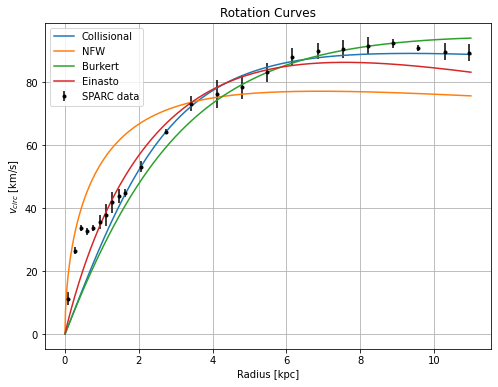}
\caption{The predicted rotation curves after using an optimization
for the collisional DM model (\ref{tanhmodel}), versus the SPARC
observational data for the galaxy NGC5585. We also plotted the
optimized curves for the NFW model, the Burkert model and the
Einasto model.} \label{NGC5585}
\end{figure}

\begin{figure}[h!]
\centering
\includegraphics[width=20pc]{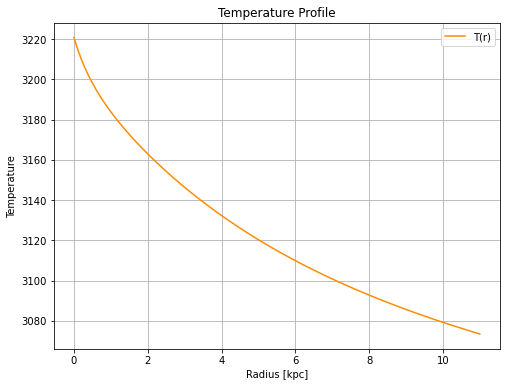}
\caption{The temperature as a function of the radius for the
collisional DM model (\ref{tanhmodel}) for the galaxy NGC5585.}
\label{NGC5585temp}
\end{figure}

\begin{figure}[h!]
\centering
\includegraphics[width=20pc]{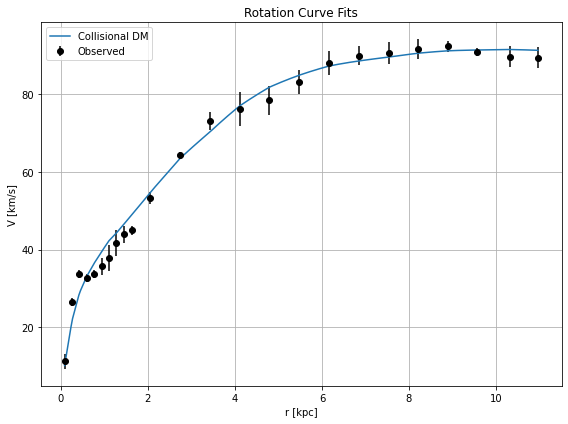}
\caption{The predicted rotation curves after using an optimization
for the collisional DM model (\ref{tanhmodel}), versus the
extended SPARC observational data for the galaxy NGC5585. The
model includes the rotation curves from all the components of the
galaxy, including gas and disk velocities, along with the
collisional DM model.} \label{extendedNGC5585}
\end{figure}

For this galaxy, we shall choose $\rho_0=5\times
10^7$$M_{\odot}/\mathrm{Kpc}^{3}$. NGC5585 is a late--type,
low--surface-brightness, dark-halo-dominated spiral galaxy
(SAB(s)d) in Ursa Major. Its distance is \(D \sim
6.2\,\mathrm{Mpc}\). In Figs. \ref{NGC5585dens}, \ref{NGC5585} and
\ref{NGC5585temp} we present the density of the collisional DM
model, the predicted rotation curves after using an optimization
for the collisional DM model (\ref{tanhmodel}), versus the SPARC
observational data and the temperature parameter as a function of
the radius respectively. As it can be seen, the SIDM model
produces non-viable rotation curves incompatible with the SPARC
data. Also in Tables \ref{collNGC5585}, \ref{NavaroNGC5585},
\ref{BuckertNGC5585} and \ref{EinastoNGC5585} we present the
optimization values for the SIDM model, and the other DM profiles.
Also in Table \ref{EVALUATIONNGC5585} we present the overall
evaluation of the SIDM model for the galaxy at hand. The resulting
phenomenology is non-viable.
\begin{table}[h!]
  \begin{center}
    \caption{Collisional Dark Matter Optimization Values}
    \label{collNGC5585}
     \begin{tabular}{|r|r|}
     \hline
      \textbf{Parameter}   & \textbf{Optimization Values}
      \\  \hline
     $\delta_{\gamma} $ & 0.0000000012
\\  \hline
$\gamma_0 $ & 1.0001 \\ \hline $K_0$ ($M_{\odot} \,
\mathrm{Kpc}^{-3} \, (\mathrm{km/s})^{2}$)& 3000  \\ \hline
    \end{tabular}
  \end{center}
\end{table}
\begin{table}[h!]
  \begin{center}
    \caption{NFW  Optimization Values}
    \label{NavaroNGC5585}
     \begin{tabular}{|r|r|}
     \hline
      \textbf{Parameter}   & \textbf{Optimization Values}
      \\  \hline
   $\rho_s$   & $5\times 10^7$
\\  \hline
$r_s$&  3.19
\\  \hline
    \end{tabular}
  \end{center}
\end{table}
\begin{table}[h!]
  \begin{center}
    \caption{Burkert Optimization Values}
    \label{BuckertNGC5585}
     \begin{tabular}{|r|r|}
     \hline
      \textbf{Parameter}   & \textbf{Optimization Values}
      \\  \hline
     $\rho_0^B$  & $1\times 10^8$
\\  \hline
$r_0$&  3.91
\\  \hline
    \end{tabular}
  \end{center}
\end{table}
\begin{table}[h!]
  \begin{center}
    \caption{Einasto Optimization Values}
    \label{EinastoNGC5585}
    \begin{tabular}{|r|r|}
     \hline
      \textbf{Parameter}   & \textbf{Optimization Values}
      \\  \hline
     $\rho_e$  &$1\times 10^7$
\\  \hline
$r_e$ & 4.28
\\  \hline
$n_e$ & 0.75
\\  \hline
    \end{tabular}
  \end{center}
\end{table}
\begin{table}[h!]
\centering \caption{Physical assessment of collisional DM
parameters for NGC5585.}
\begin{tabular}{lcc}
\hline
Parameter & Value & Physical Verdict \\
\hline
$\gamma_0$ & 1.004 & Essentially isothermal \\
$\delta_\gamma$ & $9\times10^{-6}$ & Negligible variation \\
$r_\gamma$ & 1.5 Kpc & Inner-halo scale\\
$K_0$ & $3.0\times10^{3}$ & Moderate entropy scale \\
$r_c$ & 0.5 Kpc & Compact core, consistent with low-mass halo \\
$p$ & 0.01 & Very shallow $K(r)$ decline, nearly constant \\
\hline
Overall &-& Physically consistent \\
\hline
\end{tabular}
\label{EVALUATIONNGC5585}
\end{table}
Now the extended picture including the rotation velocity from the
other components of the galaxy, such as the disk and gas, makes
the collisional DM model viable for this galaxy. In Fig.
\ref{extendedNGC5585} we present the combined rotation curves
including the other components of the galaxy along with the
collisional matter. As it can be seen, the extended collisional DM
model is non-viable.
Also in Table \ref{evaluationextendedNGC5585} we present the
values of the free parameters of the collisional DM model for
which the maximum compatibility with the SPARC data comes for the
galaxy NGC5585.
\begin{table}[h!]
\centering \caption{Physical assessment of Extended collisional DM
parameters for NGC5585.}
\begin{tabular}{lcc}
\hline
Parameter & Value & Physical Verdict \\
\hline
$\gamma_0$ & 1.01348530 & Extremely close to isothermal \\
$\delta_\gamma$ & 0.02116810 & Small but noticeable radial variation \\
$K_0$ & 3000 & Moderate entropy scale \\
$M/L_{\text{disk}}$ & 0.79086180 & Reasonable disk mass-to-light ratio, compatible with stellar population expectations \\
$M/L_{\text{bulge}}$ & 0.00000000 & No bulge contribution assumed; disk-dominated system \\
\hline
Overall &-& Physically plausible \\
\hline
\end{tabular}
\label{evaluationextendedNGC5585}
\end{table}

\subsection{The Galaxy NGC5907 Non-viable}

\begin{figure}[h!]
\centering
\includegraphics[width=20pc]{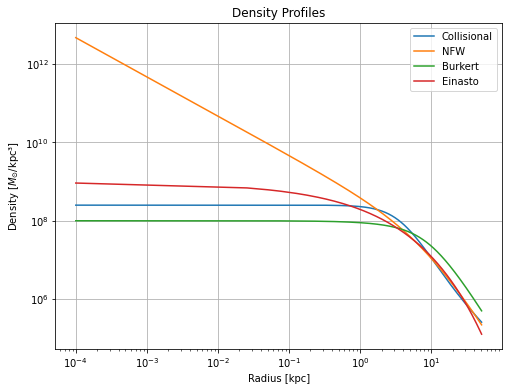}
\caption{The density of the collisional DM model (\ref{tanhmodel})
for the galaxy NGC5907, as a function of the radius.}
\label{NGC5907dens}
\end{figure}

\begin{figure}[h!]
\centering
\includegraphics[width=20pc]{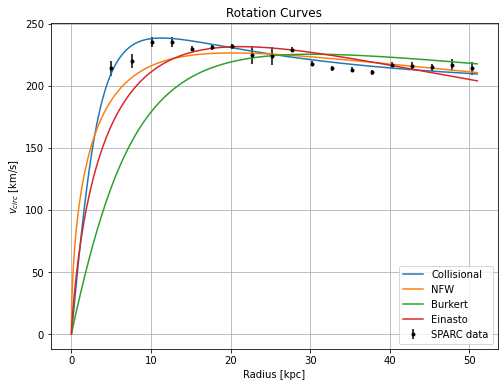}
\caption{The predicted rotation curves after using an optimization
for the collisional DM model (\ref{tanhmodel}), versus the SPARC
observational data for the galaxy NGC5907. We also plotted the
optimized curves for the NFW model, the Burkert model and the
Einasto model.} \label{NGC5907}
\end{figure}

\begin{figure}[h!]
\centering
\includegraphics[width=20pc]{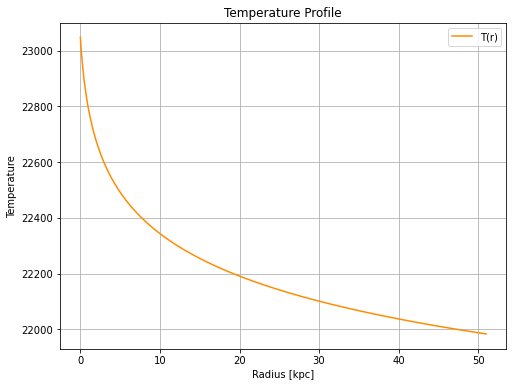}
\caption{The temperature as a function of the radius for the
collisional DM model (\ref{tanhmodel}) for the galaxy NGC5907.}
\label{NGC5907temp}
\end{figure}

\begin{figure}[h!]
\centering
\includegraphics[width=20pc]{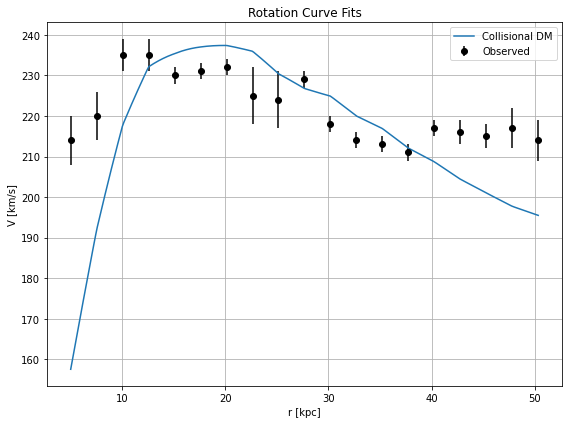}
\caption{The predicted rotation curves after using an optimization
for the collisional DM model (\ref{tanhmodel}), versus the
extended SPARC observational data for the galaxy NGC5907. The
model includes the rotation curves from all the components of the
galaxy, including gas and disk velocities, along with the
collisional DM model.} \label{extendedNGC5907}
\end{figure}

For this galaxy, we shall choose $\rho_0=2.5\times
10^8$$M_{\odot}/\mathrm{Kpc}^{3}$. NGC5907, also known as the
Splinter Galaxy or Knife Edge Galaxy, is a well-studied, edge-on
spiral galaxy located in the constellation Draco. It is classified
as SA(s)c, an unbarred spiral galaxy of late type with a very thin
disk. The galaxy is observed edge-on, revealing a remarkably thin
stellar disk surrounded by a faint, extended halo. Its estimated
distance is approximately $14-17$ Mpc. In Figs. \ref{NGC5907dens},
\ref{NGC5907} and \ref{NGC5907temp} we present the density of the
collisional DM model, the predicted rotation curves after using an
optimization for the collisional DM model (\ref{tanhmodel}),
versus the SPARC observational data and the temperature parameter
as a function of the radius respectively. As it can be seen, the
SIDM model produces non-viable rotation curves incompatible with
the SPARC data. Also in Tables \ref{collNGC5907},
\ref{NavaroNGC5907}, \ref{BuckertNGC5907} and \ref{EinastoNGC5907}
we present the optimization values for the SIDM model, and the
other DM profiles. Also in Table \ref{EVALUATIONNGC5907} we
present the overall evaluation of the SIDM model for the galaxy at
hand. The resulting phenomenology is non-viable.
\begin{table}[h!]
  \begin{center}
    \caption{Collisional Dark Matter Optimization Values}
    \label{collNGC5907}
     \begin{tabular}{|r|r|}
     \hline
      \textbf{Parameter}   & \textbf{Optimization Values}
      \\  \hline
     $\delta_{\gamma} $ & 0.0000000012
\\  \hline
$\gamma_0 $ & 1.0001 \\ \hline $K_0$ ($M_{\odot} \,
\mathrm{Kpc}^{-3} \, (\mathrm{km/s})^{2}$)& 23000  \\ \hline
    \end{tabular}
  \end{center}
\end{table}
\begin{table}[h!]
  \begin{center}
    \caption{NFW  Optimization Values}
    \label{NavaroNGC5907}
     \begin{tabular}{|r|r|}
     \hline
      \textbf{Parameter}   & \textbf{Optimization Values}
      \\  \hline
   $\rho_s$   & $5\times 10^7$
\\  \hline
$r_s$&  9.37
\\  \hline
    \end{tabular}
  \end{center}
\end{table}
\begin{table}[h!]
  \begin{center}
    \caption{Burkert Optimization Values}
    \label{BuckertNGC5907}
     \begin{tabular}{|r|r|}
     \hline
      \textbf{Parameter}   & \textbf{Optimization Values}
      \\  \hline
     $\rho_0^B$  & $1\times 10^8$
\\  \hline
$r_0$& 9.35
\\  \hline
    \end{tabular}
  \end{center}
\end{table}
\begin{table}[h!]
  \begin{center}
    \caption{Einasto Optimization Values}
    \label{EinastoNGC5907}
    \begin{tabular}{|r|r|}
     \hline
      \textbf{Parameter}   & \textbf{Optimization Values}
      \\  \hline
     $\rho_e$  &$1\times 10^7$
\\  \hline
$r_e$ & 11.06
\\  \hline
$n_e$ & 0.44
\\  \hline
    \end{tabular}
  \end{center}
\end{table}
\begin{table}[h!]
\centering \caption{Physical assessment of collisional DM
parameters for NGC5907.}
\begin{tabular}{lcc}
\hline
Parameter & Value & Physical Verdict \\
\hline
$\gamma_0$ & 1.0001 & Essentially isothermal, minimal pressure support \\
$\delta_\gamma$ & $9\times10^{-6}$ & Negligible variation \\
$r_\gamma$ & 1.5 Kpc & Transition radius irrelevant \\
$K_0$ & $2.3\times10^{4}$ & Very high entropy scale\\
$r_c$ & 0.5 Kpc & Small core scale, consistent with inner halo flattening \\
$p$ & 0.01 & Extremely shallow $K(r)$ decline, nearly constant \\
\hline
Overall &-& Model is physically marginal \\
\hline
\end{tabular}
\label{EVALUATIONNGC5907}
\end{table}
Now the extended picture including the rotation velocity from the
other components of the galaxy, such as the disk and gas, makes
the collisional DM model viable for this galaxy. In Fig.
\ref{extendedNGC5907} we present the combined rotation curves
including the other components of the galaxy along with the
collisional matter. As it can be seen, the extended collisional DM
model is non-viable.
Also in Table \ref{evaluationextendedNGC5907} we present the
values of the free parameters of the collisional DM model for
which the maximum compatibility with the SPARC data comes for the
galaxy NGC5907.
\begin{table}[h!]
\centering \caption{Physical assessment of Extended collisional DM
parameters for NGC5907.}
\begin{tabular}{lcc}
\hline
Parameter & Value & Physical Verdict \\
\hline
$\gamma_0$ & 1.09109620 & Slightly above isothermal \\
$\delta_\gamma$ & 0.0001 & Negligible radial variation \\
$K_0$ & 3000 & Moderate entropy   \\
$M/L_{\text{disk}}$ & 0.77715858 & Reasonable disk mass-to-light\\
$M/L_{\text{bulge}}$ & 0.00000000 & No bulge contribution assumed; disk-dominated morphology \\
\hline
Overall &-& Physically plausible\\
\hline
\end{tabular}
\label{evaluationextendedNGC5907}
\end{table}

\subsection{The Galaxy NGC5985 Marginally}

\begin{figure}[h!]
\centering
\includegraphics[width=20pc]{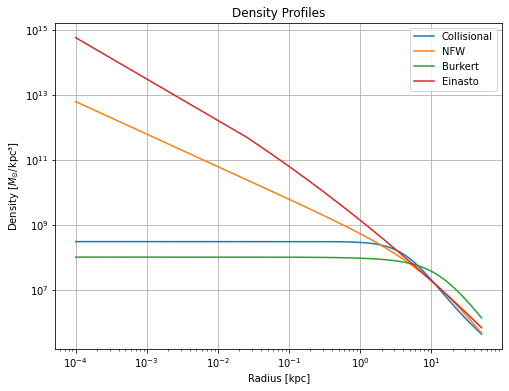}
\caption{The density of the collisional DM model (\ref{tanhmodel})
for the galaxy NGC5985, as a function of the radius.}
\label{NGC5985dens}
\end{figure}
\begin{figure}[h!]
\centering
\includegraphics[width=20pc]{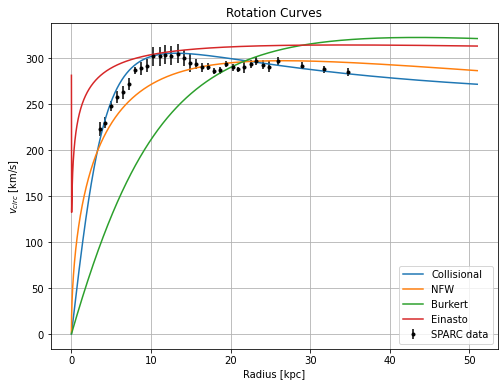}
\caption{The predicted rotation curves after using an optimization
for the collisional DM model (\ref{tanhmodel}), versus the SPARC
observational data for the galaxy NGC5985. We also plotted the
optimized curves for the NFW model, the Burkert model and the
Einasto model.} \label{NGC5985}
\end{figure}
\begin{figure}[h!]
\centering
\includegraphics[width=20pc]{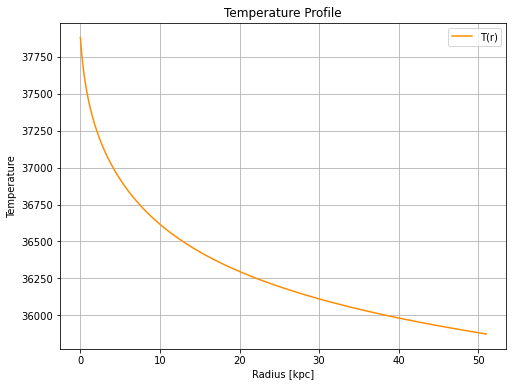}
\caption{The temperature as a function of the radius for the
collisional DM model (\ref{tanhmodel}) for the galaxy NGC5985.}
\label{NGC5985temp}
\end{figure}
\begin{figure}[h!]
\centering
\includegraphics[width=20pc]{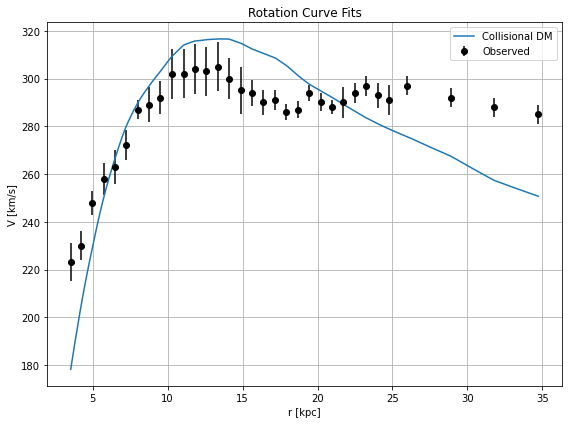}
\caption{The predicted rotation curves after using an optimization
for the collisional DM model (\ref{tanhmodel}), versus the
extended SPARC observational data for the galaxy NGC5985. The
model includes the rotation curves from all the components of the
galaxy, including gas and disk velocities, along with the
collisional DM model.} \label{extendedNGC5985}
\end{figure}

For this galaxy, we shall choose $\rho_0=3\times
10^8$$M_{\odot}/\mathrm{Kpc}^{3}$. NGC5985 is a barred spiral
galaxy of type SAB(r)b located in Draco at a distance of about
\(D\sim43\pm11\;\mathrm{Mpc}\). In Figs. \ref{NGC5985dens},
\ref{NGC5985} and \ref{NGC5985temp} we present the density of the
collisional DM model, the predicted rotation curves after using an
optimization for the collisional DM model (\ref{tanhmodel}),
versus the SPARC observational data and the temperature parameter
as a function of the radius respectively. As it can be seen, the
SIDM model produces marginally viable rotation curves compatible
with the SPARC data. Also in Tables \ref{collNGC5985},
\ref{NavaroNGC5985}, \ref{BuckertNGC5985} and \ref{EinastoNGC5985}
we present the optimization values for the SIDM model, and the
other DM profiles. Also in Table \ref{EVALUATIONNGC5985} we
present the overall evaluation of the SIDM model for the galaxy at
hand. The resulting phenomenology is marginally viable.
\begin{table}[h!]
  \begin{center}
    \caption{Collisional Dark Matter Optimization Values}
    \label{collNGC5985}
     \begin{tabular}{|r|r|}
     \hline
      \textbf{Parameter}   & \textbf{Optimization Values}
      \\  \hline
     $\delta_{\gamma} $ & 0.0000000012
\\  \hline
$\gamma_0 $ & 1.0001 \\ \hline $K_0$ ($M_{\odot} \,
\mathrm{Kpc}^{-3} \, (\mathrm{km/s})^{2}$)& 37000  \\ \hline
    \end{tabular}
  \end{center}
\end{table}
\begin{table}[h!]
  \begin{center}
    \caption{NFW  Optimization Values}
    \label{NavaroNGC5985}
     \begin{tabular}{|r|r|}
     \hline
      \textbf{Parameter}   & \textbf{Optimization Values}
      \\  \hline
   $\rho_s$   & $5\times 10^7$
\\  \hline
$r_s$& 12.30
\\  \hline
    \end{tabular}
  \end{center}
\end{table}
\begin{table}[h!]
  \begin{center}
    \caption{Burkert Optimization Values}
    \label{BuckertNGC5985}
     \begin{tabular}{|r|r|}
     \hline
      \textbf{Parameter}   & \textbf{Optimization Values}
      \\  \hline
     $\rho_0^B$  & $1\times 10^8$
\\  \hline
$r_0$&  13.38
\\  \hline
    \end{tabular}
  \end{center}
\end{table}
\begin{table}[h!]
  \begin{center}
    \caption{Einasto Optimization Values}
    \label{EinastoNGC5985}
    \begin{tabular}{|r|r|}
     \hline
      \textbf{Parameter}   & \textbf{Optimization Values}
      \\  \hline
     $\rho_e$  &$1\times 10^7$
\\  \hline
$r_e$ & 13.81
\\  \hline
$n_e$ & 0.05
\\  \hline
    \end{tabular}
  \end{center}
\end{table}
\begin{table}[h!]
\centering \caption{Physical assessment of collisional DM
parameters for NGC5985.}
\begin{tabular}{lcc}
\hline
Parameter & Value & Physical Verdict \\
\hline
$\gamma_0$ & 1.0012 & Nearly isothermal \\
$\delta_\gamma$ & $9\times10^{-6}$ & Negligible variation \\
$r_\gamma$ & 1.5 Kpc & Transition radius unimportant \\
$K_0$ & $3.7\times10^{4}$ & Very high entropy scale \\
$r_c$ & 0.5 Kpc & Small core scale \\
$p$ & 0.01 & Extremely shallow $K(r)$ decline, nearly constant \\
\hline
Overall &-& Physically marginal \\
\hline
\end{tabular}
\label{EVALUATIONNGC5985}
\end{table}
Now the extended picture including the rotation velocity from the
other components of the galaxy, such as the disk and gas, makes
the collisional DM model viable for this galaxy. In Fig.
\ref{extendedNGC5985} we present the combined rotation curves
including the other components of the galaxy along with the
collisional matter. As it can be seen, the extended collisional DM
model is viable.
Also in Table \ref{evaluationextendedNGC5985} we present the
values of the free parameters of the collisional DM model for
which the maximum compatibility with the SPARC data comes for the
galaxy NGC5985.
\begin{table}[h!]
\centering \caption{Physical assessment of Extended collisional DM
parameters for galaxy NGC5985.}
\begin{tabular}{lcc}
\hline
Parameter & Value & Physical Verdict \\
\hline
$\gamma_0$ & 1.25307927 & Above isothermal \\
$\delta_\gamma$ & 0.15214168 & Moderate-to-large radial variation \\
$K_0$ & 3000 & Moderate entropy scale \\
$ml_{disk}$ & 1.00000000 & High disk M/L \\
$ml_{bulge}$ & 0.00000000 & No bulge contribution  \\
\hline
Overall &-& Physically plausible but mixed \\
\hline
\end{tabular}
\label{evaluationextendedNGC5985}
\end{table}

\subsection{The Galaxy NGC6015 Non-viable, Extended non-viable Too}

\begin{figure}[h!]
\centering
\includegraphics[width=20pc]{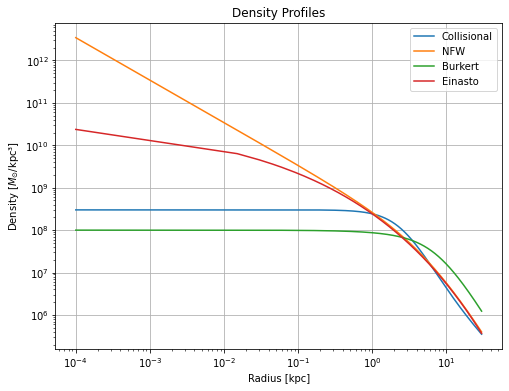}
\caption{The density of the collisional DM model (\ref{tanhmodel})
for the galaxy NGC6015, as a function of the radius.}
\label{NGC6015dens}
\end{figure}
\begin{figure}[h!]
\centering
\includegraphics[width=20pc]{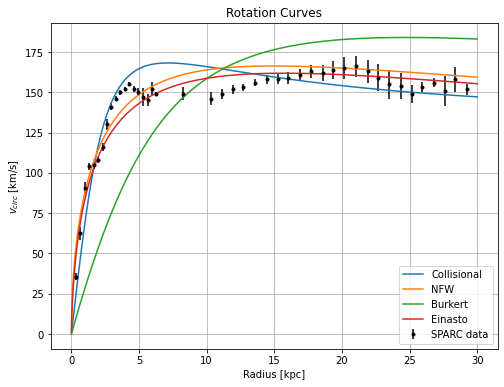}
\caption{The predicted rotation curves after using an optimization
for the collisional DM model (\ref{tanhmodel}), versus the SPARC
observational data for the galaxy NGC6015. We also plotted the
optimized curves for the NFW model, the Burkert model and the
Einasto model.} \label{NGC6015}
\end{figure}
\begin{figure}[h!]
\centering
\includegraphics[width=20pc]{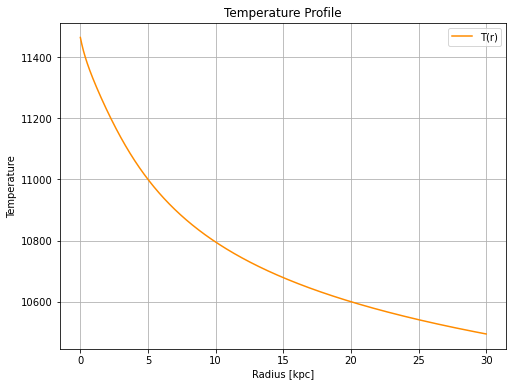}
\caption{The temperature as a function of the radius for the
collisional DM model (\ref{tanhmodel}) for the galaxy NGC6015.}
\label{NGC6015temp}
\end{figure}
\begin{figure}[h!]
\centering
\includegraphics[width=20pc]{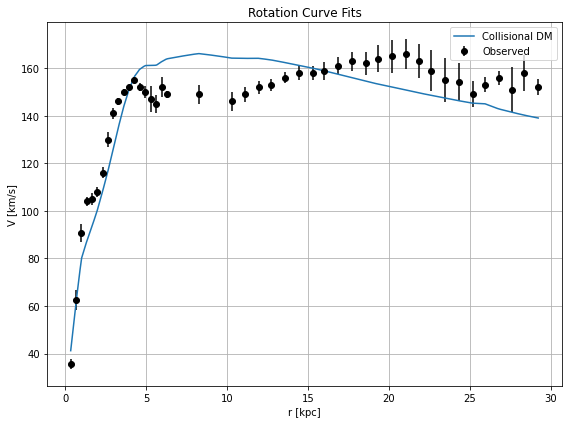}
\caption{The predicted rotation curves after using an optimization
for the collisional DM model (\ref{tanhmodel}), versus the
extended SPARC observational data for the galaxy NGC6015. The
model includes the rotation curves from all the components of the
galaxy, including gas and disk velocities, along with the
collisional DM model.} \label{extendedNGC6015}
\end{figure}

For this galaxy, we shall choose $\rho_0=3\times
10^8$$M_{\odot}/\mathrm{Kpc}^{3}$. NGC6015 is a late-type spiral
galaxy, unbarred and somewhat warped, located in the constellation
Draco, at a distance of about $13.7\,$Mpc. In Figs.
\ref{NGC6015dens}, \ref{NGC6015} and \ref{NGC6015temp} we present
the density of the collisional DM model, the predicted rotation
curves after using an optimization for the collisional DM model
(\ref{tanhmodel}), versus the SPARC observational data and the
temperature parameter as a function of the radius respectively. As
it can be seen, the SIDM model produces non-viable rotation curves
incompatible with the SPARC data. Also in Tables
\ref{collNGC6015}, \ref{NavaroNGC6015}, \ref{BuckertNGC6015} and
\ref{EinastoNGC6015} we present the optimization values for the
SIDM model, and the other DM profiles. Also in Table
\ref{EVALUATIONNGC6015} we present the overall evaluation of the
SIDM model for the galaxy at hand. The resulting phenomenology is
non-viable.
\begin{table}[h!]
  \begin{center}
    \caption{Collisional Dark Matter Optimization Values}
    \label{collNGC6015}
     \begin{tabular}{|r|r|}
     \hline
      \textbf{Parameter}   & \textbf{Optimization Values}
      \\  \hline
     $\delta_{\gamma} $ & 0.0000000012
\\  \hline
$\gamma_0 $ & 1.0001 \\ \hline $K_0$ ($M_{\odot} \,
\mathrm{Kpc}^{-3} \, (\mathrm{km/s})^{2}$)& 10000  \\ \hline
    \end{tabular}
  \end{center}
\end{table}
\begin{table}[h!]
  \begin{center}
    \caption{NFW  Optimization Values}
    \label{NavaroNGC6015}
     \begin{tabular}{|r|r|}
     \hline
      \textbf{Parameter}   & \textbf{Optimization Values}
      \\  \hline
   $\rho_s$   & $5\times 10^7$
\\  \hline
$r_s$&  6.88
\\  \hline
    \end{tabular}
  \end{center}
\end{table}
\begin{table}[h!]
  \begin{center}
    \caption{Burkert Optimization Values}
    \label{BuckertNGC6015}
     \begin{tabular}{|r|r|}
     \hline
      \textbf{Parameter}   & \textbf{Optimization Values}
      \\  \hline
     $\rho_0^B$  & $1\times 10^8$
\\  \hline
$r_0$& 7.63
\\  \hline
    \end{tabular}
  \end{center}
\end{table}
\begin{table}[h!]
  \begin{center}
    \caption{Einasto Optimization Values}
    \label{EinastoNGC6015}
    \begin{tabular}{|r|r|}
     \hline
      \textbf{Parameter}   & \textbf{Optimization Values}
      \\  \hline
     $\rho_e$  &$1\times 10^7$
\\  \hline
$r_e$ & 7.47
\\  \hline
$n_e$ &0.24
\\  \hline
    \end{tabular}
  \end{center}
\end{table}
\begin{table}[h!]
\centering \caption{Physical assessment of collisional DM
parameters (NGC6015).}
\begin{tabular}{lcc}
\hline
Parameter & Value & Physical Verdict \\
\hline
$\gamma_0$ & $1.001$ & Essentially isothermal \\
$\delta_\gamma$ & $9\times10^{-12}$ & Practically zero: $\gamma(r)$ is constant \\
$r_\gamma$ & $1.5\ \mathrm{Kpc}$ & Reasonable transition radius \\
$K_0$ & $1.0\times10^{4}$ & Enough pressure support \\
$r_c$ & $0.5\ \mathrm{Kpc}$ & Small core scale  \\
$p$ & $0.01$ & Very shallow decline of $K(r)$ \\
\hline
Overall &-& Physically consistent \\
\hline
\end{tabular}
\label{EVALUATIONNGC6015}
\end{table}
Now the extended picture including the rotation velocity from the
other components of the galaxy, such as the disk and gas, makes
the collisional DM model viable for this galaxy. In Fig.
\ref{extendedNGC6015} we present the combined rotation curves
including the other components of the galaxy along with the
collisional matter. As it can be seen, the extended collisional DM
model is non-viable.
Also in Table \ref{evaluationextendedNGC6015} we present the
values of the free parameters of the collisional DM model for
which the maximum compatibility with the SPARC data comes for the
galaxy NGC6015.
\begin{table}[h!]
\centering \caption{Physical assessment of Extended collisional DM
parameters for NGC6015.}
\begin{tabular}{lcc}
\hline
Parameter & Value & Physical Verdict \\
\hline
$\gamma_0$ & 1.04362085 & Very close to isothermal \\
$\delta_\gamma$ & 0.0001 & Negligible radial variation \\
$K_0$ & 3000 & Moderate entropy scale \\
$M/L_{\text{disk}}$ & 0.97264036 & Realistic disk mass-to-light ratio \\
$M/L_{\text{bulge}}$ & 0.00000000 & No bulge contribution assumed \\
\hline
Overall &-& Physically plausible \\
\hline
\end{tabular}
\label{evaluationextendedNGC6015}
\end{table}


\subsection{The Galaxy NGC6195 Non-viable, Extended non-viable too}

\begin{figure}[h!]
\centering
\includegraphics[width=20pc]{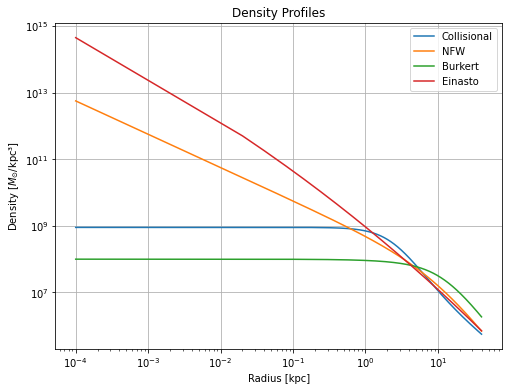}
\caption{The density of the collisional DM model (\ref{tanhmodel})
for the galaxy NGC6195, as a function of the radius.}
\label{NGC6195dens}
\end{figure}

\begin{figure}[h!]
\centering
\includegraphics[width=20pc]{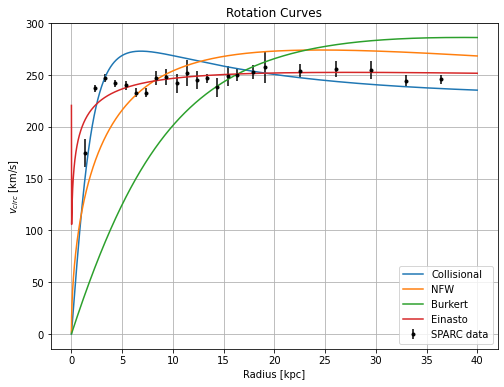}
\caption{The predicted rotation curves after using an optimization
for the collisional DM model (\ref{tanhmodel}), versus the SPARC
observational data for the galaxy NGC6195. We also plotted the
optimized curves for the NFW model, the Burkert model and the
Einasto model.} \label{NGC6195}
\end{figure}

\begin{figure}[h!]
\centering
\includegraphics[width=20pc]{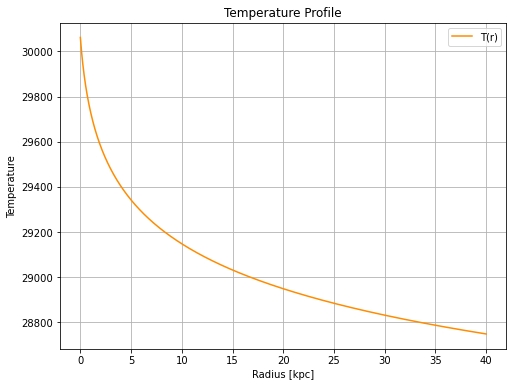}
\caption{The temperature as a function of the radius for the
collisional DM model (\ref{tanhmodel}) for the galaxy NGC6195.}
\label{NGC6195temp}
\end{figure}

\begin{figure}[h!]
\centering
\includegraphics[width=20pc]{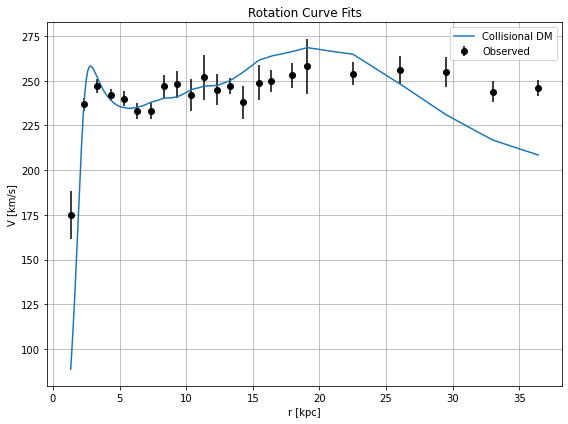}
\caption{The predicted rotation curves after using an optimization
for the collisional DM model (\ref{tanhmodel}), versus the
extended SPARC observational data for the galaxy NGC6195. The
model includes the rotation curves from all the components of the
galaxy, including gas and disk velocities, along with the
collisional DM model.} \label{extendedNGC6195}
\end{figure}

For this galaxy, we shall choose $\rho_0=9\times
10^8$$M_{\odot}/\mathrm{Kpc}^{3}$. NGC6195 is a luminous Sb spiral
galaxy in the constellation Hercules. In Figs. \ref{NGC6195dens},
\ref{NGC6195} and \ref{NGC6195temp} we present the density of the
collisional DM model, the predicted rotation curves after using an
optimization for the collisional DM model (\ref{tanhmodel}),
versus the SPARC observational data and the temperature parameter
as a function of the radius respectively. As it can be seen, the
SIDM model produces non-viable rotation curves incompatible with
the SPARC data. Also in Tables \ref{collNGC6195},
\ref{NavaroNGC6195}, \ref{BuckertNGC6195} and \ref{EinastoNGC6195}
we present the optimization values for the SIDM model, and the
other DM profiles. Also in Table \ref{EVALUATIONNGC6195} we
present the overall evaluation of the SIDM model for the galaxy at
hand. The resulting phenomenology is non-viable.
\begin{table}[h!]
  \begin{center}
    \caption{Collisional Dark Matter Optimization Values}
    \label{collNGC6195}
     \begin{tabular}{|r|r|}
     \hline
      \textbf{Parameter}   & \textbf{Optimization Values}
      \\  \hline
     $\delta_{\gamma} $ &  0.0000000012
\\  \hline
$\gamma_0 $ & 1.0001 \\ \hline $K_0$ ($M_{\odot} \,
\mathrm{Kpc}^{-3} \, (\mathrm{km/s})^{2}$)& 30000  \\ \hline
    \end{tabular}
  \end{center}
\end{table}
\begin{table}[h!]
  \begin{center}
    \caption{NFW  Optimization Values}
    \label{NavaroNGC6195}
     \begin{tabular}{|r|r|}
     \hline
      \textbf{Parameter}   & \textbf{Optimization Values}
      \\  \hline
   $\rho_s$   & $5\times 10^7$
\\  \hline
$r_s$&  11.34
\\  \hline
    \end{tabular}
  \end{center}
\end{table}
\begin{table}[h!]
  \begin{center}
    \caption{Burkert Optimization Values}
    \label{BuckertNGC6195}
     \begin{tabular}{|r|r|}
     \hline
      \textbf{Parameter}   & \textbf{Optimization Values}
      \\  \hline
     $\rho_0^B$  & $1\times 10^8$
\\  \hline
$r_0$&  11.87
\\  \hline
    \end{tabular}
  \end{center}
\end{table}
\begin{table}[h!]
  \begin{center}
    \caption{Einasto Optimization Values}
    \label{EinastoNGC6195}
    \begin{tabular}{|r|r|}
     \hline
      \textbf{Parameter}   & \textbf{Optimization Values}
      \\  \hline
     $\rho_e$  &$1\times 10^7$
\\  \hline
$r_e$ & 11.09
\\  \hline
$n_e$ & 0.05
\\  \hline
    \end{tabular}
  \end{center}
\end{table}
\begin{table}[h!]
\centering \caption{Physical assessment of collisional DM
parameters (NGC6195).}
\begin{tabular}{lcc}
\hline
Parameter & Value & Physical Verdict \\
\hline
$\gamma_0$ & $1.0001$ & Essentially isothermal\\
$\delta_\gamma$ & $9\times10^{-12}$ & Negligible; $\gamma(r)$ constant \\
$r_\gamma$ & $1.5\ \mathrm{Kpc}$ & Inner transition scale \\
$K_0$ & $3.0\times10^{4}$ & Large entropy scale \\
$r_c$ & $0.5\ \mathrm{Kpc}$ & Small core radius, physically reasonable \\
$p$ & $0.01$ & Very shallow $K(r)$ variation \\
\hline
Overall &-& Plausible\\
\hline
\end{tabular}
\label{EVALUATIONNGC6195}
\end{table}
Now the extended picture including the rotation velocity from the
other components of the galaxy, such as the disk and gas, makes
the collisional DM model viable for this galaxy. In Fig.
\ref{extendedNGC6195} we present the combined rotation curves
including the other components of the galaxy along with the
collisional matter. As it can be seen, the extended collisional DM
model is non-viable.
Also in Table \ref{evaluationextendedNGC6195} we present the
values of the free parameters of the collisional DM model for
which the maximum compatibility with the SPARC data comes for the
galaxy NGC6195.
\begin{table}[h!]
\centering \caption{Physical assessment of Extended collisional DM
parameters for NGC6195.}
\begin{tabular}{lcc}
\hline
Parameter & Value & Physical Verdict \\
\hline
$\gamma_0$ & 1.34373174 & Moderately above isothermal \\
$\delta_\gamma$ & 0.37007047 & Strong variation; $\gamma(r)$ rises significantly with radius, indicating thermal gradient \\
$K_0$ & 3000 & Moderate entropy scale; typical of large spiral DM halos \\
$ml_{disk}$ & 1.00000000 & High stellar mass-to-light ratio; disk dominates baryonic component \\
$ml_{bulge}$ & 0.00027831 & Negligible bulge contribution; consistent with disk-dominated morphology \\
\hline
Overall &-& Physically acceptable; halo moderately polytropic with strong outer heating and stable disk structure \\
\hline
\end{tabular}
\label{evaluationextendedNGC6195}
\end{table}

\subsection{The Galaxy NGC6503 Non-viable, Extended non-viable too}

\begin{figure}[h!]
\centering
\includegraphics[width=20pc]{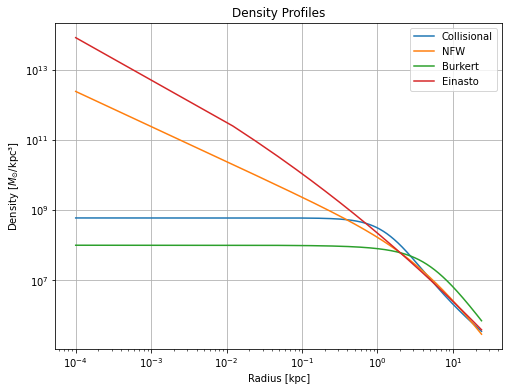}
\caption{The density of the collisional DM model (\ref{tanhmodel})
for the galaxy NGC6503, as a function of the radius.}
\label{NGC6503dens}
\end{figure}
\begin{figure}[h!]
\centering
\includegraphics[width=20pc]{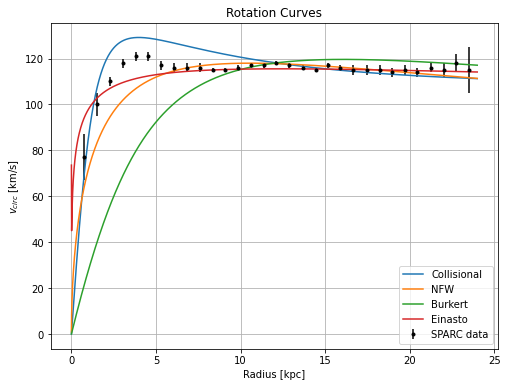}
\caption{The predicted rotation curves after using an optimization
for the collisional DM model (\ref{tanhmodel}), versus the SPARC
observational data for the galaxy NGC6503. We also plotted the
optimized curves for the NFW model, the Burkert model and the
Einasto model.} \label{NGC6503}
\end{figure}
\begin{figure}[h!]
\centering
\includegraphics[width=20pc]{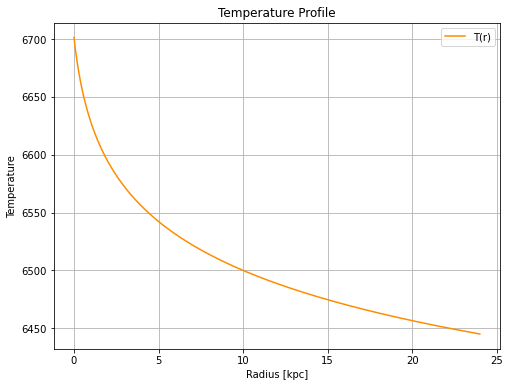}
\caption{The temperature as a function of the radius for the
collisional DM model (\ref{tanhmodel}) for the galaxy NGC6503.}
\label{NGC6503temp}
\end{figure}
\begin{figure}[h!]
\centering
\includegraphics[width=20pc]{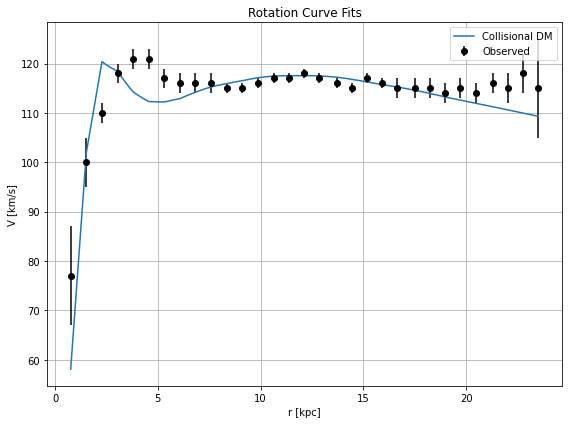}
\caption{The predicted rotation curves after using an optimization
for the collisional DM model (\ref{tanhmodel}), versus the
extended SPARC observational data for the galaxy NGC6503. The
model includes the rotation curves from all the components of the
galaxy, including gas and disk velocities, along with the
collisional DM model.} \label{extendedNGC6503}
\end{figure}

For this galaxy, we shall choose $\rho_0=6\times
10^8$$M_{\odot}/\mathrm{Kpc}^{3}$. NGC6503 is an isolated field
spiral galaxy of morphological type SA(s)cd, lying on the edge of
the Local Void in the constellation Draco, at a distance of about
$5.3$--$6.0\,$Mpc. In Figs. \ref{NGC6503dens}, \ref{NGC6503} and
\ref{NGC6503temp} we present the density of the collisional DM
model, the predicted rotation curves after using an optimization
for the collisional DM model (\ref{tanhmodel}), versus the SPARC
observational data and the temperature parameter as a function of
the radius respectively. As it can be seen, the SIDM model
produces non-viable rotation curves incompatible with the SPARC
data. Also in Tables \ref{collNGC6503}, \ref{NavaroNGC6503},
\ref{BuckertNGC6503} and \ref{EinastoNGC6503} we present the
optimization values for the SIDM model, and the other DM profiles.
Also in Table \ref{EVALUATIONNGC6503} we present the overall
evaluation of the SIDM model for the galaxy at hand. The resulting
phenomenology is non-viable.
\begin{table}[h!]
  \begin{center}
    \caption{Collisional Dark Matter Optimization Values}
    \label{collNGC6503}
     \begin{tabular}{|r|r|}
     \hline
      \textbf{Parameter}   & \textbf{Optimization Values}
      \\  \hline
     $\delta_{\gamma} $ & 0.0000000012
\\  \hline
$\gamma_0 $ & 1.0001 \\ \hline $K_0$ ($M_{\odot} \,
\mathrm{Kpc}^{-3} \, (\mathrm{km/s})^{2}$)& 6700  \\ \hline
    \end{tabular}
  \end{center}
\end{table}
\begin{table}[h!]
  \begin{center}
    \caption{NFW  Optimization Values}
    \label{NavaroNGC6503}
     \begin{tabular}{|r|r|}
     \hline
      \textbf{Parameter}   & \textbf{Optimization Values}
      \\  \hline
   $\rho_s$   & $5\times 10^7$
\\  \hline
$r_s$& 4.88
\\  \hline
    \end{tabular}
  \end{center}
\end{table}
\begin{table}[h!]
  \begin{center}
    \caption{Burkert Optimization Values}
    \label{BuckertNGC6503}
     \begin{tabular}{|r|r|}
     \hline
      \textbf{Parameter}   & \textbf{Optimization Values}
      \\  \hline
     $\rho_0^B$  & $1\times 10^8$
\\  \hline
$r_0$&  4.96
\\  \hline
    \end{tabular}
  \end{center}
\end{table}
\begin{table}[h!]
  \begin{center}
    \caption{Einasto Optimization Values}
    \label{EinastoNGC6503}
    \begin{tabular}{|r|r|}
     \hline
      \textbf{Parameter}   & \textbf{Optimization Values}
      \\  \hline
     $\rho_e$  &$1\times 10^7$
\\  \hline
$r_e$ & 5.09
\\  \hline
$n_e$ & 0.06
\\  \hline
    \end{tabular}
  \end{center}
\end{table}
\begin{table}[h!]
\centering \caption{Physical assessment of collisional DM
parameters (NGC6503).}
\begin{tabular}{lcc}
\hline
Parameter & Value & Physical Verdict \\
\hline
$\gamma_0$ & 1.0001 & Essentially isothermal \\
$\delta_\gamma$ & $9\times10^{-12}$ & Negligible \\
$r_\gamma$ & 1.5 Kpc & Inner-halo scale\\
$K_0$ & $6.7\times10^{3}$ & Moderate entropy scale \\
$r_c$ & 0.5 Kpc & Small, realistic core radius \\
$p$ & 0.01 & Almost constant $K(r)$ profile \\
\hline
Overall &-& Plausible; globally isothermal\\
\hline
\end{tabular}
\label{EVALUATIONNGC6503}
\end{table}
Now the extended picture including the rotation velocity from the
other components of the galaxy, such as the disk and gas, makes
the collisional DM model viable for this galaxy. In Fig.
\ref{extendedNGC6503} we present the combined rotation curves
including the other components of the galaxy along with the
collisional matter. As it can be seen, the extended collisional DM
model is non-viable.
Also in Table \ref{evaluationextendedNGC6503} we present the
values of the free parameters of the collisional DM model for
which the maximum compatibility with the SPARC data comes for the
galaxy NGC6503.
\begin{table}[h!]
\centering \caption{Physical assessment of Extended collisional DM
parameters for NGC6503.}
\begin{tabular}{lcc}
\hline
Parameter & Value & Physical Verdict \\
\hline
$\gamma_0$ & 1.03570556 & Very close to isothermal \\
$\delta_\gamma$ & 0.01310298 & Extremely small variation  across the halo \\
$K_0$ & 3000 & Moderate entropy scale \\
$ml_{disk}$ & 0.80149860 & Reasonable disk mass-to-light ratio  \\
$ml_{bulge}$ & 0.00000000 & Negligible bulge contribution \\
\hline
Overall &-& Physically plausible; inner halo nearly isothermal \\
\hline
\end{tabular}
\label{evaluationextendedNGC6503}
\end{table}


\subsection{The Galaxy NGC6674 Marginally}

\begin{figure}[h!]
\centering
\includegraphics[width=20pc]{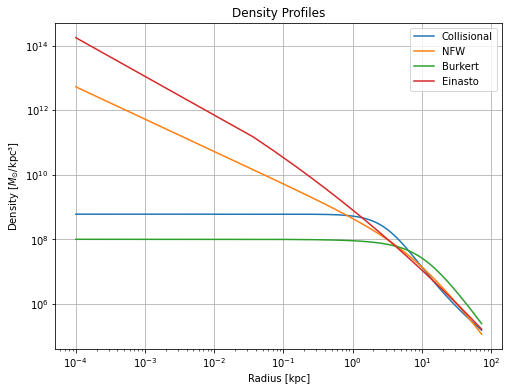}
\caption{The density of the collisional DM model (\ref{tanhmodel})
for the galaxy NGC6674, as a function of the radius.}
\label{NGC6674dens}
\end{figure}

\begin{figure}[h!]
\centering
\includegraphics[width=20pc]{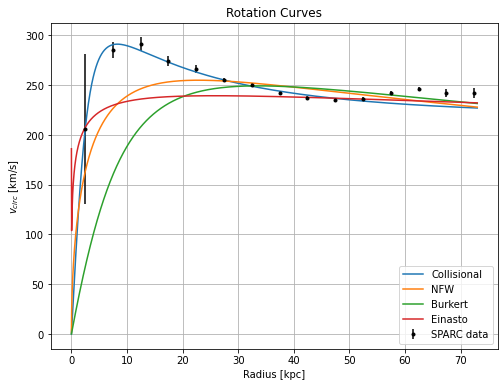}
\caption{The predicted rotation curves after using an optimization
for the collisional DM model (\ref{tanhmodel}), versus the SPARC
observational data for the galaxy NGC6674. We also plotted the
optimized curves for the NFW model, the Burkert model and the
Einasto model.} \label{NGC6674}
\end{figure}

\begin{figure}[h!]
\centering
\includegraphics[width=20pc]{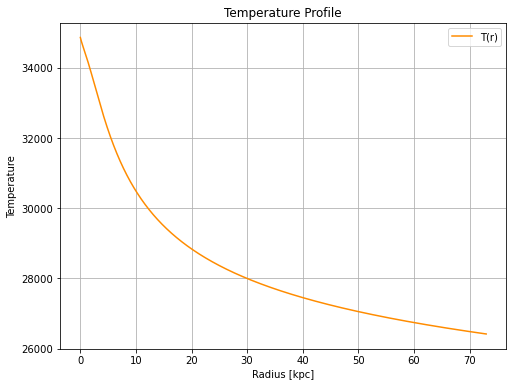}
\caption{The temperature as a function of the radius for the
collisional DM model (\ref{tanhmodel}) for the galaxy NGC6674.}
\label{NGC6674temp}
\end{figure}

\begin{figure}[h!]
\centering
\includegraphics[width=20pc]{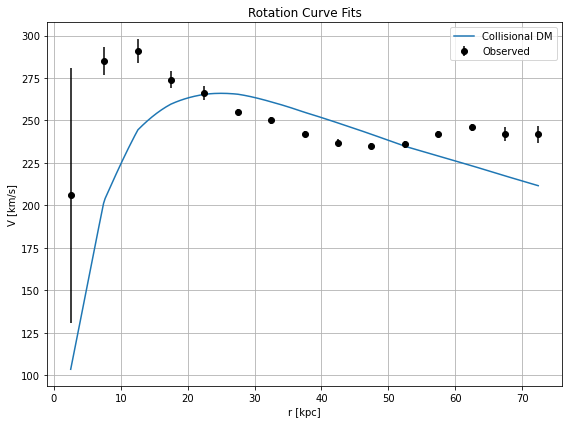}
\caption{The predicted rotation curves after using an optimization
for the collisional DM model (\ref{tanhmodel}), versus the
extended SPARC observational data for the galaxy NGC6674. The
model includes the rotation curves from all the components of the
galaxy, including gas and disk velocities, along with the
collisional DM model.} \label{extendedNGC6674}
\end{figure}

For this galaxy, we shall choose $\rho_0=6\times
10^8$$M_{\odot}/\mathrm{Kpc}^{3}$. NGC6674 is a barred spiral
galaxy of Hubble type SBb in Hercules, at a distance of about
$50\,$Mpc. In Figs. \ref{NGC6674dens}, \ref{NGC6674} and
\ref{NGC6674temp} we present the density of the collisional DM
model, the predicted rotation curves after using an optimization
for the collisional DM model (\ref{tanhmodel}), versus the SPARC
observational data and the temperature parameter as a function of
the radius respectively. As it can be seen, the SIDM model
produces marginally viable rotation curves compatible with the
SPARC data. Also in Tables \ref{collNGC6674}, \ref{NavaroNGC6674},
\ref{BuckertNGC6674} and \ref{EinastoNGC6674} we present the
optimization values for the SIDM model, and the other DM profiles.
Also in Table \ref{EVALUATIONNGC6674} we present the overall
evaluation of the SIDM model for the galaxy at hand. The resulting
phenomenology is marginally viable.
\begin{table}[h!]
  \begin{center}
    \caption{Collisional Dark Matter Optimization Values}
    \label{collNGC6674}
     \begin{tabular}{|r|r|}
     \hline
      \textbf{Parameter}   & \textbf{Optimization Values}
      \\  \hline
     $\delta_{\gamma} $ & 0.0000000012
\\  \hline
$\gamma_0 $ & 1.0001 \\ \hline $K_0$ ($M_{\odot} \,
\mathrm{Kpc}^{-3} \, (\mathrm{km/s})^{2}$)& 20000  \\ \hline
    \end{tabular}
  \end{center}
\end{table}
\begin{table}[h!]
  \begin{center}
    \caption{NFW  Optimization Values}
    \label{NavaroNGC6674}
     \begin{tabular}{|r|r|}
     \hline
      \textbf{Parameter}   & \textbf{Optimization Values}
      \\  \hline
   $\rho_s$   & $5\times 10^7$
\\  \hline
$r_s$&  10.55
\\  \hline
    \end{tabular}
  \end{center}
\end{table}
\begin{table}[h!]
  \begin{center}
    \caption{Burkert Optimization Values}
    \label{BuckertNGC6674}
     \begin{tabular}{|r|r|}
     \hline
      \textbf{Parameter}   & \textbf{Optimization Values}
      \\  \hline
     $\rho_0^B$  & $1\times 10^8$
\\  \hline
$r_0$&  10.34
\\  \hline
    \end{tabular}
  \end{center}
\end{table}
\begin{table}[h!]
  \begin{center}
    \caption{Einasto Optimization Values}
    \label{EinastoNGC6674}
    \begin{tabular}{|r|r|}
     \hline
      \textbf{Parameter}   & \textbf{Optimization Values}
      \\  \hline
     $\rho_e$  &$1\times 10^7$
\\  \hline
$r_e$ & 10.55
\\  \hline
$n_e$ & 0.06
\\  \hline
    \end{tabular}
  \end{center}
\end{table}
\begin{table}[h!]
\centering \caption{Physical assessment of collisional DM
parameters (NGC6674).}
\begin{tabular}{lcc}
\hline
Parameter & Value & Physical Verdict \\
\hline
$\gamma_0$ & 1.0275 & Slightly above isothermal\\
$\delta_\gamma$ & $9\times10^{-8}$ & Negligible \\
$r_\gamma$ & 1.5 Kpc & Inner-halo transition \\
$K_0$ & $2.0\times10^{4}$ & High entropy scale; fits massive spiral halo \\
$r_c$ & 0.5 Kpc & Small but reasonable core radius \\
$p$ & 0.01 & Almost constant $K(r)$ profile \\
\hline
Overall &-& Plausible; near-isothermal halo \\
\hline
\end{tabular}
\label{EVALUATIONNGC6674}
\end{table}
Now the extended picture including the rotation velocity from the
other components of the galaxy, such as the disk and gas, makes
the collisional DM model viable for this galaxy. In Fig.
\ref{extendedNGC6674} we present the combined rotation curves
including the other components of the galaxy along with the
collisional matter. As it can be seen, the extended collisional DM
model is marginally viable.
Also in Table \ref{evaluationextendedNGC6674} we present the
values of the free parameters of the collisional DM model for
which the maximum compatibility with the SPARC data comes for the
galaxy NGC6674.
\begin{table}[h!]
\centering \caption{Physical assessment of Extended collisional DM
parameters for galaxy NGC6674.}
\begin{tabular}{lcc}
\hline
Parameter & Value & Physical Verdict \\
\hline
$\gamma_0$ & 1.12183685 & Slightly above isothermal \\
$\delta_\gamma$ & 0.000000001 & No radial variation \\
$K_0$ & 3000 & Moderate entropy scale \\
$ml_{disk}$ & 0.78293269 & Moderate-to-high disk M/L \\
$ml_{bulge}$ & 0.00000000 & No bulge contribution \\
\hline
Overall &-& Physically plausible\\
\hline
\end{tabular}
\label{evaluationextendedNGC6674}
\end{table}


\subsection{The Galaxy NGC6789}

\begin{figure}[h!]
\centering
\includegraphics[width=20pc]{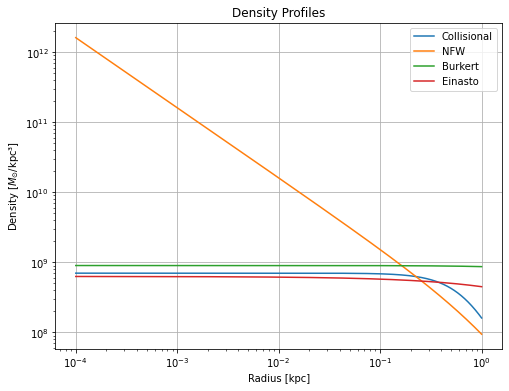}
\caption{The density of the collisional DM model (\ref{tanhmodel})
for the galaxy NGC6789, as a function of the radius.}
\label{NGC6789dens}
\end{figure}

\begin{figure}[h!]
\centering
\includegraphics[width=20pc]{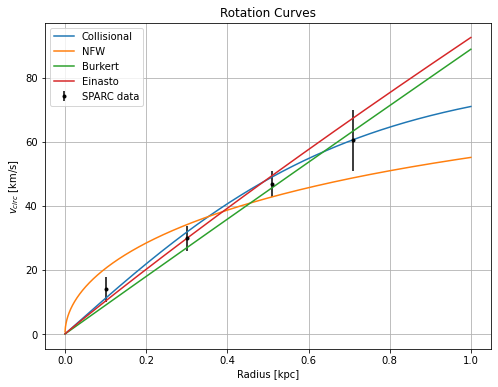}
\caption{The predicted rotation curves after using an optimization
for the collisional DM model (\ref{tanhmodel}), versus the SPARC
observational data for the galaxy NGC6789. We also plotted the
optimized curves for the NFW model, the Burkert model and the
Einasto model.} \label{NGC6789}
\end{figure}

\begin{figure}[h!]
\centering
\includegraphics[width=20pc]{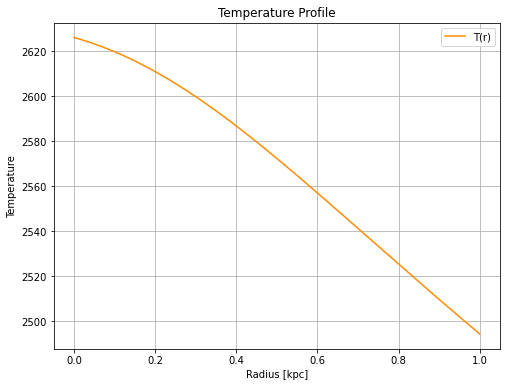}
\caption{The temperature as a function of the radius for the
collisional DM model (\ref{tanhmodel}) for the galaxy NGC6789.}
\label{NGC6789temp}
\end{figure}

For this galaxy, we shall choose $\rho_0=7\times
10^8$$M_{\odot}/\mathrm{Kpc}^{3}$. NGC6789 is a blue compact dwarf
irregular galaxy in the constellation Draco, located in the Local
Void. Its distance is about \(3.6\) Mpc. In Figs.
\ref{NGC6789dens}, \ref{NGC6789} and \ref{NGC6789temp} we present
the density of the collisional DM model, the predicted rotation
curves after using an optimization for the collisional DM model
(\ref{tanhmodel}), versus the SPARC observational data and the
temperature parameter as a function of the radius respectively. As
it can be seen, the SIDM model produces viable rotation curves
compatible with the SPARC data. Also in Tables \ref{collNGC6789},
\ref{NavaroNGC6789}, \ref{BuckertNGC6789} and \ref{EinastoNGC6789}
we present the optimization values for the SIDM model, and the
other DM profiles. Also in Table \ref{EVALUATIONNGC6789} we
present the overall evaluation of the SIDM model for the galaxy at
hand. The resulting phenomenology is viable.
\begin{table}[h!]
  \begin{center}
    \caption{Collisional Dark Matter Optimization Values}
    \label{collNGC6789}
     \begin{tabular}{|r|r|}
     \hline
      \textbf{Parameter}   & \textbf{Optimization Values}
      \\  \hline
     $\delta_{\gamma} $ & 0.0000000012
\\  \hline
$\gamma_0 $ & 1.0001 \\ \hline $K_0$ ($M_{\odot} \,
\mathrm{Kpc}^{-3} \, (\mathrm{km/s})^{2}$)& 1500  \\ \hline
    \end{tabular}
  \end{center}
\end{table}
\begin{table}[h!]
  \begin{center}
    \caption{NFW  Optimization Values}
    \label{NavaroNGC6789}
     \begin{tabular}{|r|r|}
     \hline
      \textbf{Parameter}   & \textbf{Optimization Values}
      \\  \hline
   $\rho_s$   & $5\times 10^7$
\\  \hline
$r_s$&  3.23
\\  \hline
    \end{tabular}
  \end{center}
\end{table}
\begin{table}[h!]
  \begin{center}
    \caption{Burkert Optimization Values}
    \label{BuckertNGC6789}
     \begin{tabular}{|r|r|}
     \hline
      \textbf{Parameter}   & \textbf{Optimization Values}
      \\  \hline
     $\rho_0^B$  & $9\times 10^8$
\\  \hline
$r_0$&  27.89
\\  \hline
    \end{tabular}
  \end{center}
\end{table}
\begin{table}[h!]
  \begin{center}
    \caption{Einasto Optimization Values}
    \label{EinastoNGC6789}
    \begin{tabular}{|r|r|}
     \hline
      \textbf{Parameter}   & \textbf{Optimization Values}
      \\  \hline
     $\rho_e$  &$2\times 10^7$
\\  \hline
$r_e$ & 55
\\  \hline
$n_e$ & 0.58
\\  \hline
    \end{tabular}
  \end{center}
\end{table}
\begin{table}[h!]
\centering \caption{Physical assessment of collisional DM
parameters (NGC 6789).} \label{EVALUATIONNGC6789}
\begin{tabular}{lcc}
\hline
\textbf{Parameter} & \textbf{Value} & \textbf{Physical Verdict} \\
\hline
$\gamma_0$ & $1.0001$ & Nearly isothermal \\
$\delta_\gamma$ & $9\times10^{-12}$ & Practically zero  \\
$r_\gamma$ & $1.5\ \mathrm{Kpc}$ & Transition radius  \\
$K_0$ & $1.5\times10^{3}$ & Enough pressure support  \\
$r_c$ & $0.5\ \mathrm{Kpc}$ & Small core scale \\
$p$ & $0.01$ & Extremely shallow decline of $K(r)$ \\
\hline
\textbf{Overall} & -- & Physically consistent \\
\hline
\end{tabular}
\end{table}


\subsection{The Galaxy NGC6946 Non-viable}

\begin{figure}[h!]
\centering
\includegraphics[width=20pc]{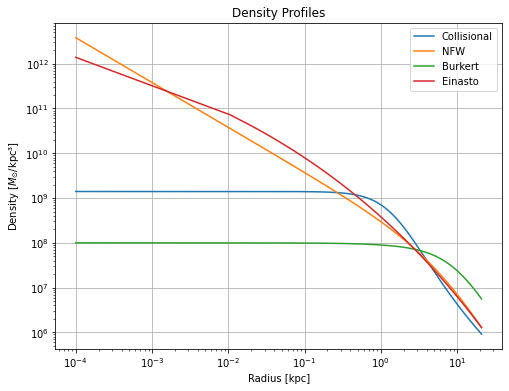}
\caption{The density of the collisional DM model (\ref{tanhmodel})
for the galaxy NGC6946, as a function of the radius.}
\label{NGC6946dens}
\end{figure}

\begin{figure}[h!]
\centering
\includegraphics[width=20pc]{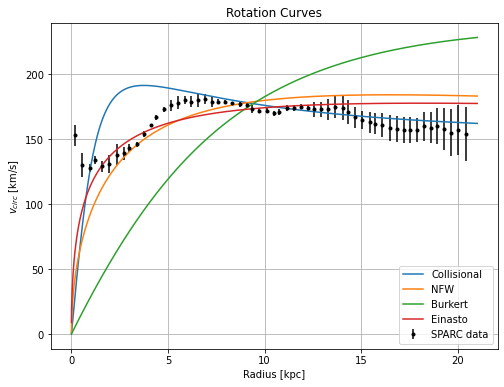}
\caption{The predicted rotation curves after using an optimization
for the collisional DM model (\ref{tanhmodel}), versus the SPARC
observational data for the galaxy NGC6946. We also plotted the
optimized curves for the NFW model, the Burkert model and the
Einasto model.} \label{NGC6946}
\end{figure}

\begin{figure}[h!]
\centering
\includegraphics[width=20pc]{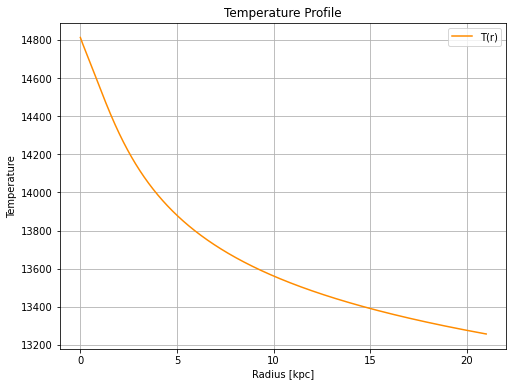}
\caption{The temperature as a function of the radius for the
collisional DM model (\ref{tanhmodel}) for the galaxy NGC6946.}
\label{NGC6946temp}
\end{figure}

\begin{figure}[h!]
\centering
\includegraphics[width=20pc]{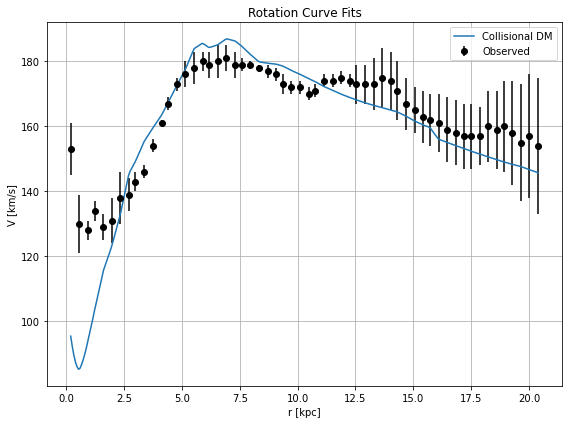}
\caption{The predicted rotation curves after using an optimization
for the collisional DM model (\ref{tanhmodel}), versus the
extended SPARC observational data for the galaxy NGC6946. The
model includes the rotation curves from all the components of the
galaxy, including gas and disk velocities, along with the
collisional DM model.} \label{extendedNGC6946}
\end{figure}

For this galaxy, we shall choose $\rho_0=1.4\times
10^9$$M_{\odot}/\mathrm{Kpc}^{3}$. NGC6946 (the ''Fireworks
Galaxy'') is a nearby, grand-design, face-on intermediate spiral
of type SAB(rs)cd at a distance of about (\(\sim 7.72\) Mpc). In
Figs. \ref{NGC6946dens}, \ref{NGC6946} and \ref{NGC6946temp} we
present the density of the collisional DM model, the predicted
rotation curves after using an optimization for the collisional DM
model (\ref{tanhmodel}), versus the SPARC observational data and
the temperature parameter as a function of the radius
respectively. As it can be seen, the SIDM model produces
non-viable rotation curves incompatible with the SPARC data. Also
in Tables \ref{collNGC6946}, \ref{NavaroNGC6946},
\ref{BuckertNGC6946} and \ref{EinastoNGC6946} we present the
optimization values for the SIDM model, and the other DM profiles.
Also in Table \ref{EVALUATIONNGC6946} we present the overall
evaluation of the SIDM model for the galaxy at hand. The resulting
phenomenology is non-viable.
\begin{table}[h!]
  \begin{center}
    \caption{Collisional Dark Matter Optimization Values}
    \label{collNGC6946}
     \begin{tabular}{|r|r|}
     \hline
      \textbf{Parameter}   & \textbf{Optimization Values}
      \\  \hline
     $\delta_{\gamma} $ & 0.0000000012
\\  \hline
$\gamma_0 $ & 1.0001 \\ \hline $K_0$ ($M_{\odot} \,
\mathrm{Kpc}^{-3} \, (\mathrm{km/s})^{2}$)& 12000  \\ \hline
    \end{tabular}
  \end{center}
\end{table}
\begin{table}[h!]
  \begin{center}
    \caption{NFW  Optimization Values}
    \label{NavaroNGC6946}
     \begin{tabular}{|r|r|}
     \hline
      \textbf{Parameter}   & \textbf{Optimization Values}
      \\  \hline
   $\rho_s$   & $5\times 10^7$
\\  \hline
$r_s$&  7.62
\\  \hline
    \end{tabular}
  \end{center}
\end{table}
\begin{table}[h!]
  \begin{center}
    \caption{Burkert Optimization Values}
    \label{BuckertNGC6946}
     \begin{tabular}{|r|r|}
     \hline
      \textbf{Parameter}   & \textbf{Optimization Values}
      \\  \hline
     $\rho_0^B$  & $1\times 10^8$
\\  \hline
$r_0$&  9.73
\\  \hline
    \end{tabular}
  \end{center}
\end{table}
\begin{table}[h!]
  \begin{center}
    \caption{Einasto Optimization Values}
    \label{EinastoNGC6946}
    \begin{tabular}{|r|r|}
     \hline
      \textbf{Parameter}   & \textbf{Optimization Values}
      \\  \hline
     $\rho_e$  &$1\times 10^7$
\\  \hline
$r_e$ & 8
\\  \hline
$n_e$ & 0.13
\\  \hline
    \end{tabular}
  \end{center}
\end{table}
\begin{table}[h!]
\centering \caption{Physical assessment of collisional DM
parameters (NGC 6946).} \label{EVALUATIONNGC6946}
\begin{tabular}{lcc}
\hline
\textbf{Parameter} & \textbf{Value} & \textbf{Physical Verdict} \\
\hline
$\gamma_0$ & $1.0001$ & Essentially isothermal \\
$\delta_\gamma$ & $9\times10^{-12}$ & Negligible \\
$r_\gamma$ & $1.5\ \mathrm{Kpc}$ & Within halo but transition not effective \\
$K_0$ ($M_{\odot}\,\mathrm{Kpc}^{-3}\,(\mathrm{km/s})^{2}$) & $1.2\times10^{3}$ & Enough pressure support\\
$r_c$ & $0.5\ \mathrm{Kpc}$ & Reasonable core scale \\
$p$ & $0.01$ & Very shallow entropy slope \\
\hline
\textbf{Overall} & -- & Physically consistent but nearly isothermal \\
\hline
\end{tabular}
\end{table}
Now the extended picture including the rotation velocity from the
other components of the galaxy, such as the disk and gas, makes
the collisional DM model viable for this galaxy. In Fig.
\ref{extendedNGC6946} we present the combined rotation curves
including the other components of the galaxy along with the
collisional matter. As it can be seen, the extended collisional DM
model is non-viable.
Also in Table \ref{evaluationextendedNGC6946} we present the
values of the free parameters of the collisional DM model for
which the maximum compatibility with the SPARC data comes for the
galaxy NGC6946.
\begin{table}[h!]
\centering \caption{Physical assessment of Extended collisional DM
parameters for NGC6946.}
\begin{tabular}{lcc}
\hline
Parameter & Value & Physical Verdict \\
\hline
$\gamma_0$ & 1.02506072 & Extremely close to isothermal \\
$\delta_\gamma$ & 0.000000001 & No variation\\
$K_0$ & 3000 & Moderate entropy/pressure scale\\
$ml_{disk}$ & 0.81683508 & Plausible disk mass-to-light ratio \\
$ml_{bulge}$ & 0.40000000 & Substantial bulge M/L \\
\hline
Overall &-& Physically plausible \\
\hline
\end{tabular}
\label{evaluationextendedNGC6946}
\end{table}

\subsection{The Galaxy NGC7331  Non-viable, Extended non-viable too}

\begin{figure}[h!]
\centering
\includegraphics[width=20pc]{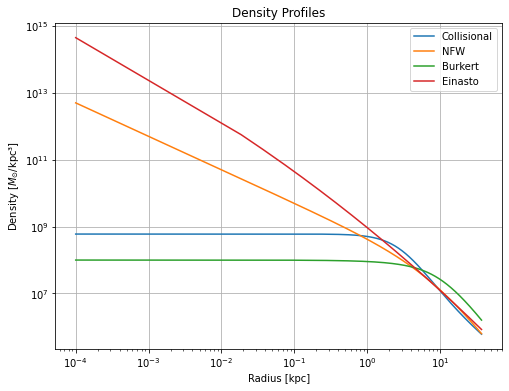}
\caption{The density of the collisional DM model (\ref{tanhmodel})
for the galaxy NGC7331, as a function of the radius.}
\label{NGC7331dens}
\end{figure}

\begin{figure}[h!]
\centering
\includegraphics[width=20pc]{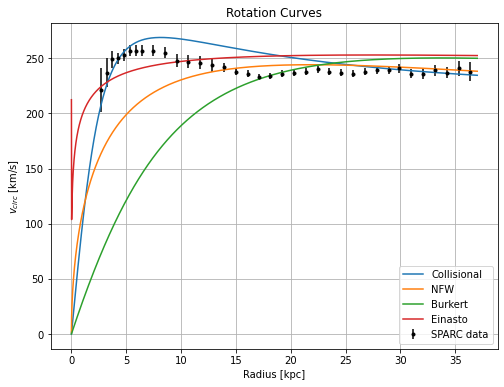}
\caption{The predicted rotation curves after using an optimization
for the collisional DM model (\ref{tanhmodel}), versus the SPARC
observational data for the galaxy NGC7331. We also plotted the
optimized curves for the NFW model, the Burkert model and the
Einasto model.} \label{NGC7331}
\end{figure}

\begin{figure}[h!]
\centering
\includegraphics[width=20pc]{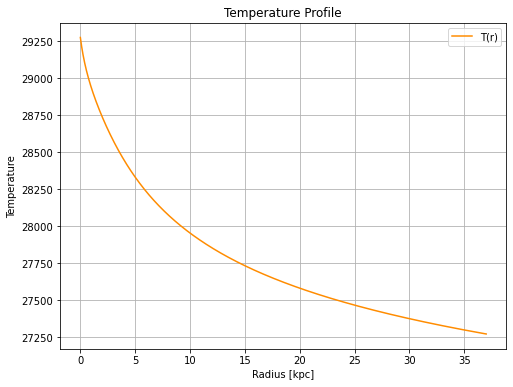}
\caption{The temperature as a function of the radius for the
collisional DM model (\ref{tanhmodel}) for the galaxy NGC7331.}
\label{NGC7331temp}
\end{figure}

\begin{figure}[h!]
\centering
\includegraphics[width=20pc]{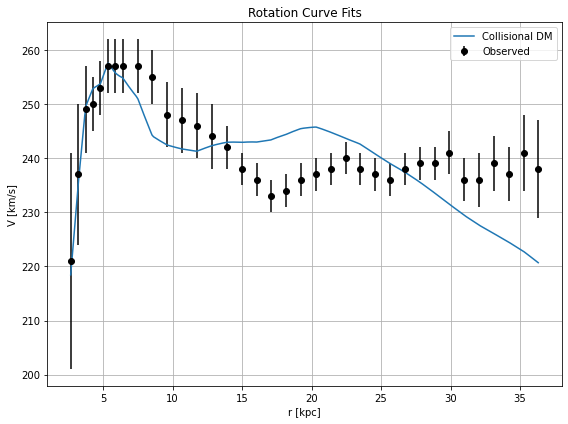}
\caption{The predicted rotation curves after using an optimization
for the collisional DM model (\ref{tanhmodel}), versus the
extended SPARC observational data for the galaxy NGC7331. The
model includes the rotation curves from all the components of the
galaxy, including gas and disk velocities, along with the
collisional DM model.} \label{extendedNGC7331}
\end{figure}

For this galaxy, we shall choose $\rho_0=6\times
10^8$$M_{\odot}/\mathrm{Kpc}^{3}$. NGC7331 is a large, unbarred
spiral galaxy of type SA(s)b in the constellation Pegasus, located
at a distance of about \(13.4\pm 2.7\) Mpc. In Figs.
\ref{NGC7331dens}, \ref{NGC7331} and \ref{NGC7331temp} we present
the density of the collisional DM model, the predicted rotation
curves after using an optimization for the collisional DM model
(\ref{tanhmodel}), versus the SPARC observational data and the
temperature parameter as a function of the radius respectively. As
it can be seen, the SIDM model produces non-viable rotation curves
incompatible with the SPARC data. Also in Tables
\ref{collNGC7331}, \ref{NavaroNGC7331}, \ref{BuckertNGC7331} and
\ref{EinastoNGC7331} we present the optimization values for the
SIDM model, and the other DM profiles. Also in Table
\ref{EVALUATIONNGC7331} we present the overall evaluation of the
SIDM model for the galaxy at hand. The resulting phenomenology is
non-viable.
\begin{table}[h!]
  \begin{center}
    \caption{Collisional Dark Matter Optimization Values}
    \label{collNGC7331}
     \begin{tabular}{|r|r|}
     \hline
      \textbf{Parameter}   & \textbf{Optimization Values}
      \\  \hline
     $\delta_{\gamma} $ & 0.0000000012
\\  \hline
$\gamma_0 $ & 1.0001 \\ \hline $K_0$ ($M_{\odot} \,
\mathrm{Kpc}^{-3} \, (\mathrm{km/s})^{2}$)& 27000  \\ \hline
    \end{tabular}
  \end{center}
\end{table}
\begin{table}[h!]
  \begin{center}
    \caption{NFW  Optimization Values}
    \label{NavaroNGC7331}
     \begin{tabular}{|r|r|}
     \hline
      \textbf{Parameter}   & \textbf{Optimization Values}
      \\  \hline
   $\rho_s$   & $5\times 10^7$
\\  \hline
$r_s$&  10.10
\\  \hline
    \end{tabular}
  \end{center}
\end{table}
\begin{table}[h!]
  \begin{center}
    \caption{Burkert Optimization Values}
    \label{BuckertNGC7331}
     \begin{tabular}{|r|r|}
     \hline
      \textbf{Parameter}   & \textbf{Optimization Values}
      \\  \hline
     $\rho_0^B$  & $1\times 10^8$
\\  \hline
$r_0$&  10.38
\\  \hline
    \end{tabular}
  \end{center}
\end{table}
\begin{table}[h!]
  \begin{center}
    \caption{Einasto Optimization Values}
    \label{EinastoNGC7331}
    \begin{tabular}{|r|r|}
     \hline
      \textbf{Parameter}   & \textbf{Optimization Values}
      \\  \hline
     $\rho_e$  &$1\times 10^7$
\\  \hline
$r_e$ & 11.11
\\  \hline
$n_e$ &0.05
\\  \hline
    \end{tabular}
  \end{center}
\end{table}
\begin{table}[h!]
\centering \caption{Physical assessment of collisional DM
parameters (NGC7331).}
\begin{tabular}{lcc}
\hline
Parameter & Value & Physical Verdict \\
\hline
$\gamma_0$ & 1.0001 & Nearly isothermal, minimal deviation \\
$\delta_\gamma$ & $12\times10^{-12}$ & Negligible variation, $\gamma(r)\sim$ const. \\
$r_\gamma$ & 1.5 Kpc & Irrelevant due to tiny $\delta_\gamma$ \\
$K_0$ & $2.7\times 10^4$ & Very high entropy scale, supports massive halo \\
$r_c$ & 0.5 Kpc & Compact entropy core radius \\
$p$ & 0.01 & Extremely shallow $K(r)$ decline, almost flat \\
\hline
Overall &-& Physically stable but rigid; halo nearly isothermal \\
\hline
\end{tabular}
\label{EVALUATIONNGC7331}
\end{table}
Now the extended picture including the rotation velocity from the
other components of the galaxy, such as the disk and gas, makes
the collisional DM model viable for this galaxy. In Fig.
\ref{extendedNGC7331} we present the combined rotation curves
including the other components of the galaxy along with the
collisional matter. As it can be seen, the extended collisional DM
model is non-viable.
Also in Table \ref{evaluationextendedNGC7331} we present the
values of the free parameters of the collisional DM model for
which the maximum compatibility with the SPARC data comes for the
galaxy NGC7331.
\begin{table}[h!]
\centering \caption{Physical assessment of Extended collisional DM
parameters for NGC7331.}
\begin{tabular}{lcc}
\hline
Parameter & Value & Physical Verdict \\
\hline
$\gamma_0$ & 1.10122973 & Slightly above isothermal \\
$\delta_\gamma$ & 0.001 & Slight variation \\
$K_0$ & 3000 & Moderate entropy/pressure scale \\
$ml_{disk}$ & 0.65156135 & Reasonable disk mass-to-light ratio \\
$ml_{bulge}$ & 0.40000000 & Substantial bulge M/L \\
\hline
Overall &-& Physically plausible \\
\hline
\end{tabular}
\label{evaluationextendedNGC7331}
\end{table}

\subsection{The Galaxy NGC7793 Non-viable, Extended Marginally Viable}

\begin{figure}[h!]
\centering
\includegraphics[width=20pc]{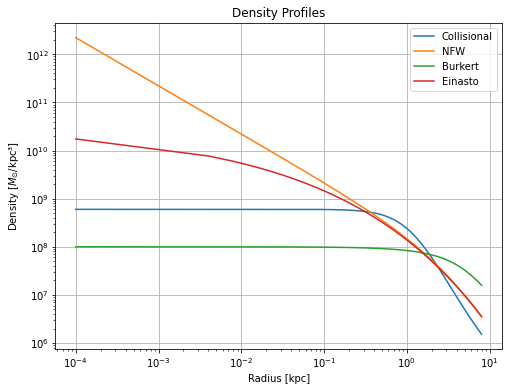}
\caption{The density of the collisional DM model (\ref{tanhmodel})
for the galaxy NGC7793, as a function of the radius.}
\label{NGC7793dens}
\end{figure}

\begin{figure}[h!]
\centering
\includegraphics[width=20pc]{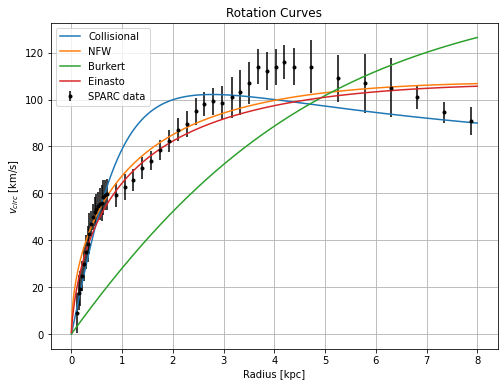}
\caption{The predicted rotation curves after using an optimization
for the collisional DM model (\ref{tanhmodel}), versus the SPARC
observational data for the galaxy NGC7793. We also plotted the
optimized curves for the NFW model, the Burkert model and the
Einasto model.} \label{NGC7793}
\end{figure}

\begin{figure}[h!]
\centering
\includegraphics[width=20pc]{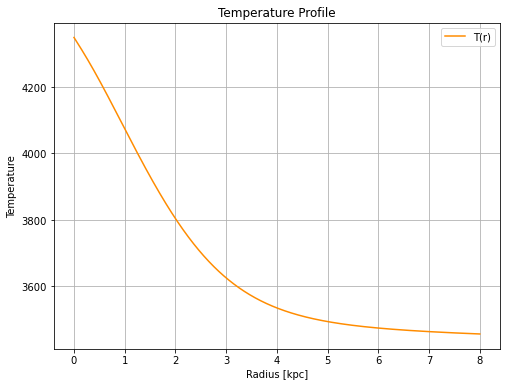}
\caption{The temperature as a function of the radius for the
collisional DM model (\ref{tanhmodel}) for the galaxy NGC7793.}
\label{NGC7793temp}
\end{figure}

\begin{figure}[h!]
\centering
\includegraphics[width=20pc]{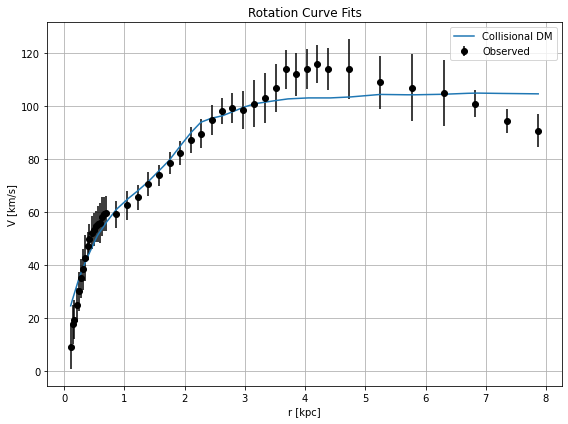}
\caption{The predicted rotation curves after using an optimization
for the collisional DM model (\ref{tanhmodel}), versus the
extended SPARC observational data for the galaxy NGC7793. The
model includes the rotation curves from all the components of the
galaxy, including gas and disk velocities, along with the
collisional DM model.} \label{extendedNGC7793}
\end{figure}

For this galaxy, we shall choose $\rho_0=6\times
10^8$$M_{\odot}/\mathrm{Kpc}^{3}$. NGC7793 is a late-type spiral
galaxy located in the Sculptor Group at a distance of about
$3.9\,\text{Mpc}$. In Figs. \ref{NGC7793dens}, \ref{NGC7793} and
\ref{NGC7793temp} we present the density of the collisional DM
model, the predicted rotation curves after using an optimization
for the collisional DM model (\ref{tanhmodel}), versus the SPARC
observational data and the temperature parameter as a function of
the radius respectively. As it can be seen, the SIDM model
produces non-viable rotation curves incompatible with the SPARC
data. Also in Tables \ref{collNGC7793}, \ref{NavaroNGC7793},
\ref{BuckertNGC7793} and \ref{EinastoNGC7793} we present the
optimization values for the SIDM model, and the other DM profiles.
Also in Table \ref{EVALUATIONNGC7793} we present the overall
evaluation of the SIDM model for the galaxy at hand. The resulting
phenomenology is non-viable.
\begin{table}[h!]
  \begin{center}
    \caption{Collisional Dark Matter Optimization Values}
    \label{collNGC7793}
     \begin{tabular}{|r|r|}
     \hline
      \textbf{Parameter}   & \textbf{Optimization Values}
      \\  \hline
     $\delta_{\gamma} $ & 0.0000000012
\\  \hline
$\gamma_0 $ & 1.0001 \\ \hline $K_0$ ($M_{\odot} \,
\mathrm{Kpc}^{-3} \, (\mathrm{km/s})^{2}$)& 40000  \\ \hline
    \end{tabular}
  \end{center}
\end{table}
\begin{table}[h!]
  \begin{center}
    \caption{NFW  Optimization Values}
    \label{NavaroNGC7793}
     \begin{tabular}{|r|r|}
     \hline
      \textbf{Parameter}   & \textbf{Optimization Values}
      \\  \hline
   $\rho_s$   & $5\times 10^7$
\\  \hline
$r_s$&  4.43
\\  \hline
    \end{tabular}
  \end{center}
\end{table}
\begin{table}[h!]
  \begin{center}
    \caption{Burkert Optimization Values}
    \label{BuckertNGC7793}
     \begin{tabular}{|r|r|}
     \hline
      \textbf{Parameter}   & \textbf{Optimization Values}
      \\  \hline
     $\rho_0^B$  & $1\times 10^8$
\\  \hline
$r_0$&  6.08
\\  \hline
    \end{tabular}
  \end{center}
\end{table}
\begin{table}[h!]
  \begin{center}
    \caption{Einasto Optimization Values}
    \label{EinastoNGC7793}
    \begin{tabular}{|r|r|}
     \hline
      \textbf{Parameter}   & \textbf{Optimization Values}
      \\  \hline
     $\rho_e$  &$1\times 10^7$
\\  \hline
$r_e$ & 4.92
\\  \hline
$n_e$ & 0.25
\\  \hline
    \end{tabular}
  \end{center}
\end{table}
\begin{table}[h!]
\centering \caption{Physical assessment of collisional DM
parameters for NGC7793.}
\begin{tabular}{lcc}
\hline
Parameter & Value & Physical Verdict \\
\hline
$\gamma_0$ & $1.0001$ & Nearly isothermal \\
$\delta_\gamma$ & $12\times10^{-9}$ & Extremely small variation \\
$r_\gamma$ & $1.5\ \mathrm{Kpc}$ & Transition radius inside inner halo \\
$K_0$ & $4.0\times10^{3}$ & Enough pressure support \\
$r_c$ & $0.5\ \mathrm{Kpc}$ & Small core scale \\
$p$ & $0.01$ & Nearly flat $K(r)$ \\
\hline
Overall &-& Physically consistent \\
\hline
\end{tabular}
\label{EVALUATIONNGC7793}
\end{table}
Now the extended picture including the rotation velocity from the
other components of the galaxy, such as the disk and gas, makes
the collisional DM model viable for this galaxy. In Fig.
\ref{extendedNGC7793} we present the combined rotation curves
including the other components of the galaxy along with the
collisional matter. As it can be seen, the extended collisional DM
model is non-viable.
Also in Table \ref{evaluationextendedNGC7793} we present the
values of the free parameters of the collisional DM model for
which the maximum compatibility with the SPARC data comes for the
galaxy NGC7793.
\begin{table}[h!]
\centering \caption{Physical assessment of Extended collisional DM
parameters for NGC7793.}
\begin{tabular}{lcc}
\hline
Parameter & Value & Physical Verdict \\
\hline
$\gamma_0$ & 1.01 & Extremely close to isothermal \\
$\delta_\gamma$ & 0.0020000000 & Minimal variation \\
$K_0$ & 3000 & Moderate entropy/pressure scale \\
$ml_{disk}$ & 0.9 & Reasonable disk mass-to-light ratio \\
$ml_{bulge}$ & 0.00000000 & No bulge contribution \\
\hline
Overall &-& Physically plausible \\
\hline
\end{tabular}
\label{evaluationextendedNGC7793}
\end{table}

\subsection{The Galaxy NGC7814 Marginally, Extended Marginal}

\begin{figure}[h!]
\centering
\includegraphics[width=20pc]{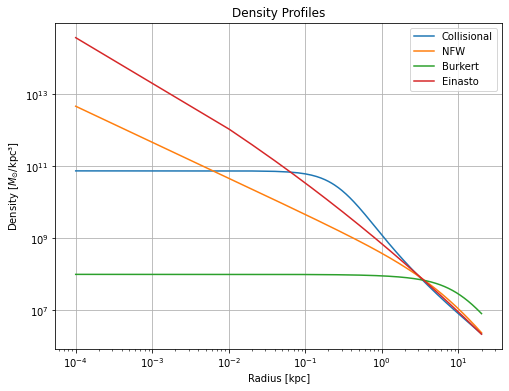}
\caption{The density of the collisional DM model (\ref{tanhmodel})
for the galaxy NGC7814, as a function of the radius.}
\label{NGC7814dens}
\end{figure}

\begin{figure}[h!]
\centering
\includegraphics[width=20pc]{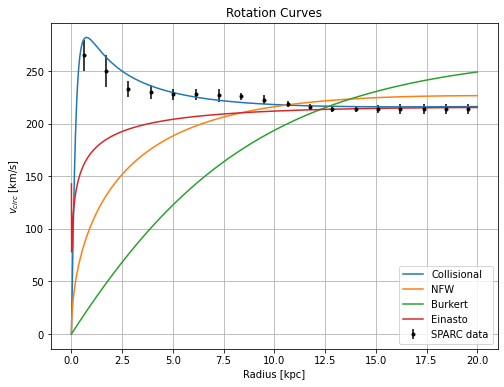}
\caption{The predicted rotation curves after using an optimization
for the collisional DM model (\ref{tanhmodel}), versus the SPARC
observational data for the galaxy NGC7814. We also plotted the
optimized curves for the NFW model, the Burkert model and the
Einasto model.} \label{NGC7814}
\end{figure}

\begin{figure}[h!]
\centering
\includegraphics[width=20pc]{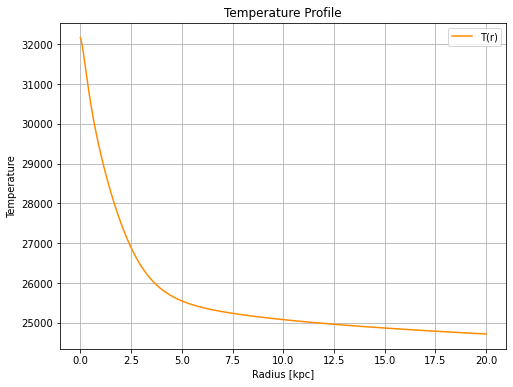}
\caption{The temperature as a function of the radius for the
collisional DM model (\ref{tanhmodel}) for the galaxy NGC7814.}
\label{NGC7814temp}
\end{figure}

\begin{figure}[h!]
\centering
\includegraphics[width=20pc]{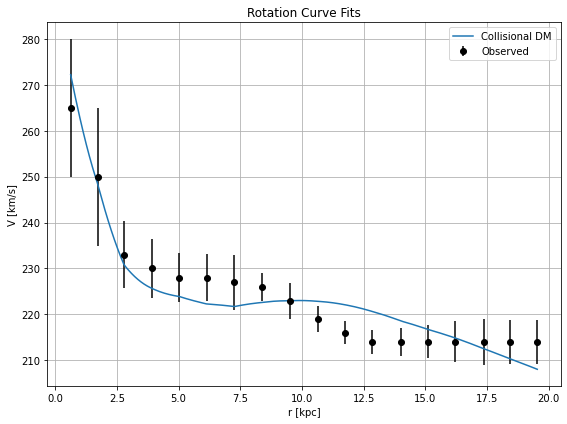}
\caption{The predicted rotation curves after using an optimization
for the collisional DM model (\ref{tanhmodel}), versus the
extended SPARC observational data for the galaxy NGC7814. The
model includes the rotation curves from all the components of the
galaxy, including gas and disk velocities, along with the
collisional DM model.} \label{extendedNGC7814}
\end{figure}

For this galaxy, we shall choose $\rho_0=7.5\times
10^{10}$$M_{\odot}/\mathrm{Kpc}^{3}$. NGC7814 is an edge-on Sab
(or SA(s)ab) spiral galaxy with a very prominent bulge, sometimes
called the ''Little Sombrero''. Its distance is about
$12.2\pm0.8\;\mathrm{Mpc}$. In Figs. \ref{NGC7814dens},
\ref{NGC7814} and \ref{NGC7814temp} we present the density of the
collisional DM model, the predicted rotation curves after using an
optimization for the collisional DM model (\ref{tanhmodel}),
versus the SPARC observational data and the temperature parameter
as a function of the radius respectively. As it can be seen, the
SIDM model produces viable rotation curves compatible with the
SPARC data. Also in Tables \ref{collNGC7814}, \ref{NavaroNGC7814},
\ref{BuckertNGC7814} and \ref{EinastoNGC7814} we present the
optimization values for the SIDM model, and the other DM profiles.
Also in Table \ref{EVALUATIONNGC7814} we present the overall
evaluation of the SIDM model for the galaxy at hand. The resulting
phenomenology is marginally viable.
\begin{table}[h!]
  \begin{center}
    \caption{Collisional Dark Matter Optimization Values}
    \label{collNGC7814}
     \begin{tabular}{|r|r|}
     \hline
      \textbf{Parameter}   & \textbf{Optimization Values}
      \\  \hline
     $\delta_{\gamma} $ & 0.0000000012
\\  \hline
$\gamma_0 $ & 1.0001\\ \hline $K_0$ ($M_{\odot} \,
\mathrm{Kpc}^{-3} \, (\mathrm{km/s})^{2}$)& 30000  \\ \hline
    \end{tabular}
  \end{center}
\end{table}
\begin{table}[h!]
  \begin{center}
    \caption{NFW  Optimization Values}
    \label{NavaroNGC7814}
     \begin{tabular}{|r|r|}
     \hline
      \textbf{Parameter}   & \textbf{Optimization Values}
      \\  \hline
   $\rho_s$   & $5\times 10^7$
\\  \hline
$r_s$&  9.38
\\  \hline
    \end{tabular}
  \end{center}
\end{table}
\begin{table}[h!]
  \begin{center}
    \caption{Burkert Optimization Values}
    \label{BuckertNGC7814}
     \begin{tabular}{|r|r|}
     \hline
      \textbf{Parameter}   & \textbf{Optimization Values}
      \\  \hline
     $\rho_0^B$  & $1\times 10^8$
\\  \hline
$r_0$& 3.91
\\  \hline
    \end{tabular}
  \end{center}
\end{table}
\begin{table}[h!]
  \begin{center}
    \caption{Einasto Optimization Values}
    \label{EinastoNGC7814}
    \begin{tabular}{|r|r|}
     \hline
      \textbf{Parameter}   & \textbf{Optimization Values}
      \\  \hline
     $\rho_e$  &$1\times 10^7$
\\  \hline
$r_e$ & 9.47
\\  \hline
$n_e$ & 0.05
\\  \hline
    \end{tabular}
  \end{center}
\end{table}
\begin{table}[h!]
\centering \caption{Physical assessment of collisional DM
parameters for NGC7814}
\begin{tabular}{lcc}
\hline
Parameter & Value & Physical Verdict \\
\hline
$\gamma_0$ & $1.0001$ & Nearly isothermal \\
$\delta_\gamma$ & $12\times 10^{-9}$ & Extremely small variation \\
$r_\gamma$ & $1.5\ \mathrm{Kpc}$ & Reasonable inner transition radius for a large spiral \\
$K_0$ & $2.35\times10^{4}$ & Plausible for a massive spiral \\
$r_c$ & $0.5\ \mathrm{Kpc}$ & Small core scale  \\
$p$ & $0.01$ & Very shallow decline of $K(r)$  \\
\hline
Overall &-& Nearly isothermal, inner halo physically plausible due to effective central density \\
\hline
\end{tabular}
\label{EVALUATIONNGC7814}
\end{table}
Now the extended picture including the rotation velocity from the
other components of the galaxy, such as the disk and gas, makes
the collisional DM model viable for this galaxy. In Fig.
\ref{extendedNGC7814} we present the combined rotation curves
including the other components of the galaxy along with the
collisional matter. As it can be seen, the extended collisional DM
model is marginally viable.
Also in Table \ref{evaluationextendedNGC7814} we present the
values of the free parameters of the collisional DM model for
which the maximum compatibility with the SPARC data comes for the
galaxy NGC7814.
\begin{table}[h!]
\centering \caption{Physical assessment of Extended collisional DM
parameters for galaxy NGC7814.}
\begin{tabular}{lcc}
\hline
Parameter & Value & Physical Verdict \\
\hline
$\gamma_0$ & 1.139 & Slightly above isothermal \\
$\delta_\gamma$ & 0.053 & Small-to-moderate radial variation \\
$K_0$ & 3000 & Moderate entropy scale \\
$ml_{disk}$ & 0.9700000000 & High disk M/L \\
$ml_{bulge}$ & 0.7800000000 & Large bulge M/L \\
\hline
Overall &-& Physically plausible but degenerate \\
\hline
\end{tabular}
\label{evaluationextendedNGC7814}
\end{table}

\subsection{The Galaxy PGC51017}

\begin{figure}[h!]
\centering
\includegraphics[width=20pc]{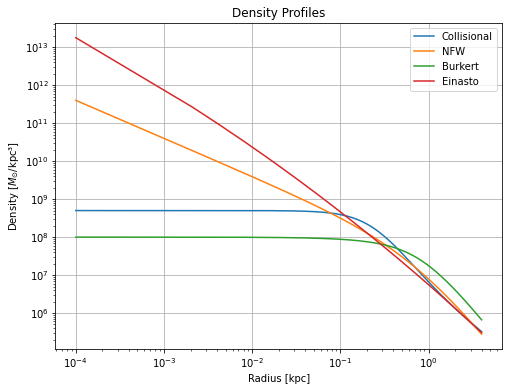}
\caption{The density of the collisional DM model (\ref{tanhmodel})
for the galaxy PGC51017, as a function of the radius.}
\label{PGC51017dens}
\end{figure}
\begin{figure}[h!]
\centering
\includegraphics[width=20pc]{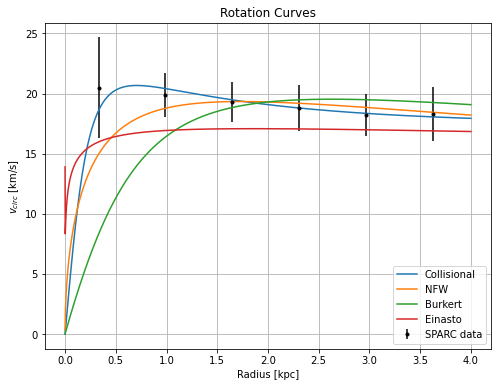}
\caption{The predicted rotation curves after using an optimization
for the collisional DM model (\ref{tanhmodel}), versus the SPARC
observational data for the galaxy PGC51017. We also plotted the
optimized curves for the NFW model, the Burkert model and the
Einasto model.} \label{PGC51017}
\end{figure}
\begin{figure}[h!]
\centering
\includegraphics[width=20pc]{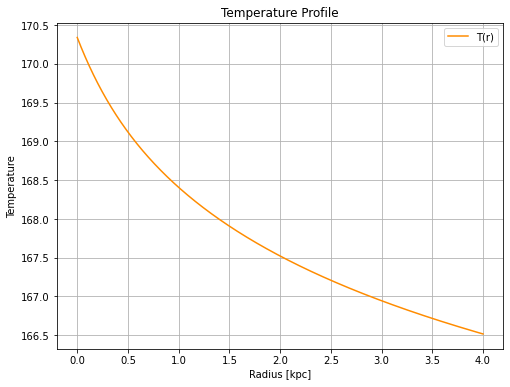}
\caption{The temperature as a function of the radius for the
collisional DM model (\ref{tanhmodel}) for the galaxy PGC51017.}
\label{PGC51017temp}
\end{figure}

For this galaxy, we shall choose $\rho_0=5\times
10^8$$M_{\odot}/\mathrm{Kpc}^{3}$. PGC51017, is a blue compact
dwarf galaxy located approximately $\sim13.9$ Mpc from the Milky
Way. In Figs. \ref{PGC51017dens}, \ref{PGC51017} and
\ref{PGC51017temp} we present the density of the collisional DM
model, the predicted rotation curves after using an optimization
for the collisional DM model (\ref{tanhmodel}), versus the SPARC
observational data and the temperature parameter as a function of
the radius respectively. As it can be seen, the SIDM model
produces viable rotation curves compatible with the SPARC data.
Also in Tables \ref{collPGC51017}, \ref{NavaroPGC51017},
\ref{BuckertPGC51017} and \ref{EinastoPGC51017} we present the
optimization values for the SIDM model, and the other DM profiles.
Also in Table \ref{EVALUATIONPGC51017} we present the overall
evaluation of the SIDM model for the galaxy at hand. The resulting
phenomenology is viable.
\begin{table}[h!]
  \begin{center}
    \caption{Collisional Dark Matter Optimization Values}
    \label{collPGC51017}
     \begin{tabular}{|r|r|}
     \hline
      \textbf{Parameter}   & \textbf{Optimization Values}
      \\  \hline
     $\delta_{\gamma} $ & 0.0000000012
\\  \hline
$\gamma_0 $ & 1.0001  \\ \hline $K_0$ ($M_{\odot} \,
\mathrm{Kpc}^{-3} \, (\mathrm{km/s})^{2}$)& 170  \\ \hline
    \end{tabular}
  \end{center}
\end{table}
\begin{table}[h!]
  \begin{center}
    \caption{NFW  Optimization Values}
    \label{NavaroPGC51017}
     \begin{tabular}{|r|r|}
     \hline
      \textbf{Parameter}   & \textbf{Optimization Values}
      \\  \hline
   $\rho_s$   & $5\times 10^7$
\\  \hline
$r_s$&  0.8
\\  \hline
    \end{tabular}
  \end{center}
\end{table}
\begin{table}[h!]
  \begin{center}
    \caption{Burkert Optimization Values}
    \label{BuckertPGC51017}
     \begin{tabular}{|r|r|}
     \hline
      \textbf{Parameter}   & \textbf{Optimization Values}
      \\  \hline
     $\rho_0^B$  & $1\times 10^8$
\\  \hline
$r_0$&  0.81
\\  \hline
    \end{tabular}
  \end{center}
\end{table}
\begin{table}[h!]
  \begin{center}
    \caption{Einasto Optimization Values}
    \label{EinastoPGC51017}
    \begin{tabular}{|r|r|}
     \hline
      \textbf{Parameter}   & \textbf{Optimization Values}
      \\  \hline
     $\rho_e$  &$1\times 10^7$
\\  \hline
$r_e$ & 0.75
\\  \hline
$n_e$ & 0.05
\\  \hline
    \end{tabular}
  \end{center}
\end{table}
\begin{table}[h!]
\centering \caption{Physical assessment of collisional DM
parameters for PGC51017.}
\begin{tabular}{lcc}
\hline
Parameter & Value & Physical Verdict \\
\hline
$\gamma_0$ & 1.0001 & Nearly isothermal \\
$\delta_\gamma$ & 0.0000000012 & Extremely small variation \\
$r_\gamma$ & 1.5 Kpc & Transition radius for $\gamma(r)$ \\
$K_0$ & 170 & Low entropy scale; weak pressure support, inner density moderately high \\
$r_c$ & 0.5 Kpc & Small core scale; K(r) nearly constant in inner halo \\
$p$ & 0.01 & Very shallow decline of K(r), inner pressure nearly uniform \\
\hline
Overall &-& Model physically plausible \\
\hline
\end{tabular}
\label{EVALUATIONPGC51017}
\end{table}


\subsection{The Galaxy UGC00128 Non-viable, Extended non-viable too}

\begin{figure}[h!]
\centering
\includegraphics[width=20pc]{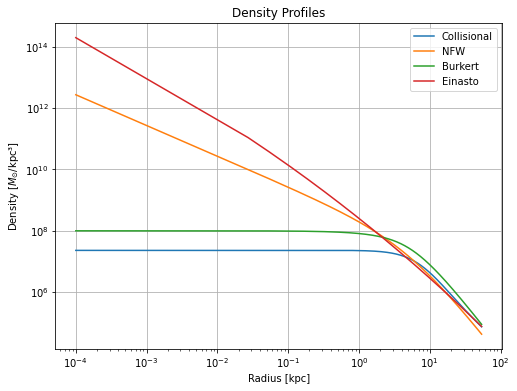}
\caption{The density of the collisional DM model (\ref{tanhmodel})
for the galaxy UGC00128, as a function of the radius.}
\label{UGC00128dens}
\end{figure}

\begin{figure}[h!]
\centering
\includegraphics[width=20pc]{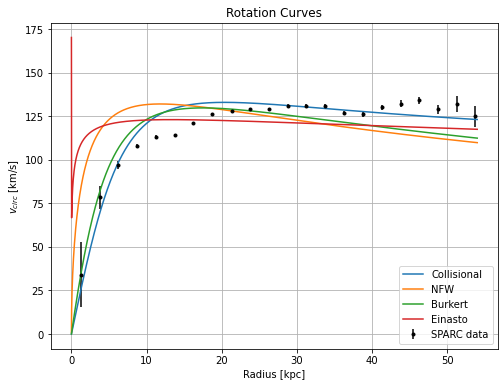}
\caption{The predicted rotation curves after using an optimization
for the collisional DM model (\ref{tanhmodel}), versus the SPARC
observational data for the galaxy UGC00128. We also plotted the
optimized curves for the NFW model, the Burkert model and the
Einasto model.} \label{UGC00128}
\end{figure}

\begin{figure}[h!]
\centering
\includegraphics[width=20pc]{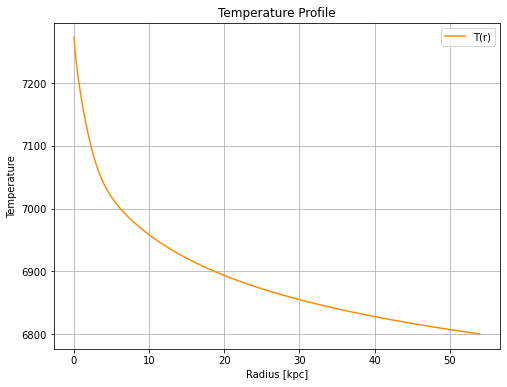}
\caption{The temperature as a function of the radius for the
collisional DM model (\ref{tanhmodel}) for the galaxy UGC00128.}
\label{UGC00128temp}
\end{figure}

\begin{figure}[h!]
\centering
\includegraphics[width=20pc]{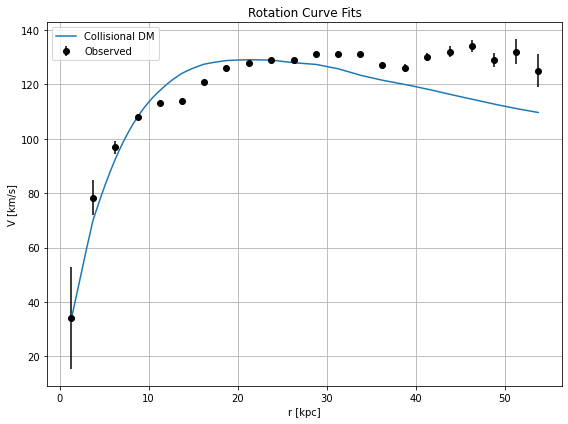}
\caption{The predicted rotation curves after using an optimization
for the collisional DM model (\ref{tanhmodel}), versus the
extended SPARC observational data for the galaxy UGC00128. The
model includes the rotation curves from all the components of the
galaxy, including gas and disk velocities, along with the
collisional DM model.} \label{extendedUGC00128}
\end{figure}

For this galaxy, we shall choose $\rho_0=2.3\times
10^7$$M_{\odot}/\mathrm{Kpc}^{3}$. UGC128 is a
low-surface-brightness spiral galaxy located approximately 10.5
Mpc from the Milky Way in the constellation Pegasus. In Figs.
\ref{UGC00128dens}, \ref{UGC00128} and \ref{UGC00128temp} we
present the density of the collisional DM model, the predicted
rotation curves after using an optimization for the collisional DM
model (\ref{tanhmodel}), versus the SPARC observational data and
the temperature parameter as a function of the radius
respectively. As it can be seen, the SIDM model produces
non-viable rotation curves incompatible with the SPARC data. Also
in Tables \ref{collUGC00128}, \ref{NavaroUGC00128},
\ref{BuckertUGC00128} and \ref{EinastoUGC00128} we present the
optimization values for the SIDM model, and the other DM profiles.
Also in Table \ref{EVALUATIONUGC00128} we present the overall
evaluation of the SIDM model for the galaxy at hand. The resulting
phenomenology is non-viable.
\begin{table}[h!]
  \begin{center}
    \caption{Collisional Dark Matter Optimization Values}
    \label{collUGC00128}
     \begin{tabular}{|r|r|}
     \hline
      \textbf{Parameter}   & \textbf{Optimization Values}
      \\  \hline
     $\delta_{\gamma} $ & 0.0000000012
\\  \hline
$\gamma_0 $ & 1.0001 \\ \hline $K_0$ ($M_{\odot} \,
\mathrm{Kpc}^{-3} \, (\mathrm{km/s})^{2}$)& 7000 \\ \hline
    \end{tabular}
  \end{center}
\end{table}
\begin{table}[h!]
  \begin{center}
    \caption{NFW  Optimization Values}
    \label{NavaroUGC00128}
     \begin{tabular}{|r|r|}
     \hline
      \textbf{Parameter}   & \textbf{Optimization Values}
      \\  \hline
   $\rho_s$   & $5\times 10^7$
\\  \hline
$r_s$&  5.46
\\  \hline
    \end{tabular}
  \end{center}
\end{table}
\begin{table}[h!]
  \begin{center}
    \caption{Burkert Optimization Values}
    \label{BuckertUGC00128}
     \begin{tabular}{|r|r|}
     \hline
      \textbf{Parameter}   & \textbf{Optimization Values}
      \\  \hline
     $\rho_0^B$  & $1\times 10^8$
\\  \hline
$r_0$& 5.38
\\  \hline
    \end{tabular}
  \end{center}
\end{table}
\begin{table}[h!]
  \begin{center}
    \caption{Einasto Optimization Values}
    \label{EinastoUGC00128}
    \begin{tabular}{|r|r|}
     \hline
      \textbf{Parameter}   & \textbf{Optimization Values}
      \\  \hline
     $\rho_e$  &$1\times 10^7$
\\  \hline
$r_e$ & 5.40
\\  \hline
$n_e$ & 0.05
\\  \hline
    \end{tabular}
  \end{center}
\end{table}
\begin{table}[h!]
\centering \caption{Physical assessment of collisional DM
parameters for UGC00128.}
\begin{tabular}{lcc}
\hline
Parameter & Value & Physical Verdict \\
\hline
$\gamma_0$ & 1.0001 & Almost perfectly isothermal \\
$\delta_\gamma$ & 0.0000000012 & Extremely small variation  across halo \\
$r_\gamma$ & 1.5 Kpc & Transition radius inside inner halo \\
$K_0$ & 700 & Moderate entropy scale \\
$r_c$ & 0.5 Kpc & Small core scale \\
$p$ & 0.01 & Very shallow K(r) decrease \\
\hline
Overall &-& Physically plausible\\
\hline
\end{tabular}
\label{EVALUATIONUGC00128}
\end{table}
Now the extended picture including the rotation velocity from the
other components of the galaxy, such as the disk and gas, makes
the collisional DM model viable for this galaxy. In Fig.
\ref{extendedUGC00128} we present the combined rotation curves
including the other components of the galaxy along with the
collisional matter. As it can be seen, the extended collisional DM
model is non-viable.
Also in Table \ref{evaluationextendedUGC00128} we present the
values of the free parameters of the collisional DM model for
which the maximum compatibility with the SPARC data comes for the
galaxy UGC00128.
\begin{table}[h!]
\centering \caption{Physical assessment of Extended collisional DM
parameters for UGC00128.}
\begin{tabular}{lcc}
\hline
Parameter & Value & Physical Verdict \\
\hline
$\gamma_0$ & 1.04 & Very close to isothermal \\
$\delta_\gamma$ & 0.0002435025 & Extremely small variation \\
$K_0$ & 3000 & Moderate entropy/pressure scale \\
$ml_{disk}$ & 0.9 & Reasonable disk mass-to-light ratio \\
$ml_{bulge}$ & 0.00000000 & No bulge contribution; consistent with a pure disk system \\
\hline
Overall &-& Physically plausible \\
\hline
\end{tabular}
\label{evaluationextendedUGC00128}
\end{table}


\subsection{The Galaxy UGC00191 Marginally}

\begin{figure}[h!]
\centering
\includegraphics[width=20pc]{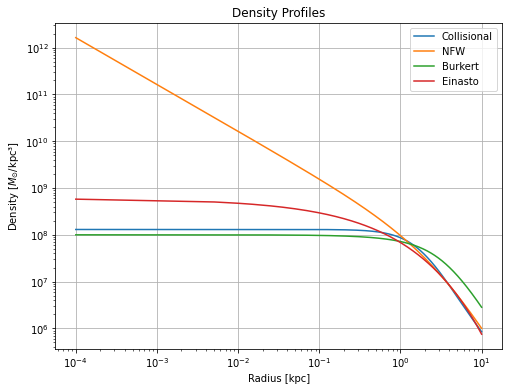}
\caption{The density of the collisional DM model (\ref{tanhmodel})
for the galaxy UGC00191, as a function of the radius.}
\label{UGC00191dens}
\end{figure}

\begin{figure}[h!]
\centering
\includegraphics[width=20pc]{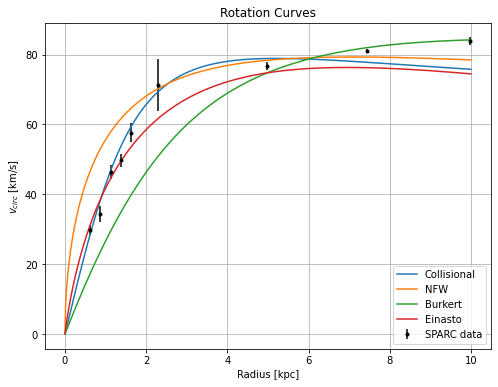}
\caption{The predicted rotation curves after using an optimization
for the collisional DM model (\ref{tanhmodel}), versus the SPARC
observational data for the galaxy UGC00191. We also plotted the
optimized curves for the NFW model, the Burkert model and the
Einasto model.} \label{UGC00191}
\end{figure}

\begin{figure}[h!]
\centering
\includegraphics[width=20pc]{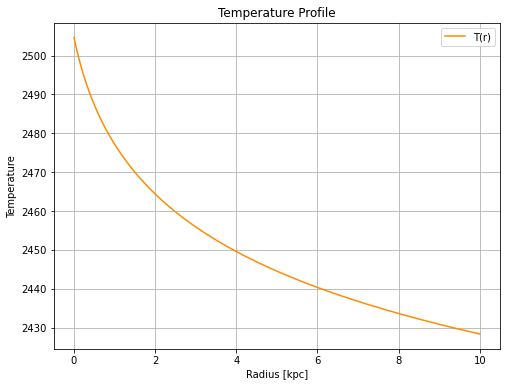}
\caption{The temperature as a function of the radius for the
collisional DM model (\ref{tanhmodel}) for the galaxy UGC00191.}
\label{UGC00191temp}
\end{figure}

\begin{figure}[h!]
\centering
\includegraphics[width=20pc]{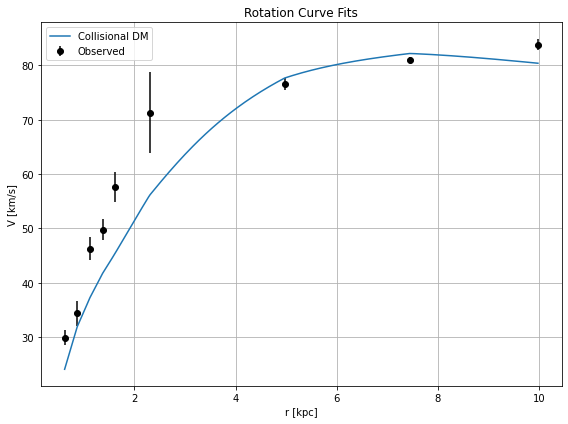}
\caption{The predicted rotation curves after using an optimization
for the collisional DM model (\ref{tanhmodel}), versus the
extended SPARC observational data for the galaxy UGC00191. The
model includes the rotation curves from all the components of the
galaxy, including gas and disk velocities, along with the
collisional DM model.} \label{extendedUGC00191}
\end{figure}

For this galaxy, we shall choose $\rho_0=1.3\times
10^8$$M_{\odot}/\mathrm{Kpc}^{3}$. UGC00191 is a relatively faint,
late-type (dwarf/spiral or irregular-spiral) system. In Figs.
\ref{UGC00191dens}, \ref{UGC00191} and \ref{UGC00191temp} we
present the density of the collisional DM model, the predicted
rotation curves after using an optimization for the collisional DM
model (\ref{tanhmodel}), versus the SPARC observational data and
the temperature parameter as a function of the radius
respectively. As it can be seen, the SIDM model produces
marginally viable rotation curves compatible with the SPARC data.
Also in Tables \ref{collUGC00191}, \ref{NavaroUGC00191},
\ref{BuckertUGC00191} and \ref{EinastoUGC00191} we present the
optimization values for the SIDM model, and the other DM profiles.
Also in Table \ref{EVALUATIONUGC00191} we present the overall
evaluation of the SIDM model for the galaxy at hand. The resulting
phenomenology is marginally viable.
\begin{table}[h!]
  \begin{center}
    \caption{Collisional Dark Matter Optimization Values}
    \label{collUGC00191}
     \begin{tabular}{|r|r|}
     \hline
      \textbf{Parameter}   & \textbf{Optimization Values}
      \\  \hline
     $\delta_{\gamma} $ & 0.0000000012
\\  \hline
$\gamma_0 $ & 1.0001 \\ \hline $K_0$ ($M_{\odot} \,
\mathrm{Kpc}^{-3} \, (\mathrm{km/s})^{2}$)& 2500 \\ \hline
    \end{tabular}
  \end{center}
\end{table}
\begin{table}[h!]
  \begin{center}
    \caption{NFW  Optimization Values}
    \label{NavaroUGC00191}
     \begin{tabular}{|r|r|}
     \hline
      \textbf{Parameter}   & \textbf{Optimization Values}
      \\  \hline
   $\rho_s$   & $5\times 10^7$
\\  \hline
$r_s$&  3.28
\\  \hline
    \end{tabular}
  \end{center}
\end{table}
\begin{table}[h!]
  \begin{center}
    \caption{Burkert Optimization Values}
    \label{BuckertUGC00191}
     \begin{tabular}{|r|r|}
     \hline
      \textbf{Parameter}   & \textbf{Optimization Values}
      \\  \hline
     $\rho_0^B$  & $1\times 10^8$
\\  \hline
$r_0$& 3.50
\\  \hline
    \end{tabular}
  \end{center}
\end{table}
\begin{table}[h!]
  \begin{center}
    \caption{Einasto Optimization Values}
    \label{EinastoUGC00191}
    \begin{tabular}{|r|r|}
     \hline
      \textbf{Parameter}   & \textbf{Optimization Values}
      \\  \hline
     $\rho_e$  &$1\times 10^7$
\\  \hline
$r_e$ & 3.67
\\  \hline
$n_e$ & 0.49
\\  \hline
    \end{tabular}
  \end{center}
\end{table}
\begin{table}[h!]
\centering \caption{Physical assessment of collisional DM
parameters for UGC00191.}
\begin{tabular}{lcc}
\hline
Parameter & Value & Physical Verdict \\
\hline
$\gamma_0$ & 1.0001 & Extremely close to isothermal\\
$\delta_\gamma$ & $1.2\times10^{-9}$ & Negligible variation \\
$r_\gamma$ & 1.5 Kpc & Transition radius \\
$K_0$ & $2.5\times10^3$ & Enough pressure support \\
$r_c$ & 0.5 Kpc & Physically reasonable for inner halo region \\
$p$ & 0.01 & Extremely shallow decrease of $K(r)$, almost constant across halo \\
\hline
Overall &-& Model is physically viable \\
\hline
\end{tabular}
\label{EVALUATIONUGC00191}
\end{table}
Now the extended picture including the rotation velocity from the
other components of the galaxy, such as the disk and gas, makes
the collisional DM model viable for this galaxy. In Fig.
\ref{extendedUGC00191} we present the combined rotation curves
including the other components of the galaxy along with the
collisional matter. As it can be seen, the extended collisional DM
model is marginally viable.
Also in Table \ref{evaluationextendedUGC00191} we present the
values of the free parameters of the collisional DM model for
which the maximum compatibility with the SPARC data comes for the
galaxy UGC00191.
\begin{table}[h!]
\centering \caption{Physical assessment of Extended collisional DM
parameters for galaxy UGC00191.}
\begin{tabular}{lcc}
\hline
Parameter & Value & Physical Verdict \\
\hline
$\gamma_0$ & 0.99547161 & Nearly isothermal core \\
$\delta_\gamma$ & 0.03188285 & Mild variation; moderate rise of $\gamma(r)$ with radius \\
$K_0$ & 3000 & Moderate entropy \\
$ml_{disk}$ & 1.00000000 & Upper viable limit; disk-dominated stellar component \\
$ml_{bulge}$ & 0.00000000 & Negligible bulge contribution, consistent with morphology \\
\hline
Overall &-& Physically viable \\
\hline
\end{tabular}
\label{evaluationextendedUGC00191}
\end{table}

\subsection{The Galaxy UGC00634}

\begin{figure}[h!]
\centering
\includegraphics[width=20pc]{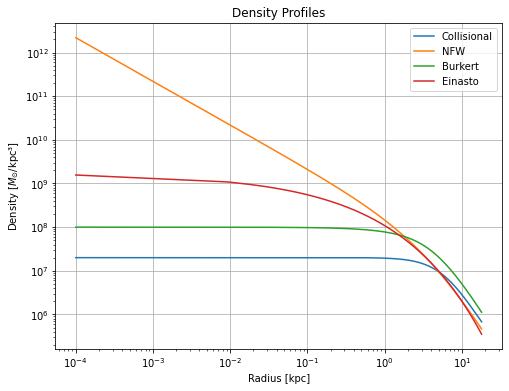}
\caption{The density of the collisional DM model (\ref{tanhmodel})
for the galaxy UGC00634, as a function of the radius.}
\label{UGC00634dens}
\end{figure}
\begin{figure}[h!]
\centering
\includegraphics[width=20pc]{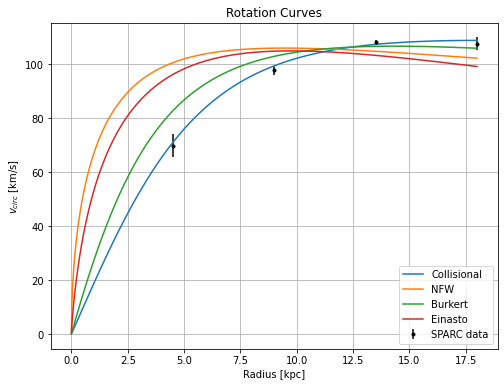}
\caption{The predicted rotation curves after using an optimization
for the collisional DM model (\ref{tanhmodel}), versus the SPARC
observational data for the galaxy UGC00634. We also plotted the
optimized curves for the NFW model, the Burkert model and the
Einasto model.} \label{UGC00634}
\end{figure}
\begin{figure}[h!]
\centering
\includegraphics[width=20pc]{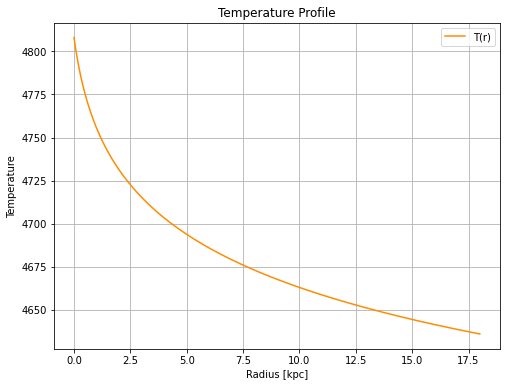}
\caption{The temperature as a function of the radius for the
collisional DM model (\ref{tanhmodel}) for the galaxy UGC00634.}
\label{UGC00634temp}
\end{figure}

For this galaxy, we shall choose $\rho_0=2\times
10^7$$M_{\odot}/\mathrm{Kpc}^{3}$. UGC00634 is classified in
catalogues as a dwarf galaxy, dwarf, weak, possibly irregular or
low-spiral system. In Figs. \ref{UGC00634dens}, \ref{UGC00634} and
\ref{UGC00634temp} we present the density of the collisional DM
model, the predicted rotation curves after using an optimization
for the collisional DM model (\ref{tanhmodel}), versus the SPARC
observational data and the temperature parameter as a function of
the radius respectively. As it can be seen, the SIDM model
produces viable rotation curves compatible with the SPARC data.
Also in Tables \ref{collUGC00634}, \ref{NavaroUGC00634},
\ref{BuckertUGC00634} and \ref{EinastoUGC00634} we present the
optimization values for the SIDM model, and the other DM profiles.
Also in Table \ref{EVALUATIONUGC00634} we present the overall
evaluation of the SIDM model for the galaxy at hand. The resulting
phenomenology is viable.
\begin{table}[h!]
  \begin{center}
    \caption{Collisional Dark Matter Optimization Values}
    \label{collUGC00634}
     \begin{tabular}{|r|r|}
     \hline
      \textbf{Parameter}   & \textbf{Optimization Values}
      \\  \hline
     $\delta_{\gamma} $ & 0.0000000012
\\  \hline
$\gamma_0 $ & 1.0001 \\ \hline $K_0$ ($M_{\odot} \,
\mathrm{Kpc}^{-3} \, (\mathrm{km/s})^{2}$)& 4800  \\ \hline
    \end{tabular}
  \end{center}
\end{table}
\begin{table}[h!]
  \begin{center}
    \caption{NFW  Optimization Values}
    \label{NavaroUGC00634}
     \begin{tabular}{|r|r|}
     \hline
      \textbf{Parameter}   & \textbf{Optimization Values}
      \\  \hline
   $\rho_s$   & $5\times 10^7$
\\  \hline
$r_s$&  4.38
\\  \hline
    \end{tabular}
  \end{center}
\end{table}
\begin{table}[h!]
  \begin{center}
    \caption{Burkert Optimization Values}
    \label{BuckertUGC00634}
     \begin{tabular}{|r|r|}
     \hline
      \textbf{Parameter}   & \textbf{Optimization Values}
      \\  \hline
     $\rho_0^B$  & $1\times 10^8$
\\  \hline
$r_0$&  4.42
\\  \hline
    \end{tabular}
  \end{center}
\end{table}
\begin{table}[h!]
  \begin{center}
    \caption{Einasto Optimization Values}
    \label{EinastoUGC00634}
    \begin{tabular}{|r|r|}
     \hline
      \textbf{Parameter}   & \textbf{Optimization Values}
      \\  \hline
     $\rho_e$  &$1\times 10^7$
\\  \hline
$r_e$ & 4.97
\\  \hline
$n_e$ & 0.39
\\  \hline
    \end{tabular}
  \end{center}
\end{table}
\begin{table}[h!]
\centering \caption{Physical assessment of collisional DM
parameters for UGC00634.}
\begin{tabular}{lcc}
\hline
Parameter & Value & Physical Verdict \\
\hline
$\gamma_0$ & 1.0001 & Extremely close to isothermal \\
$\delta_\gamma$ & $1.2\times10^{-9}$ & Negligible variation \\
$r_\gamma$ & 1.5 Kpc & Transition radius \\
$K_0$ & $4.8\times10^3$ & Enough pressure support \\
$r_c$ & 0.5 Kpc & Small core scale \\
$p$ & 0.01 & Extremely shallow decrease of $K(r)$, almost constant across halo \\
\hline
Overall &-& Model is physically acceptable \\
\hline
\end{tabular}
\label{EVALUATIONUGC00634}
\end{table}


\subsection{The Galaxy UGC00731 Non-viable, Extended Marginally Viable}

\begin{figure}[h!]
\centering
\includegraphics[width=20pc]{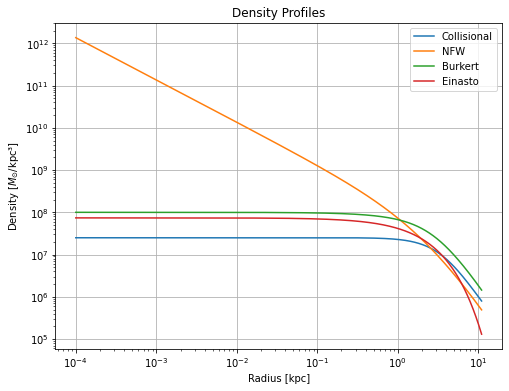}
\caption{The density of the collisional DM model (\ref{tanhmodel})
for the galaxy UGC00731, as a function of the radius.}
\label{UGC00731dens}
\end{figure}
\begin{figure}[h!]
\centering
\includegraphics[width=20pc]{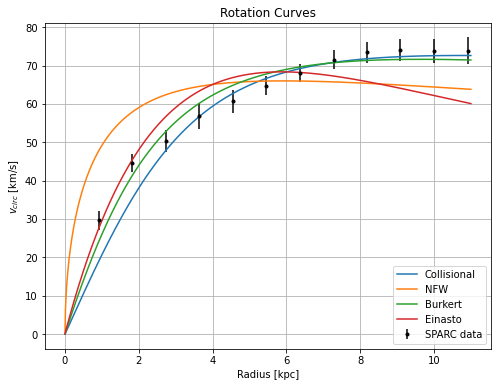}
\caption{The predicted rotation curves after using an optimization
for the collisional DM model (\ref{tanhmodel}), versus the SPARC
observational data for the galaxy UGC00731. We also plotted the
optimized curves for the NFW model, the Burkert model and the
Einasto model.} \label{UGC00731}
\end{figure}
\begin{figure}[h!]
\centering
\includegraphics[width=20pc]{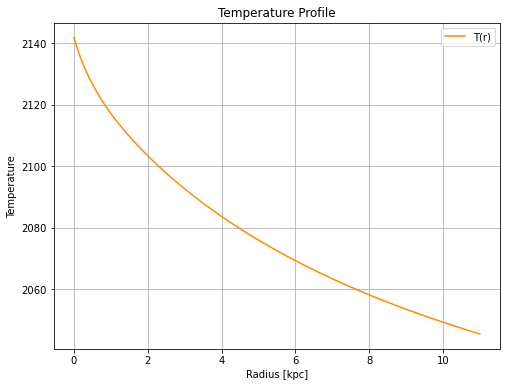}
\caption{The temperature as a function of the radius for the
collisional DM model (\ref{tanhmodel}) for the galaxy UGC00731.}
\label{UGC00731temp}
\end{figure}
\begin{figure}[h!]
\centering
\includegraphics[width=20pc]{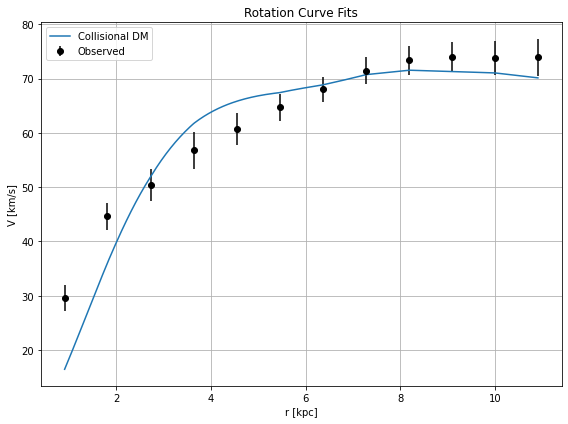}
\caption{The predicted rotation curves after using an optimization
for the collisional DM model (\ref{tanhmodel}), versus the
extended SPARC observational data for the galaxy UGC00731. The
model includes the rotation curves from all the components of the
galaxy, including gas and disk velocities, along with the
collisional DM model.} \label{extendedUGC00731}
\end{figure}

For this galaxy, we shall choose $\rho_0=2.5\times
10^7$$M_{\odot}/\mathrm{Kpc}^{3}$. UGC731 is a late-type,
low-luminosity disc system (often treated as a small/late spiral
or dwarf-spiral). In Figs. \ref{UGC00731dens}, \ref{UGC00731} and
\ref{UGC00731temp} we present the density of the collisional DM
model, the predicted rotation curves after using an optimization
for the collisional DM model (\ref{tanhmodel}), versus the SPARC
observational data and the temperature parameter as a function of
the radius respectively. As it can be seen, the SIDM model
produces non-viable rotation curves incompatible with the SPARC
data. Also in Tables \ref{collUGC00731}, \ref{NavaroUGC00731},
\ref{BuckertUGC00731} and \ref{EinastoUGC00731} we present the
optimization values for the SIDM model, and the other DM profiles.
Also in Table \ref{EVALUATIONUGC00731} we present the overall
evaluation of the SIDM model for the galaxy at hand. The resulting
phenomenology is non-viable.
\begin{table}[h!]
  \begin{center}
    \caption{Collisional Dark Matter Optimization Values}
    \label{collUGC00731}
     \begin{tabular}{|r|r|}
     \hline
      \textbf{Parameter}   & \textbf{Optimization Values}
      \\  \hline
     $\delta_{\gamma} $ & 0.0000000012
\\  \hline
$\gamma_0 $ & 1.0001 \\ \hline $K_0$ ($M_{\odot} \,
\mathrm{Kpc}^{-3} \, (\mathrm{km/s})^{2}$)& 2000 \\ \hline
    \end{tabular}
  \end{center}
\end{table}
\begin{table}[h!]
  \begin{center}
    \caption{NFW  Optimization Values}
    \label{NavaroUGC00731}
     \begin{tabular}{|r|r|}
     \hline
      \textbf{Parameter}   & \textbf{Optimization Values}
      \\  \hline
   $\rho_s$   & $5\times 10^7$
\\  \hline
$r_s$&  2.73
\\  \hline
    \end{tabular}
  \end{center}
\end{table}
\begin{table}[h!]
  \begin{center}
    \caption{Burkert Optimization Values}
    \label{BuckertUGC00731}
     \begin{tabular}{|r|r|}
     \hline
      \textbf{Parameter}   & \textbf{Optimization Values}
      \\  \hline
     $\rho_0^B$  & $1\times 10^8$
\\  \hline
$r_0$&  2.97
\\  \hline
    \end{tabular}
  \end{center}
\end{table}
\begin{table}[h!]
  \begin{center}
    \caption{Einasto Optimization Values}
    \label{EinastoUGC00731}
    \begin{tabular}{|r|r|}
     \hline
      \textbf{Parameter}   & \textbf{Optimization Values}
      \\  \hline
     $\rho_e$  &$1\times 10^7$
\\  \hline
$r_e$ & 3.47
\\  \hline
$n_e$ & 1
\\  \hline
    \end{tabular}
  \end{center}
\end{table}
\begin{table}[h!]
\centering \caption{Physical assessment of collisional DM
parameters (UGC00731).}
\begin{tabular}{lcc}
\hline
Parameter & Value & Physical Verdict \\
\hline
$\gamma_0$ & 1.0001 & Practically isothermal \\
$\delta_\gamma$ & 0.0000000012 & Negligible  \\
$r_\gamma$ & 1.5 Kpc & Transition radius inside inner halo \\
$K_0$ & $2\times10^{3}$ & Moderate entropy scale; reasonable for low-mass halo \\
$r_c$ & 0.5 Kpc & Small core radius; physically acceptable \\
$p$ & 0.01 & Nearly constant $K(r)$; almost no radial entropy gradient \\
\hline
Overall &-& Physically consistent \\
\hline
\end{tabular}
\label{EVALUATIONUGC00731}
\end{table}
Now the extended picture including the rotation velocity from the
other components of the galaxy, such as the disk and gas, makes
the collisional DM model viable for this galaxy. In Fig.
\ref{extendedUGC00731} we present the combined rotation curves
including the other components of the galaxy along with the
collisional matter. As it can be seen, the extended collisional DM
model is marginally viable.
Also in Table \ref{evaluationextendedUGC00731} we present the
values of the free parameters of the collisional DM model for
which the maximum compatibility with the SPARC data comes for the
galaxy UGC00731.
\begin{table}[h!]
\centering \caption{Physical assessment of Extended collisional DM
parameters (second set).}
\begin{tabular}{lcc}
\hline
Parameter & Value & Physical Verdict \\
\hline
$\gamma_0$ & 1.01558728 & Nearly isothermal.\\
$\delta_\gamma$ & 0.06366026 & Small but non-negligible radial variation of $\gamma(r)$ \\
$K_0$ & 3000 & Moderate entropy/pressure scale for galaxy-scale halo \\
ml\_disk & 0.50000000 & Stellar disk $M/L\sim0.5$ reasonable \\
ml\_bulge & 0.00000000 & Zero bulge mass-to-light \\
\hline
Overall &-& Physically plausible parameter set\\
\hline
\end{tabular}
\label{evaluationextendedUGC00731}
\end{table}


\subsection{The Galaxy UGC00891}

\begin{figure}[h!]
\centering
\includegraphics[width=20pc]{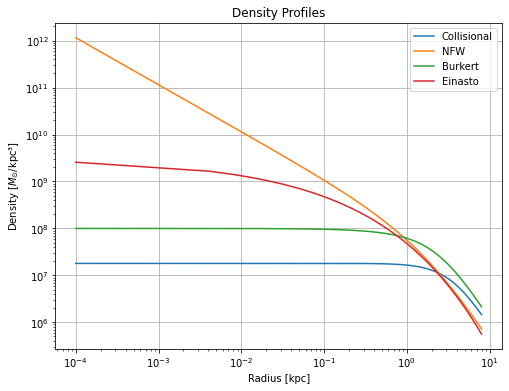}
\caption{The density of the collisional DM model (\ref{tanhmodel})
for the galaxy UGC00891, as a function of the radius.}
\label{UGC00891dens}
\end{figure}
\begin{figure}[h!]
\centering
\includegraphics[width=20pc]{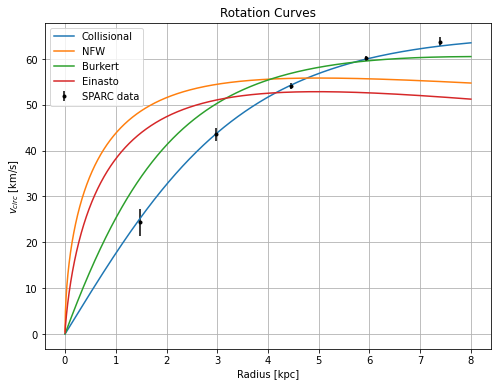}
\caption{The predicted rotation curves after using an optimization
for the collisional DM model (\ref{tanhmodel}), versus the SPARC
observational data for the galaxy UGC00891. We also plotted the
optimized curves for the NFW model, the Burkert model and the
Einasto model.} \label{UGC00891}
\end{figure}
\begin{figure}[h!]
\centering
\includegraphics[width=20pc]{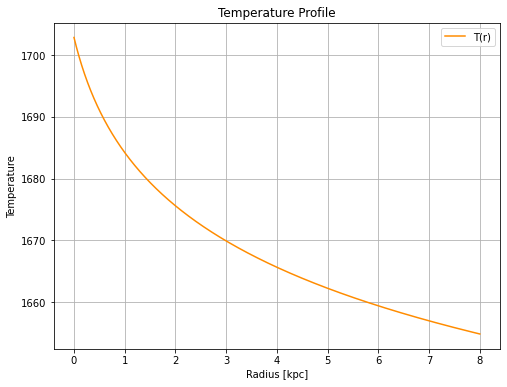}
\caption{The temperature as a function of the radius for the
collisional DM model (\ref{tanhmodel}) for the galaxy UGC00891.}
\label{UGC00891temp}
\end{figure}

For this galaxy, we shall choose $\rho_0=1.8\times
10^7$$M_{\odot}/\mathrm{Kpc}^{3}$. UGC00891 is an edge-on, large
SA(s)b spiral galaxy in the constellation Andromeda at a distance
of order $8.4\text{--}10\ \mathrm{Mpc}$. In Figs.
\ref{UGC00891dens}, \ref{UGC00891} and \ref{UGC00891temp} we
present the density of the collisional DM model, the predicted
rotation curves after using an optimization for the collisional DM
model (\ref{tanhmodel}), versus the SPARC observational data and
the temperature parameter as a function of the radius
respectively. As it can be seen, the SIDM model produces viable
rotation curves compatible with the SPARC data. Also in Tables
\ref{collUGC00891}, \ref{NavaroUGC00891}, \ref{BuckertUGC00891}
and \ref{EinastoUGC00891} we present the optimization values for
the SIDM model, and the other DM profiles. Also in Table
\ref{EVALUATIONUGC00891} we present the overall evaluation of the
SIDM model for the galaxy at hand. The resulting phenomenology is
viable.
\begin{table}[h!]
  \begin{center}
    \caption{Collisional Dark Matter Optimization Values}
    \label{collUGC00891}
     \begin{tabular}{|r|r|}
     \hline
      \textbf{Parameter}   & \textbf{Optimization Values}
      \\  \hline
     $\delta_{\gamma} $ & 0.0000000012
\\  \hline
$\gamma_0 $ & 1.0001  \\ \hline $K_0$ ($M_{\odot} \,
\mathrm{Kpc}^{-3} \, (\mathrm{km/s})^{2}$)& 1700  \\ \hline
    \end{tabular}
  \end{center}
\end{table}
\begin{table}[h!]
  \begin{center}
    \caption{NFW  Optimization Values}
    \label{NavaroUGC00891}
     \begin{tabular}{|r|r|}
     \hline
      \textbf{Parameter}   & \textbf{Optimization Values}
      \\  \hline
   $\rho_s$   & $5\times 10^7$
\\  \hline
$r_s$& 2.31
\\  \hline
    \end{tabular}
  \end{center}
\end{table}
\begin{table}[h!]
  \begin{center}
    \caption{Burkert Optimization Values}
    \label{BuckertUGC00891}
     \begin{tabular}{|r|r|}
     \hline
      \textbf{Parameter}   & \textbf{Optimization Values}
      \\  \hline
     $\rho_0^B$  & $1\times 10^8$
\\  \hline
$r_0$&  2.51
\\  \hline
    \end{tabular}
  \end{center}
\end{table}
\begin{table}[h!]
  \begin{center}
    \caption{Einasto Optimization Values}
    \label{EinastoUGC00891}
    \begin{tabular}{|r|r|}
     \hline
      \textbf{Parameter}   & \textbf{Optimization Values}
      \\  \hline
     $\rho_e$  &$1\times 10^7$
\\  \hline
$r_e$ & 2.49
\\  \hline
$n_e$ & 0.35
\\  \hline
    \end{tabular}
  \end{center}
\end{table}
\begin{table}[h!]
\centering \caption{Physical assessment of collisional DM
parameters for UGC00891.}
\begin{tabular}{lcc}
\hline
Parameter & Value & Physical Verdict \\
\hline
$\gamma_0$ & 1.0001 & Extremely close to isothermal, nearly uniform central pressure \\
$\delta_\gamma$ & $1.2\times10^{-9}$ & Negligible variation, $\gamma(r)$ essentially constant \\
$r_\gamma$ & 1.5 Kpc & Transition radius \\
$K_0$ & $1.7\times10^3$ & High entropy scale\\
$r_c$ & 0.5 Kpc & Small core scale \\
$p$ & 0.01 & Extremely shallow decrease of $K(r)$ \\
\hline
Overall &-& Model is physically acceptable \\
\hline
\end{tabular}
\label{EVALUATIONUGC00891}
\end{table}


\subsection{The Galaxy UGC01230}

\begin{figure}[h!]
\centering
\includegraphics[width=20pc]{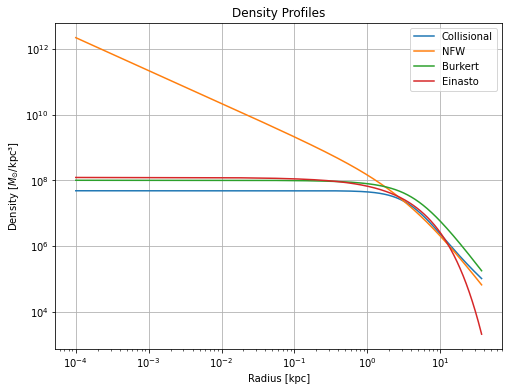}
\caption{The density of the collisional DM model (\ref{tanhmodel})
for the galaxy UGC01230, as a function of the radius.}
\label{UGC01230dens}
\end{figure}

\begin{figure}[h!]
\centering
\includegraphics[width=20pc]{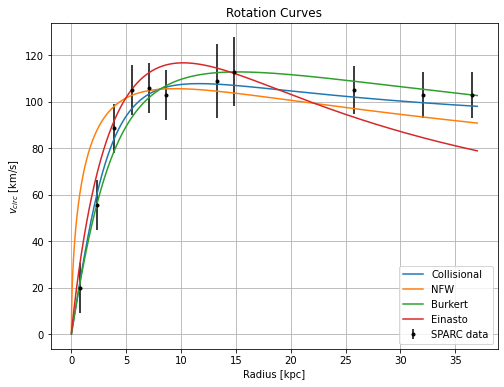}
\caption{The predicted rotation curves after using an optimization
for the collisional DM model (\ref{tanhmodel}), versus the SPARC
observational data for the galaxy UGC01230. We also plotted the
optimized curves for the NFW model, the Burkert model and the
Einasto model.} \label{UGC01230}
\end{figure}

\begin{figure}[h!]
\centering
\includegraphics[width=20pc]{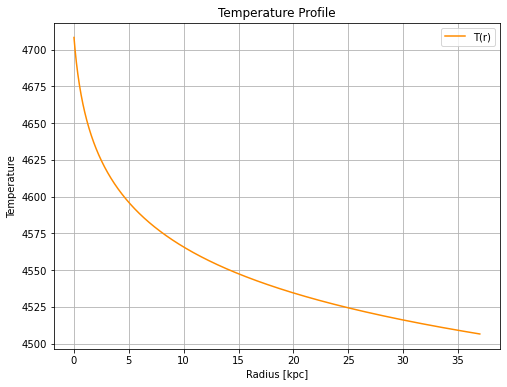}
\caption{The temperature as a function of the radius for the
collisional DM model (\ref{tanhmodel}) for the galaxy UGC01230.}
\label{UGC01230temp}
\end{figure}

For this galaxy, we shall choose $\rho_0=4.8\times
10^7$$M_{\odot}/\mathrm{Kpc}^{3}$. UGC01230 is a
low-surface-brightness, late-type spiral galaxy at a distance of
order $\sim 40\text{-}50\ \mathrm{Mpc}$, classified among the
low-surface-brightness systems with an unusually large, diffuse
stellar disk. It is a large low-surface-brightness spiral,
sometimes called a ''giant low-surface-brightness.''In Figs.
\ref{UGC01230dens}, \ref{UGC01230} and \ref{UGC01230temp} we
present the density of the collisional DM model, the predicted
rotation curves after using an optimization for the collisional DM
model (\ref{tanhmodel}), versus the SPARC observational data and
the temperature parameter as a function of the radius
respectively. As it can be seen, the SIDM model produces viable
rotation curves compatible with the SPARC data. Also in Tables
\ref{collUGC01230}, \ref{NavaroUGC01230}, \ref{BuckertUGC01230}
and \ref{EinastoUGC01230} we present the optimization values for
the SIDM model, and the other DM profiles. Also in Table
\ref{EVALUATIONUGC01230} we present the overall evaluation of the
SIDM model for the galaxy at hand. The resulting phenomenology is
viable.
\begin{table}[h!]
  \begin{center}
    \caption{Collisional Dark Matter Optimization Values}
    \label{collUGC01230}
     \begin{tabular}{|r|r|}
     \hline
      \textbf{Parameter}   & \textbf{Optimization Values}
      \\  \hline
     $\delta_{\gamma} $ & 0.0000000012
\\  \hline
$\gamma_0 $ &1.0001  \\ \hline $K_0$ ($M_{\odot} \,
\mathrm{Kpc}^{-3} \, (\mathrm{km/s})^{2}$)& 4700  \\ \hline
    \end{tabular}
  \end{center}
\end{table}
\begin{table}[h!]
  \begin{center}
    \caption{NFW  Optimization Values}
    \label{NavaroUGC01230}
     \begin{tabular}{|r|r|}
     \hline
      \textbf{Parameter}   & \textbf{Optimization Values}
      \\  \hline
   $\rho_s$   & $5\times 10^7$
\\  \hline
$r_s$&  4.37
\\  \hline
    \end{tabular}
  \end{center}
\end{table}
\begin{table}[h!]
  \begin{center}
    \caption{Burkert Optimization Values}
    \label{BuckertUGC01230}
     \begin{tabular}{|r|r|}
     \hline
      \textbf{Parameter}   & \textbf{Optimization Values}
      \\  \hline
     $\rho_0^B$  & $1\times 10^8$
\\  \hline
$r_0$&  4.68
\\  \hline
    \end{tabular}
  \end{center}
\end{table}
\begin{table}[h!]
  \begin{center}
    \caption{Einasto Optimization Values}
    \label{EinastoUGC01230}
    \begin{tabular}{|r|r|}
     \hline
      \textbf{Parameter}   & \textbf{Optimization Values}
      \\  \hline
     $\rho_e$  &$1\times 10^7$
\\  \hline
$r_e$ & 5.82
\\  \hline
$n_e$ & 1
\\  \hline
    \end{tabular}
  \end{center}
\end{table}
\begin{table}[h!]
\centering \caption{Physical assessment of collisional DM
parameters for UGC01230.}
\begin{tabular}{lcc}
\hline
Parameter & Value & Physical Verdict \\
\hline
$\gamma_0$ & 1.0001 & Practically isothermal\\
$\delta_\gamma$ & $1.2\times10^{-9}$ & Negligible variation \\
$r_\gamma$ & 1.5 Kpc & Transition radius \\
$K_0$ & $4.7\times10^3$ & High entropy scale \\
$r_c$ & 0.5 Kpc & Small core scale, physically reasonable for inner halo region \\
$p$ & 0.01 & Extremely shallow K(r) decrease \\
\hline
Overall &-& Model is physically simple and nearly isothermal \\
\hline
\end{tabular}
\label{EVALUATIONUGC01230}
\end{table}

\subsection{The Galaxy UGC02023  Remarkably has the Same Model Parameter Values as UGC01281}

\begin{figure}[h!]
\centering
\includegraphics[width=20pc]{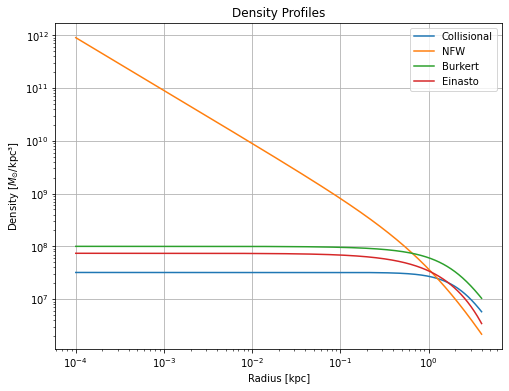}
\caption{The density of the collisional DM model (\ref{tanhmodel})
for the galaxy UGC02023, as a function of the radius.}
\label{UGC02023dens}
\end{figure}

\begin{figure}[h!]
\centering
\includegraphics[width=20pc]{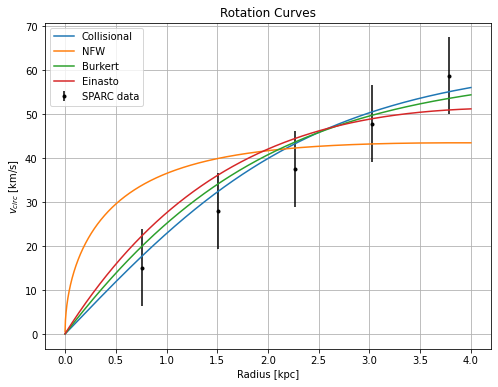}
\caption{The predicted rotation curves after using an optimization
for the collisional DM model (\ref{tanhmodel}), versus the SPARC
observational data for the galaxy UGC02023. We also plotted the
optimized curves for the NFW model, the Burkert model and the
Einasto model.} \label{UGC02023}
\end{figure}

\begin{figure}[h!]
\centering
\includegraphics[width=20pc]{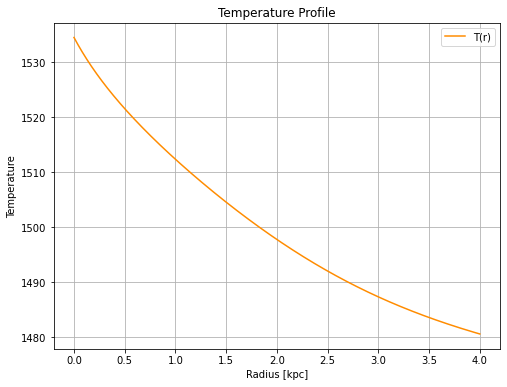}
\caption{The temperature as a function of the radius for the
collisional DM model (\ref{tanhmodel}) for the galaxy UGC02023.}
\label{UGC02023temp}
\end{figure}

For this galaxy, we shall choose $\rho_0=2.8\times
10^7$$M_{\odot}/\mathrm{Kpc}^{3}$. The galaxy UGC 02023 may be
considered a late-type dwarf/low-surface-brightness spiral at a
distance of about 15 Mpc, and is thus an ordinary (not giant)
spiral with a modest stellar disk and extended HI envelope. In
Figs. \ref{UGC02023dens}, \ref{UGC02023} and \ref{UGC02023temp} we
present the density of the collisional DM model, the predicted
rotation curves after using an optimization for the collisional DM
model (\ref{tanhmodel}), versus the SPARC observational data and
the temperature parameter as a function of the radius
respectively. As it can be seen, the SIDM model produces viable
rotation curves compatible with the SPARC data. Also in Tables
\ref{collUGC02023}, \ref{NavaroUGC02023}, \ref{BuckertUGC02023}
and \ref{EinastoUGC02023} we present the optimization values for
the SIDM model, and the other DM profiles. Also in Table
\ref{EVALUATIONUGC02023} we present the overall evaluation of the
SIDM model for the galaxy at hand. The resulting phenomenology is
viable.
\begin{table}[h!]
  \begin{center}
    \caption{Collisional Dark Matter Optimization Values}
    \label{collUGC02023}
     \begin{tabular}{|r|r|}
     \hline
      \textbf{Parameter}   & \textbf{Optimization Values}
      \\  \hline
     $\delta_{\gamma} $ & 0.0000000012
\\  \hline
$\gamma_0 $ & 1.0001  \\ \hline $K_0$ ($M_{\odot} \,
\mathrm{Kpc}^{-3} \, (\mathrm{km/s})^{2}$)& 1600  \\ \hline
    \end{tabular}
  \end{center}
\end{table}
\begin{table}[h!]
  \begin{center}
    \caption{NFW  Optimization Values}
    \label{NavaroUGC02023}
     \begin{tabular}{|r|r|}
     \hline
      \textbf{Parameter}   & \textbf{Optimization Values}
      \\  \hline
   $\rho_s$   & $5\times 10^7$
\\  \hline
$r_s$&  1.80
\\  \hline
    \end{tabular}
  \end{center}
\end{table}
\begin{table}[h!]
  \begin{center}
    \caption{Burkert Optimization Values}
    \label{BuckertUGC02023}
     \begin{tabular}{|r|r|}
     \hline
      \textbf{Parameter}   & \textbf{Optimization Values}
      \\  \hline
     $\rho_0^B$  & $1\times 10^8$
\\  \hline
$r_0$& 2.45
\\  \hline
    \end{tabular}
  \end{center}
\end{table}
\begin{table}[h!]
  \begin{center}
    \caption{Einasto Optimization Values}
    \label{EinastoUGC02023}
    \begin{tabular}{|r|r|}
     \hline
      \textbf{Parameter}   & \textbf{Optimization Values}
      \\  \hline
     $\rho_e$  &$1\times 10^7$
\\  \hline
$r_e$ & 2.61
\\  \hline
$n_e$ & 1
\\  \hline
    \end{tabular}
  \end{center}
\end{table}
\begin{table}[h!]
\centering \caption{Physical assessment of collisional DM
parameters (UGC02023).}
\begin{tabular}{lcc}
\hline
\textbf{Parameter} & \textbf{Value} & \textbf{Physical Verdict} \\
\hline
$\gamma_0$ & $1.0001$ & Practically isothermal \\
$\delta_\gamma$ & $1.2\times10^{-9}$ & Extremely tiny variation \\
$r_\gamma$ & $1.5\ \mathrm{Kpc}$ & Transition radius \\
$K_0$ & $1.6\times10^{3}$ & Moderate entropy/temperature scale \\
$r_c$ & $0.5\ \mathrm{Kpc}$ & Small core scale \\
$p$ & $0.01$ & Very shallow radial decline of $K(r)$ \\
\hline
\textbf{Overall} & -- & Physically consistent and numerically stable \\
\hline
\end{tabular}
\label{EVALUATIONUGC02023}
\end{table}


\subsection{The Galaxy UGC02259 Non-Viable, Extended non-viable too}

\begin{figure}[h!]
\centering
\includegraphics[width=20pc]{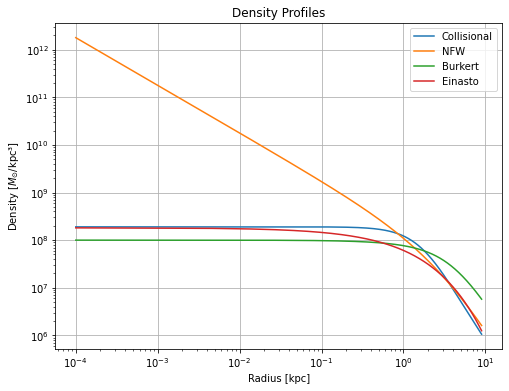}
\caption{The density of the collisional DM model (\ref{tanhmodel})
for the galaxy UGC02259, as a function of the radius.}
\label{UGC02259dens}
\end{figure}
\begin{figure}[h!]
\centering
\includegraphics[width=20pc]{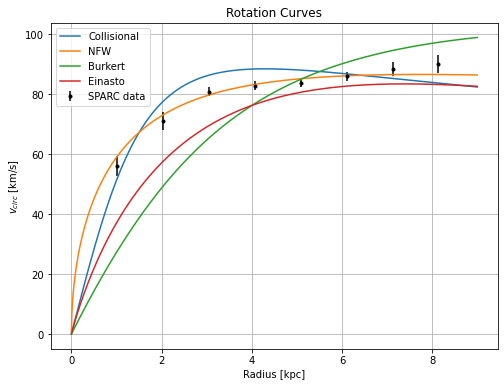}
\caption{The predicted rotation curves after using an optimization
for the collisional DM model (\ref{tanhmodel}), versus the SPARC
observational data for the galaxy UGC02259. We also plotted the
optimized curves for the NFW model, the Burkert model and the
Einasto model.} \label{UGC02259}
\end{figure}
\begin{figure}[h!]
\centering
\includegraphics[width=20pc]{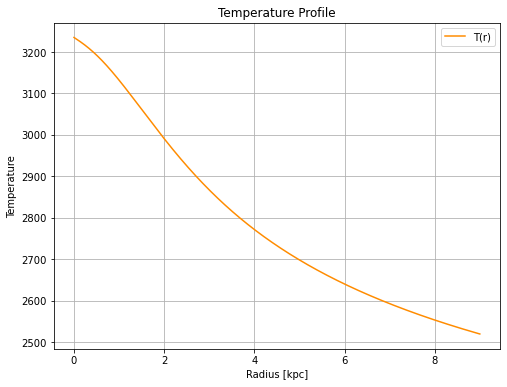}
\caption{The temperature as a function of the radius for the
collisional DM model (\ref{tanhmodel}) for the galaxy UGC02259.}
\label{UGC02259temp}
\end{figure}
\begin{figure}[h!]
\centering
\includegraphics[width=20pc]{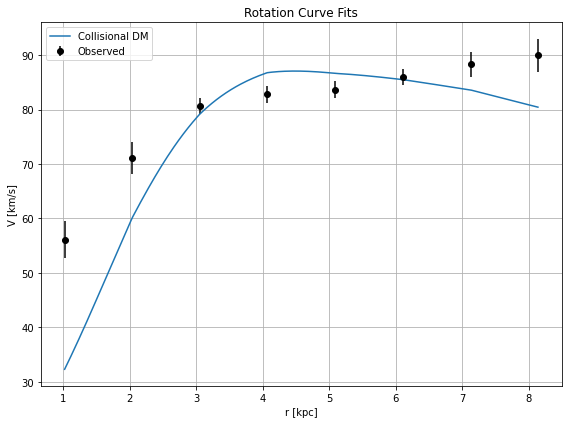}
\caption{The predicted rotation curves after using an optimization
for the collisional DM model (\ref{tanhmodel}), versus the
extended SPARC observational data for the galaxy UGC02259. The
model includes the rotation curves from all the components of the
galaxy, including gas and disk velocities, along with the
collisional DM model.} \label{extendedUGC02259}
\end{figure}

For this galaxy, we shall choose $\rho_0=1.9\times
10^8$$M_{\odot}/\mathrm{Kpc}^{3}$. The galaxy UGC02259 is a nearby
late-type, low-luminosity spiral (classified SB(s)cd / dwarf
''regular'' spiral) at a distance of about $D\sim 10.0\pm1.3\
\mathrm{Mpc}$. In Figs. \ref{UGC02259dens}, \ref{UGC02259} and
\ref{UGC02259temp} we present the density of the collisional DM
model, the predicted rotation curves after using an optimization
for the collisional DM model (\ref{tanhmodel}), versus the SPARC
observational data and the temperature parameter as a function of
the radius respectively. As it can be seen, the SIDM model
produces non-viable rotation curves incompatible with the SPARC
data. Also in Tables \ref{collUGC02259}, \ref{NavaroUGC02259},
\ref{BuckertUGC02259} and \ref{EinastoUGC02259} we present the
optimization values for the SIDM model, and the other DM profiles.
Also in Table \ref{EVALUATIONUGC02259} we present the overall
evaluation of the SIDM model for the galaxy at hand. The resulting
phenomenology is non-viable.
\begin{table}[h!]
  \begin{center}
    \caption{Collisional Dark Matter Optimization Values}
    \label{collUGC02259}
     \begin{tabular}{|r|r|}
     \hline
      \textbf{Parameter}   & \textbf{Optimization Values}
      \\  \hline
     $\delta_{\gamma} $ &0.0000000012
\\  \hline
$\gamma_0 $ & 1.0001  \\ \hline $K_0$ ($M_{\odot} \,
\mathrm{Kpc}^{-3} \, (\mathrm{km/s})^{2}$)& 1500  \\ \hline
    \end{tabular}
  \end{center}
\end{table}
\begin{table}[h!]
  \begin{center}
    \caption{NFW  Optimization Values}
    \label{NavaroUGC02259}
     \begin{tabular}{|r|r|}
     \hline
      \textbf{Parameter}   & \textbf{Optimization Values}
      \\  \hline
   $\rho_s$   & $5\times 10^7$
\\  \hline
$r_s$&  3.58
\\  \hline
    \end{tabular}
  \end{center}
\end{table}
\begin{table}[h!]
  \begin{center}
    \caption{Burkert Optimization Values}
    \label{BuckertUGC02259}
     \begin{tabular}{|r|r|}
     \hline
      \textbf{Parameter}   & \textbf{Optimization Values}
      \\  \hline
     $\rho_0^B$  & $1\times 10^8$
\\  \hline
$r_0$&  4.22
\\  \hline
    \end{tabular}
  \end{center}
\end{table}
\begin{table}[h!]
  \begin{center}
    \caption{Einasto Optimization Values}
    \label{EinastoUGC02259}
    \begin{tabular}{|r|r|}
     \hline
      \textbf{Parameter}   & \textbf{Optimization Values}
      \\  \hline
     $\rho_e$  &$1\times 10^7$
\\  \hline
$r_e$ & 4.11
\\  \hline
$n_e$ & 0.69
\\  \hline
    \end{tabular}
  \end{center}
\end{table}
\begin{table}[h!]
\centering \caption{Physical assessment of collisional DM
parameters (UGC02259).}
\begin{tabular}{lcc}
\hline
Parameter & Value & Physical Verdict \\
\hline
$\gamma_0$ & $1.0001$ & Slightly above isothermal \\
$\delta_\gamma$ & $1.2\times10^{-9}$ & Extremely tiny  \\
$r_\gamma$ & $1.5\ \mathrm{Kpc}$ & Transition radius inside inner halo   \\
$K_0$ ($M_{\odot}\,\mathrm{Kpc}^{-3}\,(\mathrm{km/s})^{2}$) & $1.5\times10^{3}$ & Moderate entropy \\
$r_c$ & $0.5\ \mathrm{Kpc}$ & Small core scale  \\
$p$ & $0.01$ & Very shallow radial decline of $K(r)$ \\
\hline
Overall & -- & Numerically stable and physically consistent \\
\hline
\end{tabular}
\label{EVALUATIONUGC02259}
\end{table}
Now the extended picture including the rotation velocity from the
other components of the galaxy, such as the disk and gas, makes
the collisional DM model viable for this galaxy. In Fig.
\ref{extendedUGC02259} we present the combined rotation curves
including the other components of the galaxy along with the
collisional matter. As it can be seen, the extended collisional DM
model is non-viable.
Also in Table \ref{evaluationextendedUGC02259} we present the
values of the free parameters of the collisional DM model for
which the maximum compatibility with the SPARC data comes for the
galaxy UGC02259.
\begin{table}[h!]
\centering \caption{Physical assessment of Extended collisional DM
parameters (second set).}
\begin{tabular}{lcc}
\hline
Parameter & Value & Physical Verdict \\
\hline
$\gamma_0$ & 1.05236091 & Slightly above isothermal \\
$\delta_\gamma$ & 0.09519684 & Moderate radial variation of $\gamma(r)$ \\
$K_0$ & 3000 & Moderate entropy/pressure scale for a galaxy-scale halo \\
ml\_disk & 1.00000000 & Stellar disk $M/L\sim1.0$ is on the high side   \\
ml\_bulge & 0.00000000 & Zero bulge mass-to-light (no bulge) \\
\hline
Overall &-& Physically plausible parameter set \\
\hline
\end{tabular}
\label{evaluationextendedUGC02259}
\end{table}

\subsection{The Galaxy UGC02487 Non-Viable Late-time Spiral, Barred, Extended non-viable too}

\begin{figure}[h!]
\centering
\includegraphics[width=20pc]{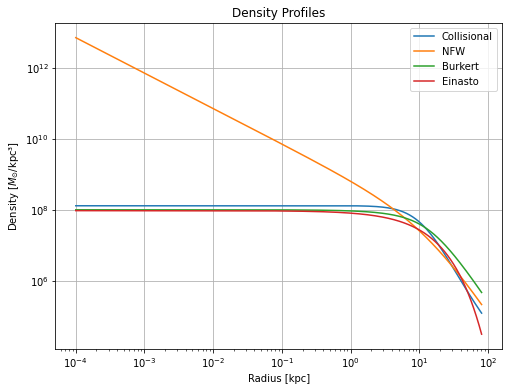}
\caption{The density of the collisional DM model (\ref{tanhmodel})
for the galaxy UGC02487, as a function of the radius.}
\label{UGC02487dens}
\end{figure}

\begin{figure}[h!]
\centering
\includegraphics[width=20pc]{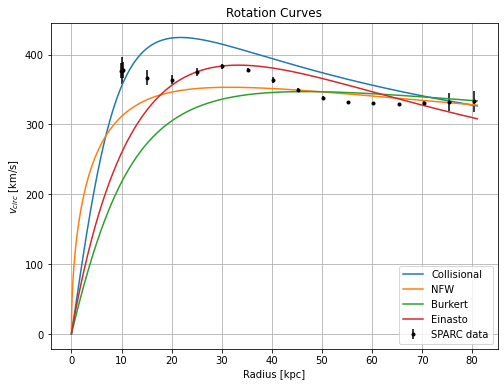}
\caption{The predicted rotation curves after using an optimization
for the collisional DM model (\ref{tanhmodel}), versus the SPARC
observational data for the galaxy UGC02487. We also plotted the
optimized curves for the NFW model, the Burkert model and the
Einasto model.} \label{UGC02487}
\end{figure}

\begin{figure}[h!]
\centering
\includegraphics[width=20pc]{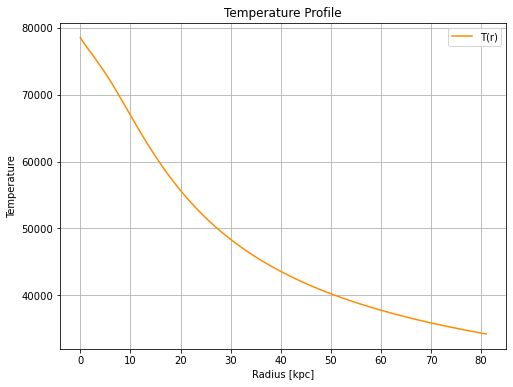}
\caption{The temperature as a function of the radius for the
collisional DM model (\ref{tanhmodel}) for the galaxy UGC02487.}
\label{UGC02487temp}
\end{figure}

\begin{figure}[h!]
\centering
\includegraphics[width=20pc]{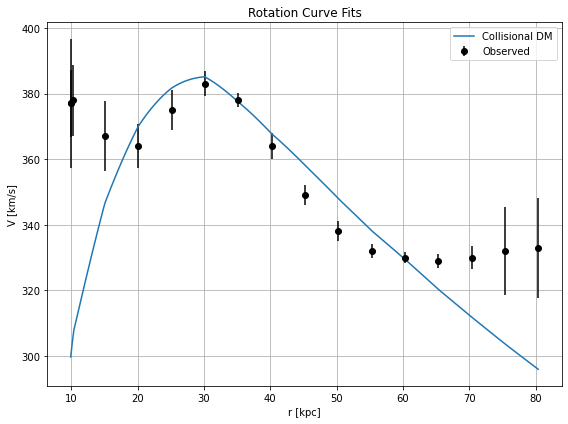}
\caption{The predicted rotation curves after using an optimization
for the collisional DM model (\ref{tanhmodel}), versus the
extended SPARC observational data for the galaxy UGC02487. The
model includes the rotation curves from all the components of the
galaxy, including gas and disk velocities, along with the
collisional DM model.} \label{extendedUGC02487}
\end{figure}

For this galaxy, we shall choose $\rho_0=1.3\times
10^8$$M_{\odot}/\mathrm{Kpc}^{3}$. The galaxy UGC 02487 is
classified as a large barred spiral of Hubble type SB(r)c / SBb to
a distance \(D\sim 65\text{-}70\ \mathrm{Mpc}\). In Figs.
\ref{UGC02487dens}, \ref{UGC02487} and \ref{UGC02487temp} we
present the density of the collisional DM model, the predicted
rotation curves after using an optimization for the collisional DM
model (\ref{tanhmodel}), versus the SPARC observational data and
the temperature parameter as a function of the radius
respectively. As it can be seen, the SIDM model produces
non-viable rotation curves incompatible with the SPARC data. Also
in Tables \ref{collUGC02487}, \ref{NavaroUGC02487},
\ref{BuckertUGC02487} and \ref{EinastoUGC02487} we present the
optimization values for the SIDM model, and the other DM profiles.
Also in Table \ref{EVALUATIONUGC02487} we present the overall
evaluation of the SIDM model for the galaxy at hand. The resulting
phenomenology is non-viable.
\begin{table}[h!]
  \begin{center}
    \caption{Collisional Dark Matter Optimization Values}
    \label{collUGC02487}
     \begin{tabular}{|r|r|}
     \hline
      \textbf{Parameter}   & \textbf{Optimization Values}
      \\  \hline
     $\delta_{\gamma} $ & 0.0005
\\  \hline
$\gamma_0 $ & 1.11  \\ \hline $K_0$ ($M_{\odot} \,
\mathrm{Kpc}^{-3} \, (\mathrm{km/s})^{2}$)& 10000  \\ \hline
    \end{tabular}
  \end{center}
\end{table}
\begin{table}[h!]
  \begin{center}
    \caption{NFW  Optimization Values}
    \label{NavaroUGC02487}
     \begin{tabular}{|r|r|}
     \hline
      \textbf{Parameter}   & \textbf{Optimization Values}
      \\  \hline
   $\rho_s$   & $5\times 10^7$
\\  \hline
$r_s$& 14.61
\\  \hline
    \end{tabular}
  \end{center}
\end{table}
\begin{table}[h!]
  \begin{center}
    \caption{Burkert Optimization Values}
    \label{BuckertUGC02487}
     \begin{tabular}{|r|r|}
     \hline
      \textbf{Parameter}   & \textbf{Optimization Values}
      \\  \hline
     $\rho_0^B$  & $1\times 10^8$
\\  \hline
$r_0$&  14.39
\\  \hline
    \end{tabular}
  \end{center}
\end{table}
\begin{table}[h!]
  \begin{center}
    \caption{Einasto Optimization Values}
    \label{EinastoUGC02487}
    \begin{tabular}{|r|r|}
     \hline
      \textbf{Parameter}   & \textbf{Optimization Values}
      \\  \hline
     $\rho_e$  &$1\times 10^7$
\\  \hline
$r_e$ & 19.35
\\  \hline
$n_e$ & 0.89
\\  \hline
    \end{tabular}
  \end{center}
\end{table}
\begin{table}[h!]
\centering \caption{Physical assessment of collisional DM
parameters (UGC02487).}
\begin{tabular}{lcc}
\hline
Parameter & Value & Physical Verdict \\
\hline
$\gamma_0$ & $1.0001$ & Nearly isothermal \\
$\delta_\gamma$ & $1.2\times10^{-9}$ & Practically zero \\
$r_\gamma$ & $1.5\ \mathrm{Kpc}$ & Transition radius inside halo but irrelevant with tiny $\delta_\gamma$ \\
$K_0$ ($M_{\odot}\,\mathrm{Kpc}^{-3}\,(\mathrm{km/s})^{2}$) & $1.0\times10^{4}$ & Moderate\\
$r_c$ & $0.5\ \mathrm{Kpc}$ & Small core scale - reasonable for inner halo \\
$p$ & $0.01$ & Very shallow radial decline of $K(r)$  \\
\hline
Overall & -- & Numerically stable and internally consistent \\
\hline
\end{tabular}
\label{EVALUATIONUGC02487}
\end{table}
Now the extended picture including the rotation velocity from the
other components of the galaxy, such as the disk and gas, makes
the collisional DM model viable for this galaxy. In Fig.
\ref{extendedUGC02487} we present the combined rotation curves
including the other components of the galaxy along with the
collisional matter. As it can be seen, the extended collisional DM
model is non-viable.
Also in Table \ref{evaluationextendedUGC02487} we present the
values of the free parameters of the collisional DM model for
which the maximum compatibility with the SPARC data comes for the
galaxy UGC02487.
\begin{table}[h!]
\centering \caption{Physical assessment of Extended collisional DM
parameters (second set).}
\begin{tabular}{lcc}
\hline
Parameter & Value & Physical Verdict \\
\hline
$\gamma_0$ & 1.16824443 & Noticeably above isothermal \\
$\delta_\gamma$ & 0.01056752 & Extremely small radial variation of $\gamma(r)$\\
$K_0$ & 3000 & Moderate entropy\\
ml\_disk & 1.00000000 & Disk $M/L\sim1.0$ is relatively high  \\
ml\_bulge & 0.00000000 & Zero bulge mass-to-light (no bulge) \\
\hline
Overall &-& Physically plausible parameter set\\
\hline
\end{tabular}
\label{evaluationextendedUGC02487}
\end{table}

\subsection{The Galaxy UGC02885 Non-viable Extraordinary Large Late-type Spiral, Extended non-viable too}

\begin{figure}[h!]
\centering
\includegraphics[width=20pc]{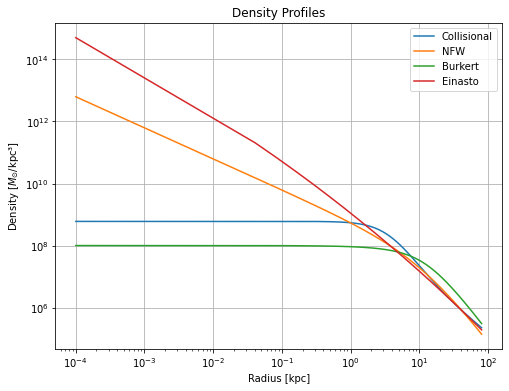}
\caption{The density of the collisional DM model (\ref{tanhmodel})
for the galaxy UGC02885, as a function of the radius.}
\label{UGC02885dens}
\end{figure}

\begin{figure}[h!]
\centering
\includegraphics[width=20pc]{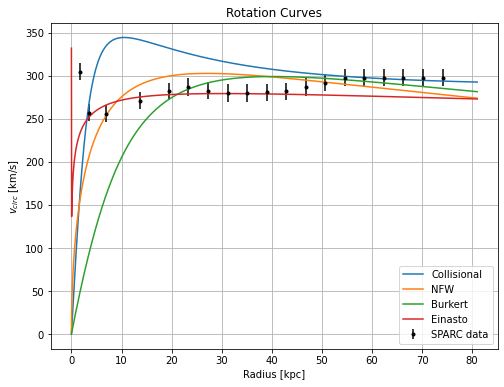}
\caption{The predicted rotation curves after using an optimization
for the collisional DM model (\ref{tanhmodel}), versus the SPARC
observational data for the galaxy UGC02885. We also plotted the
optimized curves for the NFW model, the Burkert model and the
Einasto model.} \label{UGC02885}
\end{figure}

\begin{figure}[h!]
\centering
\includegraphics[width=20pc]{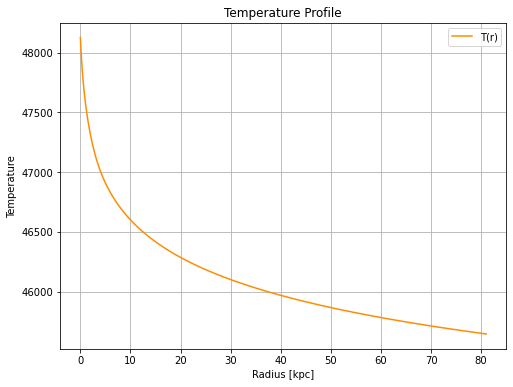}
\caption{The temperature as a function of the radius for the
collisional DM model (\ref{tanhmodel}) for the galaxy UGC02885.}
\label{UGC02885temp}
\end{figure}

\begin{figure}[h!]
\centering
\includegraphics[width=20pc]{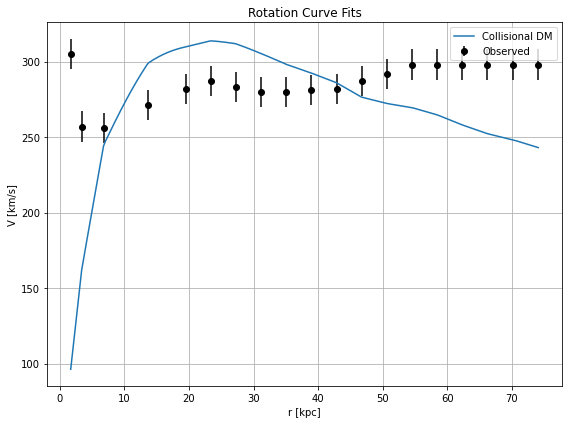}
\caption{The predicted rotation curves after using an optimization
for the collisional DM model (\ref{tanhmodel}), versus the
extended SPARC observational data for the galaxy UGC02885. The
model includes the rotation curves from all the components of the
galaxy, including gas and disk velocities, along with the
collisional DM model.} \label{extendedUGC02885}
\end{figure}

For this galaxy, we shall choose $\rho_0=6\times
10^8$$M_{\odot}/\mathrm{Kpc}^{3}$. UGC02885 (commonly referred to
as UGC2885 or ''Rubin's Galaxy'') is an extraordinarily large,
late-type barred spiral (SA(rs)c) in Perseus at a distance of
roughly $\sim 70-85\mathrm{Mpc}$, making it one of the most
massive nearby disk galaxies. In Figs. \ref{UGC02885dens},
\ref{UGC02885} and \ref{UGC02885temp} we present the density of
the collisional DM model, the predicted rotation curves after
using an optimization for the collisional DM model
(\ref{tanhmodel}), versus the SPARC observational data and the
temperature parameter as a function of the radius respectively. As
it can be seen, the SIDM model produces non-viable rotation curves
incompatible with the SPARC data. Also in Tables
\ref{collUGC02885}, \ref{NavaroUGC02885}, \ref{BuckertUGC02885}
and \ref{EinastoUGC02885} we present the optimization values for
the SIDM model, and the other DM profiles. Also in Table
\ref{EVALUATIONUGC02885} we present the overall evaluation of the
SIDM model for the galaxy at hand. The resulting phenomenology is
non-viable.
\begin{table}[h!]
  \begin{center}
    \caption{Collisional Dark Matter Optimization Values}
    \label{collUGC02885}
     \begin{tabular}{|r|r|}
     \hline
      \textbf{Parameter}   & \textbf{Optimization Values}
      \\  \hline
     $\delta_{\gamma} $ & 0.0000000012
\\  \hline
$\gamma_0 $ & 1.0001 \\ \hline $K_0$ ($M_{\odot} \,
\mathrm{Kpc}^{-3} \, (\mathrm{km/s})^{2}$)& 48000  \\ \hline
    \end{tabular}
  \end{center}
\end{table}
\begin{table}[h!]
  \begin{center}
    \caption{NFW  Optimization Values}
    \label{NavaroUGC02885}
     \begin{tabular}{|r|r|}
     \hline
      \textbf{Parameter}   & \textbf{Optimization Values}
      \\  \hline
   $\rho_s$   & $5\times 10^7$
\\  \hline
$r_s$&  12.53
\\  \hline
    \end{tabular}
  \end{center}
\end{table}
\begin{table}[h!]
  \begin{center}
    \caption{Burkert Optimization Values}
    \label{BuckertUGC02885}
     \begin{tabular}{|r|r|}
     \hline
      \textbf{Parameter}   & \textbf{Optimization Values}
      \\  \hline
     $\rho_0^B$  & $1\times 10^8$
\\  \hline
$r_0$&  12.41
\\  \hline
    \end{tabular}
  \end{center}
\end{table}
\begin{table}[h!]
  \begin{center}
    \caption{Einasto Optimization Values}
    \label{EinastoUGC02885}
    \begin{tabular}{|r|r|}
     \hline
      \textbf{Parameter}   & \textbf{Optimization Values}
      \\  \hline
     $\rho_e$  &$1\times 10^7$
\\  \hline
$r_e$ & 12.27
\\  \hline
$n_e$ & 0.05
\\  \hline
    \end{tabular}
  \end{center}
\end{table}
\begin{table}[h!]
\centering \caption{Physical assessment of collisional DM
parameters (UGC02885).}
\begin{tabular}{lcc}
\hline
Parameter & Value & Physical Verdict \\
\hline
$\gamma_0$ & $1.0001$ & Practically isothermal; negligible polytropic stiffness \\
$\delta_\gamma$ & $1.2\times10^{-9}$ & Vanishingly small  \\
$r_\gamma$ & $1.5\ \mathrm{Kpc}$ & Transition radius inside inner halo but irrelevant with tiny $\delta_\gamma$ \\
$K_0$ ($M_{\odot}\,\mathrm{Kpc}^{-3}\,(\mathrm{km/s})^{2}$) & $4.8\times10^{4}$ & Large entropy \\
$r_c$ & $0.5\ \mathrm{Kpc}$ & Small core scale - acceptable for inner halo; effect muted by tiny $p$ \\
$p$ & $0.01$ & Very shallow radial decline of $K(r)$  \\
\hline
Overall & -- & Numerically stable and internally consistent \\
\hline
\end{tabular}
\label{EVALUATIONUGC02885}
\end{table}
Now the extended picture including the rotation velocity from the
other components of the galaxy, such as the disk and gas, makes
the collisional DM model viable for this galaxy. In Fig.
\ref{extendedUGC02885} we present the combined rotation curves
including the other components of the galaxy along with the
collisional matter. As it can be seen, the extended collisional DM
model is non-viable.
Also in Table \ref{evaluationextendedUGC02885} we present the
values of the free parameters of the collisional DM model for
which the maximum compatibility with the SPARC data comes for the
galaxy UGC02885.
\begin{table}[h!]
\centering \caption{Physical assessment of Extended collisional DM
parameters (second set) for UGC02885.}
\begin{tabular}{lcc}
\hline
Parameter & Value & Physical Verdict \\
\hline
$\gamma_0$ & 1.12133906 & Near-isothermal core, mildly pressure-supported \\
$\delta_\gamma$ & 0.01 & Very small variation, nearly constant $\gamma(r)$ \\
$K_0$ & 3000 & Moderate entropy  , stable configuration \\
$ml_{disk}$ & 1.00000000 & Disk-dominated mass, typical for large spirals \\
$ml_{bulge}$ & 0.00006916 & Negligible bulge contribution, dynamically unimportant \\
\hline
Overall &-& Physically viable; quasi-isothermal inner halo, disk-dominated system \\
\hline
\end{tabular}
\label{evaluationextendedUGC02885}
\end{table}

\subsection{The Galaxy UGC02953 Non-viable Late type spiral, Extended non-viable too}

\begin{figure}[h!]
\centering
\includegraphics[width=20pc]{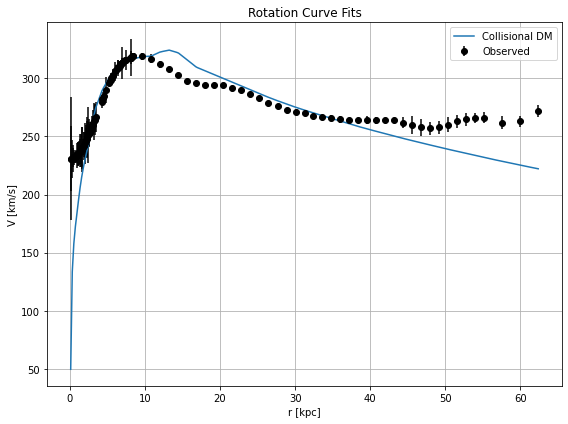}
\caption{The predicted rotation curves after using an optimization
for the collisional DM model (\ref{tanhmodel}), versus the
extended SPARC observational data for the galaxy UGC02953. The
model includes the rotation curves from all the components of the
galaxy, including gas and disk velocities, along with the
collisional DM model.} \label{extendedUGC02953}
\end{figure}

For this galaxy, we shall choose $\rho_0=9\times
10^9$$M_{\odot}/\mathrm{Kpc}^{3}$. UGC2953 (catalogued also as
IC\,356 / PGC\,14508) is classified as an Sb spiral galaxy in
Camelopardalis at a distance of about \(16.6\ \mathrm{Mpc}\).
\begin{figure}[h!]
\centering
\includegraphics[width=20pc]{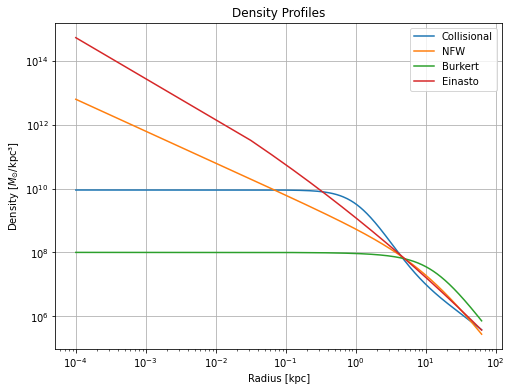}
\caption{The density of the collisional DM model (\ref{tanhmodel})
for the galaxy UGC02953, as a function of the radius.}
\label{UGC02953dens}
\end{figure}
\begin{figure}[h!]
\centering
\includegraphics[width=20pc]{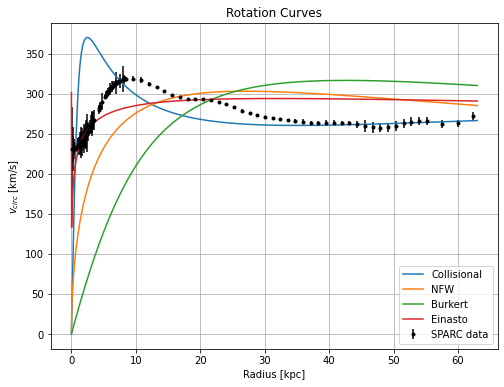}
\caption{The predicted rotation curves after using an optimization
for the collisional DM model (\ref{tanhmodel}), versus the SPARC
observational data for the galaxy UGC02953. We also plotted the
optimized curves for the NFW model, the Burkert model and the
Einasto model.} \label{UGC02953}
\end{figure}
\begin{table}[h!]
  \begin{center}
    \caption{Collisional Dark Matter Optimization Values}
    \label{collUGC02953}
     \begin{tabular}{|r|r|}
     \hline
      \textbf{Parameter}   & \textbf{Optimization Values}
      \\  \hline
     $\delta_{\gamma} $ & 0.0000000012
\\  \hline
$\gamma_0 $ & 1.0001  \\ \hline $K_0$ ($M_{\odot} \,
\mathrm{Kpc}^{-3} \, (\mathrm{km/s})^{2}$)& 41000  \\ \hline
    \end{tabular}
  \end{center}
\end{table}
\begin{table}[h!]
  \begin{center}
    \caption{NFW  Optimization Values}
    \label{NavaroUGC02953}
     \begin{tabular}{|r|r|}
     \hline
      \textbf{Parameter}   & \textbf{Optimization Values}
      \\  \hline
   $\rho_s$   & $5\times 10^7$
\\  \hline
$r_s$&  12.54
\\  \hline
    \end{tabular}
  \end{center}
\end{table}
\begin{figure}[h!]
\centering
\includegraphics[width=20pc]{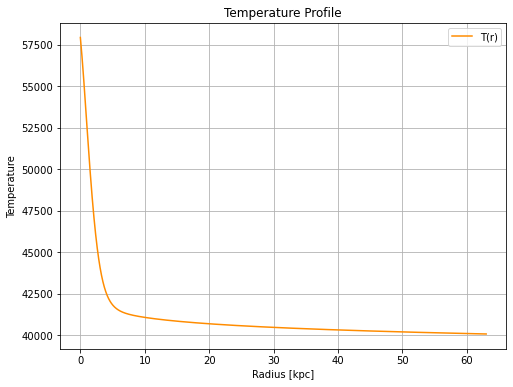}
\caption{The temperature as a function of the radius for the
collisional DM model (\ref{tanhmodel}) for the galaxy UGC02953.}
\label{UGC02953temp}
\end{figure}
\begin{table}[h!]
  \begin{center}
    \caption{Burkert Optimization Values}
    \label{BuckertUGC02953}
     \begin{tabular}{|r|r|}
     \hline
      \textbf{Parameter}   & \textbf{Optimization Values}
      \\  \hline
     $\rho_0^B$  & $1\times 10^8$
\\  \hline
$r_0$& 13.13
\\  \hline
    \end{tabular}
  \end{center}
\end{table}
\begin{table}[h!]
  \begin{center}
    \caption{Einasto Optimization Values}
    \label{EinastoUGC02953}
    \begin{tabular}{|r|r|}
     \hline
      \textbf{Parameter}   & \textbf{Optimization Values}
      \\  \hline
     $\rho_e$  &$1\times 10^7$
\\  \hline
$r_e$ & 12.91
\\  \hline
$n_e$ & 0.05
\\  \hline
    \end{tabular}
  \end{center}
\end{table}
\begin{table}[h!]
\centering \caption{Physical assessment of collisional DM
parameters (UGC02953).}
\begin{tabular}{lcc}
\hline
Parameter & Value & Physical Verdict \\
\hline
$\gamma_0$ & $1.0001$ & Nearly isothermal; very soft EoS, favors shallow inner slope \\
$\delta_\gamma$ & $0.0000000012$ & Negligible radial variation  \\
$r_\gamma$ & $1.5\ \mathrm{Kpc}$ & Transition radius inside inner halo, but minimal effect \\
$K_0$ & $4.10\times10^{4}$ & Entropy large\\
$r_c$ & $0.5\ \mathrm{Kpc}$ & Small core scale; reasonable for compact inner core \\
$p$ & $0.01$ & Practically constant $K(r)$; no significant entropy decline \\
\hline
Overall &-& Physically plausible as a nearly-isothermal, cored halo; limited radial flexibility \\
\hline
\end{tabular}
\label{EVALUATIONUGC02953}
\end{table}
Now the extended picture including the rotation velocity from the
other components of the galaxy, such as the disk and gas, makes
the collisional DM model viable for this galaxy. In Fig.
\ref{extendedUGC02953} we present the combined rotation curves
including the other components of the galaxy along with the
collisional matter. As it can be seen, the extended collisional DM
model is non-viable.
Also in Table \ref{evaluationextendedUGC02953} we present the
values of the free parameters of the collisional DM model for
which the maximum compatibility with the SPARC data comes for the
galaxy UGC02953.
\begin{table}[h!]
\centering \caption{Physical assessment of Extended collisional DM
parameters (second set) for UGC02953.}
\begin{tabular}{lcc}
\hline
Parameter & Value & Physical Verdict \\
\hline
$\gamma_0$ & 1.115 & Near-isothermal core, low-to-moderate central pressure/support \\
$\delta_\gamma$ & 0.0001 & Negligible variation -effectively constant $\gamma(r)$ \\
$K_0$ & 3000 & Moderate entropy    \\
$ml_{disk}$ & 0.92405534 & Sub-maximal to near-maximal disk; substantial disk contribution \\
$ml_{bulge}$ & 0.60000000 & Significant bulge mass-to-light; strong central baryonic potential \\
\hline
Overall &-& Physically plausible\\
\hline
\end{tabular}
\label{evaluationextendedUGC02953}
\end{table}

\subsection{The Galaxy UGC03205 Marginal maximum compatibility Late-time Spiral}

\begin{figure}[h!]
\centering
\includegraphics[width=20pc]{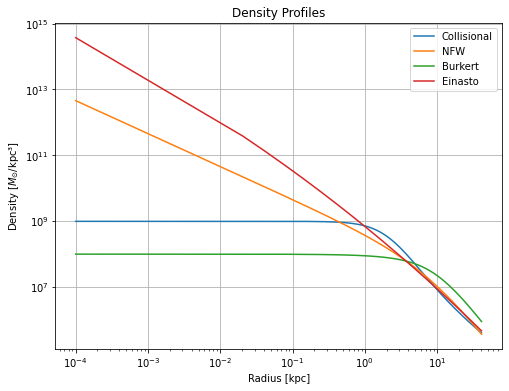}
\caption{The density of the collisional DM model (\ref{tanhmodel})
for the galaxy UGC03205, as a function of the radius.}
\label{UGC03205dens}
\end{figure}
\begin{figure}[h!]
\centering
\includegraphics[width=20pc]{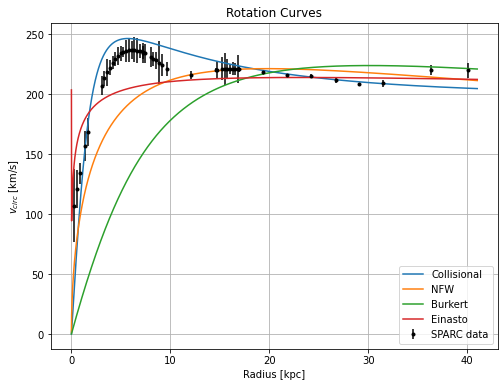}
\caption{The predicted rotation curves after using an optimization
for the collisional DM model (\ref{tanhmodel}), versus the SPARC
observational data for the galaxy UGC03205. We also plotted the
optimized curves for the NFW model, the Burkert model and the
Einasto model.} \label{UGC03205}
\end{figure}
\begin{figure}[h!]
\centering
\includegraphics[width=20pc]{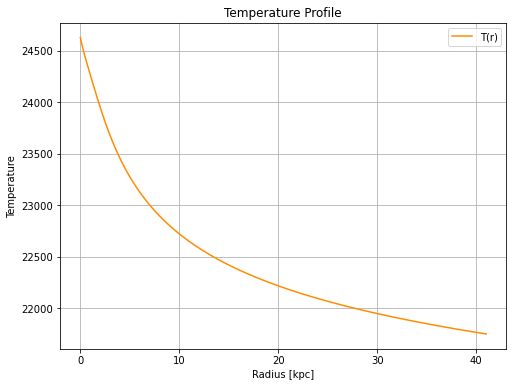}
\caption{The temperature as a function of the radius for the
collisional DM model (\ref{tanhmodel}) for the galaxy UGC03205.}
\label{UGC03205temp}
\end{figure}
\begin{figure}[h!]
\centering
\includegraphics[width=20pc]{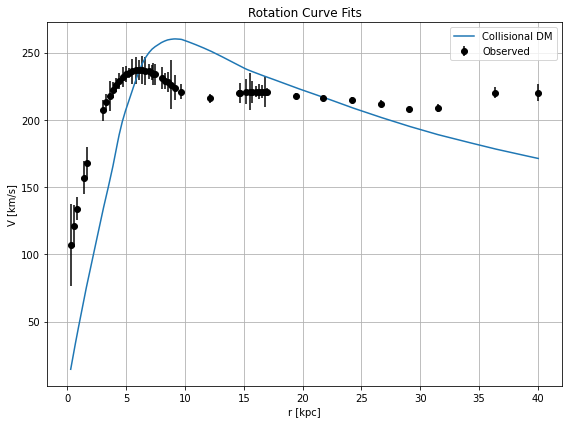}
\caption{The predicted rotation curves after using an optimization
for the collisional DM model (\ref{tanhmodel}), versus the
extended SPARC observational data for the galaxy UGC03205. The
model includes the rotation curves from all the components of the
galaxy, including gas and disk velocities, along with the
collisional DM model.} \label{extendedUGC03205}
\end{figure}

For this galaxy, we shall choose $\rho_0=9.9\times
10^8$$M_{\odot}/\mathrm{Kpc}^{3}$. UGC03205 is a late-type spiral
galaxy (Sc or Sbc class) at a distance of roughly \(40\!-\!50\
\mathrm{Mpc}\). In Figs. \ref{UGC03205dens}, \ref{UGC03205} and
\ref{UGC03205temp} we present the density of the collisional DM
model, the predicted rotation curves after using an optimization
for the collisional DM model (\ref{tanhmodel}), versus the SPARC
observational data and the temperature parameter as a function of
the radius respectively. As it can be seen, the SIDM model
produces non-viable rotation curves incompatible with the SPARC
data. Also in Tables \ref{collUGC03205}, \ref{NavaroUGC03205},
\ref{BuckertUGC03205} and \ref{EinastoUGC03205} we present the
optimization values for the SIDM model, and the other DM profiles.
Also in Table \ref{EVALUATIONUGC03205} we present the overall
evaluation of the SIDM model for the galaxy at hand. The resulting
phenomenology is non-viable.
\begin{table}[h!]
  \begin{center}
    \caption{Collisional Dark Matter Optimization Values}
    \label{collUGC03205}
     \begin{tabular}{|r|r|}
     \hline
      \textbf{Parameter}   & \textbf{Optimization Values}
      \\  \hline
     $\delta_{\gamma} $ & 0.0000000012
\\  \hline
$\gamma_0 $ & 1.0001  \\ \hline $K_0$ ($M_{\odot} \,
\mathrm{Kpc}^{-3} \, (\mathrm{km/s})^{2}$)& 20000  \\ \hline
    \end{tabular}
  \end{center}
\end{table}
\begin{table}[h!]
  \begin{center}
    \caption{NFW  Optimization Values}
    \label{NavaroUGC03205}
     \begin{tabular}{|r|r|}
     \hline
      \textbf{Parameter}   & \textbf{Optimization Values}
      \\  \hline
   $\rho_s$   & $5\times 10^7$
\\  \hline
$r_s$&  9.15
\\  \hline
    \end{tabular}
  \end{center}
\end{table}
\begin{table}[h!]
  \begin{center}
    \caption{Burkert Optimization Values}
    \label{BuckertUGC03205}
     \begin{tabular}{|r|r|}
     \hline
      \textbf{Parameter}   & \textbf{Optimization Values}
      \\  \hline
     $\rho_0^B$  & $1\times 10^8$
\\  \hline
$r_0$&  9.28
\\  \hline
    \end{tabular}
  \end{center}
\end{table}
\begin{table}[h!]
  \begin{center}
    \caption{Einasto Optimization Values}
    \label{EinastoUGC03205}
    \begin{tabular}{|r|r|}
     \hline
      \textbf{Parameter}   & \textbf{Optimization Values}
      \\  \hline
     $\rho_e$  &$1\times 10^7$
\\  \hline
$r_e$ & 9.39
\\  \hline
$n_e$ & 0.05
\\  \hline
    \end{tabular}
  \end{center}
\end{table}
\begin{table}[h!]
\centering \caption{Physical assessment of collisional DM
parameters (UGC03205).}
\begin{tabular}{lcc}
\hline
Parameter & Value & Physical Verdict \\
\hline
$\gamma_0$ & $1.0001$ & Nearly isothermal; soft EoS, shallow inner slope \\
$\delta_\gamma$ & $0.0000000012$ & Negligible variation  \\
$r_\gamma$ & $1.5\ \mathrm{Kpc}$ & Transition radius inside inner halo, no effect \\
$K_0$ & $2.0\times10^{4}$ & Large entropy \\
$r_c$ & $0.5\ \mathrm{Kpc}$ & Small core scale, reasonable \\
$p$ & $0.01$ & Practically constant $K(r)$, no entropy decline \\
\hline
Overall &-& Physically plausible; nearly isothermal, cored halo \\
\hline
\end{tabular}
\label{EVALUATIONUGC03205}
\end{table}
Now the extended picture including the rotation velocity from the
other components of the galaxy, such as the disk and gas, makes
the collisional DM model viable for this galaxy. In Fig.
\ref{extendedUGC03205} we present the combined rotation curves
including the other components of the galaxy along with the
collisional matter. As it can be seen, the extended collisional DM
model is non-viable.
Also in Table \ref{evaluationextendedUGC03205} we present the
values of the free parameters of the collisional DM model for
which the maximum compatibility with the SPARC data comes for the
galaxy UGC03205.
\begin{table}[h!]
\centering \caption{Physical assessment of Extended collisional DM
parameters for galaxy UGC03205.}
\begin{tabular}{lcc}
\hline
Parameter & Value & Physical Verdict \\
\hline
$\gamma_0$ & 1.13060309 & Slightly above isothermal; modest central stiffness, reasonable for a spiral halo \\
$\delta_\gamma$ & 0.06133000 & Moderate variation; noticeable radial increase of $\gamma(r)$ giving extra flexibility in outer halo \\
$K_0$ & 3000 & Moderate entropy \\
$ml_{disk}$ & 1.00000000 & At the upper end (disk-dominated) \\
$ml_{bulge}$ & 0.00071888 & Practically negligible bulge mass \\
\hline
Overall &-& Physically plausible \\
\hline
\end{tabular}
\label{evaluationextendedUGC03205}
\end{table}

\subsection{The Galaxy UGC03580 Non-viable Late-time spiral, Multi-parameter model is viable extended}

\begin{figure}[h!]
\centering
\includegraphics[width=20pc]{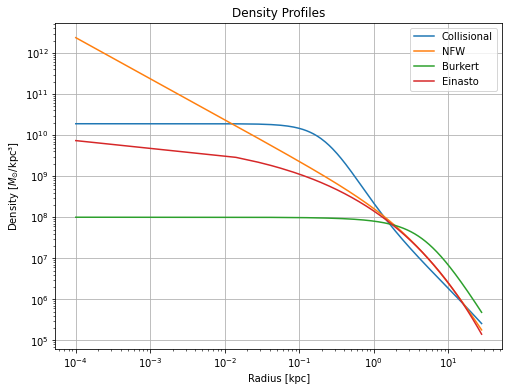}
\caption{The density of the collisional DM model (\ref{tanhmodel})
for the galaxy UGC03580, as a function of the radius.}
\label{UGC03580dens}
\end{figure}

\begin{figure}[h!]
\centering
\includegraphics[width=20pc]{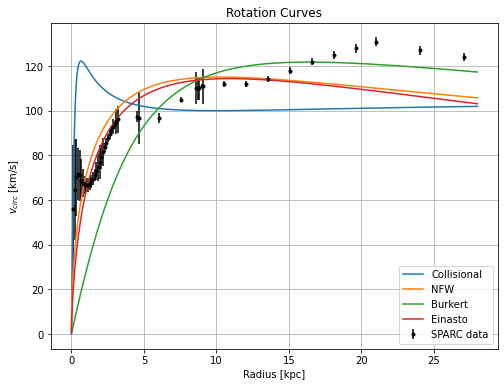}
\caption{The predicted rotation curves after using an optimization
for the collisional DM model (\ref{tanhmodel}), versus the SPARC
observational data for the galaxy UGC03580. We also plotted the
optimized curves for the NFW model, the Burkert model and the
Einasto model.} \label{UGC03580}
\end{figure}

\begin{figure}[h!]
\centering
\includegraphics[width=20pc]{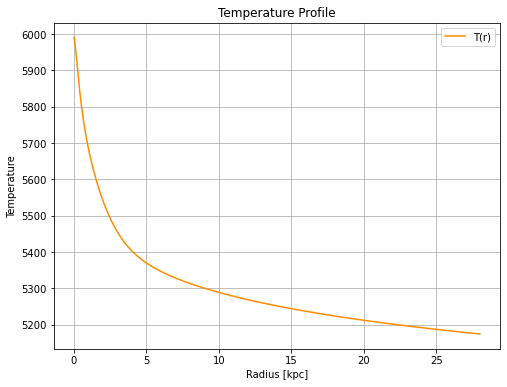}
\caption{The temperature as a function of the radius for the
collisional DM model (\ref{tanhmodel}) for the galaxy UGC03580.}
\label{UGC03580temp}
\end{figure}

\begin{figure}[h!]
\centering
\includegraphics[width=20pc]{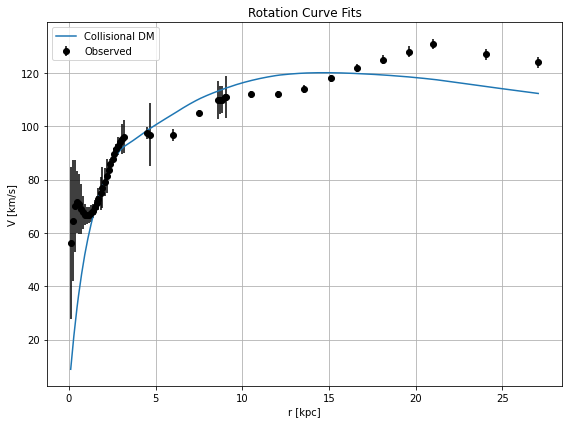}
\caption{The predicted rotation curves after using an optimization
for the collisional DM model (\ref{tanhmodel}), versus the
extended SPARC observational data for the galaxy UGC03580. The
model includes the rotation curves from all the components of the
galaxy, including gas and disk velocities, along with the
collisional DM model.} \label{extendedUGC03580}
\end{figure}

\begin{figure}[h!]
\centering
\includegraphics[width=20pc]{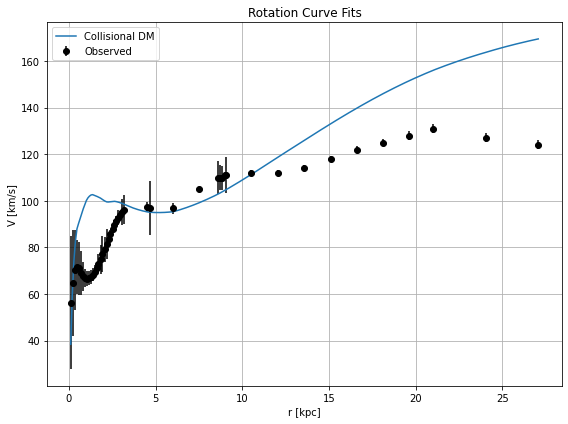}
\caption{The predicted rotation curves after using an optimization
for the collisional DM model (\ref{tanhmodel}), using all the
parameters, versus the extended SPARC observational data for the
galaxy UGC03580. The model includes the rotation curves from all
the components of the galaxy, including gas and disk velocities,
along with the collisional DM model.}
\label{extendedmultiUGC03580}
\end{figure}

For this galaxy, we shall choose $\rho_0=1.9\times
10^{10}$$M_{\odot}/\mathrm{Kpc}^{3}$. UGC3580 (also catalogued as
PGC\,19867) is listed in galaxy catalogues as an early-type spiral
(Sa-S0/Sa). In Figs. \ref{UGC03580dens}, \ref{UGC03580} and
\ref{UGC03580temp} we present the density of the collisional DM
model, the predicted rotation curves after using an optimization
for the collisional DM model (\ref{tanhmodel}), versus the SPARC
observational data and the temperature parameter as a function of
the radius respectively. As it can be seen, the SIDM model
produces non-viable rotation curves incompatible with the SPARC
data. Also in Tables \ref{collUGC03580}, \ref{NavaroUGC03580},
\ref{BuckertUGC03580} and \ref{EinastoUGC03580} we present the
optimization values for the SIDM model, and the other DM profiles.
Also in Table \ref{EVALUATIONUGC03580} we present the overall
evaluation of the SIDM model for the galaxy at hand. The resulting
phenomenology is non-viable.
\begin{table}[h!]
  \begin{center}
    \caption{Collisional Dark Matter Optimization Values}
    \label{collUGC03580}
     \begin{tabular}{|r|r|}
     \hline
      \textbf{Parameter}   & \textbf{Optimization Values}
      \\  \hline
     $\delta_{\gamma} $ & 0.001
\\  \hline
$\gamma_0 $ & 1.007 \\ \hline $K_0$ ($M_{\odot} \,
\mathrm{Kpc}^{-3} \, (\mathrm{km/s})^{2}$)& 5000 \\ \hline
    \end{tabular}
  \end{center}
\end{table}
\begin{table}[h!]
  \begin{center}
    \caption{NFW  Optimization Values}
    \label{NavaroUGC03580}
     \begin{tabular}{|r|r|}
     \hline
      \textbf{Parameter}   & \textbf{Optimization Values}
      \\  \hline
   $\rho_s$   & $5\times 10^7$
\\  \hline
$r_s$&  4.76
\\  \hline
    \end{tabular}
  \end{center}
\end{table}
\begin{table}[h!]
  \begin{center}
    \caption{Burkert Optimization Values}
    \label{BuckertUGC03580}
     \begin{tabular}{|r|r|}
     \hline
      \textbf{Parameter}   & \textbf{Optimization Values}
      \\  \hline
     $\rho_0^B$  & $1\times 10^8$
\\  \hline
$r_0$&  5.05
\\  \hline
    \end{tabular}
  \end{center}
\end{table}
\begin{table}[h!]
  \begin{center}
    \caption{Einasto Optimization Values}
    \label{EinastoUGC03580}
    \begin{tabular}{|r|r|}
     \hline
      \textbf{Parameter}   & \textbf{Optimization Values}
      \\  \hline
     $\rho_e$  &$1\times 10^7$
\\  \hline
$r_e$ & 5.33
\\  \hline
$n_e$ & 0.29
\\  \hline
    \end{tabular}
  \end{center}
\end{table}
\begin{table}[h!]
\centering \caption{Physical assessment of collisional DM
parameters (UGC03580).}
\begin{tabular}{lcc}
\hline
Parameter & Value & Physical Verdict \\
\hline
$\gamma_0$ & $1.007$ & Nearly isothermal; negligible polytropic stratification \\
$\delta_\gamma$ & $0.001$ & Effectively zero variation \\
$r_\gamma$ & $1.5\ \mathrm{Kpc}$ & Inner transition radius  \\
$K_0$ & $5.0\times10^{3}$ & Large but plausible \\
$r_c$ & $0.5\ \mathrm{Kpc}$ & Small core scale \\
$p$ & $0.01$ & Almost flat $K(r)$; very weak radial dependence \\
\hline
Overall &-& Physically consistent as an almost-isothermal halo \\
\hline
\end{tabular}
\label{EVALUATIONUGC03580}
\end{table}
Now the extended picture including the rotation velocity from the
other components of the galaxy, such as the disk and gas, makes
the collisional DM model viable for this galaxy. In Fig.
\ref{extendedUGC03580} we present the combined rotation curves
including the other components of the galaxy along with the
collisional matter. As it can be seen, the extended collisional DM
model is non-viable.
Also in Table \ref{evaluationextendedUGC03580} we present the
values of the free parameters of the collisional DM model for
which the maximum compatibility with the SPARC data comes for the
galaxy UGC03580.
\begin{table}[h!]
\centering \caption{Physical assessment of Extended collisional DM
parameters (second set) for UGC03580.}
\begin{tabular}{lcc}
\hline
Parameter & Value & Physical Verdict \\
\hline
$\gamma_0$ & 1.02792070 & Noticeably above isothermal  \\
$\delta_\gamma$ & 0.000001 & Negligible radial variation in $\gamma(r)$ \\
$K_0$ & 3000 & Moderate entropy   \\
$ml_{disk}$ & 0.84166058 & Large disk assumption \\
$ml_{bulge}$ & 0.00003008 & Minimally bulge mass-to-light \\
\hline
Overall &-& Physically plausible\\
\hline
\end{tabular}
\label{evaluationextendedUGC03580}
\end{table}
The multi-parameter model is viable though, see Fig.
\ref{extendedmultiUGC03580} and Table
\ref{evaluationextendedmultiUGC03580} for the final verdict.
\begin{table}[h!]
\centering \caption{Physical assessment of Extended collisional DM
parameters (second set) for UGC05253.}
\begin{tabular}{lcc}
\hline
Parameter & Value & Physical Verdict \\
\hline
$\gamma_0$ & 1.10297799 & Near-isothermal core; low-to-moderate central pressure/support \\
$\delta_\gamma$ & 0.00204817 & Negligible radial variation  \\
$K_0$ & 2986.77280485 & Moderate entropy \\
$r_\gamma$ & 13.60690745 & Large transition radius \\
$\alpha_K$ & 0.10000000 & Small regularization term in $K(r)$  \\
$ml_{disk}$ & 0.80047294 & Sub-maximal disk \\
$ml_{bulge}$ & 0.84983971 & Large bulge mass-to-light \\
\hline
Overall &-& Physically plausible \\
\hline
\end{tabular}
\label{evaluationextendedmultiUGC03580}
\end{table}

\subsection{The Galaxy UGC04278 Marginally, Extended Viable}

\begin{figure}[h!]
\centering
\includegraphics[width=20pc]{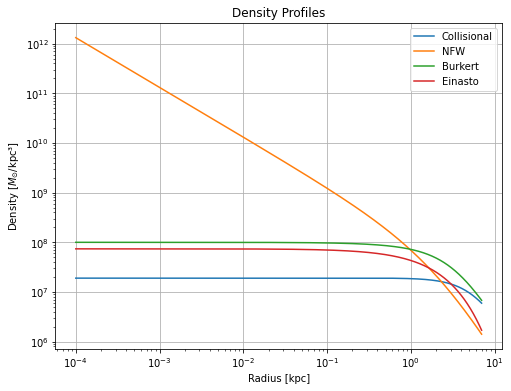}
\caption{The density of the collisional DM model (\ref{tanhmodel})
for the galaxy UGC04278, as a function of the radius.}
\label{UGC04278dens}
\end{figure}

\begin{figure}[h!]
\centering
\includegraphics[width=20pc]{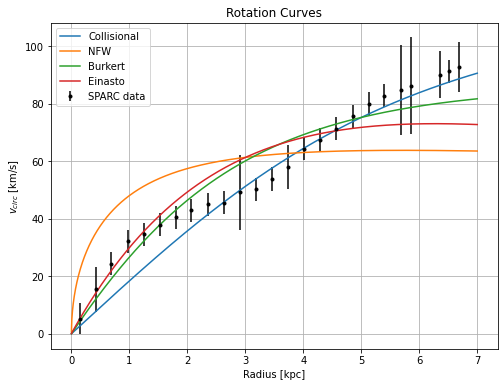}
\caption{The predicted rotation curves after using an optimization
for the collisional DM model (\ref{tanhmodel}), versus the SPARC
observational data for the galaxy UGC04278. We also plotted the
optimized curves for the NFW model, the Burkert model and the
Einasto model.} \label{UGC04278}
\end{figure}

\begin{figure}[h!]
\centering
\includegraphics[width=20pc]{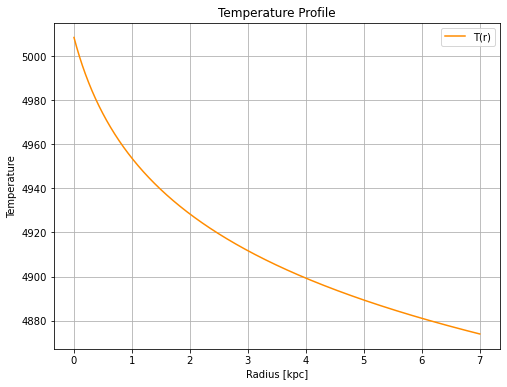}
\caption{The temperature as a function of the radius for the
collisional DM model (\ref{tanhmodel}) for the galaxy UGC04278.}
\label{UGC04278temp}
\end{figure}

\begin{figure}[h!]
\centering
\includegraphics[width=20pc]{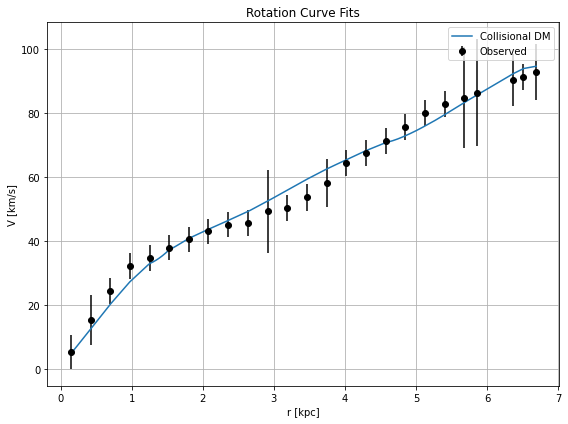}
\caption{The predicted rotation curves after using an optimization
for the collisional DM model (\ref{tanhmodel}), versus the
extended SPARC observational data for the galaxy UGC04278. The
model includes the rotation curves from all the components of the
galaxy, including gas and disk velocities, along with the
collisional DM model.} \label{extendedUGC04278}
\end{figure}

For this galaxy, we shall choose $\rho_0=1.9\times
10^7$$M_{\odot}/\mathrm{Kpc}^{3}$. UGC04278 galaxy is classified
as a late-type, low-surface-brightness   edge-on spiral (often
studied in HI rotation-curve work). Its distance is typically
adopted at $\sim 15.4\,$Mpc. In Figs. \ref{UGC04278dens},
\ref{UGC04278} and \ref{UGC04278temp} we present the density of
the collisional DM model, the predicted rotation curves after
using an optimization for the collisional DM model
(\ref{tanhmodel}), versus the SPARC observational data and the
temperature parameter as a function of the radius respectively. As
it can be seen, the SIDM model produces marginally viable rotation
curves compatible with the SPARC data. Also in Tables
\ref{collUGC04278}, \ref{NavaroUGC04278}, \ref{BuckertUGC04278}
and \ref{EinastoUGC04278} we present the optimization values for
the SIDM model, and the other DM profiles. Also in Table
\ref{EVALUATIONUGC04278} we present the overall evaluation of the
SIDM model for the galaxy at hand. The resulting phenomenology is
marginally viable.
\begin{table}[h!]
  \begin{center}
    \caption{Collisional Dark Matter Optimization Values}
    \label{collUGC04278}
     \begin{tabular}{|r|r|}
     \hline
      \textbf{Parameter}   & \textbf{Optimization Values}
      \\  \hline
     $\delta_{\gamma} $ & 0.0000000012
\\  \hline
$\gamma_0 $ & 1.0001  \\ \hline $K_0$ ($M_{\odot} \,
\mathrm{Kpc}^{-3} \, (\mathrm{km/s})^{2}$)& 5000  \\ \hline
    \end{tabular}
  \end{center}
\end{table}
\begin{table}[h!]
  \begin{center}
    \caption{NFW  Optimization Values}
    \label{NavaroUGC04278}
     \begin{tabular}{|r|r|}
     \hline
      \textbf{Parameter}   & \textbf{Optimization Values}
      \\  \hline
   $\rho_s$   & $5\times 10^7$
\\  \hline
$r_s$&  2.64
\\  \hline
    \end{tabular}
  \end{center}
\end{table}
\begin{table}[h!]
  \begin{center}
    \caption{Burkert Optimization Values}
    \label{BuckertUGC04278}
     \begin{tabular}{|r|r|}
     \hline
      \textbf{Parameter}   & \textbf{Optimization Values}
      \\  \hline
     $\rho_0^B$  & $1\times 10^8$
\\  \hline
$r_0$&  3.53
\\  \hline
    \end{tabular}
  \end{center}
\end{table}
\begin{table}[h!]
  \begin{center}
    \caption{Einasto Optimization Values}
    \label{EinastoUGC04278}
    \begin{tabular}{|r|r|}
     \hline
      \textbf{Parameter}   & \textbf{Optimization Values}
      \\  \hline
     $\rho_e$  &$1\times 10^7$
\\  \hline
$r_e$ & 3.711
\\  \hline
$n_e$ & 1
\\  \hline
    \end{tabular}
  \end{center}
\end{table}
\begin{table}[h!]
\centering \caption{Physical assessment of collisional DM
parameters for UGC04278.}
\begin{tabular}{lcc}
\hline
Parameter & Value & Physical Verdict \\
\hline
$\gamma_0$ & $1.0001$ & Almost exactly isothermal \\
$\delta_\gamma$ & $1.2\times10^{-9}$ & Practically zero  \\
$r_\gamma$ & $1.5~\mathrm{Kpc}$ & Reasonable transition scale \\
$K_0$ & $5.0\times10^{3}$ & Moderate energy scale  \\
$r_c$ & $0.5~\mathrm{Kpc}$ & Small core scale \\
$p$ & $0.01$ & Extremely shallow decline  \\
\hline
Overall & --- & Nearly isothermal\\
\hline
\end{tabular}
\label{EVALUATIONUGC04278}
\end{table}
Now the extended picture including the rotation velocity from the
other components of the galaxy, such as the disk and gas, makes
the collisional DM model viable for this galaxy. In Fig.
\ref{extendedUGC04278} we present the combined rotation curves
including the other components of the galaxy along with the
collisional matter. As it can be seen, the extended collisional DM
model is viable.
Also in Table \ref{evaluationextendedUGC04278} we present the
values of the free parameters of the collisional DM model for
which the maximum compatibility with the SPARC data comes for the
galaxy UGC04278.
\begin{table}[h!]
\centering \caption{Physical assessment of Extended collisional DM
parameters for galaxy UGC04278.}
\begin{tabular}{lcc}
\hline
Parameter & Value & Physical Verdict \\
\hline
$\gamma_0$ & 1.02771159 & Very close to isothermal \\
$\delta_\gamma$ & 0.00000001 & No radial variation \\
$K_0$ & 3000 & Moderate entropy  \\
$ml_{disk}$ & 0.94211218 & High but realistic disk M/L \\
$ml_{bulge}$ & 0.00000000 & Negligible bulge contribution \\
\hline
Overall &-& Physically plausible\\
\hline
\end{tabular}
\label{evaluationextendedUGC04278}
\end{table}

\subsection{The Galaxy UGC04305 Marginally Dwarf! Extended is Viable}

\begin{figure}[h!]
\centering
\includegraphics[width=20pc]{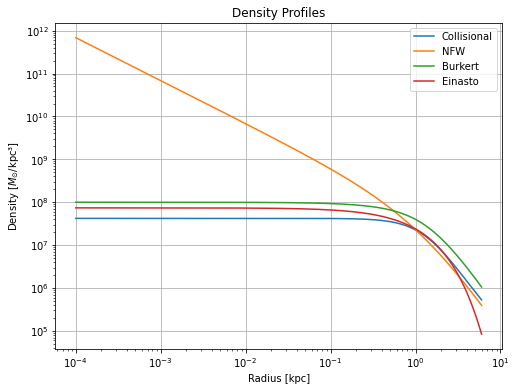}
\caption{The density of the collisional DM model (\ref{tanhmodel})
for the galaxy UGC04305, as a function of the radius.}
\label{UGC04305dens}
\end{figure}

\begin{figure}[h!]
\centering
\includegraphics[width=20pc]{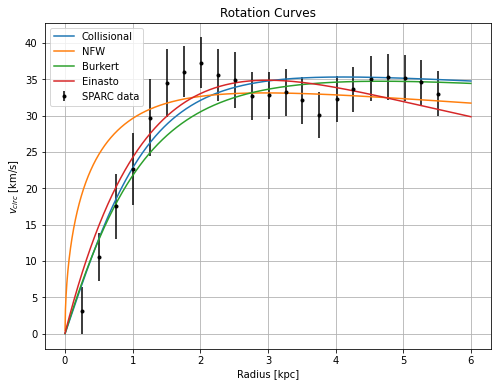}
\caption{The predicted rotation curves after using an optimization
for the collisional DM model (\ref{tanhmodel}), versus the SPARC
observational data for the galaxy UGC04305. We also plotted the
optimized curves for the NFW model, the Burkert model and the
Einasto model.} \label{UGC04305}
\end{figure}

\begin{figure}[h!]
\centering
\includegraphics[width=20pc]{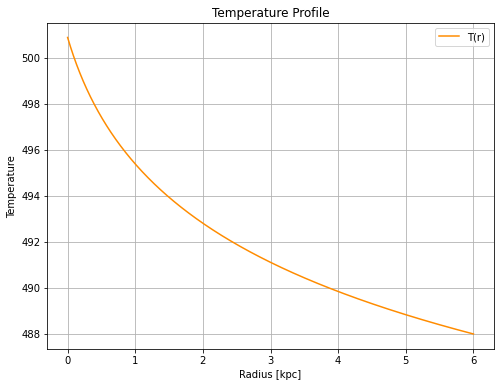}
\caption{The temperature as a function of the radius for the
collisional DM model (\ref{tanhmodel}) for the galaxy UGC04305.}
\label{UGC04305temp}
\end{figure}

\begin{figure}[h!]
\centering
\includegraphics[width=20pc]{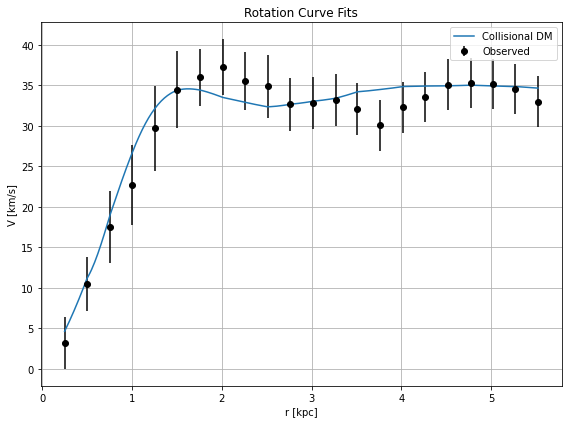}
\caption{The predicted rotation curves after using an optimization
for the collisional DM model (\ref{tanhmodel}), versus the
extended SPARC observational data for the galaxy UGC04305. The
model includes the rotation curves from all the components of the
galaxy, including gas and disk velocities, along with the
collisional DM model.} \label{extendedUGC04305}
\end{figure}

For this galaxy, we shall choose $\rho_0=4.2\times
10^7$$M_{\odot}/\mathrm{Kpc}^{3}$. UGC4305 is a nearby gas-rich
dwarf irregular galaxy in the M81 group at a distance of order
$\sim 3.4\,$Mpc. In Figs. \ref{UGC04305dens}, \ref{UGC04305} and
\ref{UGC04305temp} we present the density of the collisional DM
model, the predicted rotation curves after using an optimization
for the collisional DM model (\ref{tanhmodel}), versus the SPARC
observational data and the temperature parameter as a function of
the radius respectively. As it can be seen, the SIDM model
produces marginally viable rotation curves compatible with the
SPARC data. Also in Tables \ref{collUGC04305},
\ref{NavaroUGC04305}, \ref{BuckertUGC04305} and
\ref{EinastoUGC04305} we present the optimization values for the
SIDM model, and the other DM profiles. Also in Table
\ref{EVALUATIONUGC04305} we present the overall evaluation of the
SIDM model for the galaxy at hand. The resulting phenomenology is
marginally viable.
\begin{table}[h!]
  \begin{center}
    \caption{Collisional Dark Matter Optimization Values}
    \label{collUGC04305}
     \begin{tabular}{|r|r|}
     \hline
      \textbf{Parameter}   & \textbf{Optimization Values}
      \\  \hline
     $\delta_{\gamma} $ & 0.0000000012
\\  \hline
$\gamma_0 $ & 1.0001  \\ \hline $K_0$ ($M_{\odot} \,
\mathrm{Kpc}^{-3} \, (\mathrm{km/s})^{2}$)& 500  \\ \hline
    \end{tabular}
  \end{center}
\end{table}
\begin{table}[h!]
  \begin{center}
    \caption{NFW  Optimization Values}
    \label{NavaroUGC04305}
     \begin{tabular}{|r|r|}
     \hline
      \textbf{Parameter}   & \textbf{Optimization Values}
      \\  \hline
   $\rho_s$   & $5\times 10^7$
\\  \hline
$r_s$&  1.37
\\  \hline
    \end{tabular}
  \end{center}
\end{table}
\begin{table}[h!]
  \begin{center}
    \caption{Burkert Optimization Values}
    \label{BuckertUGC04305}
     \begin{tabular}{|r|r|}
     \hline
      \textbf{Parameter}   & \textbf{Optimization Values}
      \\  \hline
     $\rho_0^B$  & $1\times 10^8$
\\  \hline
$r_0$&  1.44
\\  \hline
    \end{tabular}
  \end{center}
\end{table}
\begin{table}[h!]
  \begin{center}
    \caption{Einasto Optimization Values}
    \label{EinastoUGC04305}
    \begin{tabular}{|r|r|}
     \hline
      \textbf{Parameter}   & \textbf{Optimization Values}
      \\  \hline
     $\rho_e$  &$1\times 10^7$
\\  \hline
$r_e$ & 1.77
\\  \hline
$n_e$ & 1
\\  \hline
    \end{tabular}
  \end{center}
\end{table}
\begin{table}[h!]
\centering \caption{Physical assessment of collisional DM
parameters for UGC04305.}
\begin{tabular}{lcc}
\hline
Parameter & Value & Physical Verdict \\
\hline
$\gamma_0$ & $1.0001$ & Almost exactly isothermal  \\
$\delta_\gamma$ & $1.2\times10^{-9}$ & Practically zero  \\
$r_\gamma$ & $1.5~\mathrm{Kpc}$ & Reasonable transition scale \\
$K_0$ & $5.0\times10^{2}$ & Enough pressure support \\
$r_c$ & $0.5~\mathrm{Kpc}$ & Small core scale\\
$p$ & $0.01$ & Extremely shallow decline\\
\hline
Overall & --- & Nearly isothermal\\
\hline
\end{tabular}
\label{EVALUATIONUGC04305}
\end{table}
Now the extended picture including the rotation velocity from the
other components of the galaxy, such as the disk and gas, makes
the collisional DM model viable for this galaxy. In Fig.
\ref{extendedUGC04305} we present the combined rotation curves
including the other components of the galaxy along with the
collisional matter. As it can be seen, the extended collisional DM
model is viable.
Also in Table \ref{evaluationextendedUGC04305} we present the
values of the free parameters of the collisional DM model for
which the maximum compatibility with the SPARC data comes for the
galaxy UGC04305.
\begin{table}[h!]
\centering \caption{Physical assessment of Extended collisional DM
parameters for galaxy UGC04305.}
\begin{tabular}{lcc}
\hline
Parameter & Value & Physical Verdict \\
\hline
$\gamma_0$ & 0.96424578 & Slightly sub-isothermal \\
$\delta_\gamma$ & 0.20696634 & Large radial variation \\
$K_0$ & 3000 & Moderate entropy   \\
$ml_{disk}$ & 0.00000000 & Zero disk M/L  \\
$ml_{bulge}$ & 0.00000000 & Zero bulge M/L  \\
\hline
Overall &-& Physically plausible  \\
\hline
\end{tabular}
\label{evaluationextendedUGC04305}
\end{table}

\subsection{The Galaxy UGC04325}

\begin{figure}[h!]
\centering
\includegraphics[width=20pc]{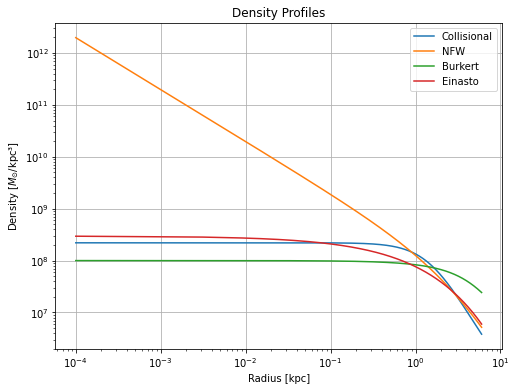}
\caption{The density of the collisional DM model (\ref{tanhmodel})
for the galaxy UGC04325, as a function of the radius.}
\label{UGC04325dens}
\end{figure}
\begin{figure}[h!]
\centering
\includegraphics[width=20pc]{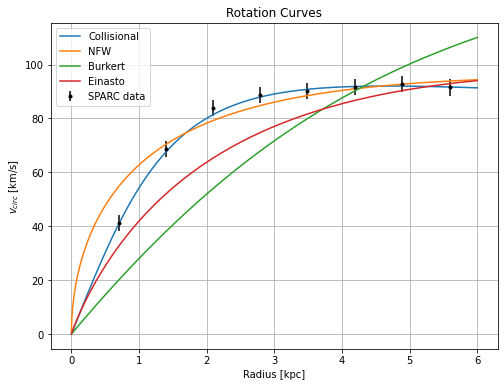}
\caption{The predicted rotation curves after using an optimization
for the collisional DM model (\ref{tanhmodel}), versus the SPARC
observational data for the galaxy UGC04325. We also plotted the
optimized curves for the NFW model, the Burkert model and the
Einasto model.} \label{UGC04325}
\end{figure}
\begin{figure}[h!]
\centering
\includegraphics[width=20pc]{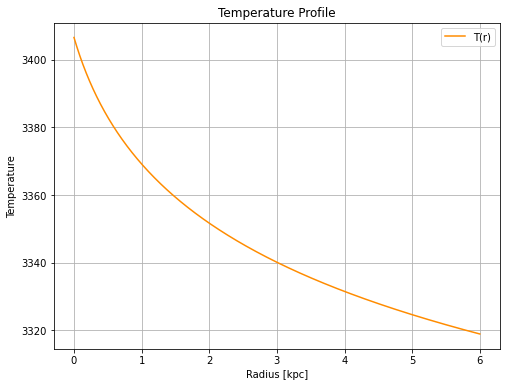}
\caption{The temperature as a function of the radius for the
collisional DM model (\ref{tanhmodel}) for the galaxy UGC04325.}
\label{UGC04325temp}
\end{figure}

For this galaxy, we shall choose $\rho_0=2.2\times
10^8$$M_{\odot}/\mathrm{Kpc}^{3}$. UGC04325  is a nearby
Magellanic-type dwarf/late-type spiral galaxy (SA(s)m / Sm) in
Lynx at a distance $\sim$7--10\,Mpc. In Figs. \ref{UGC04325dens},
\ref{UGC04325} and \ref{UGC04325temp} we present the density of
the collisional DM model, the predicted rotation curves after
using an optimization for the collisional DM model
(\ref{tanhmodel}), versus the SPARC observational data and the
temperature parameter as a function of the radius respectively. As
it can be seen, the SIDM model produces viable rotation curves
compatible with the SPARC data. Also in Tables \ref{collUGC04325},
\ref{NavaroUGC04325}, \ref{BuckertUGC04325} and
\ref{EinastoUGC04325} we present the optimization values for the
SIDM model, and the other DM profiles. Also in Table
\ref{EVALUATIONUGC04325} we present the overall evaluation of the
SIDM model for the galaxy at hand. The resulting phenomenology is
viable.
\begin{table}[h!]
  \begin{center}
    \caption{Collisional Dark Matter Optimization Values}
    \label{collUGC04325}
     \begin{tabular}{|r|r|}
     \hline
      \textbf{Parameter}   & \textbf{Optimization Values}
      \\  \hline
     $\delta_{\gamma} $ & 0.0000000012
\\  \hline
$\gamma_0 $ & 1.0001  \\ \hline $K_0$ ($M_{\odot} \,
\mathrm{Kpc}^{-3} \, (\mathrm{km/s})^{2}$)& 3400  \\ \hline
    \end{tabular}
  \end{center}
\end{table}
\begin{table}[h!]
  \begin{center}
    \caption{NFW  Optimization Values}
    \label{NavaroUGC04325}
     \begin{tabular}{|r|r|}
     \hline
      \textbf{Parameter}   & \textbf{Optimization Values}
      \\  \hline
   $\rho_s$   & $5\times 10^7$
\\  \hline
$r_s$&  3.95
\\  \hline
    \end{tabular}
  \end{center}
\end{table}
\begin{table}[h!]
  \begin{center}
    \caption{Burkert Optimization Values}
    \label{BuckertUGC04325}
     \begin{tabular}{|r|r|}
     \hline
      \textbf{Parameter}   & \textbf{Optimization Values}
      \\  \hline
     $\rho_0^B$  & $1\times 10^8$
\\  \hline
$r_0$&  5.88
\\  \hline
    \end{tabular}
  \end{center}
\end{table}
\begin{table}[h!]
  \begin{center}
    \caption{Einasto Optimization Values}
    \label{EinastoUGC04325}
    \begin{tabular}{|r|r|}
     \hline
      \textbf{Parameter}   & \textbf{Optimization Values}
      \\  \hline
     $\rho_e$  &$1\times 10^7$
\\  \hline
$r_e$ & 4.72
\\  \hline
$n_e$ & 0.59
\\  \hline
    \end{tabular}
  \end{center}
\end{table}
\begin{table}[h!]
\centering \caption{Physical assessment of collisional DM
parameters for UGC04325.}
\begin{tabular}{lcc}
\hline
Parameter & Value & Physical Verdict \\
\hline
$\gamma_0$ & $1.0001$ & Almost exactly isothermal \\
$\delta_\gamma$ & $1.2\times10^{-9}$ & Practically zero \\
$r_\gamma$ & $1.5~\mathrm{Kpc}$ & Reasonable transition scale \\
$K_0$ & $3.4\times10^{3}$ & Enough pressure support for a dwarf \\
$r_c$ & $0.5~\mathrm{Kpc}$ & Small core scale \\
$p$ & $0.01$ & Extremely shallow decline  \\
\hline
Overall & --- & Nearly isothermal\\
\hline
\end{tabular}
\label{EVALUATIONUGC04325}
\end{table}

\subsection{The Galaxy UGC04499}

\begin{figure}[h!]
\centering
\includegraphics[width=20pc]{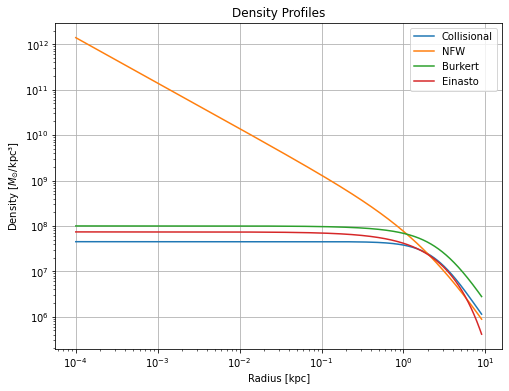}
\caption{The density of the collisional DM model (\ref{tanhmodel})
for the galaxy UGC04499, as a function of the radius.}
\label{UGC04499dens}
\end{figure}

\begin{figure}[h!]
\centering
\includegraphics[width=20pc]{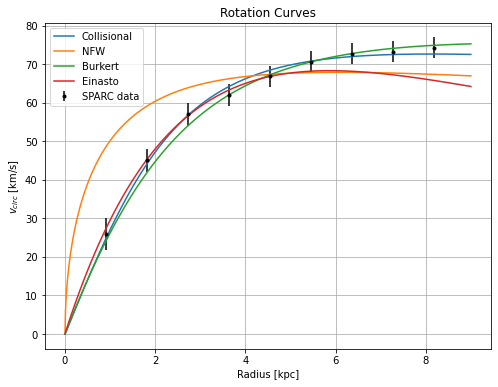}
\caption{The predicted rotation curves after using an optimization
for the collisional DM model (\ref{tanhmodel}), versus the SPARC
observational data for the galaxy UGC04499. We also plotted the
optimized curves for the NFW model, the Burkert model and the
Einasto model.} \label{UGC04499}
\end{figure}

\begin{figure}[h!]
\centering
\includegraphics[width=20pc]{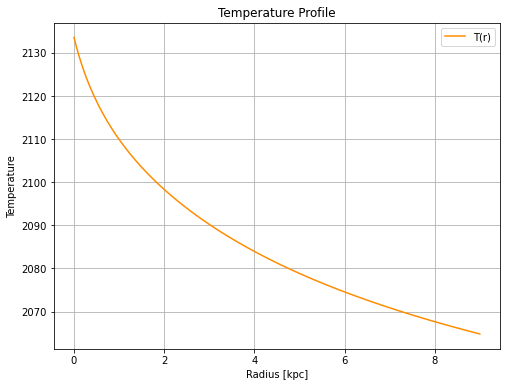}
\caption{The temperature as a function of the radius for the
collisional DM model (\ref{tanhmodel}) for the galaxy UGC04499.}
\label{UGC04499temp}
\end{figure}

For this galaxy, we shall choose $\rho_0=4.5\times
10^7$$M_{\odot}/\mathrm{Kpc}^{3}$. UGC04499 is a relatively
obscure galaxy of uncertain morphological classification, likely a
low-mass late-type or irregular spiral. In Figs.
\ref{UGC04499dens}, \ref{UGC04499} and \ref{UGC04499temp} we
present the density of the collisional DM model, the predicted
rotation curves after using an optimization for the collisional DM
model (\ref{tanhmodel}), versus the SPARC observational data and
the temperature parameter as a function of the radius
respectively. As it can be seen, the SIDM model produces viable
rotation curves compatible with the SPARC data. Also in Tables
\ref{collUGC04499}, \ref{NavaroUGC04499}, \ref{BuckertUGC04499}
and \ref{EinastoUGC04499} we present the optimization values for
the SIDM model, and the other DM profiles. Also in Table
\ref{EVALUATIONUGC04499} we present the overall evaluation of the
SIDM model for the galaxy at hand. The resulting phenomenology is
viable.
\begin{table}[h!]
  \begin{center}
    \caption{Collisional Dark Matter Optimization Values}
    \label{collUGC04499}
     \begin{tabular}{|r|r|}
     \hline
      \textbf{Parameter}   & \textbf{Optimization Values}
      \\  \hline
     $\delta_{\gamma} $ & 0.0000000012
\\  \hline
$\gamma_0 $ & 1.0001  \\ \hline $K_0$ ($M_{\odot} \,
\mathrm{Kpc}^{-3} \, (\mathrm{km/s})^{2}$)& 2100  \\ \hline
    \end{tabular}
  \end{center}
\end{table}
\begin{table}[h!]
  \begin{center}
    \caption{NFW  Optimization Values}
    \label{NavaroUGC04499}
     \begin{tabular}{|r|r|}
     \hline
      \textbf{Parameter}   & \textbf{Optimization Values}
      \\  \hline
   $\rho_s$   & $5\times 10^7$
\\  \hline
$r_s$&  2.81
\\  \hline
    \end{tabular}
  \end{center}
\end{table}
\begin{table}[h!]
  \begin{center}
    \caption{Burkert Optimization Values}
    \label{BuckertUGC04499}
     \begin{tabular}{|r|r|}
     \hline
      \textbf{Parameter}   & \textbf{Optimization Values}
      \\  \hline
     $\rho_0^B$  & $1\times 10^8$
\\  \hline
$r_0$&  3.13
\\  \hline
    \end{tabular}
  \end{center}
\end{table}
\begin{table}[h!]
  \begin{center}
    \caption{Einasto Optimization Values}
    \label{EinastoUGC04499}
    \begin{tabular}{|r|r|}
     \hline
      \textbf{Parameter}   & \textbf{Optimization Values}
      \\  \hline
     $\rho_e$  &$1\times 10^7$
\\  \hline
$r_e$ & 3.47
\\  \hline
$n_e$ & 1
\\  \hline
    \end{tabular}
  \end{center}
\end{table}
\begin{table}[h!]
\centering \caption{Physical assessment of collisional DM
parameters (UGC04499).}
\begin{tabular}{lcc}
\hline
Parameter & Value   & Physical Verdict \\
\hline
$\gamma_0$ & $1.0001$ & Practically isothermal; \\
$\delta_\gamma$ & $0.0000000012$ & Essentially zero  \\
$r_\gamma$ & $1.5\ \mathrm{Kpc}$ & Transition radius placed inside inner halo but irrelevant with tiny $\delta_\gamma$ \\
$K_0$ & $2.1\times10^{3}$ & Enough pressure support \\
$r_c$ & $0.5\ \mathrm{Kpc}$ & Small core radius  \\
$p$ & $0.01$ & Almost flat $K(r)$  \\
\hline Overall &-& Numerically stable yet physically
near-isothermal \\ \hline
\end{tabular}
\label{EVALUATIONUGC04499}
\end{table}


\subsection{The Galaxy UGC05005}

\begin{figure}[h!]
\centering
\includegraphics[width=20pc]{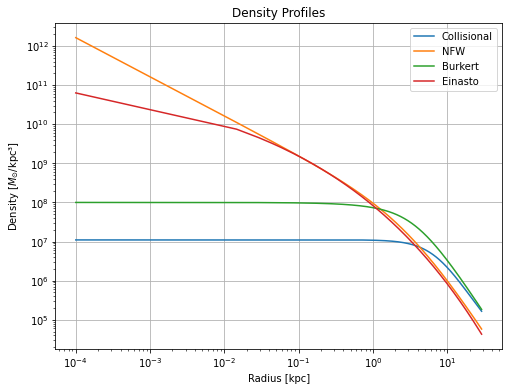}
\caption{The density of the collisional DM model (\ref{tanhmodel})
for the galaxy UGC05005, as a function of the radius.}
\label{UGC05005dens}
\end{figure}

\begin{figure}[h!]
\centering
\includegraphics[width=20pc]{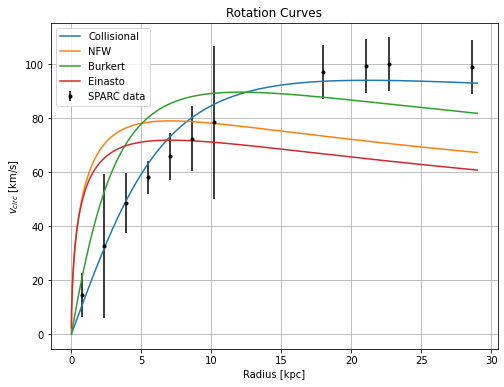}
\caption{The predicted rotation curves after using an optimization
for the collisional DM model (\ref{tanhmodel}), versus the SPARC
observational data for the galaxy UGC05005. We also plotted the
optimized curves for the NFW model, the Burkert model and the
Einasto model.} \label{UGC05005}
\end{figure}

\begin{figure}[h!]
\centering
\includegraphics[width=20pc]{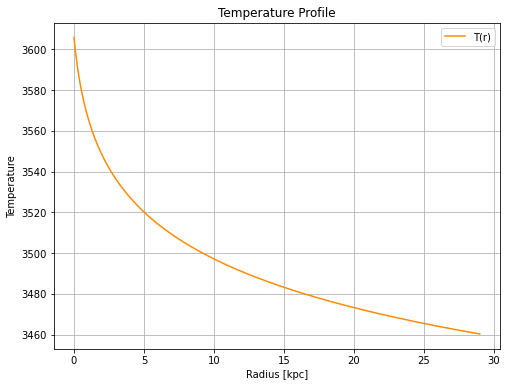}
\caption{The temperature as a function of the radius for the
collisional DM model (\ref{tanhmodel}) for the galaxy UGC05005.}
\label{UGC05005temp}
\end{figure}

For this galaxy, we shall choose $\rho_0=1\times
10^7$$M_{\odot}/\mathrm{Kpc}^{3}$. UGC05005 (commonly catalogued
as UGC5005) is a nearby, low-to-intermediate-mass, late-type disk
galaxy often treated in studies of rotation curves and HI
kinematics; it is located in the local Universe at a distance of
order tens of Mpc. In Figs. \ref{UGC05005dens}, \ref{UGC05005} and
\ref{UGC05005temp} we present the density of the collisional DM
model, the predicted rotation curves after using an optimization
for the collisional DM model (\ref{tanhmodel}), versus the SPARC
observational data and the temperature parameter as a function of
the radius respectively. As it can be seen, the SIDM model
produces viable rotation curves compatible with the SPARC data.
Also in Tables \ref{collUGC05005}, \ref{NavaroUGC05005},
\ref{BuckertUGC05005} and \ref{EinastoUGC05005} we present the
optimization values for the SIDM model, and the other DM profiles.
Also in Table \ref{EVALUATIONUGC05005} we present the overall
evaluation of the SIDM model for the galaxy at hand. The resulting
phenomenology is viable.
\begin{table}[h!]
  \begin{center}
    \caption{Collisional Dark Matter Optimization Values}
    \label{collUGC05005}
     \begin{tabular}{|r|r|}
     \hline
      \textbf{Parameter}   & \textbf{Optimization Values}
      \\  \hline
     $\delta_{\gamma} $ & 0.0000000012
\\  \hline
$\gamma_0 $ & 1.0001  \\ \hline $K_0$ ($M_{\odot} \,
\mathrm{Kpc}^{-3} \, (\mathrm{km/s})^{2}$)& 3600  \\ \hline
    \end{tabular}
  \end{center}
\end{table}
\begin{table}[h!]
  \begin{center}
    \caption{NFW  Optimization Values}
    \label{NavaroUGC05005}
     \begin{tabular}{|r|r|}
     \hline
      \textbf{Parameter}   & \textbf{Optimization Values}
      \\  \hline
   $\rho_s$   & $5\times 10^7$
\\  \hline
$r_s$&  3.27
\\  \hline
    \end{tabular}
  \end{center}
\end{table}
\begin{table}[h!]
  \begin{center}
    \caption{Burkert Optimization Values}
    \label{BuckertUGC05005}
     \begin{tabular}{|r|r|}
     \hline
      \textbf{Parameter}   & \textbf{Optimization Values}
      \\  \hline
     $\rho_0^B$  & $1\times 10^8$
\\  \hline
$r_0$&  3.72
\\  \hline
    \end{tabular}
  \end{center}
\end{table}
\begin{table}[h!]
  \begin{center}
    \caption{Einasto Optimization Values}
    \label{EinastoUGC05005}
    \begin{tabular}{|r|r|}
     \hline
      \textbf{Parameter}   & \textbf{Optimization Values}
      \\  \hline
     $\rho_e$  &$1\times 10^7$
\\  \hline
$r_e$ & 3.29
\\  \hline
$n_e$ & 0.2
\\  \hline
    \end{tabular}
  \end{center}
\end{table}
\begin{table}[h!]
\centering \caption{Physical assessment of collisional DM
parameters for UGC05005.}
\begin{tabular}{lcc}
\hline
Parameter & Value & Physical Verdict \\
\hline
$\gamma_0$ & $1.0001$ & Almost exactly isothermal \\
$\delta_\gamma$ & $1.2\times10^{-9}$ & Practically zero  \\
$r_\gamma$ & $1.5~\mathrm{Kpc}$ & Reasonable transition scale \\
$K_0$ & $3.6\times10^{3}$ & Enough pressure support \\
$r_c$ & $0.5~\mathrm{Kpc}$ & Small core scale \\
$p$ & $0.01$ & Extremely shallow decline  \\
\hline
Overall & --- & Nearly isothermal\\
\hline
\end{tabular}
\label{EVALUATIONUGC05005}
\end{table}

\subsection{The Galaxy UGC05253 Non Viable}

\begin{figure}[h!]
\centering
\includegraphics[width=20pc]{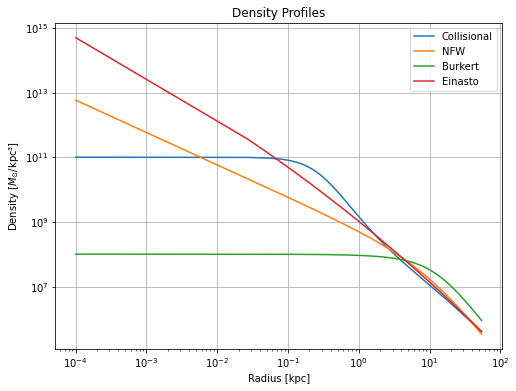}
\caption{The density of the collisional DM model (\ref{tanhmodel})
for the galaxy UGC05253, as a function of the radius.}
\label{UGC05253dens}
\end{figure}

\begin{figure}[h!]
\centering
\includegraphics[width=20pc]{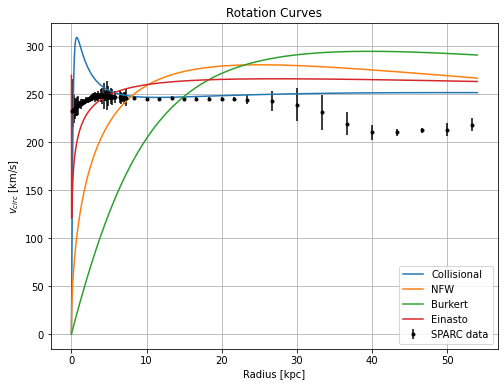}
\caption{The predicted rotation curves after using an optimization
for the collisional DM model (\ref{tanhmodel}), versus the SPARC
observational data for the galaxy UGC05253. We also plotted the
optimized curves for the NFW model, the Burkert model and the
Einasto model.} \label{UGC05253}
\end{figure}

\begin{figure}[h!]
\centering
\includegraphics[width=20pc]{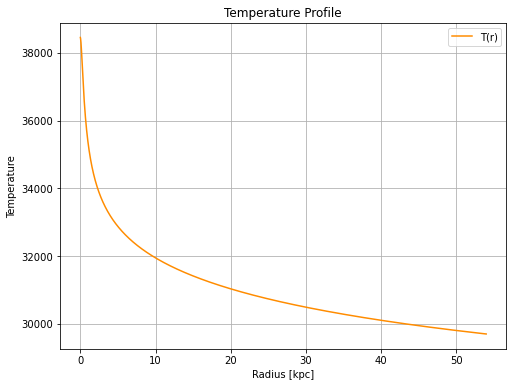}
\caption{The temperature as a function of the radius for the
collisional DM model (\ref{tanhmodel}) for the galaxy UGC05253.}
\label{UGC05253temp}
\end{figure}

For this galaxy, we shall choose $\rho_0=1\times
10^{11}$$M_{\odot}/\mathrm{Kpc}^{3}$. UGC05253 is a relatively
large spiral galaxy of morphological type SA(rs)ab, located at a
distance of about 21.5\,Mpc. In Figs. \ref{UGC05253dens},
\ref{UGC05253} and \ref{UGC05253temp} we present the density of
the collisional DM model, the predicted rotation curves after
using an optimization for the collisional DM model
(\ref{tanhmodel}), versus the SPARC observational data and the
temperature parameter as a function of the radius respectively. As
it can be seen, the SIDM model produces non-viable rotation curves
incompatible with the SPARC data. Also in Tables
\ref{collUGC05253}, \ref{NavaroUGC05253}, \ref{BuckertUGC05253}
and \ref{EinastoUGC05253} we present the optimization values for
the SIDM model, and the other DM profiles. Also in Table
\ref{EVALUATIONUGC05253} we present the overall evaluation of the
SIDM model for the galaxy at hand. The resulting phenomenology is
non-viable.
\begin{table}[h!]
  \begin{center}
    \caption{Collisional Dark Matter Optimization Values}
    \label{collUGC05253}
     \begin{tabular}{|r|r|}
     \hline
      \textbf{Parameter}   & \textbf{Optimization Values}
      \\  \hline
     $\delta_{\gamma} $ & 0.00000093
\\  \hline
$\gamma_0 $ & 1.017  \\ \hline $K_0$ ($M_{\odot} \,
\mathrm{Kpc}^{-3} \, (\mathrm{km/s})^{2}$)& 25000  \\ \hline
    \end{tabular}
  \end{center}
\end{table}
\begin{table}[h!]
  \begin{center}
    \caption{NFW  Optimization Values}
    \label{NavaroUGC05253}
     \begin{tabular}{|r|r|}
     \hline
      \textbf{Parameter}   & \textbf{Optimization Values}
      \\  \hline
   $\rho_s$   & $5\times 10^7$
\\  \hline
$r_s$&  11.60
\\  \hline
    \end{tabular}
  \end{center}
\end{table}
\begin{table}[h!]
  \begin{center}
    \caption{Burkert Optimization Values}
    \label{BuckertUGC05253}
     \begin{tabular}{|r|r|}
     \hline
      \textbf{Parameter}   & \textbf{Optimization Values}
      \\  \hline
     $\rho_0^B$  & $1\times 10^8$
\\  \hline
$r_0$&  12.21
\\  \hline
    \end{tabular}
  \end{center}
\end{table}
\begin{table}[h!]
  \begin{center}
    \caption{Einasto Optimization Values}
    \label{EinastoUGC05253}
    \begin{tabular}{|r|r|}
     \hline
      \textbf{Parameter}   & \textbf{Optimization Values}
      \\  \hline
     $\rho_e$  &$1\times 10^7$
\\  \hline
$r_e$ & 11.13
\\  \hline
$n_e$ & 0.05
\\  \hline
    \end{tabular}
  \end{center}
\end{table}
\begin{table}[h!]
\centering \caption{Physical assessment of collisional DM
parameters (UGC05253).}
\begin{tabular}{lcc}
\hline
Parameter & Value & Physical verdict \\
\hline
$\gamma_0$ & $1.017$ & Near-isothermal \\
$\delta_\gamma$ & $9.3\times10^{-7}$ & Practically zero \\
$r_\gamma$ & $1.5\ \mathrm{Kpc}$ & Reasonable inner-halo transition radius \\
$K_0$ & $2.5\times10^{4}$ & Large numerical scale  \\
$r_c$ & $0.5\ \mathrm{Kpc}$ & Small core scale \\
$p$ & $0.01$ & Extremely shallow decline \\
\hline Overall &-& Model is numerically consistent  \\
\hline
\end{tabular}
\label{EVALUATIONUGC05253}
\end{table}


\subsection{The Galaxy UGC05414}

\begin{figure}[h!]
\centering
\includegraphics[width=20pc]{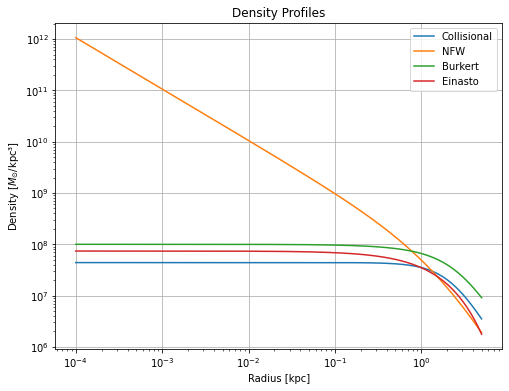}
\caption{The density of the collisional DM model (\ref{tanhmodel})
for the galaxy UGC05414, as a function of the radius.}
\label{UGC05414dens}
\end{figure}

\begin{figure}[h!]
\centering
\includegraphics[width=20pc]{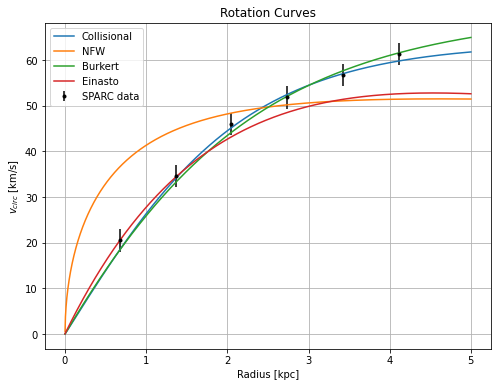}
\caption{The predicted rotation curves after using an optimization
for the collisional DM model (\ref{tanhmodel}), versus the SPARC
observational data for the galaxy UGC05414. We also plotted the
optimized curves for the NFW model, the Burkert model and the
Einasto model.} \label{UGC05414}
\end{figure}

\begin{figure}[h!]
\centering
\includegraphics[width=20pc]{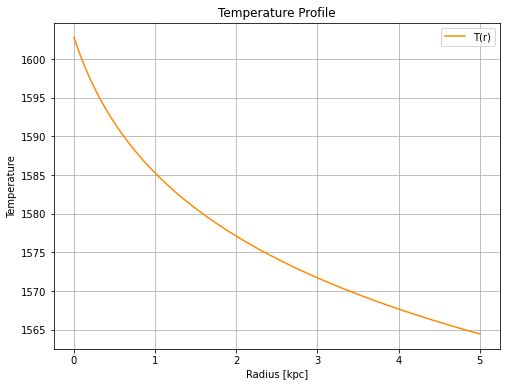}
\caption{The temperature as a function of the radius for the
collisional DM model (\ref{tanhmodel}) for the galaxy UGC05414.}
\label{UGC05414temp}
\end{figure}

For this galaxy, we shall choose $\rho_0=4.4\times
10^7$$M_{\odot}/\mathrm{Kpc}^{3}$. UGC05414 is a poorly studied
galaxy within the UGC catalog whose precise morphological
classification and distance remain uncertain; in modelling work it
has been treated akin to a late-type disk system owing to its
appearance and gas content. Its distance is sometimes assumed of
order 20-30 Mpc in the absence of precise redshift data. In Figs.
\ref{UGC05414dens}, \ref{UGC05414} and \ref{UGC05414temp} we
present the density of the collisional DM model, the predicted
rotation curves after using an optimization for the collisional DM
model (\ref{tanhmodel}), versus the SPARC observational data and
the temperature parameter as a function of the radius
respectively. As it can be seen, the SIDM model produces viable
rotation curves compatible with the SPARC data. Also in Tables
\ref{collUGC05414}, \ref{NavaroUGC05414}, \ref{BuckertUGC05414}
and \ref{EinastoUGC05414} we present the optimization values for
the SIDM model, and the other DM profiles. Also in Table
\ref{EVALUATIONUGC05414} we present the overall evaluation of the
SIDM model for the galaxy at hand. The resulting phenomenology is
viable.
\begin{table}[h!]
  \begin{center}
    \caption{Collisional Dark Matter Optimization Values}
    \label{collUGC05414}
     \begin{tabular}{|r|r|}
     \hline
      \textbf{Parameter}   & \textbf{Optimization Values}
      \\  \hline
     $\delta_{\gamma} $ & 0.0000000012
\\  \hline
$\gamma_0 $ & 1.0001  \\ \hline $K_0$ ($M_{\odot} \,
\mathrm{Kpc}^{-3} \, (\mathrm{km/s})^{2}$)& 1600  \\ \hline
    \end{tabular}
  \end{center}
\end{table}
\begin{table}[h!]
  \begin{center}
    \caption{NFW  Optimization Values}
    \label{NavaroUGC05414}
     \begin{tabular}{|r|r|}
     \hline
      \textbf{Parameter}   & \textbf{Optimization Values}
      \\  \hline
   $\rho_s$   & $5\times 10^7$
\\  \hline
$r_s$&  2.13
\\  \hline
    \end{tabular}
  \end{center}
\end{table}
\begin{table}[h!]
  \begin{center}
    \caption{Burkert Optimization Values}
    \label{BuckertUGC05414}
     \begin{tabular}{|r|r|}
     \hline
      \textbf{Parameter}   & \textbf{Optimization Values}
      \\  \hline
     $\rho_0^B$  & $1\times 10^8$
\\  \hline
$r_0$&  2.88
\\  \hline
    \end{tabular}
  \end{center}
\end{table}
\begin{table}[h!]
  \begin{center}
    \caption{Einasto Optimization Values}
    \label{EinastoUGC05414}
    \begin{tabular}{|r|r|}
     \hline
      \textbf{Parameter}   & \textbf{Optimization Values}
      \\  \hline
     $\rho_e$  &$1\times 10^7$
\\  \hline
$r_e$ & 2.68
\\  \hline
$n_e$ & 1
\\  \hline
    \end{tabular}
  \end{center}
\end{table}
\begin{table}[h!]
\centering \caption{Physical assessment of collisional DM
parameters for UGC05414.}
\begin{tabular}{lcc}
\hline
Parameter & Value & Physical Verdict \\
\hline
$\gamma_0$ & $1.0001$ & Almost exactly isothermal \\
$\delta_\gamma$ & $1.2\times10^{-9}$ & Practically zero \\
$r_\gamma$ & $1.5~\mathrm{Kpc}$ & Reasonable transition scale \\
$K_0$ & $1.6\times10^{3}$ & Moderate  scale, but enough pressure support \\
$r_c$ & $0.5~\mathrm{Kpc}$ & Small core scale \\
$p$ & $0.01$ & Extremely shallow decline\\
\hline
Overall & --- & Nearly isothermal\\
\hline
\end{tabular}
\label{EVALUATIONUGC05414}
\end{table}


\subsection{The Galaxy UGC05716 Non-viable, Extended Viable but 3 Parameter Model}

\begin{figure}[h!]
\centering
\includegraphics[width=20pc]{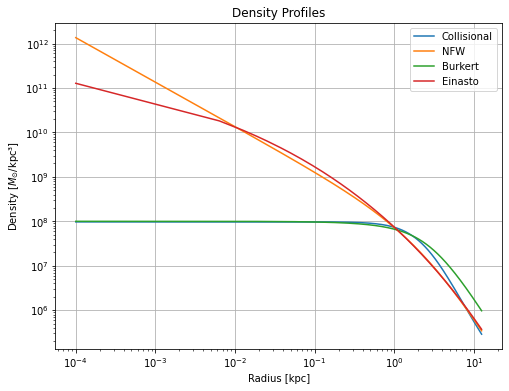}
\caption{The density of the collisional DM model (\ref{tanhmodel})
for the galaxy UGC05716, as a function of the radius.}
\label{UGC05716dens}
\end{figure}

\begin{figure}[h!]
\centering
\includegraphics[width=20pc]{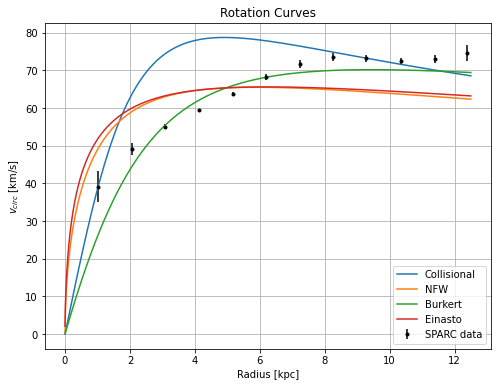}
\caption{The predicted rotation curves after using an optimization
for the collisional DM model (\ref{tanhmodel}), versus the SPARC
observational data for the galaxy UGC05716. We also plotted the
optimized curves for the NFW model, the Burkert model and the
Einasto model.} \label{UGC05716}
\end{figure}

\begin{figure}[h!]
\centering
\includegraphics[width=20pc]{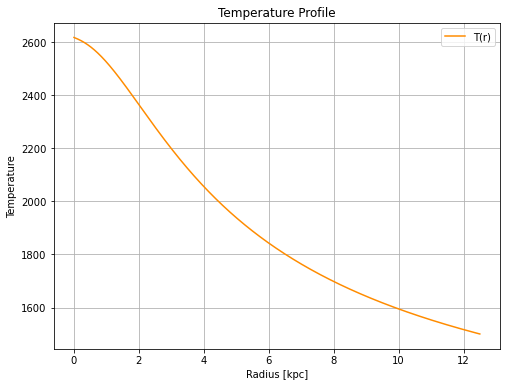}
\caption{The temperature as a function of the radius for the
collisional DM model (\ref{tanhmodel}) for the galaxy UGC05716.}
\label{UGC05716temp}
\end{figure}

\begin{figure}[h!]
\centering
\includegraphics[width=20pc]{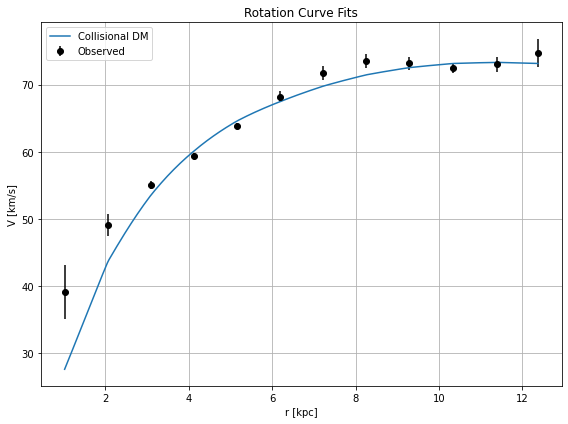}
\caption{The predicted rotation curves after using an optimization
for the collisional DM model (\ref{tanhmodel}), versus the
extended SPARC observational data for the galaxy UGC05716. The
model includes the rotation curves from all the components of the
galaxy, including gas and disk velocities, along with the
collisional DM model.} \label{extendedUGC05716}
\end{figure}

For this galaxy, we shall choose $\rho_0=0.5\times
10^7$$M_{\odot}/\mathrm{Kpc}^{3}$. UGC05716 is a sparsely
documented object in the UGC catalogue whose precise morphological
type and distance are not available in major databases; by analogy
with poorly studied UGC entries it is commonly treated as a
late-type disk or small spiral/dwarf system with an assumed
distance of order tens of Mpc. In Figs. \ref{UGC05716dens},
\ref{UGC05716} and \ref{UGC05716temp} we present the density of
the collisional DM model, the predicted rotation curves after
using an optimization for the collisional DM model
(\ref{tanhmodel}), versus the SPARC observational data and the
temperature parameter as a function of the radius respectively. As
it can be seen, the SIDM model produces non-viable rotation curves
incompatible with the SPARC data. Also in Tables
\ref{collUGC05716}, \ref{NavaroUGC05716}, \ref{BuckertUGC05716}
and \ref{EinastoUGC05716} we present the optimization values for
the SIDM model, and the other DM profiles. Also in Table
\ref{EVALUATIONUGC05716} we present the overall evaluation of the
SIDM model for the galaxy at hand. The resulting phenomenology is
non-viable.
\begin{table}[h!]
  \begin{center}
    \caption{Collisional Dark Matter Optimization Values}
    \label{collUGC05716}
     \begin{tabular}{|r|r|}
     \hline
      \textbf{Parameter}   & \textbf{Optimization Values}
      \\  \hline
     $\delta_{\gamma} $ & 0.0000000012
\\  \hline
$\gamma_0 $ & 1.0001  \\ \hline $K_0$ ($M_{\odot} \,
\mathrm{Kpc}^{-3} \, (\mathrm{km/s})^{2}$)& 500  \\ \hline
    \end{tabular}
  \end{center}
\end{table}
\begin{table}[h!]
  \begin{center}
    \caption{NFW  Optimization Values}
    \label{NavaroUGC05716}
     \begin{tabular}{|r|r|}
     \hline
      \textbf{Parameter}   & \textbf{Optimization Values}
      \\  \hline
   $\rho_s$   & $5\times 10^7$
\\  \hline
$r_s$&  2.71
\\  \hline
    \end{tabular}
  \end{center}
\end{table}
\begin{table}[h!]
  \begin{center}
    \caption{Burkert Optimization Values}
    \label{BuckertUGC05716}
     \begin{tabular}{|r|r|}
     \hline
      \textbf{Parameter}   & \textbf{Optimization Values}
      \\  \hline
     $\rho_0^B$  & $1\times 10^8$
\\  \hline
$r_0$&  2.91
\\  \hline
    \end{tabular}
  \end{center}
\end{table}
\begin{table}[h!]
  \begin{center}
    \caption{Einasto Optimization Values}
    \label{EinastoUGC05716}
    \begin{tabular}{|r|r|}
     \hline
      \textbf{Parameter}   & \textbf{Optimization Values}
      \\  \hline
     $\rho_e$  &$1\times 10^7$
\\  \hline
$r_e$ & 2.85
\\  \hline
$n_e$ & 0.18
\\  \hline
    \end{tabular}
  \end{center}
\end{table}
\begin{table}[h!]
\centering \caption{Physical assessment of collisional DM
parameters (UGC05716).}
\begin{tabular}{lcc}
\hline
Parameter & Value & Physical verdict \\
\hline
$\gamma_0$ & $1.0001$ & Slightly above isothermal; moderate central pressure support \\
$\delta_\gamma$ & $0.0000000012$ & Practically zero  \\
$r_\gamma$ & $1.5\ \mathrm{Kpc}$ & Plausible inner-halo transition radius \\
$K_0$ & $5\times10^{2}$ & Moderate numerical scale  \\
$r_c$ & $0.5\ \mathrm{Kpc}$ & Small core scale \\
$p$ & $0.01$ & Extremely shallow decline; $K(r)\sim$ constant \\
\hline
Overall &-& Numerically stable and plausible\\
\hline
\end{tabular}
\label{EVALUATIONUGC05716}
\end{table}
Now the extended picture including the rotation velocity from the
other components of the galaxy, such as the disk and gas, makes
the collisional DM model viable for this galaxy. In Fig.
\ref{extendedUGC05716} we present the combined rotation curves
including the other components of the galaxy along with the
collisional matter. As it can be seen, the extended collisional DM
model is viable.
Also in Table \ref{evaluationextendedUGC05716} we present the
values of the free parameters of the collisional DM model for
which the maximum compatibility with the SPARC data comes for the
galaxy UGC05716.
\begin{table}[h!]
\centering \caption{Physical assessment of Extended collisional DM
parameters (second set) for UGC05716.}
\begin{tabular}{lcc}
\hline
Parameter & Value & Physical Verdict \\
\hline
$\gamma_0$ & 0.81495132 & Sub-isothermal may be thermodynamically suspicious or indicate fitting pathology \\
$\delta_\gamma$ & 0.02155816 & Small radial variation \\
$K_0$ & 47803.27607890 & Very large entropy    \\
$ml_{disk}$ & 1.00000000 & Maximal-disk assumption \\
$ml_{bulge}$ & 0.00000000 & No bulge contribution \\
\hline
Overall &-& Exceptional Case\\
\hline
\end{tabular}
\label{evaluationextendedUGC05716}
\end{table}

\subsection{The Galaxy UGC05721 Marginally Viable}

\begin{figure}[h!]
\centering
\includegraphics[width=20pc]{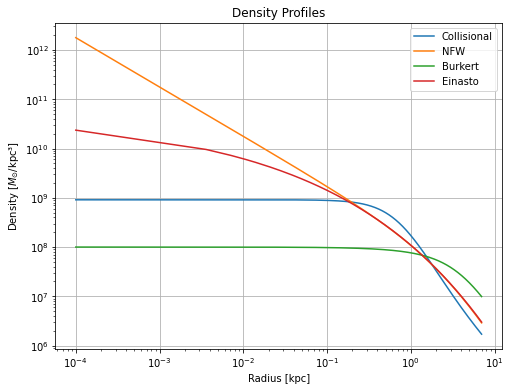}
\caption{The density of the collisional DM model (\ref{tanhmodel})
for the galaxy UGC05721, as a function of the radius.}
\label{UGC05721dens}
\end{figure}

\begin{figure}[h!]
\centering
\includegraphics[width=20pc]{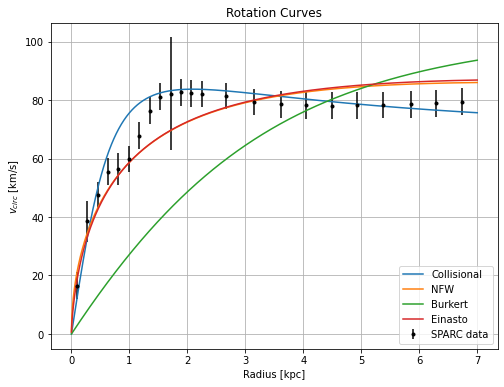}
\caption{The predicted rotation curves after using an optimization
for the collisional DM model (\ref{tanhmodel}), versus the SPARC
observational data for the galaxy UGC05721. We also plotted the
optimized curves for the NFW model, the Burkert model and the
Einasto model.} \label{UGC05721}
\end{figure}

\begin{figure}[h!]
\centering
\includegraphics[width=20pc]{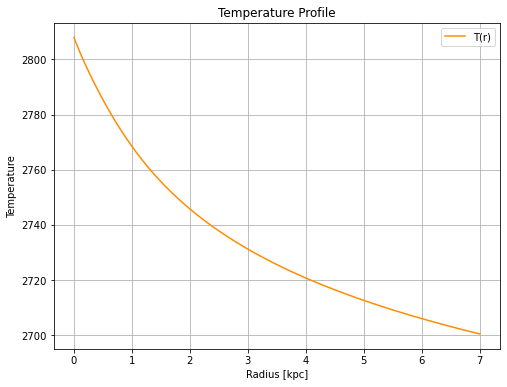}
\caption{The temperature as a function of the radius for the
collisional DM model (\ref{tanhmodel}) for the galaxy UGC05721.}
\label{UGC05721temp}
\end{figure}

\begin{figure}[h!]
\centering
\includegraphics[width=20pc]{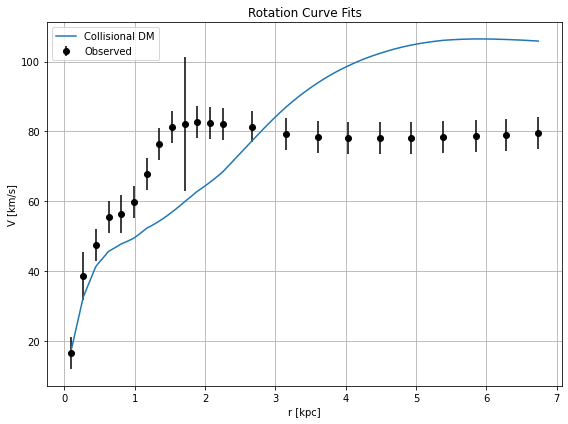}
\caption{The predicted rotation curves after using an optimization
for the collisional DM model (\ref{tanhmodel}), versus the
extended SPARC observational data for the galaxy UGC05721. The
model includes the rotation curves from all the components of the
galaxy, including gas and disk velocities, along with the
collisional DM model.} \label{extendedUGC05721}
\end{figure}

For this galaxy, we shall choose $\rho_0=9.1\times
10^8$$M_{\odot}/\mathrm{Kpc}^{3}$. UGC05721  is a low-mass spiral
(SAB / SBcd) galaxy in the field, at a distance of approximately
\(13.0\) Mpc.  In Figs. \ref{UGC05721dens}, \ref{UGC05721} and
\ref{UGC05721temp} we present the density of the collisional DM
model, the predicted rotation curves after using an optimization
for the collisional DM model (\ref{tanhmodel}), versus the SPARC
observational data and the temperature parameter as a function of
the radius respectively. As it can be seen, the SIDM model
produces marginally viable rotation curves compatible with the
SPARC data. Also in Tables \ref{collUGC05721},
\ref{NavaroUGC05721}, \ref{BuckertUGC05721} and
\ref{EinastoUGC05721} we present the optimization values for the
SIDM model, and the other DM profiles. Also in Table
\ref{EVALUATIONUGC05721} we present the overall evaluation of the
SIDM model for the galaxy at hand. The resulting phenomenology is
marginally viable.
\begin{table}[h!]
  \begin{center}
    \caption{Collisional Dark Matter Optimization Values}
    \label{collUGC05721}
     \begin{tabular}{|r|r|}
     \hline
      \textbf{Parameter}   & \textbf{Optimization Values}
      \\  \hline
     $\delta_{\gamma} $ & 0.0000000012
\\  \hline
$\gamma_0 $ & 1.0001 \\ \hline $K_0$ ($M_{\odot} \,
\mathrm{Kpc}^{-3} \, (\mathrm{km/s})^{2}$)& 1500  \\ \hline
    \end{tabular}
  \end{center}
\end{table}
\begin{table}[h!]
  \begin{center}
    \caption{NFW  Optimization Values}
    \label{NavaroUGC05721}
     \begin{tabular}{|r|r|}
     \hline
      \textbf{Parameter}   & \textbf{Optimization Values}
      \\  \hline
   $\rho_s$   & $5\times 10^7$
\\  \hline
$r_s$&  3.56
\\  \hline
    \end{tabular}
  \end{center}
\end{table}
\begin{table}[h!]
  \begin{center}
    \caption{Burkert Optimization Values}
    \label{BuckertUGC05721}
     \begin{tabular}{|r|r|}
     \hline
      \textbf{Parameter}   & \textbf{Optimization Values}
      \\  \hline
     $\rho_0^B$  & $1\times 10^8$
\\  \hline
$r_0$& 4.19
\\  \hline
    \end{tabular}
  \end{center}
\end{table}
\begin{table}[h!]
  \begin{center}
    \caption{Einasto Optimization Values}
    \label{EinastoUGC05721}
    \begin{tabular}{|r|r|}
     \hline
      \textbf{Parameter}   & \textbf{Optimization Values}
      \\  \hline
     $\rho_e$  &$1\times 10^7$
\\  \hline
$r_e$ & 3.83
\\  \hline
$n_e$ & 0.24
\\  \hline
    \end{tabular}
  \end{center}
\end{table}
\begin{table}[h!]
\centering \caption{Physical assessment of collisional DM
parameters for UGC05721.}
\begin{tabular}{lcc}
\hline
Parameter & Value & Physical Verdict \\
\hline
$\gamma_0$ & $1.0001$ & Practically isothermal \\
$\delta_\gamma$ & $1.2\times10^{-9}$ & Negligible variation \\
$r_\gamma$ & $1.5\ \mathrm{Kpc}$ & Transition radius nominally inside inner halo\\
$K_0$ ($M_{\odot}\,\mathrm{Kpc}^{-3}\,(\mathrm{km/s})^{2}$) & $2.7\times10^{3}$ & Entropy/pressure scale is large in absolute terms\\
$r_c$ & $0.5\ \mathrm{Kpc}$ & Small core length . \\
$p$ & $0.01$ & Very shallow decline of $K(r)$ \\
Overall & -- & Model effectively reduces to an almost-isothermal \\
\hline
\end{tabular}
\label{EVALUATIONUGC05721}
\end{table}
Now the extended picture including the rotation velocity from the
other components of the galaxy, such as the disk and gas, makes
the collisional DM model viable for this galaxy. In Fig.
\ref{extendedUGC05721} we present the combined rotation curves
including the other components of the galaxy along with the
collisional matter. As it can be seen, the extended collisional DM
model is marginally viable.
Also in Table \ref{evaluationextendedUGC05721} we present the
values of the free parameters of the collisional DM model for
which the maximum compatibility with the SPARC data comes for the
galaxy UGC05721.
\begin{table}[h!]
\centering \caption{Physical assessment of Extended collisional DM
parameters for galaxy UGC05721.}
\begin{tabular}{lcc}
\hline
Parameter & Value & Physical Verdict \\
\hline
$\gamma_0$ & 1.10109807 & Slightly above isothermal \\
$\delta_\gamma$ & 0.1 & Noticeable radial variation \\
$K_0$ & 3000 & Moderate entropy   \\
$ml_{disk}$ & 1.00000000 & At the upper plausible limit \\
$ml_{bulge}$ & 0.00000000 & Negligible bulge \\
\hline
Overall &-& Physically plausible but borderline \\
\hline
\end{tabular}
\label{evaluationextendedUGC05721}
\end{table}


\subsection{The Galaxy UGC05764   Marginally}

\begin{figure}[h!]
\centering
\includegraphics[width=20pc]{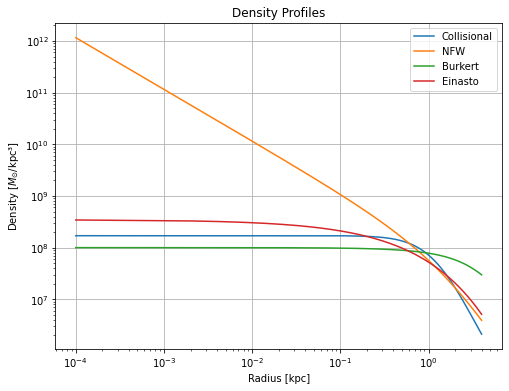}
\caption{The density of the collisional DM model (\ref{tanhmodel})
for the galaxy UGC05764, as a function of the radius.}
\label{UGC05764dens}
\end{figure}

\begin{figure}[h!]
\centering
\includegraphics[width=20pc]{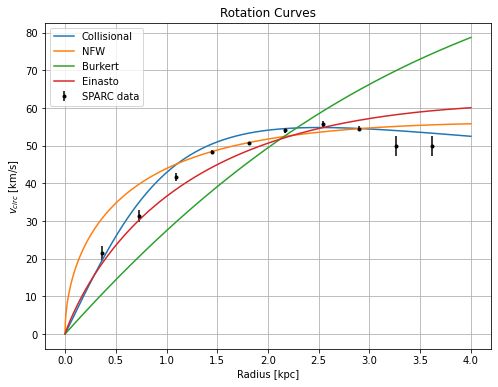}
\caption{The predicted rotation curves after using an optimization
for the collisional DM model (\ref{tanhmodel}), versus the SPARC
observational data for the galaxy UGC05764. We also plotted the
optimized curves for the NFW model, the Burkert model and the
Einasto model.} \label{UGC05764}
\end{figure}

\begin{figure}[h!]
\centering
\includegraphics[width=20pc]{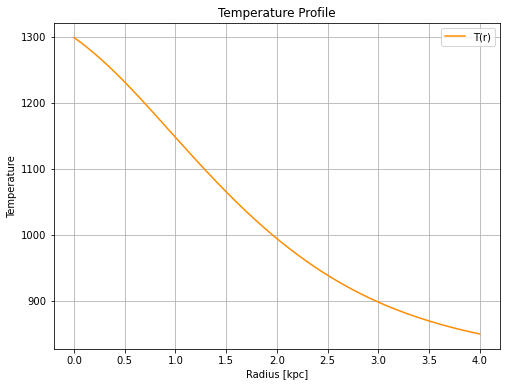}
\caption{The temperature as a function of the radius for the
collisional DM model (\ref{tanhmodel}) for the galaxy UGC05764.}
\label{UGC05764temp}
\end{figure}

\begin{figure}[h!]
\centering
\includegraphics[width=20pc]{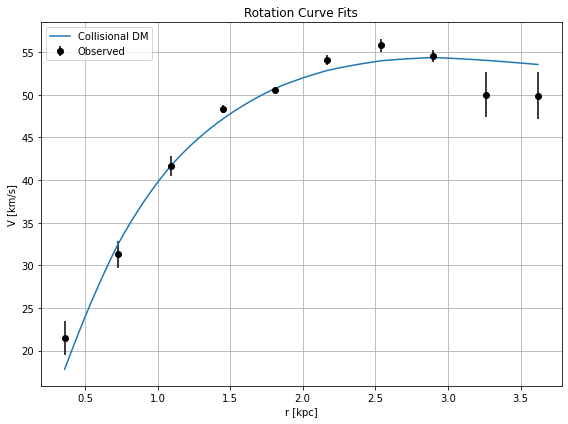}
\caption{The predicted rotation curves after using an optimization
for the collisional DM model (\ref{tanhmodel}), versus the
extended SPARC observational data for the galaxy UGC05764. The
model includes the rotation curves from all the components of the
galaxy, including gas and disk velocities, along with the
collisional DM model.} \label{extendedUGC05764}
\end{figure}

For this galaxy, we shall choose $\rho_0=1.7\times
10^8$$M_{\odot}/\mathrm{Kpc}^{3}$. UGC05764 is a gas-rich,
late-type dwarf/irregular (sometimes classified as blue
compact/irregular) galaxy rather than a classical spiral. Distance
estimates used in kinematic and HI studies place it at roughly
\(D\sim 8\!-\!10\) Mpc, so it is a nearby dwarf galaxy. In Figs.
\ref{UGC05764dens}, \ref{UGC05764} and \ref{UGC05764temp} we
present the density of the collisional DM model, the predicted
rotation curves after using an optimization for the collisional DM
model (\ref{tanhmodel}), versus the SPARC observational data and
the temperature parameter as a function of the radius
respectively. As it can be seen, the SIDM model produces
marginally viable rotation curves compatible with the SPARC data.
Also in Tables \ref{collUGC05764}, \ref{NavaroUGC05764},
\ref{BuckertUGC05764} and \ref{EinastoUGC05764} we present the
optimization values for the SIDM model, and the other DM profiles.
Also in Table \ref{EVALUATIONUGC05764} we present the overall
evaluation of the SIDM model for the galaxy at hand. The resulting
phenomenology is marginally viable.
\begin{table}[h!]
  \begin{center}
    \caption{Collisional Dark Matter Optimization Values}
    \label{collUGC05764}
     \begin{tabular}{|r|r|}
     \hline
      \textbf{Parameter}   & \textbf{Optimization Values}
      \\  \hline
     $\delta_{\gamma} $ & 0.0000000012
\\  \hline
$\gamma_0 $ & 1.0001  \\ \hline $K_0$ ($M_{\odot} \,
\mathrm{Kpc}^{-3} \, (\mathrm{km/s})^{2}$)& 1200  \\ \hline
    \end{tabular}
  \end{center}
\end{table}
\begin{table}[h!]
  \begin{center}
    \caption{NFW  Optimization Values}
    \label{NavaroUGC05764}
     \begin{tabular}{|r|r|}
     \hline
      \textbf{Parameter}   & \textbf{Optimization Values}
      \\  \hline
   $\rho_s$   & $5\times 10^7$
\\  \hline
$r_s$&  2.32
\\  \hline
    \end{tabular}
  \end{center}
\end{table}
\begin{table}[h!]
  \begin{center}
    \caption{Burkert Optimization Values}
    \label{BuckertUGC05764}
     \begin{tabular}{|r|r|}
     \hline
      \textbf{Parameter}   & \textbf{Optimization Values}
      \\  \hline
     $\rho_0^B$  & $1\times 10^8$
\\  \hline
$r_0$&  4.52
\\  \hline
    \end{tabular}
  \end{center}
\end{table}
\begin{table}[h!]
  \begin{center}
    \caption{Einasto Optimization Values}
    \label{EinastoUGC05764}
    \begin{tabular}{|r|r|}
     \hline
      \textbf{Parameter}   & \textbf{Optimization Values}
      \\  \hline
     $\rho_e$  &$1\times 10^7$
\\  \hline
$r_e$ & 2.83
\\  \hline
$n_e$ & 0.58
\\  \hline
    \end{tabular}
  \end{center}
\end{table}
\begin{table}[h!]
\centering \caption{Physical assessment of collisional DM
parameters for UGC05764.}
\begin{tabular}{lcc}
\hline
Parameter & Value & Physical Verdict \\
\hline
$\gamma_0$ & $1.0001$ & Slightly above isothermal \\
$\delta_\gamma$ & $1.2\times10^{-9}$ & Modest variation \\
$r_\gamma$ & $1.5\ \mathrm{Kpc}$ & Transition inside inner halo \\
$K_0$ ($M_{\odot}\,\mathrm{Kpc}^{-3}\,(\mathrm{km/s})^{2}$) & $7.0\times10^{2}$ & Enough pressure support \\
$r_c$ & $0.5\ \mathrm{Kpc}$ & Small core length \\
$p$ & $0.01$ & Very shallow radial decline of $K(r)$ \\
Overall & -- & Physically consistent and numerically stable\\
\hline
\end{tabular}
\label{EVALUATIONUGC05764}
\end{table}
Now the extended picture including the rotation velocity from the
other components of the galaxy, such as the disk and gas, makes
the collisional DM model viable for this galaxy. In Fig.
\ref{extendedUGC05764} we present the combined rotation curves
including the other components of the galaxy along with the
collisional matter. As it can be seen, the extended collisional DM
model is marginally viable.
Also in Table \ref{evaluationextendedUGC05764} we present the
values of the free parameters of the collisional DM model for
which the maximum compatibility with the SPARC data comes for the
galaxy UGC05764.
\begin{table}[h!]
\centering \caption{Physical assessment of Extended collisional DM
parameters for galaxy UGC05764.}
\begin{tabular}{lcc}
\hline
Parameter & Value & Physical Verdict \\
\hline
$\gamma_0$ & 1.02713240 & Very close to isothermal \\
$\delta_\gamma$ & 0.0020206815 & Extremely small variation \\
$K_0$ & 600 & Low entropy  \\
$ml_{disk}$ & 1.00000000 & At the high end of plausible disk M/L \\
$ml_{bulge}$ & 0.00000000 & Negligible bulge contribution \\
\hline
Overall &-& Physically possible\\
\hline
\end{tabular}
\label{evaluationextendedUGC05764}
\end{table}

\subsection{The Galaxy UGC05829}

\begin{figure}[h!]
\centering
\includegraphics[width=20pc]{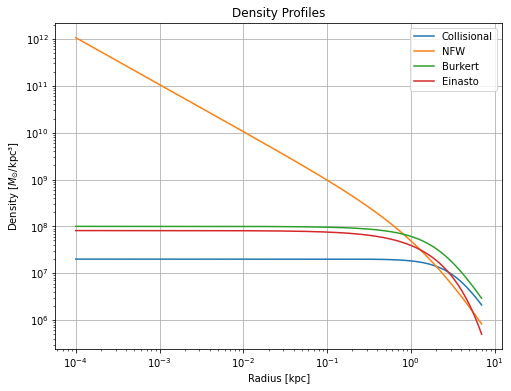}
\caption{The density of the collisional DM model (\ref{tanhmodel})
for the galaxy UGC05829, as a function of the radius.}
\label{UGC05829dens}
\end{figure}

\begin{figure}[h!]
\centering
\includegraphics[width=20pc]{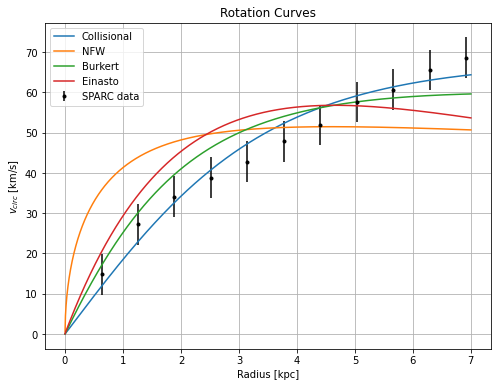}
\caption{The predicted rotation curves after using an optimization
for the collisional DM model (\ref{tanhmodel}), versus the SPARC
observational data for the galaxy UGC05829. We also plotted the
optimized curves for the NFW model, the Burkert model and the
Einasto model.} \label{UGC05829}
\end{figure}

\begin{figure}[h!]
\centering
\includegraphics[width=20pc]{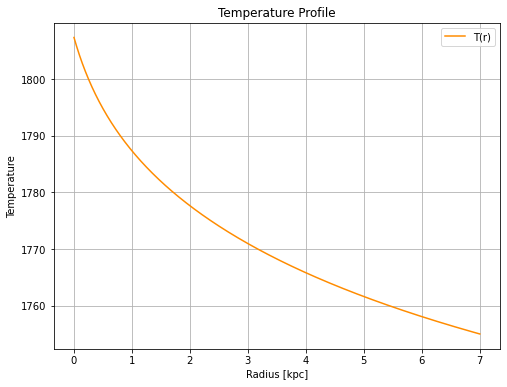}
\caption{The temperature as a function of the radius for the
collisional DM model (\ref{tanhmodel}) for the galaxy UGC05829.}
\label{UGC05829temp}
\end{figure}

For this galaxy, we shall choose $\rho_0=2\times
10^7$$M_{\odot}/\mathrm{Kpc}^{3}$ UGC5829 (the ''Spider Galaxy'')
is a nearby irregular dwarf  Magellanic--type galaxy located in
the direction of Leo. The distance to UGC 5829 is approximately
3.75 Mpc. In Figs. \ref{UGC05829dens}, \ref{UGC05829} and
\ref{UGC05829temp} we present the density of the collisional DM
model, the predicted rotation curves after using an optimization
for the collisional DM model (\ref{tanhmodel}), versus the SPARC
observational data and the temperature parameter as a function of
the radius respectively. As it can be seen, the SIDM model
produces viable rotation curves compatible with the SPARC data.
Also in Tables \ref{collUGC05829}, \ref{NavaroUGC05829},
\ref{BuckertUGC05829} and \ref{EinastoUGC05829} we present the
optimization values for the SIDM model, and the other DM profiles.
Also in Table \ref{EVALUATIONUGC05829} we present the overall
evaluation of the SIDM model for the galaxy at hand. The resulting
phenomenology is viable.
\begin{table}[h!]
  \begin{center}
    \caption{Collisional Dark Matter Optimization Values}
    \label{collUGC05829}
     \begin{tabular}{|r|r|}
     \hline
      \textbf{Parameter}   & \textbf{Optimization Values}
      \\  \hline
     $\delta_{\gamma} $ & 0.0000000012
\\  \hline
$\gamma_0 $ & 1.0001  \\ \hline $K_0$ ($M_{\odot} \,
\mathrm{Kpc}^{-3} \, (\mathrm{km/s})^{2}$)& 1780 \\ \hline
    \end{tabular}
  \end{center}
\end{table}
\begin{table}[h!]
  \begin{center}
    \caption{NFW  Optimization Values}
    \label{NavaroUGC05829}
     \begin{tabular}{|r|r|}
     \hline
      \textbf{Parameter}   & \textbf{Optimization Values}
      \\  \hline
   $\rho_s$   & $5\times 10^7$
\\  \hline
$r_s$&  2.13
\\  \hline
    \end{tabular}
  \end{center}
\end{table}
\begin{table}[h!]
  \begin{center}
    \caption{Burkert Optimization Values}
    \label{BuckertUGC05829}
     \begin{tabular}{|r|r|}
     \hline
      \textbf{Parameter}   & \textbf{Optimization Values}
      \\  \hline
     $\rho_0^B$  & $1\times 10^8$
\\  \hline
$r_0$&  2.48
\\  \hline
    \end{tabular}
  \end{center}
\end{table}

\begin{table}[h!]
  \begin{center}
    \caption{Einasto Optimization Values}
    \label{EinastoUGC05829}
    \begin{tabular}{|r|r|}
     \hline
      \textbf{Parameter}   & \textbf{Optimization Values}
      \\  \hline
     $\rho_e$  &$1\times 10^7$
\\  \hline
$r_e$ & 2.75
\\  \hline
$n_e$ & 1
\\  \hline
    \end{tabular}
  \end{center}
\end{table}
\begin{table}[h!]
\centering \caption{Physical assessment of collisional DM
parameters (UGC05829).}
\begin{tabular}{lcc}
\hline
Parameter & Value & Physical Verdict \\
\hline
$\gamma_0$ & $1.0001$ & Essentially isothermal \\
$\delta_\gamma$ & $0.0000000012$ & Variation negligible \\
$r_\gamma$ & $1.5\ \mathrm{Kpc}$ & Transition radius inside inner halo \\
$K_0$ & $1.78\times10^{3}$ & Large entropy/pressure scale \\
$r_c$ & $0.5\ \mathrm{Kpc}$ & Small core scale \\
$p$ & $0.01$ & Very shallow decline \\
\hline
Overall &-& Physically consistent  \\
\hline
\end{tabular}
\label{EVALUATIONUGC05829}
\end{table}

\subsection{The Galaxy UGC05918}

\begin{figure}[h!]
\centering
\includegraphics[width=20pc]{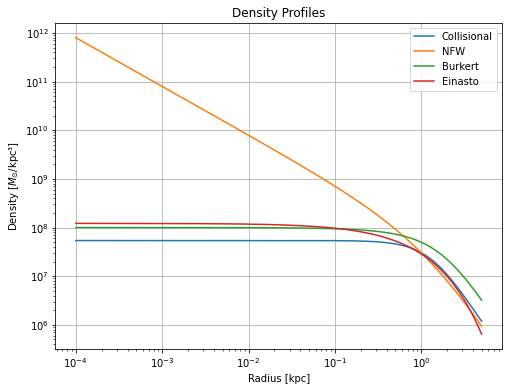}
\caption{The density of the collisional DM model (\ref{tanhmodel})
for the galaxy UGC05918, as a function of the radius.}
\label{UGC05918dens}
\end{figure}
\begin{figure}[h!]
\centering
\includegraphics[width=20pc]{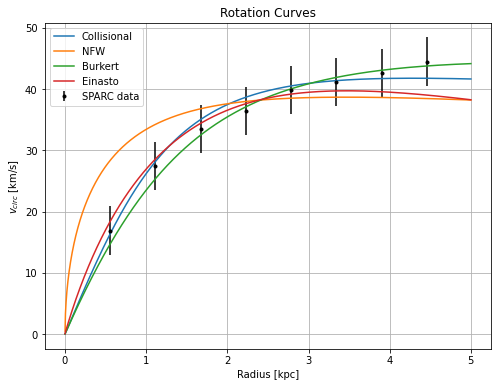}
\caption{The predicted rotation curves after using an optimization
for the collisional DM model (\ref{tanhmodel}), versus the SPARC
observational data for the galaxy UGC05918. We also plotted the
optimized curves for the NFW model, the Burkert model and the
Einasto model.} \label{UGC05918}
\end{figure}
\begin{figure}[h!]
\centering
\includegraphics[width=20pc]{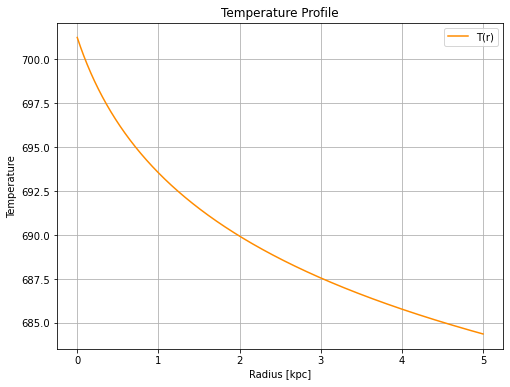}
\caption{The temperature as a function of the radius for the
collisional DM model (\ref{tanhmodel}) for the galaxy UGC05918.}
\label{UGC05918temp}
\end{figure}

For this galaxy, we shall choose $\rho_0=4\times
10^7$$M_{\odot}/\mathrm{Kpc}^{3}$. UGC5918 is cataloged as a dwarf
galaxy (likely of irregular / low-surface-brightness type) with
very low optical brightness located in Ursa Major. Its precise
distance is uncertain, but in catalogs it is linked to DDO 87 and
placed among isolated low-luminosity systems with distances of
order 3-10 Mpc. In Figs. \ref{UGC05918dens}, \ref{UGC05918} and
\ref{UGC05918temp} we present the density of the collisional DM
model, the predicted rotation curves after using an optimization
for the collisional DM model (\ref{tanhmodel}), versus the SPARC
observational data and the temperature parameter as a function of
the radius respectively. As it can be seen, the SIDM model
produces viable rotation curves compatible with the SPARC data.
Also in Tables \ref{collUGC05918}, \ref{NavaroUGC05918},
\ref{BuckertUGC05918} and \ref{EinastoUGC05918} we present the
optimization values for the SIDM model, and the other DM profiles.
Also in Table \ref{EVALUATIONUGC05918} we present the overall
evaluation of the SIDM model for the galaxy at hand. The resulting
phenomenology is viable.
\begin{table}[h!]
  \begin{center}
    \caption{Collisional Dark Matter Optimization Values}
    \label{collUGC05918}
     \begin{tabular}{|r|r|}
     \hline
      \textbf{Parameter}   & \textbf{Optimization Values}
      \\  \hline
     $\delta_{\gamma} $ & 0.0000000012
\\  \hline
$\gamma_0 $ & 1.0001 \\ \hline $K_0$ ($M_{\odot} \,
\mathrm{Kpc}^{-3} \, (\mathrm{km/s})^{2}$)& 700  \\ \hline
    \end{tabular}
  \end{center}
\end{table}
\begin{table}[h!]
  \begin{center}
    \caption{NFW  Optimization Values}
    \label{NavaroUGC05918}
     \begin{tabular}{|r|r|}
     \hline
      \textbf{Parameter}   & \textbf{Optimization Values}
      \\  \hline
   $\rho_s$   & $5\times 10^7$
\\  \hline
$r_s$&  1.60
\\  \hline
    \end{tabular}
  \end{center}
\end{table}
\begin{table}[h!]
  \begin{center}
    \caption{Burkert Optimization Values}
    \label{BuckertUGC05918}
     \begin{tabular}{|r|r|}
     \hline
      \textbf{Parameter}   & \textbf{Optimization Values}
      \\  \hline
     $\rho_0^B$  & $1\times 10^8$
\\  \hline
$r_0$&  1.84
\\  \hline
    \end{tabular}
  \end{center}
\end{table}
\begin{table}[h!]
  \begin{center}
    \caption{Einasto Optimization Values}
    \label{EinastoUGC05918}
    \begin{tabular}{|r|r|}
     \hline
      \textbf{Parameter}   & \textbf{Optimization Values}
      \\  \hline
     $\rho_e$  &$1\times 10^7$
\\  \hline
$r_e$ & 1.98
\\  \hline
$n_e$ & 0.8
\\  \hline
    \end{tabular}
  \end{center}
\end{table}
\begin{table}[h!]
\centering \caption{Physical assessment of collisional DM
parameters for UGC05918.}
\begin{tabular}{lcc}
\hline
Parameter & Value & Physical Verdict \\
\hline
$\gamma_0$ & $1.0001$ & Almost exactly isothermal  \\
$\delta_\gamma$ & $1.2\times10^{-9}$ & Practically zero  \\
$r_\gamma$ & $1.5~\mathrm{Kpc}$ & Plausible transition scale  \\
$K_0$ & $7.0\times10^{2}$ & Enough pressure support suited to very low-mass systems \\
$r_c$ & $0.5~\mathrm{Kpc}$ & Small core scale \\
$p$ & $0.01$ & Extremely shallow decline \\
\hline
Overall & --- & Nearly isothermal \\
\hline
\end{tabular}
\label{EVALUATIONUGC05918}
\end{table}

\subsection{The Galaxy UGC05986 Peculiar maximum compatibility 4 Parameter Model}

\begin{figure}[h!]
\centering
\includegraphics[width=20pc]{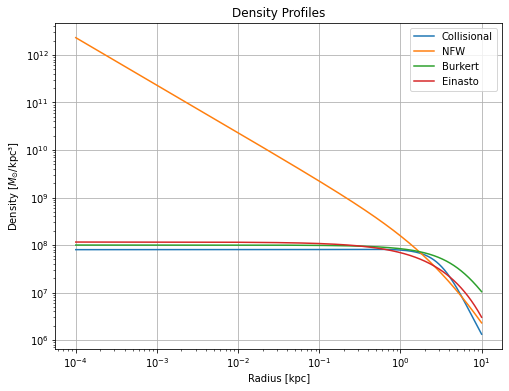}
\caption{The density of the collisional DM model (\ref{tanhmodel})
for the galaxy UGC05986, as a function of the radius.}
\label{UGC05986dens}
\end{figure}

\begin{figure}[h!]
\centering
\includegraphics[width=20pc]{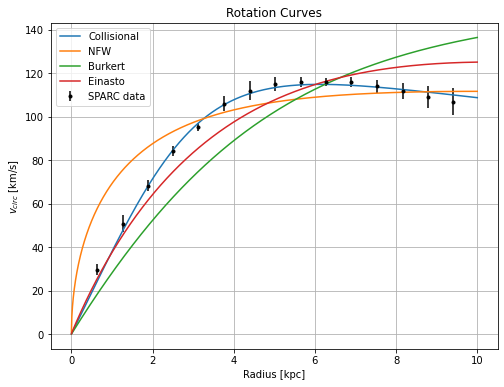}
\caption{The predicted rotation curves after using an optimization
for the collisional DM model (\ref{tanhmodel}), versus the SPARC
observational data for the galaxy UGC05986. We also plotted the
optimized curves for the NFW model, the Burkert model and the
Einasto model.} \label{UGC05986}
\end{figure}

\begin{figure}[h!]
\centering
\includegraphics[width=20pc]{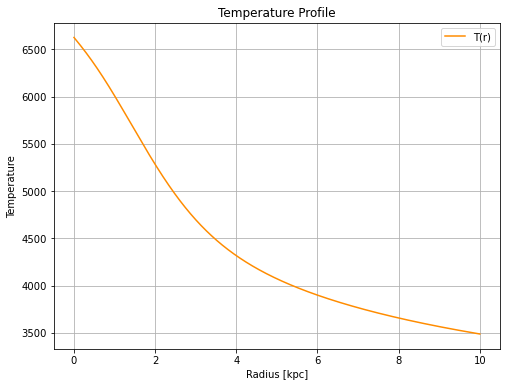}
\caption{The temperature as a function of the radius for the
collisional DM model (\ref{tanhmodel}) for the galaxy UGC05986.}
\label{UGC05986temp}
\end{figure}

For this galaxy, we shall choose $\rho_0=8\times
10^7$$M_{\odot}/\mathrm{Kpc}^{3}$. UGC5986 is an edge-on,
late-type/irregular spiral with prominent star-forming regions and
tidal features from a nearby companion; it lies at an estimated
distance of $\sim 12$--$13$\,Mpc. In Figs. \ref{UGC05986dens},
\ref{UGC05986} and \ref{UGC05986temp} we present the density of
the collisional DM model, the predicted rotation curves after
using an optimization for the collisional DM model
(\ref{tanhmodel}), versus the SPARC observational data and the
temperature parameter as a function of the radius respectively. As
it can be seen, the SIDM model produces viable rotation curves
compatible with the SPARC data. Also in Tables \ref{collUGC05986},
\ref{NavaroUGC05986}, \ref{BuckertUGC05986} and
\ref{EinastoUGC05986} we present the optimization values for the
SIDM model, and the other DM profiles. Also in Table
\ref{EVALUATIONUGC05986} we present the overall evaluation of the
SIDM model for the galaxy at hand. The resulting phenomenology is
viable.
\begin{table}[h!]
  \begin{center}
    \caption{Collisional Dark Matter Optimization Values}
    \label{collUGC05986}
     \begin{tabular}{|r|r|}
     \hline
      \textbf{Parameter}   & \textbf{Optimization Values}
      \\  \hline
     $\delta_{\gamma} $ & 0.0000000012
\\  \hline
$\gamma_0 $ & 1.0001  \\ \hline $K_0$ ($M_{\odot} \,
\mathrm{Kpc}^{-3} \, (\mathrm{km/s})^{2}$)& 1500 \\ \hline
    \end{tabular}
  \end{center}
\end{table}
\begin{table}[h!]
  \begin{center}
    \caption{NFW  Optimization Values}
    \label{NavaroUGC05986}
     \begin{tabular}{|r|r|}
     \hline
      \textbf{Parameter}   & \textbf{Optimization Values}
      \\  \hline
   $\rho_s$   & $5\times 10^7$
\\  \hline
$r_s$& 4.62
\\  \hline
    \end{tabular}
  \end{center}
\end{table}
\begin{table}[h!]
  \begin{center}
    \caption{Burkert Optimization Values}
    \label{BuckertUGC05986}
     \begin{tabular}{|r|r|}
     \hline
      \textbf{Parameter}   & \textbf{Optimization Values}
      \\  \hline
     $\rho_0^B$  & $1\times 10^8$
\\  \hline
$r_0$&  6.15
\\  \hline
    \end{tabular}
  \end{center}
\end{table}
\begin{table}[h!]
  \begin{center}
    \caption{Einasto Optimization Values}
    \label{EinastoUGC05986}
    \begin{tabular}{|r|r|}
     \hline
      \textbf{Parameter}   & \textbf{Optimization Values}
      \\  \hline
     $\rho_e$  &$1\times 10^7$
\\  \hline
$r_e$ & 5.98
\\  \hline
$n_e$ & 0.85
\\  \hline
    \end{tabular}
  \end{center}
\end{table}
\begin{table}[h!]
\centering \caption{Physical assessment of collisional DM
parameters (UGC05986).}
\begin{tabular}{lcc}
\hline
Parameter & Value & Physical Verdict \\
\hline
$\gamma_0$ & $1.074$ & Mildly super-isothermal \\
$\delta_\gamma$ & $1.2\times10^{-2}$ & Small but measurable variation\\
$r_\gamma$ & $1.5\ \mathrm{Kpc}$ & Transition inside inner halo; affects inner slope slightly \\
$K_0$ & $1.5\times10^{3}$ & Large entropy/pressure scale \\
$r_c$ & $0.5\ \mathrm{Kpc}$ & Small core radius \\
$p$ & $0.01$ & Very shallow decline \\
\hline
Overall &-& Physically consistent and modestly flexible\\
\hline
\end{tabular}
\label{EVALUATIONUGC05986}
\end{table}


\subsection{The Galaxy UGC05999}

\begin{figure}[h!]
\centering
\includegraphics[width=20pc]{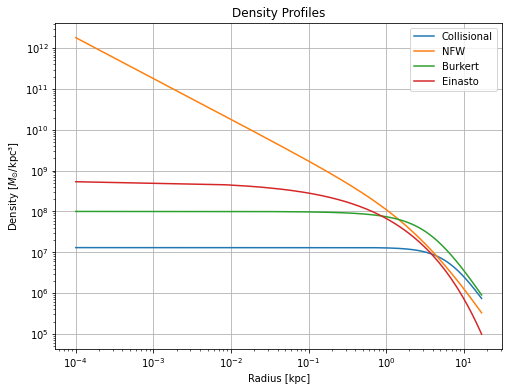}
\caption{The density of the collisional DM model (\ref{tanhmodel})
for the galaxy UGC05999, as a function of the radius.}
\label{UGC05999dens}
\end{figure}

\begin{figure}[h!]
\centering
\includegraphics[width=20pc]{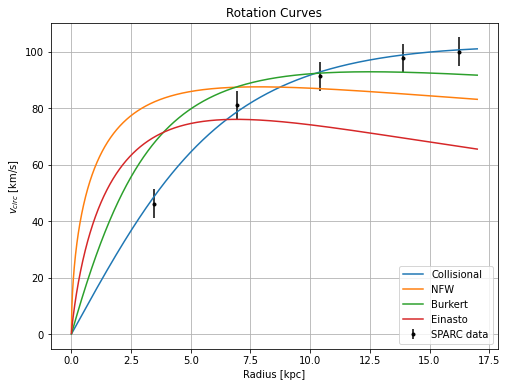}
\caption{The predicted rotation curves after using an optimization
for the collisional DM model (\ref{tanhmodel}), versus the SPARC
observational data for the galaxy UGC05999. We also plotted the
optimized curves for the NFW model, the Burkert model and the
Einasto model.} \label{UGC05999}
\end{figure}

\begin{figure}[h!]
\centering
\includegraphics[width=20pc]{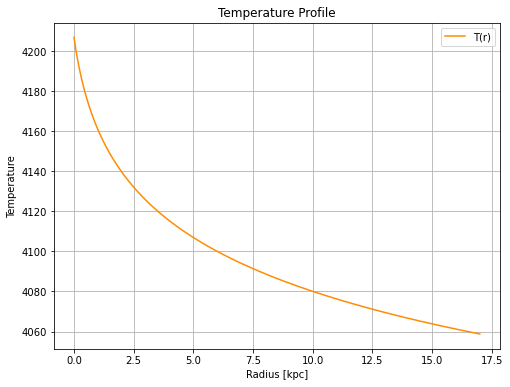}
\caption{The temperature as a function of the radius for the
collisional DM model (\ref{tanhmodel}) for the galaxy UGC05999.}
\label{UGC05999temp}
\end{figure}

For this galaxy, we shall choose $\rho_0=0.9\times
10^7$$M_{\odot}/\mathrm{Kpc}^{3}$. UGC5999 is a
low-surface-brightness   spiral galaxy located at an estimated
distance of approximately 10--20\,Mpc. It is characterized by a
large, extended H\,I disk and a low central surface brightness,
classifying it as a low-luminosity spiral galaxy. In Figs.
\ref{UGC05999dens}, \ref{UGC05999} and \ref{UGC05999temp} we
present the density of the collisional DM model, the predicted
rotation curves after using an optimization for the collisional DM
model (\ref{tanhmodel}), versus the SPARC observational data and
the temperature parameter as a function of the radius
respectively. As it can be seen, the SIDM model produces viable
rotation curves compatible with the SPARC data. Also in Tables
\ref{collUGC05999}, \ref{NavaroUGC05999}, \ref{BuckertUGC05999}
and \ref{EinastoUGC05999} we present the optimization values for
the SIDM model, and the other DM profiles. Also in Table
\ref{EVALUATIONUGC05999} we present the overall evaluation of the
SIDM model for the galaxy at hand. The resulting phenomenology is
viable.
\begin{table}[h!]
  \begin{center}
    \caption{Collisional Dark Matter Optimization Values}
    \label{collUGC05999}
     \begin{tabular}{|r|r|}
     \hline
      \textbf{Parameter}   & \textbf{Optimization Values}
      \\  \hline
     $\delta_{\gamma} $ & 0.012
\\  \hline
$\gamma_0 $ & 1.073  \\ \hline $K_0$ ($M_{\odot} \,
\mathrm{Kpc}^{-3} \, (\mathrm{km/s})^{2}$)& 1500  \\ \hline
    \end{tabular}
  \end{center}
\end{table}
\begin{table}[h!]
  \begin{center}
    \caption{NFW  Optimization Values}
    \label{NavaroUGC05999}
     \begin{tabular}{|r|r|}
     \hline
      \textbf{Parameter}   & \textbf{Optimization Values}
      \\  \hline
   $\rho_s$   & $5\times 10^7$
\\  \hline
$r_s$&  3.62
\\  \hline
    \end{tabular}
  \end{center}
\end{table}
\begin{table}[h!]
  \begin{center}
    \caption{Burkert Optimization Values}
    \label{BuckertUGC05999}
     \begin{tabular}{|r|r|}
     \hline
      \textbf{Parameter}   & \textbf{Optimization Values}
      \\  \hline
     $\rho_0^B$  & $1\times 10^8$
\\  \hline
$r_0$&  3.85
\\  \hline
    \end{tabular}
  \end{center}
\end{table}
\begin{table}[h!]
  \begin{center}
    \caption{Einasto Optimization Values}
    \label{EinastoUGC05999}
    \begin{tabular}{|r|r|}
     \hline
      \textbf{Parameter}   & \textbf{Optimization Values}
      \\  \hline
     $\rho_e$  &$1\times 10^7$
\\  \hline
$r_e$ & 3.66
\\  \hline
$n_e$ & 0.5
\\  \hline
    \end{tabular}
  \end{center}
\end{table}
\begin{table}[h!]
\centering \caption{Physical assessment of collisional DM
parameters for UGC05999.}
\begin{tabular}{lcc}
\hline
Parameter & Value & Physical Verdict \\
\hline
$\gamma_0$ & $1.0001$ & Almost exactly isothermal \\
$\delta_\gamma$ & $1.2\times10^{-9}$ & Practically zero \\
$r_\gamma$ & $1.5~\mathrm{Kpc}$ & Plausible transition scale \\
$K_0$ & $4.2\times10^{3}$ & Moderate scale \\
$r_c$ & $0.5~\mathrm{Kpc}$ & Small core scale \\
$p$ & $0.01$ & Extremely shallow decline  \\
\hline
Overall & --- & Nearly isothermal \\
\hline
\end{tabular}
\label{EVALUATIONUGC05999}
\end{table}

\subsection{The Galaxy UGC06399 low-surface-brightness seem compatible}

\begin{figure}[h!]
\centering
\includegraphics[width=20pc]{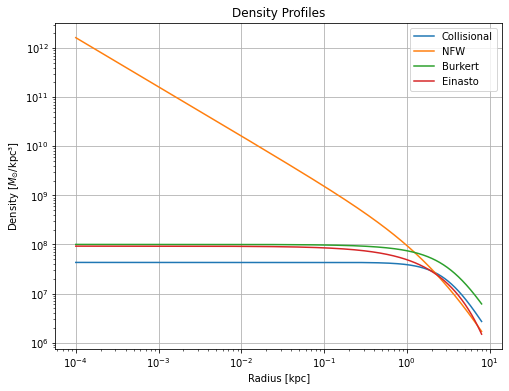}
\caption{The density of the collisional DM model (\ref{tanhmodel})
for the galaxy UGC06399, as a function of the radius.}
\label{UGC06399dens}
\end{figure}

\begin{figure}[h!]
\centering
\includegraphics[width=20pc]{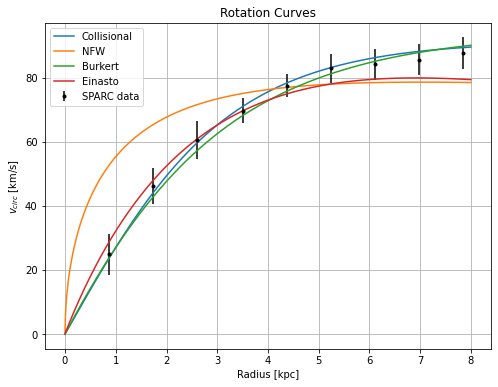}
\caption{The predicted rotation curves after using an optimization
for the collisional DM model (\ref{tanhmodel}), versus the SPARC
observational data for the galaxy UGC06399. We also plotted the
optimized curves for the NFW model, the Burkert model and the
Einasto model.} \label{UGC06399}
\end{figure}

\begin{figure}[h!]
\centering
\includegraphics[width=20pc]{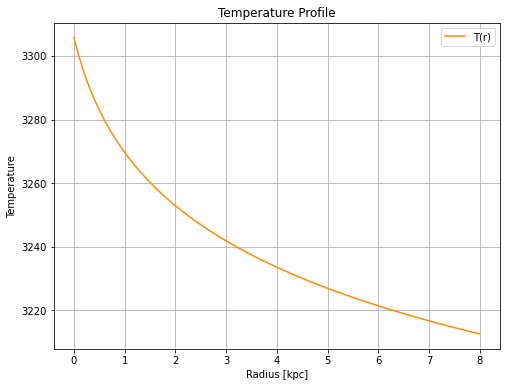}
\caption{The temperature as a function of the radius for the
collisional DM model (\ref{tanhmodel}) for the galaxy UGC06399.}
\label{UGC06399temp}
\end{figure}

For this galaxy, we shall choose $\rho_0=4.3\times
10^7$$M_{\odot}/\mathrm{Kpc}^{3}$. UGC6399 is a
low-surface-brightness spiral galaxy located in the constellation
Ursa Major. Its distance is approximately 10-20\,Mpc. In Figs.
\ref{UGC06399dens}, \ref{UGC06399} and \ref{UGC06399temp} we
present the density of the collisional DM model, the predicted
rotation curves after using an optimization for the collisional DM
model (\ref{tanhmodel}), versus the SPARC observational data and
the temperature parameter as a function of the radius
respectively. As it can be seen, the SIDM model produces viable
rotation curves compatible with the SPARC data. Also in Tables
\ref{collUGC06399}, \ref{NavaroUGC06399}, \ref{BuckertUGC06399}
and \ref{EinastoUGC06399} we present the optimization values for
the SIDM model, and the other DM profiles. Also in Table
\ref{EVALUATIONUGC06399} we present the overall evaluation of the
SIDM model for the galaxy at hand. The resulting phenomenology is
viable.
\begin{table}[h!]
  \begin{center}
    \caption{Collisional Dark Matter Optimization Values}
    \label{collUGC06399}
     \begin{tabular}{|r|r|}
     \hline
      \textbf{Parameter}   & \textbf{Optimization Values}
      \\  \hline
     $\delta_{\gamma} $ & 0.0000000012
\\  \hline
$\gamma_0 $ & 1.0001  \\ \hline $K_0$ ($M_{\odot} \,
\mathrm{Kpc}^{-3} \, (\mathrm{km/s})^{2}$)& 3300  \\ \hline
    \end{tabular}
  \end{center}
\end{table}
\begin{table}[h!]
  \begin{center}
    \caption{NFW  Optimization Values}
    \label{NavaroUGC06399}
     \begin{tabular}{|r|r|}
     \hline
      \textbf{Parameter}   & \textbf{Optimization Values}
      \\  \hline
   $\rho_s$   & $5\times 10^7$
\\  \hline
$r_s$&  3.25
\\  \hline
    \end{tabular}
  \end{center}
\end{table}
\begin{table}[h!]
  \begin{center}
    \caption{Burkert Optimization Values}
    \label{BuckertUGC06399}
     \begin{tabular}{|r|r|}
     \hline
      \textbf{Parameter}   & \textbf{Optimization Values}
      \\  \hline
     $\rho_0^B$  & $1\times 10^8$
\\  \hline
$r_0$&  3.86
\\  \hline
    \end{tabular}
  \end{center}
\end{table}
\begin{table}[h!]
  \begin{center}
    \caption{Einasto Optimization Values}
    \label{EinastoUGC06399}
    \begin{tabular}{|r|r|}
     \hline
      \textbf{Parameter}   & \textbf{Optimization Values}
      \\  \hline
     $\rho_e$  &$1\times 10^7$
\\  \hline
$r_e$ & 4.02
\\  \hline
$n_e$ & 0.9
\\  \hline
    \end{tabular}
  \end{center}
\end{table}
\begin{table}[h!]
\centering \caption{Physical assessment of collisional DM
parameters for UGC06399.}
\begin{tabular}{lcc}
\hline
Parameter & Value & Physical Verdict \\
\hline
$\gamma_0$ & $1.0001$ & Nearly isothermal\\
$\delta_\gamma$ & $1.2\times10^{-9}$ & Practically zero  \\
$r_\gamma$ & $1.5~\mathrm{Kpc}$ & Transition radius plausible but inactive \\
$K_0$ & $3.3\times10^{3}$ & Moderate scale \\
$r_c$ & $0.5~\mathrm{Kpc}$ & Small core \\
$p$ & $0.01$ & Very shallow decline \\
\hline
Overall & --- & Physically plausible \\
\hline
\end{tabular}
\label{EVALUATIONUGC06399}
\end{table}

\subsection{The Galaxy UGC06446 Marginal, Extended Marginal too}

\begin{figure}[h!]
\centering
\includegraphics[width=20pc]{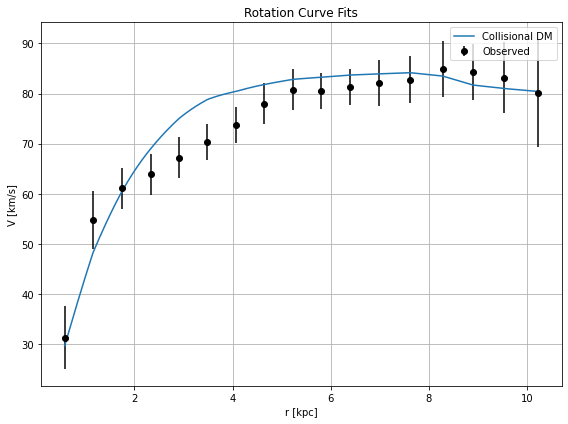}
\caption{The predicted rotation curves after using an optimization
for the collisional DM model (\ref{tanhmodel}), versus the
extended SPARC observational data for the galaxy UGC06446. The
model includes the rotation curves from all the components of the
galaxy, including gas and disk velocities, along with the
collisional DM model.} \label{extendedUGC06446}
\end{figure}

For this galaxy, we shall choose $\rho_0=1.4\times
10^8$$M_{\odot}/\mathrm{Kpc}^{3}$. UGC06446 is a nearby dwarf
irregular galaxy located at a distance of approximately \(D \sim
9.1\ \mathrm{Mpc}\). In Figs. \ref{UGC05764dens}, \ref{UGC05764}
and \ref{UGC05764temp} we present the density of the collisional
DM model, the predicted rotation curves after using an
optimization for the collisional DM model (\ref{tanhmodel}),
versus the SPARC observational data and the temperature parameter
as a function of the radius respectively. As it can be seen, the
SIDM model produces marginally viable rotation curves compatible
with the SPARC data. Also in Tables \ref{collCamB},
\ref{NavaroUGC05764}, \ref{BuckertUGC05764} and
\ref{EinastoUGC05764} we present the optimization values for the
SIDM model, and the other DM profiles. Also in Table
\ref{EVALUATIONUGC05764} we present the overall evaluation of the
SIDM model for the galaxy at hand. The resulting phenomenology is
marginally viable.
\begin{figure}[h!]
\centering
\includegraphics[width=20pc]{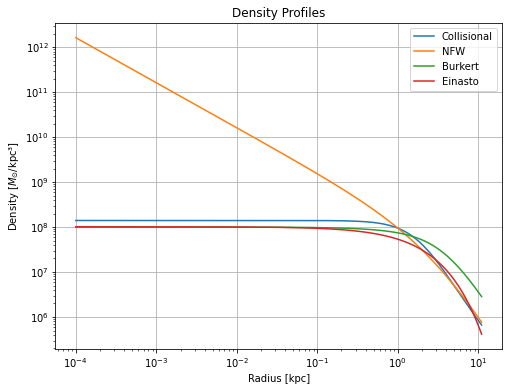}
\caption{The density of the collisional DM model (\ref{tanhmodel})
for the galaxy UGC06446, as a function of the radius.}
\label{UGC06446dens}
\end{figure}
\begin{figure}[h!]
\centering
\includegraphics[width=20pc]{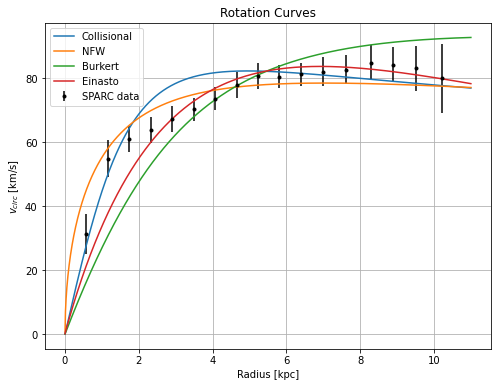}
\caption{The predicted rotation curves after using an optimization
for the collisional DM model (\ref{tanhmodel}), versus the SPARC
observational data for the galaxy UGC06446. We also plotted the
optimized curves for the NFW model, the Burkert model and the
Einasto model.} \label{UGC06446}
\end{figure}
\begin{table}[h!]
  \begin{center}
    \caption{Collisional Dark Matter Optimization Values}
    \label{collUGC06446}
     \begin{tabular}{|r|r|}
     \hline
      \textbf{Parameter}   & \textbf{Optimization Values}
      \\  \hline
     $\delta_{\gamma} $ & 0.0000000012
\\  \hline
$\gamma_0 $ & 1.0001 \\ \hline $K_0$ ($M_{\odot} \,
\mathrm{Kpc}^{-3} \, (\mathrm{km/s})^{2}$)& 1900 \\ \hline
    \end{tabular}
  \end{center}
\end{table}
\begin{table}[h!]
  \begin{center}
    \caption{NFW  Optimization Values}
    \label{NavaroUGC06446}
     \begin{tabular}{|r|r|}
     \hline
      \textbf{Parameter}   & \textbf{Optimization Values}
      \\  \hline
   $\rho_s$   & $5\times 10^7$
\\  \hline
$r_s$& 3.25
\\  \hline
    \end{tabular}
  \end{center}
\end{table}
\begin{figure}[h!]
\centering
\includegraphics[width=20pc]{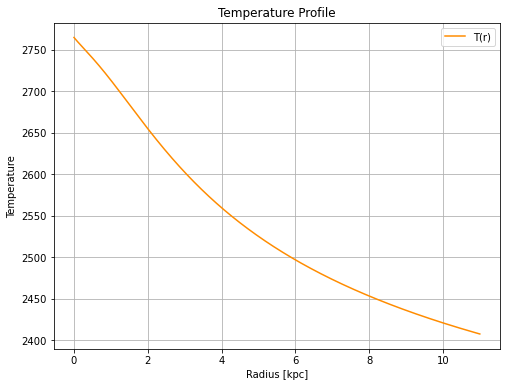}
\caption{The temperature as a function of the radius for the
collisional DM model (\ref{tanhmodel}) for the galaxy UGC06446.}
\label{UGC06446temp}
\end{figure}
\begin{table}[h!]
  \begin{center}
    \caption{Burkert Optimization Values}
    \label{BuckertUGC06446}
     \begin{tabular}{|r|r|}
     \hline
      \textbf{Parameter}   & \textbf{Optimization Values}
      \\  \hline
     $\rho_0^B$  & $1\times 10^8$
\\  \hline
$r_0$&  3.86
\\  \hline
    \end{tabular}
  \end{center}
\end{table}
\begin{table}[h!]
  \begin{center}
    \caption{Einasto Optimization Values}
    \label{EinastoUGC06446}
    \begin{tabular}{|r|r|}
     \hline
      \textbf{Parameter}   & \textbf{Optimization Values}
      \\  \hline
     $\rho_e$  &$1\times 10^7$
\\  \hline
$r_e$ & 4.02
\\  \hline
$n_e$ & 0.9
\\  \hline
    \end{tabular}
  \end{center}
\end{table}
\begin{table}[h!]
\centering \caption{Physical assessment of collisional DM
parameters (UGC06446).}
\begin{tabular}{lcc}
\hline
Parameter & Value & Physical Verdict \\
\hline
$\gamma_0$ & $1.0001$ & Nearly isothermal \\
$\delta_\gamma$ & $0.0000000012$ & Negligible variation \\
$r_\gamma$ & $1.5\ \mathrm{Kpc}$ & Reasonable transition scale \\
$K_0$ & $1.9\times10^{3}$ (km$^2$/s$^2$) & Enough pressure support \\
$r_c$ & $0.5\ \mathrm{Kpc}$ & Small core radius \\
$p$ & $0.01$ & Almost constant $K(r)$ \\
\hline
Overall &-& Physically plausible for a low-mass system \\
\hline
\end{tabular}
\label{EVALUATIONUGC06446}
\end{table}
Now the extended picture including the rotation velocity from the
other components of the galaxy, such as the disk and gas, makes
the collisional DM model viable for this galaxy. In Fig.
\ref{extendedUGC06446} we present the combined rotation curves
including the other components of the galaxy along with the
collisional matter. As it can be seen, the extended collisional DM
model is marginally viable.
Also in Table \ref{evaluationextendedUGC06446} we present the
values of the free parameters of the collisional DM model for
which the maximum compatibility with the SPARC data comes for the
galaxy UGC06446.
\begin{table}[h!]
\centering \caption{Physical assessment of Extended collisional DM
parameters for galaxy UGC06446.}
\begin{tabular}{lcc}
\hline
Parameter & Value & Physical Verdict \\
\hline
$\gamma_0$ & 1.03555248 & Very close to isothermal \\
$\delta_\gamma$ & 0.0007364302 & Extremely small variation \\
$K_0$ & 1200 & Lower entropy normalization than the 3000 benchmark\\
$ml_{disk}$ & 1.00000000 & At the upper plausible limit\\
$ml_{bulge}$ & 0.00000000 & Negligible bulge contribution \\
\hline
Overall &-& Physically plausible but borderline\\
\hline
\end{tabular}
\label{evaluationextendedUGC06446}
\end{table}

\subsection{The Galaxy UGC06614 Non-viable, Extended non-viable too}

\begin{figure}[h!]
\centering
\includegraphics[width=20pc]{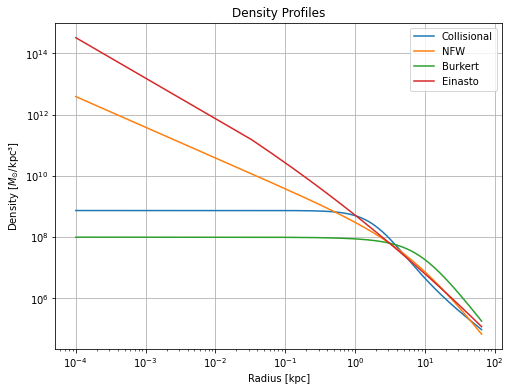}
\caption{The density of the collisional DM model (\ref{tanhmodel})
for the galaxy UGC06614, as a function of the radius.}
\label{UGC06614dens}
\end{figure}
\begin{figure}[h!]
\centering
\includegraphics[width=20pc]{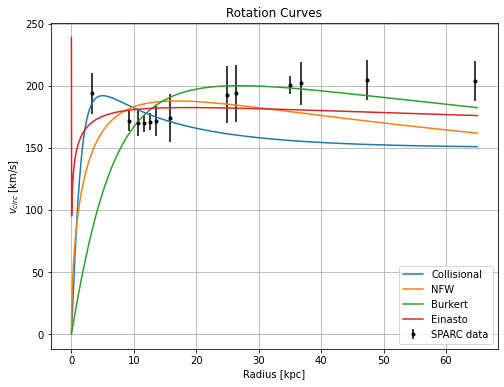}
\caption{The predicted rotation curves after using an optimization
for the collisional DM model (\ref{tanhmodel}), versus the SPARC
observational data for the galaxy UGC06614. We also plotted the
optimized curves for the NFW model, the Burkert model and the
Einasto model.} \label{UGC06614}
\end{figure}
\begin{figure}[h!]
\centering
\includegraphics[width=20pc]{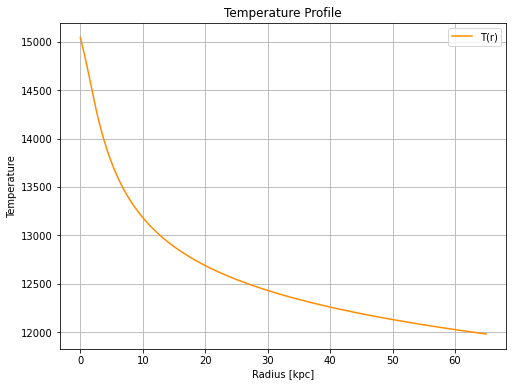}
\caption{The temperature as a function of the radius for the
collisional DM model (\ref{tanhmodel}) for the galaxy UGC06614.}
\label{UGC06614temp}
\end{figure}
\begin{figure}[h!]
\centering
\includegraphics[width=20pc]{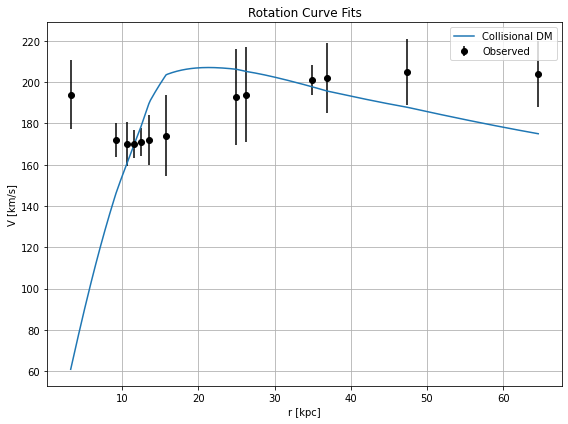}
\caption{The predicted rotation curves after using an optimization
for the collisional DM model (\ref{tanhmodel}), versus the
extended SPARC observational data for the galaxy UGC06614. The
model includes the rotation curves from all the components of the
galaxy, including gas and disk velocities, along with the
collisional DM model.} \label{extendedUGC06614}
\end{figure}

For this galaxy, we shall choose $\rho_0=7.4\times
10^8$$M_{\odot}/\mathrm{Kpc}^{3}$. UGC06614  is classified as a
giant low-surface-brightness spiral galaxy at a distance of order
\(D \sim 98\text{-}100\ \mathrm{Mpc}\). In Figs.
\ref{UGC06614dens}, \ref{UGC06614} and \ref{UGC06614temp} we
present the density of the collisional DM model, the predicted
rotation curves after using an optimization for the collisional DM
model (\ref{tanhmodel}), versus the SPARC observational data and
the temperature parameter as a function of the radius
respectively. As it can be seen, the SIDM model produces
non-viable rotation curves incompatible with the SPARC data. Also
in Tables \ref{collUGC06614}, \ref{NavaroUGC06614},
\ref{BuckertUGC06614} and \ref{EinastoUGC06614} we present the
optimization values for the SIDM model, and the other DM profiles.
Also in Table \ref{EVALUATIONUGC06614} we present the overall
evaluation of the SIDM model for the galaxy at hand. The resulting
phenomenology is non-viable.
\begin{table}[h!]
  \begin{center}
    \caption{Collisional Dark Matter Optimization Values}
    \label{collUGC06614}
     \begin{tabular}{|r|r|}
     \hline
      \textbf{Parameter}   & \textbf{Optimization Values}
      \\  \hline
     $\delta_{\gamma} $ & 0.0000000012
\\  \hline
$\gamma_0 $ & 1.0001  \\ \hline $K_0$ ($M_{\odot} \,
\mathrm{Kpc}^{-3} \, (\mathrm{km/s})^{2}$)& 10000  \\ \hline
    \end{tabular}
  \end{center}
\end{table}
\begin{table}[h!]
  \begin{center}
    \caption{NFW  Optimization Values}
    \label{NavaroUGC06614}
     \begin{tabular}{|r|r|}
     \hline
      \textbf{Parameter}   & \textbf{Optimization Values}
      \\  \hline
   $\rho_s$   & $5\times 10^7$
\\  \hline
$r_s$&  7.77
\\  \hline
    \end{tabular}
  \end{center}
\end{table}
\begin{table}[h!]
  \begin{center}
    \caption{Burkert Optimization Values}
    \label{BuckertUGC06614}
     \begin{tabular}{|r|r|}
     \hline
      \textbf{Parameter}   & \textbf{Optimization Values}
      \\  \hline
     $\rho_0^B$  & $1\times 10^8$
\\  \hline
$r_0$&  8.30
\\  \hline
    \end{tabular}
  \end{center}
\end{table}
\begin{table}[h!]
  \begin{center}
    \caption{Einasto Optimization Values}
    \label{EinastoUGC06614}
    \begin{tabular}{|r|r|}
     \hline
      \textbf{Parameter}   & \textbf{Optimization Values}
      \\  \hline
     $\rho_e$  &$1\times 10^7$
\\  \hline
$r_e$ & 7.64
\\  \hline
$n_e$ & 0.05
\\  \hline
    \end{tabular}
  \end{center}
\end{table}
\begin{table}[h!]
\centering \caption{Physical assessment of collisional DM
parameters (UGC06614).}
\begin{tabular}{lcc}
\hline
Parameter & Value & Physical Verdict \\
\hline
$\gamma_0$ & $1.0001$ & Nearly isothermal \\
$\delta_\gamma$ & $0.0000000012$ & Negligible variation \\
$r_\gamma$ & $1.5\ \mathrm{Kpc}$ & Reasonable transition radius  \\
$K_0$ & $1.0\times10^{4}$ (km$^2$/s$^2$) & Enough pressure support \\
$r_c$ & $0.5\ \mathrm{Kpc}$ & Small core radius \\
$p$ & $0.01$ & Almost constant $K(r)$ \\
\hline
Overall &-& Model behaves effectively isothermally \\
\hline
\end{tabular}
\label{EVALUATIONUGC06614}
\end{table}
Now the extended picture including the rotation velocity from the
other components of the galaxy, such as the disk and gas, makes
the collisional DM model viable for this galaxy. In Fig.
\ref{extendedUGC06614} we present the combined rotation curves
including the other components of the galaxy along with the
collisional matter. As it can be seen, the extended collisional DM
model is non-viable.
Also in Table \ref{evaluationextendedUGC06614} we present the
values of the free parameters of the collisional DM model for
which the maximum compatibility with the SPARC data comes for the
galaxy UGC06614.
\begin{table}[h!]
\centering \caption{Physical assessment of Extended collisional DM
parameters for galaxy UGC06614.}
\begin{tabular}{lcc}
\hline
Parameter & Value & Physical Verdict \\
\hline
$\gamma_0$ & 1.1 & Near-isothermal core \\
$\delta_\gamma$ & 0.0037 & Very small gradient\\
$K_0$ & 3000 & Moderate entropy  \\
$ml_{\rm disk}$ & 0.90 & Physically reasonable \\
$ml_{\rm bulge}$ & 0.0000268 & Negligible bulge mass, consistent with disk-dominated morphology \\
\hline
Overall &-& Viable parameter set \\
\hline
\end{tabular}
\label{evaluationextendedUGC06614}
\end{table}

\subsection{The Galaxy UGC06628}

\begin{figure}[h!]
\centering
\includegraphics[width=20pc]{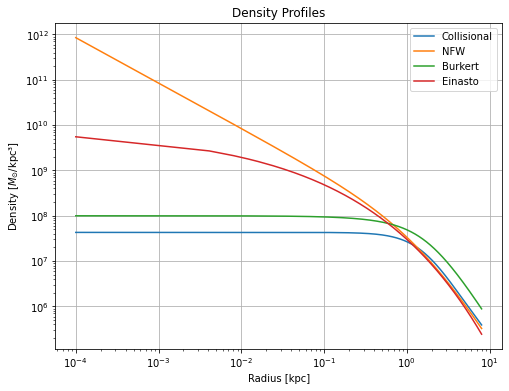}
\caption{The density of the collisional DM model (\ref{tanhmodel})
for the galaxy UGC06628, as a function of the radius.}
\label{UGC06628dens}
\end{figure}

\begin{figure}[h!]
\centering
\includegraphics[width=20pc]{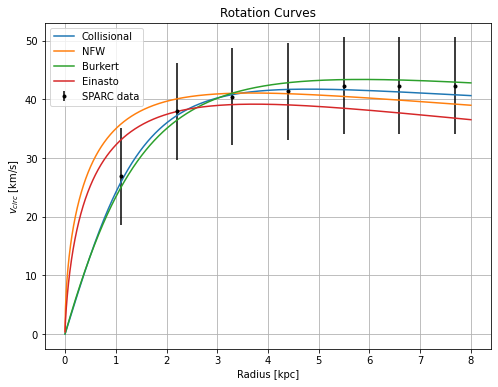}
\caption{The predicted rotation curves after using an optimization
for the collisional DM model (\ref{tanhmodel}), versus the SPARC
observational data for the galaxy UGC06628. We also plotted the
optimized curves for the NFW model, the Burkert model and the
Einasto model.} \label{UGC06628}
\end{figure}

\begin{figure}[h!]
\centering
\includegraphics[width=20pc]{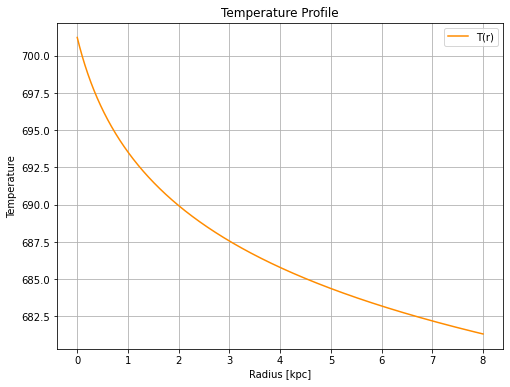}
\caption{The temperature as a function of the radius for the
collisional DM model (\ref{tanhmodel}) for the galaxy UGC06628.}
\label{UGC06628temp}
\end{figure}

For this galaxy, we shall choose $\rho_0=4.3\times
10^7$$M_{\odot}/\mathrm{Kpc}^{3}$. UGC06628 is a late-type,
gas-rich dwarf/spiral galaxy located at an approximate distance
\(D\sim 34.7\ \mathrm{Mpc}\). In Figs. \ref{UGC06628dens},
\ref{UGC06628} and \ref{UGC06628temp} we present the density of
the collisional DM model, the predicted rotation curves after
using an optimization for the collisional DM model
(\ref{tanhmodel}), versus the SPARC observational data and the
temperature parameter as a function of the radius respectively. As
it can be seen, the SIDM model produces viable rotation curves
compatible with the SPARC data. Also in Tables \ref{collUGC06628},
\ref{NavaroUGC06628}, \ref{BuckertUGC06628} and
\ref{EinastoUGC06628} we present the optimization values for the
SIDM model, and the other DM profiles. Also in Table
\ref{EVALUATIONUGC06628} we present the overall evaluation of the
SIDM model for the galaxy at hand. The resulting phenomenology is
viable.
\begin{table}[h!]
  \begin{center}
    \caption{Collisional Dark Matter Optimization Values}
    \label{collUGC06628}
     \begin{tabular}{|r|r|}
     \hline
      \textbf{Parameter}   & \textbf{Optimization Values}
      \\  \hline
     $\delta_{\gamma} $ &  0.0000000012
\\  \hline
$\gamma_0 $ & 1.0001  \\ \hline $K_0$ ($M_{\odot} \,
\mathrm{Kpc}^{-3} \, (\mathrm{km/s})^{2}$)& 700 \\ \hline
    \end{tabular}
  \end{center}
\end{table}
\begin{table}[h!]
  \begin{center}
    \caption{NFW  Optimization Values}
    \label{NavaroUGC06628}
     \begin{tabular}{|r|r|}
     \hline
      \textbf{Parameter}   & \textbf{Optimization Values}
      \\  \hline
   $\rho_s$   & $5\times 10^7$
\\  \hline
$r_s$&  1.70
\\  \hline
    \end{tabular}
  \end{center}
\end{table}
\begin{table}[h!]
  \begin{center}
    \caption{Burkert Optimization Values}
    \label{BuckertUGC06628}
     \begin{tabular}{|r|r|}
     \hline
      \textbf{Parameter}   & \textbf{Optimization Values}
      \\  \hline
     $\rho_0^B$  & $1\times 10^8$
\\  \hline
$r_0$& 1.80
\\  \hline
    \end{tabular}
  \end{center}
\end{table}
\begin{table}[h!]
  \begin{center}
    \caption{Einasto Optimization Values}
    \label{EinastoUGC06628}
    \begin{tabular}{|r|r|}
     \hline
      \textbf{Parameter}   & \textbf{Optimization Values}
      \\  \hline
     $\rho_e$  &$1\times 10^7$
\\  \hline
$r_e$ & 1.83
\\  \hline
$n_e$ & 0.3
\\  \hline
    \end{tabular}
  \end{center}
\end{table}
\begin{table}[h!]
\centering \caption{Physical assessment of collisional DM
parameters for UGC06628.}
\begin{tabular}{lcc}
\hline
Parameter & Value & Physical Verdict \\
\hline
$\gamma_0$ & $1.0001$ & Nearly isothermal \\
$\delta_\gamma$ & $1.2\times10^{-9}$ & Practically zero \\
$r_\gamma$ & $1.5~\mathrm{Kpc}$ & Transition radius plausible \\
$K_0$ & $700$ & Low pressure support consistent with dwarf galaxy halo \\
$r_c$ & $0.5~\mathrm{Kpc}$ & Small core; $K(r)$ effectively constant \\
$p$ & $0.01$ & Very shallow decline; nearly constant $K(r)$ \\
\hline
Overall & --- & Physically plausible \\
\hline
\end{tabular}
\label{EVALUATIONUGC06628}
\end{table}

\subsection{The Galaxy UGC06667}

\begin{figure}[h!]
\centering
\includegraphics[width=20pc]{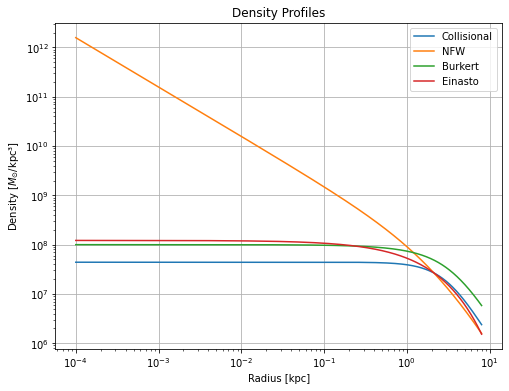}
\caption{The density of the collisional DM model (\ref{tanhmodel})
for the galaxy UGC06667, as a function of the radius.}
\label{UGC06667dens}
\end{figure}
\begin{figure}[h!]
\centering
\includegraphics[width=20pc]{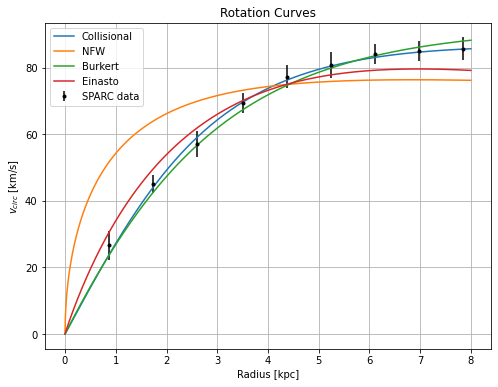}
\caption{The predicted rotation curves after using an optimization
for the collisional DM model (\ref{tanhmodel}), versus the SPARC
observational data for the galaxy UGC06667. We also plotted the
optimized curves for the NFW model, the Burkert model and the
Einasto model.} \label{UGC06667}
\end{figure}
\begin{figure}[h!]
\centering
\includegraphics[width=20pc]{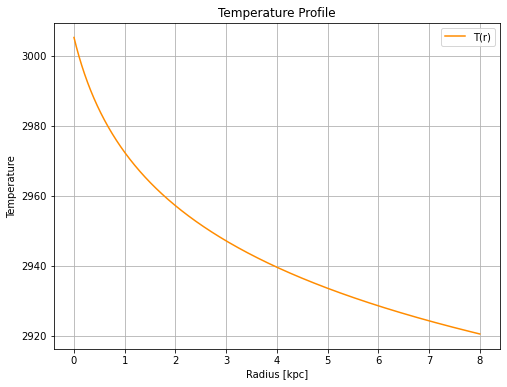}
\caption{The temperature as a function of the radius for the
collisional DM model (\ref{tanhmodel}) for the galaxy UGC06667.}
\label{UGC06667temp}
\end{figure}

For this galaxy, we shall choose $\rho_0=4.4\times
10^7$$M_{\odot}/\mathrm{Kpc}^{3}$. UGC06667 is a late-type,
gas-rich disk galaxy included in the SPARC sample; it is located
at a distance of approximately \(D \sim 23.8\pm1.3\ \mathrm{Mpc}\)
and observed nearly edge-on. In Figs. \ref{UGC06667dens},
\ref{UGC06667} and \ref{UGC06667temp} we present the density of
the collisional DM model, the predicted rotation curves after
using an optimization for the collisional DM model
(\ref{tanhmodel}), versus the SPARC observational data and the
temperature parameter as a function of the radius respectively. As
it can be seen, the SIDM model produces viable rotation curves
compatible with the SPARC data. Also in Tables \ref{collUGC06667},
\ref{NavaroUGC06667}, \ref{BuckertUGC06667} and
\ref{EinastoUGC06667} we present the optimization values for the
SIDM model, and the other DM profiles. Also in Table
\ref{EVALUATIONUGC06667} we present the overall evaluation of the
SIDM model for the galaxy at hand. The resulting phenomenology is
viable.
\begin{table}[h!]
  \begin{center}
    \caption{Collisional Dark Matter Optimization Values}
    \label{collUGC06667}
     \begin{tabular}{|r|r|}
     \hline
      \textbf{Parameter}   & \textbf{Optimization Values}
      \\  \hline
     $\delta_{\gamma} $ & 0.0000000012
\\  \hline
$\gamma_0 $ & 1.0001  \\ \hline $K_0$ ($M_{\odot} \,
\mathrm{Kpc}^{-3} \, (\mathrm{km/s})^{2}$)& 3000  \\ \hline
    \end{tabular}
  \end{center}
\end{table}
\begin{table}[h!]
  \begin{center}
    \caption{NFW  Optimization Values}
    \label{NavaroUGC06667}
     \begin{tabular}{|r|r|}
     \hline
      \textbf{Parameter}   & \textbf{Optimization Values}
      \\  \hline
   $\rho_s$   & $5\times 10^7$
\\  \hline
$r_s$&  3.16
\\  \hline
    \end{tabular}
  \end{center}
\end{table}
\begin{table}[h!]
  \begin{center}
    \caption{Burkert Optimization Values}
    \label{BuckertUGC06667}
     \begin{tabular}{|r|r|}
     \hline
      \textbf{Parameter}   & \textbf{Optimization Values}
      \\  \hline
     $\rho_0^B$  & $1\times 10^8$
\\  \hline
$r_0$&  3.77
\\  \hline
    \end{tabular}
  \end{center}
\end{table}
\begin{table}[h!]
  \begin{center}
    \caption{Einasto Optimization Values}
    \label{EinastoUGC06667}
    \begin{tabular}{|r|r|}
     \hline
      \textbf{Parameter}   & \textbf{Optimization Values}
      \\  \hline
     $\rho_e$  &$1\times 10^7$
\\  \hline
$r_e$ & 3.97
\\  \hline
$n_e$ & 0.8
\\  \hline
    \end{tabular}
  \end{center}
\end{table}
\begin{table}[h!]
\centering \caption{Physical assessment of collisional DM
parameters for UGC06667.}
\begin{tabular}{lcc}
\hline
Parameter & Value & Physical Verdict \\
\hline
$\gamma_0$ & $1.0001$ & Nearly isothermal \\
$\delta_\gamma$ & $1.2\times10^{-9}$ & Practically zero  \\
$r_\gamma$ & $1.5~\mathrm{Kpc}$ & Transition radius plausible but inactive \\
$K_0$ & $3000$ & Moderately high pressure support \\
$r_c$ & $0.5~\mathrm{Kpc}$ & Small core; $K(r)$ effectively constant \\
$p$ & $0.01$ & Very shallow decline; nearly constant $K(r)$ \\
\hline
Overall & --- & Physically plausible; inner halo nearly isothermal \\
\hline
\end{tabular}
\label{EVALUATIONUGC06667}
\end{table}

\subsection{The Galaxy UGC06786 Non-viable}

\begin{figure}[h!]
\centering
\includegraphics[width=20pc]{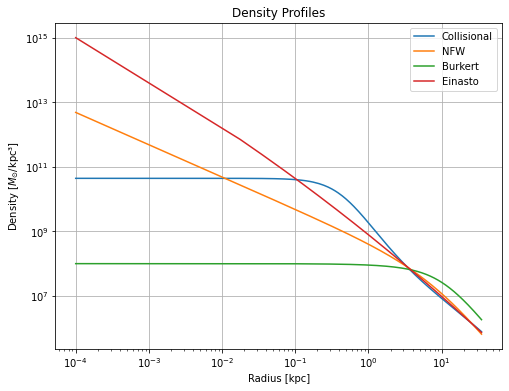}
\caption{The density of the collisional DM model (\ref{tanhmodel})
for the galaxy UGC06786, as a function of the radius.}
\label{UGC06786dens}
\end{figure}
\begin{figure}[h!]
\centering
\includegraphics[width=20pc]{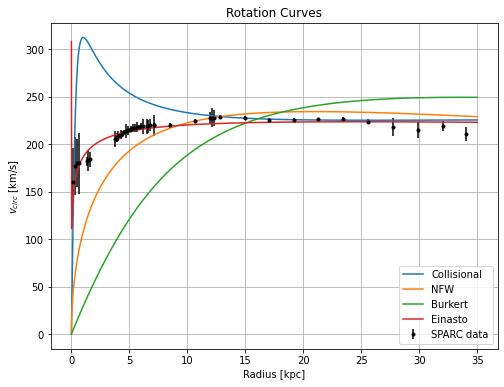}
\caption{The predicted rotation curves after using an optimization
for the collisional DM model (\ref{tanhmodel}), versus the SPARC
observational data for the galaxy UGC06786. We also plotted the
optimized curves for the NFW model, the Burkert model and the
Einasto model.} \label{UGC06786}
\end{figure}
\begin{figure}[h!]
\centering
\includegraphics[width=20pc]{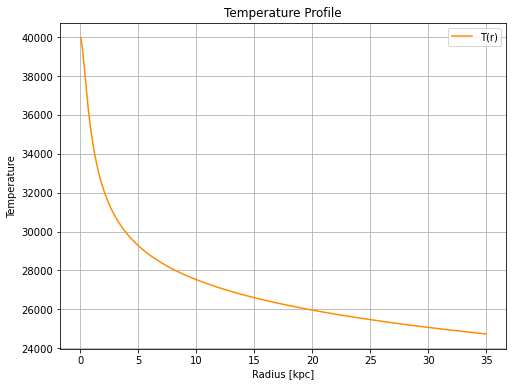}
\caption{The temperature as a function of the radius for the
collisional DM model (\ref{tanhmodel}) for the galaxy UGC06786.}
\label{UGC06786temp}
\end{figure}
\begin{figure}[h!]
\centering
\includegraphics[width=20pc]{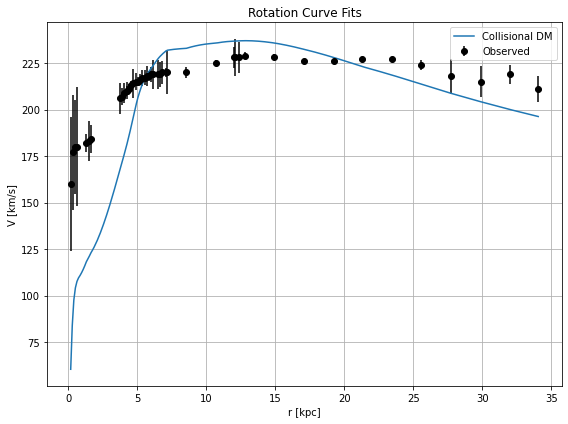}
\caption{The predicted rotation curves after using an optimization
for the collisional DM model (\ref{tanhmodel}), versus the
extended SPARC observational data for the galaxy UGC06786. The
model includes the rotation curves from all the components of the
galaxy, including gas and disk velocities, along with the
collisional DM model.} \label{extendedUGC06786}
\end{figure}

For this galaxy, we shall choose $\rho_0=4.4\times
10^{10}$$M_{\odot}/\mathrm{Kpc}^{3}$. The galaxy UGC06786 is
identified with the interacting barred spiral NGC6786 located in
the constellation Draco at a Hubble distance of about $110\
\mathrm{Mpc}$. In Figs. \ref{UGC06786dens}, \ref{UGC06786} and
\ref{UGC06786temp} we present the density of the collisional DM
model, the predicted rotation curves after using an optimization
for the collisional DM model (\ref{tanhmodel}), versus the SPARC
observational data and the temperature parameter as a function of
the radius respectively. As it can be seen, the SIDM model
produces non-viable rotation curves incompatible with the SPARC
data. Also in Tables \ref{collUGC06786}, \ref{NavaroUGC06786},
\ref{BuckertUGC06786} and \ref{EinastoUGC06786} we present the
optimization values for the SIDM model, and the other DM profiles.
Also in Table \ref{EVALUATIONUGC06786} we present the overall
evaluation of the SIDM model for the galaxy at hand. The resulting
phenomenology is non-viable.
\begin{table}[h!]
  \begin{center}
    \caption{Collisional Dark Matter Optimization Values}
    \label{collUGC06786}
     \begin{tabular}{|r|r|}
     \hline
      \textbf{Parameter}   & \textbf{Optimization Values}
      \\  \hline
     $\delta_{\gamma} $ & 0.0000000012
\\  \hline
$\gamma_0 $ & 1.0001  \\ \hline $K_0$ ($M_{\odot} \,
\mathrm{Kpc}^{-3} \, (\mathrm{km/s})^{2}$)& 15000 \\ \hline
    \end{tabular}
  \end{center}
\end{table}
\begin{table}[h!]
  \begin{center}
    \caption{NFW  Optimization Values}
    \label{NavaroUGC06786}
     \begin{tabular}{|r|r|}
     \hline
      \textbf{Parameter}   & \textbf{Optimization Values}
      \\  \hline
   $\rho_s$   & $5\times 10^7$
\\  \hline
$r_s$& 9.70
\\  \hline
    \end{tabular}
  \end{center}
\end{table}
\begin{table}[h!]
  \begin{center}
    \caption{Burkert Optimization Values}
    \label{BuckertUGC06786}
     \begin{tabular}{|r|r|}
     \hline
      \textbf{Parameter}   & \textbf{Optimization Values}
      \\  \hline
     $\rho_0^B$  & $1\times 10^8$
\\  \hline
$r_0$&  10.35
\\  \hline
    \end{tabular}
  \end{center}
\end{table}
\begin{table}[h!]
  \begin{center}
    \caption{Einasto Optimization Values}
    \label{EinastoUGC06786}
    \begin{tabular}{|r|r|}
     \hline
      \textbf{Parameter}   & \textbf{Optimization Values}
      \\  \hline
     $\rho_e$  &$1\times 10^7$
\\  \hline
$r_e$ & 9.79
\\  \hline
$n_e$ &  0.04
\\  \hline
    \end{tabular}
  \end{center}
\end{table}
\begin{table}[h!]
\centering \caption{Physical assessment of collisional DM
parameters (UGC06786).}
\begin{tabular}{lcc}
\hline
Parameter & Value & Physical Verdict \\
\hline
$\gamma_0$ & $1.0001$ & Nearly isothermal \\
$\delta_\gamma$ & $1.2\times10^{-9}$ & Practically zero  \\
$r_\gamma$ & $1.5\ \mathrm{Kpc}$ & Reasonable location for a transition\\
$K_0$ ($M_{\odot}\,\mathrm{Kpc}^{-3}\,(\mathrm{km/s})^{2}$) & $1.5\times10^{4}$ & Large entropy/pressure scale given $\rho_0$ \\
$r_c$ & $0.5\ \mathrm{Kpc}$ & Small core scale \\
$p$ & $0.01$ & Very shallow decline -  \\
\hline
Overall & -- & Implementation is algebraically consistent \\
\hline
\end{tabular}
\label{EVALUATIONUGC06786}
\end{table}
Now the extended picture including the rotation velocity from the
other components of the galaxy, such as the disk and gas, makes
the collisional DM model viable for this galaxy. In Fig.
\ref{extendedUGC06786} we present the combined rotation curves
including the other components of the galaxy along with the
collisional matter. As it can be seen, the extended collisional DM
model is non-viable.
Also in Table \ref{evaluationextendedUGC06786} we present the
values of the free parameters of the collisional DM model for
which the maximum compatibility with the SPARC data comes for the
galaxy UGC06786.
\begin{table}[h!]
\centering \caption{Physical assessment of Extended collisional DM
parameters for galaxy UGC06786.}
\begin{tabular}{lcc}
\hline
Parameter & Value & Physical Verdict \\
\hline
$\gamma_0$ & 1.15433412 & Slightly above isothermal \\
$\delta_\gamma$ & 0.05939432 & Small but non-negligible radial variation \\
$K_0$ & 3000 & Moderate entropy   \\
$ml_{\rm disk}$ & 1.00 & Maximal disk assumption \\
$ml_{\rm bulge}$ & 0.50 & Significant bulge mass fraction\\
\hline
Overall &-& Physically plausible \\
\hline
\end{tabular}
\label{evaluationextendedUGC06786}
\end{table}

\subsection{The Galaxy UGC06787 Non-viable}

\begin{figure}[h!]
\centering
\includegraphics[width=20pc]{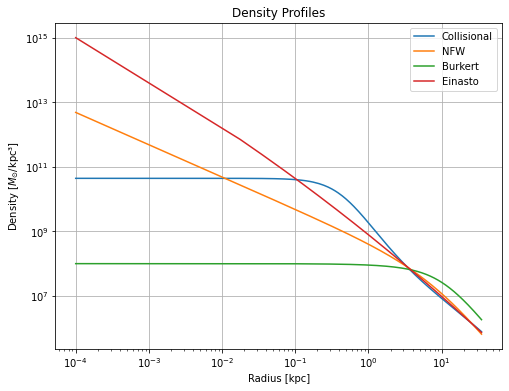}
\caption{The density of the collisional DM model (\ref{tanhmodel})
for the galaxy UGC06787, as a function of the radius.}
\label{UGC06787dens}
\end{figure}

\begin{figure}[h!]
\centering
\includegraphics[width=20pc]{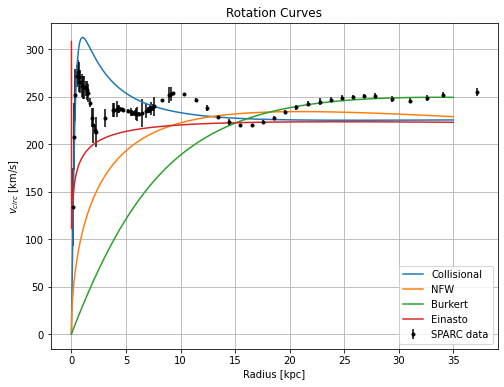}
\caption{The predicted rotation curves after using an optimization
for the collisional DM model (\ref{tanhmodel}), versus the SPARC
observational data for the galaxy UGC06787. We also plotted the
optimized curves for the NFW model, the Burkert model and the
Einasto model.} \label{UGC06787}
\end{figure}

\begin{figure}[h!]
\centering
\includegraphics[width=20pc]{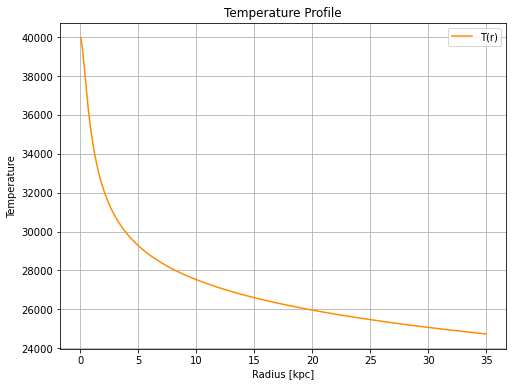}
\caption{The temperature as a function of the radius for the
collisional DM model (\ref{tanhmodel}) for the galaxy UGC06787.}
\label{UGC06787temp}
\end{figure}

\begin{figure}[h!]
\centering
\includegraphics[width=20pc]{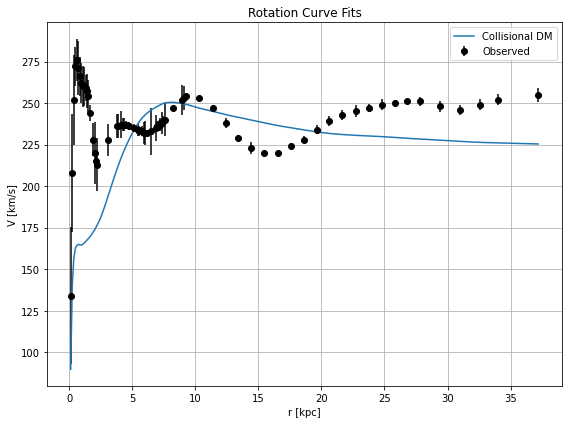}
\caption{The predicted rotation curves after using an optimization
for the collisional DM model (\ref{tanhmodel}), versus the
extended SPARC observational data for the galaxy UGC06787. The
model includes the rotation curves from all the components of the
galaxy, including gas and disk velocities, along with the
collisional DM model.} \label{extendedUGC06787}
\end{figure}

For this galaxy, we shall choose $\rho_0=4.4\times
10^{10}$$M_{\odot}/\mathrm{Kpc}^{3}$. The galaxy UGC06787 is an
intermediate spiral (type SAB(s)b), located in the constellation
Draco at a distance of roughly $164\ \mathrm{Mpc}$. In Figs.
\ref{UGC06787dens}, \ref{UGC06787} and \ref{UGC06787temp} we
present the density of the collisional DM model, the predicted
rotation curves after using an optimization for the collisional DM
model (\ref{tanhmodel}), versus the SPARC observational data and
the temperature parameter as a function of the radius
respectively. As it can be seen, the SIDM model produces
non-viable rotation curves incompatible with the SPARC data. Also
in Tables \ref{collUGC06787}, \ref{NavaroUGC06787},
\ref{BuckertUGC06787} and \ref{EinastoUGC06787} we present the
optimization values for the SIDM model, and the other DM profiles.
Also in Table \ref{EVALUATIONUGC06787} we present the overall
evaluation of the SIDM model for the galaxy at hand. The resulting
phenomenology is non-viable.
\begin{table}[h!]
  \begin{center}
    \caption{Collisional Dark Matter Optimization Values}
    \label{collUGC06787}
     \begin{tabular}{|r|r|}
     \hline
      \textbf{Parameter}   & \textbf{Optimization Values}
      \\  \hline
     $\delta_{\gamma} $ & 0.0000000012
\\  \hline
$\gamma_0 $ & 1.0001  \\ \hline $K_0$ ($M_{\odot} \,
\mathrm{Kpc}^{-3} \, (\mathrm{km/s})^{2}$)& 15000  \\ \hline
    \end{tabular}
  \end{center}
\end{table}
\begin{table}[h!]
  \begin{center}
    \caption{NFW  Optimization Values}
    \label{NavaroUGC06787}
     \begin{tabular}{|r|r|}
     \hline
      \textbf{Parameter}   & \textbf{Optimization Values}
      \\  \hline
   $\rho_s$   & $5\times 10^7$
\\  \hline
$r_s$&  9.70
\\  \hline
    \end{tabular}
  \end{center}
\end{table}
\begin{table}[h!]
  \begin{center}
    \caption{Burkert Optimization Values}
    \label{BuckertUGC06787}
     \begin{tabular}{|r|r|}
     \hline
      \textbf{Parameter}   & \textbf{Optimization Values}
      \\  \hline
     $\rho_0^B$  & $1\times 10^8$
\\  \hline
$r_0$&  10.35
\\  \hline
    \end{tabular}
  \end{center}
\end{table}
\begin{table}[h!]
  \begin{center}
    \caption{Einasto Optimization Values}
    \label{EinastoUGC06787}
    \begin{tabular}{|r|r|}
     \hline
      \textbf{Parameter}   & \textbf{Optimization Values}
      \\  \hline
     $\rho_e$  &$1\times 10^7$
\\  \hline
$r_e$ & 9.79
\\  \hline
$n_e$ & 0.04
\\  \hline
    \end{tabular}
  \end{center}
\end{table}
\begin{table}[h!]
\centering \caption{Physical assessment of collisional DM
parameters (UGC06787).}
\begin{tabular}{lcc}
\hline
Parameter & Value & Physical Verdict \\
\hline
$\gamma_0$ & $1.0001$ & Nearly isothermal \\
$\delta_\gamma$ & $1.2\times10^{-9}$ & Practically zero  \\
$r_\gamma$ & $1.5\ \mathrm{Kpc}$ & Reasonable placement for a transition\\
$K_0$ ($M_{\odot}\,\mathrm{Kpc}^{-3}\,(\mathrm{km/s})^{2}$) & $1.5\times10^{4}$ & Large entropy/pressure scale given $\rho_0$ \\
$r_c$ & $0.5\ \mathrm{Kpc}$ & Small core scale \\
$p$ & $0.01$ & Very shallow decline . \\
\hline
Overall & - & Algebraically consistent\\
\hline
\end{tabular}
\label{EVALUATIONUGC06787}
\end{table}
Now the extended picture including the rotation velocity from the
other components of the galaxy, such as the disk and gas, makes
the collisional DM model viable for this galaxy. In Fig.
\ref{extendedUGC06787} we present the combined rotation curves
including the other components of the galaxy along with the
collisional matter. As it can be seen, the extended collisional DM
model is non-viable.
Also in Table \ref{evaluationextendedUGC06787} we present the
values of the free parameters of the collisional DM model for
which the maximum compatibility with the SPARC data comes for the
galaxy UGC06787.
\begin{table}[h!]
\centering \caption{Physical assessment of Extended collisional DM
parameters for galaxy UGC06787.}
\begin{tabular}{lcc}
\hline
Parameter & Value & Physical Verdict \\
\hline
$\gamma_0$ & 0.93266444 & Sub-isothermal ($\gamma_0<1$) \\
$\delta_\gamma$ & 0.11992891 & Moderate radial variation\\
$K_0$ & $4.80788\times10^{5}$ & Very large entropy compared to typical values \\
$ml_{\rm disk}$ & 0.76321172 & Reasonable sub-maximal disk \\
$ml_{\rm bulge}$ & 0.50 & Substantial bulge mass \\
\hline
Overall &-& Marginally viable  \\
\hline
\end{tabular}
\label{evaluationextendedUGC06787}
\end{table}

\subsection{The Galaxy  UGC06818 Non-viable, Extended Viable}

\begin{figure}[h!]
\centering
\includegraphics[width=20pc]{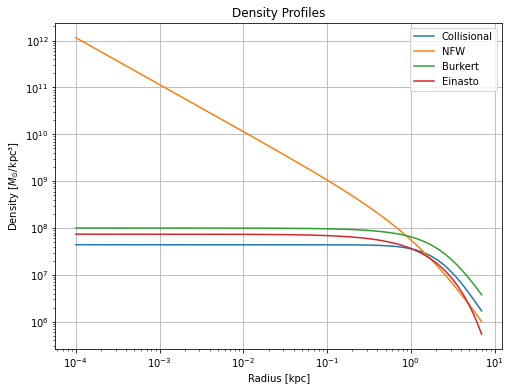}
\caption{The density of the collisional DM model (\ref{tanhmodel})
for the galaxy UGC06818, as a function of the radius.}
\label{UGC06818dens}
\end{figure}

\begin{figure}[h!]
\centering
\includegraphics[width=20pc]{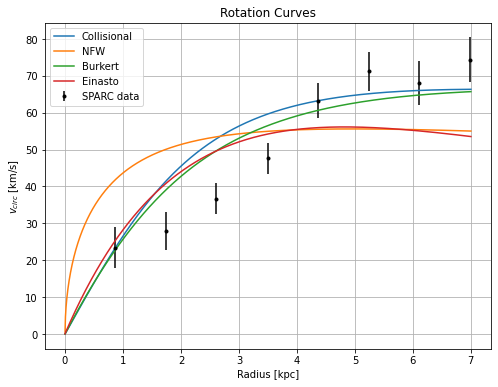}
\caption{The predicted rotation curves after using an optimization
for the collisional DM model (\ref{tanhmodel}), versus the SPARC
observational data for the galaxy UGC06818. We also plotted the
optimized curves for the NFW model, the Burkert model and the
Einasto model.} \label{UGC06818}
\end{figure}

\begin{figure}[h!]
\centering
\includegraphics[width=20pc]{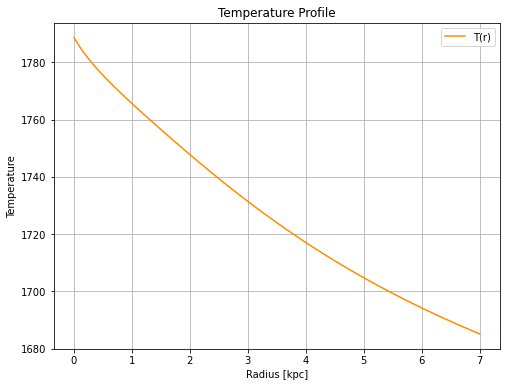}
\caption{The temperature as a function of the radius for the
collisional DM model (\ref{tanhmodel}) for the galaxy UGC06818.}
\label{UGC06818temp}
\end{figure}

\begin{figure}[h!]
\centering
\includegraphics[width=20pc]{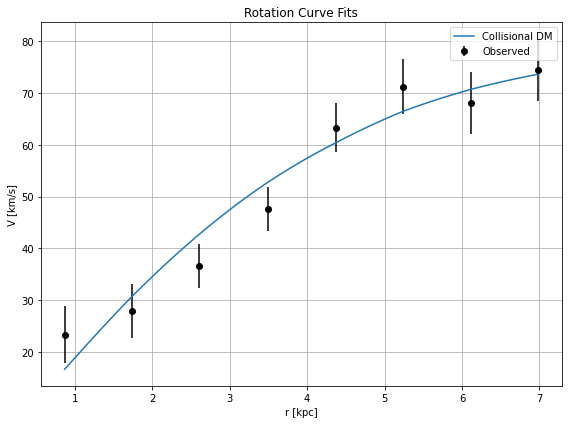}
\caption{The predicted rotation curves after using an optimization
for the collisional DM model (\ref{tanhmodel}), versus the
extended SPARC observational data for the galaxy UGC06818. The
model includes the rotation curves from all the components of the
galaxy, including gas and disk velocities, along with the
collisional DM model.} \label{extendedUGC06818}
\end{figure}

For this galaxy, we shall choose $\rho_0=4.4\times
10^7$$M_{\odot}/\mathrm{Kpc}^{3}$. The galaxy UGC06818 is presumed
to be a late-type disk (spiral or irregular), at a redshift
corresponding to a Hubble distance of order $\sim100$-$200\
\mathrm{Mpc}$. In Figs. \ref{UGC06818dens}, \ref{UGC06818} and
\ref{UGC06818temp} we present the density of the collisional DM
model, the predicted rotation curves after using an optimization
for the collisional DM model (\ref{tanhmodel}), versus the SPARC
observational data and the temperature parameter as a function of
the radius respectively. As it can be seen, the SIDM model
produces non-viable rotation curves incompatible with the SPARC
data. Also in Tables \ref{collUGC06818}, \ref{NavaroUGC06818},
\ref{BuckertUGC06818} and \ref{EinastoUGC06818} we present the
optimization values for the SIDM model, and the other DM profiles.
Also in Table \ref{EVALUATIONUGC06818} we present the overall
evaluation of the SIDM model for the galaxy at hand. The resulting
phenomenology is non-viable.
\begin{table}[h!]
  \begin{center}
    \caption{Collisional Dark Matter Optimization Values}
    \label{collUGC06818}
     \begin{tabular}{|r|r|}
     \hline
      \textbf{Parameter}   & \textbf{Optimization Values}
      \\  \hline
     $\delta_{\gamma} $ & 0.0000000012
\\  \hline
$\gamma_0 $ & 1.0001 \\ \hline $K_0$ ($M_{\odot} \,
\mathrm{Kpc}^{-3} \, (\mathrm{km/s})^{2}$)& 1500 \\ \hline
    \end{tabular}
  \end{center}
\end{table}
\begin{table}[h!]
  \begin{center}
    \caption{NFW  Optimization Values}
    \label{NavaroUGC06818}
     \begin{tabular}{|r|r|}
     \hline
      \textbf{Parameter}   & \textbf{Optimization Values}
      \\  \hline
   $\rho_s$   & $5\times 10^7$
\\  \hline
$r_s$&  2.30
\\  \hline
    \end{tabular}
  \end{center}
\end{table}
\begin{table}[h!]
  \begin{center}
    \caption{Burkert Optimization Values}
    \label{BuckertUGC06818}
     \begin{tabular}{|r|r|}
     \hline
      \textbf{Parameter}   & \textbf{Optimization Values}
      \\  \hline
     $\rho_0^B$  & $1\times 10^8$
\\  \hline
$r_0$&  2.75
\\  \hline
    \end{tabular}
  \end{center}
\end{table}
\begin{table}[h!]
  \begin{center}
    \caption{Einasto Optimization Values}
    \label{EinastoUGC06818}
    \begin{tabular}{|r|r|}
     \hline
      \textbf{Parameter}   & \textbf{Optimization Values}
      \\  \hline
     $\rho_e$  &$1\times 10^7$
\\  \hline
$r_e$ & 2.85
\\  \hline
$n_e$ & 1
\\  \hline
    \end{tabular}
  \end{center}
\end{table}
\begin{table}[h!]
\centering \caption{Physical assessment of collisional DM
parameters (UGC06818).}
\begin{tabular}{lcc}
\hline
Parameter & Value & Physical Verdict \\
\hline
$\gamma_0$ & $1.0001$ & Extremely close to isothermal \\
$\delta_\gamma$ & $1.2\times10^{-9}$ & Practically zero  \\
$r_\gamma$ & $1.5\ \mathrm{Kpc}$ & Reasonable transition radius  \\
$K_0$ ($M_{\odot}\,\mathrm{Kpc}^{-3}\,(\mathrm{km/s})^{2}$) & $1.5\times10^{3}$ & Modest entropy/pressure scale \\
$r_c$ & $0.5\ \mathrm{Kpc}$ & Small core scale  \\
$p$ & $0.01$ & Very shallow decline  \\
\hline
Overall & - & Algebraically correct \\
\hline
\end{tabular}
\label{EVALUATIONUGC06818}
\end{table}
Now the extended picture including the rotation velocity from the
other components of the galaxy, such as the disk and gas, makes
the collisional DM model viable for this galaxy. In Fig.
\ref{extendedUGC06818} we present the combined rotation curves
including the other components of the galaxy along with the
collisional matter. As it can be seen, the extended collisional DM
model is viable.
Also in Table \ref{evaluationextendedUGC06818} we present the
values of the free parameters of the collisional DM model for
which the maximum compatibility with the SPARC data comes for the
galaxy UGC06818.
\begin{table}[h!]
\centering \caption{Physical assessment of Extended collisional DM
parameters for galaxy UGC06818.}
\begin{tabular}{lcc}
\hline
Parameter & Value & Physical Verdict \\
\hline
$\gamma_0$ & 1.01 & Very near-isothermal \\
$\delta_\gamma$ & 0.0001 & Essentially zero gradient \\
$K_0$ & 2000 & Moderate entropy   \\
$ml_{\rm disk}$ & 0.23457151 & Low stellar mass-to-light \\
$ml_{\rm bulge}$ & 0.00000000 & Negligible bulge \\
\hline
Overall &-& Physically plausible and self-consistent \\
\hline
\end{tabular}
\label{evaluationextendedUGC06818}
\end{table}

\subsection{The Galaxy UGC06917}

\begin{figure}[h!]
\centering
\includegraphics[width=20pc]{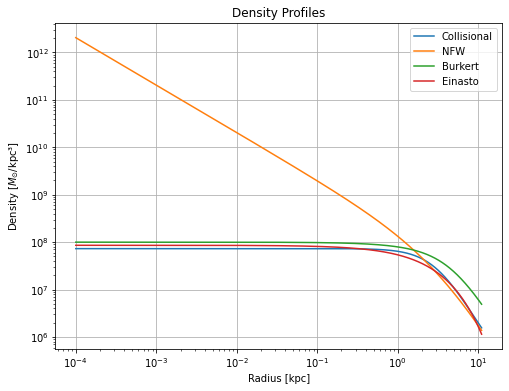}
\caption{The density of the collisional DM model (\ref{tanhmodel})
for the galaxy UGC06917, as a function of the radius.}
\label{UGC06917dens}
\end{figure}

\begin{figure}[h!]
\centering
\includegraphics[width=20pc]{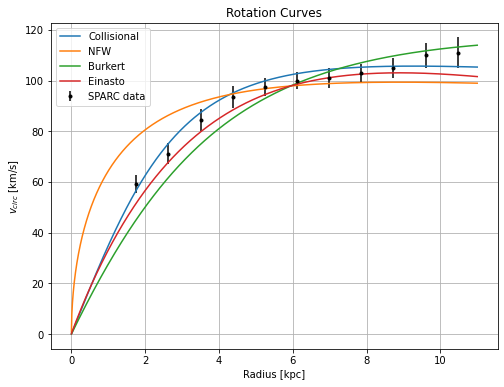}
\caption{The predicted rotation curves after using an optimization
for the collisional DM model (\ref{tanhmodel}), versus the SPARC
observational data for the galaxy UGC06917. We also plotted the
optimized curves for the NFW model, the Burkert model and the
Einasto model.} \label{UGC06917}
\end{figure}

\begin{figure}[h!]
\centering
\includegraphics[width=20pc]{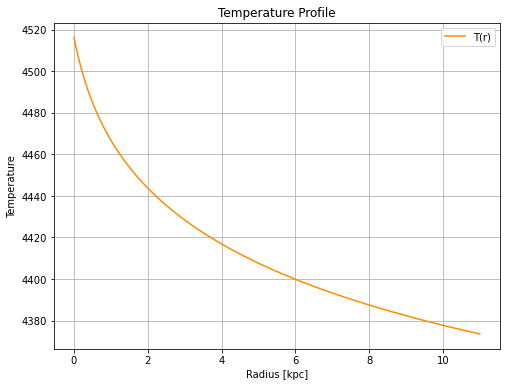}
\caption{The temperature as a function of the radius for the
collisional DM model (\ref{tanhmodel}) for the galaxy UGC06917.}
\label{UGC06917temp}
\end{figure}

For this galaxy, we shall choose $\rho_0=7.3\times
10^7$$M_{\odot}/\mathrm{Kpc}^{3}$. UGC06917 is a late-type spiral
(member of the Ursa Major / NGC3992 group) located at a distance
of order $18.4\ \mathrm{Mpc}$. In Figs. \ref{UGC06917dens},
\ref{UGC06917} and \ref{UGC06917temp} we present the density of
the collisional DM model, the predicted rotation curves after
using an optimization for the collisional DM model
(\ref{tanhmodel}), versus the SPARC observational data and the
temperature parameter as a function of the radius respectively. As
it can be seen, the SIDM model produces viable rotation curves
compatible with the SPARC data. Also in Tables \ref{collUGC06917},
\ref{NavaroUGC06917}, \ref{BuckertUGC06917} and
\ref{EinastoUGC06917} we present the optimization values for the
SIDM model, and the other DM profiles. Also in Table
\ref{EVALUATIONUGC06917} we present the overall evaluation of the
SIDM model for the galaxy at hand. The resulting phenomenology is
viable.
\begin{table}[h!]
  \begin{center}
    \caption{Collisional Dark Matter Optimization Values}
    \label{collUGC06917}
     \begin{tabular}{|r|r|}
     \hline
      \textbf{Parameter}   & \textbf{Optimization Values}
      \\  \hline
     $\delta_{\gamma} $ & 0.000000012
\\  \hline
$\gamma_0 $ & 1.0001 \\ \hline $K_0$ ($M_{\odot} \,
\mathrm{Kpc}^{-3} \, (\mathrm{km/s})^{2}$)& 4500 \\ \hline
    \end{tabular}
  \end{center}
\end{table}
\begin{table}[h!]
  \begin{center}
    \caption{NFW  Optimization Values}
    \label{NavaroUGC06917}
     \begin{tabular}{|r|r|}
     \hline
      \textbf{Parameter}   & \textbf{Optimization Values}
      \\  \hline
   $\rho_s$   & $5\times 10^7$
\\  \hline
$r_s$&  4.11
\\  \hline
    \end{tabular}
  \end{center}
\end{table}
\begin{table}[h!]
  \begin{center}
    \caption{Burkert Optimization Values}
    \label{BuckertUGC06917}
     \begin{tabular}{|r|r|}
     \hline
      \textbf{Parameter}   & \textbf{Optimization Values}
      \\  \hline
     $\rho_0^B$  & $1\times 10^8$
\\  \hline
$r_0$&  4.82
\\  \hline
    \end{tabular}
  \end{center}
\end{table}
\begin{table}[h!]
  \begin{center}
    \caption{Einasto Optimization Values}
    \label{EinastoUGC06917}
    \begin{tabular}{|r|r|}
     \hline
      \textbf{Parameter}   & \textbf{Optimization Values}
      \\  \hline
     $\rho_e$  &$1\times 10^7$
\\  \hline
$r_e$ & 5.20
\\  \hline
$n_e$ & 0.93
\\  \hline
    \end{tabular}
  \end{center}
\end{table}
\begin{table}[h!]
\centering \caption{Physical assessment of collisional DM
parameters (UGC06917).}
\begin{tabular}{lcc}
\hline
Parameter & Value & Physical Verdict \\
\hline
$\gamma_0$ & $1.0001$ & Practically isothermal \\
$\delta_\gamma$ & $1.2\times10^{-8}$ & Negligible  \\
$r_\gamma$ & $1.5\ \mathrm{Kpc}$ & Reasonable transition radius \\
$K_0$ ($M_{\odot}\,\mathrm{Kpc}^{-3}\,(\mathrm{km/s})^{2}$) & $4.5\times10^{3}$ & Moderate entropy scale  \\
$r_c$ & $0.5\ \mathrm{Kpc}$ & Small core scale \\
$p$ & $0.01$ & Very shallow decline - $K(r)\sim\mathrm{const}$ \\
\hline
Overall & - & Algebraically consistent\\
\hline
\end{tabular}
\label{EVALUATIONUGC06917}
\end{table}

\subsection{The Galaxy UGC06923}

\begin{figure}[h!]
\centering
\includegraphics[width=20pc]{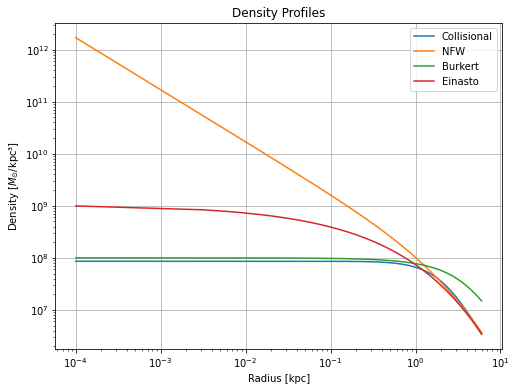}
\caption{The density of the collisional DM model (\ref{tanhmodel})
for the galaxy UGC06923, as a function of the radius.}
\label{UGC06923dens}
\end{figure}
\begin{figure}[h!]
\centering
\includegraphics[width=20pc]{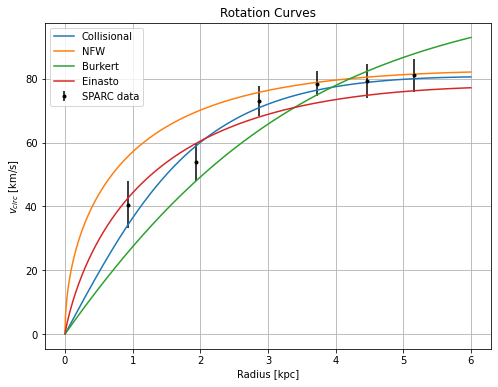}
\caption{The predicted rotation curves after using an optimization
for the collisional DM model (\ref{tanhmodel}), versus the SPARC
observational data for the galaxy UGC06923. We also plotted the
optimized curves for the NFW model, the Burkert model and the
Einasto model.} \label{UGC06923}
\end{figure}
\begin{figure}[h!]
\centering
\includegraphics[width=20pc]{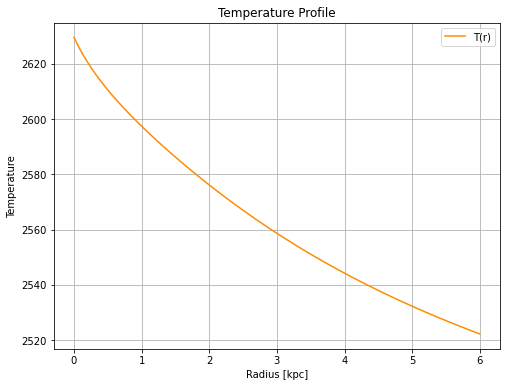}
\caption{The temperature as a function of the radius for the
collisional DM model (\ref{tanhmodel}) for the galaxy UGC06923.}
\label{UGC06923temp}
\end{figure}

For this galaxy, we shall choose $\rho_0=8.7\times
10^7$$M_{\odot}/\mathrm{Kpc}^{3}$. The galaxy UGC,06923 is a small
companion of the large barred spiral NGC3992 (Messier 109) and is
usually classified as a dwarf spiral/irregular type. The system
lies in the M109 group (Ursa Major cloud) at a distance of about
$\sim15$-$18\ \mathrm{Mpc}$. In Figs. \ref{UGC06923dens},
\ref{UGC06923} and \ref{UGC06923temp} we present the density of
the collisional DM model, the predicted rotation curves after
using an optimization for the collisional DM model
(\ref{tanhmodel}), versus the SPARC observational data and the
temperature parameter as a function of the radius respectively. As
it can be seen, the SIDM model produces viable rotation curves
compatible with the SPARC data. Also in Tables \ref{collUGC06923},
\ref{NavaroUGC06923}, \ref{BuckertUGC06923} and
\ref{EinastoUGC06923} we present the optimization values for the
SIDM model, and the other DM profiles. Also in Table
\ref{EVALUATIONUGC06923} we present the overall evaluation of the
SIDM model for the galaxy at hand. The resulting phenomenology is
viable.
\begin{table}[h!]
  \begin{center}
    \caption{Collisional Dark Matter Optimization Values}
    \label{collUGC06923}
     \begin{tabular}{|r|r|}
     \hline
      \textbf{Parameter}   & \textbf{Optimization Values}
      \\  \hline
     $\delta_{\gamma} $ & 0.000000012
\\  \hline
$\gamma_0 $ & 1.0001 \\ \hline $K_0$ ($M_{\odot} \,
\mathrm{Kpc}^{-3} \, (\mathrm{km/s})^{2}$)& 2400  \\ \hline
    \end{tabular}
  \end{center}
\end{table}
\begin{table}[h!]
  \begin{center}
    \caption{NFW  Optimization Values}
    \label{NavaroUGC06923}
     \begin{tabular}{|r|r|}
     \hline
      \textbf{Parameter}   & \textbf{Optimization Values}
      \\  \hline
   $\rho_s$   & $5\times 10^7$
\\  \hline
$r_s$& 3.41
\\  \hline
    \end{tabular}
  \end{center}
\end{table}
\begin{table}[h!]
  \begin{center}
    \caption{Burkert Optimization Values}
    \label{BuckertUGC06923}
     \begin{tabular}{|r|r|}
     \hline
      \textbf{Parameter}   & \textbf{Optimization Values}
      \\  \hline
     $\rho_0^B$  & $1\times 10^8$
\\  \hline
$r_0$&  4.42
\\  \hline
    \end{tabular}
  \end{center}
\end{table}
\begin{table}[h!]
  \begin{center}
    \caption{Einasto Optimization Values}
    \label{EinastoUGC06923}
    \begin{tabular}{|r|r|}
     \hline
      \textbf{Parameter}   & \textbf{Optimization Values}
      \\  \hline
     $\rho_e$  &$1\times 10^7$
\\  \hline
$r_e$ & 3.7
\\  \hline
$n_e$ & 0.43
\\  \hline
    \end{tabular}
  \end{center}
\end{table}
\begin{table}[h!]
\centering \caption{Physical assessment of collisional DM
parameters (UGC06923).}
\begin{tabular}{lcc}
\hline
Parameter & Value & Physical Verdict \\
\hline
$\gamma_0$ & $1.0001$ & Essentially isothermal \\
$\delta_\gamma$ & $1.2\times10^{-8}$ & Practically zero  \\
$r_\gamma$ & $1.5\ \mathrm{Kpc}$ & Transition radius nominal  \\
$K_0$ ($M_{\odot}\,\mathrm{Kpc}^{-3}\,(\mathrm{km/s})^{2}$) & $2.4\times10^{3}$ & Relatively large entropy scale \\
$r_c$ & $0.5\ \mathrm{Kpc}$ & Small core radius \\
$p$ & $0.01$ & Extremely shallow decline of $K(r)$ \\
\hline
Overall & - & Physically plausible  \\
\hline
\end{tabular}
\label{EVALUATIONUGC06923}
\end{table}

\subsection{The Galaxy UGC06930}

\begin{figure}[h!]
\centering
\includegraphics[width=20pc]{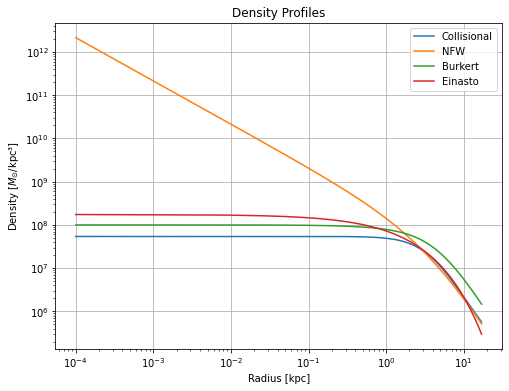}
\caption{The density of the collisional DM model (\ref{tanhmodel})
for the galaxy UGC06930, as a function of the radius.}
\label{UGC06930dens}
\end{figure}

\begin{figure}[h!]
\centering
\includegraphics[width=20pc]{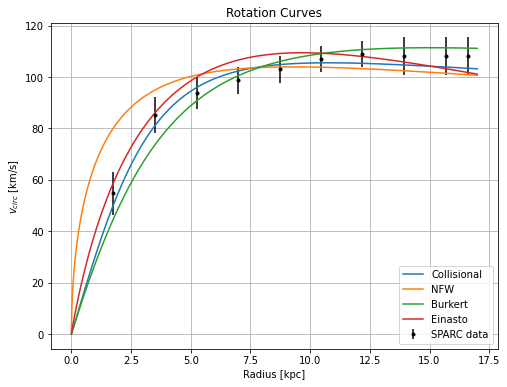}
\caption{The predicted rotation curves after using an optimization
for the collisional DM model (\ref{tanhmodel}), versus the SPARC
observational data for the galaxy UGC06930. We also plotted the
optimized curves for the NFW model, the Burkert model and the
Einasto model.} \label{UGC06930}
\end{figure}

\begin{figure}[h!]
\centering
\includegraphics[width=20pc]{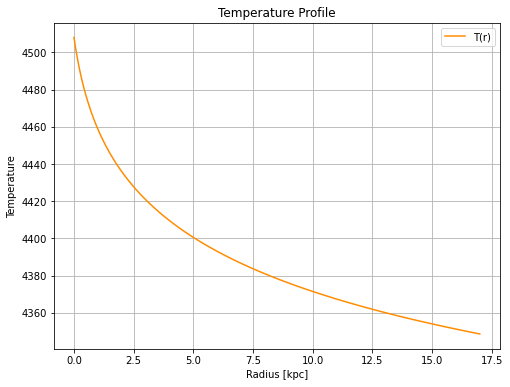}
\caption{The temperature as a function of the radius for the
collisional DM model (\ref{tanhmodel}) for the galaxy UGC06930.}
\label{UGC06930temp}
\end{figure}

For this galaxy, we shall choose $\rho_0=5.2\times
10^7$$M_{\odot}/\mathrm{Kpc}^{3}$. UGC,06930 is a spiral, small
late-type spiral system with an extended HI disc, classified among
galaxies with substantial gas content. In Figs.
\ref{UGC06930dens}, \ref{UGC06930} and \ref{UGC06930temp} we
present the density of the collisional DM model, the predicted
rotation curves after using an optimization for the collisional DM
model (\ref{tanhmodel}), versus the SPARC observational data and
the temperature parameter as a function of the radius
respectively. As it can be seen, the SIDM model produces viable
rotation curves compatible with the SPARC data. Also in Tables
\ref{collUGC06930}, \ref{NavaroUGC06930}, \ref{BuckertUGC06930}
and \ref{EinastoUGC06930} we present the optimization values for
the SIDM model, and the other DM profiles. Also in Table
\ref{EVALUATIONUGC06930} we present the overall evaluation of the
SIDM model for the galaxy at hand. The resulting phenomenology is
viable.
\begin{table}[h!]
  \begin{center}
    \caption{Collisional Dark Matter Optimization Values}
    \label{collUGC06930}
     \begin{tabular}{|r|r|}
     \hline
      \textbf{Parameter}   & \textbf{Optimization Values}
      \\  \hline
     $\delta_{\gamma} $ & 0.0000000012
\\  \hline
$\gamma_0 $ & 1.0001 \\ \hline $K_0$ ($M_{\odot} \,
\mathrm{Kpc}^{-3} \, (\mathrm{km/s})^{2}$)& 4500  \\ \hline
    \end{tabular}
  \end{center}
\end{table}
\begin{table}[h!]
  \begin{center}
    \caption{NFW  Optimization Values}
    \label{NavaroUGC06930}
     \begin{tabular}{|r|r|}
     \hline
      \textbf{Parameter}   & \textbf{Optimization Values}
      \\  \hline
   $\rho_s$   & $5\times 10^7$
\\  \hline
$r_s$&  4.3
\\  \hline
    \end{tabular}
  \end{center}
\end{table}
\begin{table}[h!]
  \begin{center}
    \caption{Burkert Optimization Values}
    \label{BuckertUGC06930}
     \begin{tabular}{|r|r|}
     \hline
      \textbf{Parameter}   & \textbf{Optimization Values}
      \\  \hline
     $\rho_0^B$  & $1\times 10^8$
\\  \hline
$r_0$&  4.62
\\  \hline
    \end{tabular}
  \end{center}
\end{table}
\begin{table}[h!]
  \begin{center}
    \caption{Einasto Optimization Values}
    \label{EinastoUGC06930}
    \begin{tabular}{|r|r|}
     \hline
      \textbf{Parameter}   & \textbf{Optimization Values}
      \\  \hline
     $\rho_e$  &$1\times 10^7$
\\  \hline
$r_e$ &  5.4
\\  \hline
$n_e$ & 0.7
\\  \hline
    \end{tabular}
  \end{center}
\end{table}
\begin{table}[h!]
\centering \caption{Physical assessment of collisional DM
parameters for UGC06930.}
\begin{tabular}{lcc}
\hline
Parameter & Value & Physical Verdict \\
\hline
$\gamma_0$ & $1.0001$ & Nearly isothermal \\
$\delta_\gamma$ & $1.2\times10^{-9}$ & Practically zero  \\
$r_\gamma$ & $1.5~\mathrm{Kpc}$ & Transition radius plausible \\
$K_0$ & $4500$ & Enough and large pressure support \\
$r_c$ & $0.5~\mathrm{Kpc}$ & Small core \\
$p$ & $0.01$ & Very shallow decline \\
\hline
Overall & --- & Physically plausible \\
\hline
\end{tabular}
\label{EVALUATIONUGC06930}
\end{table}

\subsection{The Galaxy UGC06973 Non-viable, Extended Viable}

\begin{figure}[h!]
\centering
\includegraphics[width=20pc]{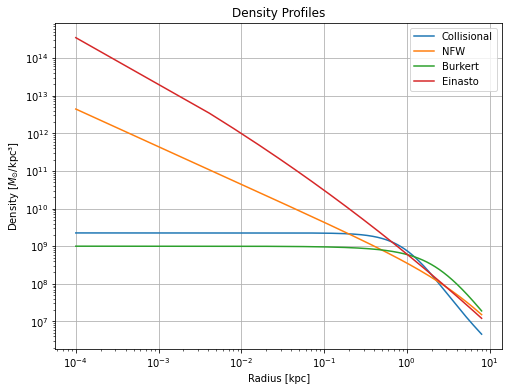}
\caption{The density of the collisional DM model (\ref{tanhmodel})
for the galaxy UGC06973, as a function of the radius.}
\label{UGC06973dens}
\end{figure}
\begin{figure}[h!]
\centering
\includegraphics[width=20pc]{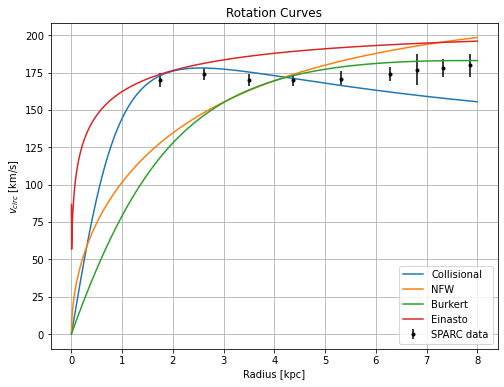}
\caption{The predicted rotation curves after using an optimization
for the collisional DM model (\ref{tanhmodel}), versus the SPARC
observational data for the galaxy UGC06973. We also plotted the
optimized curves for the NFW model, the Burkert model and the
Einasto model.} \label{UGC06973}
\end{figure}
\begin{figure}[h!]
\centering
\includegraphics[width=20pc]{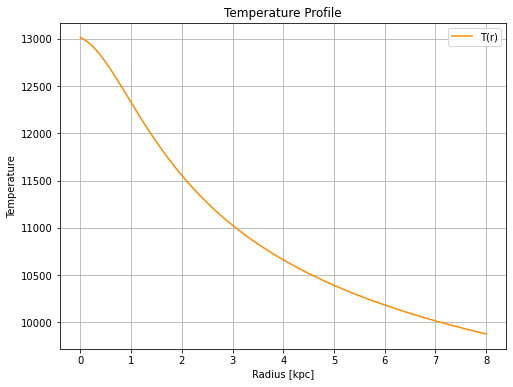}
\caption{The temperature as a function of the radius for the
collisional DM model (\ref{tanhmodel}) for the galaxy UGC06973.}
\label{UGC06973temp}
\end{figure}
\begin{figure}[h!]
\centering
\includegraphics[width=20pc]{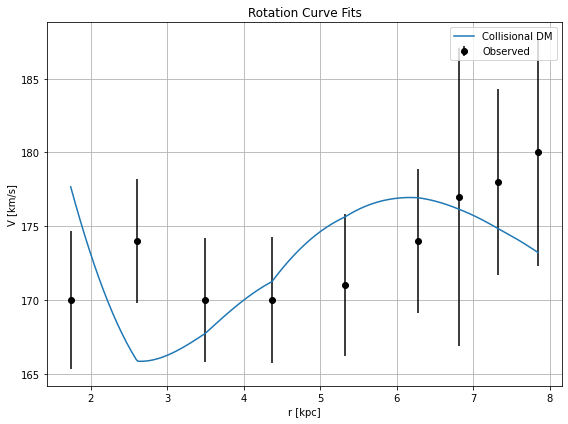}
\caption{The predicted rotation curves after using an optimization
for the collisional DM model (\ref{tanhmodel}), versus the
extended SPARC observational data for the galaxy UGC06973. The
model includes the rotation curves from all the components of the
galaxy, including gas and disk velocities, along with the
collisional DM model.} \label{extendedUGC06973}
\end{figure}

For this galaxy, we shall choose $\rho_0=2.25\times
10^9$$M_{\odot}/\mathrm{Kpc}^{3}$. The galaxy UGC06973 is a
late-type disc system in the SPARC sample, located at a distance
of $D\simeq23.0\pm1.6\ \mathrm{Mpc}$. In Figs. \ref{UGC06973dens},
\ref{UGC06973} and \ref{UGC06973temp} we present the density of
the collisional DM model, the predicted rotation curves after
using an optimization for the collisional DM model
(\ref{tanhmodel}), versus the SPARC observational data and the
temperature parameter as a function of the radius respectively. As
it can be seen, the SIDM model produces non-viable rotation curves
incompatible with the SPARC data. Also in Tables
\ref{collUGC06973}, \ref{NavaroUGC06973}, \ref{BuckertUGC06973}
and \ref{EinastoUGC06973} we present the optimization values for
the SIDM model, and the other DM profiles. Also in Table
\ref{EVALUATIONUGC06973} we present the overall evaluation of the
SIDM model for the galaxy at hand. The resulting phenomenology is
non-viable.
\begin{table}[h!]
  \begin{center}
    \caption{Collisional Dark Matter Optimization Values}
    \label{collUGC06973}
     \begin{tabular}{|r|r|}
     \hline
      \textbf{Parameter}   & \textbf{Optimization Values}
      \\  \hline
     $\delta_{\gamma} $ & 0.0000000012
\\  \hline
$\gamma_0 $ & 1.0001  \\ \hline $K_0$ ($M_{\odot} \,
\mathrm{Kpc}^{-3} \, (\mathrm{km/s})^{2}$)& 5500  \\ \hline
    \end{tabular}
  \end{center}
\end{table}
\begin{table}[h!]
  \begin{center}
    \caption{NFW  Optimization Values}
    \label{NavaroUGC06973}
     \begin{tabular}{|r|r|}
     \hline
      \textbf{Parameter}   & \textbf{Optimization Values}
      \\  \hline
   $\rho_s$   & $5\times 10^7$
\\  \hline
$r_s$&  8.82
\\  \hline
    \end{tabular}
  \end{center}
\end{table}
\begin{table}[h!]
  \begin{center}
    \caption{Burkert Optimization Values}
    \label{BuckertUGC06973}
     \begin{tabular}{|r|r|}
     \hline
      \textbf{Parameter}   & \textbf{Optimization Values}
      \\  \hline
     $\rho_0^B$  & $1\times 10^8$
\\  \hline
$r_0$&  2.4
\\  \hline
    \end{tabular}
  \end{center}
\end{table}
\begin{table}[h!]
  \begin{center}
    \caption{Einasto Optimization Values}
    \label{EinastoUGC06973}
    \begin{tabular}{|r|r|}
     \hline
      \textbf{Parameter}   & \textbf{Optimization Values}
      \\  \hline
     $\rho_e$  &$1\times 10^7$
\\  \hline
$r_e$ & 8.81
\\  \hline
$n_e$ & 0.05
\\  \hline
    \end{tabular}
  \end{center}
\end{table}
\begin{table}[h!]
\centering \caption{Physical assessment of collisional DM
parameters (UGC06973).}
\begin{tabular}{lcc}
\hline
Parameter & Value & Physical Verdict \\
\hline
$\gamma_0$ & $1.0001$ & Mildly super-isothermal \\
$\delta_\gamma$ & $0.0000000012$ & Negligible \\
$r_\gamma$ & $1.5\ \mathrm{Kpc}$ & Nominal transition radius \\
$K_0$ & $5.5\times10^{3}$ & Large entropy scale \\
$r_c$ & $0.5\ \mathrm{Kpc}$ & Small core scale \\
$p$ & $0.01$ & Extremely shallow decline of $K(r)$ \\
Overall &-& Physically consistent as an almost-isothermal \\
\hline
\end{tabular}
\label{EVALUATIONUGC06973}
\end{table}
Now the extended picture including the rotation velocity from the
other components of the galaxy, such as the disk and gas, makes
the collisional DM model viable for this galaxy. In Fig.
\ref{extendedUGC06973} we present the combined rotation curves
including the other components of the galaxy along with the
collisional matter. As it can be seen, the extended collisional DM
model is viable.
Also in Table \ref{evaluationextendedUGC06973} we present the
values of the free parameters of the collisional DM model for
which the maximum compatibility with the SPARC data comes for the
galaxy UGC06973.
\begin{table}[h!]
\centering \caption{Physical assessment of Extended collisional DM
parameters (UGC06973).}
\begin{tabular}{lcc}
\hline
Parameter & Value & Physical Verdict \\
\hline
$\gamma_0$ & 1.16350014 & Slightly above isothermal \\
$\delta_\gamma$ & 0.11481604 & Moderate radial variation in $\gamma(r)$ \\
$K_0$ & 3000 & Moderate entropy/pressure scale\\
ml\_disk & 0.52051799 & Reasonable stellar mass-to-light for a disk  \\
ml\_bulge & 0.00000164 & Effectively zero bulge contribution  \\
\hline
Overall &-& Physically plausible \\
\hline
\end{tabular}
\label{evaluationextendedUGC06973}
\end{table}

\subsection{The Galaxy UGC06983}

\begin{figure}[h!]
\centering
\includegraphics[width=20pc]{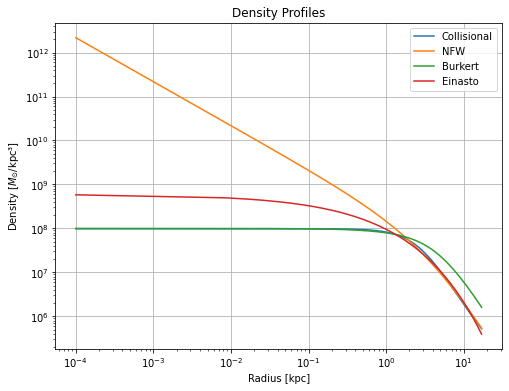}
\caption{The density of the collisional DM model (\ref{tanhmodel})
for the galaxy UGC06983, as a function of the radius.}
\label{UGC06983dens}
\end{figure}

\begin{figure}[h!]
\centering
\includegraphics[width=20pc]{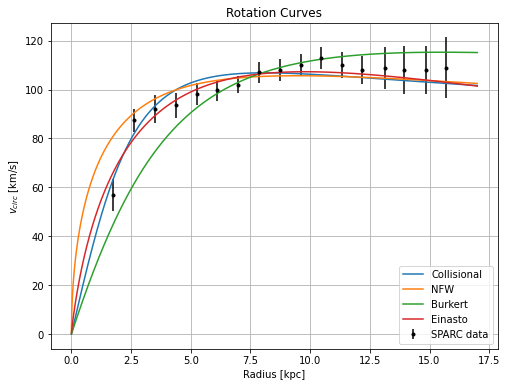}
\caption{The predicted rotation curves after using an optimization
for the collisional DM model (\ref{tanhmodel}), versus the SPARC
observational data for the galaxy UGC06983. We also plotted the
optimized curves for the NFW model, the Burkert model and the
Einasto model.} \label{UGC06983}
\end{figure}

\begin{figure}[h!]
\centering
\includegraphics[width=20pc]{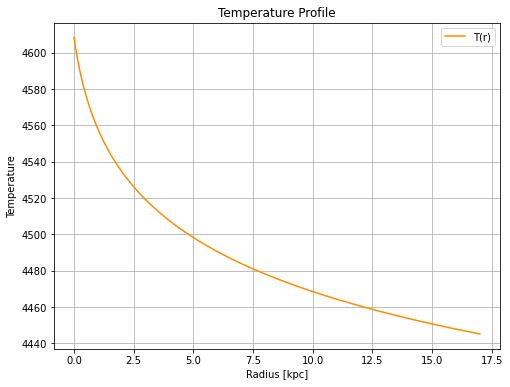}
\caption{The temperature as a function of the radius for the
collisional DM model (\ref{tanhmodel}) for the galaxy UGC06983.}
\label{UGC06983temp}
\end{figure}

For this galaxy, we shall choose $\rho_0=9.8\times
10^7$$M_{\odot}/\mathrm{Kpc}^{3}$. The galaxy UGC06983 is
classified as a low-surface-brightness or late-type disc galaxy
rather than a classic dwarf. From redshift-based scaling it likely
lies at tens of Mpc distance. In Figs. \ref{UGC06983dens},
\ref{UGC06983} and \ref{UGC06983temp} we present the density of
the collisional DM model, the predicted rotation curves after
using an optimization for the collisional DM model
(\ref{tanhmodel}), versus the SPARC observational data and the
temperature parameter as a function of the radius respectively. As
it can be seen, the SIDM model produces viable rotation curves
compatible with the SPARC data. Also in Tables \ref{collUGC06983},
\ref{NavaroUGC06983}, \ref{BuckertUGC06983} and
\ref{EinastoUGC06983} we present the optimization values for the
SIDM model, and the other DM profiles. Also in Table
\ref{EVALUATIONUGC06983} we present the overall evaluation of the
SIDM model for the galaxy at hand. The resulting phenomenology is
viable.
\begin{table}[h!]
  \begin{center}
    \caption{Collisional Dark Matter Optimization Values}
    \label{collUGC06983}
     \begin{tabular}{|r|r|}
     \hline
      \textbf{Parameter}   & \textbf{Optimization Values}
      \\  \hline
     $\delta_{\gamma} $ & 0.0000000012
\\  \hline
$\gamma_0 $ & 1.0001\\ \hline $K_0$ ($M_{\odot} \,
\mathrm{Kpc}^{-3} \, (\mathrm{km/s})^{2}$)& 4600  \\ \hline
    \end{tabular}
  \end{center}
\end{table}
\begin{table}[h!]
  \begin{center}
    \caption{NFW  Optimization Values}
    \label{NavaroUGC06983}
     \begin{tabular}{|r|r|}
     \hline
      \textbf{Parameter}   & \textbf{Optimization Values}
      \\  \hline
   $\rho_s$   & $5\times 10^7$
\\  \hline
$r_s$&  4.37
\\  \hline
    \end{tabular}
  \end{center}
\end{table}
\begin{table}[h!]
  \begin{center}
    \caption{Burkert Optimization Values}
    \label{BuckertUGC06983}
     \begin{tabular}{|r|r|}
     \hline
      \textbf{Parameter}   & \textbf{Optimization Values}
      \\  \hline
     $\rho_0^B$  & $1\times 10^8$
\\  \hline
$r_0$&  4.78
\\  \hline
    \end{tabular}
  \end{center}
\end{table}
\begin{table}[h!]
  \begin{center}
    \caption{Einasto Optimization Values}
    \label{EinastoUGC06983}
    \begin{tabular}{|r|r|}
     \hline
      \textbf{Parameter}   & \textbf{Optimization Values}
      \\  \hline
     $\rho_e$  &$1\times 10^7$
\\  \hline
$r_e$ & 5.16
\\  \hline
$n_e$ & 0.49
\\  \hline
    \end{tabular}
  \end{center}
\end{table}
\begin{table}[h!]
\centering \caption{Physical assessment of collisional DM
parameters (UGC06983).}
\begin{tabular}{lcc}
\hline
Parameter & Value & Physical Verdict \\
\hline
$\gamma_0$ & $1.0001$ & Nearly isothermal \\
$\delta_\gamma$ & $ 0.0000000012$ & Negligible \\
$r_\gamma$ & $1.5\ \mathrm{Kpc}$ & Transition radius inactive due to tiny $\delta_\gamma$ \\
$K_0$ & $4.5\times10^{3}$ & Moderate entropy scale \\
$r_c$ & $0.5\ \mathrm{Kpc}$ & Small core \\
$p$ & $0.01$ & Extremely shallow decline of $K(r)$ \\
\hline
Overall &-& Physically plausible \\
\hline
\end{tabular}
\label{EVALUATIONUGC06983}
\end{table}

\subsection{The Galaxy UGC07089}

\begin{figure}[h!]
\centering
\includegraphics[width=20pc]{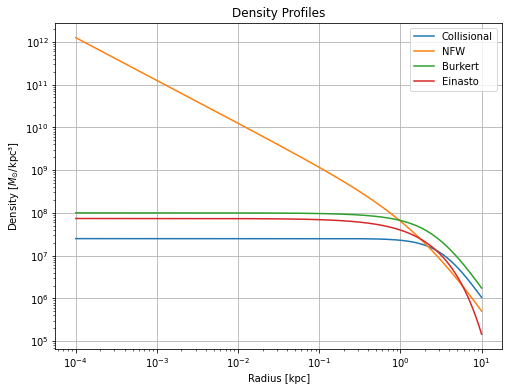}
\caption{The density of the collisional DM model (\ref{tanhmodel})
for the galaxy UGC07089, as a function of the radius.}
\label{UGC07089dens}
\end{figure}

\begin{figure}[h!]
\centering
\includegraphics[width=20pc]{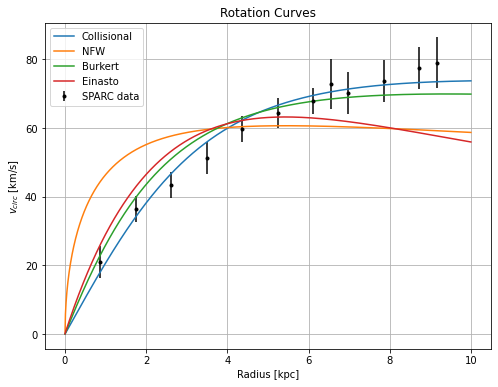}
\caption{The predicted rotation curves after using an optimization
for the collisional DM model (\ref{tanhmodel}), versus the SPARC
observational data for the galaxy UGC07089. We also plotted the
optimized curves for the NFW model, the Burkert model and the
Einasto model.} \label{UGC07089}
\end{figure}

\begin{figure}[h!]
\centering
\includegraphics[width=20pc]{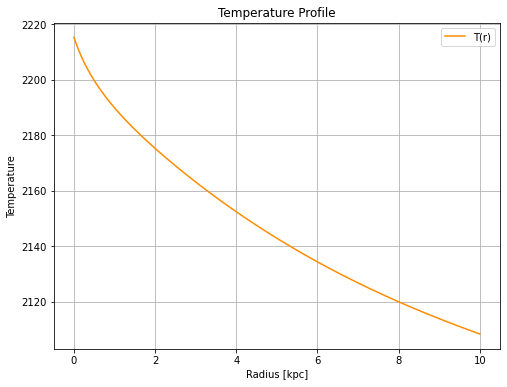}
\caption{The temperature as a function of the radius for the
collisional DM model (\ref{tanhmodel}) for the galaxy UGC07089.}
\label{UGC07089temp}
\end{figure}

For this galaxy, we shall choose $\rho_0=2.5\times
10^7$$M_{\odot}/\mathrm{Kpc}^{3}$. UGC07089 is classified as a
late-type spiral (Sd) galaxy. Its distance is about \(13.3\pm 1.2\
\mathrm{Mpc}\). In Figs. \ref{UGC07089dens}, \ref{UGC07089} and
\ref{UGC07089temp} we present the density of the collisional DM
model, the predicted rotation curves after using an optimization
for the collisional DM model (\ref{tanhmodel}), versus the SPARC
observational data and the temperature parameter as a function of
the radius respectively. As it can be seen, the SIDM model
produces viable rotation curves compatible with the SPARC data.
Also in Tables \ref{collUGC07089}, \ref{NavaroUGC07089},
\ref{BuckertUGC07089} and \ref{EinastoUGC07089} we present the
optimization values for the SIDM model, and the other DM profiles.
Also in Table \ref{EVALUATIONUGC07089} we present the overall
evaluation of the SIDM model for the galaxy at hand. The resulting
phenomenology is viable.
\begin{table}[h!]
  \begin{center}
    \caption{Collisional Dark Matter Optimization Values}
    \label{collUGC07089}
     \begin{tabular}{|r|r|}
     \hline
      \textbf{Parameter}   & \textbf{Optimization Values}
      \\  \hline
     $\delta_{\gamma} $ &  0.0000000012
\\  \hline
$\gamma_0 $ & 1.0001 \\ \hline $K_0$ ($M_{\odot} \,
\mathrm{Kpc}^{-3} \, (\mathrm{km/s})^{2}$)& 2000 \\ \hline
    \end{tabular}
  \end{center}
\end{table}
\begin{table}[h!]
  \begin{center}
    \caption{NFW  Optimization Values}
    \label{NavaroUGC07089}
     \begin{tabular}{|r|r|}
     \hline
      \textbf{Parameter}   & \textbf{Optimization Values}
      \\  \hline
   $\rho_s$   & $5\times 10^7$
\\  \hline
$r_s$&  2.51
\\  \hline
    \end{tabular}
  \end{center}
\end{table}
\begin{table}[h!]
  \begin{center}
    \caption{Burkert Optimization Values}
    \label{BuckertUGC07089}
     \begin{tabular}{|r|r|}
     \hline
      \textbf{Parameter}   & \textbf{Optimization Values}
      \\  \hline
     $\rho_0^B$  & $1\times 10^8$
\\  \hline
$r_0$&  2.90
\\  \hline
    \end{tabular}
  \end{center}
\end{table}
\begin{table}[h!]
  \begin{center}
    \caption{Einasto Optimization Values}
    \label{EinastoUGC07089}
    \begin{tabular}{|r|r|}
     \hline
      \textbf{Parameter}   & \textbf{Optimization Values}
      \\  \hline
     $\rho_e$  &$1\times 10^7$
\\  \hline
$r_e$ & 3.21
\\  \hline
$n_e$ & 1
\\  \hline
    \end{tabular}
  \end{center}
\end{table}
\begin{table}[h!]
\centering \caption{Physical assessment of collisional DM
parameters for \texttt{UGC07089}.}
\begin{tabular}{lcc}
\hline
Parameter & Value   & Physical Verdict \\
\hline
$\gamma_0$ & $1.0001$ & Nearly isothermal \\
$\delta_\gamma$ & $1.2\times10^{-9}$ & Practically zero  \\
$r_\gamma$ & $1.5\ \mathrm{Kpc}$ & Transition radius lies within the inner halo\\
$K_0$ ($M_{\odot}\,\mathrm{Kpc}^{-3}\,(\mathrm{km/s})^{2}$) & $2.0\times10^{3}$ & Enough pressure support \\
$r_c$ & $0.5\ \mathrm{Kpc}$ & Small core scale \\
$p$ & $0.01$ & Very shallow decline \\
\hline
Overall & - & Physically self-consistent \\
\hline
\end{tabular}
\label{EVALUATIONUGC07089}
\end{table}

\subsection{The Galaxy UGC07125 Marginally Viable, Extended Viable}

\begin{figure}[h!]
\centering
\includegraphics[width=20pc]{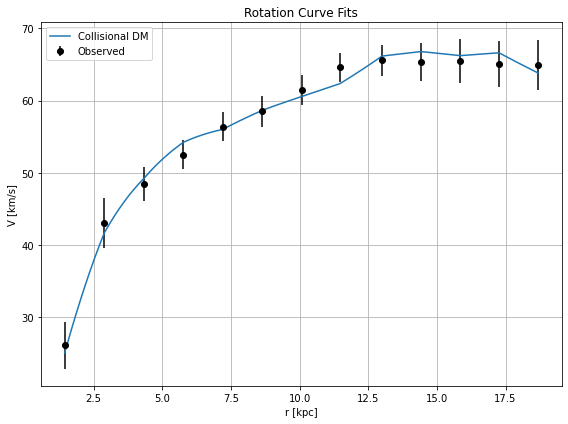}
\caption{The predicted rotation curves after using an optimization
for the collisional DM model (\ref{tanhmodel}), versus the
extended SPARC observational data for the galaxy UGC07125. The
model includes the rotation curves from all the components of the
galaxy, including gas and disk velocities, along with the
collisional DM model.} \label{extendedUGC07125}
\end{figure}

For this galaxy, we shall choose $\rho_0=1.5\times
10^7$$M_{\odot}/\mathrm{Kpc}^{3}$. UGC07125 is a faint, late-type
disk galaxy located at an estimated distance of order 15.5
$\mathrm{Mpc}$. In Figs. \ref{UGC05764dens}, \ref{UGC05764} and
\ref{UGC05764temp} we present the density of the collisional DM
model, the predicted rotation curves after using an optimization
for the collisional DM model (\ref{tanhmodel}), versus the SPARC
observational data and the temperature parameter as a function of
the radius respectively. As it can be seen, the SIDM model
produces marginally viable rotation curves compatible with the
SPARC data. Also in Tables \ref{collUGC05764},
\ref{NavaroUGC05764}, \ref{BuckertUGC05764} and
\ref{EinastoUGC05764} we present the optimization values for the
SIDM model, and the other DM profiles. Also in Table
\ref{EVALUATIONUGC05764} we present the overall evaluation of the
SIDM model for the galaxy at hand. The resulting phenomenology is
marginally viable.
\begin{figure}[h!]
\centering
\includegraphics[width=20pc]{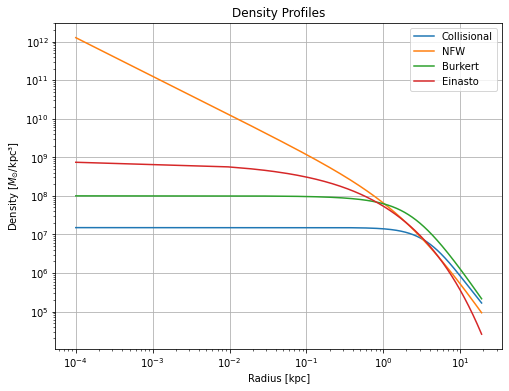}
\caption{The density of the collisional DM model (\ref{tanhmodel})
for the galaxy UGC07125, as a function of the radius.}
\label{UGC07125dens}
\end{figure}
\begin{figure}[h!]
\centering
\includegraphics[width=20pc]{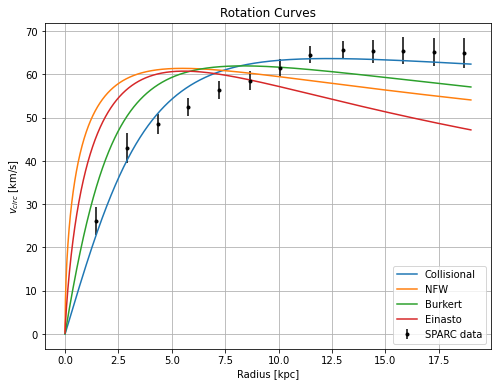}
\caption{The predicted rotation curves after using an optimization
for the collisional DM model (\ref{tanhmodel}), versus the SPARC
observational data for the galaxy UGC07125. We also plotted the
optimized curves for the NFW model, the Burkert model and the
Einasto model.} \label{UGC07125}
\end{figure}
\begin{table}[h!]
  \begin{center}
    \caption{Collisional Dark Matter Optimization Values}
    \label{collUGC07125}
     \begin{tabular}{|r|r|}
     \hline
      \textbf{Parameter}   & \textbf{Optimization Values}
      \\  \hline
     $\delta_{\gamma} $ & 0.000000000012
\\  \hline
$\gamma_0 $ & 1.0001 \\ \hline $K_0$ ($M_{\odot} \,
\mathrm{Kpc}^{-3} \, (\mathrm{km/s})^{2}$)& 1640  \\ \hline
    \end{tabular}
  \end{center}
\end{table}
\begin{table}[h!]
  \begin{center}
    \caption{NFW  Optimization Values}
    \label{NavaroUGC07125}
     \begin{tabular}{|r|r|}
     \hline
      \textbf{Parameter}   & \textbf{Optimization Values}
      \\  \hline
   $\rho_s$   & $5\times 10^7$
\\  \hline
$r_s$& 2.54
\\  \hline
    \end{tabular}
  \end{center}
\end{table}
\begin{figure}[h!]
\centering
\includegraphics[width=20pc]{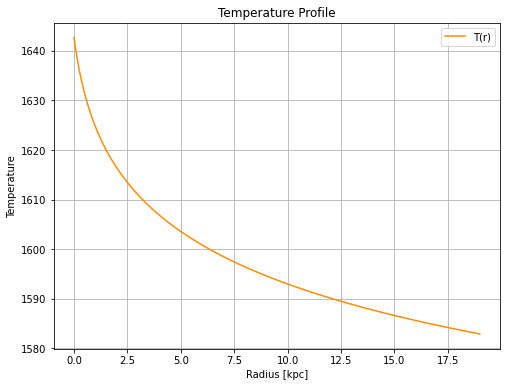}
\caption{The temperature as a function of the radius for the
collisional DM model (\ref{tanhmodel}) for the galaxy UGC07125.}
\label{UGC07125temp}
\end{figure}
\begin{table}[h!]
  \begin{center}
    \caption{Burkert Optimization Values}
    \label{BuckertUGC07125}
     \begin{tabular}{|r|r|}
     \hline
      \textbf{Parameter}   & \textbf{Optimization Values}
      \\  \hline
     $\rho_0^B$  & $1\times 10^8$
\\  \hline
$r_0$&  2.57
\\  \hline
    \end{tabular}
  \end{center}
\end{table}
\begin{table}[h!]
  \begin{center}
    \caption{Einasto Optimization Values}
    \label{EinastoUGC07125}
    \begin{tabular}{|r|r|}
     \hline
      \textbf{Parameter}   & \textbf{Optimization Values}
      \\  \hline
     $\rho_e$  &$1\times 10^7$
\\  \hline
$r_e$ & 2.91
\\  \hline
$n_e$ & 0.46
\\  \hline
    \end{tabular}
  \end{center}
\end{table}
\begin{table}[h!]
\centering \caption{Physical assessment of collisional DM
parameters for \texttt{UGC07125}.}
\begin{tabular}{lcc}
\hline
Parameter & Value   & Physical Verdict \\
\hline
$\gamma_0$ & $1.0001$ & Practically isothermal \\
$\delta_\gamma$ & $1.2\times10^{-11}$ & Essentially zero  \\
$r_\gamma$ & $1.5\ \mathrm{Kpc}$ & Transition radius placed within the inner halo \\
$K_0$ ($M_{\odot}\,\mathrm{Kpc}^{-3}\,(\mathrm{km/s})^{2}$) & $1.64\times10^{3}$ & Entropy/pressure scale moderately large \\
$r_c$ & $0.5\ \mathrm{Kpc}$ & Small core scale  \\
$p$ & $0.01$ & Nearly flat $K(r)$  \\
\hline
Overall & - & Model is self-consistent \\
\hline
\end{tabular}
\label{EVALUATIONUGC07125}
\end{table}
Now the extended picture including the rotation velocity from the
other components of the galaxy, such as the disk and gas, makes
the collisional DM model viable for this galaxy. In Fig.
\ref{extendedUGC07125} we present the combined rotation curves
including the other components of the galaxy along with the
collisional matter. As it can be seen, the extended collisional DM
model is viable.
Also in Table \ref{evaluationextendedUGC07125} we present the
values of the free parameters of the collisional DM model for
which the maximum compatibility with the SPARC data comes for the
galaxy UGC07125.
\begin{table}[h!]
\centering \caption{Physical assessment of Extended collisional DM
parameters for galaxy UGC07125.}
\begin{tabular}{lcc}
\hline
Parameter & Value & Physical Verdict \\
\hline
$\gamma_0$ & 0.94049240 & Slightly sub-isothermal\\
$\delta_\gamma$ & 0.00001 & Effectively zero variation \\
$K_0$ & 3000 & Moderate entropy   \\
$ml_{disk}$ & 0.38010194 & Low-to-moderate disk M/L \\
$ml_{bulge}$ & 0.00000000 & Negligible bulge contribution \\
\hline
Overall &-& Physically plausible \\
\hline
\end{tabular}
\label{evaluationextendedUGC07125}
\end{table}

\subsection{The Galaxy UGC07151}

\begin{figure}[h!]
\centering
\includegraphics[width=20pc]{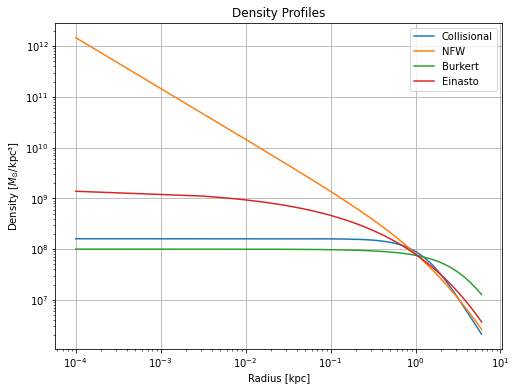}
\caption{The density of the collisional DM model (\ref{tanhmodel})
for the galaxy UGC07151, as a function of the radius.}
\label{UGC07151dens}
\end{figure}
\begin{figure}[h!]
\centering
\includegraphics[width=20pc]{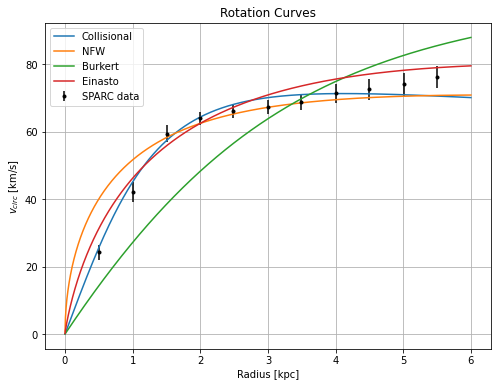}
\caption{The predicted rotation curves after using an optimization
for the collisional DM model (\ref{tanhmodel}), versus the SPARC
observational data for the galaxy UGC07151. We also plotted the
optimized curves for the NFW model, the Burkert model and the
Einasto model.} \label{UGC07151}
\end{figure}
\begin{figure}[h!]
\centering
\includegraphics[width=20pc]{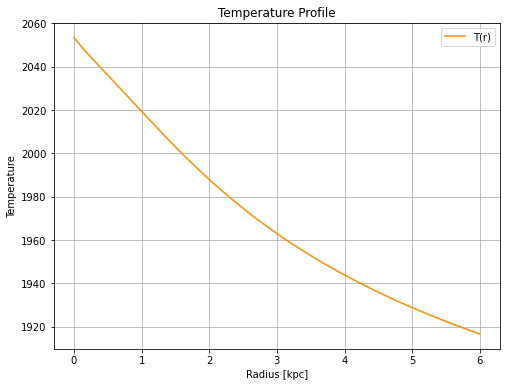}
\caption{The temperature as a function of the radius for the
collisional DM model (\ref{tanhmodel}) for the galaxy UGC07151.}
\label{UGC07151temp}
\end{figure}

For this galaxy, we shall choose $\rho_0=1.6\times
10^7$$M_{\odot}/\mathrm{Kpc}^{3}$. UGC07151 is classified as an
irregular/late-type spiral galaxy, located at an approximate
distance of  $12$Mpc. In Figs. \ref{UGC07151dens}, \ref{UGC07151}
and \ref{UGC07151temp} we present the density of the collisional
DM model, the predicted rotation curves after using an
optimization for the collisional DM model (\ref{tanhmodel}),
versus the SPARC observational data and the temperature parameter
as a function of the radius respectively. As it can be seen, the
SIDM model produces viable rotation curves compatible with the
SPARC data. Also in Tables \ref{collUGC07151},
\ref{NavaroUGC07151}, \ref{BuckertUGC07151} and
\ref{EinastoUGC07151} we present the optimization values for the
SIDM model, and the other DM profiles. Also in Table
\ref{EVALUATIONUGC07151} we present the overall evaluation of the
SIDM model for the galaxy at hand. The resulting phenomenology is
viable.
\begin{table}[h!]
  \begin{center}
    \caption{Collisional Dark Matter Optimization Values}
    \label{collUGC07151}
     \begin{tabular}{|r|r|}
     \hline
      \textbf{Parameter}   & \textbf{Optimization Values}
      \\  \hline
     $\delta_{\gamma} $ & 0.000000000012
\\  \hline
$\gamma_0 $ & 1.0001 \\ \hline $K_0$ ($M_{\odot} \,
\mathrm{Kpc}^{-3} \, (\mathrm{km/s})^{2}$)& 1700  \\ \hline
    \end{tabular}
  \end{center}
\end{table}
\begin{table}[h!]
  \begin{center}
    \caption{NFW  Optimization Values}
    \label{NavaroUGC07151}
     \begin{tabular}{|r|r|}
     \hline
      \textbf{Parameter}   & \textbf{Optimization Values}
      \\  \hline
   $\rho_s$   & $5\times 10^7$
\\  \hline
$r_s$&  2.93
\\  \hline
    \end{tabular}
  \end{center}
\end{table}
\begin{table}[h!]
  \begin{center}
    \caption{Burkert Optimization Values}
    \label{BuckertUGC07151}
     \begin{tabular}{|r|r|}
     \hline
      \textbf{Parameter}   & \textbf{Optimization Values}
      \\  \hline
     $\rho_0^B$  & $1\times 10^8$
\\  \hline
$r_0$&  4.07
\\  \hline
    \end{tabular}
  \end{center}
\end{table}
\begin{table}[h!]
  \begin{center}
    \caption{Einasto Optimization Values}
    \label{EinastoUGC07151}
    \begin{tabular}{|r|r|}
     \hline
      \textbf{Parameter}   & \textbf{Optimization Values}
      \\  \hline
     $\rho_e$  &$1\times 10^7$
\\  \hline
$r_e$ & 3.8
\\  \hline
$n_e$ & 0.40
\\  \hline
    \end{tabular}
  \end{center}
\end{table}
\begin{table}[h!]
\centering \caption{Physical assessment of collisional DM
parameters for \texttt{UGC07151}.}
\begin{tabular}{lcc}
\hline
Parameter & Value   & Physical Verdict \\
\hline
$\gamma_0$ & $1.0001$ & Very close to isothermal\\
$\delta_\gamma$ & $1.2\times10^{-9}$ & Practically zero \\
$r_\gamma$ & $1.5\ \mathrm{Kpc}$ & Transition radius set inside the inner halo \\
$K_0$ ($M_{\odot}\,\mathrm{Kpc}^{-3}\,(\mathrm{km/s})^{2}$) & $1.7\times10^{3}$ & Enough pressure support \\
$r_c$ & $0.5\ \mathrm{Kpc}$ & Small core scale  \\
$p$ & $0.01$ & Extremely shallow decline of $K(r)$ \\
\hline
Overall & - & Self-consistent numerically  \\
\hline
\end{tabular}
\label{EVALUATIONUGC07151}
\end{table}


\subsection{The Galaxy UGC07232}

\begin{figure}[h!]
\centering
\includegraphics[width=20pc]{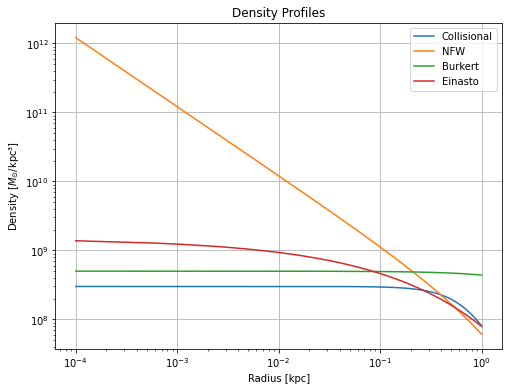}
\caption{The density of the collisional DM model (\ref{tanhmodel})
for the galaxy UGC07232, as a function of the radius.}
\label{UGC07232dens}
\end{figure}
\begin{figure}[h!]
\centering
\includegraphics[width=20pc]{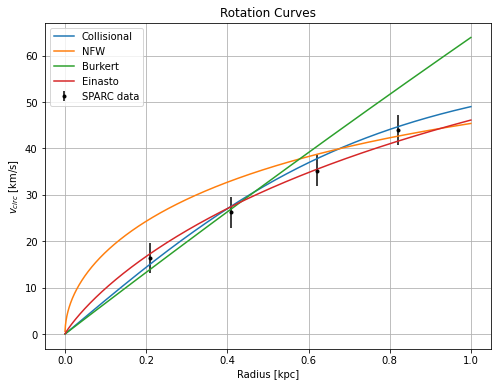}
\caption{The predicted rotation curves after using an optimization
for the collisional DM model (\ref{tanhmodel}), versus the SPARC
observational data for the galaxy UGC07232. We also plotted the
optimized curves for the NFW model, the Burkert model and the
Einasto model.} \label{UGC07232}
\end{figure}
\begin{figure}[h!]
\centering
\includegraphics[width=20pc]{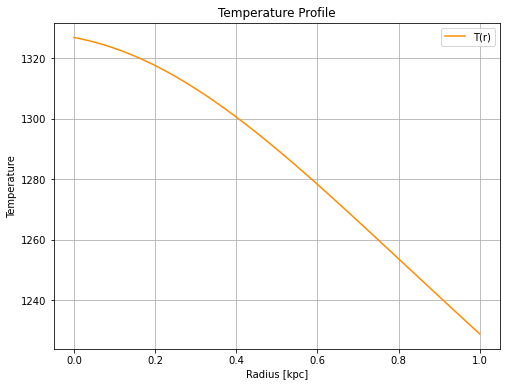}
\caption{The temperature as a function of the radius for the
collisional DM model (\ref{tanhmodel}) for the galaxy UGC07232.}
\label{UGC07232temp}
\end{figure}

For this galaxy, we shall choose $\rho_0=3\times
10^8$$M_{\odot}/\mathrm{Kpc}^{3}$. UGC07232 is a dwarf irregular/
late-type dwarfish disk galaxy, lying at a distance of
approximately \(4.3\)-\(4.5\) Mpc. In Figs. \ref{UGC07232dens},
\ref{UGC07232} and \ref{UGC07232temp} we present the density of
the collisional DM model, the predicted rotation curves after
using an optimization for the collisional DM model
(\ref{tanhmodel}), versus the SPARC observational data and the
temperature parameter as a function of the radius respectively. As
it can be seen, the SIDM model produces viable rotation curves
compatible with the SPARC data. Also in Tables \ref{collUGC07232},
\ref{NavaroUGC07232}, \ref{BuckertUGC07232} and
\ref{EinastoUGC07232} we present the optimization values for the
SIDM model, and the other DM profiles. Also in Table
\ref{EVALUATIONUGC07232} we present the overall evaluation of the
SIDM model for the galaxy at hand. The resulting phenomenology is
viable.
\begin{table}[h!]
  \begin{center}
    \caption{Collisional Dark Matter Optimization Values}
    \label{collUGC07232}
     \begin{tabular}{|r|r|}
     \hline
      \textbf{Parameter}   & \textbf{Optimization Values}
      \\  \hline
     $\delta_{\gamma} $ & 0.0000000012
\\  \hline
$\gamma_0 $ & 1.0001 \\ \hline $K_0$ ($M_{\odot} \,
\mathrm{Kpc}^{-3} \, (\mathrm{km/s})^{2}$)& 1500 \\ \hline
    \end{tabular}
  \end{center}
\end{table}
\begin{table}[h!]
  \begin{center}
    \caption{NFW  Optimization Values}
    \label{NavaroUGC07232}
     \begin{tabular}{|r|r|}
     \hline
      \textbf{Parameter}   & \textbf{Optimization Values}
      \\  \hline
   $\rho_s$   & $5\times 10^7$
\\  \hline
$r_s$& 2.43
\\  \hline
    \end{tabular}
  \end{center}
\end{table}
\begin{table}[h!]
  \begin{center}
    \caption{Burkert Optimization Values}
    \label{BuckertUGC07232}
     \begin{tabular}{|r|r|}
     \hline
      \textbf{Parameter}   & \textbf{Optimization Values}
      \\  \hline
     $\rho_0^B$  & $1\times 10^8$
\\  \hline
$r_0$& 8.07
\\  \hline
    \end{tabular}
  \end{center}
\end{table}
\begin{table}[h!]
  \begin{center}
    \caption{Einasto Optimization Values}
    \label{EinastoUGC07232}
    \begin{tabular}{|r|r|}
     \hline
      \textbf{Parameter}   & \textbf{Optimization Values}
      \\  \hline
     $\rho_e$  &$1\times 10^7$
\\  \hline
$r_e$ & 3.8
\\  \hline
$n_e$ & 0.40
\\  \hline
    \end{tabular}
  \end{center}
\end{table}
\begin{table}[h!]
\centering \caption{Physical assessment of collisional DM
parameters for \texttt{UGC07232}.}
\begin{tabular}{lcc}
\hline
Parameter & Value   & Physical Verdict \\
\hline
$\gamma_0$ & $1.0001$ & Slightly above isothermal \\
$\delta_\gamma$ & $1.2\times10^{-9}$ & Extremely small  \\
$r_\gamma$ & $1.5\ \mathrm{Kpc}$ & Transition radius placed inside the inner halo \\
$K_0$ ($M_{\odot}\,\mathrm{Kpc}^{-3}\,(\mathrm{km/s})^{2}$) & $1.5\times10^{3}$ & Modest entropy/pressure scale.\\
$r_c$ & $0.5\ \mathrm{Kpc}$ & Small core scale \\
$p$ & $0.01$ & Very shallow decline of $K(r)$ \\
\hline
Overall & - & Numerically self-consistent  \\
\hline
\end{tabular}
\label{EVALUATIONUGC07232}
\end{table}

\subsection{The Galaxy UGC07261}

\begin{figure}[h!]
\centering
\includegraphics[width=20pc]{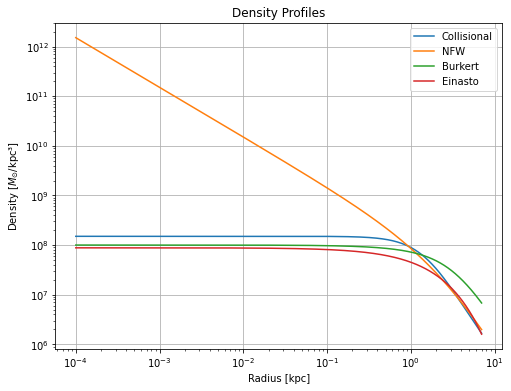}
\caption{The density of the collisional DM model (\ref{tanhmodel})
for the galaxy UGC07261, as a function of the radius.}
\label{UGC07261dens}
\end{figure}
\begin{figure}[h!]
\centering
\includegraphics[width=20pc]{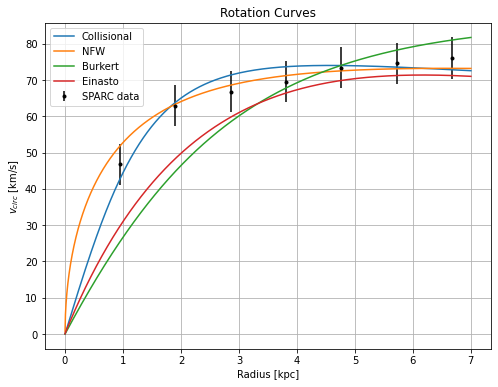}
\caption{The predicted rotation curves after using an optimization
for the collisional DM model (\ref{tanhmodel}), versus the SPARC
observational data for the galaxy UGC07261. We also plotted the
optimized curves for the NFW model, the Burkert model and the
Einasto model.} \label{UGC07261}
\end{figure}
\begin{figure}[h!]
\centering
\includegraphics[width=20pc]{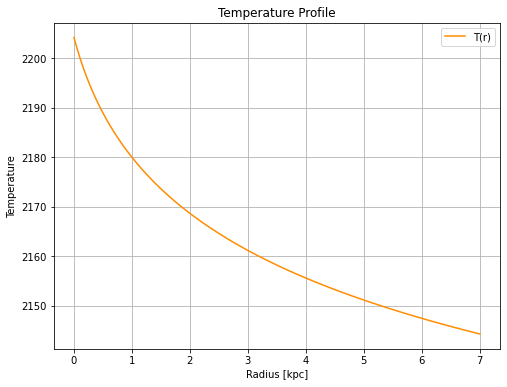}
\caption{The temperature as a function of the radius for the
collisional DM model (\ref{tanhmodel}) for the galaxy UGC07261.}
\label{UGC07261temp}
\end{figure}

For this galaxy, we shall choose $\rho_0=1.5\times
10^8$$M_{\odot}/\mathrm{Kpc}^{3}$. UGC07261 is a nearby, small
barred small, late-type, gas-rich disk system with a noticeable
bar structure embedded in its faint spiral disk. Its distance
estimates is $\simeq 12\ \mathrm{Mpc}$. In Figs.
\ref{UGC07261dens}, \ref{UGC07261} and \ref{UGC07261temp} we
present the density of the collisional DM model, the predicted
rotation curves after using an optimization for the collisional DM
model (\ref{tanhmodel}), versus the SPARC observational data and
the temperature parameter as a function of the radius
respectively. As it can be seen, the SIDM model produces viable
rotation curves compatible with the SPARC data. Also in Tables
\ref{collUGC07261}, \ref{NavaroUGC07261}, \ref{BuckertUGC07261}
and \ref{EinastoUGC07261} we present the optimization values for
the SIDM model, and the other DM profiles. Also in Table
\ref{EVALUATIONUGC07261} we present the overall evaluation of the
SIDM model for the galaxy at hand. The resulting phenomenology is
viable.
\begin{table}[h!]
  \begin{center}
    \caption{Collisional Dark Matter Optimization Values}
    \label{collUGC07261}
     \begin{tabular}{|r|r|}
     \hline
      \textbf{Parameter}   & \textbf{Optimization Values}
      \\  \hline
     $\delta_{\gamma} $ &  0.0000000012
\\  \hline
$\gamma_0 $ & 1.0001 \\ \hline $K_0$ ($M_{\odot} \,
\mathrm{Kpc}^{-3} \, (\mathrm{km/s})^{2}$)& 2200 \\ \hline
    \end{tabular}
  \end{center}
\end{table}
\begin{table}[h!]
  \begin{center}
    \caption{NFW  Optimization Values}
    \label{NavaroUGC07261}
     \begin{tabular}{|r|r|}
     \hline
      \textbf{Parameter}   & \textbf{Optimization Values}
      \\  \hline
   $\rho_s$   & $5\times 10^7$
\\  \hline
$r_s$&  3.03
\\  \hline
    \end{tabular}
  \end{center}
\end{table}
\begin{table}[h!]
  \begin{center}
    \caption{Burkert Optimization Values}
    \label{BuckertUGC07261}
     \begin{tabular}{|r|r|}
     \hline
      \textbf{Parameter}   & \textbf{Optimization Values}
      \\  \hline
     $\rho_0^B$  & $1\times 10^8$
\\  \hline
$r_0$&  3.53
\\  \hline
    \end{tabular}
  \end{center}
\end{table}
\begin{table}[h!]
  \begin{center}
    \caption{Einasto Optimization Values}
    \label{EinastoUGC07261}
    \begin{tabular}{|r|r|}
     \hline
      \textbf{Parameter}   & \textbf{Optimization Values}
      \\  \hline
     $\rho_e$  &$1\times 10^7$
\\  \hline
$r_e$ & 3.60
\\  \hline
$n_e$ & 0.92
\\  \hline
    \end{tabular}
  \end{center}
\end{table}
\begin{table}[h!]
\centering \caption{Physical assessment of collisional DM
parameters (UGC07261).}
\begin{tabular}{lcc}
\hline
Parameter & Value & Physical Verdict \\
\hline
$\gamma_0$ & $1.0001$ & Essentially isothermal \\
$\delta_\gamma$ & $1.2\times10^{-9}$ & Negligible \\
$r_\gamma$ & $1.5\ \mathrm{Kpc}$ & Transition radius inside inner halo \\
$K_0$ ($M_{\odot}\, \mathrm{Kpc}^{-3}\, (\mathrm{km/s})^{2}$) & $2.2\times10^{3}$ & Moderate entropy/pressure scale \\
$r_c$ & $0.5\ \mathrm{Kpc}$ & Small core scale  \\
$p$ & $0.01$ & Very shallow radial decline of $K(r)$ \\
\hline
Overall & - & Physically consistent \\
\hline
\end{tabular}
\label{EVALUATIONUGC07261}
\end{table}


\subsection{The Galaxy UGC07323 Non-viable, Extended is viable}

\begin{figure}[h!]
\centering
\includegraphics[width=20pc]{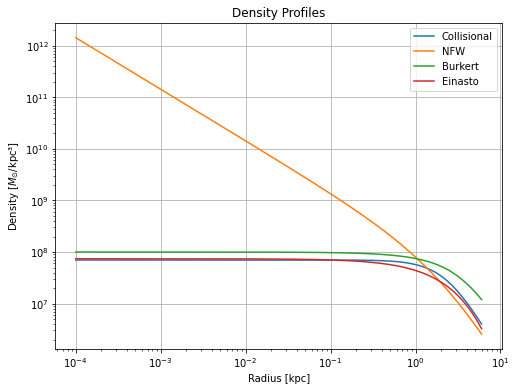}
\caption{The density of the collisional DM model (\ref{tanhmodel})
for the galaxy UGC07323, as a function of the radius.}
\label{UGC07323dens}
\end{figure}
\begin{figure}[h!]
\centering
\includegraphics[width=20pc]{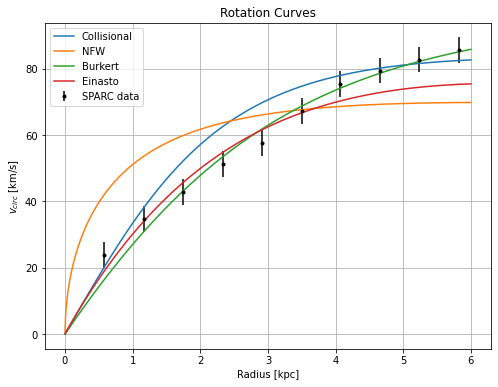}
\caption{The predicted rotation curves after using an optimization
for the collisional DM model (\ref{tanhmodel}), versus the SPARC
observational data for the galaxy UGC07323. We also plotted the
optimized curves for the NFW model, the Burkert model and the
Einasto model.} \label{UGC07323}
\end{figure}
\begin{figure}[h!]
\centering
\includegraphics[width=20pc]{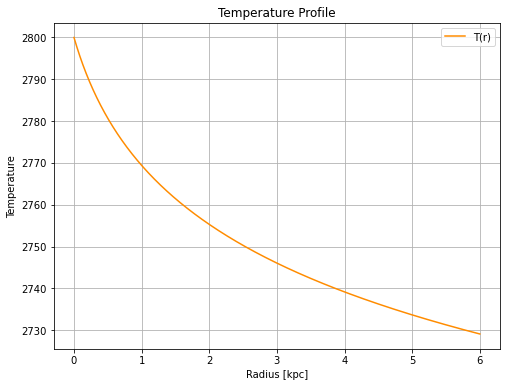}
\caption{The temperature as a function of the radius for the
collisional DM model (\ref{tanhmodel}) for the galaxy UGC07323.}
\label{UGC07323temp}
\end{figure}
\begin{figure}[h!]
\centering
\includegraphics[width=20pc]{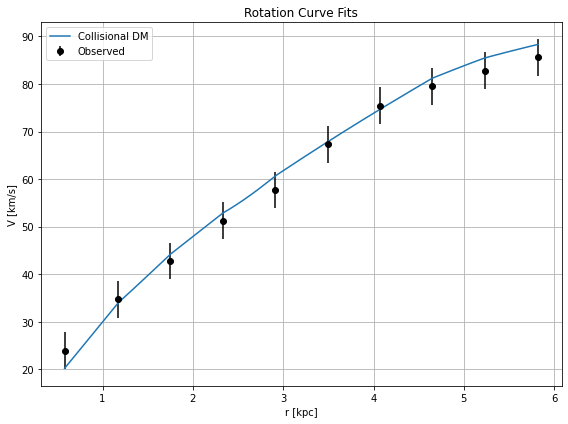}
\caption{The predicted rotation curves after using an optimization
for the collisional DM model (\ref{tanhmodel}), versus the
extended SPARC observational data for the galaxy UGC07323. The
model includes the rotation curves from all the components of the
galaxy, including gas and disk velocities, along with the
collisional DM model.} \label{extendedUGC07323}
\end{figure}

For this galaxy, we shall choose $\rho_0=7\times
10^7$$M_{\odot}/\mathrm{Kpc}^{3}$. UGC07323 is classified as a
late-type spiral (Sd) galaxy, relatively small but not strictly
''dwarf'' with a redshift $z\sim 0.0016$ and an inferred
Hubble-flow distance of order $\sim 10\ \mathrm{Mpc}$ (though with
possible peculiar-velocity corrections). In Figs.
\ref{UGC07323dens}, \ref{UGC07323} and \ref{UGC07323temp} we
present the density of the collisional DM model, the predicted
rotation curves after using an optimization for the collisional DM
model (\ref{tanhmodel}), versus the SPARC observational data and
the temperature parameter as a function of the radius
respectively. As it can be seen, the SIDM model produces
non-viable rotation curves incompatible with the SPARC data. Also
in Tables \ref{collUGC07323}, \ref{NavaroUGC07323},
\ref{BuckertUGC07323} and \ref{EinastoUGC07323} we present the
optimization values for the SIDM model, and the other DM profiles.
Also in Table \ref{EVALUATIONUGC07323} we present the overall
evaluation of the SIDM model for the galaxy at hand. The resulting
phenomenology is non-viable.
\begin{table}[h!]
  \begin{center}
    \caption{Collisional Dark Matter Optimization Values}
    \label{collUGC07323}
     \begin{tabular}{|r|r|}
     \hline
      \textbf{Parameter}   & \textbf{Optimization Values}
      \\  \hline
     $\delta_{\gamma} $ & 0.0000000012
\\  \hline
$\gamma_0 $ & 1.0001 \\ \hline $K_0$ ($M_{\odot} \,
\mathrm{Kpc}^{-3} \, (\mathrm{km/s})^{2}$)& 2800 \\ \hline
    \end{tabular}
  \end{center}
\end{table}
\begin{table}[h!]
  \begin{center}
    \caption{NFW  Optimization Values}
    \label{NavaroUGC07323}
     \begin{tabular}{|r|r|}
     \hline
      \textbf{Parameter}   & \textbf{Optimization Values}
      \\  \hline
   $\rho_s$   & $5\times 10^7$
\\  \hline
$r_s$& 2.89
\\  \hline
    \end{tabular}
  \end{center}
\end{table}
\begin{table}[h!]
  \begin{center}
    \caption{Burkert Optimization Values}
    \label{BuckertUGC07323}
     \begin{tabular}{|r|r|}
     \hline
      \textbf{Parameter}   & \textbf{Optimization Values}
      \\  \hline
     $\rho_0^B$  & $1\times 10^8$
\\  \hline
$r_0$& 3.94
\\  \hline
    \end{tabular}
  \end{center}
\end{table}
\begin{table}[h!]
  \begin{center}
    \caption{Einasto Optimization Values}
    \label{EinastoUGC07323}
    \begin{tabular}{|r|r|}
     \hline
      \textbf{Parameter}   & \textbf{Optimization Values}
      \\  \hline
     $\rho_e$  &$1\times 10^7$
\\  \hline
$r_e$ & 3.84
\\  \hline
$n_e$ & 1
\\  \hline
    \end{tabular}
  \end{center}
\end{table}
\begin{table}[h!]
\centering \caption{Physical assessment of collisional DM
parameters (UGC07323).}
\begin{tabular}{lcc}
\hline
Parameter & Value & Physical Verdict \\
\hline
$\gamma_0$ & $1.0001$ & Practically isothermal \\
$\delta_\gamma$ & $1.2\times10^{-9}$ & Negligible  \\
$r_\gamma$ & $1.5\ \mathrm{Kpc}$ & Transition radius placed inside inner halo \\
$K_0$ ($M_{\odot}\, \mathrm{Kpc}^{-3}\, (\mathrm{km/s})^{2}$) & $2.8\times10^{3}$ & Enough pressure support \\
$r_c$ & $0.5\ \mathrm{Kpc}$ & Small core radius  \\
$p$ & $0.01$ & Very shallow radial decline \\
\hline
Overall & - & Physically consistent \\
\hline
\end{tabular}
\label{EVALUATIONUGC07323}
\end{table}
Now the extended picture including the rotation velocity from the
other components of the galaxy, such as the disk and gas, makes
the collisional DM model viable for this galaxy. In Fig.
\ref{extendedUGC07323} we present the combined rotation curves
including the other components of the galaxy along with the
collisional matter. As it can be seen, the extended collisional DM
model is viable.
Also in Table \ref{evaluationextendedUGC07323} we present the
values of the free parameters of the collisional DM model for
which the maximum compatibility with the SPARC data comes for the
galaxy UGC07323.
\begin{table}[h!]
\centering \caption{Physical assessment of Extended collisional DM
parameters (UGC07323).}
\begin{tabular}{lcc}
\hline
Parameter & Value & Physical Verdict \\
\hline
$\gamma_0$ & 1.00001 & Essentially isothermal \\
$\delta_\gamma$ & 0.00001 & Practically zero radial variation  \\
$K_0$ & 3000 & Moderate entropy/pressure scale\\
ml\_disk & 0.82536793 & Realistic disk mass-to-light  \\
ml\_bulge & 0.00000000 & Zero bulge contribution  \\
\hline
Overall &-& Physically plausible \\
\hline
\end{tabular}
\label{evaluationextendedUGC07323}
\end{table}

\subsection{The Galaxy UGC07399 Non-viable}

\begin{figure}[h!]
\centering
\includegraphics[width=20pc]{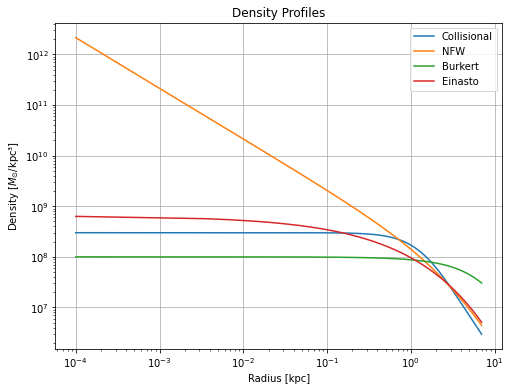}
\caption{The density of the collisional DM model (\ref{tanhmodel})
for the galaxy UGC07399, as a function of the radius.}
\label{UGC07399dens}
\end{figure}

\begin{figure}[h!]
\centering
\includegraphics[width=20pc]{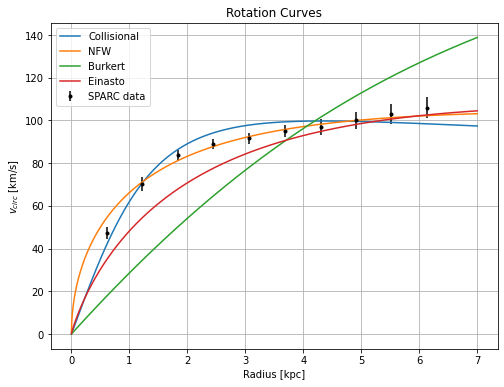}
\caption{The predicted rotation curves after using an optimization
for the collisional DM model (\ref{tanhmodel}), versus the SPARC
observational data for the galaxy UGC07399. We also plotted the
optimized curves for the NFW model, the Burkert model and the
Einasto model.} \label{UGC07399}
\end{figure}

\begin{figure}[h!]
\centering
\includegraphics[width=20pc]{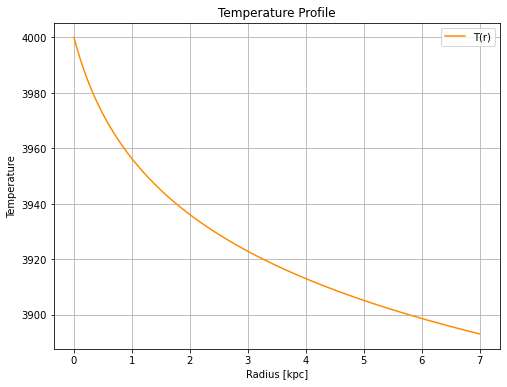}
\caption{The temperature as a function of the radius for the
collisional DM model (\ref{tanhmodel}) for the galaxy UGC07399.}
\label{UGC07399temp}
\end{figure}

\begin{figure}[h!]
\centering
\includegraphics[width=20pc]{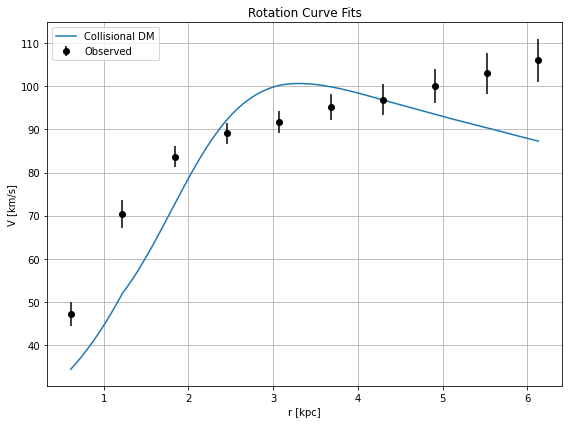}
\caption{The predicted rotation curves after using an optimization
for the collisional DM model (\ref{tanhmodel}), versus the
extended SPARC observational data for the galaxy UGC07399. The
model includes the rotation curves from all the components of the
galaxy, including gas and disk velocities, along with the
collisional DM model.} \label{extendedUGC07399}
\end{figure}

For this galaxy, we shall choose $\rho_0=3\times
10^8$$M_{\odot}/\mathrm{Kpc}^{3}$. UGC07399 is a small, late-type
barred spiral (SBcd / SBm) in Canes Venatici, catalogued as a
gas-rich, low-luminosity disk often treated as a
dwarf/Magellanic-type spiral. It lies at a nearby distance of
order $\sim 8\text{--}10\ \mathrm{Mpc}$. In Figs.
\ref{UGC07399dens}, \ref{UGC07399} and \ref{UGC07399temp} we
present the density of the collisional DM model, the predicted
rotation curves after using an optimization for the collisional DM
model (\ref{tanhmodel}), versus the SPARC observational data and
the temperature parameter as a function of the radius
respectively. As it can be seen, the SIDM model produces
non-viable rotation curves incompatible with the SPARC data. Also
in Tables \ref{collUGC07399}, \ref{NavaroUGC07399},
\ref{BuckertUGC07399} and \ref{EinastoUGC07399} we present the
optimization values for the SIDM model, and the other DM profiles.
Also in Table \ref{EVALUATIONUGC07399} we present the overall
evaluation of the SIDM model for the galaxy at hand. The resulting
phenomenology is non-viable.
\begin{table}[h!]
  \begin{center}
    \caption{Collisional Dark Matter Optimization Values}
    \label{collUGC07399}
     \begin{tabular}{|r|r|}
     \hline
      \textbf{Parameter}   & \textbf{Optimization Values}
      \\  \hline
     $\delta_{\gamma} $ & 0.0000000012
\\  \hline
$\gamma_0 $ & 1.0001 \\ \hline $K_0$ ($M_{\odot} \,
\mathrm{Kpc}^{-3} \, (\mathrm{km/s})^{2}$)& 4000 \\ \hline
    \end{tabular}
  \end{center}
\end{table}
\begin{table}[h!]
  \begin{center}
    \caption{NFW  Optimization Values}
    \label{NavaroUGC07399}
     \begin{tabular}{|r|r|}
     \hline
      \textbf{Parameter}   & \textbf{Optimization Values}
      \\  \hline
   $\rho_s$   & $5\times 10^7$
\\  \hline
$r_s$&  4.30
\\  \hline
    \end{tabular}
  \end{center}
\end{table}
\begin{table}[h!]
  \begin{center}
    \caption{Burkert Optimization Values}
    \label{BuckertUGC07399}
     \begin{tabular}{|r|r|}
     \hline
      \textbf{Parameter}   & \textbf{Optimization Values}
      \\  \hline
     $\rho_0^B$  & $1\times 10^8$
\\  \hline
$r_0$& 8.05
\\  \hline
    \end{tabular}
  \end{center}
\end{table}
\begin{table}[h!]
  \begin{center}
    \caption{Einasto Optimization Values}
    \label{EinastoUGC07399}
    \begin{tabular}{|r|r|}
     \hline
      \textbf{Parameter}   & \textbf{Optimization Values}
      \\  \hline
     $\rho_e$  &$1\times 10^7$
\\  \hline
$r_e$ & 5.12
\\  \hline
$n_e$ & 0.48
\\  \hline
    \end{tabular}
  \end{center}
\end{table}
\begin{table}[h!]
\centering \caption{Physical assessment of collisional DM
parameters (UGC07399).}
\begin{tabular}{lcc}
\hline
Parameter & Value & Physical Verdict \\
\hline
$\gamma_0$ & $1.0001$ & Practically isothermal \\
$\delta_\gamma$ & $1.2\times10^{-9}$ & Negligible \\
$r_\gamma$ & $1.5\ \mathrm{Kpc}$ & Transition radius inside inner halo \\
$K_0$ ($M_{\odot}\, \mathrm{Kpc}^{-3}\, (\mathrm{km/s})^{2}$) & $4.0\times10^{3}$ & Enough pressure support \\
$r_c$ & $0.5\ \mathrm{Kpc}$ & Small core radius  \\
$p$ & $0.01$ & Very shallow radial decline  \\
\hline
Overall & - & Physically consistent \\
\hline
\end{tabular}
\label{EVALUATIONUGC07399}
\end{table}
Now the extended picture including the rotation velocity from the
other components of the galaxy, such as the disk and gas, makes
the collisional DM model viable for this galaxy. In Fig.
\ref{extendedUGC07399} we present the combined rotation curves
including the other components of the galaxy along with the
collisional matter. As it can be seen, the extended collisional DM
model is non-viable.
Also in Table \ref{evaluationextendedUGC07399} we present the
values of the free parameters of the collisional DM model for
which the maximum compatibility with the SPARC data comes for the
galaxy UGC07399.
\begin{table}[h!]
\centering \caption{Physical assessment of Extended collisional DM
parameters (UGC07399).}
\begin{tabular}{lcc}
\hline
Parameter & Value & Physical Verdict \\
\hline
$\gamma_0$ & 1.11799952 & Slightly above isothermal \\
$\delta_\gamma$ & 0.17148054 & Significant radial variation in $\gamma(r)$  \\
$K_0$ & 3000 & Moderate entropy/pressure scale \\
ml\_disk & 1.00000000 & High disk mass-to-light  \\
ml\_bulge & 0.00000000 & Zero bulge contribution \\
\hline
Overall &-& Physically plausible \\
\hline
\end{tabular}
\label{evaluationextendedUGC07399}
\end{table}

\subsection{The Galaxy UGC07524 Non-viable Dwarf, Extended Viable, 3 parameter model}

\begin{figure}[h!]
\centering
\includegraphics[width=20pc]{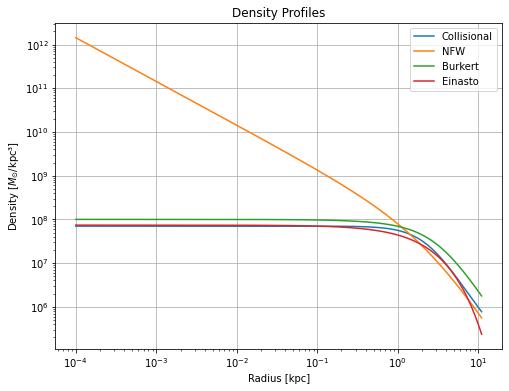}
\caption{The density of the collisional DM model (\ref{tanhmodel})
for the galaxy UGC07524, as a function of the radius.}
\label{UGC07524dens}
\end{figure}

\begin{figure}[h!]
\centering
\includegraphics[width=20pc]{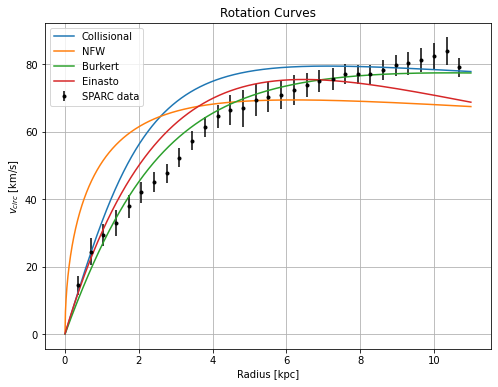}
\caption{The predicted rotation curves after using an optimization
for the collisional DM model (\ref{tanhmodel}), versus the SPARC
observational data for the galaxy UGC07524. We also plotted the
optimized curves for the NFW model, the Burkert model and the
Einasto model.} \label{UGC07524}
\end{figure}

\begin{figure}[h!]
\centering
\includegraphics[width=20pc]{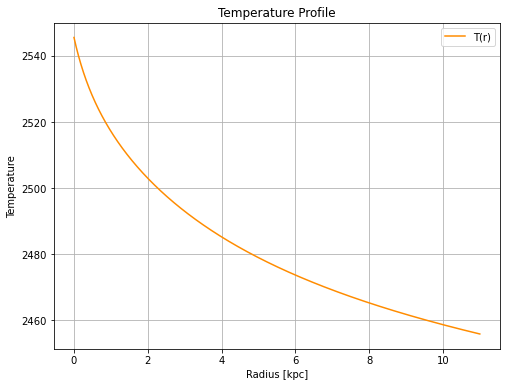}
\caption{The temperature as a function of the radius for the
collisional DM model (\ref{tanhmodel}) for the galaxy UGC07524.}
\label{UGC07524temp}
\end{figure}

\begin{figure}[h!]
\centering
\includegraphics[width=20pc]{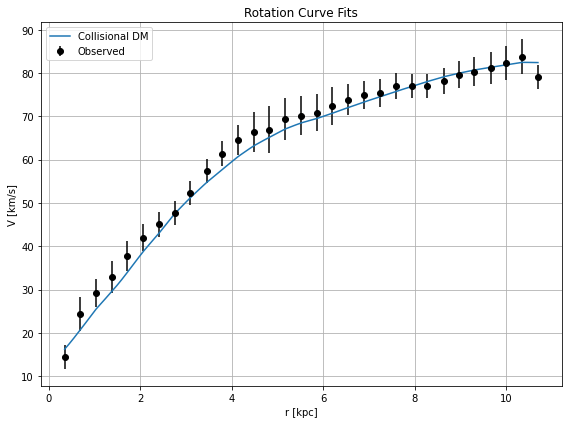}
\caption{The predicted rotation curves after using an optimization
for the collisional DM model (\ref{tanhmodel}), versus the
extended SPARC observational data for the galaxy UGC07524. The
model includes the rotation curves from all the components of the
galaxy, including gas and disk velocities, along with the
collisional DM model.} \label{extendedUGC07524}
\end{figure}

For this galaxy, we shall choose $\rho_0=7\times
10^7$$M_{\odot}/\mathrm{Kpc}^{3}$. UGC07524 is a nearby,
low-surface-brightness barred spiral galaxy (SA(s)m) in Canes
Venatici, catalogued as a gas-rich, low-luminosity disk often
treated as a dwarf/Magellanic-type spiral; it lies at a distance
of order $\sim 4.3\ \mathrm{Mpc}$. In Figs. \ref{UGC07524dens},
\ref{UGC07524} and \ref{UGC07524temp} we present the density of
the collisional DM model, the predicted rotation curves after
using an optimization for the collisional DM model
(\ref{tanhmodel}), versus the SPARC observational data and the
temperature parameter as a function of the radius respectively. As
it can be seen, the SIDM model produces non-viable rotation curves
incompatible with the SPARC data. Also in Tables
\ref{collUGC07524}, \ref{NavaroUGC07524}, \ref{BuckertUGC07524}
and \ref{EinastoUGC07524} we present the optimization values for
the SIDM model, and the other DM profiles. Also in Table
\ref{EVALUATIONUGC07524} we present the overall evaluation of the
SIDM model for the galaxy at hand. The resulting phenomenology is
non-viable.
\begin{table}[h!]
  \begin{center}
    \caption{Collisional Dark Matter Optimization Values}
    \label{collUGC07524}
     \begin{tabular}{|r|r|}
     \hline
      \textbf{Parameter}   & \textbf{Optimization Values}
      \\  \hline
     $\delta_{\gamma} $ & 0.0000000012
\\  \hline
$\gamma_0 $ & 1.0001 \\ \hline $K_0$ ($M_{\odot} \,
\mathrm{Kpc}^{-3} \, (\mathrm{km/s})^{2}$)& 2500  \\ \hline
    \end{tabular}
  \end{center}
\end{table}
\begin{table}[h!]
  \begin{center}
    \caption{NFW  Optimization Values}
    \label{NavaroUGC07524}
     \begin{tabular}{|r|r|}
     \hline
      \textbf{Parameter}   & \textbf{Optimization Values}
      \\  \hline
   $\rho_s$   & $5\times 10^7$
\\  \hline
$r_s$&   2.87
\\  \hline
    \end{tabular}
  \end{center}
\end{table}
\begin{table}[h!]
  \begin{center}
    \caption{Burkert Optimization Values}
    \label{BuckertUGC07524}
     \begin{tabular}{|r|r|}
     \hline
      \textbf{Parameter}   & \textbf{Optimization Values}
      \\  \hline
     $\rho_0^B$  & $1\times 10^8$
\\  \hline
$r_0$& 3.21
\\  \hline
    \end{tabular}
  \end{center}
\end{table}
\begin{table}[h!]
  \begin{center}
    \caption{Einasto Optimization Values}
    \label{EinastoUGC07524}
    \begin{tabular}{|r|r|}
     \hline
      \textbf{Parameter}   & \textbf{Optimization Values}
      \\  \hline
     $\rho_e$  &$1\times 10^7$
\\  \hline
$r_e$ & 3.83
\\  \hline
$n_e$ & 1
\\  \hline
    \end{tabular}
  \end{center}
\end{table}
\begin{table}[h!]
\centering \caption{Physical assessment of collisional DM
parameters (UGC07524).}
\begin{tabular}{lcc}
\hline
Parameter & Value & Physical Verdict \\
\hline
$\gamma_0$ & $1.0001$ & Nearly isothermal \\
$\delta_\gamma$ & $1.2\times10^{-9}$ & Negligible \\
$r_\gamma$ & $1.5\ \mathrm{Kpc}$ & Transition radius inside inner halo \\
$K_0$ ($M_{\odot}\, \mathrm{Kpc}^{-3}\, (\mathrm{km/s})^{2}$) & $2.5\times10^{3}$ & Enough pressure support \\
$r_c$ & $0.5\ \mathrm{Kpc}$ & Small core radius \\
$p$ & $0.01$ & Very shallow radial decline of $K(r)$ \\
\hline
Overall & - & Physically plausible \\
\hline
\end{tabular}
\label{EVALUATIONUGC07524}
\end{table}
Now the extended picture including the rotation velocity from the
other components of the galaxy, such as the disk and gas, makes
the collisional DM model viable for this galaxy. In Fig.
\ref{extendedUGC07524} we present the combined rotation curves
including the other components of the galaxy along with the
collisional matter. As it can be seen, the extended collisional DM
model is viable.
Also in Table \ref{evaluationextendedUGC07524} we present the
values of the free parameters of the collisional DM model for
which the maximum compatibility with the SPARC data comes for the
galaxy UGC07524.
\begin{table}[h!]
\centering \caption{Physical assessment of Extended collisional DM
parameters (UGC07524).}
\begin{tabular}{lcc}
\hline
Parameter & Value & Physical Verdict \\
\hline
$\gamma_0$ & 1.0001 & Essentially isothermal \\
$\delta_\gamma$ & 0.00001681718 & Practically zero radial variation \\
$K_0$ & 1750 & Lower entropy/pressure scale than the 3000 benchmark \\
ml\_disk & 0.99000000 & Disk mass-to-light near unity \\
ml\_bulge & 0.00000000 & Zero bulge contribution  \\
\hline
Overall &-& Physically plausible \\
\hline
\end{tabular}
\label{evaluationextendedUGC07524}
\end{table}
\begin{table}[h!]
\centering \caption{Physical assessment of collisional DM
parameters (UGC07559).}
\begin{tabular}{lcc}
\hline
Parameter & Value & Physical Verdict \\
\hline
$\gamma_0$ & $1.0001$ & Nearly isothermal \\
$\delta_\gamma$ & $1.2\times10^{-9}$ & Negligible \\
$r_\gamma$ & $1.5\ \mathrm{Kpc}$ & Transition radius inside inner halo \\
$K_0$ ($M_{\odot}\, \mathrm{Kpc}^{-3}\, (\mathrm{km/s})^{2}$) & $5.0\times10^{2}$ & Enough pressure support for dwarf system \\
$r_c$ & $0.5\ \mathrm{Kpc}$ & Small core radius  \\
$p$ & $0.01$ & Very shallow radial decline of $K(r)$\\
\hline
Overall & - & Physically plausible \\
\hline
\end{tabular}
\label{EVALUATIONUGC07559}
\end{table}

\subsection{The Galaxy UGC07559}

\begin{figure}[h!]
\centering
\includegraphics[width=20pc]{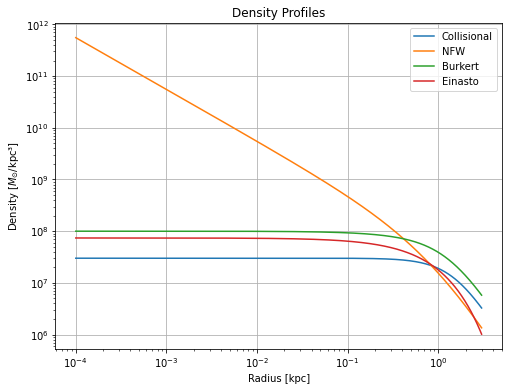}
\caption{The density of the collisional DM model (\ref{tanhmodel})
for the galaxy UGC07559, as a function of the radius.}
\label{UGC07559dens}
\end{figure}

\begin{figure}[h!]
\centering
\includegraphics[width=20pc]{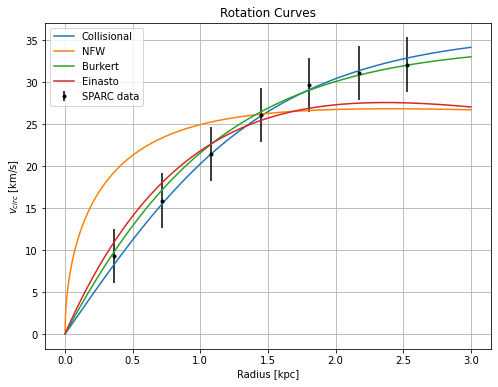}
\caption{The predicted rotation curves after using an optimization
for the collisional DM model (\ref{tanhmodel}), versus the SPARC
observational data for the galaxy UGC07559. We also plotted the
optimized curves for the NFW model, the Burkert model and the
Einasto model.} \label{UGC07559}
\end{figure}

\begin{figure}[h!]
\centering
\includegraphics[width=20pc]{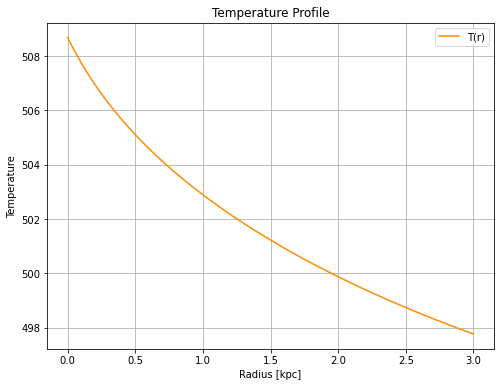}
\caption{The temperature as a function of the radius for the
collisional DM model (\ref{tanhmodel}) for the galaxy UGC07559.}
\label{UGC07559temp}
\end{figure}

For this galaxy, we shall choose $\rho_0=3\times
10^7$$M_{\odot}/\mathrm{Kpc}^{3}$. UGC07559 is a nearby,
low-luminosity barred spiral galaxy located in the Coma Berenices
constellation. Despite having a small disk and a weak bar, its low
luminosity, low mass, and small optical radius (a few Kpc)
classify it firmly as a dwarf system. It lies at a distance of
approximately $2.7\ \mathrm{Mpc}$ (redshift $z \sim 0.00043$). In
Figs. \ref{UGC07559dens}, \ref{UGC07559} and \ref{UGC07559temp} we
present the density of the collisional DM model, the predicted
rotation curves after using an optimization for the collisional DM
model (\ref{tanhmodel}), versus the SPARC observational data and
the temperature parameter as a function of the radius
respectively. As it can be seen, the SIDM model produces viable
rotation curves compatible with the SPARC data. Also in Tables
\ref{collUGC07559}, \ref{NavaroUGC07559}, \ref{BuckertUGC07559}
and \ref{EinastoUGC07559} we present the optimization values for
the SIDM model, and the other DM profiles. Also in Table
\ref{EVALUATIONUGC07559} we present the overall evaluation of the
SIDM model for the galaxy at hand. The resulting phenomenology is
viable.
\begin{table}[h!]
  \begin{center}
    \caption{Collisional Dark Matter Optimization Values}
    \label{collUGC07559}
     \begin{tabular}{|r|r|}
     \hline
      \textbf{Parameter}   & \textbf{Optimization Values}
      \\  \hline
     $\delta_{\gamma} $ & 0.0000000012
\\  \hline
$\gamma_0 $ & 1.0001 \\ \hline $K_0$ ($M_{\odot} \,
\mathrm{Kpc}^{-3} \, (\mathrm{km/s})^{2}$)& 500  \\ \hline
    \end{tabular}
  \end{center}
\end{table}
\begin{table}[h!]
  \begin{center}
    \caption{NFW  Optimization Values}
    \label{NavaroUGC07559}
     \begin{tabular}{|r|r|}
     \hline
      \textbf{Parameter}   & \textbf{Optimization Values}
      \\  \hline
   $\rho_s$   & $5\times 10^7$
\\  \hline
$r_s$&  1.11
\\  \hline
    \end{tabular}
  \end{center}
\end{table}
\begin{table}[h!]
  \begin{center}
    \caption{Burkert Optimization Values}
    \label{BuckertUGC07559}
     \begin{tabular}{|r|r|}
     \hline
      \textbf{Parameter}   & \textbf{Optimization Values}
      \\  \hline
     $\rho_0^B$  & $1\times 10^8$
\\  \hline
$r_0$&  1.41
\\  \hline
    \end{tabular}
  \end{center}
\end{table}
\begin{table}[h!]
  \begin{center}
    \caption{Einasto Optimization Values}
    \label{EinastoUGC07559}
    \begin{tabular}{|r|r|}
     \hline
      \textbf{Parameter}   & \textbf{Optimization Values}
      \\  \hline
     $\rho_e$  &$1\times 10^7$
\\  \hline
$r_e$ & 1.40
\\  \hline
$n_e$ & 1
\\  \hline
    \end{tabular}
  \end{center}
\end{table}
\begin{table}[h!]
\centering \caption{Physical assessment of collisional DM
parameters (UGC07559).}
\begin{tabular}{lcc}
\hline
Parameter & Value & Physical Verdict \\
\hline
$\gamma_0$ & $1.0001$ & Nearly isothermal \\
$\delta_\gamma$ & $1.2\times10^{-9}$ & Negligible  \\
$r_\gamma$ & $1.5\ \mathrm{Kpc}$ & Transition radius inside inner halo \\
$K_0$ ($M_{\odot}\, \mathrm{Kpc}^{-3}\, (\mathrm{km/s})^{2}$) & $5.0\times10^{2}$ & Enough pressure support for dwarf system \\
$r_c$ & $0.5\ \mathrm{Kpc}$ & Small core radius \\
$p$ & $0.01$ & Very shallow radial decline of $K(r)$ \\
\hline
Overall & - & Physically plausible \\
\hline
\end{tabular}
\label{EVALUATIONUGC07559}
\end{table}

\subsection{The Galaxy UGC07577}

For this galaxy, we shall choose $\rho_0=1\times
10^7$$M_{\odot}/\mathrm{Kpc}^{3}$. UGC07577 is a nearby,
low-luminosity barred spiral galaxy located in the constellation
Coma Berenices. It lies at a distance of approximately $2.7\
\mathrm{Mpc}$ (redshift $z \sim 0.00043$).
\begin{figure}[h!]
\centering
\includegraphics[width=20pc]{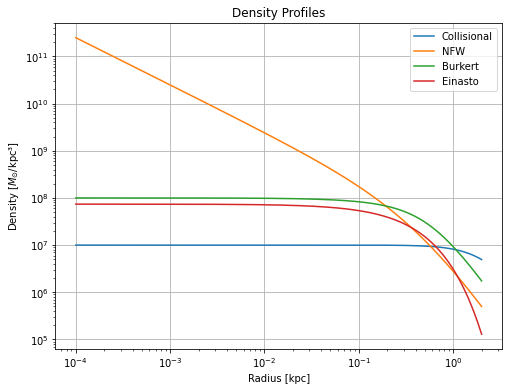}
\caption{The density of the collisional DM model (\ref{tanhmodel})
for the galaxy UGC07577, as a function of the radius.}
\label{UGC07577dens}
\end{figure}
\begin{figure}[h!]
\centering
\includegraphics[width=20pc]{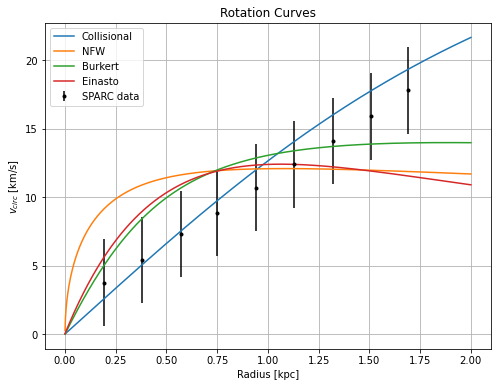}
\caption{The predicted rotation curves after using an optimization
for the collisional DM model (\ref{tanhmodel}), versus the SPARC
observational data for the galaxy UGC07577. We also plotted the
optimized curves for the NFW model, the Burkert model and the
Einasto model.} \label{UGC07577}
\end{figure}
\begin{table}[h!]
  \begin{center}
    \caption{Collisional Dark Matter Optimization Values}
    \label{collUGC07577}
     \begin{tabular}{|r|r|}
     \hline
      \textbf{Parameter}   & \textbf{Optimization Values}
      \\  \hline
     $\delta_{\gamma} $ & 0.0000000012
\\  \hline
$\gamma_0 $ & 1.0001 \\ \hline $K_0$ ($M_{\odot} \,
\mathrm{Kpc}^{-3} \, (\mathrm{km/s})^{2}$)& 400  \\ \hline
    \end{tabular}
  \end{center}
\end{table}
\begin{table}[h!]
  \begin{center}
    \caption{NFW  Optimization Values}
    \label{NavaroUGC07577}
     \begin{tabular}{|r|r|}
     \hline
      \textbf{Parameter}   & \textbf{Optimization Values}
      \\  \hline
   $\rho_s$   & $5\times 10^7$
\\  \hline
$r_s$&  0.50
\\  \hline
    \end{tabular}
  \end{center}
\end{table}
\begin{figure}[h!]
\centering
\includegraphics[width=20pc]{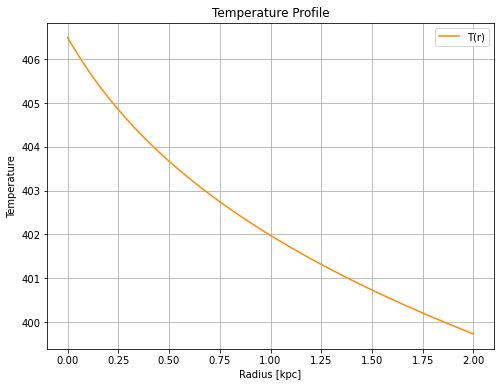}
\caption{The temperature as a function of the radius for the
collisional DM model (\ref{tanhmodel}) for the galaxy UGC07577.}
\label{UGC07577temp}
\end{figure}
\begin{table}[h!]
  \begin{center}
    \caption{Burkert Optimization Values}
    \label{BuckertUGC07577}
     \begin{tabular}{|r|r|}
     \hline
      \textbf{Parameter}   & \textbf{Optimization Values}
      \\  \hline
     $\rho_0^B$  & $1\times 10^8$
\\  \hline
$r_0$& 0.58
\\  \hline
    \end{tabular}
  \end{center}
\end{table}
\begin{table}[h!]
  \begin{center}
    \caption{Einasto Optimization Values}
    \label{EinastoUGC07577}
    \begin{tabular}{|r|r|}
     \hline
      \textbf{Parameter}   & \textbf{Optimization Values}
      \\  \hline
     $\rho_e$  &$1\times 10^7$
\\  \hline
$r_e$ & 0.63
\\  \hline
$n_e$ & 1
\\  \hline
    \end{tabular}
  \end{center}
\end{table}
\begin{table}[h!]
\centering \caption{Physical assessment of collisional DM
parameters (UGC07577).}
\begin{tabular}{lcc}
\hline
Parameter & Value & Physical Verdict \\
\hline
$\gamma_0$ & $1.0001$ & Nearly isothermal \\
$\delta_\gamma$ & $1.2\times10^{-9}$ & Negligible variation \\
$r_\gamma$ & $1.5\ \mathrm{Kpc}$ & Transition radius within inner halo \\
$K_0$ ($M_{\odot}\, \mathrm{Kpc}^{-3}\, (\mathrm{km/s})^{2}$) & $4.0\times10^{2}$ & Enough pressure support \\
$r_c$ & $0.5\ \mathrm{Kpc}$ & Small core radius \\
$p$ & $0.01$ & Very shallow radial decline of $K(r)$ \\
\hline
Overall & - & Physically plausible  \\
\hline
\end{tabular}
\label{EVALUATIONUGC07577}
\end{table}

\subsection{The Galaxy UGC07603}

\begin{figure}[h!]
\centering
\includegraphics[width=20pc]{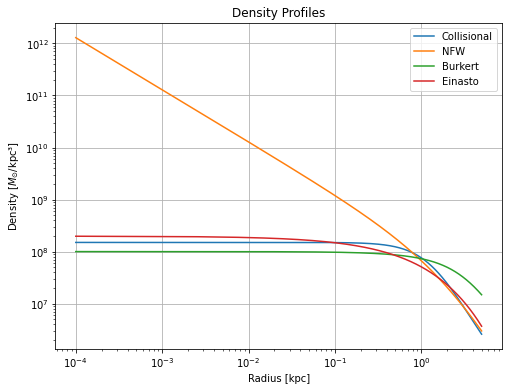}
\caption{The density of the collisional DM model (\ref{tanhmodel})
for the galaxy UGC07603, as a function of the radius.}
\label{UGC07603dens}
\end{figure}
\begin{figure}[h!]
\centering
\includegraphics[width=20pc]{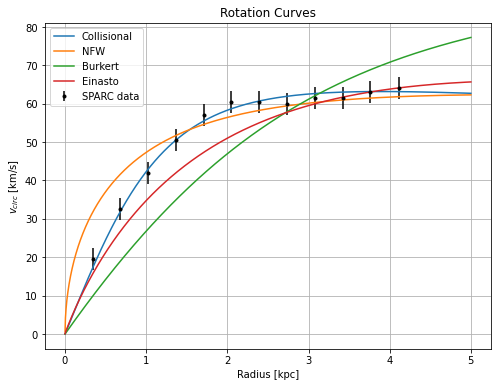}
\caption{The predicted rotation curves after using an optimization
for the collisional DM model (\ref{tanhmodel}), versus the SPARC
observational data for the galaxy UGC07603. We also plotted the
optimized curves for the NFW model, the Burkert model and the
Einasto model.} \label{UGC07603}
\end{figure}
\begin{figure}[h!]
\centering
\includegraphics[width=20pc]{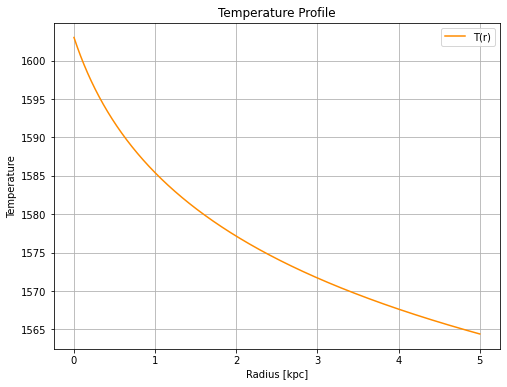}
\caption{The temperature as a function of the radius for the
collisional DM model (\ref{tanhmodel}) for the galaxy UGC07603.}
\label{UGC07603temp}
\end{figure}

For this galaxy, we shall choose $\rho_0=1.5\times
10^8$$M_{\odot}/\mathrm{Kpc}^{3}$. UGC07603 is a nearby,
low-luminosity barred spiral galaxy (SBcd) located in the
constellation Coma Berenices. It lies at a distance of
approximately $2.7\ \mathrm{Mpc}$ (redshift $z \sim 0.00043$). In
Figs. \ref{UGC07603dens}, \ref{UGC07603} and \ref{UGC07603temp} we
present the density of the collisional DM model, the predicted
rotation curves after using an optimization for the collisional DM
model (\ref{tanhmodel}), versus the SPARC observational data and
the temperature parameter as a function of the radius
respectively. As it can be seen, the SIDM model produces viable
rotation curves compatible with the SPARC data. Also in Tables
\ref{collUGC07603}, \ref{NavaroUGC07603}, \ref{BuckertUGC07603}
and \ref{EinastoUGC07603} we present the optimization values for
the SIDM model, and the other DM profiles. Also in Table
\ref{EVALUATIONUGC07603} we present the overall evaluation of the
SIDM model for the galaxy at hand. The resulting phenomenology is
viable.
\begin{table}[h!]
  \begin{center}
    \caption{Collisional Dark Matter Optimization Values}
    \label{collUGC07603}
     \begin{tabular}{|r|r|}
     \hline
      \textbf{Parameter}   & \textbf{Optimization Values}
      \\  \hline
     $\delta_{\gamma} $ & 0.0000000012
\\  \hline
$\gamma_0 $ & 1.0001 \\ \hline $K_0$ ($M_{\odot} \,
\mathrm{Kpc}^{-3} \, (\mathrm{km/s})^{2}$)& 1600  \\ \hline
    \end{tabular}
  \end{center}
\end{table}
\begin{table}[h!]
  \begin{center}
    \caption{NFW  Optimization Values}
    \label{NavaroUGC07603}
     \begin{tabular}{|r|r|}
     \hline
      \textbf{Parameter}   & \textbf{Optimization Values}
      \\  \hline
   $\rho_s$   & $5\times 10^7$
\\  \hline
$r_s$&  2.58
\\  \hline
    \end{tabular}
  \end{center}
\end{table}
\begin{table}[h!]
  \begin{center}
    \caption{Burkert Optimization Values}
    \label{BuckertUGC07603}
     \begin{tabular}{|r|r|}
     \hline
      \textbf{Parameter}   & \textbf{Optimization Values}
      \\  \hline
     $\rho_0^B$  & $1\times 10^8$
\\  \hline
$r_0$& 3.67
\\  \hline
    \end{tabular}
  \end{center}
\end{table}
\begin{table}[h!]
  \begin{center}
    \caption{Einasto Optimization Values}
    \label{EinastoUGC07603}
    \begin{tabular}{|r|r|}
     \hline
      \textbf{Parameter}   & \textbf{Optimization Values}
      \\  \hline
     $\rho_e$  &$1\times 10^7$
\\  \hline
$r_e$ & 3.25
\\  \hline
$n_e$ & 0.67
\\  \hline
    \end{tabular}
  \end{center}
\end{table}
\begin{table}[h!]
\centering \caption{Physical assessment of collisional DM
parameters (UGC07603).}
\begin{tabular}{lcc}
\hline
Parameter & Value & Physical Verdict \\
\hline
$\gamma_0$ & $1.0001$ & Nearly perfectly isothermal \\
$\delta_\gamma$ & $1.2\times10^{-9}$ & Negligible variation \\
$r_\gamma$ & $1.5\ \mathrm{Kpc}$ & Transition radius within inner halo\\
$K_0$ ($M_{\odot}\, \mathrm{Kpc}^{-3}\, (\mathrm{km/s})^{2}$) & $1.6\times10^{3}$ & Higher internal dispersion/entropy scale\\
$r_c$ & $0.5\ \mathrm{Kpc}$ & Small core radius  \\
$p$ & $0.01$ & Very shallow radial decline of $K(r)$ \\
\hline
Overall & - & Physically plausible \\
\hline
\end{tabular}
\label{EVALUATIONUGC07603}
\end{table}


\subsection{The Galaxy UGC07608}

\begin{figure}[h!]
\centering
\includegraphics[width=20pc]{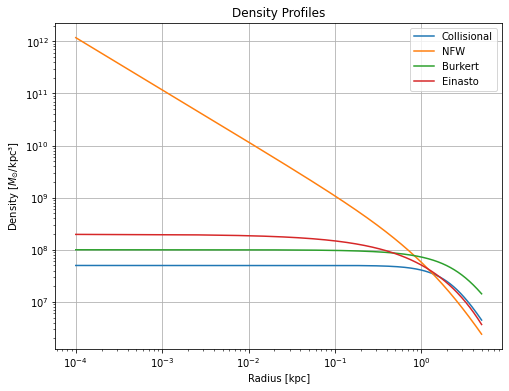}
\caption{The density of the collisional DM model (\ref{tanhmodel})
for the galaxy UGC07608, as a function of the radius.}
\label{UGC07608dens}
\end{figure}
\begin{figure}[h!]
\centering
\includegraphics[width=20pc]{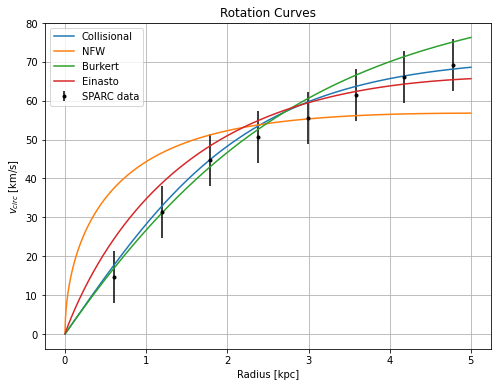}
\caption{The predicted rotation curves after using an optimization
for the collisional DM model (\ref{tanhmodel}), versus the SPARC
observational data for the galaxy UGC07608. We also plotted the
optimized curves for the NFW model, the Burkert model and the
Einasto model.} \label{UGC07608}
\end{figure}
\begin{figure}[h!]
\centering
\includegraphics[width=20pc]{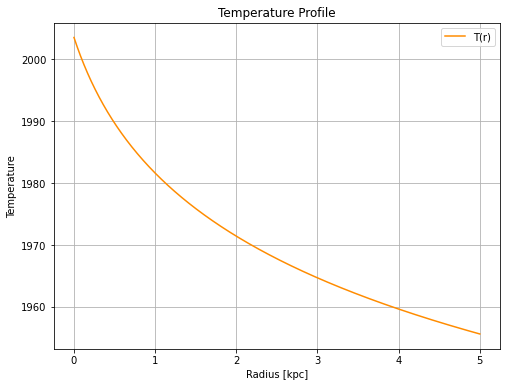}
\caption{The temperature as a function of the radius for the
collisional DM model (\ref{tanhmodel}) for the galaxy UGC07608.}
\label{UGC07608temp}
\end{figure}

For this galaxy, we shall choose $\rho_0=5\times
10^7$$M_{\odot}/\mathrm{Kpc}^{3}$. The galaxy UGC7608 is a dwarf
irregular galaxy situated approximately $1.7\ \mathrm{Mpc}$ from
the Milky Way. It is characterized by a low-surface-brightness and
lacks a well-defined spiral structure, distinguishing it from
typical spiral galaxies. The optical radius of UGC 7608 is
approximately $2.5\ \mathrm{Kpc}$, while its HI radius extends to
about $5.5\ \mathrm{Kpc}$, indicating a significant amount of
neutral hydrogen extending beyond the optical disk. In Figs.
\ref{UGC07608dens}, \ref{UGC07608} and \ref{UGC07608temp} we
present the density of the collisional DM model, the predicted
rotation curves after using an optimization for the collisional DM
model (\ref{tanhmodel}), versus the SPARC observational data and
the temperature parameter as a function of the radius
respectively. As it can be seen, the SIDM model produces viable
rotation curves compatible with the SPARC data. Also in Tables
\ref{collUGC07608}, \ref{NavaroUGC07608}, \ref{BuckertUGC07608}
and \ref{EinastoUGC07608} we present the optimization values for
the SIDM model, and the other DM profiles. Also in Table
\ref{EVALUATIONUGC07608} we present the overall evaluation of the
SIDM model for the galaxy at hand. The resulting phenomenology is
viable.
\begin{table}[h!]
  \begin{center}
    \caption{Collisional Dark Matter Optimization Values}
    \label{collUGC07608}
     \begin{tabular}{|r|r|}
     \hline
      \textbf{Parameter}   & \textbf{Optimization Values}
      \\  \hline
     $\delta_{\gamma} $ & 0.0000000012
\\  \hline
$\gamma_0 $ & 1.0001 \\ \hline $K_0$ ($M_{\odot} \,
\mathrm{Kpc}^{-3} \, (\mathrm{km/s})^{2}$)& 2000  \\ \hline
    \end{tabular}
  \end{center}
\end{table}
\begin{table}[h!]
  \begin{center}
    \caption{NFW  Optimization Values}
    \label{NavaroUGC07608}
     \begin{tabular}{|r|r|}
     \hline
      \textbf{Parameter}   & \textbf{Optimization Values}
      \\  \hline
   $\rho_s$   & $5\times 10^7$
\\  \hline
$r_s$&  2.35
\\  \hline
    \end{tabular}
  \end{center}
\end{table}
\begin{table}[h!]
  \begin{center}
    \caption{Burkert Optimization Values}
    \label{BuckertUGC07608}
     \begin{tabular}{|r|r|}
     \hline
      \textbf{Parameter}   & \textbf{Optimization Values}
      \\  \hline
     $\rho_0^B$  & $1\times 10^8$
\\  \hline
$r_0$&  3.6
\\  \hline
    \end{tabular}
  \end{center}
\end{table}
\begin{table}[h!]
  \begin{center}
    \caption{Einasto Optimization Values}
    \label{EinastoUGC07608}
    \begin{tabular}{|r|r|}
     \hline
      \textbf{Parameter}   & \textbf{Optimization Values}
      \\  \hline
     $\rho_e$  &$1\times 10^7$
\\  \hline
$r_e$ & 3.25
\\  \hline
$n_e$ & 0.67
\\  \hline
    \end{tabular}
  \end{center}
\end{table}
\begin{table}[h!]
\centering \caption{Physical assessment of collisional DM
parameters for UGC07608.}
\begin{tabular}{lcc}
\hline
Parameter & Value & Physical Verdict \\
\hline
$\gamma_0$ & 1.0001 & Almost perfectly isothermal \\
$\delta_\gamma$ & $1.2 \times 10^{-9}$ & Essentially no variation \\
$r_\gamma$ & 1.5 Kpc & Transition radius irrelevant due to negligible $\delta_\gamma$ \\
$K_0$ & 2000 & High entropy scale \\
$r_c$ & 0.5 Kpc & Small core radius, reasonable for inner halo \\
$p$ & 0.01 & Nearly constant $K(r)$, minimal radial variation \\
\hline
Overall &-& Essentially isothermal \\
\hline
\end{tabular}
\label{EVALUATIONUGC07608}
\end{table}

\subsection{The Galaxy UGC07690}

\begin{figure}[h!]
\centering
\includegraphics[width=20pc]{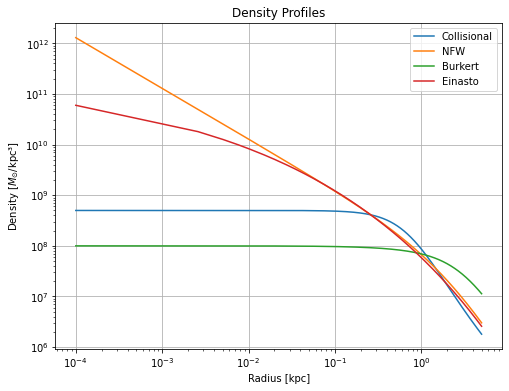}
\caption{The density of the collisional DM model (\ref{tanhmodel})
for the galaxy UGC07690, as a function of the radius.}
\label{UGC07690dens}
\end{figure}

\begin{figure}[h!]
\centering
\includegraphics[width=20pc]{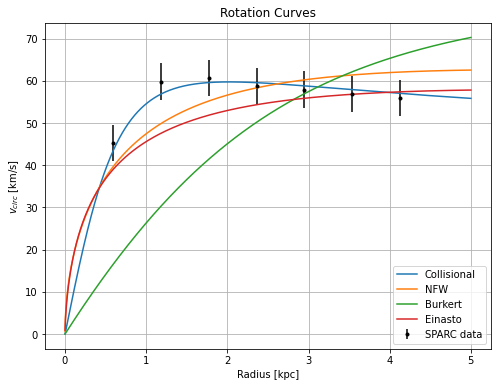}
\caption{The predicted rotation curves after using an optimization
for the collisional DM model (\ref{tanhmodel}), versus the SPARC
observational data for the galaxy UGC07690. We also plotted the
optimized curves for the NFW model, the Burkert model and the
Einasto model.} \label{UGC07690}
\end{figure}

\begin{figure}[h!]
\centering
\includegraphics[width=20pc]{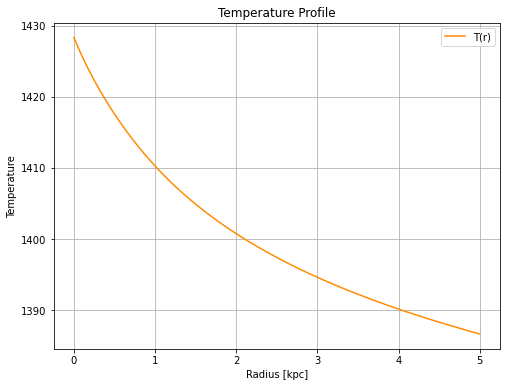}
\caption{The temperature as a function of the radius for the
collisional DM model (\ref{tanhmodel}) for the galaxy UGC07690.}
\label{UGC07690temp}
\end{figure}

For this galaxy, we shall choose $\rho_0=5\times
10^8$$M_{\odot}/\mathrm{Kpc}^{3}$. The galaxy UGC07690 is a barred
spiral galaxy located approximately $12.2$ Mpc from the Milky Way,
situated in the Virgo Cluster. It exhibits a well-defined spiral
structure and is classified as type Sb. The galaxy's optical
radius is approximately $6.5$ Kpc, while its HI radius extends to
about $15$ Kpc, indicating a significant amount of neutral
hydrogen beyond the optical disk. Overall, UGC07690 exemplifies a
moderate-mass, gas-rich spiral galaxy with a large HI extent
relative to its optical size. In Figs. \ref{UGC07690dens},
\ref{UGC07690} and \ref{UGC07690temp} we present the density of
the collisional DM model, the predicted rotation curves after
using an optimization for the collisional DM model
(\ref{tanhmodel}), versus the SPARC observational data and the
temperature parameter as a function of the radius respectively. As
it can be seen, the SIDM model produces viable rotation curves
compatible with the SPARC data. Also in Tables \ref{collUGC07690},
\ref{NavaroUGC07690}, \ref{BuckertUGC07690} and
\ref{EinastoUGC07690} we present the optimization values for the
SIDM model, and the other DM profiles. Also in Table
\ref{EVALUATIONUGC07690} we present the overall evaluation of the
SIDM model for the galaxy at hand. The resulting phenomenology is
viable.
\begin{table}[h!]
  \begin{center}
    \caption{Collisional Dark Matter Optimization Values}
    \label{collUGC07690}
     \begin{tabular}{|r|r|}
     \hline
      \textbf{Parameter}   & \textbf{Optimization Values}
      \\  \hline
     $\delta_{\gamma} $ & 0.0000000012
\\  \hline
$\gamma_0 $ & 1.0001  \\ \hline $K_0$ ($M_{\odot} \,
\mathrm{Kpc}^{-3} \, (\mathrm{km/s})^{2}$)& 1400  \\ \hline
    \end{tabular}
  \end{center}
\end{table}
\begin{table}[h!]
  \begin{center}
    \caption{NFW  Optimization Values}
    \label{NavaroUGC07690}
     \begin{tabular}{|r|r|}
     \hline
      \textbf{Parameter}   & \textbf{Optimization Values}
      \\  \hline
   $\rho_s$   & $5\times 10^7$
\\  \hline
$r_s$&  2.59
\\  \hline
    \end{tabular}
  \end{center}
\end{table}
\begin{table}[h!]
  \begin{center}
    \caption{Burkert Optimization Values}
    \label{BuckertUGC07690}
     \begin{tabular}{|r|r|}
     \hline
      \textbf{Parameter}   & \textbf{Optimization Values}
      \\  \hline
     $\rho_0^B$  & $1\times 10^8$
\\  \hline
$r_0$&  3.20
\\  \hline
    \end{tabular}
  \end{center}
\end{table}
\begin{table}[h!]
  \begin{center}
    \caption{Einasto Optimization Values}
    \label{EinastoUGC07690}
    \begin{tabular}{|r|r|}
     \hline
      \textbf{Parameter}   & \textbf{Optimization Values}
      \\  \hline
     $\rho_e$  &$1\times 10^7$
\\  \hline
$r_e$ & 2.65
\\  \hline
$n_e$ & 0.20
\\  \hline
    \end{tabular}
  \end{center}
\end{table}
\begin{table}[h!]
\centering \caption{Physical assessment of collisional DM
parameters for UGC07690.}
\begin{tabular}{lcc}
\hline
Parameter & Value & Physical Verdict \\
\hline
$\gamma_0$ & 1.001 & Almost perfectly isothermal \\
$\delta_\gamma$ & $1.2 \times 10^{-9}$ & Essentially no variation, $\gamma(r)$ constant \\
$r_\gamma$ & 1.5 Kpc & Transition radius irrelevant due to negligible $\delta_\gamma$ \\
$K_0$ & 1400 & Enough pressure support \\
$r_c$ & 0.5 Kpc & Small core radius, reasonable for inner halo \\
$p$ & 0.01 & Nearly constant $K(r)$, minimal radial variation \\
\hline
Overall &-& Physically acceptable \\
\hline
\end{tabular}
\label{EVALUATIONUGC07690}
\end{table}

\subsection{The Galaxy UGC07866}

\begin{figure}[h!]
\centering
\includegraphics[width=20pc]{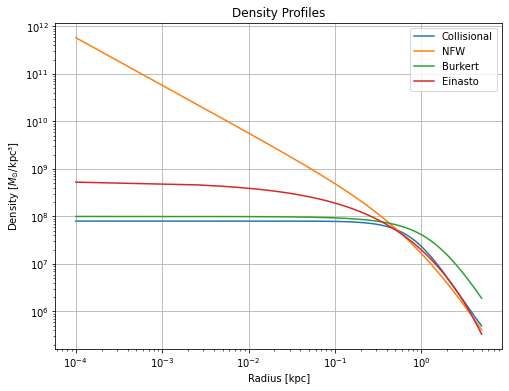}
\caption{The density of the collisional DM model (\ref{tanhmodel})
for the galaxy UGC07866, as a function of the radius.}
\label{UGC07866dens}
\end{figure}

\begin{figure}[h!]
\centering
\includegraphics[width=20pc]{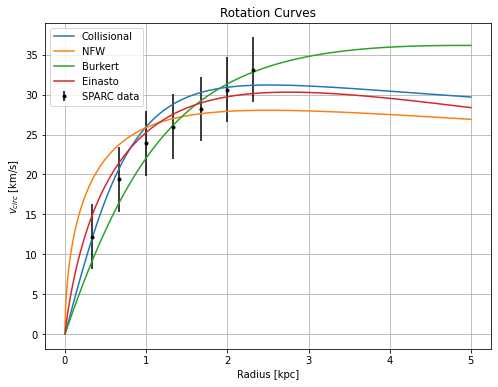}
\caption{The predicted rotation curves after using an optimization
for the collisional DM model (\ref{tanhmodel}), versus the SPARC
observational data for the galaxy UGC07866. We also plotted the
optimized curves for the NFW model, the Burkert model and the
Einasto model.} \label{UGC07866}
\end{figure}

\begin{figure}[h!]
\centering
\includegraphics[width=20pc]{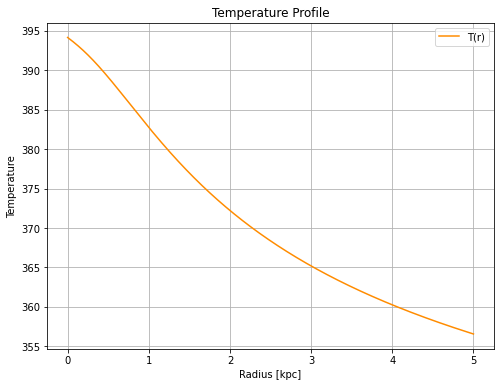}
\caption{The temperature as a function of the radius for the
collisional DM model (\ref{tanhmodel}) for the galaxy UGC07866.}
\label{UGC07866temp}
\end{figure}

For this galaxy, we shall choose $\rho_0=8\times
10^7$$M_{\odot}/\mathrm{Kpc}^{3}$. The galaxy UGC07866 is a dwarf
irregular galaxy situated approximately $1.7\ \mathrm{Mpc}$ from
the Milky Way, located in the M94 group. It is characterized by a
low-surface-brightness and lacks a well-defined spiral structure,
distinguishing it from typical spiral galaxies. In Figs.
\ref{UGC07866dens}, \ref{UGC07866} and \ref{UGC07866temp} we
present the density of the collisional DM model, the predicted
rotation curves after using an optimization for the collisional DM
model (\ref{tanhmodel}), versus the SPARC observational data and
the temperature parameter as a function of the radius
respectively. As it can be seen, the SIDM model produces viable
rotation curves compatible with the SPARC data. Also in Tables
\ref{collUGC07866}, \ref{NavaroUGC07866}, \ref{BuckertUGC07866}
and \ref{EinastoUGC07866} we present the optimization values for
the SIDM model, and the other DM profiles. Also in Table
\ref{EVALUATIONUGC07866} we present the overall evaluation of the
SIDM model for the galaxy at hand. The resulting phenomenology is
viable.
\begin{table}[h!]
  \begin{center}
    \caption{Collisional Dark Matter Optimization Values}
    \label{collUGC07866}
     \begin{tabular}{|r|r|}
     \hline
      \textbf{Parameter}   & \textbf{Optimization Values}
      \\  \hline
     $\delta_{\gamma} $ & 0.0000000012
\\  \hline
$\gamma_0 $ & 1.0001 \\ \hline $K_0$ ($M_{\odot} \,
\mathrm{Kpc}^{-3} \, (\mathrm{km/s})^{2}$)& 360 \\ \hline
    \end{tabular}
  \end{center}
\end{table}
\begin{table}[h!]
  \begin{center}
    \caption{NFW  Optimization Values}
    \label{NavaroUGC07866}
     \begin{tabular}{|r|r|}
     \hline
      \textbf{Parameter}   & \textbf{Optimization Values}
      \\  \hline
   $\rho_s$   & $5\times 10^7$
\\  \hline
$r_s$&  1.16
\\  \hline
    \end{tabular}
  \end{center}
\end{table}
\begin{table}[h!]
  \begin{center}
    \caption{Burkert Optimization Values}
    \label{BuckertUGC07866}
     \begin{tabular}{|r|r|}
     \hline
      \textbf{Parameter}   & \textbf{Optimization Values}
      \\  \hline
     $\rho_0^B$  & $1\times 10^8$
\\  \hline
$r_0$&  1.50
\\  \hline
    \end{tabular}
  \end{center}
\end{table}
\begin{table}[h!]
  \begin{center}
    \caption{Einasto Optimization Values}
    \label{EinastoUGC07866}
    \begin{tabular}{|r|r|}
     \hline
      \textbf{Parameter}   & \textbf{Optimization Values}
      \\  \hline
     $\rho_e$  &$1\times 10^7$
\\  \hline
$r_e$ & 1.46
\\  \hline
$n_e$ & 0.5
\\  \hline
    \end{tabular}
  \end{center}
\end{table}
\begin{table}[h!]
\centering \caption{Physical assessment of collisional DM
parameters for UGC07866.}
\begin{tabular}{lcc}
\hline
Parameter & Value & Physical Verdict \\
\hline
$\gamma_0$ & 1.0001 & Slightly above isothermal \\
$\delta_\gamma$ & $1.2 \times 10^{-9}$ & Essentially no variation, $\gamma(r)$ constant \\
$r_\gamma$ & 1.5 Kpc & Transition radius irrelevant due to negligible $\delta_\gamma$ \\
$K_0$ & 360 & Enough pressure support \\
$r_c$ & 0.5 Kpc & Small core radius, typical inner halo scale \\
$p$ & 0.01 & Very shallow $K(r)$ decrease, nearly constant \\
\hline
Overall &-& Physically plausible \\
\hline
\end{tabular}
\label{EVALUATIONUGC07866}
\end{table}

\subsection{The Galaxy UGC08286 Marginally Viable}

\begin{figure}[h!]
\centering
\includegraphics[width=20pc]{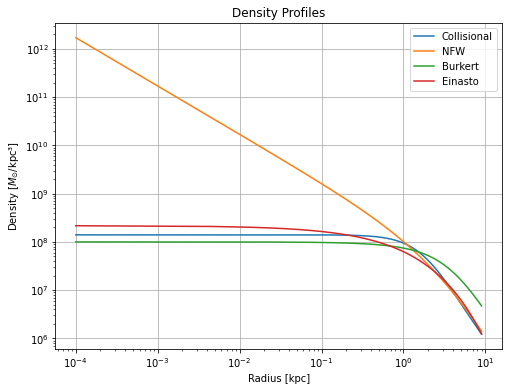}
\caption{The density of the collisional DM model (\ref{tanhmodel})
for the galaxy UGC08286, as a function of the radius.}
\label{UGC08286dens}
\end{figure}

\begin{figure}[h!]
\centering
\includegraphics[width=20pc]{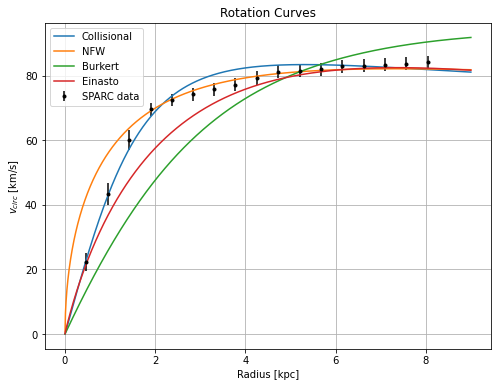}
\caption{The predicted rotation curves after using an optimization
for the collisional DM model (\ref{tanhmodel}), versus the SPARC
observational data for the galaxy UGC08286. We also plotted the
optimized curves for the NFW model, the Burkert model and the
Einasto model.} \label{UGC08286}
\end{figure}

\begin{figure}[h!]
\centering
\includegraphics[width=20pc]{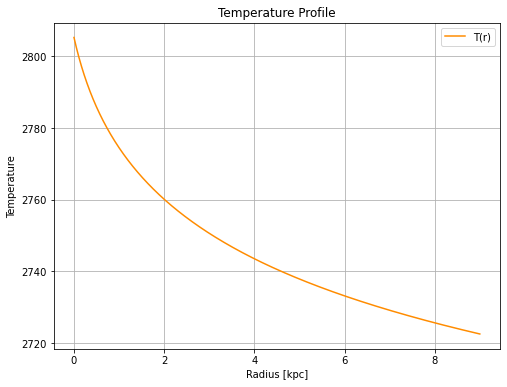}
\caption{The temperature as a function of the radius for the
collisional DM model (\ref{tanhmodel}) for the galaxy UGC08286.}
\label{UGC08286temp}
\end{figure}

\begin{figure}[h!]
\centering
\includegraphics[width=20pc]{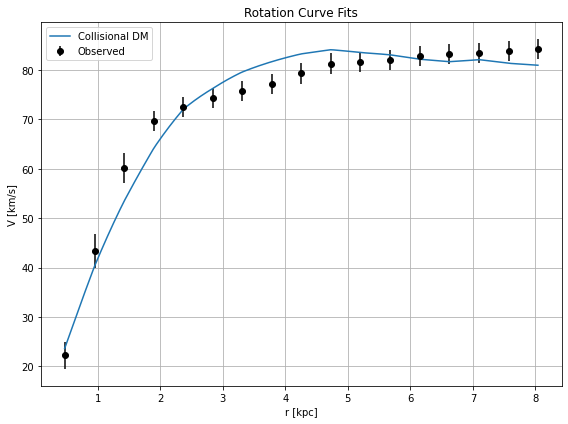}
\caption{The predicted rotation curves after using an optimization
for the collisional DM model (\ref{tanhmodel}), versus the
extended SPARC observational data for the galaxy UGC08286. The
model includes the rotation curves from all the components of the
galaxy, including gas and disk velocities, along with the
collisional DM model.} \label{extendedUGC08286}
\end{figure}

For this galaxy, we shall choose $\rho_0=1.4\times
10^8$$M_{\odot}/\mathrm{Kpc}^{3}$. The galaxy UGC08286 is a
low-surface-brightness dwarf irregular galaxy located at a
distance of approximately 10 Mpc, featuring a chaotic structure
without prominent spiral arms or a central bulge typical of
ordinary spirals. In Figs. \ref{UGC08286dens}, \ref{UGC08286} and
\ref{UGC08286temp} we present the density of the collisional DM
model, the predicted rotation curves after using an optimization
for the collisional DM model (\ref{tanhmodel}), versus the SPARC
observational data and the temperature parameter as a function of
the radius respectively. As it can be seen, the SIDM model
produces marginally viable rotation curves compatible with the
SPARC data. Also in Tables \ref{collUGC08286},
\ref{NavaroUGC08286}, \ref{BuckertUGC08286} and
\ref{EinastoUGC08286} we present the optimization values for the
SIDM model, and the other DM profiles. Also in Table
\ref{EVALUATIONUGC08286} we present the overall evaluation of the
SIDM model for the galaxy at hand. The resulting phenomenology is
marginally viable.
\begin{table}[h!]
  \begin{center}
    \caption{Collisional Dark Matter Optimization Values}
    \label{collUGC08286}
     \begin{tabular}{|r|r|}
     \hline
      \textbf{Parameter}   & \textbf{Optimization Values}
      \\  \hline
     $\delta_{\gamma} $ & 0.0000000012
\\  \hline
$\gamma_0 $ & 1.0001 \\ \hline $K_0$ ($M_{\odot} \,
\mathrm{Kpc}^{-3} \, (\mathrm{km/s})^{2}$)& 2800  \\ \hline
    \end{tabular}
  \end{center}
\end{table}
\begin{table}[h!]
  \begin{center}
    \caption{NFW  Optimization Values}
    \label{NavaroUGC08286}
     \begin{tabular}{|r|r|}
     \hline
      \textbf{Parameter}   & \textbf{Optimization Values}
      \\  \hline
   $\rho_s$   & $5\times 10^7$
\\  \hline
$r_s$&  3.40
\\  \hline
    \end{tabular}
  \end{center}
\end{table}
\begin{table}[h!]
  \begin{center}
    \caption{Burkert Optimization Values}
    \label{BuckertUGC08286}
     \begin{tabular}{|r|r|}
     \hline
      \textbf{Parameter}   & \textbf{Optimization Values}
      \\  \hline
     $\rho_0^B$  & $1\times 10^8$
\\  \hline
$r_0$&  3.88
\\  \hline
    \end{tabular}
  \end{center}
\end{table}
\begin{table}[h!]
  \begin{center}
    \caption{Einasto Optimization Values}
    \label{EinastoUGC08286}
    \begin{tabular}{|r|r|}
     \hline
      \textbf{Parameter}   & \textbf{Optimization Values}
      \\  \hline
     $\rho_e$  &$1\times 10^7$
\\  \hline
$r_e$ & 4.05
\\  \hline
$n_e$ & 0.65
\\  \hline
    \end{tabular}
  \end{center}
\end{table}
\begin{table}[h!]
\centering \caption{Physical assessment of collisional DM
parameters for UGC08286.}
\begin{tabular}{lcc}
\hline
Parameter & Value & Physical Verdict \\
\hline
$\gamma_0$ & 1.0001 & Essentially isothermal \\
$\delta_\gamma$ & $1.2 \times 10^{-9}$ & Negligible variation \\
$r_\gamma$ & 1.5 Kpc & Transition radius irrelevant due to negligible $\delta_\gamma$ \\
$K_0$ & 2800 &  Reasonable for a dwarf/low-mass galaxy core \\
$r_c$ & 0.5 Kpc & Small core radius, typical for inner halo \\
$p$ & 0.01 & Very shallow $K(r)$ decrease, nearly constant \\
\hline
Overall &-& Physically plausible \\
\hline
\end{tabular}
\label{EVALUATIONUGC08286}
\end{table}
Now the extended picture including the rotation velocity from the
other components of the galaxy, such as the disk and gas, makes
the collisional DM model viable for this galaxy. In Fig.
\ref{extendedUGC08286} we present the combined rotation curves
including the other components of the galaxy along with the
collisional matter. As it can be seen, the extended collisional DM
model is marginally viable.
Also in Table \ref{evaluationextendedUGC08286} we present the
values of the free parameters of the collisional DM model for
which the maximum compatibility with the SPARC data comes for the
galaxy UGC08286.
\begin{table}[h!]
\centering \caption{Physical assessment of Extended collisional DM
parameters for galaxy UGC08286.}
\begin{tabular}{lcc}
\hline
Parameter & Value & Physical Verdict \\
\hline
$\gamma_0$ & 1.003768997 & Extremely close to isothermal \\
$\delta_\gamma$ & 0.00008925896 & Negligible radial variation \\
$K_0$ & 2000 & Plausible moderate \\
$ml_{disk}$ & 1.00000000 & Maximal disk M/L; disk-dominated scenario \\
$ml_{bulge}$ & 0.00000000 & Negligible bulge contribution \\
\hline
Overall &-& Physically plausible \\
\hline
\end{tabular}
\label{evaluationextendedUGC08286}
\end{table}

\subsection{The Galaxy UGC08490 Marginally Viable by One Data Point}

\begin{figure}[h!]
\centering
\includegraphics[width=20pc]{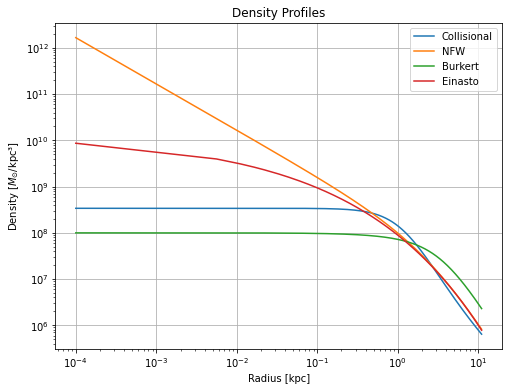}
\caption{The density of the collisional DM model (\ref{tanhmodel})
for the galaxy UGC08490, as a function of the radius.}
\label{UGC08490dens}
\end{figure}

\begin{figure}[h!]
\centering
\includegraphics[width=20pc]{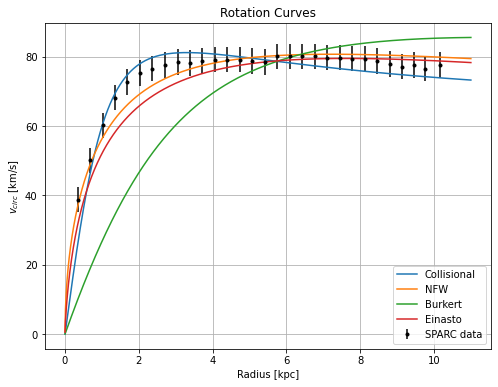}
\caption{The predicted rotation curves after using an optimization
for the collisional DM model (\ref{tanhmodel}), versus the SPARC
observational data for the galaxy UGC08490. We also plotted the
optimized curves for the NFW model, the Burkert model and the
Einasto model.} \label{UGC08490}
\end{figure}

\begin{figure}[h!]
\centering
\includegraphics[width=20pc]{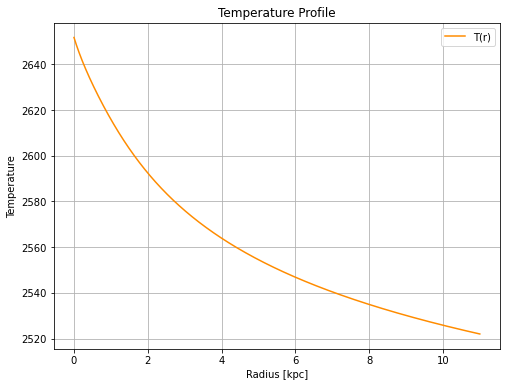}
\caption{The temperature as a function of the radius for the
collisional DM model (\ref{tanhmodel}) for the galaxy UGC08490.}
\label{UGC08490temp}
\end{figure}

\begin{figure}[h!]
\centering
\includegraphics[width=20pc]{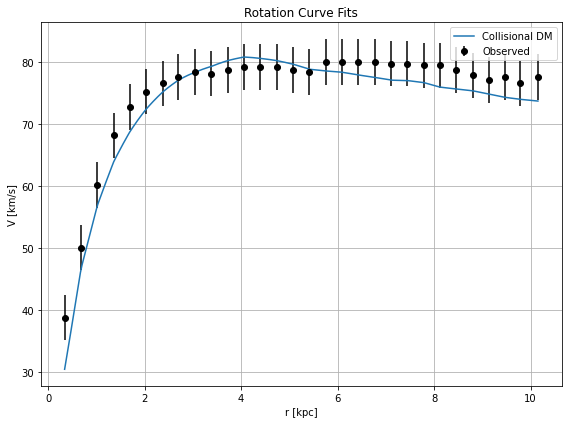}
\caption{The predicted rotation curves after using an optimization
for the collisional DM model (\ref{tanhmodel}), versus the
extended SPARC observational data for the galaxy UGC08490. The
model includes the rotation curves from all the components of the
galaxy, including gas and disk velocities, along with the
collisional DM model.} \label{extendedUGC08490}
\end{figure}

For this galaxy, we shall choose $\rho_0=3.4\times
10^8$$M_{\odot}/\mathrm{Kpc}^{3}$. The galaxy UGC08490 is a
low-surface-brightness dwarf irregular galaxy at a distance of
approximately 10 Mpc. In Figs. \ref{UGC08490dens}, \ref{UGC08490}
and \ref{UGC08490temp} we present the density of the collisional
DM model, the predicted rotation curves after using an
optimization for the collisional DM model (\ref{tanhmodel}),
versus the SPARC observational data and the temperature parameter
as a function of the radius respectively. As it can be seen, the
SIDM model produces marginally viable (by one miss in data points)
rotation curves compatible with the SPARC data. Also in Tables
\ref{collUGC08490}, \ref{NavaroUGC08490}, \ref{BuckertUGC08490}
and \ref{EinastoUGC08490} we present the optimization values for
the SIDM model, and the other DM profiles. Also in Table
\ref{EVALUATIONUGC08490} we present the overall evaluation of the
SIDM model for the galaxy at hand. The resulting phenomenology is
marginally viable.
\begin{table}[h!]
  \begin{center}
    \caption{Collisional Dark Matter Optimization Values}
    \label{collUGC08490}
     \begin{tabular}{|r|r|}
     \hline
      \textbf{Parameter}   & \textbf{Optimization Values}
      \\  \hline
     $\delta_{\gamma} $ & 0.0000000012
\\  \hline
$\gamma_0 $ & 1.0001 \\ \hline $K_0$ ($M_{\odot} \,
\mathrm{Kpc}^{-3} \, (\mathrm{km/s})^{2}$)& 2500  \\ \hline
    \end{tabular}
  \end{center}
\end{table}
\begin{table}[h!]
  \begin{center}
    \caption{NFW  Optimization Values}
    \label{NavaroUGC08490}
     \begin{tabular}{|r|r|}
     \hline
      \textbf{Parameter}   & \textbf{Optimization Values}
      \\  \hline
   $\rho_s$   & $5\times 10^7$
\\  \hline
$r_s$&  3.34
\\  \hline
    \end{tabular}
  \end{center}
\end{table}
\begin{table}[h!]
  \begin{center}
    \caption{Burkert Optimization Values}
    \label{BuckertUGC08490}
     \begin{tabular}{|r|r|}
     \hline
      \textbf{Parameter}   & \textbf{Optimization Values}
      \\  \hline
     $\rho_0^B$  & $1\times 10^8$
\\  \hline
$r_0$&  3.55
\\  \hline
    \end{tabular}
  \end{center}
\end{table}
\begin{table}[h!]
  \begin{center}
    \caption{Einasto Optimization Values}
    \label{EinastoUGC08490}
    \begin{tabular}{|r|r|}
     \hline
      \textbf{Parameter}   & \textbf{Optimization Values}
      \\  \hline
     $\rho_e$  &$1\times 10^7$
\\  \hline
$r_e$ & 3.70
\\  \hline
$n_e$ & 0.28
\\  \hline
    \end{tabular}
  \end{center}
\end{table}
\begin{table}[h!]
\centering \caption{Physical assessment of collisional DM
parameters for UGC08490.}
\begin{tabular}{lcc}
\hline
Parameter & Value & Physical Verdict \\
\hline
$\gamma_0$ & 1.0001 & Nearly isothermal, very low central pressure \\
$\delta_\gamma$ & $1.2 \times 10^{-9}$ & Negligible variation \\
$r_\gamma$ & 1.5 Kpc & Transition radius irrelevant due to tiny $\delta_\gamma$ \\
$K_0$ & 2500 & Reasonable for a dwarf/low-mass galaxy core \\
$r_c$ & 0.5 Kpc & Small core radius, typical inner halo scale \\
$p$ & 0.01 & Very shallow $K(r)$ decrease, nearly constant \\
\hline
Overall &-& Physically plausible\\
\hline
\end{tabular}
\label{EVALUATIONUGC08490}
\end{table}
Now the extended picture including the rotation velocity from the
other components of the galaxy, such as the disk and gas, makes
the collisional DM model viable for this galaxy. In Fig.
\ref{extendedUGC08490} we present the combined rotation curves
including the other components of the galaxy along with the
collisional matter. As it can be seen, the extended collisional DM
model is marginally viable.
Also in Table \ref{evaluationextendedUGC08490} we present the
values of the free parameters of the collisional DM model for
which the maximum compatibility with the SPARC data comes for the
galaxy UGC08490.
\begin{table}[h!]
\centering \caption{Physical assessment of Extended collisional DM
parameters for galaxy UGC08490.}
\begin{tabular}{lcc}
\hline
Parameter & Value & Physical Verdict \\
\hline
$\gamma_0$ & 1.0001478264 & Essentially isothermal \\
$\delta_\gamma$ & 0.001365974 & Negligible radial variation \\
$K_0$ & 2000 & Moderate \\
$ml_{disk}$ & 1.00000000 & Maximal disk M/L \\
$ml_{bulge}$ & 0.00000000 & Negligible bulge contribution \\
\hline
Overall &-& Physically plausible \\
\hline
\end{tabular}
\label{evaluationextendedUGC08490}
\end{table}


\subsection{The Galaxy UGC08550 Marginally Viable}

\begin{figure}[h!]
\centering
\includegraphics[width=20pc]{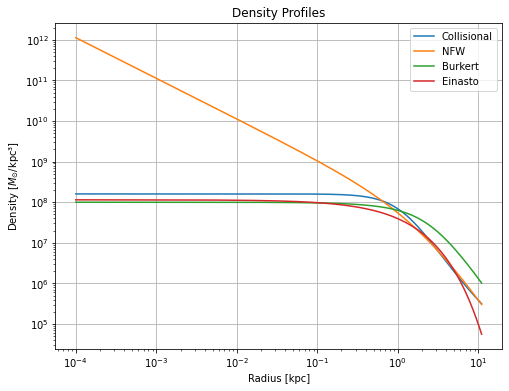}
\caption{The density of the collisional DM model (\ref{tanhmodel})
for the galaxy UGC08550, as a function of the radius.}
\label{UGC08550dens}
\end{figure}

\begin{figure}[h!]
\centering
\includegraphics[width=20pc]{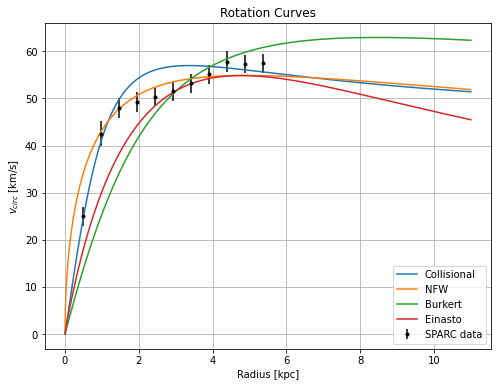}
\caption{The predicted rotation curves after using an optimization
for the collisional DM model (\ref{tanhmodel}), versus the SPARC
observational data for the galaxy UGC08550. We also plotted the
optimized curves for the NFW model, the Burkert model and the
Einasto model.} \label{UGC08550}
\end{figure}

\begin{figure}[h!]
\centering
\includegraphics[width=20pc]{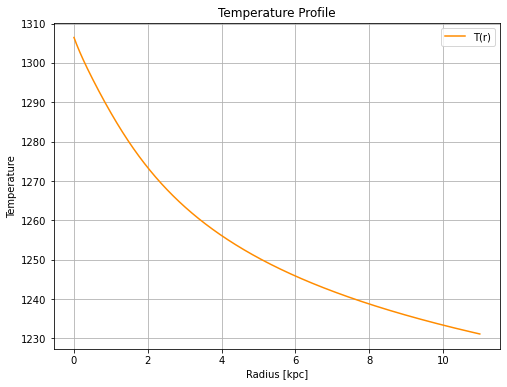}
\caption{The temperature as a function of the radius for the
collisional DM model (\ref{tanhmodel}) for the galaxy UGC08550.}
\label{UGC08550temp}
\end{figure}

\begin{figure}[h!]
\centering
\includegraphics[width=20pc]{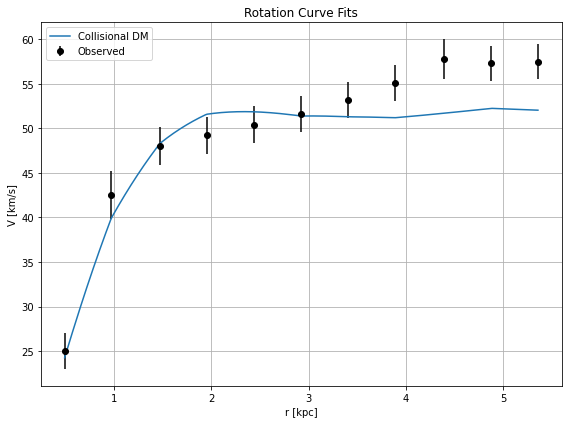}
\caption{The predicted rotation curves after using an optimization
for the collisional DM model (\ref{tanhmodel}), versus the
extended SPARC observational data for the galaxy UGC08550. The
model includes the rotation curves from all the components of the
galaxy, including gas and disk velocities, along with the
collisional DM model.} \label{extendedUGC08550}
\end{figure}

For this galaxy, we shall choose $\rho_0=1.6\times
10^8$$M_{\odot}/\mathrm{Kpc}^{3}$. UGC2885 is a massive, isolated
giant spiral galaxy at a distance of about $71 \mathrm{Mpc}$. In
Figs. \ref{UGC08550dens}, \ref{UGC08550} and \ref{UGC08550temp} we
present the density of the collisional DM model, the predicted
rotation curves after using an optimization for the collisional DM
model (\ref{tanhmodel}), versus the SPARC observational data and
the temperature parameter as a function of the radius
respectively. As it can be seen, the SIDM model produces
marginally viable rotation curves compatible with the SPARC data.
Also in Tables \ref{collUGC08550}, \ref{NavaroUGC08550},
\ref{BuckertUGC08550} and \ref{EinastoUGC08550} we present the
optimization values for the SIDM model, and the other DM profiles.
Also in Table \ref{EVALUATIONUGC08550} we present the overall
evaluation of the SIDM model for the galaxy at hand. The resulting
phenomenology is marginally viable.
\begin{table}[h!]
  \begin{center}
    \caption{Collisional Dark Matter Optimization Values}
    \label{collUGC08550}
     \begin{tabular}{|r|r|}
     \hline
      \textbf{Parameter}   & \textbf{Optimization Values}
      \\  \hline
     $\delta_{\gamma} $ & 0.0000000012
\\  \hline
$\gamma_0 $ & 1.0001\\ \hline $K_0$ ($M_{\odot} \,
\mathrm{Kpc}^{-3} \, (\mathrm{km/s})^{2}$)& 1200  \\ \hline
    \end{tabular}
  \end{center}
\end{table}
\begin{table}[h!]
  \begin{center}
    \caption{NFW  Optimization Values}
    \label{NavaroUGC08550}
     \begin{tabular}{|r|r|}
     \hline
      \textbf{Parameter}   & \textbf{Optimization Values}
      \\  \hline
   $\rho_s$   & $5\times 10^7$
\\  \hline
$r_s$&  2.27
\\  \hline
    \end{tabular}
  \end{center}
\end{table}
\begin{table}[h!]
  \begin{center}
    \caption{Burkert Optimization Values}
    \label{BuckertUGC08550}
     \begin{tabular}{|r|r|}
     \hline
      \textbf{Parameter}   & \textbf{Optimization Values}
      \\  \hline
     $\rho_0^B$  & $1\times 10^8$
\\  \hline
$r_0$&  2.61
\\  \hline
    \end{tabular}
  \end{center}
\end{table}
\begin{table}[h!]
  \begin{center}
    \caption{Einasto Optimization Values}
    \label{EinastoUGC08550}
    \begin{tabular}{|r|r|}
     \hline
      \textbf{Parameter}   & \textbf{Optimization Values}
      \\  \hline
     $\rho_e$  &$1\times 10^7$
\\  \hline
$r_e$ & 2.74
\\  \hline
$n_e$ & 0.82
\\  \hline
    \end{tabular}
  \end{center}
\end{table}
\begin{table}[h!]
\centering \caption{Physical assessment of collisional DM
parameters for UGC08550  .}
\begin{tabular}{lcc}
\hline
Parameter & Value   & Physical Verdict \\
\hline
$\gamma_0$ & $1.0001$ & Almost isothermal \\
$\delta_\gamma$ & $1.2\times10^{-9}$ & Effectively zero \\
$r_\gamma$ & $1.5\ \mathrm{Kpc}$ & Transition radius set outside very inner core \\
$K_0$ & $1.2\times10^{3}$ & Plausible moderate high \\
$r_c$ & $0.5\ \mathrm{Kpc}$ & Small core scale  \\
$p$ & $0.01$ & Very shallow decline of $K(r)$ \\
\hline
Overall &-& Physically consistent \\
\hline
\end{tabular}
\label{EVALUATIONUGC08550}
\end{table}
Now the extended picture including the rotation velocity from the
other components of the galaxy, such as the disk and gas, makes
the collisional DM model viable for this galaxy. In Fig.
\ref{extendedUGC08550} we present the combined rotation curves
including the other components of the galaxy along with the
collisional matter. As it can be seen, the extended collisional DM
model is marginally viable.
Also in Table \ref{evaluationextendedUGC08550} we present the
values of the free parameters of the collisional DM model for
which the maximum compatibility with the SPARC data comes for the
galaxy UGC08550.
\begin{table}[h!]
\centering \caption{Physical assessment of Extended collisional DM
parameters for galaxy UGC08550.}
\begin{tabular}{lcc}
\hline
Parameter & Value & Physical Verdict \\
\hline
$\gamma_0$ & 1.01 & Essentially isothermal \\
$\delta_\gamma$ & 0.0007429340 & Negligible radial variation\\
$K_0$ & 700 & Low entropy \\
$ml_{disk}$ & 1.00000000 & Maximal disk M/L \\
$ml_{bulge}$ & 0.00000000 & Negligible bulge contribution \\
\hline
Overall &-& Physically plausible  \\
\hline
\end{tabular}
\label{evaluationextendedUGC08550}
\end{table}

\subsection{The Galaxy UGC08699 Non-viable}

\begin{figure}[h!]
\centering
\includegraphics[width=20pc]{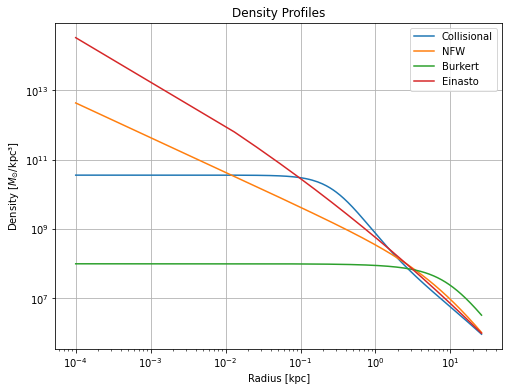}
\caption{The density of the collisional DM model (\ref{tanhmodel})
for the galaxy UGC08699, as a function of the radius.}
\label{UGC08699dens}
\end{figure}
\begin{figure}[h!]
\centering
\includegraphics[width=20pc]{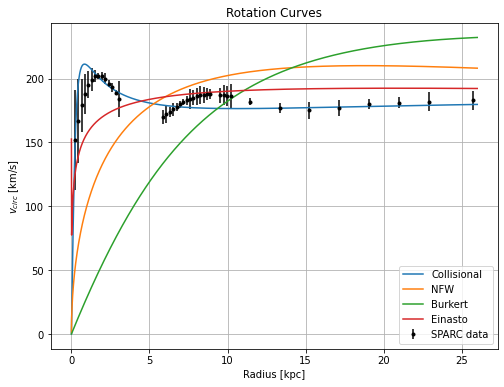}
\caption{The predicted rotation curves after using an optimization
for the collisional DM model (\ref{tanhmodel}), versus the SPARC
observational data for the galaxy UGC08699. We also plotted the
optimized curves for the NFW model, the Burkert model and the
Einasto model.} \label{UGC08699}
\end{figure}
\begin{figure}[h!]
\centering
\includegraphics[width=20pc]{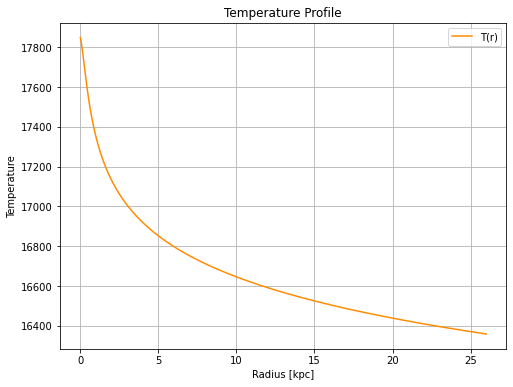}
\caption{The temperature as a function of the radius for the
collisional DM model (\ref{tanhmodel}) for the galaxy UGC08699.}
\label{UGC08699temp}
\end{figure}
\begin{figure}[h!]
\centering
\includegraphics[width=20pc]{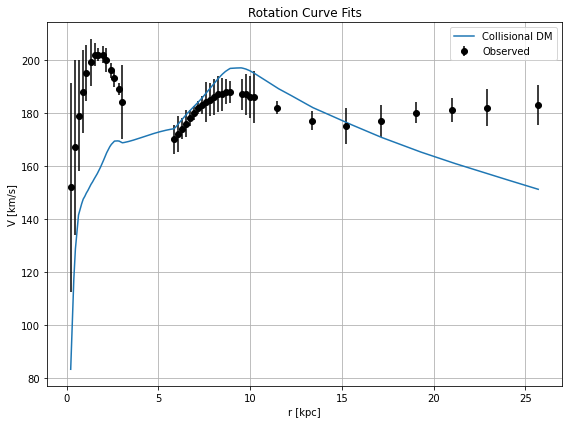}
\caption{The predicted rotation curves after using an optimization
for the collisional DM model (\ref{tanhmodel}), versus the
extended SPARC observational data for the galaxy UGC08699. The
model includes the rotation curves from all the components of the
galaxy, including gas and disk velocities, along with the
collisional DM model.} \label{extendedUGC08699}
\end{figure}

For this galaxy, we shall choose $\rho_0=3.6\times
10^{10}$$M_{\odot}/\mathrm{Kpc}^{3}$. UGC8699 is a highly inclined
early-type spiral galaxy, a normal/large disk system. In Figs.
\ref{UGC08699dens}, \ref{UGC08699} and \ref{UGC08699temp} we
present the density of the collisional DM model, the predicted
rotation curves after using an optimization for the collisional DM
model (\ref{tanhmodel}), versus the SPARC observational data and
the temperature parameter as a function of the radius
respectively. As it can be seen, the SIDM model produces
non-viable rotation curves incompatible with the SPARC data. Also
in Tables \ref{collUGC08699}, \ref{NavaroUGC08699},
\ref{BuckertUGC08699} and \ref{EinastoUGC08699} we present the
optimization values for the SIDM model, and the other DM profiles.
Also in Table \ref{EVALUATIONUGC08699} we present the overall
evaluation of the SIDM model for the galaxy at hand. The resulting
phenomenology is non-viable.
\begin{table}[h!]
  \begin{center}
    \caption{Collisional Dark Matter Optimization Values}
    \label{collUGC08699}
     \begin{tabular}{|r|r|}
     \hline
      \textbf{Parameter}   & \textbf{Optimization Values}
      \\  \hline
     $\delta_{\gamma} $ & 0.0000000012
\\  \hline
$\gamma_0 $ & 1.0001 \\ \hline $K_0$ ($M_{\odot} \,
\mathrm{Kpc}^{-3} \, (\mathrm{km/s})^{2}$)& 1500 \\ \hline
    \end{tabular}
  \end{center}
\end{table}
\begin{table}[h!]
  \begin{center}
    \caption{NFW  Optimization Values}
    \label{NavaroUGC08699}
     \begin{tabular}{|r|r|}
     \hline
      \textbf{Parameter}   & \textbf{Optimization Values}
      \\  \hline
   $\rho_s$   & $5\times 10^7$
\\  \hline
$r_s$&  8.69
\\  \hline
    \end{tabular}
  \end{center}
\end{table}
\begin{table}[h!]
  \begin{center}
    \caption{Burkert Optimization Values}
    \label{BuckertUGC08699}
     \begin{tabular}{|r|r|}
     \hline
      \textbf{Parameter}   & \textbf{Optimization Values}
      \\  \hline
     $\rho_0^B$  & $1\times 10^8$
\\  \hline
$r_0$&  9.68
\\  \hline
    \end{tabular}
  \end{center}
\end{table}
\begin{table}[h!]
  \begin{center}
    \caption{Einasto Optimization Values}
    \label{EinastoUGC08699}
    \begin{tabular}{|r|r|}
     \hline
      \textbf{Parameter}   & \textbf{Optimization Values}
      \\  \hline
     $\rho_e$  &$1\times 10^7$
\\  \hline
$r_e$ & 8.45
\\  \hline
$n_e$ & 0.05
\\  \hline
    \end{tabular}
  \end{center}
\end{table}
\begin{table}[h!]
\centering \caption{Physical assessment of collisional DM
parameters for UGC08699.}
\begin{tabular}{lcc}
\hline
Parameter & Value   & Physical Verdict \\
\hline
$\gamma_0$ & $1.0001$ & Nearly isothermal \\
$\delta_\gamma$ & $1.2\times10^{-9}$ & Practically zero \\
$r_\gamma$ & $1.5\ \mathrm{Kpc}$ & Transition radius irrelevant with tiny $\delta_\gamma$ \\
$K_0$ & $1.6\times10^{4}$ & Enough pressure support \\
$r_c$ & $0.5\ \mathrm{Kpc}$ & Small core; plausible for compact inner halo \\
$p$ & $0.01$ & Very shallow decline\\
\hline
Overall &-& Numerically stable  \\
\hline
\end{tabular}
\label{EVALUATIONUGC08699}
\end{table}
Now the extended picture including the rotation velocity from the
other components of the galaxy, such as the disk and gas, makes
the collisional DM model viable for this galaxy. In Fig.
\ref{extendedUGC08699} we present the combined rotation curves
including the other components of the galaxy along with the
collisional matter. As it can be seen, the extended collisional DM
model is non-viable.
Also in Table \ref{evaluationextendedUGC08699} we present the
values of the free parameters of the collisional DM model for
which the maximum compatibility with the SPARC data comes for the
galaxy UGC08699.
\begin{table}[h!]
\centering \caption{Physical assessment of Extended collisional DM
parameters (UGC08699).}
\begin{tabular}{lcc}
\hline
Parameter & Value & Physical Verdict \\
\hline
$\gamma_0$ & 1.10645390 & Slightly above isothermal \\
$\delta_\gamma$ & 0.05676139 & Small-to-moderate radial variation in $\gamma(r)$ \\
$K_0$ & 3000 & Moderate entropy/pressure scale \\
ml\_disk & 1.00000000 & High disk mass-to-light \\
ml\_bulge & 0.50000000 & Substantial bulge mass-to-light\\
\hline
Overall &-& Physically plausible \\
\hline
\end{tabular}
\label{evaluationextendedUGC08699}
\end{table}

\subsection{The Galaxy UGC09037}

\begin{figure}[h!]
\centering
\includegraphics[width=20pc]{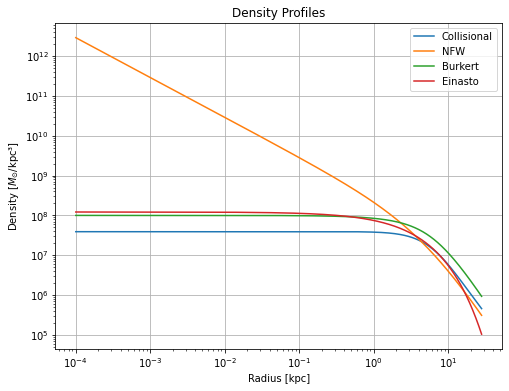}
\caption{The density of the collisional DM model (\ref{tanhmodel})
for the galaxy UGC09037, as a function of the radius.}
\label{UGC09037dens}
\end{figure}

\begin{figure}[h!]
\centering
\includegraphics[width=20pc]{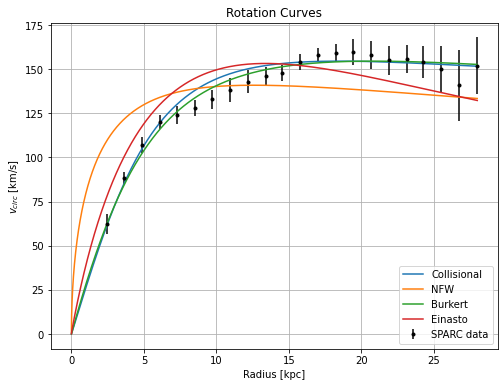}
\caption{The predicted rotation curves after using an optimization
for the collisional DM model (\ref{tanhmodel}), versus the SPARC
observational data for the galaxy UGC09037. We also plotted the
optimized curves for the NFW model, the Burkert model and the
Einasto model.} \label{UGC09037}
\end{figure}

\begin{figure}[h!]
\centering
\includegraphics[width=20pc]{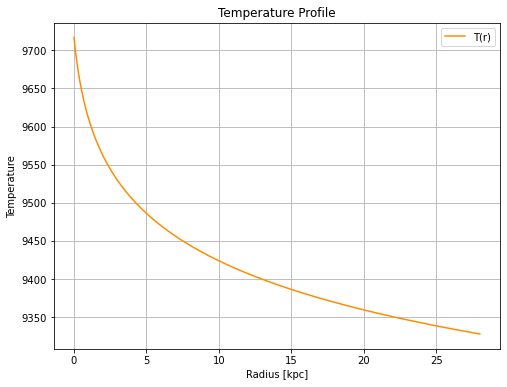}
\caption{The temperature as a function of the radius for the
collisional DM model (\ref{tanhmodel}) for the galaxy UGC09037.}
\label{UGC09037temp}
\end{figure}

For this galaxy, we shall choose $\rho_0=3.9\times
10^7$$M_{\odot}/\mathrm{Kpc}^{3}$. UGC09037 is a gas-rich,
late-type spiral galaxy at a distance $\sim 120 \mathrm{Mpc}$. In
Figs. \ref{UGC09037dens}, \ref{UGC09037} and \ref{UGC09037temp} we
present the density of the collisional DM model, the predicted
rotation curves after using an optimization for the collisional DM
model (\ref{tanhmodel}), versus the SPARC observational data and
the temperature parameter as a function of the radius
respectively. As it can be seen, the SIDM model produces viable
rotation curves compatible with the SPARC data. Also in Tables
\ref{collUGC09037}, \ref{NavaroUGC09037}, \ref{BuckertUGC09037}
and \ref{EinastoUGC09037} we present the optimization values for
the SIDM model, and the other DM profiles. Also in Table
\ref{EVALUATIONUGC09037} we present the overall evaluation of the
SIDM model for the galaxy at hand. The resulting phenomenology is
viable.
\begin{table}[h!]
  \begin{center}
    \caption{Collisional Dark Matter Optimization Values}
    \label{collUGC09037}
     \begin{tabular}{|r|r|}
     \hline
      \textbf{Parameter}   & \textbf{Optimization Values}
      \\  \hline
     $\delta_{\gamma} $ & 0.0000000012
\\  \hline
$\gamma_0 $ & 1.0001 \\ \hline $K_0$ ($M_{\odot} \,
\mathrm{Kpc}^{-3} \, (\mathrm{km/s})^{2}$)& 9700  \\ \hline
    \end{tabular}
  \end{center}
\end{table}
\begin{table}[h!]
  \begin{center}
    \caption{NFW  Optimization Values}
    \label{NavaroUGC09037}
     \begin{tabular}{|r|r|}
     \hline
      \textbf{Parameter}   & \textbf{Optimization Values}
      \\  \hline
   $\rho_s$   & $5\times 10^7$
\\  \hline
$r_s$&  5.83
\\  \hline
    \end{tabular}
  \end{center}
\end{table}
\begin{table}[h!]
  \begin{center}
    \caption{Burkert Optimization Values}
    \label{BuckertUGC09037}
     \begin{tabular}{|r|r|}
     \hline
      \textbf{Parameter}   & \textbf{Optimization Values}
      \\  \hline
     $\rho_0^B$  & $1\times 10^8$
\\  \hline
$r_0$&  6.41
\\  \hline
    \end{tabular}
  \end{center}
\end{table}
\begin{table}[h!]
  \begin{center}
    \caption{Einasto Optimization Values}
    \label{EinastoUGC09037}
    \begin{tabular}{|r|r|}
     \hline
      \textbf{Parameter}   & \textbf{Optimization Values}
      \\  \hline
     $\rho_e$  &$1\times 10^7$
\\  \hline
$r_e$ & 7.64
\\  \hline
$n_e$ & 1
\\  \hline
    \end{tabular}
  \end{center}
\end{table}
\begin{table}[h!]
\centering \caption{Physical assessment of collisional DM
parameters for UGC09037  .}
\begin{tabular}{lcc}
\hline
Parameter & Value   & Physical Verdict \\
\hline
$\gamma_0$ & $1.0001$ & Practically isothermal \\
$\delta_\gamma$ & $1.2\times10^{-9}$ & Negligible\\
$r_\gamma$ & $1.5\ \mathrm{Kpc}$ & Transition radius irrelevant  \\
$K_0$ & $9.7\times10^{3}$ & Enough pressure support \\
$r_c$ & $0.5\ \mathrm{Kpc}$ & Small core scale  \\
$p$ & $0.01$ & Very shallow decline \\
\hline
Overall &-& Numerically stable and physically plausible \\
\hline
\end{tabular}
\label{EVALUATIONUGC09037}
\end{table}


\subsection{The Galaxy UGC09133 Non-viable, Extended Marginal, 4 parameter Model}

\begin{figure}[h!]
\centering
\includegraphics[width=20pc]{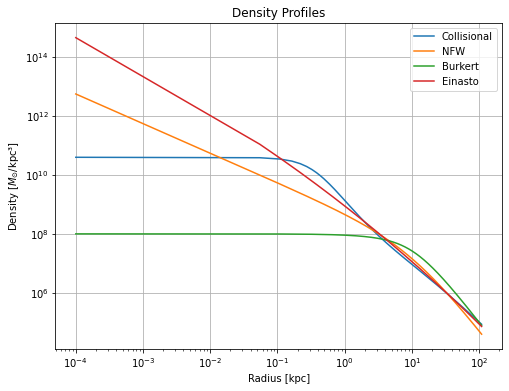}
\caption{The density of the collisional DM model (\ref{tanhmodel})
for the galaxy UGC09133, as a function of the radius.}
\label{UGC09133dens}
\end{figure}

\begin{figure}[h!]
\centering
\includegraphics[width=20pc]{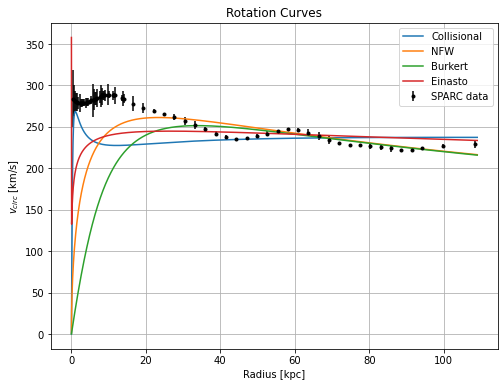}
\caption{The predicted rotation curves after using an optimization
for the collisional DM model (\ref{tanhmodel}), versus the SPARC
observational data for the galaxy UGC09133. We also plotted the
optimized curves for the NFW model, the Burkert model and the
Einasto model.} \label{UGC09133}
\end{figure}

\begin{figure}[h!]
\centering
\includegraphics[width=20pc]{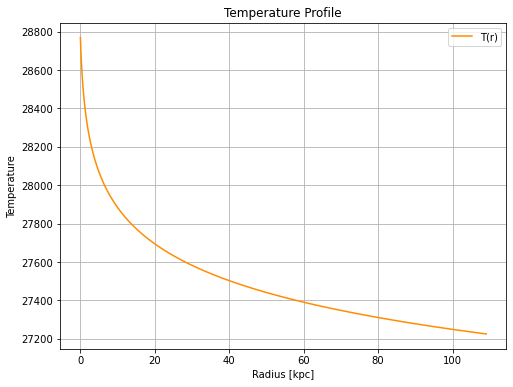}
\caption{The temperature as a function of the radius for the
collisional DM model (\ref{tanhmodel}) for the galaxy UGC09133.}
\label{UGC09133temp}
\end{figure}

\begin{figure}[h!]
\centering
\includegraphics[width=20pc]{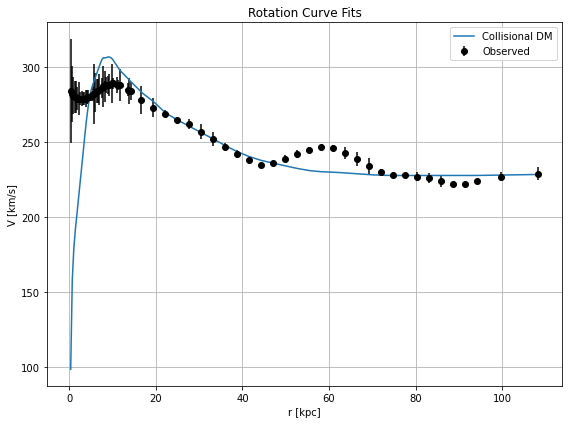}
\caption{The predicted rotation curves after using an optimization
for the collisional DM model (\ref{tanhmodel}), versus the
extended SPARC observational data for the galaxy UGC09133. The
model includes the rotation curves from all the components of the
galaxy, including gas and disk velocities, along with the
collisional DM model.} \label{extendedUGC09133}
\end{figure}

For this galaxy, we shall choose $\rho_0=3.9\times
10^{10}$$M_{\odot}/\mathrm{Kpc}^{3}$. UGC09133 is a large,
unbarred spiral galaxy at a distance of about $54\ \mathrm{Mpc}$.
In Figs. \ref{UGC09133dens}, \ref{UGC09133} and \ref{UGC09133temp}
we present the density of the collisional DM model, the predicted
rotation curves after using an optimization for the collisional DM
model (\ref{tanhmodel}), versus the SPARC observational data and
the temperature parameter as a function of the radius
respectively. As it can be seen, the SIDM model produces
non-viable rotation curves incompatible with the SPARC data. Also
in Tables \ref{collUGC09133}, \ref{NavaroUGC09133},
\ref{BuckertUGC09133} and \ref{EinastoUGC09133} we present the
optimization values for the SIDM model, and the other DM profiles.
Also in Table \ref{EVALUATIONUGC09133} we present the overall
evaluation of the SIDM model for the galaxy at hand. The resulting
phenomenology is non-viable.
\begin{table}[h!]
  \begin{center}
    \caption{Collisional Dark Matter Optimization Values}
    \label{collUGC09133}
     \begin{tabular}{|r|r|}
     \hline
      \textbf{Parameter}   & \textbf{Optimization Values}
      \\  \hline
     $\delta_{\gamma} $ & 0.0000000012
\\  \hline
$\gamma_0 $ & 1.0001 \\ \hline $K_0$ ($M_{\odot} \,
\mathrm{Kpc}^{-3} \, (\mathrm{km/s})^{2}$)& 28700  \\ \hline
    \end{tabular}
  \end{center}
\end{table}
\begin{table}[h!]
  \begin{center}
    \caption{NFW  Optimization Values}
    \label{NavaroUGC09133}
     \begin{tabular}{|r|r|}
     \hline
      \textbf{Parameter}   & \textbf{Optimization Values}
      \\  \hline
   $\rho_s$   & $5\times 10^7$
\\  \hline
$r_s$&  10.81
\\  \hline
    \end{tabular}
  \end{center}
\end{table}
\begin{table}[h!]
  \begin{center}
    \caption{Burkert Optimization Values}
    \label{BuckertUGC09133}
     \begin{tabular}{|r|r|}
     \hline
      \textbf{Parameter}   & \textbf{Optimization Values}
      \\  \hline
     $\rho_0^B$  & $1\times 10^8$
\\  \hline
$r_0$&  10.43
\\  \hline
    \end{tabular}
  \end{center}
\end{table}
\begin{table}[h!]
  \begin{center}
    \caption{Einasto Optimization Values}
    \label{EinastoUGC09133}
    \begin{tabular}{|r|r|}
     \hline
      \textbf{Parameter}   & \textbf{Optimization Values}
      \\  \hline
     $\rho_e$  &$1\times 10^7$
\\  \hline
$r_e$ & 10.75
\\  \hline
$n_e$ & 0.05
\\  \hline
    \end{tabular}
  \end{center}
\end{table}
\begin{table}[h!]
\centering \caption{Physical assessment of collisional DM
parameters for UGC09133  .}
\begin{tabular}{lcc}
\hline
Parameter & Value   & Physical Verdict \\
\hline
$\gamma_0$ & $1.0001$ & Practically isothermal\\
$\delta_\gamma$ & $1.2\times10^{-9}$ & Negligible \\
$r_\gamma$ & $1.5\ \mathrm{Kpc}$ & Transition radius irrelevant \\
$K_0$ & $2.87\times10^{4}$ & Large Pressure support \\
$r_c$ & $0.5\ \mathrm{Kpc}$ & Small core scale  \\
$p$ & $0.01$ & Very shallow decline \\
\hline
Overall &-& Numerically stable and plausible \\
\hline
\end{tabular}
\label{EVALUATIONUGC09133}
\end{table}
Now the extended picture including the rotation velocity from the
other components of the galaxy, such as the disk and gas, makes
the collisional DM model viable for this galaxy. In Fig.
\ref{extendedUGC09133} we present the combined rotation curves
including the other components of the galaxy along with the
collisional matter. As it can be seen, the extended collisional DM
model is marginally viable.
Also in Table \ref{evaluationextendedUGC09133} we present the
values of the free parameters of the collisional DM model for
which the maximum compatibility with the SPARC data comes for the
galaxy UGC09133.
\begin{table}[h!]
\centering \caption{Physical assessment of Extended collisional DM
parameters (UGC09133).}
\begin{tabular}{lcc}
\hline
Parameter & Value & Physical Verdict \\
\hline
$\gamma_0$ & 0.88624373 & Below isothermal ($\gamma<1$)\\
$\delta_\gamma$ & 0.09463253 & Moderate radial variation  \\
$K_0$ & 425867.08386591 & Extremely large entropy/pressure scale compared with typical model values  \\
ml\_disk & 0.96441896 & Disk mass-to-light near unity  \\
ml\_bulge & 0.50000000 & Significant bulge M/L  \\
\hline
Overall &-& Mixed plausibility: stellar M/L values are reasonable but extremely large $K_0$  \\
\hline
\end{tabular}
\label{evaluationextendedUGC09133}
\end{table}

\subsection{The Galaxy UGC09992}

\begin{figure}[h!]
\centering
\includegraphics[width=20pc]{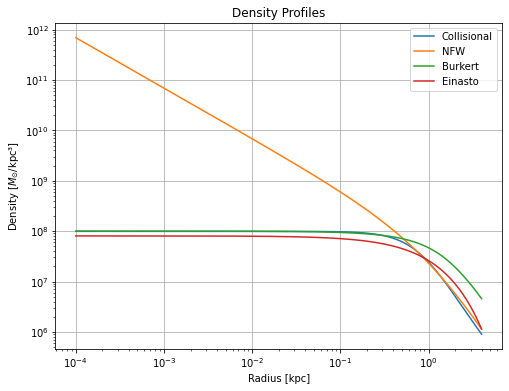}
\caption{The density of the collisional DM model (\ref{tanhmodel})
for the galaxy UGC09992, as a function of the radius.}
\label{UGC09992dens}
\end{figure}

\begin{figure}[h!]
\centering
\includegraphics[width=20pc]{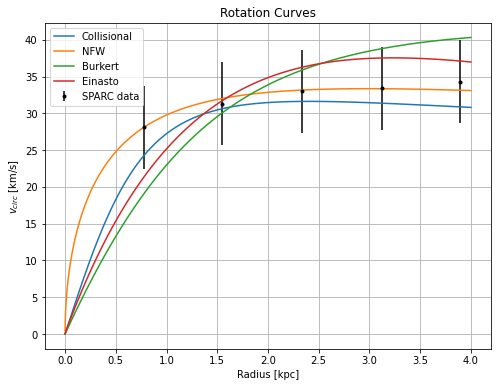}
\caption{The predicted rotation curves after using an optimization
for the collisional DM model (\ref{tanhmodel}), versus the SPARC
observational data for the galaxy UGC09992. We also plotted the
optimized curves for the NFW model, the Burkert model and the
Einasto model.} \label{UGC09992}
\end{figure}

\begin{figure}[h!]
\centering
\includegraphics[width=20pc]{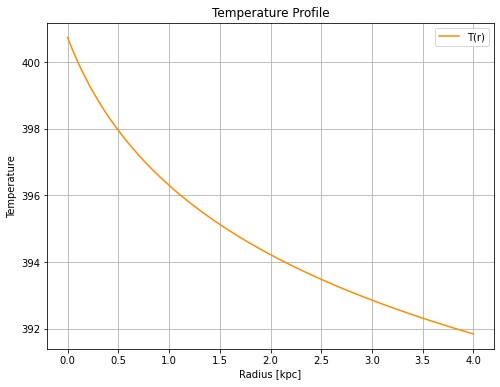}
\caption{The temperature as a function of the radius for the
collisional DM model (\ref{tanhmodel}) for the galaxy UGC09992.}
\label{UGC09992temp}
\end{figure}

For this galaxy, we shall choose $\rho_0=9.9\times
10^7$$M_{\odot}/\mathrm{Kpc}^{3}$. UGC09992 is a late-type spiral
galaxy (type Sd) located at a distance of approximately  $\sim 10
\text{ Mpc}$. In Figs. \ref{UGC09992dens}, \ref{UGC09992} and
\ref{UGC09992temp} we present the density of the collisional DM
model, the predicted rotation curves after using an optimization
for the collisional DM model (\ref{tanhmodel}), versus the SPARC
observational data and the temperature parameter as a function of
the radius respectively. As it can be seen, the SIDM model
produces viable rotation curves compatible with the SPARC data.
Also in Tables \ref{collUGC09992}, \ref{NavaroUGC09992},
\ref{BuckertUGC09992} and \ref{EinastoUGC09992} we present the
optimization values for the SIDM model, and the other DM profiles.
Also in Table \ref{EVALUATIONUGC09992} we present the overall
evaluation of the SIDM model for the galaxy at hand. The resulting
phenomenology is viable.
\begin{table}[h!]
  \begin{center}
    \caption{Collisional Dark Matter Optimization Values}
    \label{collUGC09992}
     \begin{tabular}{|r|r|}
     \hline
      \textbf{Parameter}   & \textbf{Optimization Values}
      \\  \hline
     $\delta_{\gamma} $ & 0.0000000012
\\  \hline
$\gamma_0 $ & 1.0001 \\ \hline $K_0$ ($M_{\odot} \,
\mathrm{Kpc}^{-3} \, (\mathrm{km/s})^{2}$)& 400 \\ \hline
    \end{tabular}
  \end{center}
\end{table}
\begin{table}[h!]
  \begin{center}
    \caption{NFW  Optimization Values}
    \label{NavaroUGC09992}
     \begin{tabular}{|r|r|}
     \hline
      \textbf{Parameter}   & \textbf{Optimization Values}
      \\  \hline
   $\rho_s$   & $5\times 10^7$
\\  \hline
$r_s$&  1.38
\\  \hline
    \end{tabular}
  \end{center}
\end{table}
\begin{table}[h!]
  \begin{center}
    \caption{Burkert Optimization Values}
    \label{BuckertUGC09992}
     \begin{tabular}{|r|r|}
     \hline
      \textbf{Parameter}   & \textbf{Optimization Values}
      \\  \hline
     $\rho_0^B$  & $1\times 10^8$
\\  \hline
$r_0$&  1.7
\\  \hline
    \end{tabular}
  \end{center}
\end{table}
\begin{table}[h!]
  \begin{center}
    \caption{Einasto Optimization Values}
    \label{EinastoUGC09992}
    \begin{tabular}{|r|r|}
     \hline
      \textbf{Parameter}   & \textbf{Optimization Values}
      \\  \hline
     $\rho_e$  &$1\times 10^7$
\\  \hline
$r_e$ & 1.90
\\  \hline
$n_e$ & 0.96
\\  \hline
    \end{tabular}
  \end{center}
\end{table}
\begin{table}[h!]
\centering \caption{Physical assessment of collisional DM
parameters for UGC09992.}
\begin{tabular}{lcc}
\hline
Parameter & Value   & Physical Verdict \\
\hline
$\gamma_0$ & $1.0001$ & Practically isothermal \\
$\delta_\gamma$ & $1.2\times10^{-9}$ & Negligible radial variation in $\gamma(r)$ \\
$r_\gamma$ & $1.5\ \mathrm{Kpc}$ & Transition radius irrelevant \\
$K_0$ & $4\times10^{2}$ & Low entropy scale \\
$r_c$ & $0.5\ \mathrm{Kpc}$ & Small core scale, plausible \\
$p$ & $0.01$ & Very shallow decline; $K(r)$ nearly constant \\
\hline
Overall &-& Physically plausible \\
\hline
\end{tabular}
\label{EVALUATIONUGC09992}
\end{table}

\subsection{The Galaxy UGC10310}

\begin{figure}[h!]
\centering
\includegraphics[width=20pc]{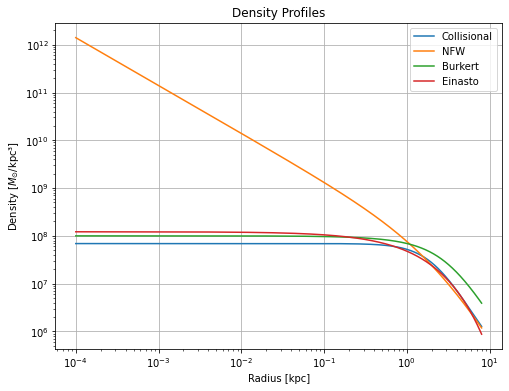}
\caption{The density of the collisional DM model (\ref{tanhmodel})
for the galaxy UGC10310, as a function of the radius.}
\label{UGC10310dens}
\end{figure}

\begin{figure}[h!]
\centering
\includegraphics[width=20pc]{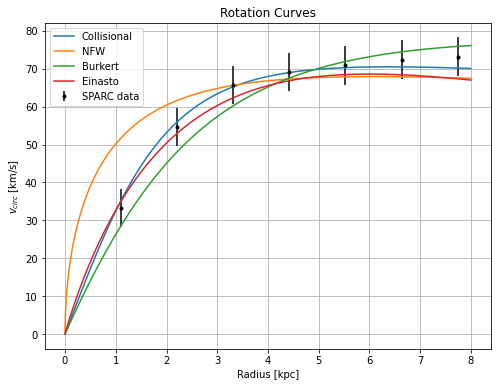}
\caption{The predicted rotation curves after using an optimization
for the collisional DM model (\ref{tanhmodel}), versus the SPARC
observational data for the galaxy UGC10310. We also plotted the
optimized curves for the NFW model, the Burkert model and the
Einasto model.} \label{UGC10310}
\end{figure}

\begin{figure}[h!]
\centering
\includegraphics[width=20pc]{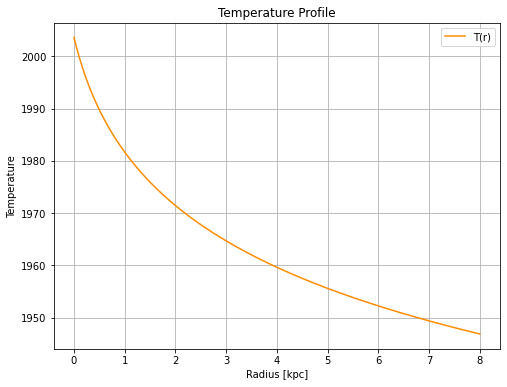}
\caption{The temperature as a function of the radius for the
collisional DM model (\ref{tanhmodel}) for the galaxy UGC10310.}
\label{UGC10310temp}
\end{figure}

For this galaxy, we shall choose $\rho_0=6.9\times
10^7$$M_{\odot}/\mathrm{Kpc}^{3}$. UGC10310 is a dwarf spiral
galaxy of Magellanic class located in the constellation Hercules,
featuring a prominent bright HII region. Its distance from the
Milky Way is approximately 220 Mpc. The galaxy exhibits an
ordinary spiral structure but qualifies as a dwarf due to its
relatively low mass and size compared to large spirals. In Figs.
\ref{UGC10310dens}, \ref{UGC10310} and \ref{UGC10310temp} we
present the density of the collisional DM model, the predicted
rotation curves after using an optimization for the collisional DM
model (\ref{tanhmodel}), versus the SPARC observational data and
the temperature parameter as a function of the radius
respectively. As it can be seen, the SIDM model produces viable
rotation curves compatible with the SPARC data. Also in Tables
\ref{collUGC10310}, \ref{NavaroUGC10310}, \ref{BuckertUGC10310}
and \ref{EinastoUGC10310} we present the optimization values for
the SIDM model, and the other DM profiles. Also in Table
\ref{EVALUATIONUGC10310} we present the overall evaluation of the
SIDM model for the galaxy at hand. The resulting phenomenology is
viable.
\begin{table}[h!]
  \begin{center}
    \caption{Collisional Dark Matter Optimization Values}
    \label{collUGC10310}
     \begin{tabular}{|r|r|}
     \hline
      \textbf{Parameter}   & \textbf{Optimization Values}
      \\  \hline
     $\delta_{\gamma} $ & 0.0000000012
\\  \hline
$\gamma_0 $ & 1.0001 \\ \hline $K_0$ ($M_{\odot} \,
\mathrm{Kpc}^{-3} \, (\mathrm{km/s})^{2}$)& 2000  \\ \hline
    \end{tabular}
  \end{center}
\end{table}
\begin{table}[h!]
  \begin{center}
    \caption{NFW  Optimization Values}
    \label{NavaroUGC10310}
     \begin{tabular}{|r|r|}
     \hline
      \textbf{Parameter}   & \textbf{Optimization Values}
      \\  \hline
   $\rho_s$   & $5\times 10^7$
\\  \hline
$r_s$&  2.81
\\  \hline
    \end{tabular}
  \end{center}
\end{table}
\begin{table}[h!]
  \begin{center}
    \caption{Burkert Optimization Values}
    \label{BuckertUGC10310}
     \begin{tabular}{|r|r|}
     \hline
      \textbf{Parameter}   & \textbf{Optimization Values}
      \\  \hline
     $\rho_0^B$  & $1\times 10^8$
\\  \hline
$r_0$&  3.19
\\  \hline
    \end{tabular}
  \end{center}
\end{table}
\begin{table}[h!]
  \begin{center}
    \caption{Einasto Optimization Values}
    \label{EinastoUGC10310}
    \begin{tabular}{|r|r|}
     \hline
      \textbf{Parameter}   & \textbf{Optimization Values}
      \\  \hline
     $\rho_e$  &$1\times 10^7$
\\  \hline
$r_e$ & 3.42
\\  \hline
$n_e$ & 0.8
\\  \hline
    \end{tabular}
  \end{center}
\end{table}
\begin{table}[h!]
\centering \caption{Physical assessment of collisional DM
parameters (second set).}
\begin{tabular}{lcc}
\hline
Parameter & Value & Physical Verdict \\
\hline
$\gamma_0$ & 1.0001 & Slightly above isothermal, low central pressure \\
$\delta_\gamma$ & 0.0000000012 & Extremely small variation \\
$r_\gamma$ & 1.5 Kpc & Transition radius inside inner halo \\
$K_0$ & 2000 & Enough pressure support \\
$r_c$ & 0.5 Kpc & Small core scale, reasonable \\
$p$ & 0.01 & Very shallow K(r) decrease, nearly constant \\
\hline
Overall &-& Physically plausible \\
\hline
\end{tabular}
\label{EVALUATIONUGC10310}
\end{table}


\subsection{The Galaxy UGC11455 Marginally, Extended Marginally presentable}

\begin{figure}[h!]
\centering
\includegraphics[width=20pc]{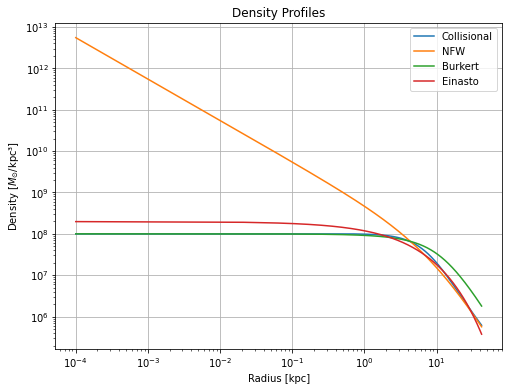}
\caption{The density of the collisional DM model (\ref{tanhmodel})
for the galaxy UGC11455, as a function of the radius.}
\label{UGC11455dens}
\end{figure}

\begin{figure}[h!]
\centering
\includegraphics[width=20pc]{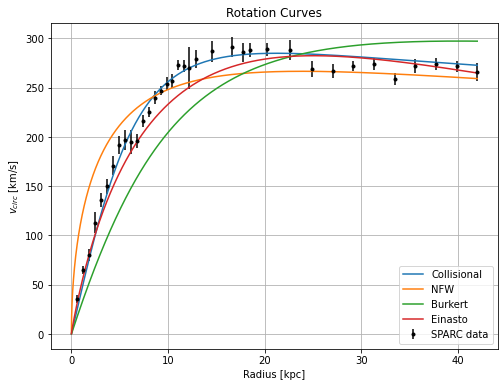}
\caption{The predicted rotation curves after using an optimization
for the collisional DM model (\ref{tanhmodel}), versus the SPARC
observational data for the galaxy UGC11455. We also plotted the
optimized curves for the NFW model, the Burkert model and the
Einasto model.} \label{UGC11455}
\end{figure}

\begin{figure}[h!]
\centering
\includegraphics[width=20pc]{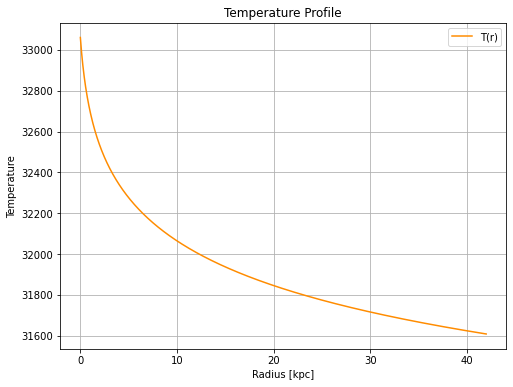}
\caption{The temperature as a function of the radius for the
collisional DM model (\ref{tanhmodel}) for the galaxy UGC11455.}
\label{UGC11455temp}
\end{figure}

\begin{figure}[h!]
\centering
\includegraphics[width=20pc]{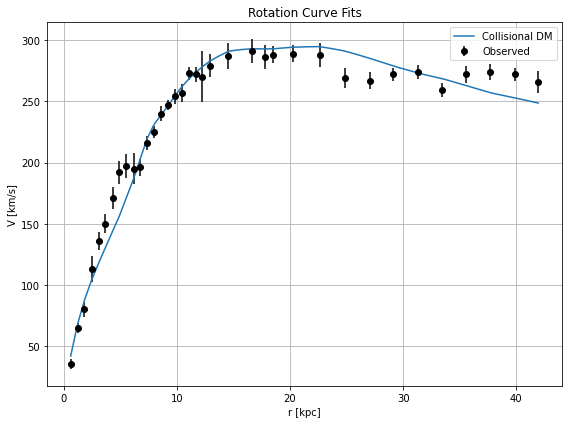}
\caption{The predicted rotation curves after using an optimization
for the collisional DM model (\ref{tanhmodel}), versus the
extended SPARC observational data for the galaxy UGC11455. The
model includes the rotation curves from all the components of the
galaxy, including gas and disk velocities, along with the
collisional DM model.} \label{extendedUGC11455}
\end{figure}

For this galaxy, we shall choose $\rho_0=1\times
10^8$$M_{\odot}/\mathrm{Kpc}^{3}$. UGC11455 is a dwarf irregular
galaxy characterized by an extended HI gas disk and a dense,
compact dark matter halo that suppresses small-scale spiral
structure. Its distance is not precisely documented but falls
within typical ranges for nearby UGC dwarfs, around 10-20 Mpc
based on similar objects. In Figs. \ref{UGC11455dens},
\ref{UGC11455} and \ref{UGC11455temp} we present the density of
the collisional DM model, the predicted rotation curves after
using an optimization for the collisional DM model
(\ref{tanhmodel}), versus the SPARC observational data and the
temperature parameter as a function of the radius respectively. As
it can be seen, the SIDM model produces marginally viable rotation
curves compatible with the SPARC data. Also in Tables
\ref{collUGC11455}, \ref{NavaroUGC11455}, \ref{BuckertUGC11455}
and \ref{EinastoUGC11455} we present the optimization values for
the SIDM model, and the other DM profiles. Also in Table
\ref{EVALUATIONUGC11455} we present the overall evaluation of the
SIDM model for the galaxy at hand. The resulting phenomenology is
marginally viable.
\begin{table}[h!]
  \begin{center}
    \caption{Collisional Dark Matter Optimization Values}
    \label{collUGC11455}
     \begin{tabular}{|r|r|}
     \hline
      \textbf{Parameter}   & \textbf{Optimization Values}
      \\  \hline
     $\delta_{\gamma} $ & 0.0000000012
\\  \hline
$\gamma_0 $ & 1.0001 \\ \hline $K_0$ ($M_{\odot} \,
\mathrm{Kpc}^{-3} \, (\mathrm{km/s})^{2}$)& 33000  \\ \hline
    \end{tabular}
  \end{center}
\end{table}
\begin{table}[h!]
  \begin{center}
    \caption{NFW  Optimization Values}
    \label{NavaroUGC11455}
     \begin{tabular}{|r|r|}
     \hline
      \textbf{Parameter}   & \textbf{Optimization Values}
      \\  \hline
   $\rho_s$   & $5\times 10^7$
\\  \hline
$r_s$&  11.03
\\  \hline
    \end{tabular}
  \end{center}
\end{table}
\begin{table}[h!]
  \begin{center}
    \caption{Burkert Optimization Values}
    \label{BuckertUGC11455}
     \begin{tabular}{|r|r|}
     \hline
      \textbf{Parameter}   & \textbf{Optimization Values}
      \\  \hline
     $\rho_0^B$  & $1\times 10^8$
\\  \hline
$r_0$&  12.33
\\  \hline
    \end{tabular}
  \end{center}
\end{table}
\begin{table}[h!]
  \begin{center}
    \caption{Einasto Optimization Values}
    \label{EinastoUGC11455}
    \begin{tabular}{|r|r|}
     \hline
      \textbf{Parameter}   & \textbf{Optimization Values}
      \\  \hline
     $\rho_e$  &$1\times 10^7$
\\  \hline
$r_e$ & 13.89
\\  \hline
$n_e$ & 0.67
\\  \hline
    \end{tabular}
  \end{center}
\end{table}
\begin{table}[h!]
\centering \caption{Physical assessment of collisional DM
parameters for UGC11455.}
\begin{tabular}{lcc}
\hline
Parameter & Value & Physical Verdict \\
\hline
$\gamma_0$ & 1.0001 & Extremely close to isothermal\\
$\delta_\gamma$ & $1.2\times10^{-9}$ & Negligible variation \\
$r_\gamma$ & 1.5 Kpc & Physically irrelevant due to tiny $\delta_\gamma$ \\
$K_0$ & 33000 & Very high entropy scale; central halo is extremely hot \\
$r_c$ & 0.5 Kpc & Small core scale; reasonable for inner halo \\
$p$ & 0.01 & Nearly flat K(r) \\
\hline
Overall &-& Physically plausible \\
\hline
\end{tabular}
\label{EVALUATIONUGC11455}
\end{table}
Now the extended picture including the rotation velocity from the
other components of the galaxy, such as the disk and gas, makes
the collisional DM model viable for this galaxy. In Fig.
\ref{extendedUGC11455} we present the combined rotation curves
including the other components of the galaxy along with the
collisional matter. As it can be seen, the extended collisional DM
model is marginally viable.
Also in Table \ref{evaluationextendedUGC11455} we present the
values of the free parameters of the collisional DM model for
which the maximum compatibility with the SPARC data comes for the
galaxy UGC11455.
\begin{table}[h!]
\centering \caption{Physical assessment of Extended collisional DM
parameters for galaxy UGC11455.}
\begin{tabular}{lcc}
\hline
Parameter & Value & Physical Verdict \\
\hline
$\gamma_0$ & 1.17946284 & Slightly above isothermal\\
$\delta_\gamma$ & 0.05691342 & Noticeable radial variation\\
$K_0$ & 3000 & Moderate entropy   \\
$ml_{disk}$ & 0.57059467 & Sub-maximal disk M/L \\
$ml_{bulge}$ & 0.00000000 & No bulge component \\
\hline
Overall &-& Physically acceptable\\
\hline
\end{tabular}
\label{evaluationextendedUGC11455}
\end{table}

\subsection{The Galaxy UGC11557}

\begin{figure}[h!]
\centering
\includegraphics[width=20pc]{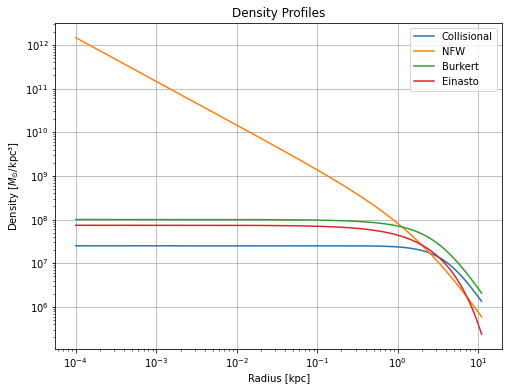}
\caption{The density of the collisional DM model (\ref{tanhmodel})
for the galaxy UGC11557, as a function of the radius.}
\label{UGC11557dens}
\end{figure}

\begin{figure}[h!]
\centering
\includegraphics[width=20pc]{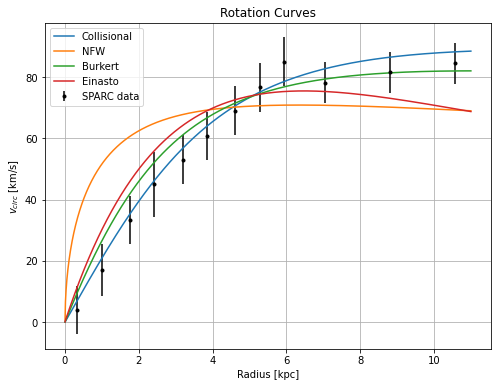}
\caption{The predicted rotation curves after using an optimization
for the collisional DM model (\ref{tanhmodel}), versus the SPARC
observational data for the galaxy UGC11557. We also plotted the
optimized curves for the NFW model, the Burkert model and the
Einasto model.} \label{UGC11557}
\end{figure}

\begin{figure}[h!]
\centering
\includegraphics[width=20pc]{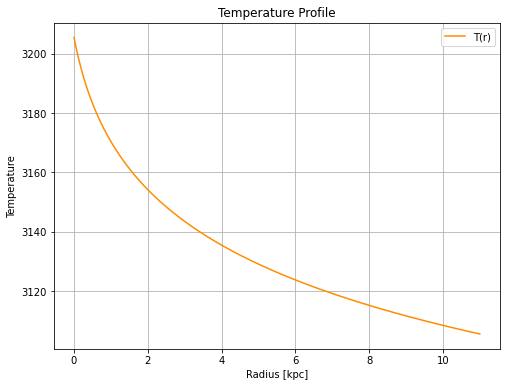}
\caption{The temperature as a function of the radius for the
collisional DM model (\ref{tanhmodel}) for the galaxy UGC11557.}
\label{UGC11557temp}
\end{figure}

For this galaxy, we shall choose $\rho_0=2.5\times
10^7$$M_{\odot}/\mathrm{Kpc}^{3}$. UGC11557 is a
low-surface-brightness galaxy located approximately 23.7 Mpc away
in the constellation Cepheus. It is classified as an Sd-type
spiral galaxy, characterized by a large HI disk extending beyond
its optical radius. In Figs. \ref{UGC11557dens}, \ref{UGC11557}
and \ref{UGC11557temp} we present the density of the collisional
DM model, the predicted rotation curves after using an
optimization for the collisional DM model (\ref{tanhmodel}),
versus the SPARC observational data and the temperature parameter
as a function of the radius respectively. As it can be seen, the
SIDM model produces viable rotation curves compatible with the
SPARC data. Also in Tables \ref{collUGC11557},
\ref{NavaroUGC11557}, \ref{BuckertUGC11557} and
\ref{EinastoUGC11557} we present the optimization values for the
SIDM model, and the other DM profiles. Also in Table
\ref{EVALUATIONUGC11557} we present the overall evaluation of the
SIDM model for the galaxy at hand. The resulting phenomenology is
viable.
\begin{table}[h!]
  \begin{center}
    \caption{Collisional Dark Matter Optimization Values}
    \label{collUGC11557}
     \begin{tabular}{|r|r|}
     \hline
      \textbf{Parameter}   & \textbf{Optimization Values}
      \\  \hline
     $\delta_{\gamma} $ & 0.0000000012
\\  \hline
$\gamma_0 $ & 1.0001 \\ \hline $K_0$ ($M_{\odot} \,
\mathrm{Kpc}^{-3} \, (\mathrm{km/s})^{2}$)& 3200 \\ \hline
    \end{tabular}
  \end{center}
\end{table}
\begin{table}[h!]
  \begin{center}
    \caption{NFW  Optimization Values}
    \label{NavaroUGC11557}
     \begin{tabular}{|r|r|}
     \hline
      \textbf{Parameter}   & \textbf{Optimization Values}
      \\  \hline
   $\rho_s$   & $5\times 10^7$
\\  \hline
$r_s$&  2.93
\\  \hline
    \end{tabular}
  \end{center}
\end{table}
\begin{table}[h!]
  \begin{center}
    \caption{Burkert Optimization Values}
    \label{BuckertUGC11557}
     \begin{tabular}{|r|r|}
     \hline
      \textbf{Parameter}   & \textbf{Optimization Values}
      \\  \hline
     $\rho_0^B$  & $1\times 10^8$
\\  \hline
$r_0$&  3.40
\\  \hline
    \end{tabular}
  \end{center}
\end{table}
\begin{table}[h!]
  \begin{center}
    \caption{Einasto Optimization Values}
    \label{EinastoUGC11557}
    \begin{tabular}{|r|r|}
     \hline
      \textbf{Parameter}   & \textbf{Optimization Values}
      \\  \hline
     $\rho_e$  &$1\times 10^7$
\\  \hline
$r_e$ & 3.83
\\  \hline
$n_e$ & 1
\\  \hline
    \end{tabular}
  \end{center}
\end{table}
\begin{table}[h!] \centering \caption{Physical assessment of
collisional DM parameters for UGC11557.}
\begin{tabular}{lcc}
\hline
Parameter & Value & Physical Verdict \\
\hline
$\gamma_0$ & 1.0001 & Extremely close to isothermal \\
$\delta_\gamma$ & $1.2\times10^{-9}$ & Negligible variation \\
$r_\gamma$ & 1.5 Kpc & Transition radius physically irrelevant\\
$K_0$ & 3200 & Enough pressure support \\
$r_c$ & 0.5 Kpc & Small core scale \\
$p$ & 0.01 & Nearly flat K(r) \\
\hline
Overall &-& Physically plausible \\
\hline
\end{tabular}
\label{EVALUATIONUGC11557}
\end{table}

\subsection{The Galaxy UGC11820 Non-viable Dwarf, One of the Very Few}

\begin{figure}[h!]
\centering
\includegraphics[width=20pc]{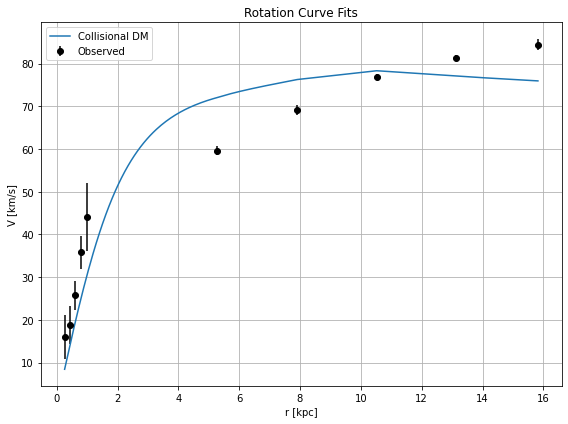}
\caption{The predicted rotation curves after using an optimization
for the collisional DM model (\ref{tanhmodel}), versus the
extended SPARC observational data for the galaxy UGC11820. The
model includes the rotation curves from all the components of the
galaxy, including gas and disk velocities, along with the
collisional DM model.} \label{extendedUGC11820}
\end{figure}

For this galaxy, we shall choose $\rho_0=1.5\times
10^8$$M_{\odot}/\mathrm{Kpc}^{3}$. UGC11820 is a nearby late-type,
low-surface-brightness/irregular disk galaxy  which is a
dwarf/late-type system with asymmetric outer kinematics; its
distance is commonly taken to be of order \( D \sim
1.9\times10^{1}\ \mathrm{Mpc}\).
\begin{figure}[h!]
\centering
\includegraphics[width=20pc]{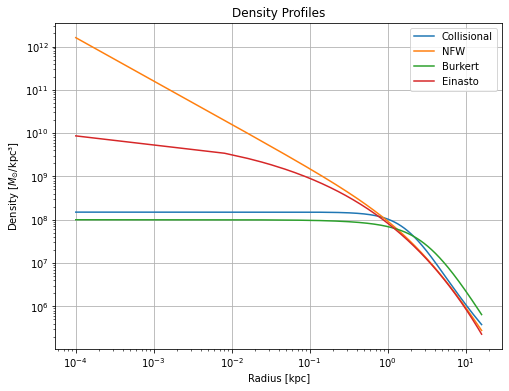}
\caption{The density of the collisional DM model (\ref{tanhmodel})
for the galaxy UGC11820, as a function of the radius.}
\label{UGC11820dens}
\end{figure}
\begin{figure}[h!]
\centering
\includegraphics[width=20pc]{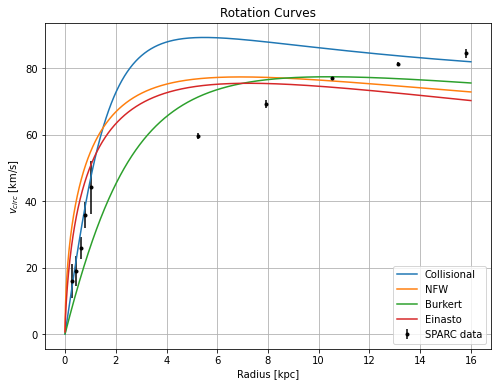}
\caption{The predicted rotation curves after using an optimization
for the collisional DM model (\ref{tanhmodel}), versus the SPARC
observational data for the galaxy UGC11820. We also plotted the
optimized curves for the NFW model, the Burkert model and the
Einasto model.} \label{UGC11820}
\end{figure}
\begin{table}[h!]
  \begin{center}
    \caption{Collisional Dark Matter Optimization Values}
    \label{collUGC11820}
     \begin{tabular}{|r|r|}
     \hline
      \textbf{Parameter}   & \textbf{Optimization Values}
      \\  \hline
     $\delta_{\gamma} $ & 0.0000000012
\\  \hline
$\gamma_0 $ & 1.0001 \\ \hline $K_0$ ($M_{\odot} \,
\mathrm{Kpc}^{-3} \, (\mathrm{km/s})^{2}$)& 3200 \\ \hline
    \end{tabular}
  \end{center}
\end{table}
\begin{table}[h!]
  \begin{center}
    \caption{NFW  Optimization Values}
    \label{NavaroUGC11820}
     \begin{tabular}{|r|r|}
     \hline
      \textbf{Parameter}   & \textbf{Optimization Values}
      \\  \hline
   $\rho_s$   & $5\times 10^7$
\\  \hline
$r_s$&  3.20
\\  \hline
    \end{tabular}
  \end{center}
\end{table}
\begin{figure}[h!]
\centering
\includegraphics[width=20pc]{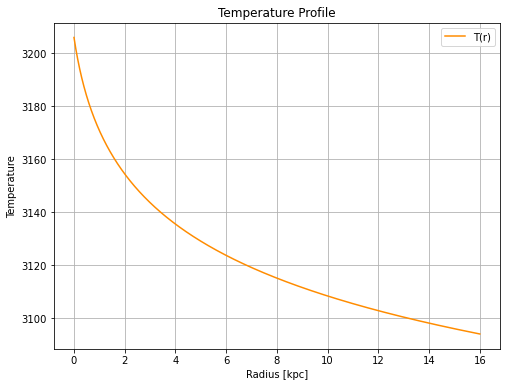}
\caption{The temperature as a function of the radius for the
collisional DM model (\ref{tanhmodel}) for the galaxy UGC11820.}
\label{UGC11820temp}
\end{figure}
\begin{table}[h!]
  \begin{center}
    \caption{Burkert Optimization Values}
    \label{BuckertUGC11820}
     \begin{tabular}{|r|r|}
     \hline
      \textbf{Parameter}   & \textbf{Optimization Values}
      \\  \hline
     $\rho_0^B$  & $1\times 10^8$
\\  \hline
$r_0$&  3.21
\\  \hline
    \end{tabular}
  \end{center}
\end{table}
\begin{table}[h!]
  \begin{center}
    \caption{Einasto Optimization Values}
    \label{EinastoUGC11820}
    \begin{tabular}{|r|r|}
     \hline
      \textbf{Parameter}   & \textbf{Optimization Values}
      \\  \hline
     $\rho_e$  &$1\times 10^7$
\\  \hline
$r_e$ & 3.51
\\  \hline
$n_e$ & 0.28
\\  \hline
    \end{tabular}
  \end{center}
\end{table}
\begin{table}[h!]
\centering \caption{Physical assessment of collisional DM
parameters for UGC11820.}
\begin{tabular}{lcc}
\hline Parameter & Value & Physical Verdict \\ \hline
$\gamma_0$ & $1.0001$ & Essentially isothermal \\
$\delta_\gamma$ & $1.2\times10^{-9}$ & Practically zero  \\
$r_\gamma$ & $1.5\ \mathrm{Kpc}$ & Transition radius inside the inner halo \\
$K_0$ ($M_{\odot}\ \mathrm{Kpc}^{-3}\ (\mathrm{km/s})^{2}$) & $3.2\times10^{3}$ & Enough pressure support \\
$r_c$ & $0.5\ \mathrm{Kpc}$ & Small core radius\\
$p$ & $0.01$ & Nearly flat $K(r)$ \\ \hline Overall & - &
Physically self-consistent \\ \hline
\end{tabular}
\label{EVALUATIONUGC11820}
\end{table}
Now the extended picture including the rotation velocity from the
other components of the galaxy, such as the disk and gas, makes
the collisional DM model viable for this galaxy. In Fig.
\ref{extendedUGC11820} we present the combined rotation curves
including the other components of the galaxy along with the
collisional matter. As it can be seen, the extended collisional DM
model is non-viable.
Also in Table \ref{evaluationextendedUGC11820} we present the
values of the free parameters of the collisional DM model for
which the maximum compatibility with the SPARC data comes for the
galaxy UGC11820.
\begin{table}[h!]
\centering \caption{Physical assessment of Extended collisional DM
parameters (UGC11820).}
\begin{tabular}{lcc}
\hline
Parameter & Value & Physical Verdict \\
\hline
$\gamma_0$ & 0.97799438 & Slightly below isothermal ($\gamma<1$): central EoS very soft \\
$\delta_\gamma$ & 0.00000001 & Exactly zero radial variation \\
$K_0$ & 3000 & Moderate entropy/pressure scale \\
ml\_disk & 0.00000041 & Effectively zero disk M/L  \\
ml\_bulge & 0.00000000 & Zero bulge contribution  \\
\hline
Overall &-& Mixed plausibility\\
\hline
\end{tabular}
\label{evaluationextendedUGC11820}
\end{table}

\subsection{The Galaxy UGC11914 Remarkably Marginally Viable Large Galaxy}

\begin{figure}[h!]
\centering
\includegraphics[width=20pc]{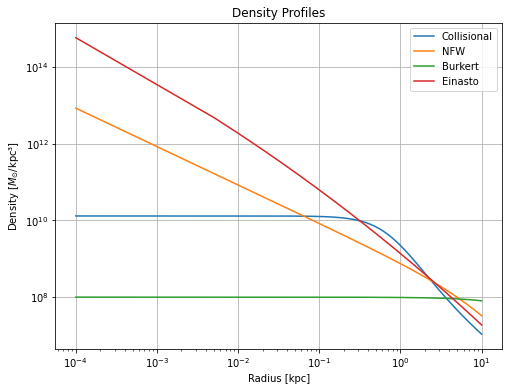}
\caption{The density of the collisional DM model (\ref{tanhmodel})
for the galaxy UGC11914, as a function of the radius.}
\label{UGC11914dens}
\end{figure}

\begin{figure}[h!]
\centering
\includegraphics[width=20pc]{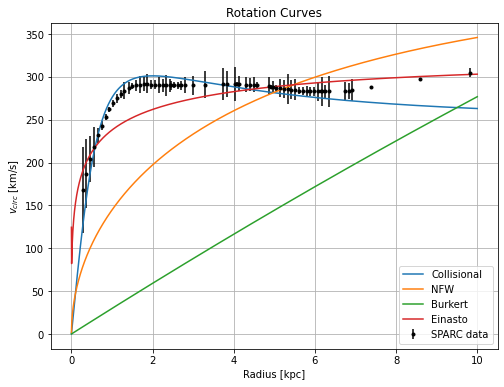}
\caption{The predicted rotation curves after using an optimization
for the collisional DM model (\ref{tanhmodel}), versus the SPARC
observational data for the galaxy UGC11914. We also plotted the
optimized curves for the NFW model, the Burkert model and the
Einasto model.} \label{UGC11914}
\end{figure}

\begin{figure}[h!]
\centering
\includegraphics[width=20pc]{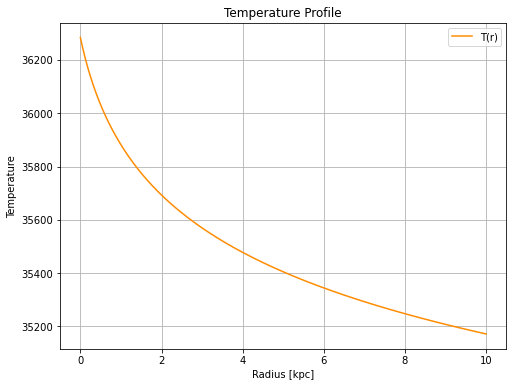}
\caption{The temperature as a function of the radius for the
collisional DM model (\ref{tanhmodel}) for the galaxy UGC11914.}
\label{UGC11914temp}
\end{figure}

For this galaxy, we shall choose $\rho_0=1.3\times
10^{10}$$M_{\odot}/\mathrm{Kpc}^{3}$. UGC11914 is an
early-type/spiral galaxy with a relatively large rotation
amplitude and a conspicuous HI morphology where the neutral gas is
concentrated in a ring. Its distance in recent observational
compilations is \(D\sim 15\ \mathrm{Mpc}\), it is morphologically
classified as an (early) spiral/SA-type system rather than a
low-mass dwarf. In Figs. \ref{UGC11914dens}, \ref{UGC11914} and
\ref{UGC11914temp} we present the density of the collisional DM
model, the predicted rotation curves after using an optimization
for the collisional DM model (\ref{tanhmodel}), versus the SPARC
observational data and the temperature parameter as a function of
the radius respectively. As it can be seen, the SIDM model
produces marginally viable rotation curves compatible with the
SPARC data. Also in Tables \ref{collUGC11914},
\ref{NavaroUGC11914}, \ref{BuckertUGC11914} and
\ref{EinastoUGC11914} we present the optimization values for the
SIDM model, and the other DM profiles. Also in Table
\ref{EVALUATIONUGC11914} we present the overall evaluation of the
SIDM model for the galaxy at hand. The resulting phenomenology is
marginally viable.
\begin{table}[h!]
  \begin{center}
    \caption{Collisional Dark Matter Optimization Values}
    \label{collUGC11914}
     \begin{tabular}{|r|r|}
     \hline
      \textbf{Parameter}   & \textbf{Optimization Values}
      \\  \hline
     $\delta_{\gamma} $ & 0.0000000012
\\  \hline
$\gamma_0 $ & 1.0001 \\ \hline $K_0$ ($M_{\odot} \,
\mathrm{Kpc}^{-3} \, (\mathrm{km/s})^{2}$)& 36200  \\ \hline
    \end{tabular}
  \end{center}
\end{table}

\begin{table}[h!]
  \begin{center}
    \caption{NFW  Optimization Values}
    \label{NavaroUGC11914}
     \begin{tabular}{|r|r|}
     \hline
      \textbf{Parameter}   & \textbf{Optimization Values}
      \\  \hline
   $\rho_s$   & $5\times 10^7$
\\  \hline
$r_s$&  16.74
\\  \hline
    \end{tabular}
  \end{center}
\end{table}
\begin{table}[h!]
  \begin{center}
    \caption{Burkert Optimization Values}
    \label{BuckertUGC11914}
     \begin{tabular}{|r|r|}
     \hline
      \textbf{Parameter}   & \textbf{Optimization Values}
      \\  \hline
     $\rho_0^B$  & $1\times 10^8$
\\  \hline
$r_0$&  50
\\  \hline
    \end{tabular}
  \end{center}
\end{table}
\begin{table}[h!]
  \begin{center}
    \caption{Einasto Optimization Values}
    \label{EinastoUGC11914}
    \begin{tabular}{|r|r|}
     \hline
      \textbf{Parameter}   & \textbf{Optimization Values}
      \\  \hline
     $\rho_e$  &$1\times 10^7$
\\  \hline
$r_e$ & 13.78
\\  \hline
$n_e$ & 0.05
\\  \hline
    \end{tabular}
  \end{center}
\end{table}
\begin{table}[h!]
\centering \caption{Physical assessment of collisional DM
parameters for UGC11914.}
\begin{tabular}{lcc}
\hline Parameter & Value   & Physical Verdict \\ \hline
$\gamma_0$ & $1.0001$ & Practically isothermal \\
$\delta_\gamma$ & $1.2\times10^{-9}$ & Negligible \\
$r_\gamma$ & $1.5\ \mathrm{Kpc}$ & Transition placed in the inner halo  \\
$K_0$ ($M_{\odot}\ \mathrm{Kpc}^{-3}\ (\mathrm{km/s})^{2}$) & $3.62\times10^{4}$ & High entropy/temperature scale\\
$r_c$ & $0.5\ \mathrm{Kpc}$ & Small core scale \\
$p$ & $0.01$ & Almost flat $K(r)$  \\ \hline Overall & - &
Physically self-consistent \\ \hline
\end{tabular}
\label{EVALUATIONUGC11914}
\end{table}

\subsection{The Galaxy UGC12506 Non-Viable Extended Marginally Viable}

\begin{figure}[h!]
\centering
\includegraphics[width=20pc]{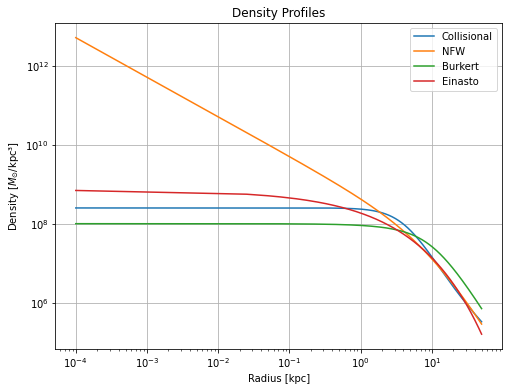}
\caption{The density of the collisional DM model (\ref{tanhmodel})
for the galaxy UGC12506, as a function of the radius.}
\label{UGC12506dens}
\end{figure}

\begin{figure}[h!]
\centering
\includegraphics[width=20pc]{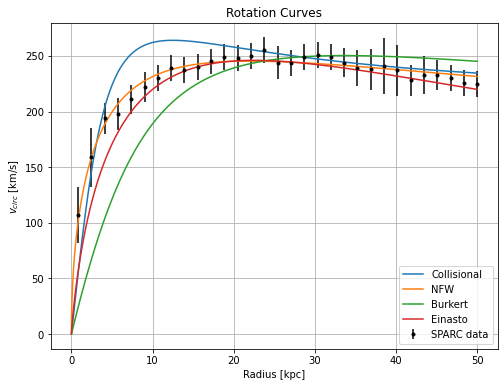}
\caption{The predicted rotation curves after using an optimization
for the collisional DM model (\ref{tanhmodel}), versus the SPARC
observational data for the galaxy UGC12506. We also plotted the
optimized curves for the NFW model, the Burkert model and the
Einasto model.} \label{UGC12506}
\end{figure}

\begin{figure}[h!]
\centering
\includegraphics[width=20pc]{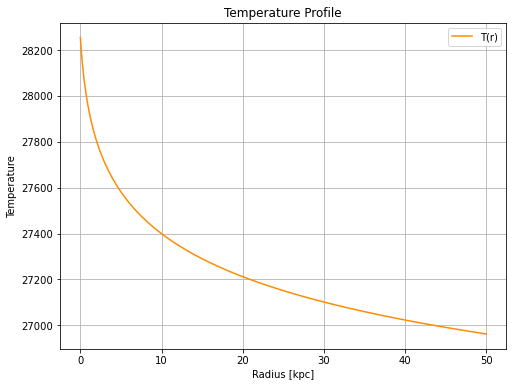}
\caption{The temperature as a function of the radius for the
collisional DM model (\ref{tanhmodel}) for the galaxy UGC12506.}
\label{UGC12506temp}
\end{figure}

\begin{figure}[h!]
\centering
\includegraphics[width=20pc]{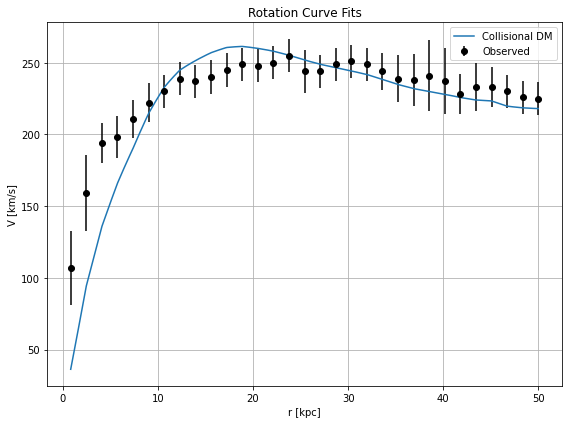}
\caption{The predicted rotation curves after using an optimization
for the collisional DM model (\ref{tanhmodel}), versus the
extended SPARC observational data for the galaxy UGC12506. The
model includes the rotation curves from all the components of the
galaxy, including gas and disk velocities, along with the
collisional DM model.} \label{extendedUGC12506}
\end{figure}

For this galaxy, we shall choose $\rho_0=2.5\times
10^8$$M_{\odot}/\mathrm{Kpc}^{3}$. UGC12506 is classified as a
spiral (Sc) galaxy, and it is included among HI-rich ''High Mass''
galaxies with extended neutral gas disks. Its distance is
estimated from redshift/HI surveys to be on the order of \(D\sim
100\text{-}150\ \mathrm{Mpc}\). In Figs. \ref{UGC12506dens},
\ref{UGC12506} and \ref{UGC12506temp} we present the density of
the collisional DM model, the predicted rotation curves after
using an optimization for the collisional DM model
(\ref{tanhmodel}), versus the SPARC observational data and the
temperature parameter as a function of the radius respectively. As
it can be seen, the SIDM model produces non-viable rotation curves
incompatible with the SPARC data. Also in Tables
\ref{collUGC12506}, \ref{NavaroUGC12506}, \ref{BuckertUGC12506}
and \ref{EinastoUGC12506} we present the optimization values for
the SIDM model, and the other DM profiles. Also in Table
\ref{EVALUATIONUGC12506} we present the overall evaluation of the
SIDM model for the galaxy at hand. The resulting phenomenology is
non-viable.
\begin{table}[h!]
  \begin{center}
    \caption{Collisional Dark Matter Optimization Values}
    \label{collUGC12506}
     \begin{tabular}{|r|r|}
     \hline
      \textbf{Parameter}   & \textbf{Optimization Values}
      \\  \hline
     $\delta_{\gamma} $ & 0.0000000012
\\  \hline
$\gamma_0 $ & 1.0001\\ \hline $K_0$ ($M_{\odot} \,
\mathrm{Kpc}^{-3} \, (\mathrm{km/s})^{2}$)& 28200  \\ \hline
    \end{tabular}
  \end{center}
\end{table}
\begin{table}[h!]
  \begin{center}
    \caption{NFW  Optimization Values}
    \label{NavaroUGC12506}
     \begin{tabular}{|r|r|}
     \hline
      \textbf{Parameter}   & \textbf{Optimization Values}
      \\  \hline
   $\rho_s$   & $5\times 10^7$
\\  \hline
$r_s$&  10.15
\\  \hline
    \end{tabular}
  \end{center}
\end{table}
\begin{table}[h!]
  \begin{center}
    \caption{Burkert Optimization Values}
    \label{BuckertUGC12506}
     \begin{tabular}{|r|r|}
     \hline
      \textbf{Parameter}   & \textbf{Optimization Values}
      \\  \hline
     $\rho_0^B$  & $1\times 10^8$
\\  \hline
$r_0$&  10.38
\\  \hline
    \end{tabular}
  \end{center}
\end{table}
\begin{table}[h!]
  \begin{center}
    \caption{Einasto Optimization Values}
    \label{EinastoUGC12506}
    \begin{tabular}{|r|r|}
     \hline
      \textbf{Parameter}   & \textbf{Optimization Values}
      \\  \hline
     $\rho_e$  &$1\times 10^7$
\\  \hline
$r_e$ & 11.80
\\  \hline
$n_e$ & 0.47
\\  \hline
    \end{tabular}
  \end{center}
\end{table}
\begin{table}[h!]
\centering \caption{Physical assessment of collisional DM
parameters for UGC12506.}
\begin{tabular}{lcc}
\hline
Parameter & Value   & Physical Verdict \\
\hline
$\gamma_0$ & $1.0001$ & Practically isothermal \\
$\delta_\gamma$ & $1.2\times10^{-9}$ & Negligible  \\
$r_\gamma$ & $1.5\ \mathrm{Kpc}$ & Transition inside inner halo \\
$K_0$ & $2.82\times10^{4}$ & Moderate-high entropy scale \\
$r_c$ & $0.5\ \mathrm{Kpc}$ & Small core radius\\
$p$ & $0.01$ & Almost flat $K(r)$  \\
\hline
Overall &-& Physically consistent \\
\hline
\end{tabular}
\label{EVALUATIONUGC12506}
\end{table}
Now the extended picture including the rotation velocity from the
other components of the galaxy, such as the disk and gas, makes
the collisional DM model viable for this galaxy. In Fig.
\ref{extendedUGC12506} we present the combined rotation curves
including the other components of the galaxy along with the
collisional matter. As it can be seen, the extended collisional DM
model is marginally viable.
Also in Table \ref{evaluationextendedUGC12506} we present the
values of the free parameters of the collisional DM model for
which the maximum compatibility with the SPARC data comes for the
galaxy UGC12506.
\begin{table}[h!]
\centering \caption{Physical assessment of Extended collisional DM
parameters (UGC12506).}
\begin{tabular}{lcc}
\hline
Parameter & Value & Physical Verdict \\
\hline
$\gamma_0$ & 1.13276520 & Slightly above isothermal \\
$\delta_\gamma$ & 0.03391874 & Small radial variation \\
$K_0$ & 3000 & Moderate entropy/pressure scale\\
ml\_disk & 1.00000000 & High disk mass-to-light \\
ml\_bulge & 0.00000000 & Zero bulge contribution  \\
\hline
Overall &-& Physically plausible \\
\hline
\end{tabular}
\label{evaluationextendedUGC12506}
\end{table}

\subsection{The Galaxy UGC12632}

\begin{figure}[h!]
\centering
\includegraphics[width=20pc]{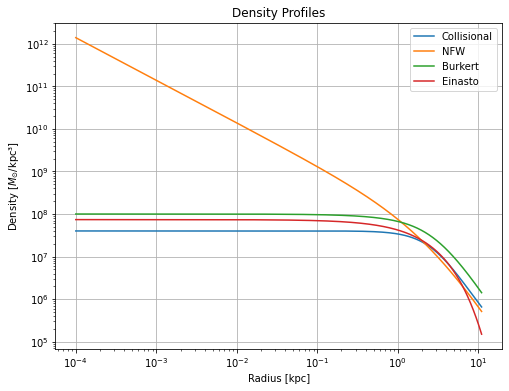}
\caption{The density of the collisional DM model (\ref{tanhmodel})
for the galaxy UGC12632, as a function of the radius.}
\label{UGC12632dens}
\end{figure}
\begin{figure}[h!]
\centering
\includegraphics[width=20pc]{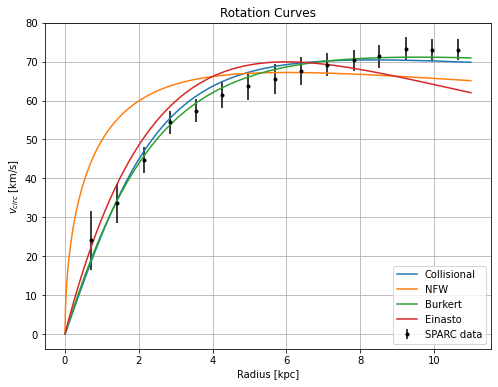}
\caption{The predicted rotation curves after using an optimization
for the collisional DM model (\ref{tanhmodel}), versus the SPARC
observational data for the galaxy UGC12632. We also plotted the
optimized curves for the NFW model, the Burkert model and the
Einasto model.} \label{UGC12632}
\end{figure}
\begin{figure}[h!]
\centering
\includegraphics[width=20pc]{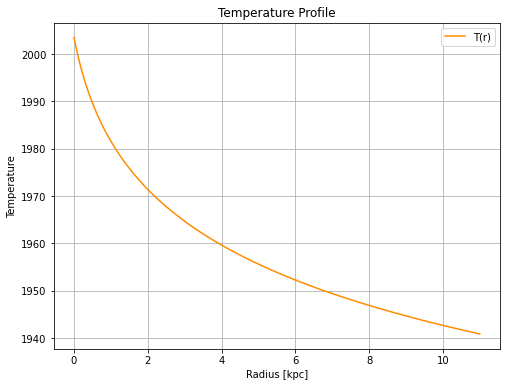}
\caption{The temperature as a function of the radius for the
collisional DM model (\ref{tanhmodel}) for the galaxy UGC12632.}
\label{UGC12632temp}
\end{figure}

For this galaxy, we shall choose $\rho_0=4\times
10^7$$M_{\odot}/\mathrm{Kpc}^{3}$. UGC12632 is a
low-surface-brightness/late-type dwarf/spiral galaxy located in
the general direction of Andromeda. Distance estimates based on
redshift and a number of observational compilations place it at of
order \(D\sim 8\text{-}12\ \mathrm{Mpc}\). In Figs.
\ref{UGC12632dens}, \ref{UGC12632} and \ref{UGC12632temp} we
present the density of the collisional DM model, the predicted
rotation curves after using an optimization for the collisional DM
model (\ref{tanhmodel}), versus the SPARC observational data and
the temperature parameter as a function of the radius
respectively. As it can be seen, the SIDM model produces viable
rotation curves compatible with the SPARC data. Also in Tables
\ref{collUGC12632}, \ref{NavaroUGC12632}, \ref{BuckertUGC12632}
and \ref{EinastoUGC12632} we present the optimization values for
the SIDM model, and the other DM profiles. Also in Table
\ref{EVALUATIONUGC12632} we present the overall evaluation of the
SIDM model for the galaxy at hand. The resulting phenomenology is
viable.
\begin{table}[h!]
  \begin{center}
    \caption{Collisional Dark Matter Optimization Values}
    \label{collUGC12632}
     \begin{tabular}{|r|r|}
     \hline
      \textbf{Parameter}   & \textbf{Optimization Values}
      \\  \hline
     $\delta_{\gamma} $ & 0.0000000012
\\  \hline
$\gamma_0 $ & 1.0001 \\ \hline $K_0$ ($M_{\odot} \,
\mathrm{Kpc}^{-3} \, (\mathrm{km/s})^{2}$)& 2000 \\ \hline
    \end{tabular}
  \end{center}
\end{table}
\begin{table}[h!]
  \begin{center}
    \caption{NFW  Optimization Values}
    \label{NavaroUGC12632}
     \begin{tabular}{|r|r|}
     \hline
      \textbf{Parameter}   & \textbf{Optimization Values}
      \\  \hline
   $\rho_s$   & $5\times 10^7$
\\  \hline
$r_s$&  2.78
\\  \hline
    \end{tabular}
  \end{center}
\end{table}
\begin{table}[h!]
  \begin{center}
    \caption{Burkert Optimization Values}
    \label{BuckertUGC12632}
     \begin{tabular}{|r|r|}
     \hline
      \textbf{Parameter}   & \textbf{Optimization Values}
      \\  \hline
     $\rho_0^B$  & $1\times 10^8$
\\  \hline
$r_0$&  2.95
\\  \hline
    \end{tabular}
  \end{center}
\end{table}
\begin{table}[h!]
  \begin{center}
    \caption{Einasto Optimization Values}
    \label{EinastoUGC12632}
    \begin{tabular}{|r|r|}
     \hline
      \textbf{Parameter}   & \textbf{Optimization Values}
      \\  \hline
     $\rho_e$  &$1\times 10^7$
\\  \hline
$r_e$ & 3.55
\\  \hline
$n_e$ & 1
\\  \hline
    \end{tabular}
  \end{center}
\end{table}
\begin{table}[h!]
\centering \caption{Physical assessment of collisional DM
parameters for UGC12632.}
\begin{tabular}{lcc}
\hline
Parameter & Value   & Physical Verdict \\
\hline
$\gamma_0$ & $1.0001$ & Effectively isothermal \\
$\delta_\gamma$ & $1.2\times10^{-9}$ & Negligible \\
$r_\gamma$ & $1.5\ \mathrm{Kpc}$ & Transition radius placed inside inner halo \\
$K_0$ & $2.0\times10^{3}$ & Enough pressure support \\
$r_c$ & $0.5\ \mathrm{Kpc}$ & Small core scale \\
$p$ & $0.01$ & Almost flat $K(r)$ \\
\hline
Overall &-& Physically consistent and numerically stable \\
\hline
\end{tabular}
\label{EVALUATIONUGC12632}
\end{table}

\subsection{The Galaxy UGC12732 Non-viable, Extended Marginally}

\begin{figure}[h!]
\centering
\includegraphics[width=20pc]{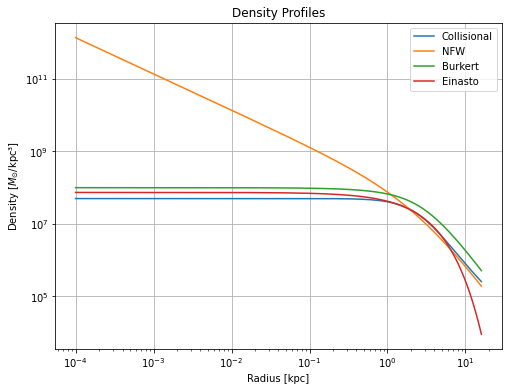}
\caption{The density of the collisional DM model (\ref{tanhmodel})
for the galaxy UGC12732, as a function of the radius.}
\label{UGC12732dens}
\end{figure}

\begin{figure}[h!]
\centering
\includegraphics[width=20pc]{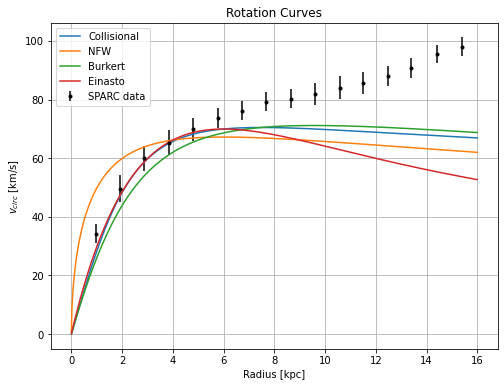}
\caption{The predicted rotation curves after using an optimization
for the collisional DM model (\ref{tanhmodel}), versus the SPARC
observational data for the galaxy UGC12732. We also plotted the
optimized curves for the NFW model, the Burkert model and the
Einasto model.} \label{UGC12732}
\end{figure}

\begin{figure}[h!]
\centering
\includegraphics[width=20pc]{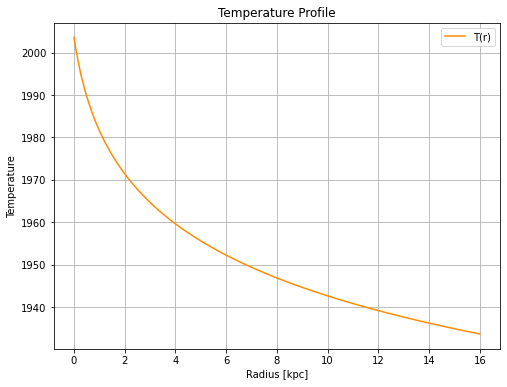}
\caption{The temperature as a function of the radius for the
collisional DM model (\ref{tanhmodel}) for the galaxy UGC12732.}
\label{UGC12732temp}
\end{figure}

\begin{figure}[h!]
\centering
\includegraphics[width=20pc]{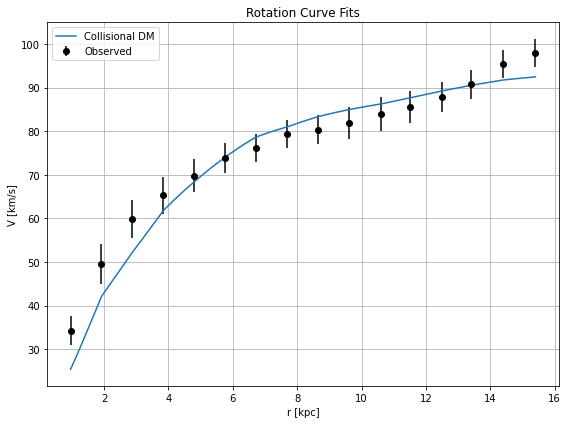}
\caption{The predicted rotation curves after using an optimization
for the collisional DM model (\ref{tanhmodel}), versus the
extended SPARC observational data for the galaxy UGC12732. The
model includes the rotation curves from all the components of the
galaxy, including gas and disk velocities, along with the
collisional DM model.} \label{extendedUGC12732}
\end{figure}

For this galaxy, we shall choose $\rho_0=5\times
10^7$$M_{\odot}/\mathrm{Kpc}^{3}$. UGC12732 is a gas-rich,
late-type spiral galaxy (classified as Scd/Sm) belonging to the
low-mass end of the SPARC sample. It is an extended, diffuse disk
system with prominent HI content and a slowly rising rotation
curve.  Its distance is typically adopted as \(D\simeq
15\text{-}17\ \mathrm{Mpc}\) from redshift-independent
measurements. In Figs. \ref{UGC12732dens}, \ref{UGC12732} and
\ref{UGC12732temp} we present the density of the collisional DM
model, the predicted rotation curves after using an optimization
for the collisional DM model (\ref{tanhmodel}), versus the SPARC
observational data and the temperature parameter as a function of
the radius respectively. As it can be seen, the SIDM model
produces non-viable rotation curves incompatible with the SPARC
data. Also in Tables \ref{collUGC12732}, \ref{NavaroUGC12732},
\ref{BuckertUGC12732} and \ref{EinastoUGC12732} we present the
optimization values for the SIDM model, and the other DM profiles.
Also in Table \ref{EVALUATIONUGC12732} we present the overall
evaluation of the SIDM model for the galaxy at hand. The resulting
phenomenology is non-viable.
\begin{table}[h!]
  \begin{center}
    \caption{Collisional Dark Matter Optimization Values}
    \label{collUGC12732}
     \begin{tabular}{|r|r|}
     \hline
      \textbf{Parameter}   & \textbf{Optimization Values}
      \\  \hline
     $\delta_{\gamma} $ & 0.0000000012
\\  \hline
$\gamma_0 $ & 1.0001  \\ \hline $K_0$ ($M_{\odot} \,
\mathrm{Kpc}^{-3} \, (\mathrm{km/s})^{2}$)& 2000  \\ \hline
    \end{tabular}
  \end{center}
\end{table}
\begin{table}[h!]
  \begin{center}
    \caption{NFW  Optimization Values}
    \label{NavaroUGC12732}
     \begin{tabular}{|r|r|}
     \hline
      \textbf{Parameter}   & \textbf{Optimization Values}
      \\  \hline
   $\rho_s$   & $5\times 10^7$
\\  \hline
$r_s$&  2.78
\\  \hline
    \end{tabular}
  \end{center}
\end{table}
\begin{table}[h!]
  \begin{center}
    \caption{Burkert Optimization Values}
    \label{BuckertUGC12732}
     \begin{tabular}{|r|r|}
     \hline
      \textbf{Parameter}   & \textbf{Optimization Values}
      \\  \hline
     $\rho_0^B$  & $1\times 10^8$
\\  \hline
$r_0$& 2.95
\\  \hline
    \end{tabular}
  \end{center}
\end{table}
\begin{table}[h!]
  \begin{center}
    \caption{Einasto Optimization Values}
    \label{EinastoUGC12732}
    \begin{tabular}{|r|r|}
     \hline
      \textbf{Parameter}   & \textbf{Optimization Values}
      \\  \hline
     $\rho_e$  &$1\times 10^7$
\\  \hline
$r_e$ & 3.55
\\  \hline
$n_e$ & 1
\\  \hline
    \end{tabular}
  \end{center}
\end{table}
\begin{table}[h!]
\centering \caption{Physical assessment of collisional DM
parameters for UGC12732.}
\begin{tabular}{lcc}
\hline Parameter & Value  & Physical Verdict \\ \hline
$\gamma_0$ & $1.0001$ & Practically isothermal \\
$\delta_\gamma$ & $1.2\times10^{-9}$ & Negligible  \\
$r_\gamma$ & $1.5\ \mathrm{Kpc}$ & Transition inside inner halo  \\
$K_0$ ($M_{\odot}\ \mathrm{Kpc}^{-3}\ (\mathrm{km/s})^{2}$) & $2.0\times10^{3}$ & Enough pressure support\\
$r_c$ & $0.5\ \mathrm{Kpc}$ & Small core radius \\
$p$ & $0.01$ & Almost flat $K(r)$  \\ \hline Overall & - &
Physically consistent \\ \hline
\end{tabular}
\label{EVALUATIONUGC12732}
\end{table}
Now the extended picture including the rotation velocity from the
other components of the galaxy, such as the disk and gas, makes
the collisional DM model viable for this galaxy. In Fig.
\ref{extendedUGC12732} we present the combined rotation curves
including the other components of the galaxy along with the
collisional matter. As it can be seen, the extended collisional DM
model is marginally viable.
Also in Table \ref{evaluationextendedUGC12732} we present the
values of the free parameters of the collisional DM model for
which the maximum compatibility with the SPARC data comes for the
galaxy UGC12732.
\begin{table}[h!]
\centering \caption{Physical assessment of Extended collisional DM
parameters (UGC12732).}
\begin{tabular}{lcc}
\hline
Parameter & Value & Physical Verdict \\
\hline
$\gamma_0$ & 0.99072155 & Essentially isothermal \\
$\delta_\gamma$ & 0.000000001 & No radial variation  \\
$K_0$ & 3000 & Moderate entropy/pressure scale \\
ml\_disk & 1.00000000 & High disk mass-to-light  \\
ml\_bulge & 0.00000000 & Zero bulge contribution  \\
\hline
Overall &-& Physically plausible \\
\hline
\end{tabular}
\label{evaluationextendedUGC12732}
\end{table}


\subsection{The Galaxy UGCA281}

\begin{figure}[h!]
\centering
\includegraphics[width=20pc]{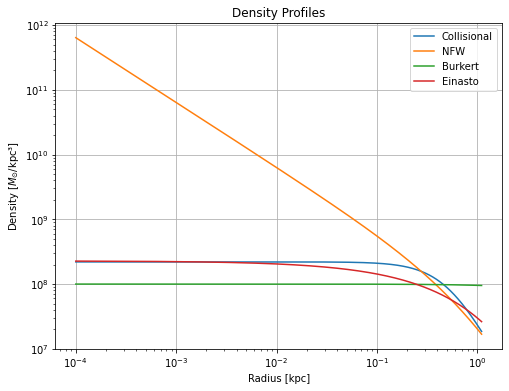}
\caption{The density of the collisional DM model (\ref{tanhmodel})
for the galaxy UGCA281, as a function of the radius.}
\label{UGCA281dens}
\end{figure}
\begin{figure}[h!]
\centering
\includegraphics[width=20pc]{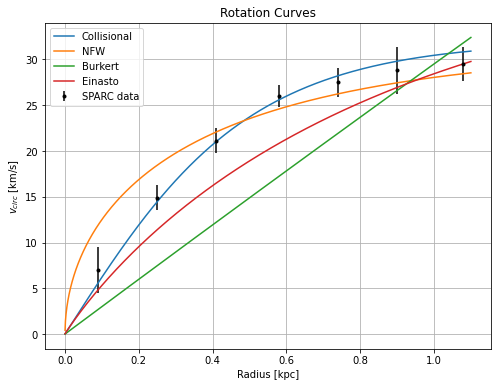}
\caption{The predicted rotation curves after using an optimization
for the collisional DM model (\ref{tanhmodel}), versus the SPARC
observational data for the galaxy UGCA281. We also plotted the
optimized curves for the NFW model, the Burkert model and the
Einasto model.} \label{UGCA281}
\end{figure}
\begin{figure}[h!]
\centering
\includegraphics[width=20pc]{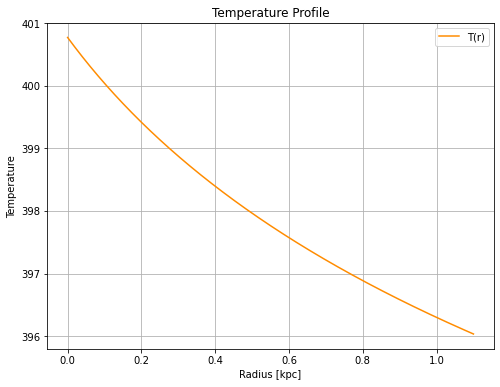}
\caption{The temperature as a function of the radius for the
collisional DM model (\ref{tanhmodel}) for the galaxy UGCA281.}
\label{UGCA281temp}
\end{figure}

For this galaxy, we shall choose $\rho_0=2.2\times
10^8$$M_{\odot}/\mathrm{Kpc}^{3}$. UGCA281 is a blue compact dwarf
galaxy located in the constellation Canes Venatici, situated
approximately 5.5 Mpc from the Milky Way. In Figs.
\ref{UGCA281dens}, \ref{UGCA281} and \ref{UGCA281temp} we present
the density of the collisional DM model, the predicted rotation
curves after using an optimization for the collisional DM model
(\ref{tanhmodel}), versus the SPARC observational data and the
temperature parameter as a function of the radius respectively. As
it can be seen, the SIDM model produces viable rotation curves
compatible with the SPARC data. Also in Tables \ref{collUGCA281},
\ref{NavaroUGCA281}, \ref{BuckertUGCA281} and \ref{EinastoUGCA281}
we present the optimization values for the SIDM model, and the
other DM profiles. Also in Table \ref{EVALUATIONUGCA281} we
present the overall evaluation of the SIDM model for the galaxy at
hand. The resulting phenomenology is viable.
\begin{table}[h!]
  \begin{center}
    \caption{Collisional Dark Matter Optimization Values}
    \label{collUGCA281}
     \begin{tabular}{|r|r|}
     \hline
      \textbf{Parameter}   & \textbf{Optimization Values}
      \\  \hline
     $\delta_{\gamma} $ & 0.0000000012
\\  \hline
$\gamma_0 $ & 1.0001  \\ \hline $K_0$ ($M_{\odot} \,
\mathrm{Kpc}^{-3} \, (\mathrm{km/s})^{2}$)& 40  \\ \hline
    \end{tabular}
  \end{center}
\end{table}
\begin{table}[h!]
  \begin{center}
    \caption{NFW  Optimization Values}
    \label{NavaroUGCA281}
     \begin{tabular}{|r|r|}
     \hline
      \textbf{Parameter}   & \textbf{Optimization Values}
      \\  \hline
   $\rho_s$   & $5\times 10^7$
\\  \hline
$r_s$& 1.28
\\  \hline
    \end{tabular}
  \end{center}
\end{table}
\begin{table}[h!]
  \begin{center}
    \caption{Burkert Optimization Values}
    \label{BuckertUGCA281}
     \begin{tabular}{|r|r|}
     \hline
      \textbf{Parameter}   & \textbf{Optimization Values}
      \\  \hline
     $\rho_0^B$  & $1\times 10^8$
\\  \hline
$r_0$& 22.38
\\  \hline
    \end{tabular}
  \end{center}
\end{table}
\begin{table}[h!]
  \begin{center}
    \caption{Einasto Optimization Values}
    \label{EinastoUGCA281}
    \begin{tabular}{|r|r|}
     \hline
      \textbf{Parameter}   & \textbf{Optimization Values}
      \\  \hline
     $\rho_e$  &$1\times 10^7$
\\  \hline
$r_e$ & 1.96
\\  \hline
$n_e$ & 0.64
\\  \hline
    \end{tabular}
  \end{center}
\end{table}
\begin{table}[h!]
\centering \caption{Physical assessment of collisional DM
parameters for UGCA281.}
\begin{tabular}{lcc}
\hline
Parameter & Value & Physical Verdict \\
\hline
$\gamma_0$ & $1.0001$ & Practically isothermal \\
$\delta_\gamma$ & $1.2\times10^{-9}$ & Negligible\\
$r_\gamma$ & $1.5\ \mathrm{Kpc}$ & Transition inside inner halo \\
$K_0$ & $4.0\times10^{2}$ & Low entropy/temperature scale; appropriate for dwarf halo \\
$r_c$ & $0.5\ \mathrm{Kpc}$ & Small core radius; nearly flat $K(r)$ \\
$p$ & $0.01$ & Very shallow $K(r)$ decrease \\
\hline
Overall &-& Physically consistent \\
\hline
\end{tabular}
\label{EVALUATIONUGCA281}
\end{table}

\subsection{The Galaxy UGCA442 Marginally viable, Extended Marginal}

\begin{figure}[h!]
\centering
\includegraphics[width=20pc]{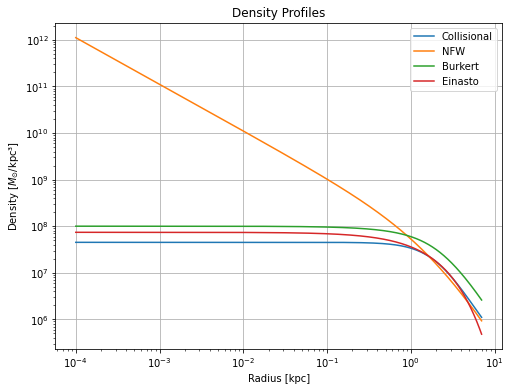}
\caption{The density of the collisional DM model (\ref{tanhmodel})
for the galaxy UGCA442, as a function of the radius.}
\label{UGCA442dens}
\end{figure}

\begin{figure}[h!]
\centering
\includegraphics[width=20pc]{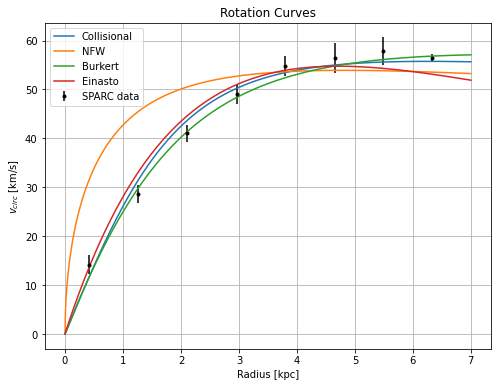}
\caption{The predicted rotation curves after using an optimization
for the collisional DM model (\ref{tanhmodel}), versus the SPARC
observational data for the galaxy UGCA442. We also plotted the
optimized curves for the NFW model, the Burkert model and the
Einasto model.} \label{UGCA442}
\end{figure}

\begin{figure}[h!]
\centering
\includegraphics[width=20pc]{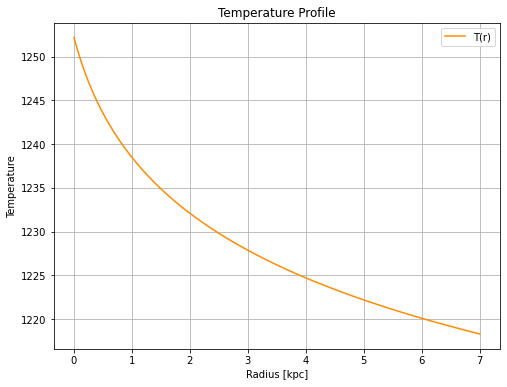}
\caption{The temperature as a function of the radius for the
collisional DM model (\ref{tanhmodel}) for the galaxy UGCA442.}
\label{UGCA442temp}
\end{figure}

\begin{figure}[h!]
\centering
\includegraphics[width=20pc]{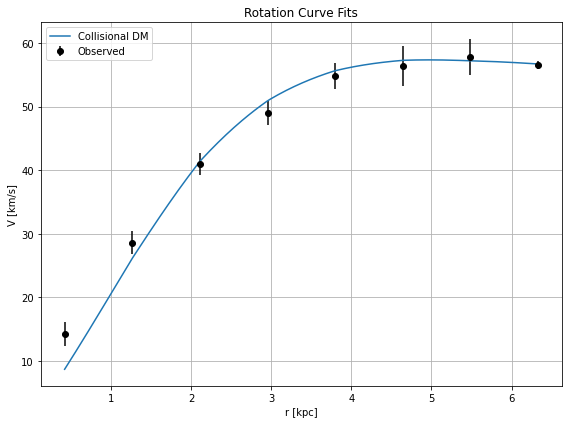}
\caption{The predicted rotation curves after using an optimization
for the collisional DM model (\ref{tanhmodel}), versus the
extended SPARC observational data for the galaxy UGCA442. The
model includes the rotation curves from all the components of the
galaxy, including gas and disk velocities, along with the
collisional DM model.} \label{extendedUGCA442}
\end{figure}

For this galaxy, we shall choose $\rho_0=4.5\times
10^7$$M_{\odot}/\mathrm{Kpc}^{3}$. UGCA442 is a
low-surface-brightness, irregular galaxy located in the Sculptor
group, approximately 4.3 Mpc from the Milky Way. In Figs.
\ref{UGCA442dens}, \ref{UGCA442} and \ref{UGCA442temp} we present
the density of the collisional DM model, the predicted rotation
curves after using an optimization for the collisional DM model
(\ref{tanhmodel}), versus the SPARC observational data and the
temperature parameter as a function of the radius respectively. As
it can be seen, the SIDM model produces marginally viable rotation
curves compatible with the SPARC data. Also in Tables
\ref{collUGCA442}, \ref{NavaroUGCA442}, \ref{BuckertUGCA442} and
\ref{EinastoUGCA442} we present the optimization values for the
SIDM model, and the other DM profiles. Also in Table
\ref{EVALUATIONUGCA442} we present the overall evaluation of the
SIDM model for the galaxy at hand. The resulting phenomenology is
marginally viable.
\begin{table}[h!]
  \begin{center}
    \caption{Collisional Dark Matter Optimization Values}
    \label{collUGCA442}
     \begin{tabular}{|r|r|}
     \hline
      \textbf{Parameter}   & \textbf{Optimization Values}
      \\  \hline
     $\delta_{\gamma} $ & 0.0000000012
\\  \hline
$\gamma_0 $ & 1.0001 \\ \hline $K_0$ ($M_{\odot} \,
\mathrm{Kpc}^{-3} \, (\mathrm{km/s})^{2}$)& 1250 \\ \hline
    \end{tabular}
  \end{center}
\end{table}
\begin{table}[h!]
  \begin{center}
    \caption{NFW  Optimization Values}
    \label{NavaroUGCA442}
     \begin{tabular}{|r|r|}
     \hline
      \textbf{Parameter}   & \textbf{Optimization Values}
      \\  \hline
   $\rho_s$   & $5\times 10^7$
\\  \hline
$r_s$&  2.23
\\  \hline
    \end{tabular}
  \end{center}
\end{table}
\begin{table}[h!]
  \begin{center}
    \caption{Burkert Optimization Values}
    \label{BuckertUGCA442}
     \begin{tabular}{|r|r|}
     \hline
      \textbf{Parameter}   & \textbf{Optimization Values}
      \\  \hline
     $\rho_0^B$  & $1\times 10^8$
\\  \hline
$r_0$&  2.37
\\  \hline
    \end{tabular}
  \end{center}
\end{table}
\begin{table}[h!]
  \begin{center}
    \caption{Einasto Optimization Values}
    \label{EinastoUGCA442}
    \begin{tabular}{|r|r|}
     \hline
      \textbf{Parameter}   & \textbf{Optimization Values}
      \\  \hline
     $\rho_e$  &$1\times 10^7$
\\  \hline
$r_e$ & 2.78
\\  \hline
$n_e$ & 1
\\  \hline
    \end{tabular}
  \end{center}
\end{table}
\begin{table}[h!]
\centering \caption{Physical assessment of collisional DM
parameters (second set).}
\begin{tabular}{lcc}
\hline
Parameter & Value & Physical Verdict \\
\hline
$\gamma_0$ & 1.0001 & Slightly above isothermal \\
$\delta_\gamma$ & 0.0000000012 & Extremely small variation \\
$r_\gamma$ & 1.5 Kpc & Transition radius inside inner halo \\
$K_0$ & 1250 & Moderate entropy scale, consistent with dwarf \\
$r_c$ & 0.5 Kpc & Small core scale \\
$p$ & 0.01 & Very shallow K(r) decrease, nearly constant \\
\hline
Overall &-& Physically plausible \\
\hline
\end{tabular}
\label{EVALUATIONUGCA442}
\end{table}
Now the extended picture including the rotation velocity from the
other components of the galaxy, such as the disk and gas, makes
the collisional DM model viable for this galaxy. In Fig.
\ref{extendedUGCA442} we present the combined rotation curves
including the other components of the galaxy along with the
collisional matter. As it can be seen, the extended collisional DM
model is marginally viable.
Also in Table \ref{evaluationextendedUGCA442} we present the
values of the free parameters of the collisional DM model for
which the maximum compatibility with the SPARC data comes for the
galaxy UGCA442.
\begin{table}[h!]
\centering \caption{Physical assessment of Extended collisional DM
parameters for galaxy UGCA442.}
\begin{tabular}{lcc}
\hline
Parameter & Value & Physical Verdict \\
\hline
$\gamma_0$ & 0.97636282 & Slightly sub-isothermal \\
$\delta_\gamma$ & 0.04269061 & Moderate radial variation \\
$K_0$ & 3000 & Moderate entropy   \\
$ml_{disk}$ & 0.50000000 & Moderate disk M/L \\
$ml_{bulge}$ & 0.00000000 & Negligible bulge contribution \\
\hline
Overall &-& Physically plausible\\
\hline
\end{tabular}
\label{evaluationextendedUGCA442}
\end{table}

\subsection{The Galaxy UGCA444}

\begin{figure}[h!]
\centering
\includegraphics[width=20pc]{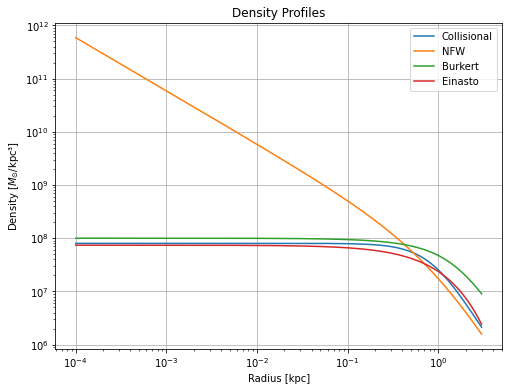}
\caption{The density of the collisional DM model (\ref{tanhmodel})
for the galaxy UGCA444, as a function of the radius.}
\label{UGCA444dens}
\end{figure}

\begin{figure}[h!]
\centering
\includegraphics[width=20pc]{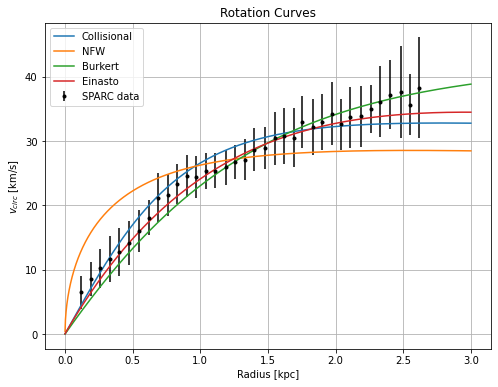}
\caption{The predicted rotation curves after using an optimization
for the collisional DM model (\ref{tanhmodel}), versus the SPARC
observational data for the galaxy UGCA444. We also plotted the
optimized curves for the NFW model, the Burkert model and the
Einasto model.} \label{UGCA444}
\end{figure}

\begin{figure}[h!]
\centering
\includegraphics[width=20pc]{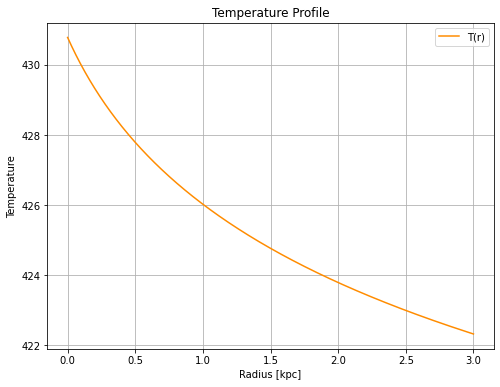}
\caption{The temperature as a function of the radius for the
collisional DM model (\ref{tanhmodel}) for the galaxy UGCA444.}
\label{UGCA444temp}
\end{figure}

For this galaxy we choose $\rho_0=8\times
10^7$$M_{\odot}/\mathrm{Kpc}^{3}$. UGCA444 is a
low-surface-brightness, irregular galaxy located in the
constellation Centaurus, approximately 3.8 Mpc from the Milky Way.
In Figs. \ref{UGCA444dens}, \ref{UGCA444} and \ref{UGCA444temp} we
present the density of the collisional DM model, the predicted
rotation curves after using an optimization for the collisional DM
model (\ref{tanhmodel}), versus the SPARC observational data and
the temperature parameter as a function of the radius
respectively. As it can be seen, the SIDM model produces viable
rotation curves compatible with the SPARC data. Also in Tables
\ref{collUGCA444}, \ref{NavaroUGCA444}, \ref{BuckertUGCA444} and
\ref{EinastoUGCA444} we present the optimization values for the
SIDM model, and the other DM profiles. Also in Table
\ref{EVALUATIONUGCA444} we present the overall evaluation of the
SIDM model for the galaxy at hand. The resulting phenomenology is
viable.
\begin{table}[h!]
  \begin{center}
    \caption{Collisional Dark Matter Optimization Values}
    \label{collUGCA444}
     \begin{tabular}{|r|r|}
     \hline
      \textbf{Parameter}   & \textbf{Optimization Values}
      \\  \hline
     $\delta_{\gamma} $ & 0.0000000012
\\  \hline
$\gamma_0 $ & 1.0001 \\ \hline $K_0$ ($M_{\odot} \,
\mathrm{Kpc}^{-3} \, (\mathrm{km/s})^{2}$)& 4300  \\ \hline
    \end{tabular}
  \end{center}
\end{table}
\begin{table}[h!]
  \begin{center}
    \caption{NFW  Optimization Values}
    \label{NavaroUGCA444}
     \begin{tabular}{|r|r|}
     \hline
      \textbf{Parameter}   & \textbf{Optimization Values}
      \\  \hline
   $\rho_s$   & $5\times 10^7$
\\  \hline
$r_s$&  1.18
\\  \hline
    \end{tabular}
  \end{center}
\end{table}
\begin{table}[h!]
  \begin{center}
    \caption{Burkert Optimization Values}
    \label{BuckertUGCA444}
     \begin{tabular}{|r|r|}
     \hline
      \textbf{Parameter}   & \textbf{Optimization Values}
      \\  \hline
     $\rho_0^B$  & $1\times 10^8$
\\  \hline
$r_0$&  1.72
\\  \hline
    \end{tabular}
  \end{center}
\end{table}
\begin{table}[h!]
  \begin{center}
    \caption{Einasto Optimization Values}
    \label{EinastoUGCA444}
    \begin{tabular}{|r|r|}
     \hline
      \textbf{Parameter}   & \textbf{Optimization Values}
      \\  \hline
     $\rho_e$  &$1\times 10^7$
\\  \hline
$r_e$ & 1.75
\\  \hline
$n_e$ & 1
\\  \hline
    \end{tabular}
  \end{center}
\end{table}
\begin{table}[h!]
\centering \caption{Physical assessment of collisional DM
parameters for UGCA444.}
\begin{tabular}{lcc}
\hline
Parameter & Value & Physical Verdict \\
\hline
$\gamma_0$ & $1.0001$ & Practically isothermal \\
$\delta_\gamma$ & $1.2\times10^{-9}$ & Negligible variation \\
$r_\gamma$ & $1.5\ \mathrm{Kpc}$ & Inner halo transition \\
$K_0$ & $430$ & Moderate entropy scale \\
$r_c$ & $0.5\ \mathrm{Kpc}$ & Small core radius \\
$p$ & $0.01$ & Very shallow radial decrease of $K(r)$ \\
\hline
Overall &-& Physically consistent \\
\hline
\end{tabular}
\label{EVALUATIONUGCA444}
\end{table}

\end{document}